\documentclass[11pt]{report}
\usepackage{amsfonts, amsmath, amssymb, amsthm,bm,bbm, stackrel}
\usepackage[utf8]{inputenc}
\usepackage{color}
\usepackage{stmaryrd}
\usepackage{graphicx}
\usepackage[babel=true]{csquotes}
\usepackage[caption=false]{subfig}
\usepackage{hyperref}
\usepackage{verbatim}
\usepackage[a4paper,width=150mm,top=25mm,bottom=25mm]{geometry}
\usepackage{cancel}
\usepackage[normalem]{ulem}
\usepackage{enumitem}
\usepackage[absolute]{textpos} 
\usepackage{multicol} 
\usepackage{calc}
\usepackage{tikz}

\DeclareMathOperator{\Tr}{Tr}

\newcommand{\be}{\begin{equation}}
\newcommand{\ee}{\end{equation}}
\newcommand{\bea}{\begin{eqnarray}}
\newcommand{\eea}{\end{eqnarray}}

\newcommand{\cA}{\mathcal{A}}
\newcommand{\cM}{\mathcal{M}}

\newcommand{\cB}{\mathcal{B}}
\newcommand{\cF}{\mathcal{F}}
\newcommand{\cC}{\mathcal{C}}
\newcommand{\cE}{\mathcal{E}}
\newcommand{\cV}{\mathcal{V}}

\newcommand{\cG}{\mathcal{G}}

\newcommand{\cO}{\mathcal{O}}
\newcommand{\cT}{\mathcal{T}}
\newcommand{\cP}{\mathcal{P}}

\newcommand{\cI}{\mathcal{I}}
\newcommand{\cJ}{\mathcal{J}}
\newcommand{\cZ}{\mathcal{Z}}

\newcommand{\cS}{\mathcal{S}}

\newcommand{\bT}{\mathbb{T}}

\newcommand{\bM}{\mathbb{M}}
\newcommand{\bG}{\mathbb{G}}
\newcommand{\bB}{\mathbb{B}}
\newcommand{\bC}{\mathbb{C}}
\newcommand{\bN}{\mathbb{N}}

\newcommand{\bS}{\mathbb{S}}
\newcommand{\bR}{\mathbb{R}}
\newcommand{\bK}{\mathbb{K}}
\newcommand{\bQ}{\mathbb{Q}}
\newcommand{\bH}{\mathbb{H}}
\newcommand{\bF}{\mathbb{F}}
\newcommand{\bX}{\mathbb{X}}

\newcommand{\un}{\mathbbm{1}}
\newcommand{\ED}{{\lfloor\frac{D}{2}\rfloor}}

\newcommand{\G}{G}
\newcommand{\nb}{b}
\newcommand{\C}{\mathcal{C}}
\newcommand{\Lc}{L}
\newcommand{\Ga}{\Gamma}

\newcommand{\Bc}{{\mathcal{C}_B}}
\newcommand{\B}{B}
\newcommand{\BC}{ { {\tilde{ \mathcal C }}_B } }
\newcommand{\BB}{\tilde B}

\newcommand{\Sb}{\mathbb{E}}
\newcommand{\Mb}{\mathbb{M}}

\newcommand{\Om}{\Omega}

\newcommand{\Omi}[1][i]{\Om^{(#1)}}

\newcommand{\BOM}[1][]{\B_{/\Om_{\rm #1}}}
\newcommand{\BCO}[1][]{\B^{\Om_{\rm #1}}}

\newcommand{\GOM}[1][]{\G_{/\Om_{\rm #1}}}
\newcommand{\Gcan}[1][]{\G_{0, \Om_{\rm #1}}}

\newcommand{\Pc}{\mathcal{P}}

\newcommand{\I}{\mathcal{I}_2}

\newcommand{\LOR }{LO }
\newcommand{\PhiM }{\Phi_0(\BCO[opt])}
\newcommand{\PhijM }{\Phi(\BCO[opt])}
\newcommand{\Opt}[1][]{{\Om_{{\rm opt} #1}} }
\newcommand{\LmM }{L_m(\BOM[opt])}

\newcommand{\Bpi}{{\B_{\bar \pi}}}

\newcommand{\Gi}[1][]{{  G_{#1}^{\hat i}  }}
\newcommand{\GI}{{G^{\hat I}}}

\newcommand{\Fint}{{F_{\rm int}}}

\newcommand{\Gae}{\Ga_e}
\newcommand{\Gai}[1][i]{\Ga^{(#1)}}
\newcommand{\Gaij}{\Ga^{(ij)}}

\newcommand{\Ps}{\Psi}

\newcommand{\lDr}{\llbracket 1,D \rrbracket}

\newcommand{\GF}{\mathcal{G}}
\newcommand{\GP}{\mathcal{P}}

\newcommand{\deltaG}{\delta_{\mathrm{Gur}}}

\newcommand{\prf}{{\noindent \bf Proof.\; }}

\DeclareMathOperator{\col}{col}

\DeclareMathOperator{\Card}{Card}

\newcommand{\dr}{\partial}
\newcommand{\cD}{{\cal D}}

\newtheorem{lemma}{Lemma}[section]
\newtheorem{definition}{Definition}[section]
\newtheorem{theorem}{Theorem}[section]
\newtheorem{coroll}{Corollary}[section]
\newtheorem{prop}{Proposition}[section]

\makeatletter
\DeclareRobustCommand{\sqcdot}{\mathbin{\mathpalette\morphic@sqcdot\relax}}
\newcommand{\morphic@sqcdot}[2]{%
  \sbox\z@{$\m@th#1\centerdot$}%
  \ht\z@=.33333\ht\z@
  \vcenter{\box\z@}%
}
\makeatother

\label{The thesis}

\newcommand{\PhDTitleEN}{{\huge Colored Discrete Spaces:}

\vspace{0.2cm}

	{\Large Higher dimensional combinatorial maps and quantum gravity}}
	
 \label{Personnal}
 \newcommand{\PhDname}{Luca Lionni} 
\newcommand{\NNT}{2017SACLS270} 
\newcommand{\ecodocnum}{564} 
\newcommand{\ecodoctitle}{Physique en \^Ile de France} 
\newcommand{\PhDspeciality}{Physique} 
\newcommand{\PhDworkingplace}{\`a l'Universit\'e Paris-Sud} 
\newcommand{\vpos}{1} 

\label{Jury}
\newcommand{\jurynameA}{Bertrand DUPLANTIER}
\newcommand{\jurygenderA}{Pr.} 
\newcommand{\juryadressA}{Commissariat \`a l'\'energie atomique}
\newcommand{\jurygradeA}{Directeur de recherche}
\newcommand{\juryroleA}{Pr\'esident}
\newcommand{\jurynameB}{Frank FERRARI}
\newcommand{\jurygenderB}{Pr.} 
\newcommand{\juryadressB}{Universit\'e Libre de Bruxelles}
\newcommand{\jurygradeB}{Professeur}
\newcommand{\juryroleB}{Rapporteur}
\newcommand{\jurynameC}{Eric FUSY}
\newcommand{\jurygenderC}{Dr.} 
\newcommand{\juryadressC}{Ecole Polytechnique}
\newcommand{\jurygradeC}{Charg\'e de recherche}
\newcommand{\juryroleC}{Rapporteur}
\newcommand{\jurynameD}{Jean-Fran\c{c}ois MARCKERT}
\newcommand{\jurygenderD}{Pr.} 
\newcommand{\juryadressD}{Universit\'e de Bordeaux}
\newcommand{\jurygradeD}{Directeur de recherche}
\newcommand{\juryroleD}{Examinateur}
\newcommand{\jurynameE}{Lionel POURNIN}
\newcommand{\jurygenderE}{Pr.} 
\newcommand{\juryadressE}{Universit\'e Paris 13}
\newcommand{\jurygradeE}{Professeur}
\newcommand{\juryroleE}{Examinateur}
\newcommand{\jurynameF}{Valentin BONZOM}
\newcommand{\jurygenderF}{Dr.} 
\newcommand{\juryadressF}{Universit\'e Paris 13}
\newcommand{\jurygradeF}{Ma\^itre de conf\'erences}
\newcommand{\juryroleF}{Co-encadrant}
\newcommand{\jurynameG}{Vincent RIVASSEAU}
\newcommand{\jurygenderG}{Pr.} 
\newcommand{\juryadressG}{Universit\'e Paris-Sud}
\newcommand{\jurygradeG}{Professeur}
\newcommand{\juryroleG}{Directeur de th\`ese}

\begin{document}

\newgeometry{textheight=150ex,textwidth=40em,top=30pt,headheight=30pt,headsep=30pt,inner=80pt, bottom = 0pt}
\label{cotutelle}

\begin{tikzpicture}[remember picture,overlay,color=blue!20!red!45!black!75!]
	\draw[very thick]
		([yshift=-160pt,xshift=45pt]current page.north west)--     
		([yshift=-160pt,xshift=-25pt]current page.north east)--    
		([yshift=35pt,xshift=-25pt]current page.south east)--      
		([yshift=35pt,xshift=45pt]current page.south west)--cycle; 
\end{tikzpicture}

\begin{textblock}{13}(1.15,3.3)
  NNT : \NNT
\end{textblock}

\begin{textblock}{1}(1.15,1)
\includegraphics[height=2.4cm]{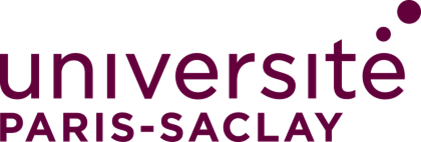} 
\label{Logo Paris Saclay}
\end{textblock}

\begin{textblock}{1}(12,\vpos)
\includegraphics[height=2.4cm]{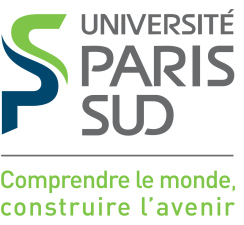} 
\label{Logo Etablissement}
\end{textblock}

\vspace{6cm}
\color{blue!20!red!45!black} 
  \begin{center}    
    \LARGE\textsc{Th\`ese de doctorat\\ de l'Universit\'e Paris-Saclay} \\
    \LARGE{\textsc{pr\'epar\'ee \PhDworkingplace}} \\ \bigskip
  \color{black} 
    \Large{Ecole doctorale n$^{\circ}\ecodocnum$}\\ 
     \Large{\ecodoctitle}  \\

     \Large{Sp\'ecialit\'e de doctorat: \PhDspeciality} 
\vspace{0.2cm}

   \Large{par}
\\
   \LARGE{\textbf{\textsc{\PhDname}}} 

\vspace{1.3cm}

    \Large{\PhDTitleEN} 
    
\vspace{0.5cm}

\end{center}
\color{black}
\begin{flushleft}
Composition du Jury :
\end{flushleft}
\begin{center}
\begin{tabular}{llll}
    \jurygenderA & \textsc{\jurynameA}  & \jurygradeA & (\juryroleA) \\
    \null & \null & \juryadressA &\\  [+1ex] 
   
    \jurygenderB & \textsc{\jurynameB}  & \jurygradeB & (\juryroleB) \\
    \null & \null & \juryadressB &\\ [+1ex] 
    
    \jurygenderC & \textsc{\jurynameC}  & \jurygradeC & (\juryroleC) \\
    \null & \null & \juryadressC &\\ [+1ex] 
    
    \jurygenderD & \textsc{\jurynameD}  & \jurygradeD & (\juryroleD) \\
    \null & \null & \juryadressD &\\ [+1ex] 
    
    \jurygenderE & \textsc{\jurynameE}  & \jurygradeE & (\juryroleE) \\
    \null & \null & \juryadressE &\\ [+1ex] 
    
    \jurygenderF & \textsc{\jurynameF}  & \jurygradeF & (\juryroleF) \\
    \null & \null & \juryadressF &\\ [+1ex] 
    
        \jurygenderG & \textsc{\jurynameG}  & \jurygradeG & (\juryroleG) \\
    \null & \null & \juryadressG &
  \end{tabular}    
\end{center}

\restoregeometry

\newpage
\thispagestyle{empty}
\mbox{}

\newpage
\thispagestyle{empty}

\

\

\begin{flushright}{\it \`A Cynthia, Alix, et Sophie.}\end{flushright}

\newpage
\thispagestyle{empty}
\mbox{}

\newpage
\thispagestyle{plain}
\noindent {{\huge \bf Abstract}

\vspace{1cm}

In two dimensions, the Euclidean Einstein-Hilbert action, which describes gravity in the absence of matter, can be discretized over random triangulations. In the physical limit of small Newton's constant, only planar triangulations survive. The limit in distribution of planar triangulations - the Brownian map - is a continuum random fractal spherical surface, which importance in the context of two-dimensional quantum gravity has been made more precise over the last years. It is interpreted as a quantum random continuum space-time, obtained in the thermodynamical limit from a statistical ensemble of random discrete surfaces, and has been shown to be equivalent to Liouville quantum gravity. The fractal properties of two-dimensional quantum gravity can therefore be studied from a discrete approach. It is well known that direct higher dimensional generalizations fail to produce appropriate quantum space-times in the continuum limit: the limit in distribution of dimension $D>2$ triangulations which survive in the limit of small Newton's constant is the continuous random tree, also called branched polymers in physics. However, while in two dimensions, discretizing the Einstein-Hilbert action over random $2p$-angulations - discrete surfaces obtained by gluing $2p$-gons together - leads to the same conclusions as for triangulations, this is not always the case in higher dimensions, as was discovered recently. Whether new continuum limit arise by considering discrete Einstein-Hilbert theories of more general random discrete spaces in dimension $D$ remains an open question.

We study discrete spaces obtained by gluing together elementary building blocks, such as polytopes with triangular facets. Such spaces generalize $2p$-angulations in higher dimensions. In the physical limit of small Newton's constant, only discrete spaces which maximize the mean curvature survive. However, identifying them is a task far too difficult in the general case, for which quantities are estimated throughout numerical computations. In order to obtain analytical results, a coloring of ($D-1$)-cells has been introduced. In any even dimension, we can find families of colored discrete spaces of maximal mean curvature in the universality classes of trees - converging towards the continuous random tree, of planar maps - converging towards the Brownian map, or of proliferating baby universes. However, it is the simple structure of the corresponding building blocks which makes it possible to obtain these results: it is similar to that of one or two dimensional objects and does not render the rich diversity of colored building blocks in dimensions three and higher. 

This work therefore aims at providing combinatorial tools which would enable a systematic study of the building blocks and of the colored discrete spaces they generate. The main result of this thesis is the derivation of a bijection between colored discrete spaces and colored combinatorial maps, which preserves the information on the local curvature. It makes it possible to use results from combinatorial maps and paves the way to a systematical study of higher dimensional colored discrete spaces. As an application, a number of blocks of small sizes are analyzed, as well as a new infinite family of building blocks. The relation to random tensor models is detailed. Emphasis is given to finding the lowest bound on the number of ($D-2$)-cells, which is equivalent to determining the correct scaling for the corresponding tensor model. We explain how the bijection can be used to identify the graphs contributing at any given order of the $1/N$ expansion of the $2n$-point functions of the colored SYK model, and apply this to the enumeration of generalized unicellular maps - discrete spaces obtained from a single building block - according to their mean curvature.  For any choice of colored building blocks, we show how to rewrite the corresponding discrete Einstein-Hilbert theory as a random matrix model with partial traces, the so-called intermediate field representation.

\newpage
\thispagestyle{empty}
\mbox{}

\newpage
\thispagestyle{plain}
\noindent {{\huge \bf Espaces discrets color\'es:} 

\

\noindent	{\Large Cartes combinatoires en dimensions sup\'erieures et gravit\'e quantique}}

\vspace{1cm}

On consid\`ere, en deux dimensions, une version euclidienne discr\`ete de l'action d'Einstein-Hilbert, qui d\'ecrit la gravit\'e en l'absence de mati\`ere. \`A l'int\'egration sur les  g\'eom\'etries se substitue une sommation sur des surfaces triangul\'ees al\'eatoires. Dans la limite physique de faible gravit\'e, seules les triangulations planaires survivent. Leur limite en distribution, la carte brownienne, est une surface fractale continue al\'eatoire dont l'importance dans le contexte de la gravit\'e quantique en deux dimensions a \'et\'e r\'ecemment pr\'ecis\'ee. Cet espace est interpr\'et\'e comme un espace-temps quantique al\'eatoire, obtenu comme limite \`a grande \'echelle d'un ensemble statistique de surfaces discr\`etes al\'eatoires. En deux dimensions, on peut donc \'etudier les propri\'et\'es fractales de la gravit\'e quantique via une approche discr\`ete. Il est bien connu que les g\'en\'eralisations directes en dimensions sup\'erieures \'echouent \`a produire des espace-temps quantiques aux propri\'et\'es ad\'equates : en dimension $D>2$, la limite en distribution des triangulations qui survivent dans la limite de faible gravit\'e est l'arbre continu al\'eatoire, ou polym\`eres branch\'es en physique. Si en deux dimensions on parvient aux m\^emes conclusions en consid\'erant non pas des triangulations, mais des surfaces discr\`etes al\'eatoires obtenues par recollements de $2p$-gones, nous savons depuis peu que ce n'est pas toujours le cas en dimension $D>2$.  L'apparition de nouvelles limites continues dans le cadre de th\'eories de gravit\'e impliquant des espaces discrets al\'eatoires reste une question ouverte. 

Nous \'etudions des espaces obtenus par recollements de blocs \'el\'ementaires,  comme des polytopes \`a facettes triangulaires. Dans la limite de faible gravit\'e, seuls les espaces qui maximisent la courbure moyenne survivent. Les identifier est cependant une t\^ache ardue dans le cas g\'en\'eral, pour lequel les r\'esultats sont obtenus num\'eriquement. Afin d'obtenir des r\'esultats analytiques, une coloration des ($D-1$) - cellules, les facettes, a \'et\'e introduite. En toute dimension paire, on peut trouver des familles d'espaces discrets color\'es de courbure moyenne maximale dans la classe d'universalit\'e des arbres - convergeant vers l'arbre continu al\'eatoire, des cartes planaires - convergeant vers la carte brownienne, ou encore dans la classe de prolif\'eration des b\'eb\'e-univers. Cependant, ces r\'esultats sont obtenus en raison de la simplicit\'e de blocs \'el\'ementaires dont la structure uni ou bidimensionnelle ne rend pas compte de la riche diversit\'e des blocs color\'es en dimensions sup\'erieures. 

Le premier objectif de cette th\`ese est donc d'\'etablir des outils combinatoires qui permettraient une \'etude syst\'ematique des blocs \'el\'ementaires color\'es et des espaces discrets qu'ils g\'en\`erent. Le principal r\'esultat de ce travail est l'\'etablissement d'une bijection entre ces espaces et des familles de cartes combinatoires, qui pr\'eserve l'information sur la courbure locale. Elle permet l'utilisation de r\'esultats sur les surfaces discr\`etes et ouvre la voie \`a une \'etude syst\'ematique des espaces discrets en dimensions sup\'erieures \`a deux. Cette bijection est appliqu\'ee \`a la caract\'erisation d'un certain nombre de blocs de petites tailles ainsi qu'\`a une nouvelle famille infinie. Le lien avec les mod\`eles de tenseurs al\'eatoires est d\'etaill\'e. Une attention particuli\`ere est donn\'ee \`a la d\'etermination du nombre maximal de ($D-2$) - cellules et de l'action appropri\'ee du mod\`ele de tenseurs correspondant. Nous montrons comment utiliser la bijection susmentionn\'ee pour identifier les contributions \`a un tout ordre du d\'eveloppement en $1/N$ des fonctions \`a $2n$ points du mod\`ele SYK color\'e, et appliquons ceci \`a l'\'enum\'eration des cartes unicellulaires g\'en\'eralis\'ees - les espaces discrets obtenus par recollement d'un unique bloc \'el\'ementaire - selon leur courbure moyenne. Pour tout choix de blocs color\'es, nous montrons comment r\'e\'ecrire la th\'eorie d'Einstein-Hilbert discr\`ete correspondante comme un mod\`ele de matrices al\'eatoires avec traces partielles, dit repr\'esentation en champs interm\'ediaires.

\newpage
\thispagestyle{empty}
\mbox{}

\newpage
\thispagestyle{empty}
\noindent {{\huge \bf Remerciements}

\vspace{1cm}

Je tiens tout d'abord \`a remercier chaudement Vincent et Valentin, pour leurs inestimables conseils et pour m'avoir donn\'e cette chance, celle d'avoir pu d\'ecouvrir l'univers de la recherche et profiter de leur impressionnante culture, le premier par ses bien connues envol\'ees aux confins de la physique ainsi que par l'intuition par laquelle il devine bien souvent ce qu'on mettra par la suite des mois \`a d\'emontrer, le second par la pr\'ecision et la pertinence de ses si pr\'ecieux conseils. Pour m'avoir guid\'e avec \'equilibre et justesse le long de ce passionnant premier voyage \`a la fronti\`ere de la physique, des math\'ematiques et de la combinatoire. Pour avoir su donner du cr\'edit \`a mes id\'ees parfois un peu trop enthousiastes mais aussi pour avoir su parfois, justement, les calmer. 

\

Je voudrais remercier Eric Fusy et Frank Ferrari, pour avoir accept\'e la t\^ache ardue qu'\'etait la relecture estivale de ce travail, relecture qui fut pourtant minutieuse. Je remercie \'egalement Bertrand Duplantier, Lionel Pournin, et Jean-Fran\c{c}ois Marckert, qui me font l'honneur de faire partie de ce jury. Je voudrais de plus exprimer chaleureusement ma gratitude \`a ce dernier, ainsi qu'\`a Razvan Gurau, pour m'avoir plusieurs fois recommand\'e. 

\

Je tiens \'egalement \`a remercier mes collaborateurs, brillants esprits, et d\'esormais amis, Vincent, Joseph, Dario, Adrian, Razvan, Thomas, Fabien, Vasily, ainsi que Johannes et St\'ephane, avec qui le travail, toujours intercal\'e de rires, est surtout une partie de plaisir. And, I want to thank Thibault, Tajron and Eduardo, for these amazing shwift revolutionary good times, all around the globe. 

Merci \`a Damir, pour les longues discussions et tous les sages conseils. Merci \`a Timoth\'ee, Andre\"i, Gabriel, Mathias, Olcyr, Antoine, Nicolas, et aux autres doctorants. Merci \`a Marie, ainsi qu'\`a tous les autres membres de ce laboratoire, sans qui la vie ici n'aurait pas \'et\'e pareille. 

Je voudrais aussi remercier Thierry, ainsi que Mario, pour tous les bons moments pass\'es ensemble, \`a Marseille ou ailleurs. Ringrazio Luigi Grasselli, Maria Rita Casali, e Paola Cristofori, per le risposte preziose a tutte le mie domande topologiche, a volte in mezzo alla notte. Merci \`a Marco Delbo, pour les bons moments pass\'es \`a chercher des grains de sables sur de lointains ast\'ero\"ides, dans le plus beau laboratoire de France.

\

Enfin, tout cela n'aurait pas \'et\'e possible sans le soutien de mes parents, Pippo et Sophie, de Richard, de ma soeur et meilleure amie, Alix, de mes oncles et tantes Philippo, Pascal et Chantale, chez qui de nombreux passages de cette th\`ese discr\`ete ont \'et\'e \'ecrits, de Rapha\"elle, de mes grands-parents, ainsi que de tous mes oncles, tantes et cousins, si chers \`a mon coeur. Merci \`a Pauline et Louis, qui relisent avec po\'esie mes papiers. Merci \`a Val\'erie, ainsi qu'\`a mon parrain, Enrico, d'\^etre venu de loin pour ce jour important. Merci \`a mes amis. Et enfin, Cynthia, pour ton amour vif et profond et ton soutien durant ces trois ann\'ees, merci.

\newpage
\thispagestyle{empty}
\mbox{}

\tableofcontents


\newpage
\thispagestyle{plain}
\newgeometry{width=150mm,top=25mm,bottom=25mm, left=20mm}
 {{\huge \bf List of symbols}

\vspace{1cm}



\noindent\begin{tabular}{rcl}



$L(\G)$& :& The circuit-rank of a graph  (Def.~\ref{def:CircuitRank}).\\

$g(M)$& :& The genus of a combinatorial map (Def.~\ref{def:genus}).\\


\\
$\cS^D$& :& The $D$-dimensional sphere.\\
$\C$ &:& A colored simplicial pseudo-complex (or triangulation), Def.~\ref{def:PseudoComp}.
\\

$n_{D}, n_{D-2}\cdots$ &: &The number of $D$-simplices, $(D-2)$-simplices... of a simplicial \\ &&pseudo-complex.
\\

\\


$\bG_D$ &: & The set of connected bipartite regular $(D+1)$-edge-colored graphs (Def. \ref{def:cG}).
\\

$\bG^q_D$ &: & The set of edge-colored graphs with $q$ valency-$D$ white vertices (Def. \ref{def:GraphWithBound}).
\\


$\Phi_{i,j}$ &:& The number of bicolored cycles $ij$ of a colored graph.\\

$\Phi$ &:& The score of a graph \eqref{eqref:score}, i.e. the  total number of bicolored cycles (Def. \ref{def:BiC}).\\

$\deltaG$ &:& Gurau's degree (Def. \ref{def:Deg}).\\

$\Phi_0$ &:& The 0-score, i.e. number of bicolored cycles $0i$ of a colored graph (Def.~\ref{def:Score0}).\\

$\delta_\bB, \delta_\B$ &:& The bubble-dependent degree (Def. \ref{def:BubDepDeg}).\\

$ a_\B, \tilde a_\B$ &: & The linear dependence of the bound on the number of $D-2$ simplices \\&&(Subsec.~\ref{subsec:BubDeg}).
\\
$ s_\B $ &: & The scaling of an enhanced tensor model \eqref{eqref:SFromTildeA}.
\\
$ \gamma_\B $ &: & The critical exponent characterizing the asymptotical behavior of the gene- \\
&&-rating functions (correlation functions) at the singularity \eqref{eqref:CritExp1} and \eqref{eqref:CritExp2}.
\\
$ \Delta_\B $ &: & The correction to Gurau's degree, the rescaled difference with the bubble \\&&-dependent degree  \eqref{eqref:CorrectionGurau}.
\\


$\Bc$& :& A bubble (Def.~\ref{def:bubbles}), and more specifically its $D-1$ dimensional boundary. \\

$\B$& :& The colored graph dual to the boundary of a bubble.\\

$n_\B(\G)$& :& The number of bubbles of type $\B$ in $\G$.\\

$b(\G)$& :& The total number of bubbles in $\G$. \\

$\bB$ &: & A set of bubbles.
\\

$\bG(\bB)$ &: & The set of graphs for which we recover copies of elements of $\bB$ when deleting \\
&&all color-0 edges (in dimension $D$ unless specified otherwise) (Def.~\ref{def:Restricted}).
\\

$\bG^q(\bB)$ &: & Graphs as in $\bG(\bB)$ but with $q$ missing color-0 edges.
\\



$\Om$& :& A pairing of the vertices of a graph (Def.~\ref{def:Pairing}).\\

$\Opt$& :& An optimal pairing (Def.~\ref{def:OptPair}).\\

$\BCO$& :& The covering of $\B$ corresponding to the pairing $\Om$ (Def.~\ref{def:Pairing}).\\


\\

$\Mb^q_D$& :& The set of combinatorial maps with edges colored in $\lDr$ and $q$ corners \\&& marked  on $q$ different vertices.\\

$\Fint$& :& The faces of a combinatorial map which do not encounter a marked corner.\\

$\BOM, \GOM$& :& The Eulerian graphs described in Def.~\ref{def:EulColGraph}.\\

$L_m$& :& The number of independent polychromatic cycles Def.~\ref{def:MultCycl}.\\


\\

$\Ps$ & :& The map described in Section~\ref{sec:StackedMaps}.\\

$\Ga$ & :& Generally a stacked map Section~\ref{sec:StackedMaps}.\\

$\bS^q_D$ &: & The set of stacked maps with $q$ marked corners on $q$ different color-0 vertices.\\

$\Gai$ & :& The color $i$  or $0i$ submap (Def.~\ref{def:BicolSubmap}).\\

$\I(e)$& :& The number of colors for which two faces run along $e$ in $\Gai$ \eqref{eqref:I2E}.\\

$\Gae$& :& The map obtained unhooking the edge $e$ (Subsec.~\ref{subsec:Unhook}).\\

\end{tabular}
\restoregeometry
\newpage
\thispagestyle{empty}
\mbox{}

\chapter*{Introduction}
\markboth{Introduction}{Introduction}

%
Gravity, on large scales, is described by general relativity, a dynamical theory of the geometry of space-time, which is a four-dimensional continuous manifold. One of its key concepts is that the presence of matter influences the curvature of space-time. 
%
%
The other fundamental forces of nature, electromagnetism,  the strong interaction, responsible for the cohesion of the proton and the neutron, and the weak interaction, responsible for radioactive decays,  are described by a unified theory, the Standard model
 \cite{Weinberg, Salam, Glashow}. It involves fields living and interacting locally on the ambient space-time. We learn from general relativity, that this background space-time is not the immovable background of Newton's gravity or Einstein's special relativity, but dynamically reacts to its matter content. 
Importantly, these fields are quantized, 
they represent quantum physical states. Measurable quantities, called observables,  are probabilities of transitions between states. These are computed from the partition function of the theory.
However, the quantities computed from these quantum high energy theories are not yet understandable from our low-energy  point of view.  Renormalization is the process that tells us how we will perceive quantum physical quantities from our scales. An electron, for instance, cannot be considered ``naked", that is without the ``cloud" of self-interactions, created and annihilated particles which ineluctably surrounds it. Renormalization takes all of this into account, and translates the bare theoretical quantum electron into a physically meaningful theory: it is the renormalized electron charge which can be compared with experimental results. Therefore, a non-renormalizable quantum theory makes no physical sense.

While gravity is by far the weakest force of all four\footnote{Relative magnitudes of forces as they act on a pair of proton in an atomic nucleus:  gravity 1, weak interaction $10^{24}$, electromagnetism $10^{35}$, and strong interaction $10^{37}$.}, it implies the existence of singularities - the black-hole singularities, the cosmological singularity - where physical quantities diverge. These singularities occur at microscopic scales shorter than the Planck scale\footnote{Approximately $10^{-35}$ meters, i.e. $10^{-20}$ times the size of the proton. Planck energy is $1.22\times10^{19} GeV$.}, and suggest that, at these scale, quantum effects should start playing a role. 
Gravity is a classical field theory.
The metric is the field describing the dynamics of space-time geometry, and quantizing gravity
therefore seems to imply quantizing the metric field. Trying to quantize it directly as it is done for other quantum field theories leads to divergences: general relativity is not a perturbatively renormalizable theory \cite{THooftGrav, Sagnotti}. 
%
%
There are many attempts to build a theory of quantum gravity, among which string theory, loop quantum gravity, causal dynamical triangulations...  However, an important problem, met in most of these theories, is to make sense of a functional integral of the kind
\be
\label{PartFunGrav1}
\cZ = \int_\cM D[g] D[X] e^{-S_\text{Gravity} - S_\text{SM}} ,
\ee 
where $S_\text{SM}$ is the action of the standard model, and the functional integration is performed over the metrics $g$ of the space-time manifold $\cM$, and over the fields $X$ involved in the standard model. In an Euclidean context, the most obvious choice for the action describing pure gravity (gravity in the absence of matter) is the Einstein-Hilbert action 
\be
\label{EH1}
S_\text{Gravity}=S_\text{EH} =\int_\cM d^Dx\sqrt {\lvert g\rvert} (\Lambda-\frac{1}{16\pi G}R),
\ee
where $\Lambda$ is the cosmological constant, $G$ is Newton's constant\footnote{$G= 6.67408(31)\times10^{-11} m^3.kg^{-1}.s^{-2}$}, and $R$ is the Ricci scalar curvature. 
One way of making sense of this partition function is by introducing a short-scale cut-off and seeing space-time on short scales as a lattice. A lattice in a general sense is a discretized manifold, a space obtained by gluing together elementary building blocks, such as tetrahedra in three dimensions. In the case where this discrete space has been given a well-defined notion of geometry, and assuming that there is enough freedom in gluing the building blocks for the discretizations of the manifold to give a good approximation of its geometries, we can try to make sense of \eqref{PartFunGrav1} by assuming that the integration over geometries can be replaced with a summation over discretizations of the manifold,
\be
\int_\cM D[g] \qquad\longleftrightarrow\quad\sum_{\text{discretizations of }\cM}.
\ee 
Topology fluctuations are not a priori excluded, and considering $\Sb$ some more general set of discrete spaces, we can choose to define the discrete partition function of pure gravity as 
\be
\label{PartFunc1}
\cZ_\Sb = \sum_{\C\in\Sb} e^{-S_\text{discrete}(\C)}, 
\ee
where $S_\text{discrete}$ should be a discrete version of \eqref{EH1}. Let us review in more details several possibilities for $\Sb$, and the corresponding discrete actions $S_\text{discrete}(\C)$, starting with dimension two.


\

Combinatorial maps are closed discrete two-dimensional surfaces obtained by gluing a finite number of polygons along segments of their boundaries. They were introduced in the 60's  by the seminal work of Tutte on discrete spheres \cite{TutteTriang, TutteBij}. Since then, their study and enumeration has grown into a rich area of research \cite{GoulJack, GraphsOnSurfaces,
Flajolet}. The study of  surfaces of higher genus was first addressed in \cite{BenCanHigher}. Let us list a few directions: rationality results \cite{Bousquet}, 
hypermaps \cite{Walsh, BernFusHyp}, 
degree-restricted maps \cite {Ben, BernFusBij, CEAMatrix, BernFusHyp},
unicellular maps \cite{ChapUnicell, BernardiHZ, ChapFusUnicell}, 
recursive counting formulae \cite{BernardiHZ,ChapRecu}, 
bijections \cite{BernChapHZ, BernFusHyp}. The bijective methods due to Cori-Vauquelin \cite{CoriVauq} and Schaeffer \cite{Schae}, which conserve the information on the geodesics in the maps, have been extended to more general cases \cite{Bou2, ChapSchaeffBij, ChapBij}, and have led to a better understanding of the metric properties of random surfaces. 
 Their connection to theoretical physics and random matrix models was highlighted by Brezin,  Itzykson, Parisi and  Zuber \cite{PlanarDiag}, following an idea by t'Hooft \cite{THooftPlanar}.  The topological recursion makes the bridge between maps and algebraic geometry \cite{EynTopo, EynOran, EynardBook}.
 
Combinatorial maps are topological surfaces in the sense that they do not carry an \emph{ad hoc} geometry. In the case of triangulations, a canonical induced geometry can be given to the map by supposing that every segment (edge) has the same length. One then has a natural notion of local curvature - the number of triangles around each point - and of geodesics - the shortest sequence of adjoining edges between two points (vertices). A bigger polygon is not fixed by only specifying that segments have the same length, and one needs to give additional information. 
However, bigger polygons can still be given a similar simple geometry by taking the star subdivision of each polygon, that is, adding a vertex in its center and lines linking it to every point of its boundary, thus dividing it into triangles, which we can take to be equilateral. The same notion of curvature is then defined. Gluings of polygons can therefore be seen as particular kind of triangulations for which there is less freedom on how the triangles can be glued together, because of the requirement that some vertices have a fixed number of triangles around them. 

\

The two-dimensional Einsten-Hilbert action can be discretized over a combinatorial map \cite{MatrixReview}. From the Gauss-Bonnet theorem, it corresponds to its Euler's characteristics. The discrete partition function therefore classifies combinatorial maps according to their genus, and assigns to  the maps some Boltzmann weights, which depends on Newton's constant and defines a probability distribution on the set of discrete two-dimensional surfaces. This distribution is uniform on maps of the same genus. 
%
 ``The smaller" the Newton constant is, the more this distribution is peaked around planar maps. Planar maps made of triangles or of polygons of even size converge in distribution towards \emph{the same} continuous random metric space called the Brownian map \cite{MarckMierm, Mie, LeGallUnivers}, first introduced by Marckert and Mokkadem \cite{MarcMok}, and which is a fractal surface homeomorphic to the 2-sphere \cite{LGPaul2Sph} but which has Hausdorff dimension 4 \cite{GallHaus}, roughly, a measure of its creaseness. 
Regarding the partition function, this continuous limit is reached at the dominant singularity. At the critical point, the area of  the maps and the number of polygons go to infinity, and the statistical system undergoes a phase transition. The area is kept finite by rescaling the length of the edges to zero. 
The Brownian map is therefore  interpreted as an Euclidean quantum emergent random space-time: not the large-scale macroscopic  two-dimensional classical  space-time, which is simply the sphere as gravity in two-dimensions is purely topological, not the microscopic  statistical ensemble of discrete surfaces, but a mesoscopic continuous random space with intriguing properties. It is seen as a thermodynamical limit of the statistical system of discrete surfaces. 
The relation to quantum gravity was studied via the KPZ relation \cite{CFTQG2, CFTQG3, DuplSheff0, DuplSheff}, conjectured in \cite{CFTQG1}.
In the recent series of papers by Miller and Sheffield \cite{MilSheff}, it is proven that the Brownian map is actually equivalent to Liouville quantum gravity \cite{Liouv}, the effective continuum gravitational theory obtained from coupling conformal matter to 2D gravity, introduced by Polyakov \cite{Polya} to describe a theory of world-sheets in string theory. 


\

In dimension $D$, the simpler choice for $\Sb$ is the set of triangulations, that is discrete spaces obtained by gluing together tetrahedra, or their higher dimensional generalizations, called simplices. This is the point of view developed in dynamical triangulations \cite{David, QuantumGeom}. As before in the two-dimensional case, assuming that all edges have the same length leads to a notion of local curvature, and a discrete version of the Einstein-Hilbert action \eqref{EH1} is obtained following Regge's prescription \cite{Regge}. 
As in the two-dimensional case, the resulting discrete partition  function classifies triangulations according to their mean curvature: the normalized sum of the number of simplices around $(D-2)$-dimensional elements. 
However, analytic computations are very difficult, and most results are numerical. This can be overcome by introducing a coloring of facets, which are the $(D-1)$-dimensional elements of the simplices. This way, the colored triangulations are encoded into edge-colored graphs, which conserve the information about their induced geometry.  This makes it possible to classify triangulations according to the Boltzmann weight assigned by the Einstein-Hilbert action, i.e. according to their mean curvature. In the case of colored triangulations, the distribution is peaked around a sub-family of triangulations, which converge \cite{MelonsAreBP} in distribution towards the continuum random tree introduced by Aldous \cite{Aldous}, also called branched polymers in physics. This continuous space gathers the properties of a continuum limit of one-dimensional discrete spaces, and cannot be interpreted as a $D$-dimensional quantum space-time. 

One therefore needs to consider some set $\Sb$ of more general -  or more constrained - discrete spaces. The same conclusion was reached numerically for dynamical triangulations (see e.g. \cite{QuantumGeom}), so the coloring does not seem to be the problem. The aim of this thesis is to investigate the situation for discrete spaces obtained by gluing other types of elementary building blocks, such as bigger polytopes, or even more singular objects. Because the coloring was the key assumption under which analytical results were obtained, we study \emph{colored discrete spaces} obtained by gluing together finitely many building blocks along their colored facets. 
As before, an induced geometry is obtained by taking the star subdivision of the building blocks - the cone - and assuming that edges all have the same length. This defines a notion of local curvature, and the mean curvature is roughly the normalized sum over $(D-2)$-cells of the number of incident building blocks.
We stress that from a combinatorial point of view, counting how many configurations have the same mean curvature is a natural and interesting problem of its own.  From the gravity perspective, the probability distribution induced by the discretized Einstein-Hilbert action 
is peaked around configurations which maximize the number of $(D-2)$-cells at fixed number of $D$-cells, and our first concern is therefore to identify this sub-family $\Sb_\text{max}$ of discrete spaces for different choices of $\Sb$. In the limit where only such maximal configurations survive, the partition function \eqref{PartFunc1} reduces to the generation function of elements of $\Sb_\text{max}$, counted according to their number of building blocks. It has a singularity, for which the volume of the discrete space diverges. It is kept finite by rescaling the volume of building blocks to zero, so that at criticality, a continuum limit is reached which we would like to characterize.  A first indicator is the critical exponent obtained from the asymptotics of the generating function near the singularity: for $1/2$ we expect the continuum limit to be the continuum random tree, for $-1/2$ we expect it to be the Brownian map... In this framework, the emergence of new critical behaviors would be an important step towards establishing a theory describing quantum gravity.

\

When I started my PhD, in October 2014, the study of colored discrete spaces in this context was at its early stages. In any dimension $D$, there are building blocks of any size called {\it melonic}, which have a tree-like structure. It is easy to show that the results of gluing such building blocks together maximize the number of $(D-2)$-cells at fixed number of building blocks if they inherit that tree-like melonic structure, leading to the continuum random tree in the continuum limit. However, V.~Bonzom had pointed out in 2013 \cite{MoreUniv} that it was possible to escape this universality class by gluing other kind of blocks, thus motivating this work.
 The case of generalized quadrangulations in dimension four was then investigated in 2015 \cite{Enhanced}. It involves two kinds of building blocks, a melonic one and a block which mimics the combinatorial structure of a two-dimensional polygon. Three critical regimes are involved, according to the balance between the two counting parameters: the universality class of trees, leading to the continuum random tree in the continuum limit, that of planar configurations, leading to the Brownian map, and a {\it transitional} one with critical exponent $\frac{1}{3}$, for which baby-universes proliferate \cite{Baby}. This universality class was known from multi-trace matrix models \cite{IndianBaby,  BabyCrit, Korchem}, that is, by gluing non-connected polygons. 
Although only known universality classes appear, in our case they are recovered in a very specific context, that of gluing connected building blocks together and selecting the spaces which maximize the number of $(D-2)$-cells at fixed number of building blocks. As mentioned previously, the only universality class which appears in this context in two dimensions is that of planar maps, referred to as the universality class of $2D$ pure gravity in physics. The first lesson we learn from these results, is that in dimension four, this very restrictive framework does not limit the universality class to a single one. Moreover, they concern one of the simplest models one can build in dimension four, a couple of building blocks which mimic lower dimensional structures and do not render the vast diversity and richness of colored building blocks in dimensions three and higher.
However, these results were precisely obtained because of the very simple structure of these building blocks.
The aim of this thesis is to provide combinatorial tools enabling a systematic study of the building blocks and of the discrete spaces they generate. The main result of this work is the derivation of  a bijection with stacked combinatorial maps that preserves the information on the number of $(D-2)$-cells, making it possible to use results from combinatorial maps and paving the way to a systematical classification of$D$-dimensional  discrete spaces according to their mean curvature.

\

Random tensor models \cite{Book}, which generalize matrix models, have been introduced in 1991 \cite{Tensors, Tensors2, Tensors3} as a non-perturbative approach to quantum gravity and an analytical tool to study random geometries in dimension three and higher.  The proof by Gurau in 2010 \cite{Gurau1NCol, Complte1N} that colored random tensor models have a well defined perturbative $1/N$ expansion opened the way to many results in the topic: $1/N$-expansions of the uncolored tensor models \cite{Uncoloring}, the multi-orientable model \cite{Tan0}, and models with $O(N)^3$  symmetry \cite{Tan}, constructive and analyticity results \cite{BeyondPert, Mutiscale, ConstrQuart, PosTens}, double-scaling limit \cite{GurauSchaeffer, DartoisDoubleScaling, DoubleScal}, higher orders \cite{GurauSchaeffer, Fusy, BLT}, enhanced models \cite{MoreUniv, Enhanced, SigmaReview}, and the topological recursion \cite{TopoTens}. References for the very recent bridge with holography and quantum black holes are given in the paragraph on the SYK model, below. There has also been very recent interest for symmetric tensor models \cite{Kleba2, GurSym}.
The Feynman graphs of their perturbative expansions are the colored graphs dual to the colored discrete spaces introduced previously, and the discrete Einstein-Hilbert partition function \eqref{PartFunc1} can be understood as the perturbative expansion of the free-energy of some random tensor model. This generalizes the link between matrix models and combinatorial maps. Solving a tensor model usually goes back to studying the combinatorial properties of its Feynman colored graphs, which is precisely the aim of this work. We stress that one of the problems we address in this thesis is to determine how to scale a tensor model interaction in $N$ to have a well-defined and non-trivial $1/N$-expansion, a question which appears in some recent publications related to holography and quantum black holes \cite{Ferrari, Kleba2}. 
The bijective methods developed in this thesis make it possible to rewrite tensor models as matrix models with partial traces - the so called intermediate field theories, which could in the future be used to prove constructive results (see \cite{Analyticity, ConstrQuart}, and our first attempt for non-quartic models \cite{PosTens}), and possibly use eigenvalues technics to solve the models, 
as was done in \cite{DartEyn}. 

\

Along this work, we detail the connections to two other areas of research.  The Italian school of Pezzana and followers studies the 
 topological properties of colored piecewise-linear manifolds and more singular colored triangulations using the dual colored graph \cite{RegImb, RegGen, OnlyGen0Man, Moves, LinsMandel, ItalianSurvey, Handle, GagliaProper, RegHeeg, Flips, GagliaSwitch, Cristo}. Recent papers investigate the topological properties of the degree introduced by Gurau in the context of colored tensor models \cite{TopoTensor1, TopoTensor2}. 
 In this thesis, we explain how certain results, such as the topological invariance under local moves on the colored graphs, translate in the bijective framework we introduce, and apply this to determine the topology of the building blocks and of the discrete spaces they generate. 
 
 \
 
 The Sachdev-Ye-Kitaev (SYK) model is a quantum mechanical model with remarkable properties \cite{kitaev}. It encounters tremendous interest as a toy model to study the quantum properties of black-holes throughout a near AdS/CFT holography \cite{PR, MS, SYK21, Fu, gurau-ultim, Gross2, SYK20, GurauSYK3, SYK26}. Introducing flavors, as done in \cite{Gross, GurSyk, BLT}, the Feynman diagrams are a particular kind of the colored graphs we study throughout this thesis. The combinatorial bijective technics we introduce can therefore be applied to the characterization of Feynman graphs contributing at any order of the $1/N$ expansion of the correlation functions, as detailed in this work. Furthermore, Witten pointed out recently \cite{Witten} that SYK-like tensor models could be considered,  that lead to the same remarkable properties but without quenched disorder. These tensor models are one-dimensional generalizations of the models introduced by Gurau, and their Feynman diagrams are therefore dual to colored triangulations \cite{GurSyk}. Since then, the bridge between tensor models and holography has been the subject of numerous papers, such as \cite{Kleba, SYK22, SYK23, SYK28, SYK27, StephSYK, SYK25}. Among these papers, some SYK-like tensor models have been considered, which Feynman diagrams are dual to the more general kind of orientable or non-orientable discrete spaces we study in this thesis. Matrix \cite{Ferrari} and vector \cite{SYK24} models have been considered which have similar diagrammatics.
 
\

The first chapter is devoted to giving a precise meaning to the statements of the introduction. 
\begin{itemize}[label=$-$]
\item In Section \ref{sec:CombMaps}, we define notions of graph theory and combinatorial maps which will be used throughout all of this thesis. In Section~\ref{sec:Simpl}, we give a precise definition of $D$-dimensional colored triangulations as pseudo-complexes obtained by gluing simplices with colored facets. 

\item In Section~\ref{sec:EdgeCol}, we explain how colored triangulations are encoded into edge-colored graphs~\ref{subsec:GEM}. Subsection~\ref{subsec:Cryst}  is a brief summary of  results  from Crystallization theory developed by the Italian school of Pezzana and followers, which we use in the rest of this work. Subsection~\ref{subsec:GurDeg} introduces the degree, which classifies triangulations according to their number of $D$-simplices and $(D-2)$-simplices, a quantity which will be central in this work. In~\ref{subsec:Melonic}, we describe the colored triangulations around which the distribution induced by the discrete Einstein-Hilbert action is peaked. 

\item In Section~\ref{sec:pAng}, we introduce colored discrete spaces obtained by gluing more general building blocks, called \emph{bubbles}. We detail the problems which we aim at solving in this thesis. 

\item Section~\ref{sec:QG} clarifies how these questions arise from our discrete approach to quantum gravity. Subsection~\ref{subsec:Guideline} is a summary of the problems which we will tackle for different sets of discrete spaces, and of the steps which we will follow to solve them. 

 In Subsection~\ref{subsec:ColTenMod}, we show that if $\Sb$ is the set of colored triangulations, the partition function \eqref{PartFunc1} can be rewritten as the free-energy of a colored random tensor model. 
 %
 In~\ref{subsec:Uncolored}, we detail this relation when $\Sb$ is a set of colored discrete spaces obtained by gluing bubbles, as described in Section~\ref{sec:pAng}. 
 
  Subsection~\ref{subsec:SYK}, is devoted to the Sachdev-Ye-Kitaev (SYK) model. We briefly explain the link with random tensor models and introduce a colored SYK model, which Feynman graphs are the colored graphs described in Section~\ref{sec:EdgeCol}.
  \end{itemize}

\

In Chapter~\ref{chap:BijMeth}, we develop a bijection between the colored graphs introduced in Section~\ref{sec:EdgeCol} and stackings of combinatorial maps we thus name stacked maps. 
\begin{itemize}[label=$-$]
\item We come to this bijection step-by-step, by first proving bijections in simpler cases of Sections~\ref{sec:SimplerBij} and~\ref{sec:Tutte}. 
\item In Section~\ref{sec:StackedMaps}, we detail the bijection with stacked maps in the general case, and in some more specific cases, corresponding to different choices of $\Sb$.

\item The simplest case, that of the so-called quartic melonic bubbles, is detailed in Section~\ref{sec:QuartMelSec} from the point of view of combinatorial maps. This has been the most studied tensor model in the last years (see e.g. \cite{ConstrQuart, DartoisDoubleScaling, DartEyn, PhaseTrans,TopoTens}). In this section we provide powerful tools to study these models, and prove a few new results concerning sub-leading orders in~\ref{subsec:SubDomQuartMel}, bounds on the degree and sub-leading orders for spaces with boundaries in~\ref{subsec:BoundQuart}, and an extension to non-orientable spaces in~\ref{subsec:LOQuart}.

\item In Section~\ref{sec:UniGraphColSYK}, we focus on self-gluings of a single building block. They generalize combinatorial maps made of a single polygon, called unicellular maps. We apply the bijection from Section~\ref{sec:StackedMaps}, along with classical methods from graph theory. These manipulations make it easy to identify all the diagrammatic contributions to the complex colored SYK model. We extend these techniques for the real colored SYK model. In~\ref{subsec:CountUni}, we apply this framework to the enumeration of generalized unicellular maps. 

\item In Section~\ref{sec:IFT}, we generalize the intermediate field approach (see e.g. \cite{DartEyn}) by  rewriting the generating functions of colored graphs as matrix models with partial traces. As mentioned previously, the intermediate field theory has proven a very powerful tool in the quartic case to obtain analyticity results \cite{Analyticity, ConstrQuart},  apply matrix-model methods \cite{DartEyn},   the topological recursion \cite{TopoTens}, and study the phase transition at criticality \cite{PhaseTrans, SymBreak}.
\end{itemize}

Chapter~\ref{chap:PropSWM} is devoted to the derivation of general results on stacked maps, which are then applied to a number of examples.
\begin{itemize}[label=$-$]
\item Section~\ref{sec:TreeBound} tackles the issue of finding the sharpest bound on the number of $(D-2)$-cells in a discrete space. An equivalent problem is determining the right scaling in $N$ for a tensor model to have a well-defined and non-trivial $1/N$ expansion.

\item In Section~\ref{sec:MaxMaps}, we prove a series of results which specify some properties of the building blocks which have a tractable impact on the configurations maximizing the number of $(D-2)$-cells at fixed number of building blocks. Subsections~\ref{subsec:hPair} and~\ref{subsec:ConecSum} make it possible to extend results on a finite number of building blocks to infinite families of building blocks. Subsection~\ref{subsec:NonConBub} treats the cases of non-connected bubbles, which generalize multi-trace matrix models. Subsection~\ref{subsec:TreeLike} is an important generalization of the results from Section~\ref{sec:TreeBound} to a more general case, encountered for non-orientable bubbles, and for orientable bubbles in dimension six or higher. A theorem is proven, which states that when it exists, there exists a unique scaling for a tensor model to have a well-defined and non-trivial $1/N$ expansion. 

\item In Section~\ref{sec:Examples}, we apply the previous results to solve a certain number of examples in dimension 3, 4 and $D$. We stress that they can in general be understood having read the two first subsections of Sec.~\ref{sec:MaxMaps}. The results are summarized in Chapter~\ref{chap:SummaryEnd}.

\item Section~\ref{sec:Fluctuations} gives an application of the intermediate field theory for certain infinite families of bubbles. We explain how to derive the effective theory describing the fluctuations around the non-necessarily melonic vacuum. 
\end{itemize}

In Chapter~\ref{chap:SummaryEnd}, we provide tables~\ref{sec:Summary} summarizing the results of Chapter 3, and argue that more exotic behaviors are a priori not excluded. 

\vspace{2cm}

 This thesis is partially based on various articles and preprints, which we list here.
 We first introduced the bijection of Subsection~\ref{subsec:BubStacked} and proved many other results from this thesis with my advisors Vincent Rivasseau and Valentin Bonzom in \\[-2ex]

 \cite{SWM} 
\emph{Colored triangulations of arbitrary dimensions are stuffed Walsh maps,} 2015, 

Electronic Journal of Combinatorics, Volume 24, Issue 1 (2017).\\[+1ex]
The results on octahedra in Section~\ref{subsec:BiPyr} were derived in  \\[-2ex]

\cite{GluOcta} \emph{Counting gluings of octahedra,} 2016, Electronic Journal of Combinatorics, \\
Volume 24, Issue 3 (2017) with V.~Bonzom.
\\[+1ex]
\noindent  The results on the characterization of sub-leading orders of the colored SYK model in Section~\ref{sec:UniGraphColSYK} were derived in a recent paper with V.~Bonzom and A.~Tanasa. In this thesis, we describe these results in the context of stacked maps. \\[-2ex]

\cite{BLT} 
\emph{Diagrammatics of a colored SYK model and of an SYK-like tensor model, leading}

\emph{and next-to-leading orders,}
Journal of Mathematical Physics {\bf 58}, 052301 (2017).\\[-2ex]

\noindent The results concerning bubbles of size 6 in the Sections~\ref{subsec:K33} and~\ref{subsec:K334} were obtained in  \\[-2ex]

\cite{Johannes} \emph{Multi-critical behaviour of 4-dimensional tensor models up to order 6,} 2017, with Johannes Th\"urigen. \\[-2ex]

Section~\ref{sec:Fluctuations} contains the very first result of a detailed study of the theory of fluctuations around the vacuums of melonic and certain infinite families of non-melonic tensor models.  \\[-2ex]

\cite{Stephane} 
\emph{Fluctuations around melonic and non-melonic vacuums,} in progress with St\'ephane Dartois. \\

And at last, some of the results have only been proven in this thesis. The simpler bijection of Section~\ref{sec:SimplerBij}, the bijection $\Ps$ of Subsection~\ref{subsec:Bij} in the case of generic colored graphs, the description of sub-leading orders of Section~\ref{sec:QuartMelSec} and the bijection extended to non-orientable quartic bubbles \ref{subsec:LOQuart}, the procedure to obtain  the contributions to the $2n$-point function of the colored SYK model at sub-leading orders for $n>2$, the counting results on unicellular graphs in~\ref{subsec:CountUni}, and many of the results from Section~\ref{sec:MaxMaps}, among which Theorem~\ref{thm:ExUnicBBD} which states a necessary and sufficient condition of existence of the scaling of an enhanced tensor model to obtain a well-defined and non-trivial $1/N$ expansion, and the unicity of this scaling when it exists. The bi-pyramids, bubbles with toroidal boundaries, and higher dimensional generalizations in~\ref{subsec:BiPyr} have not been treated before. I hope the new dual representation for colored polytopes in Subsection~\ref{subsec:ColPol} will lead to future work.

\

Moreover, together with V.~Rivasseau, we have  tackled  the question of proving the uniform Borel summability of matrix models with  interactions of order higher than four, and of certain families of tensor models. 
More precisely, the loop vertex expansion  is a powerful combinatorial constructive method. It aims at proving the analyticity of correlation functions in the Borel summability domain of the perturbative expansion at the origin \cite{BeyondPert,Mutiscale, Analyticity}. Using this method, R. Gurau and T. Krajewski showed in  \cite{Analyticity} that the planar sector is the large $N$ limit of quartic one-matrix models beyond perturbation theory. During my PhD, V.~Rivasseau and I have succeeded in applying intermediate field methods to show the non-uniform Borel-Leroy summability of positive scalar, matrix and tensor interactions, in a series of two papers \\[-1ex]

\cite{Note} \emph{Note on the intermediate field representation of $\Phi^{2k}$ theory in zero dimension,} \\[-1ex]

\cite{PosTens} \emph{Intermediate Field Representation for Positive Matrix and Tensor Interactions}.  \\[-1ex]

As made clear in the titles, these results rely on a version of the intermediate field representation adapted to exploit the positivity of the interactions.  However, we could not prove the uniform Borel-Leroy summability, needed to extend these results to the large $N$ limit.  Recent progress by Rivasseau \cite{Riv2017} should be extendable to matrix and tensor models. These technics differ in many ways  from the bijective methods developed in this thesis, and analyticity results are out of the scope of this work, as we focus on the problem of identifying spaces according to their mean curvature. I have therefore chosen not to include them.

\newpage
\thispagestyle{plain}
\noindent {{\huge \bf Introduction} 
	{\Large (en langue fran\c{c}aise)}}

\vspace{1cm}

\`A grandes \'echelles, la gravit\'e est d\'ecrite par la relativit\'e g\'en\'erale, une th\'eorie dynamique de la g\'eom\'etrie de l'espace-temps, qui est une vari\'et\'e continue en quatre dimensions. L'un de ses concepts cl\'es est que la pr\'esence de mati\`ere influe sur la courbure de l'espace-temps. Les autres forces fondamentales d\'ecrivant la nature, l'\'electromagn\'etisme, l'interaction forte - responsable de la coh\'esion du proton et du neutron, et l'interaction faible - responsable de la radioactivit\'e, sont unifi\'ees au sein d'une m\^eme th\'eorie, le mod\`ele standard \cite{Weinberg, Salam, Glashow}. Il met en jeu des champs d\'efinis et interagissant dans l'espace-temps ambient. Nous apprenons de la relativit\'e g\'en\'erale que cet espace-temps n'est pas le th\'e\^atre  immuable des interactions de la mati\`ere, comme le d\'ecrit la relativit\'e restreinte, mais r\'eagit de fa\c{c}on dynamique \`a son contenu materiel. Ces champs sont quantifi\'es, et repr\'esentent des \'etats physiques quantiques. Les quantit\'es mesurables, appel\'ees observables, sont des probabilit\'es de transitions entre ces \'etats. Elles sont calcul\'ees \`a partir de la fonction de partition de la th\'eorie. On ne peut cependant comprendre telles quelles ces quantit\'es, qui sont calcul\'ees \`a partir de th\'eories quantiques \`a hautes \'energies, depuis notre point de vue effectif \`a faible \'energie. La renormalisation est le processus qui d\'ecrit comment nous percevront des quantit\'es physiques quantiques depuis notre \'echelle macroscopique. Un \'electron, par exemple, ne peut \^etre consid\'er\'e ``nu", c'est \`a dire sans le ``nuage" d'auto-interactions, de particules cr\'ees et d\'etruites, qui l'entoure in\'eluctablement. La renormalisation prends tout ceci en compte, et traduit l'\'electron quantique th\'eorique nu en une th\'eorie ayant un sens physique: c'est la charge renormalis\'ee de l'\'electron qui peut \^etre compar\'ee aux r\'esultats exp\'erimentaux. En cons\'equence, une th\'eorie quantique non-renormalisable n'a pas de sens physique. 

Si la gravit\'e est de loin la plus faible des quatre forces fondamentales\footnote{Les magnitudes relatives des forces agissant sur une paire de protons dans un noyau atomique sont : gravit\'e 1, interaction faible $10^{24}$, \'electromagn\'etisme $10^{35}$, interaction forte $10^{37}$}, elle implique l'existence de singularit\'es - les singularit\'es des trous-noirs, la singularit\'e cosmologique - o\`u les quantit\'es physiques divergent. Ces singularit\'es apparaissent \`a des \'echelles de l'ordre de l'\'echelle de Planck\footnote{Environ $10^{35}$ m\`etres, c'est \`a dire $10^{-20}$ fois la taille du proton. L'\'energie de Planck est $1.22\times 10^{19}GeV$.}, et sugg\`erent que, \`a cette \'echelle, les effets quantiques ne sont plus n\'egligeables. La gravit\'e est une th\'eorie classique des champs. La m\'etrique est le champs d\'ecrivant la dynamique de la g\'eom\'etrie de l'espace-temps, et quantifier la gravit\'e semble donc impliquer la quantification du champs m\'etrique. Mais tenter de le quantifier de fa\c{c}on directe comme on le fait pour les autres th\'eories quantiques des champs m\`ene \`a des divergences : la relativit\'e g\'en\'erale n'est pas une th\'eorie perturbativement renormalisable  \cite{THooftGrav, Sagnotti}. Il y a de nombreuses approches \`a la gravit\'e quantique, parmi lesquelles la th\'eorie des cordes, la gravit\'e quantique \`a boucles, les triangulations dynamiques causales... Mais un probl\`eme important, qu'on rencontre dans presque toutes ces th\'eories, et de faire sens d'une int\'egrale fonctionnelle du type 
\be
\label{PartFunGrav1FR}
\cZ = \int_\cM D[g] D[X] e^{-S_\text{Gravit\'e} - S_\text{MS}} ,
\ee 
o\`u $S_{MS}$ est l'action du mod\`ele standard, et l'int\'egration fonctionnelle est faite sur les m\'etriques $g$ de la vari\'et\'e d'espace-temps $\cM$ ainsi que sur les champs $X$ qui entrent en jeu dans le mod\`ele standard. Dans un contexte euclidien, le choix le plus \'evident pour l'action d\'ecrivant la gravit\'e pure (la gravit\'e en l'absence de mati\`ere) est l'action d'Einstein-Hilbert 
\be
\label{EH1FR}
S_\text{Gravit\'e}=S_\text{EH} =\int_\cM d^Dx\sqrt {\lvert g\rvert} (\Lambda-\frac{1}{16\pi G}R),
\ee
o\`u $\Lambda$ est la constante cosmologique, $G$ is est la constante de Newton\footnote{$G= 6.67408(31)\times10^{-11} m^3.kg^{-1}.s^{-2}$}, et $R$ est la courbure scalaire de Ricci. 
Une fa\c{c}on de donner un sens \`a la fonction de partition \eqref{PartFunGrav1FR} est d'introduire une borne d'\'echelle microscopique, et de voir l'espace-temps \`a courtes \'echelles comme un r\'eseau. Au sens large, un r\'eseau est une vari\'et\'e discr\'etis\'ee, un espace obtenu en collant des blocs \'el\'ementaires, comme des t\'etra\`edres en trois dimensions. Dans le cas o\`u l'on a \'equip\'e ces espaces discrets d'une notion de g\'eom\'etrie bien d\'efinie, et \`a supposer que les espaces discrets obtenus en collant ces blocs \'el\'ementaires de toutes les fa\c{c}ons possibles fournissent une bonne approximation des g\'eom\'etries de la vari\'et\'e, nous pouvons tenter de donner un sens \`a \eqref{PartFunGrav1FR} en rempla\c{c}ant l'int\'egration sur les g\'eom\'etries par une somme sur les discr\'etisations de la vari\'et\'e,
\be
\int_\cM D[g] \qquad\longleftrightarrow\quad\sum_{\text{discr\'etisations de }\cM}.
\ee 
Les fluctuations de topologies ne sont pas a priori interdites, et consid\'erant $\Sb$ un ensemble plus g\'en\'eral d'espaces discrets, nous pouvons choisir de d\'efinir la fonction de partition discr\`ete suivante pour la gravit\'e pure 
\be
\label{PartFunc1FR}
\cZ_\Sb = \sum_{\C\in\Sb} e^{-S_\text{discret}(\C)}, 
\ee
o\`u $S_{discret}$ serait une version discr\`ete de \eqref{EH1FR}. Voyons un peu plus en d\'etail certaines possibilit\'es pour l'ensemble $\Sb$, et les actions discr\`etes $S_{discret}$ correspondantes, en commen\c{c}ant par le cas bi-dimensionnel.

\

Les cartes combinatoires sont des surfaces bidimensionnelles discr\`etes ferm\'ees obtenues en collant des polygones le long des segments de leurs bords. Introduites dans les ann\'ees soixante par les travaux fondateurs de Tutte sur les sph\`eres discr\`etes \cite{TutteTriang, TutteBij}, leur \'etude et \'enum\'eration ont donn\'e naissance \`a un domaine de recherche foisonnant  \cite{GoulJack, GraphsOnSurfaces,
Flajolet}. L'\'etude des surfaces de genres sup\'erieurs a \'et\'e initi\'ee par \cite{BenCanHigher}. Listons quelques directions~: Caract\`ere rationnel de fonctions g\'en\'eratrices \cite{Bousquet}, 
hypercartes \cite{Walsh, BernFusHyp}, 
cartes \`a degr\'es restreints \cite {Ben, BernFusBij, CEAMatrix, BernFusHyp},
cartes unicellulaires \cite{ChapUnicell, BernardiHZ, ChapFusUnicell}, 
formules de comptage r\'ecursif \cite{BernardiHZ,ChapRecu}, 
 bijections \cite{BernChapHZ, BernFusHyp}. Les m\'ethodes bijectives dues \`a Cori-Vauquelin \cite{CoriVauq} et Schaeffer \cite{Schae}, qui conservent l'information sur les g\'eod\'esiques des cartes, ont \'et\'e \'etendues \`a des cas plus g\'en\'eraux \cite{Bou2, ChapSchaeffBij, ChapBij}, et ont men\'e \`a une meilleure compr\'ehension des propri\'et\'es m\'etriques des surfaces al\'eatoires. Le lien avec la physique th\'eorique et les mod\`eles de matrices a \'et\'e mis en valeur par Brezin,  Itzykson, Parisi et  Zuber \cite{PlanarDiag}, suivant une id\'ee de t'Hooft \cite{THooftPlanar}.  La r\'ecurrence topologique fait le pont entre les cartes et la g\'eom\'etrie alg\'ebrique \cite{EynTopo, EynOran, EynardBook}. 
 
 Les cartes combinatoires sont des surfaces topologiques au sens o\`u elles n'ont pas de g\'eom\'etries \emph{ad hoc}. Les triangulations - les collages de triangles - peuvent \^etre \'equip\'ees d'une g\'eom\'etrie canonique induite en supposant que tous les segments (les ar\^etes) ont la m\^eme longueur. On a alors une notion naturelle de courbure locale - le nombre de triangles autour d'un point (un sommet), et de g\'eod\'esiques - la plus petite suite d'ar\^etes adjacentes entre deux sommets. Un polygone de taille sup\'erieure n'est pas fix\'e en requ\'erant que toutes ses ar\^etes aient la m\^eme longueur, et il faut pour cela donner d'autres informations. Cependant, on peut toujours donner une g\'eom\'etrie canonique \`a de tels polygones en prenant leur subdivision en \'etoile, c'est \`a dire en ajoutant un sommet en leur centre que l'on relie \`a tous ses sommets, ce qui les divise en triangles, que l'on peut choisir \'equilat\'eraux. La m\^eme notion de courbure est alors d\'efinie. Les collages de polygones peuvent donc \^etre vus comme des triangulations d'un type particulier, pour lesquelles on a moins de libert\'e quant au collage des triangles, puisqu'on requiert que certains sommets aient un nombre fix\'e de triangles adjacents. 

\

On peut discr\'etiser l'action d'Einstein-Hilbert en consid\'erant une carte combinatoire \cite{MatrixReview}. D'apr\`es le th\'eor\`eme de Gauss-Bonnet, il s'agit alors de sa caract\'eristique d'Euler. La fonction de partition discr\`ete classifie donc les cartes combinatoires selon leur genre, et assigne \`a chaque carte un poids de Boltzmann qui d\'epend de la contante de Newton et d\'efinit une distribution de probabilit\'e sur l'ensemble des surfaces bidimensionnelles discr\`etes. Cette distribution est uniforme pour les cartes de m\^eme genre. Plus la constante de Newton est ``petite", plus la distribution est piqu\'ee sur les cartes planaires. Les cartes planaires faites de triangles ou de polygones de taille paire convergent en distribution vers \emph{le m\^eme} espace m\'etrique continu al\'eatoire, appel\'e la carte - ou sph\`ere -  brownienne   \cite{MarckMierm, Mie, LeGallUnivers}, initialement introduite par Marckert et Mokkadem  \cite{MarcMok}. Il s'agit d'une surface fractale hom\'eomorphe \`a la 2-sph\`ere  \cite{LGPaul2Sph} mais dont la dimension de Hausdorff est 4 \cite{GallHaus}, ce qui signifie, intuitivement, qu'elle est tr\`es ``chiffonn\'ee". Du point de vue de la fonction de partition, on atteint la limite continue \`a la singularit\'e dominante. Au point critique, l'aire des cartes et le nombre de polygones diverge, et le syst\`eme statistique subit une transition de phase. On maintient une aire finie en faisant tendre l'aire des polygones vers z\'ero. La carte brownienne est donc interpr\'et\'ee comme un espace-temps al\'eatoire quantique \'emergent : non pas l'espace-temps classique \`a large-\'echelle, qui est simplement la sph\`ere car la gravit\'e en deux dimensions est purement topologique, non pas l'ensemble statistique microscopique de surfaces discr\`etes, mais un espace continu al\'eatoire mesoscopique aux propri\'et\'es intrigantes. Elle est vue comme la limite thermodynamique de l'ensemble statistique de surfaces discr\`etes. La relation \`a la gravit\'e a tout d'abord \'et\'e \'etudi\'ee via la relation KPZ \cite{CFTQG2, CFTQG3, DuplSheff0, DuplSheff}, conjectur\'ee dans \cite{CFTQG1}.
Dans la r\'ecente s\'erie de papier de Miller et Sheffield \cite{MilSheff}, il est prouv\'e que la carte brownienne est en fait \'equivalente \`a la gravit\'e quantique de Liouville  \cite{Liouv}, la th\'eorie gravitationnelle continue effective obtenue en couplant de la mati\`ere conforme \`a de la gravit\'e en deux dimensions, qu'a introduite Polyakov \cite{Polya} afin de d\'ecrire les surfaces d'univers en th\'eorie des cordes. 

\

En dimension $D$, le choix le plus simple pour  $\Sb$ est l'ensemble des triangulations, c'est \`a dire des espaces discrets obtenus en collant des t\'etra\`edres, ou leurs g\'en\'eralisations en dimensions sup\'erieures, qu'on appelle des simplexes. C'est le point de vue d\'evelopp\'e dans les triangulations dynamiques \cite{David, QuantumGeom}. Comme pr\'ec\'edemment dans le cas bidimensionnel, imposer une m\^eme longueur \`a toutes les ar\^etes m\`ene \`a une notion de courbure locale, et une version discr\`ete de l'action de Einstein-Hilbert \eqref{EH1FR} s'obtient \`a partir le l'action de Regge \cite{Regge}. 
La fonction de partition r\'esultante classifie les triangulations selon leur courbure moyenne : la somme normalis\'ee du nombre de $D$-simplexes autour de chaque \'el\'ement de dimension $D-2$. Cependant, les calculs analytiques sont tr\`es difficiles, et la plupart des r\'esultats sont obtenus num\'eriquement. On peut surmonter cet obstacle en introduisant une coloration des facettes, qui sont les sous-\'el\'ements de dimension $D-1$ des simplexes. Ainsi, les triangulations color\'ees peuvent \^etre encod\'ees par des graphes color\'es qui conservent l'information sur la g\'eom\'etrie induite. Cela rend possible la classification des triangulations selon le poids de Boltzmann que leur associe l'action d'Einstein-Hilbert, i.e. selon leur courbure moyenne. Dans le cas des triangulations color\'ees, la distribution est piqu\'ee autour d'une sous-famille de triangulations, qui converge \cite{MelonsAreBP} en distribution vers l'arbre continu al\'eatoire introduit par Aldous \cite{Aldous}, que l'on nomme aussi polym\`eres branch\'es en physique. Les propri\'et\'es de cet espace continu al\'eatoire sont celles d'un espace \`a une dimension, et il ne peut donc pas \^etre interpr\'et\'e comme un espace-temps quantique $D$-dimensionnel.

\

Il est donc n\'ecessaire de consid\'erer un ensemble $\Sb$  d'espaces discrets plus g\'en\'eraux, ou plus contraints. On parvient num\'eriquement aux m\^emes conclusions en consid\'erant les triangulations dynamiques (voir \cite{QuantumGeom} par exemple), donc la coloration ne semble pas \^etre le probl\`eme. Le but de cette th\`ese est d'\'etudier la situation dans le cas d'espaces discrets obtenus en collant d'autres types de blocs \'el\'ementaires, comme des polytopes de tailles sup\'erieures, voir m\^eme des objets plus singuliers. Comme la coloration \'etait l'hypoth\`ese cl\'e permettant l'obtention de r\'esultats analytiques, nous \'etudions des \emph{espaces discrets color\'es} obtenus en collant des blocs \'el\'ementaires le long de leurs facettes color\'ees. Comme pr\'ec\'edemment, on obtient une g\'eom\'etrie induite en prenant la subdivision en \'etoile des blocs \'el\'ementaires - le cone - et en imposant une m\^eme longueur \`a toutes les ar\^etes ainsi obtenues. Une notion de courbure locale en r\'esulte, et la courbure moyenne est approximativement la somme normalis\'ee du nombre de blocs \'el\'ementaires autour de chaque $(D-2)$-cellule. D'un point de vue combinatoire, compter le nombre de configurations qui ont la m\^eme courbure moyenne est un probl\`eme naturel et int\'eressant en soi. Du point de vue de la gravit\'e, la distribution de probabilit\'e induite par l'action d'Einstein-Hilbert discr\'etis\'ee est piqu\'ee autour des configurations qui maximisent le nombre de $(D-2)$-cellules \`a nombre fix\'e de blocs \'el\'ementaires, et notre premier objectif est donc d'identifier cette sous-famille $\Sb_\text{max}$ d'espaces discrets pour diff\'erents choix  de $\Sb$. Dans la limite o\`u seuls survivent ces espaces maximaux, la fonction de partition \eqref{PartFunc1FR} se r\'eduit \`a la fonction g\'en\'eratrice des \'el\'ements  de $\Sb_\text{max}$, compt\'es en fonction de leur nombre de blocs \'el\'ementaires. Cette derni\`ere a une singularit\'e dominante, pour laquelle le volume des espaces discrets diverge. On le maintient fini en faisant tendre le volume des blocs \'el\'ementaires vers z\'ero, et on atteint donc au point critique une limite continue, que l'on voudrait caract\'eriser. Une premi\`ere indication quant \`a la nature de cette limite est l'exposant critique obtenu \`a partir du comportement asymptotique de la fonction g\'en\'eratrice \`a la singularit\'e: si c'est $1/2$, on s'attend  \`a ce que la limite continue soit l'arbre continue al\'eatoire, si c'est $-1/2$, on s'attend \`a ce que ce soit la carte brownienne... Dans ce contexte, l'apparition de nouveaux comportements critiques serait un important progr\`es quant \`a la quantification de la gravit\'e.

\

Lorsque j'ai commenc\'e ma th\`ese, en octobre 2014, l'\'etude des espaces discrets color\'es dans le contexte pr\'esent\'e ci-dessus en \'etait \`a ses d\'ebuts. Il existe, en toute dimension $D$, des blocs \'el\'ementaires appel\'es \emph{m\'eloniques}, qui ont une structure arborescente. Il est facile de montrer que les collages de tels blocs ne maximisent la courbure moyenne que s'ils h\'eritent eux-m\^emes de cette structure m\'elonique arborescente, et ils convergent donc vers l'arbre continu al\'eatoire dans la limite continue. Cependant, V. Bonzom avait soulign\'e dans \cite{MoreUniv} qu'il \'etait possible d'\'echapper \`a cette classe d'universalit\'e en collant d'autres types de blocs \'el\'ementaires, motivant ainsi ces travaux. Le cas des quadrangulations g\'en\'eralis\'ees en dimension quatre a ensuite \'et\'e explor\'e en 2015 dans \cite{Enhanced}. Il met en jeu deux types de blocs \'el\'ementaires, un bloc m\'elonique et un bloc qui mime la structure combinatoire d'un polygone bidimensionnel. Trois r\'egimes critiques entrent en jeu, selon le rapport entre les deux variables de comptage : la classe d'universalit\'e des arbres, menant \`a l'arbre continu al\'eatoire dans la limite continue, la classe des configurations planaires, qui convergent vers la carte brownienne, et une classe \emph{transitoire} avec exposant critique  $\frac{1}{3}$, pour laquelle les b\'eb\'e-univers prolif\`erent  \cite{Baby}. Celle classe d'universalit\'e \'etait connue dans le contexte des mod\`eles multi-traces \cite{IndianBaby,  BabyCrit, Korchem}, c'est \`a dire, des espaces obtenus en collant des polygones non connexes. Si seules des classes d'universalit\'e connues \'emergent, on les retrouve ici dans un contexte bien particulier : on colle des blocs \'el\'ementaires \emph{connexes}, et on s\'electionne les espaces qui maximisent le nombre de $(D-2)$-cellules \`a nombre fix\'e de blocs. Comme mentionn\'e pr\'ec\'edemment, la seule classe qui apparait dans ce contexte en deux dimensions est celle des cartes planaires, \'egalement appel\'ee classe d'universalit\'e de la gravit\'e 2D pure en physique. On apprend donc de ces r\'esultats qu'en dimension quatre, le contexte tr\`es contraignant dans lequel on se place ne restreint pas la classe d'universalit\'e. De plus, ces r\'esultats concernent les mod\`eles les plus simples que l'on peut construire en dimension 4, deux blocs \'el\'ementaires qui miment des structures de plus faible dimensionnalit\'e et ne rendent pas compte de l'immense diversit\'e et de la richesse des blocs \'el\'ementaires color\'es en dimensions trois et sup\'erieures. Ces r\'esultats ont cependant \'et\'e obtenus pr\'ecis\'ement du fait de la simplicit\'e de la structure combinatoire de ces blocs \'el\'ementaires. L'objectif de cette th\`ese est de fournir des outils combinatoires qui permettraient une \'etude syst\'ematique des blocs \'el\'ementaires et des espaces discrets qu'ils g\'en\`erent. Le r\'esultat principal de ce travail est l'\'etablissement d'une bijection avec des cartes combinatoires empil\'ees, qui pr\'eserve l'information sur le nombre de $(D-2)$-cellules, rendant possible l'utilisation de r\'esultats sur les cartes combinatoires et ouvrant la voie \`a une classification syst\'ematique des espaces discrets $D$-dimensionnels selon leur courbure moyenne. 

\

Les mod\`eles de tenseurs al\'eatoires  \cite{Book}, qui g\'en\'eralisent les mod\`eles de matrices, ont \'et\'e introduits en 1991 \cite{Tensors, Tensors2, Tensors3} comme une approche non-perturbative \`a la gravit\'e quantique et un outil analytique pour \'etudier les g\'eom\'etries al\'eatoires en dimensions trois et sup\'erieure. La preuve par Gurau en 2010  \cite{Gurau1NCol, Complte1N} que les mod\`eles de tenseurs  al\'eatoires color\'es ont un d\'eveloppement en $1/N$ bien d\'efini a ouvert la voie \`a de nombreux r\'esultats : 
le d\'eveloppement en $1/N$ des mod\`eles d\'ecolor\'es \cite{Uncoloring}, le mod\`ele multi-orientable  \cite{Tan0}, et les mod\`eles avec sym\'etrie $O(N)^3$  \cite{Tan}, des r\'esultats constructifs et d'analycit\'e \cite{BeyondPert, Mutiscale, ConstrQuart, PosTens}, la limite de double-scaling \cite{GurauSchaeffer, DartoisDoubleScaling, DoubleScal}, l'\'etude des ordres non-dominants \cite{GurauSchaeffer, Fusy, BLT}, les mod\`eles augment\'es \cite{MoreUniv, Enhanced, SigmaReview}, et la r\'ecurrence topologique \cite{TopoTens}. Des r\'ef\'erences concernant le pont tr\`es r\'ecent avec l'holographie et les trous-noirs quantiques sont donn\'ees dans le paragraphe sur le mod\`ele SYK, ci-dessous. De tr\`es r\'ecents r\'esultats montrent un int\'er\^et croissant pour les mod\`eles de tenseurs sym\'etriques \cite{Kleba2, GurSym}.
Les graphes de Feynman de leur d\'eveloppements perturbatifs sont des graphes color\'es duaux aux espaces discrets color\'es introduits pr\'ec\'edemment, et la fonction de partition d'Einstein-Hilbert \eqref{PartFunc1FR} peut donc s'interpr\'eter comme le d\'eveloppement perturbatif de l'\'energie libre de mod\`eles de tenseurs al\'eatoires. Cela g\'en\'eralise le lien entre les mod\`eles de matrices al\'eatoires et les cartes combinatoires. R\'esoudre un mod\`ele de tenseurs al\'eatoires revient usuellement \`a \'etudier les propri\'et\'es combinatoires de ses graphes de Feynman color\'es, ce qui est pr\'ecis\'ement l'objectif de ce travail. Soulignons ici que l'un des probl\`emes centraux de cette th\`ese est la d\'etermination de la puissance de $N$ ad\'equate multipliant l'interaction dans un mod\`ele de tenseurs donn\'e afin que ce dernier ait un d\'eveloppement en $1/N$ bien d\'efini et non-trivial dans la limite continue, une question que l'on retrouve dans des publications r\'ecentes li\'ees \`a l'holographie et aux trous noirs quantiques \cite{Ferrari, Kleba2}. 
Les m\'ethodes bijectives d\'evelopp\'ees dans cette th\`ese permettent de r\'e\'ecrire les mod\`eles de tenseurs comme des mod\`eles de matrices avec des traces partielles - connues sous le nom de th\'eories de champs interm\'ediaires, qui pourraient \^etre utilis\'ees dans le futur pour \'etablir des r\'esultats constructifs  (voir \cite{Analyticity, ConstrQuart}, et notre premier essai pour les mod\`eles non-quartiques  \cite{PosTens}), et possiblement pour appliquer des techniques de valeurs propres afin de r\'esoudre les mod\`eles, comme cela fut fait dans  \cite{DartEyn}. 

\

Au long de cette th\`ese, nous d\'etaillons les connexions avec deux autres domaines de recherche. L'\'ecole italienne de Pezzana \'etudie les propri\'et\'es topologiques des vari\'et\'es triangul\'ees color\'ees \`a l'aide du graphe color\'e dual  \cite{RegImb, RegGen, OnlyGen0Man, Moves, LinsMandel, ItalianSurvey, Handle, GagliaProper, RegHeeg, Flips, GagliaSwitch, Cristo}. Des publications r\'ecentes explorent les propri\'et\'es topologiques du degr\'e introduit par Gurau dans le contexte des mod\`eles de tenseurs al\'eatoires color\'es \cite{TopoTensor1, TopoTensor2}.  Dans cette th\`ese, nous expliquons comment certains r\'esultats, tels que l'invariance topologique selon des mouvements locaux sur les graphes color\'es, se traduisent dans le langage bijectif que nous introduisons, et appliquons cela \`a la d\'etermination de la topologie des blocs \'el\'ementaires et des espaces discrets qu'ils g\'en\`erent.
 
 \
 
 Le mod\`ele de  Sachdev-Ye-Kitaev (SYK) est un mod\`ele de m\'ecanique quantique poss\'edant des propri\'et\'es remarquables \cite{kitaev}. Il rencontre  depuis peu un int\'er\^et mondial comme mod\`ele-jouet d'\'etude des propri\'et\'es quantiques des trous-noirs, \`a travers une holographie AdS/CFT approch\'ee \cite{PR, MS, SYK21, Fu, gurau-ultim, Gross2, SYK20, GurauSYK3, SYK26}. Si l'on introduit des saveurs, comme fait dans  \cite{Gross, GurSyk, BLT}, les graphes de Feynman sont un sous-ensemble des graphes color\'es \'etudi\'es dans cette th\`ese. Les techniques bijectives introduites s'appliquent donc \`a la caract\'erisation des graphes contribuant a un ordre donn\'ee du d\'eveloppement en $1/N$ des fonctions de correlation, comme d\'etaill\'e dans ce travail. Par ailleurs, Witten a r\'ecemment  soulign\'e \cite{Witten} que consid\'erer des mod\`eles de tenseurs similaires au mod\`ele SYK mais sans d\'esordre menait aux m\^emes importantes conclusions. Ces mod\`eles de tenseurs sont des g\'en\'eralisations en une dimension des mod\`eles introduits par Gurau, et leurs graphes de Feynman sont donc duaux de triangulations color\'ees \cite{GurSyk}. Depuis, le pont entre les mod\`eles de tenseurs et l'holographie est au coeur de nombreuses publications, telles que  \cite{Kleba, SYK22, SYK23, SYK28, SYK27, StephSYK, SYK25}.

\chapter{Colored simplices and edge-colored graphs}

\section{Combinatorial maps}
\label{sec:CombMaps}



In this section, we define several notions of graph and map theory which we will use throughout this thesis. There are different equivalent ways of defining combinatorial maps. A map is obtained by taking a finite collection of polygons and identifying two-by-two all of the segments of their boundaries. The result is a discretized two-dimensional surface. A map can therefore also be seen as a graph embedded in this surface such that the connected components of the complement of the embedded graph - which are its faces - are homeomorphic
 to discs. In all of this work, a graph is the following (also called multigraph or pseudograph in graph theory)
 \begin{definition}[Graph] A graph is an ordered pair $(\cV,\cE)$, where $\cV$ is a set and $\cE$ is a multiset of unordered pairs of elements of $\cV$. 
 \end{definition}
The elements of $\cV$ are called vertices, and an element  $(v,v')\in\cE$ is an edge between $v$ and $v'$. The definition allows edges between the same vertex (loops) and multiple 
 edges, which are edges that link the same two vertices, also called parallel edges.  
 A combinatorial map can also be seen as a graph with an additional constraint, which is a cyclic ordering of the half-edges around each vertex. We will use the following formal definition
\begin{definition}[labeled combinatorial maps] 
\label{def:CombMap}
A labeled combinatorial map  is a triplet $(\cD, \sigma, \alpha)$ where
$\cD$ is a set of half-edges (or darts) labeled from 1 to $2E$  such that
\begin{itemize}
\item $\sigma$ is a permutation on $\cD$ 
\item $\alpha$ is a fixed-point free involution on $\cD$
\end{itemize}
\end{definition}

Example of labeled map are shown in Fig.~\ref{fig:EquivLabMap}. Each element of the unique decomposition of $\sigma$ into disjoint cycles is interpreted as a {\it vertex} of $M$, together with a cyclic ordering or the incident half-edges: the edge following the half-edge $i\in\cD$ counterclockwise is $\sigma(i)$. A $corner$ is a pair $(i,\sigma(i))$.
The {\it edges} of $M$ are the disjoint transpositions of $\alpha$: the other half-edge of the same edge as $i$ is $\alpha(i)$. 
The {\it faces} are the disjoint cycles of $\sigma\circ\alpha$. The half-edge following $i$ clockwise in a face, $\sigma(\alpha(i))$ is the one which shares a corner with the other half-edge of the same edge. By faces of a non-connected map, we mean the sum of the faces of its connected components.

The {\it degree} (or {\it valency}) of a vertex or a face is the number of half-edges in the corresponding cycle.
An edge and a vertex, or a vertex and a face, are said to be $incident$ if the corresponding cycles contain a common half-edge. An edge and a face  are said to be incident if one or the other of its half-edges belongs to the face. The $extremities$ (or endpoints) of an edge are its incident vertices.

\begin{figure}[!h]
\centering
\includegraphics[scale=0.6]{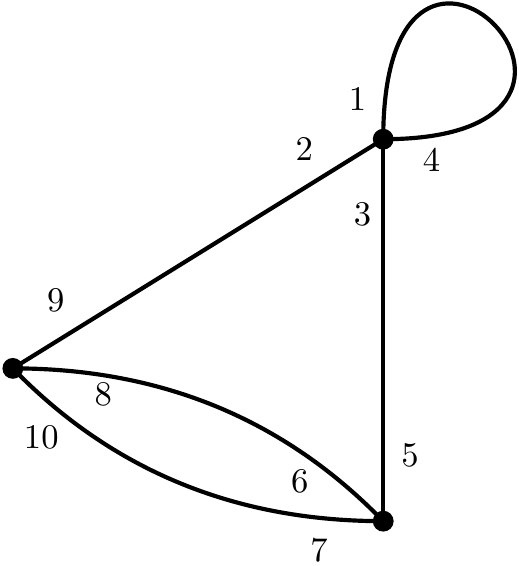} 
\hspace{1cm}\raisebox{10ex}{$\equiv$}\hspace{1cm}
\includegraphics[scale=0.6]{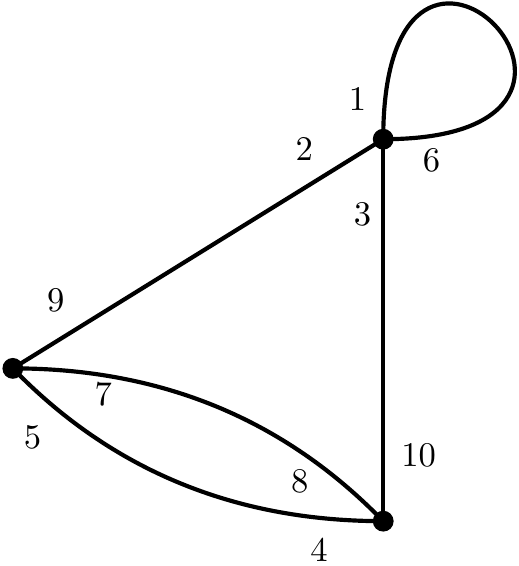} 
\caption{Equivalent labeled maps. }
\label{fig:EquivLabMap}
\end{figure}

Labeled combinatorial maps depend on the arbitrary labeling of half-edges. Combinatorial maps are the equivalence classes upon relabeling the half-edges.
\begin{definition}[Combinatorial maps] An (unlabeled) combinatorial map is 
\be
\cM=\{ (\rho\circ\sigma\circ\rho^{-1}, \rho\circ\alpha\circ\rho^{-1}) \mid \rho\in \cS_{2E} \},
\ee
where $\sigma$ and $\alpha$ are as in Def.~\ref{def:CombMap}.
\end{definition}

An example of two equivalent maps is shown in Fig.~\ref{fig:EquivLabMap}. Indeed, the labeled map on the left and on the right are respectively
\begin{align}
    \sigma  &= (1234)(567)(8910)           &  \sigma' &= (1236)(8410)(795)  \\ 
    \alpha &= (14)(35)(29)(68)(710)  &  \alpha'  &= (16)(310)(29)(78)(45),
\end{align}
and one can verify that the following permutation satisfies
\be
\rho = (4687)(510); \qquad \rho\circ\sigma\circ\rho^{-1}=\sigma', \quad\text{and}\quad \rho\circ\alpha\circ\rho^{-1}=\alpha'.
\ee
The number of labeled map corresponding to the same unlabeled map is 
\be
\frac{(2E)!}{\lvert \text{Aut}(\sigma,\alpha) \rvert}, 
\ee
where $\lvert \text{Aut}(\sigma,\alpha) \rvert$ is the number of automorphisms of the unlabeled map:
\be
\lvert \text{Aut}(\sigma,\alpha)\rvert = \Card\{ \rho\in \cS_{2E} \mid (\sigma,\alpha)=(\rho\circ\sigma\circ\rho^{-1}, \rho\circ\alpha\circ\rho^{-1})\}.
\ee

\begin{figure}[!h]
\centering
\includegraphics[scale=0.6]{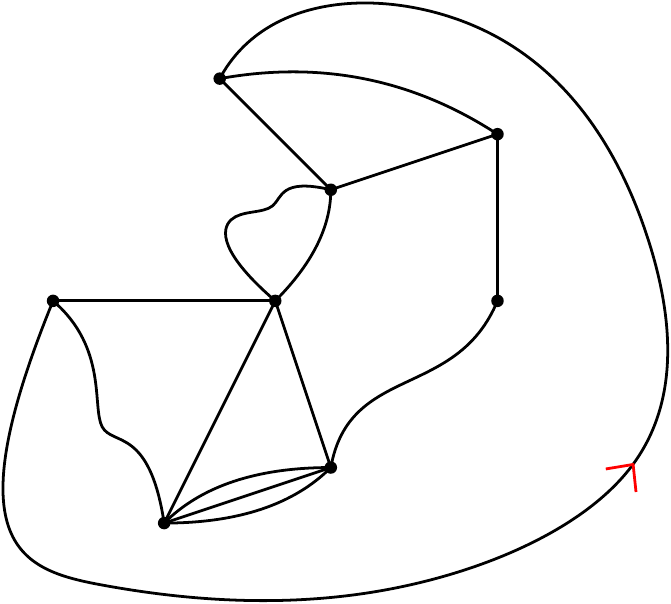} 
\qquad\raisebox{10ex}{$=$}\qquad
\raisebox{2ex}{\includegraphics[scale=0.6]{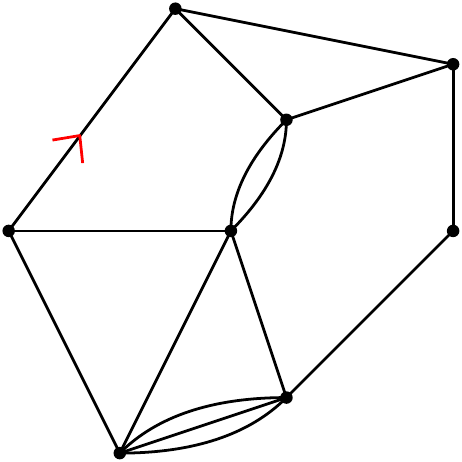} }
\qquad\raisebox{10ex}{$\neq$}\qquad
\raisebox{2ex}{\includegraphics[scale=0.6]{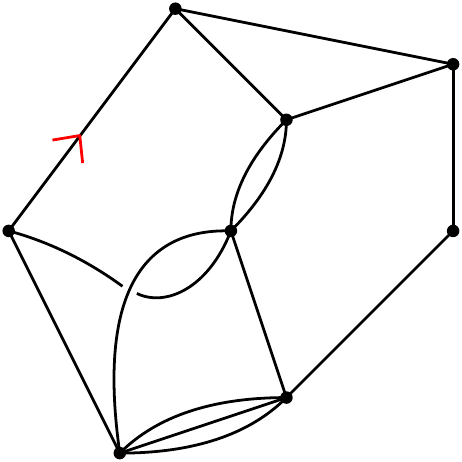}}
\caption{Isomorphic and non-isomorphic rooted maps. }
\label{fig:EquivMap}
\end{figure}

$Rooted$ maps are such that a particular edge - the root - has been oriented. This is equivalent to distinguishing a particular corner (e.g. the one before the outgoing half-edge, counterclockwise) or to distinguishing a particular face (e.g. the one on the right of the root edge).  A root edge, root corner or root face, is also called a marked edge, marked corner, marked face. The automorphism group of a rooted map is trivial, so that the number of labeled and unlabeled rooted maps coincide.

We call {\it underlying graph} the graph obtained by releasing the ordering constraint on half-edges. More precisely, its vertices are the disjoint cycles of $\sigma$ and there is an edge $(v,v')$ for each disjoint pair $(d,d')$ of $\alpha$ such that $d$ (resp. $d'$) is incident to $v$ (resp. $v'$). It is also called the  \emph{1-skeleton}. This notion generalizes to higher dimensional discrete spaces.

A $path$ in a graph is a  set of edges such that consecutive edges share an endpoint, and which never passes twice through the same vertex (such paths are also called proper paths). A graph is \emph{connected} if any two vertices have a path between them. A $cycle$ in a graph is a cyclic succession of edges which share an endpoint, i.e. a connected subgraph that only has degree two vertices.

\begin{definition}[Circuit-rank]
\label{def:CircuitRank}
The circuit-rank of a graph (or a map) with E edges, V vertices and K connected components, is defined as 
\be
L=E-V+K.
\ee
It corresponds to its number of independent cycles.
\end{definition}

The example of Figure~\ref{fig:EquivMap} has 8 independent cycles.  A \emph{forest} is a graph (or a map) which as a vanishing circuit-rank, i.e. which has no cycle, and a \emph{tree} is a connected forest. A subgraph contains all the vertices of a graph, and a subset of its edges. An \emph{isolated vertex} is a vertex with no incident edge. 
Given a graph $\G$, a \emph{spanning forest} $T$ is a subgraph of $\G$ with no isolated vertex and no cycle. If $\G$ is connected, $T$ is a spanning tree. Given a graph $\G$ and a spanning forest $T$, the number of edges which are not in $T$ is $L(\G)$.

\emph{Deleting an edge} is removing it from the edge set. A \emph{cut-edge}, or \emph{bridge}, is an edge which when deleted, raises the number of connected components by one. An \emph{edge-cut} is a set of edges which, when deleted, raises the number of connected components. A $k$-bond is a minimal edge-cut comprised of $k$ edges, i.e. a set $S$ of edges such that deleting all of them disconnects a connected graph into two connected components while deleting the edges of any proper subset of $S$ does not.

The {\it dual map} is the combinatorial map $(\cD, \phi^{-1}, \alpha^{-1})$. 
Its faces are the disjoint cycles of $\sigma^{-1}$. The dual $M^*$ of $M$ has one vertex for each face of $M$, one face for each vertex of $M$, and an edge between two non-necessarily distinct vertices if there was an edge between the corresponding face(s). One may choose to restrict the degrees of faces and/or vertices, e.g. to consider maps that have vertices of degrees 4 or 6 and only faces of degree 5. A {\it $p$-angulation} is a map that has solely faces of degree $p$. A triangulation is shown on the left of Fig.~\ref{fig:Cone}. A regular graph or map is such that all vertices have the same valency. It is said to be  {\it $p$-valent} if  all  vertices have valency $p$. The dual map of a $p$-angulation is a $p$-valent map.

A graph or a map is said to be $bipartite$ if its vertices can be partitioned into two sets A and B, such that edges  can only have an extremity in $A$ and the other in $B$. We generally color the vertices in A and those in B with two different colors. The dual of a bipartite map is face-bipartite, or face-bicolored. 


\begin{definition}[Genus]
\label{def:genus}
The genus $g$ of a combinatorial map with $E$ edges, $V$ vertices, $F$ faces and $K$ connected components is defined as 
\be
2K-2g=V-E+F.
\ee
For a connected map, it is the genus of the surface on which the underlying graph can be embedded. 
\end{definition}
\begin{figure}[!h]
\centering
\hspace{0.2cm}\raisebox{1.5ex}{\includegraphics[scale=0.8]{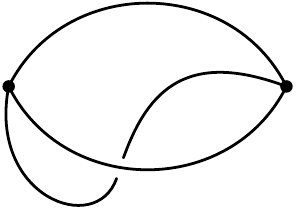}} \hspace{2cm}
\includegraphics[scale=0.8]{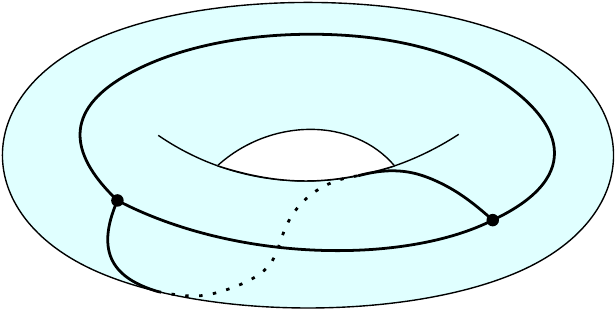}
\caption{A genus one map can be embedded on the torus. }
\label{fig:Gen1}
\end{figure}
The genus of a connected map is therefore the minimal genus of surfaces on which the map can be drawn without crossings. The map on the left of  Fig.~\ref{fig:EquivMap} is planar ($g=0$) as it is embedded on the sphere.  The map on the right of Fig.~\ref{fig:EquivMap} and the map of Fig.~\ref{fig:Gen1} have genus 1: the surface of minimum genus on which they can be drawn without crossings is the torus.

\section{Colored simplicial pseudo-complexes}
\label{sec:Simpl}


A simplicial pseudo-complex is a set of vertices, edges, triangles, and $D$-dimensional generalizations - called $D$-simplices - that satisfies additional rules. 
Before stating them, we define the $D$-simplex recursively from the $D-1$ one by taking its cone. 
 The 1-skeleton, or underlying graph, is obtained by keeping only the graph consisting of the vertices and edges. 

\begin{figure}[!h]
\centering
\includegraphics[scale=0.6]{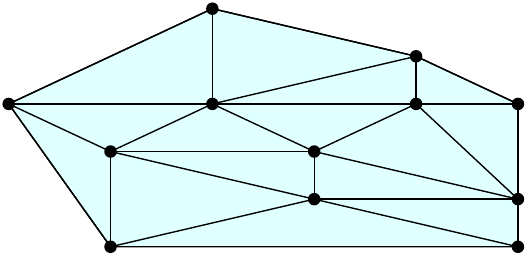} \hspace{1cm}
\raisebox{6ex}{\begin{tabular}{@{}c@{}}  cone C(X)\\$\longrightarrow$\\$\longleftarrow$\\ $X =\partial C(X)\setminus C(\partial X)$ \end{tabular} }
\hspace{1cm}\includegraphics[scale=0.6]{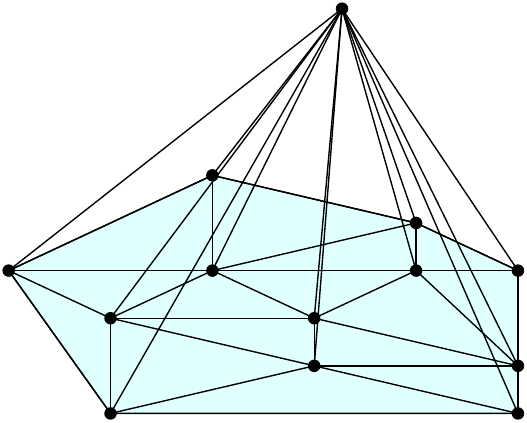}
\caption{A discrete disc and its cone. }
\label{fig:Cone}
\end{figure}
\begin{definition}[Cone] 
\label{def:Cone}
The cone of a graph is its star subdivision, obtained adding a vertex, and  edges connecting that vertex to every existing one. The cone of a discrete $D-1$ dimensional space $X$ without boundary $\partial X$ is a $D$-dimensional space $C(X)$ with boundary $X=\partial C(X)$  which 1-skeleton is the cone of the 1-skeleton of $X$. If furthermore $X$ has a boundary, the boundary of $C(X)$ is the $D-1$ dimensional space obtained by identifying $X$ and $C(\partial X)$ along $\partial X$ (Fig.~\ref{fig:Cone}). 
\end{definition}
%

 A 0-dimensional simplex is just a vertex, a 1-dimensional simplex consists of two vertices joint by an edge, 
 a 2-dimensional simplex is a
  triangle and its interior, a 3-dimensional simplex is a tetrahedron and the volume it contains, etc. A 3-simplex is pictured in Figure~\ref{fig:ColSimp}.
\begin{figure}[!h]
\centering
\includegraphics[scale=0.6]{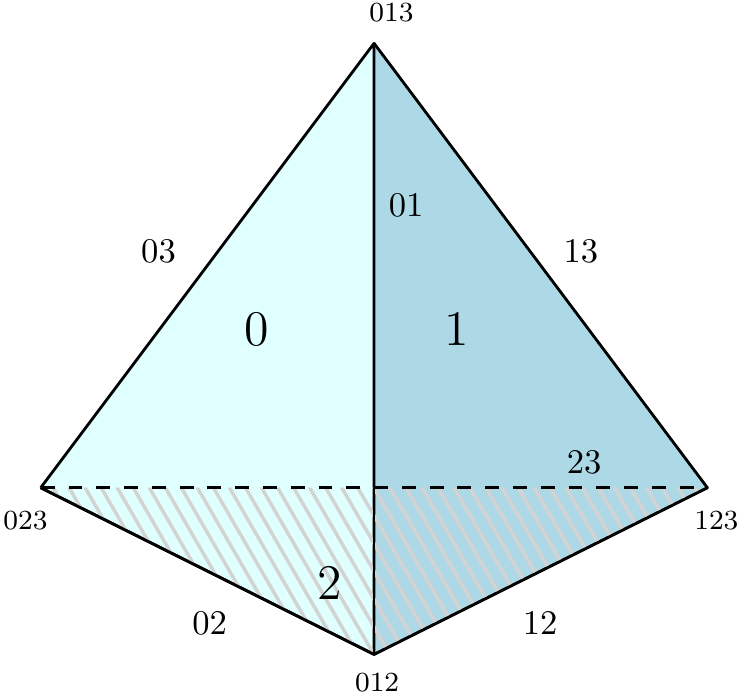}
\caption{A 3-simplex with colored facets. }
\label{fig:ColSimp}
\end{figure}
Throughout this thesis, we consider that edges have unit length, and that $D$-dimensional simplices have colored $(D-1)$-simplices, also called  \emph{facets}.  In practice, we represent the coloring of facets by arbitrarily indexing the colors from 0 to $D$. Simplices are glued together along facets of the same color. Besides avoiding singularities (such as self-gluings of a simplex), the major motivation for considering colored facets is that we can specify the attachment map so that the gluing of two facets is done in a unique way. More precisely, every $(D-k)$-simplex inherits the colors of the $k$ facets it belongs to. For instance, an edge incident to two facets respectively of color 1 and 2 will carry both colors 1 and 2. To each set of distinct colors $i_1,\cdots, i_k$ corresponds a unique $(D-k)$-simplex. We therefore require that the gluing of two $D$-simplices along facets of color 1 is done identifying the sub-simplices that have the same set of colors, as shown in Fig.~\ref{fig:ColGlu}. Again, this is done in a unique way.
\begin{figure}[!h]
\centering
\includegraphics[scale=0.9]{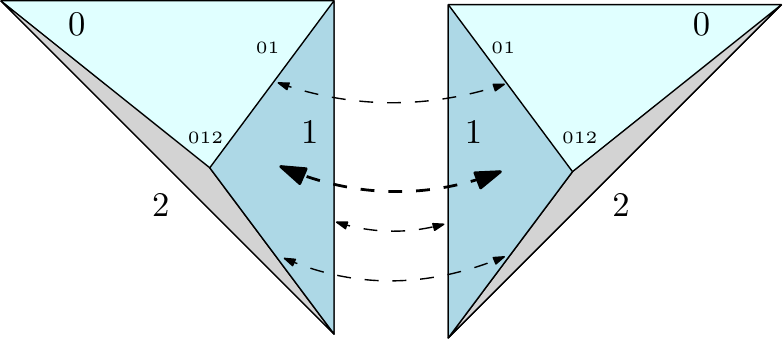}
\caption{Simplices are glued along colored facets in a unique way. }
\label{fig:ColGlu}
\end{figure}
The resulting discrete space is a simplicial pseudo-complex.
\begin{definition}
\label{def:PseudoComp}
A dimension $D$ simplicial pseudo-complex $X$ is a set of $D$-simplices satisfying
\begin{itemize}
\item Any $(D-k)$-simplex of $X$ is also in $X$.
\item The intersection of two $D$-simplices of $X$ is a subset of their subsimplices.
\end{itemize}

\end{definition}

A simplicial complex is such that two distinct $D$-simplices can at most share one $(D-k)$-simplex and its subsimplices: there is less freedom in how the simplices can be glued together. We do not make this stronger requirement. Throughout this thesis, we will sometimes refer to pseudo-complexes as {\it triangulations}. 
We stress however that this is somehow a conflictual denomination with that of generalized $p$-angulations in Section~\ref{sec:pAng}.
Topologically, the pseudo-complex  obtained by gluing a collection of simplices along all their facets is a pseudo-manifold. Intuitively, a pseudo-manifold is almost a manifold, apart for a certain number of singularities. More precisely,
\begin{definition}[Pseudo-manifold]
A topological space $X$ with triangulation $\cC$ is a $D$-dimensional pseudo-manifold if 
\begin{itemize}
\item $X$ is the union of all $D$-simplices
\item the facets belong to precisely two $D$-simplices
\item for any two $D$-simplices $\sigma$ and $\sigma'$ of $\cC$, there is a sequence $\sigma=\sigma_0, \sigma_1,\cdots,\sigma_p=\sigma'$ such that $\forall i\in\llbracket 0,p-1\rrbracket$, $\sigma_i\cap \sigma_{i+1}$ is a $(D-1)$-simplex.
\end{itemize}
\end{definition}

%
From a discrete pseudo-manifold, one can always build a colored triangulation by taking its barycentric subdivision, which is always colored.
It is obtained by adding a vertex in every sub-simplex at the barycentre of the 0-simplices, and joining all the newly added vertices.
Therefore, as long as a pseudo-manifold possesses a discretization, it should also possess a colored triangulation.

By gluing only a subset of the facets of the simplices, one obtains a colored triangulation of a pseudo-manifold with boundaries, which are lower dimensional pseudo-manifolds themselves. 
The discretization induces colored triangulations of the boundaries. In Figure~\ref{fig:Octa1}, we have represented a 3-dimensional triangulation of a ball with a connected spherical boundary of color 0. 
\begin{figure}[!h]
\centering
\includegraphics[scale=0.7]{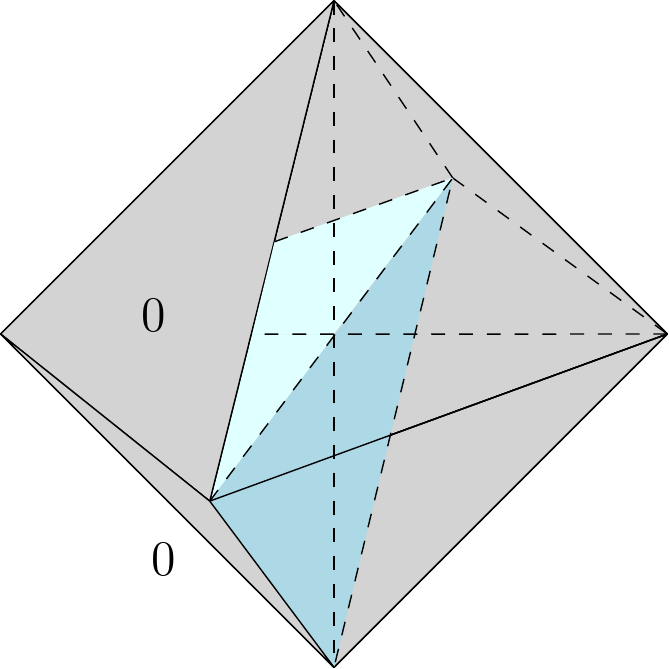}
\caption{A triangulated ball in 3 dimensions. }
\label{fig:Octa1}
\end{figure}
The boundary inherits a triangulation such that the color set of every sub-simplex contains color 0. By considering all the colors but 0, we obtain a planar colored triangulation of the boundary.

\section{Edge-colored graphs}
\label{sec:EdgeCol}
	\subsection{Graph encoded manifolds}
	\label{subsec:GEM}
	
We represent a $D$-simplex by a $(D+1)$-valent vertex. An edge is dual to a facet of the simplex, and carries the corresponding color. 
Because the gluing of two simplices along facets of the same color $i$ is done in a unique way, we can just represent this gluing by identifying the two half-edges of color $i$ incident to each vertex. This is pictured in Figure~\ref{fig:DualGluing}. 
\begin{figure}[!h]
\centering
\includegraphics[scale=0.8]{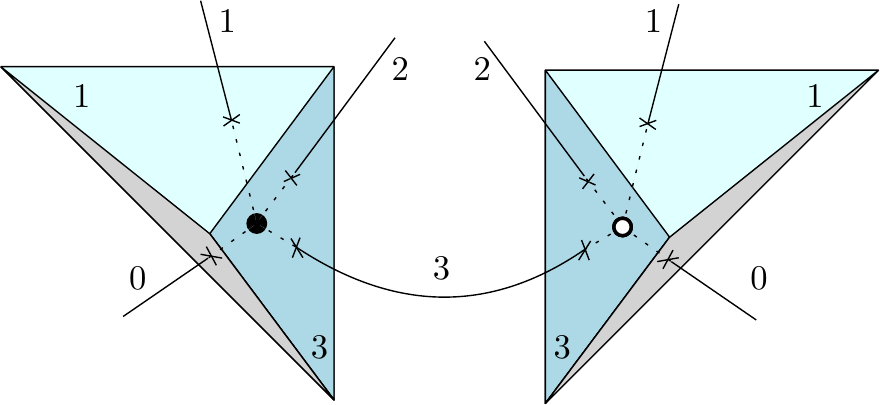}
\caption{An edge of color 3 encodes the identification of the dual facets. }
\label{fig:DualGluing}
\end{figure}
 It is the 1-skeleton of the cellular dual of the triangulation, but we will refer to it as the dual colored graph\footnote{Note that in the two dimensional case, the trivalent combinatorial map dual to the embedded triangulation is referred to as the dual {\it map}, its underlying graph being the dual colored graph.}. We say that the graph represents the corresponding pseudo-manifold.
  In the crystallization literature (see Section~\ref{subsec:Cryst} and references therein),
 it is referred to as {\it graph encoded manifold} (GEM). The corresponding graph is such that every vertex has valency $D+1$, and an edge of each color is incident to each vertex once, and only once (it is said to have a proper $(D+1)$-edge-coloring). Examples in $D=2$ are shown in Fig.~\ref{fig:Ex2D}. A $D=3$ example is shown in Fig.~\ref{fig:Melo} and both orientable and non-orientable examples in $D=4$ are shown in Fig.~\ref{fig:Handle}.
\begin{definition}
\label{def:cG}
We define $\bG_D$ as the set of connected $(D+1)$-regular properly edge-colored bipartite graphs with color set $\{0,\cdots,D\}$.
We denote $\tilde \bG_D$ 
the set obtained by dropping the bipartiteness condition. We denote $\bG_D^U$ and $\tilde \bG_D^U$ the sets obtained by dropping the connectivity. 
\end{definition}

Every pseudo-manifold has a colored triangulation, and this triangulation can be encoded into an edge-colored graph (the neighborhood of vertices has to be strongly connected, if not the triangulation cannot be reconstructed from the colored graph).  This condition is satisfied in  the case of singular-manifolds, which are such that the links of the vertices are piecewise-linear manifolds. We refer the reader to the beginning of Subsection~\ref{subsec:Cryst} for the definitions of piecewise-linear and singular manifolds.
\begin{prop}[Casali, Cristofori, Grasselli, 2017 \cite{TopoTensor2}]
\label{prop:Pezz1}
In any dimension, any singular-manifold admits a colored triangulation, and a $(D+1)$-edge-colored graph representing it.
\end{prop}

Our assumption that every edge has the same length induces notions of distance and curvature in the triangulation. The colored graph dual to a triangulation contains all the information on its induced geometry.
The $D-k$ sub-simplices carry a set of $k$ colors and are identified in the dual colored graph by $k$-edge-colored subgraphs. In particular, $(D-2)$-simplices are cycles in the graph, which alternate edges of two different colors, $i$ and $j$. We call such a cycle a \emph{color-$ij$ cycle}, and in general a \emph{bicolored cycle}. We have the following correspondences :
\be
\label{Dictionnary}
{\renewcommand{\arraystretch}{1.2}
\qquad
\begin{tabular}{rcl}
{\bf triangulation} &  & {\bf dual graph}\\
$D$-simplex & $\leftrightarrow$ & vertex\\
facet& $\leftrightarrow$ & edge\\
$(D-2)$-simplex & $\leftrightarrow$ & bicolored cycle\\
$(D-k)$-simplex & $\leftrightarrow$ &  subgraph   in $\bG_{k-1}$\\
\end{tabular}}
\ee


If the triangulation discretizes an orientable pseudo-manifold, one can choose a local orientation of each simplex which will translate to a global orientation of the manifold. More precisely, we have the following classical proposition \cite{ItalianSurvey}.
\begin{prop}[Orientability]
\label{prop:Orient2}
A  pseudo-manifold is orientable if and only if its dual colored graph is bipartite.
\end{prop}

\subsubsection*{Two dimensions}

In two dimensions, if $\G\in\bG_2$ is a colored graph dual to a triangulation $\cC$, then it is the 1-skeleton of the map dual to $\C$. The simpler colored triangulations of the 2-sphere, of the torus and of the real projective plane are represented in Figure~\ref{fig:Ex2D}.
\begin{figure}[!h]
\centering
\raisebox{5.3ex}{\includegraphics[scale=0.8]{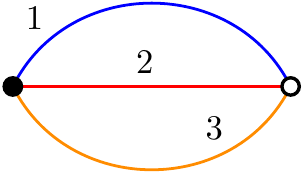}}
\hspace{1.5cm}
\includegraphics[scale=0.75]{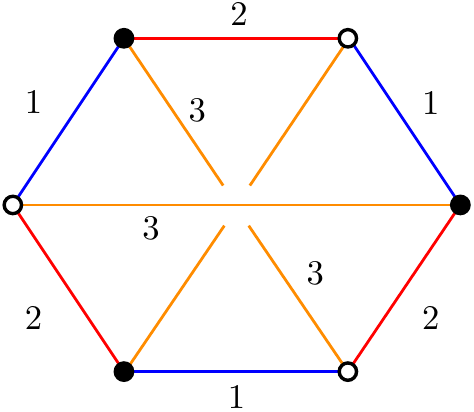}
\hspace{1.5cm}
\raisebox{1.2ex}{\includegraphics[scale=0.75]{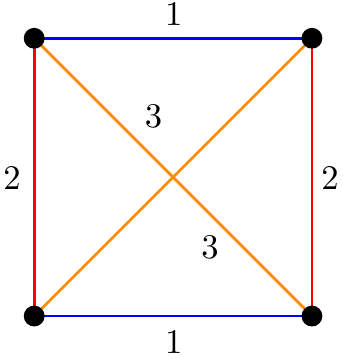}}
\caption{2D triangulations of the sphere, the torus and the real projective plane. }
\label{fig:Ex2D}
\end{figure}
We can verify this right away by computing the genus of the corresponding triangulations.  Considering a triangulation $\C$ and its colored dual graph $\G$, from the dictionary \eqref{Dictionnary}, its genus $g$ writes 
\be
2-2g(\C)=F(\G)-E(\G)+V(\G)
\ee
where we have denoted $F$ the number of bicolored cycles of $\G$ (also the number of faces of the trivalent map dual to $\C$), and $E$ and $V$ its number of edges and vertices. We respectively find genera 0, 1 and 1/2 for the examples of Figure~\ref{fig:Ex2D}.
\begin{figure}[!h]
\centering
\raisebox{0.9ex}{\includegraphics[scale=0.8]{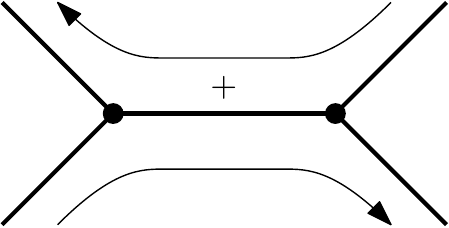}}
\hspace{3cm}
\includegraphics[scale=0.8]{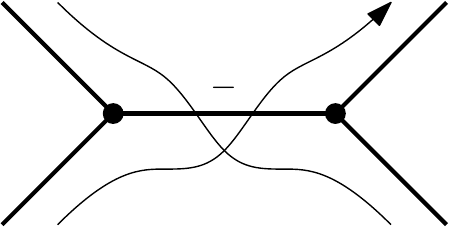}
\caption{Locally orientable maps can have twist edges. }
\label{fig:LOMaps}
\end{figure}

The orientability condition can be visualized in two dimensions. Triangles with colored edges are glued without specifying an ordering of the colors around their boundary. One may however choose a local orientation by embedding every triangle on the sphere, therefore specifying an ordering of colors around the boundaries, clockwise or counter-clockwise. The locally orientable combinatorial map dual to the triangulation therefore has two kinds of edges, which represent the attachment maps: the usual combinatorial map edges, carrying a $(+)$~label, and twisted edges, reversing the faces and carrying a $(-)$~label, also called twist factor. The way the faces behave is shown in Figure~\ref{fig:LOMaps}. To define locally orientable maps with permutations, we need two blades per half-edge, but we refer the reader to \cite{Atlas, GraphsOnSurfaces}. The gluing of two triangles with the same (resp. opposite) orientation is represented by a $(-)$ edge (resp. $(+)$ edge). An example of a colored triangulation together with a choice of orientation for each triangle, and the corresponding dual map is shown in Fig.~\ref{fig:Orient}.
%
%

\begin{figure}[!h]
\centering
\raisebox{0.9ex}{\includegraphics[scale=0.8]{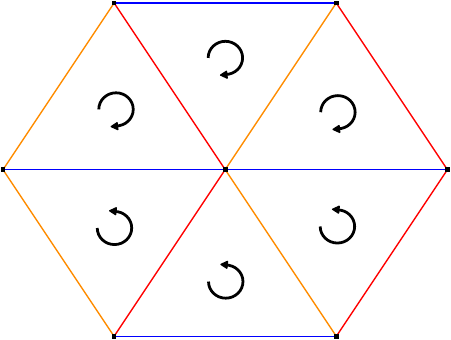}}
\hspace{1cm}
\includegraphics[scale=0.75]{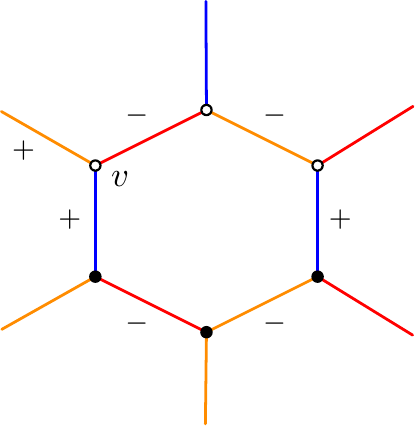}
\hspace{1cm}
\includegraphics[scale=0.75]{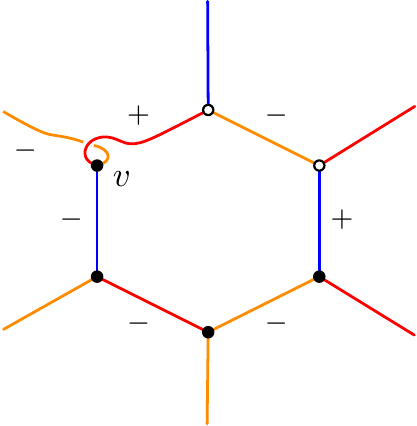}
\caption{A choice of orientation, the corresponding dual map, and a local change of orientation at vertex $v$. }
\label{fig:Orient}
\end{figure}

\begin{definition}[Local change of orientation]
\label{def:LocChange}
 A local change of orientation at a vertex is done by inverting the ordering of the edges around that vertex and exchanging the $(+)$ signs of the incident edges by $(-)$ signs. 
 \end{definition}
It is a classical result (see e.g. \cite{GraphsOnSurfaces}, Lemma 4.1.4) that the corresponding surface is orientable if and only if all twists in the map can be changed to $(+)$ signs 
by a finite number of local changes of orientation.
In two dimensions, 
bipartiteness of the colored dual graph/map therefore  implies that the  triangulated surface is orientable. One may e.g. choose  to orient black vertices clockwise and white vertices counter-clockwise, so that every edge carries a $(+)$ sign. Equivalently, one may choose to give the same orientation to every triangle, so that every edge carries a $(-)$ sign, and then perform local changes on white vertices.  In dimension 3, a similar construction also leads to a natural oriented geometric realization.

\subsubsection*{Boundaries}


Throughout this work, we will be mainly interested in triangulations with boundaries such that unglued faces all have the same color, which we usually take to be 0. They are dual to edge-colored graphs such that a certain number of white/black vertices do not have an incident color-0 edge, and therefore have valency $D$ instead of $D+1$. 
\begin{figure}[!h]
\centering
\includegraphics[scale=0.8]{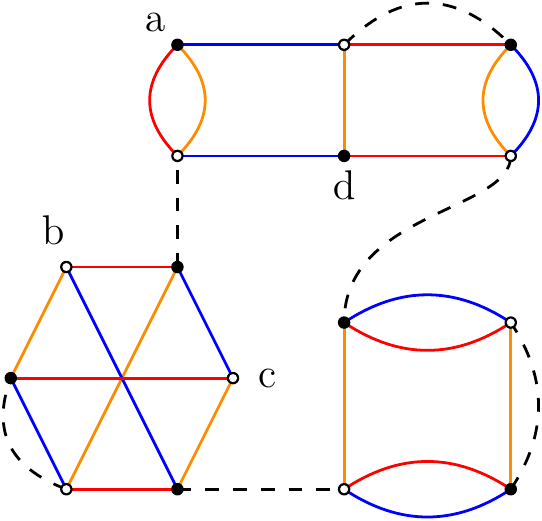}\hspace{3cm}
\raisebox{6ex}{\includegraphics[scale=0.8]{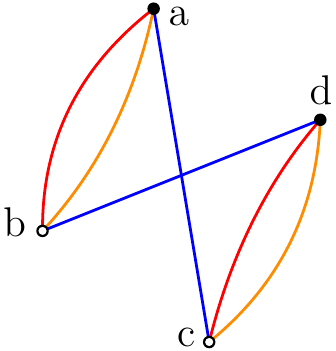}}
\caption{A colored graph in $\bG_D^2$ and its boundary graph. }
\label{fig:ExColBound}
\end{figure}
\begin{definition}
\label{def:GraphWithBound}
We denote $\bG_D^q$ the set of connected bipartite edge-colored graphs such that $q$ white vertices and $q$ black vertices have one incident edge for each color in $\lDr$, and the others vertices have one incident edge for each color in $\llbracket 0,D\rrbracket$.
\end{definition}
See the example in Fig.~\ref{fig:ExColBound}. Some paths alternating color 0 and some color $i\in\lDr$ begin on a degree-$D$ white vertex and end on a degree-$D$ black vertex. Let us define the boundary graph:
\begin{definition}[Boundary graph]
\label{eqref:Boundary}
Given a colored graph $\G\in\bG_D^q$, the boundary graph $\partial\G$ has all the degree-$D$ vertices of $\G$ and an edge of color $i$ between a black and a white vertex if there is a color $0i$ path between them in $\G$. 
\end{definition}
Remark that if there is only one degree-$D$ white vertex, the boundary graph is necessarily the elementary melon shown in Figure~\ref{fig:ElMel}. As there is only one way to add a vertex in order to obtain a graph in $\bG_D$, we can consider equivalently graphs in $\bG_D^2$ as graphs in $\bG_D$ with a distinguished edge. Another example of a graph and its boundary is shown in Fig.~\ref{fig:ExColBound}. We have the following known properties. 
%
\begin{prop}
If $\G\in\bG_D^q$, the connected components of the boundary graph $\partial\G$ belong to $\bG_{D-1}$. The triangulation dual to $\G$ induces a colored triangulation of its boundary, which dual graph is $\partial\G$.
\end{prop}
\begin{figure}[!h]
\centering
\includegraphics[scale=0.7]{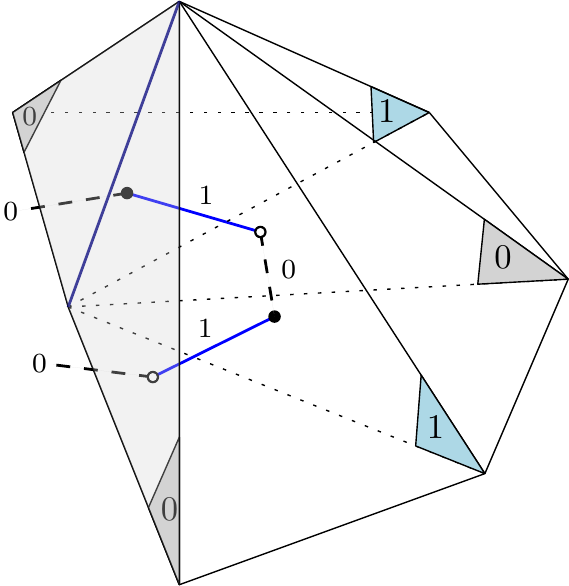}
\hspace{1.2cm}\raisebox{+12ex}{$\xrightarrow[]{\partial}$}\hspace{1.5cm}
\includegraphics[scale=0.7]{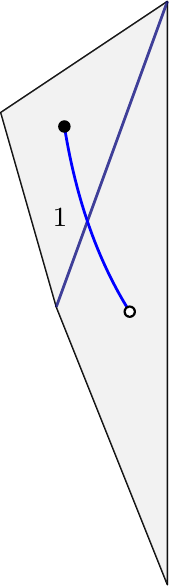}
\caption{A bicolored path between two pending color-0 half-edges dual to facets of the boundary gives a color-1 edge in the boundary graph. }
\label{fig:BrokenFace}
\end{figure}
In Figure~\ref{fig:BrokenFace}, we have represented a color-$01$ path between two pending half-edges dual to color-0 facets (shaded) of the boundary which have remained unglued (in our convention, we would not represent the pending half-edges, but we do so here for readability). As shown on the right, the boundary graph would have a color-1 edge between the two vertices incident to the pending color-0 half-edges (which would be degree-$D$ vertices in our usual convention).

	\subsection{Topology 
	}
	\label{subsec:Cryst}

	Given a colored graph $\G\in\bG_D$, the graph $\Gi$ is the graph obtained from $\G$ by deleting all the color-$i$ edges. The following result is classical (see e.g. \cite{TopoTensor1}, Prop.~3).
	
	\begin{prop}[PL-Manifolds]
	\label{prop:Manifolds}
	A colored graph $\G\in\bG_D$ represents
	a piecewise-linear manifold (PL-manifold) iff for every color $i$, the connected components of $\Gi$ are dual to triangulated spheres.
	In $D=3$, the Euler characteristics is positive and vanishes solely for manifolds
	\be
	\sum_{k=0}^4 n_k(\C)=0,
	\ee
	where $n_k(\C)$ is the number of $k$-simplices of $\cC$.
	\end{prop}
	
A colored graph represents a \emph{singular manifold} iff for every color $i$, the connected components of $\Gi$ represent $(D-1)$-dimensional  PL-manifolds.

	\begin{prop}[Regular embedding (jacket); Gagliardi, 1981 \cite{RegImb}]
	\label{prop:Jack}
	Given a colored graph $\G\in\tilde\bG_D$ and for every cyclic permutation $\mu\in\cS_{D+1}$, there exists an embedding $\G_\mu$ of $\G$ onto a surface $F_\mu$, such that the faces are bounded by edges of color $i$ and $\mu(i)$. The embedded graph is called a jacket, or a regular embedding of $\G$. $F_\mu$ is orientable if and only if $\G$ is bipartite.
	\end{prop}
	
Every graph  admits $n!/2$ regular embeddings. The regular genus of a graph is the smallest genus of the regular embeddings of the graph. In particular, if it vanishes, then the graph represents a $D$-sphere 
	
	
	\begin{prop}[Ferri, Gagliardi, 1982 \cite{OnlyGen0Man}]
	\label{prop:PlanJack}
	If a colored graph has a planar jacket, then it represents 
	a triangulated sphere. 
	\be
	g(\G_\mu)=0 \quad\Rightarrow \quad\C(\G) \cong \cS^D.
	\ee
	\end{prop}
	
The above condition is sufficient but not necessary. The regular genus of a manifold is the minimal regular genus of all the colored graphs representing that manifold. It was introduced by Gagliardi in \cite{RegGen}. In dimension two, the regular genus is simply the genus, and for 3-manifolds, the regular genus is the Heegaard genus if the manifold is orientable, and twice the Heegaard genus otherwise \cite{RegGen, RegHeeg}.
	
	\begin{figure}[!h]
\centering
\raisebox{+1ex}{\includegraphics[scale=1]{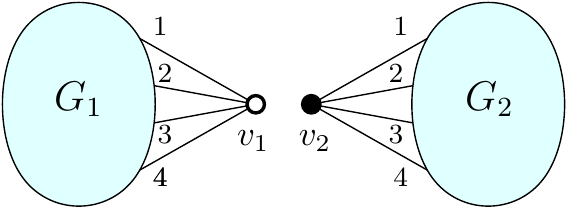}}
\hspace{1.7cm}
\includegraphics[scale=1]{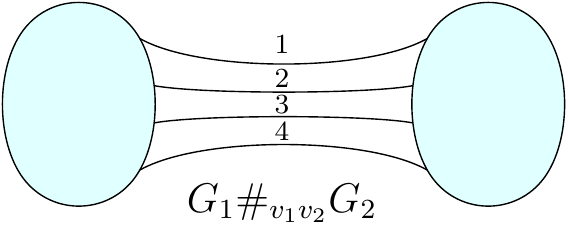}
\caption{Connected sum (1).}
\label{fig:DirSum1}
\end{figure}

\begin{prop}[Connected sum]
\label{prop:DirSum}
A graph connected sum $\G_1\#_{v_1v_2}\G_2$ of two graphs $\G_1$ and $\G_2$ representing two PL-manifolds $\cM_1$ and $\cM_2$ is obtained by deleting a vertex $v_1$ in $\G_1$, a vertex $v_2$ in  $\G_2$ and by reconnecting the pending half-edges in the unique possible way (Fig.~\ref{fig:DirSum1}). The resulting graph represents a connected sum of $\cM_1$ and $\cM_2$. If the two graphs are bipartite, then topologically there is a unique connected sum of $\cM_1$ and $\cM_2$, denoted $\cM_1\#\cM_2$, and 
\be
\forall\  v_1^\bullet \in \G_1 \text{ and } v_2^\circ \in \G_2,\quad  \G_1\#_{v_1^\bullet v_2^\circ}\G_2\ \text{ represents }\ \cM_1\#\cM_2.
\ee
This property extends to the case of pseudo-manifolds if in every $\Gi_1$ and $\Gi_2$, $v_1$ and $v_2$ belong to spheres. Furthermore, 
\be
\delta(\G_1)+\delta(\G_2) = \delta(\G_1\#_{v_1v_2}\G_2).
\ee
\end{prop}

The connected sums of triangulations with boundaries can also be done. In this case, in every $\Gi_1$ and $\Gi_2$, $v_1$ and $v_2$ must belong to spheres, or balls if they are incident to the boundaries. In the case where $v_1$ and $v_2$ are incident to connected components $\partial\G_1^a$ and $\partial\G_2^b$ of the boundaries, then one of the connected components of the boundary of the graph connected sum $\G_1\#_{v_1v_2}\G_2$ is the graph connected sum $\partial\G_1^a\#_{v_1v_2}\partial\G_2^b$. If the boundaries are connected, 
\be
\partial(\G_1\#_{v_1v_2}\G_2)=\partial(\G_1)\#_{v_1v_2}\partial(\G_2).
\ee
Furthermore, if the colored graphs $\G_1$ and $\G_2$ are bipartite, then $\G_1\#_{v_1v_2}\G_2$ is too.


	\subsubsection{Moves}
	\label{subsec:Moves}
	
We first define $h$-pairs
\begin{definition}[$h$-pair]	
\label{def:hPair}
An $h$-pair is a pair of vertices linked by $h$ parallel edges.
\end{definition}

We will be interested in the following chapters in the influence on the local curvature of the insertion or contraction of pairs, as pictured in Figure~\ref{fig:PairIns}. Topologically, an additional condition is important.
	
\begin{figure}[!h]
\centering
\includegraphics[scale=0.6]{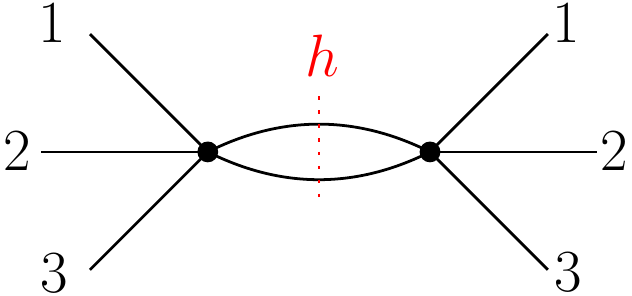}
\hspace{1cm}\raisebox{+5ex}{$ \leftrightarrow$}\hspace{1cm}
\includegraphics[scale=0.6]{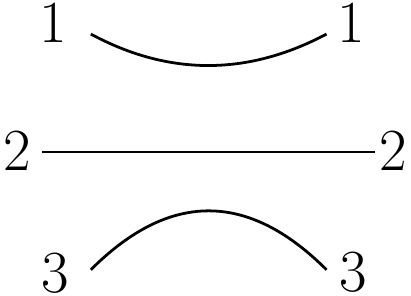}\hspace{0.5cm}
\caption{Pair insertion and contraction.}
\label{fig:PairIns}
\end{figure}
	
\begin{definition}[$h$-dipole]
\label{def:Dipole}
An $h$-dipole of a colored graph dual to a manifold is an $h$-uple of parallel edges of colors $I=\{i_1,\cdots,i_h\}$ between two vertices $v_1$ and $v_2$, such that 
in the graph $\GI$ obtained by deleting all the edges with colors in $I$, $v_1$ and $v_2$ belong to two different connected components.
The dipole is said to be proper, if at least one of the connected components of $\GI$ containing $v_1$ or $v_2$ is dual to a $(D-h)$-sphere.
\end{definition}

Conversely, the condition for a dipole insertion is that the $D-h+1$ considered edges which should all have different colors must be in the same connected component of $\GI$.

\begin{theorem}[Gagliardi, 1987 \cite{GagliaProper}]
\label{thm:GagliaProper}
Two colored graphs that are obtained one from another by a finite sequence of proper dipole insertions and contractions represent the same (pseudo)-manifold.
\end{theorem}

\begin{figure}[!h]
\centering
\includegraphics[scale=0.85]{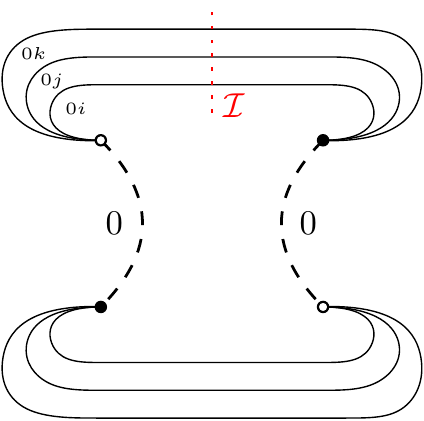}
\hspace{1cm}\raisebox{+9.5ex}{$ \longrightarrow$}\hspace{1cm}
\includegraphics[scale=0.85]{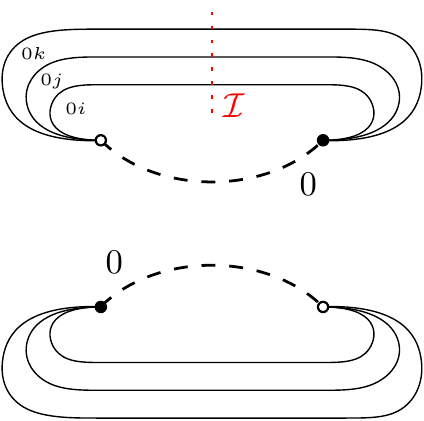}
\caption{$\rho_\cI$-pair switching.}
\label{fig:RhoISwitch}
\end{figure}

Considering two edges of the same color in a colored bipartite graph, there is a unique way to exchange them while preserving bipartiteness. If the graph is not bipartite, there are two ways to do so. We focus on the bipartite case.

\begin{definition}[$\rho$-pair switching]
\label{def:RhoHSwitch}
A $\rho_\cI$-pair switching in a graph $\G\in\bG_D$ is the exchange of two edges of the same color $i$ which originally belong to precisely $\cI$ common bicolored cycles. 
\end{definition}


\begin{prop}
\label{prop:DirSum2}
If a $\rho_0$-pair switching disconnects the graph $\G$, the original graph represents the connected sum of the PL-manifolds represented by $\G_1$ and $\G_2$. 
\end{prop}

\prf The operation decomposes into a series of $D$-dipole insertions and contractions and a connected sum, as defined in Prop.~\ref{prop:DirSum}. This sequence is shown in Figure~\ref{fig:DirSum2}. \qed 

\begin{figure}[!h]
\centering
\includegraphics[scale=0.9]{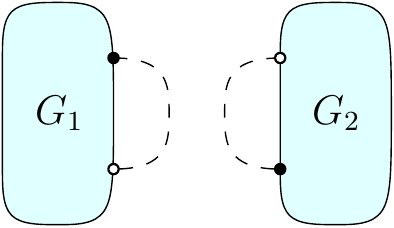}\ \ 
\includegraphics[scale=0.9]{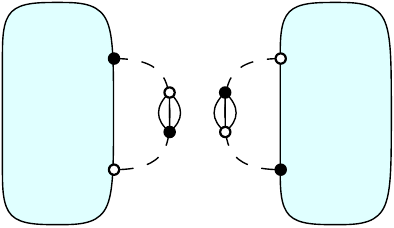}\ \ 
\includegraphics[scale=0.9]{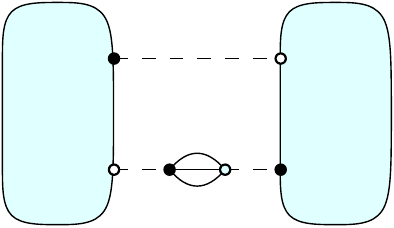}\ \ 
\includegraphics[scale=0.9]{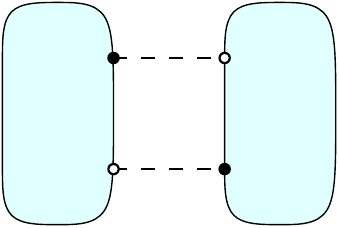}
\caption{Connected sum (2).}
\label{fig:DirSum2}
\end{figure}

It extends to pseudo-manifolds upon the same conditions as in Prop.~\ref{prop:DirSum}.
A flip is a particular kind of $\rho$-pair switching.

\begin{definition}[Flip]
\label{def:Flip}
A flip in a graph $\G\in\bG_D$  is the exchange of two edges of the same color incident to the two vertices of a proper
 $h$-dipole with $h\in\llbracket 1, D-1\rrbracket$. It is illustrated in Fig.~\ref{fig:Flip}.
\end{definition}

\begin{theorem}[Lins, Mulazzani, 2006 \cite{Flips}]
\label{prop:FlipBlop}
Two colored graphs in $\bG_D$ that are obtained one from another by a finite sequence of $D$-dipole insertions and flips are dual to the same PL-manifold.
\end{theorem}

\begin{figure}[!h]
\centering
\includegraphics[scale=1.1]{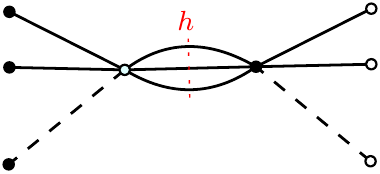}
\hspace{2cm}
\includegraphics[scale=1.1]{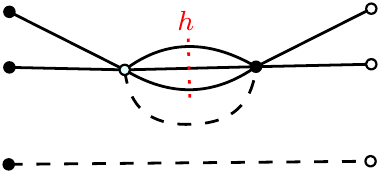}
\caption{A flip between two
edges incident to a $h$-dipole.}
\label{fig:Flip}
\end{figure}

\begin{definition}[Combinatorial handle]
\label{def:CombHandle}
A combinatorial handle is a $(D-1)$-pair such that the four edges not in the pair belong to the same bicolored cycle.
\end{definition}

\begin{theorem}[Gagliardi, Volzone, 1987 \cite{Handle}\footnote{Similar results were proven in 1982 by Gagliardi for the 3-dimensional case, ``Recognizing a 3-dimensional handle among 4-coloured graphs," Ricerche Mat. {\bf31} (1982),
389-404.}]
\label{thm:CombHandle}
If $\G\in\bG_D$ represents a PL-manifold $M$, and If $\G'$ is obtained from $\G$ by contraction of  a combinatorial handle, then
\begin{itemize}
\item If $\G'$ is connected, it represents a PL-manifold $N$ such that $M\cong_{PL} N\#\bH$
\item If the contraction separates the graph into two connected components which represent two Pl-manifolds $N_1$ and $N_2$, $\G'=\G'_1\sqcup\G_2'$, then $M\cong_{PL} N_1\#N_2$,
\end{itemize}
where $\bH$ is one of the two PL-manifolds represented by the colored graphs in Fig.~\ref{fig:Handle} (or $D$-dimensional generalization). 
\end{theorem}
\begin{figure}[!h]
\centering
\includegraphics[scale=0.45]{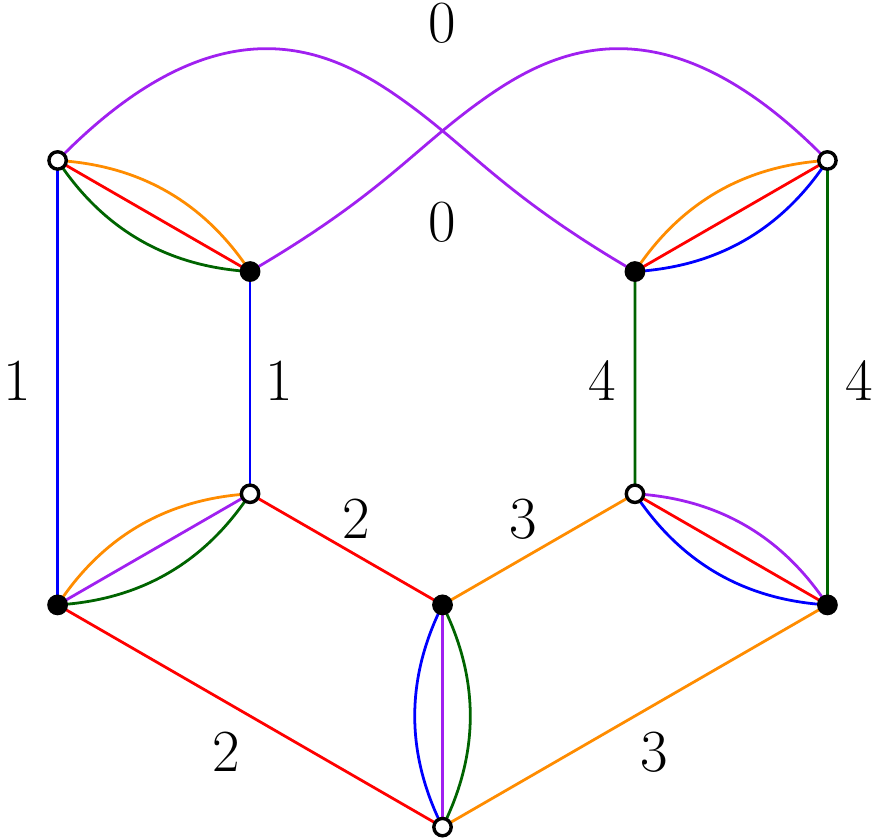}
\hspace{2cm}
\includegraphics[scale=0.45]{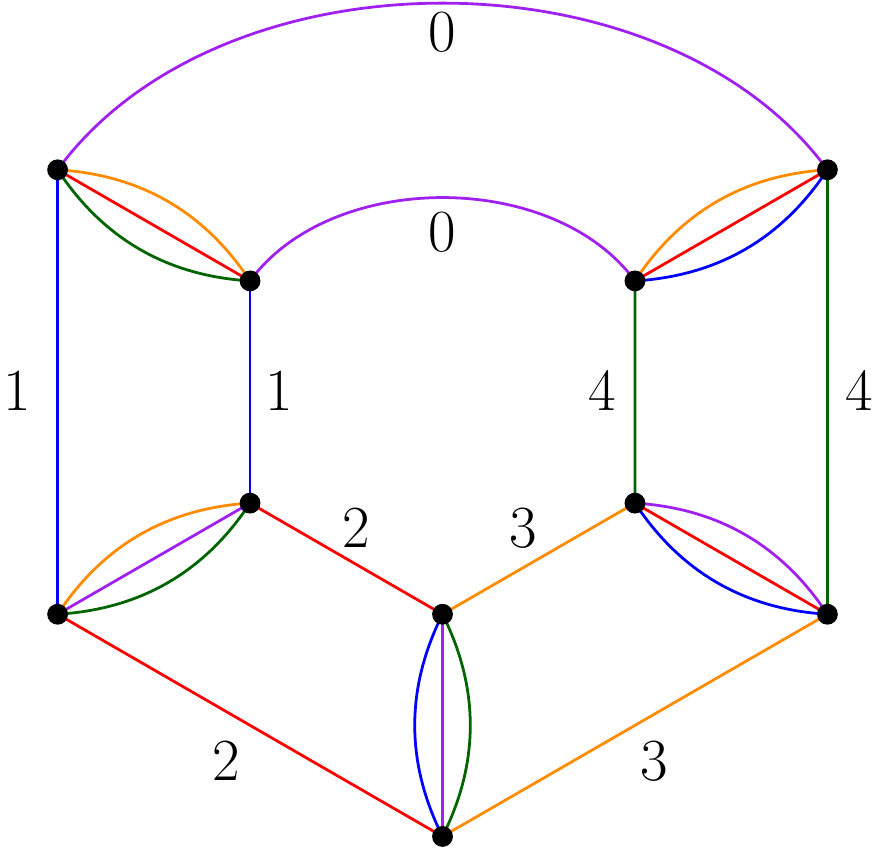}
\caption{Examples of $D$-dimensional handles (orientable $\cS^{D-1}\times\cS^1$ and non-orientable $\cS^{D-1}\tilde\times\cS^1$).}
\label{fig:Handle}
\end{figure}

	\subsection{Gurau's degree}
	\label{subsec:GurDeg}
	
	A cycle is a connected subgraph that only has degree two vertices. In a colored graph $\G\in\bG$, the subgraph containing all the vertices, and the edges of color $i$ and $j\neq i$ is a collection of disjoint bicolored cycles. 
	
	\begin{definition}[Score]
	\label{def:BiC}
	The number of bicolored cycles of a graph $\G\in\bG$ that alternate edges of colors $i$ and $j$
	is denoted $\Phi_{i,j}(\G)$. The score $\Phi$ of a colored graph is defined as its total number of 	bicolored cycles
	\be
	\label{eqref:score}
	\Phi(\G)=\sum_{i<j} \Phi_{i,j}(\G)
	\ee
	If $\G$ is the colored graph dual to a triangulation $\C$, the $(D-2)$-simplices of $\C$ are dual to bicolored cycles of $\G$, and in particular 
	\be
	n_{D-2}(\C)=\Phi(\G)
	\ee 
	\end{definition}
	We stress that bicolored cycles are generally called \textit{faces} in the random tensor literature, and $\Phi$, the number of faces. The reason is that in dimension 2, they coincide with the faces of the trivalent map dual to the triangulation. We have chosen to use \textit{bicolored cycle} and \textit{score} instead, not to be confused with the \textit{facets} of triangulations or the \textit{faces} of combinatorial maps.


	\begin{definition}[Gurau degree] 
	\label{def:Deg}
	The degree of a connected colored $D$-dimensional triangulation $\C$  is defined as
	\be
	\label{eqref:Deg}
	\deltaG(\C)=D + \frac{D(D-1)}4 n_D(\C) - n_{D-2}(\C),
	\ee
	in which $n_D$ (resp. $n_{D-2}$) is the number of $D$-simplices (resp. $(D-2)$-simplices).
	The degree of the dual colored graph $\G\in\cG_D$ is 
	\be
	\deltaG(\G)=D + \frac{D(D-1)}4 V(\G) - \Phi(\G).
	\ee
	\end{definition}
	
	
	
	\begin{theorem}[Gurau, 2011 \cite{Gurau1NCol, Complte1N}] 
	\label{thm:Gurau}
	The degree rewrites in terms of the genera of the jackets, and is therefore a positive or vanishing integer
	\be
	\deltaG(\G)=\frac 2 {(D-1)!} \sum_{\cJ} g(\G_\cJ)\ge 0,
	\ee
	the jackets $\G_\cJ$ being the embedded graphs defined in Prop.~\ref{prop:Jack}.
	\end{theorem}
	
	
	
	Remark that the degree as defined here is usually referred to as the {\it reduced} degree, the degree being commonly defined as 
	\be
	\omega_\text{Gur}(\G)= \frac{(D-1)!}2 \deltaG(\G).
	\ee

\begin{prop}[Casali, Cristofori, Dartois, Grasselli, 2017  \cite{TopoTensor1}]
Every bipartite graph $\G\in\bG_D$ with Gurau degree $\deltaG(\G)<D$ represents the sphere.
\end{prop}	

\begin{prop}[Bonzom, Lionni, Tanasa, 2017  \cite{BLT}]
No non-bipartite graph $\G\in\tilde\bG_D$ exists with Gurau degree $\deltaG(\G)<D-1$.
\end{prop}

It can also be seen from \cite{GurauSchaeffer} and Prop.~\ref{prop:FlipBlop} that graphs with degree $D-2$  represent the sphere $\cS^D$. Furthermore, it is not complicated to extend the proof of \cite{BLT} to show that graphs with degree $D-1$ represent non-orientable $D$-dimensional handles $\bS^{D-1}\tilde\times\bS^1$, such as on the graph on the right of Figure~\ref{fig:Handle}.

Considering a triangulation with boundary, we will be interested in both the degree of its boundary and the degree of its interior. We focus on the case where all missing edges of the colored dual graph are of color 0. It is therefore obtained from a $(D+1)$-colored graph by deleting some color-0 edges. In the case where there are no remaining edge of color 0, only a (connected) $D$-colored graph is left, which is interpreted as a $(D+1)$-colored graph with no color-0 edge. It is therefore  dual to a triangulation in which all the color-0 facets remain unglued. See the dedicated section~\ref{subsec:Bubbles}. 
%
 The degree of the boundary of $\G$ is the degree of the boundary graph $\partial \G$, which is a non-necessarily connected $D$-colored graph,
\be
\deltaG(\partial \G)=(D-1)K(\partial \G)+\frac{(D-1)(D-2)}4V(\partial \G) - \Phi(\partial \G),
\ee
where $K$ is the number of connected components of $\partial \G$, and $\Phi(\partial \G)$ is the score of the boundary graph. The degree of the interior of $\G$ is its degree as defined in Def.~\ref{def:Deg}, with the difference that the score only counts bicolored cycles (i.e. it does not count the bicolored open paths, dual to $(D-2)$-simplices on the boundary).


	\subsection{Melonic graphs}
	\label{subsec:Melonic}
	
	We focus on a specific family of series-parallel colored graphs called {\it melonic},  defined as the recursive insertion of $D$-dipoles on the elementary melon, which is the only graph in $\bG_D$ with two vertices (Fig.~\ref{fig:ElMel}).
	\begin{figure}
	\centering
	\includegraphics[scale=0.9]{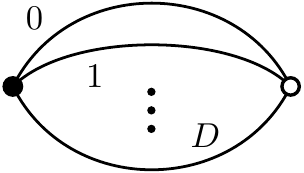}
\caption{An elementary melon in $\bG_D$. }
\label{fig:ElMel}
	\end{figure}
	
	We recall that $h$-dipoles were defined in Def.~\ref{def:Dipole}.  A $D$-pair is always a $D$-dipole. A $D$-dipole insertion is the operation illustrated below in Figure \ref{fig:DDipIns}.
	 \begin{figure}[h!]
	\centering
	\raisebox{3.7ex}{\includegraphics[scale=0.8]{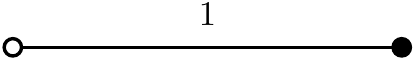}}\hspace{1cm} \raisebox{3.8ex}{$\longrightarrow$ }\hspace{1cm}\includegraphics[scale=0.8]{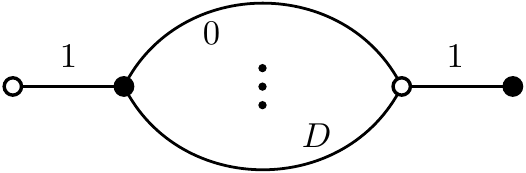}
\caption{A $D$ dipole insertion on a color 1 edge. }
\label{fig:DDipIns}
	\end{figure}

	\begin{definition}[Melonic graph]  Melonic graphs are obtained by recursive $D$-dipole insertions on the elementary melon. 	
	\end{definition}
	An example of melonic graph is shown on the left of Figure~\ref{fig:Melo}. We have the following classical result.
	\begin{theorem} [Bonzom, Gurau, Riello, Rivasseau, 2011 \cite{CritBehavior}]
	\label{thm:Melo}
	Graphs of vanishing degree are the melonic ones. 
	\end{theorem}
	
	%
	\begin{coroll}
	Melonic graphs are dual to triangulated spheres.
	\end{coroll}	
	\prf From Thm.~\ref{thm:Gurau}, graphs of vanishing degree only have planar jackets (Prop.~\ref{prop:Jack}), and we conclude with Prop.~\ref{prop:PlanJack}. \qed

	\begin{figure}[h!]
	\centering
	\raisebox{1ex}{\includegraphics[scale=0.6]{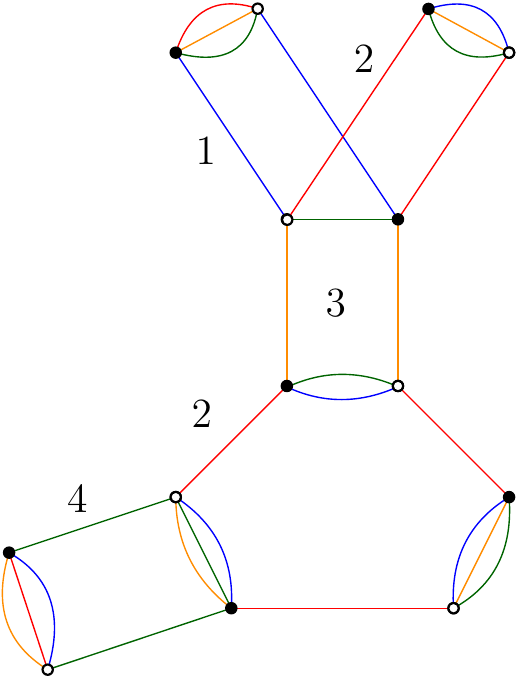}}\hspace{2cm}\includegraphics[scale=0.6]{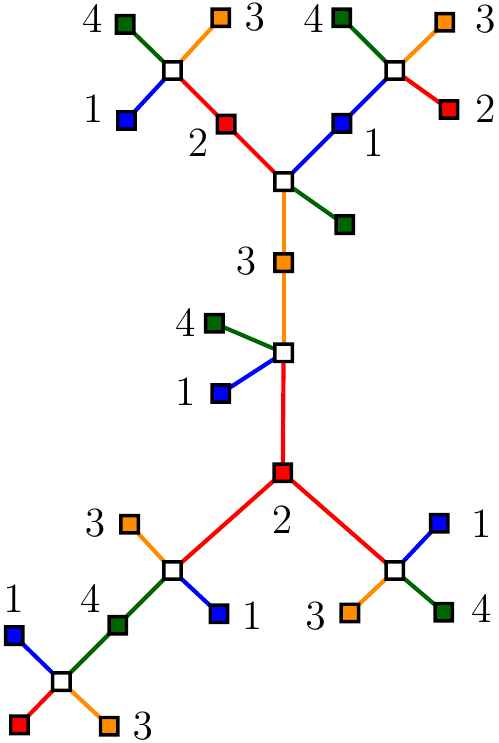}
\caption{A melonic graph and the corresponding bipartite plane tree. }
\label{fig:Melo}
	\end{figure}

	
	\begin{definition}[Canonical pairing] 	
	\label{def:CanoPairing}
	The canonical pairing of a melonic graph is defined as the (unordered) set of pairs of black and white vertices corresponding to the recursive $D$-dipole insertions. They are called canonical pairs and there is a unique such set\footnote{We do not prove this statement  here, it follows from the recursive structure. However, it is a consequence of the more general result of Lemma~\ref{lemma:TreesImplieOpt}.}. \end{definition}

	A cycle that alternates $k$ edges of color $i$ and $k$ canonical pairs is  completely separating, if when deleting those $k$ edges the number of connected components of the graph increases by $k-1$.
	Completely separating cycles and vertices of trees
	 are characterized the same way. The following proposition indicates that melonic graphs have a tree-like structure.
	 
	 \begin{prop} 
	 \label{prop:Melo2}
	 A graph in $\bG$ is melonic if and only if every edge 
	 belongs to a completely separating cycle.
	 \end{prop}
\prf (sketch)
	 It is enough to show that the two half-edges of any color $i$ incident to a canonical pair either belong to the same edge, or form a 2-cut (the number of connected components increases when we delete the two edges). This canonical pair was inserted on some color-$j$ edge in the recursive $D$-dipole insertion. At that step, the two color-$j$ edges incident to the pair do form a 2-cut. The following dipole insertions do not change this property, as if some $D$-dipole are inserted on the edges linking the vertices of the pair, then the corresponding half-edges become 2-cuts. Conversely, if every edge belongs to a completely separating cycle, then the incidence relations between the pairs is tree-like. In particular, there exists a leaf, i.e. a $D$-dipole. Contracting it, one obtains a smaller graph with the same property. This way, one recursively recovers the elementary melon, which proves that the graph is melonic. \qed

	 \
	 
	(Full) $D$-ary trees are rooted $(D+1)$-valent plane trees. The $D$-ary trees we consider are properly edge-colored, with colors in $\llbracket 1, D+1\rrbracket$.
	 
	  \begin{prop}[Gurau, Ryan, 2014 \cite{MelonsAreBP}]
	  \label{prop:MeloBij1}
	  Melonic graphs with one marked color-0 edge are in bijection with properly edge-colored $D$-ary trees. 
	  \end{prop}
	 
	 We state here the following result, which is a consequence of the bijection we will prove in Section~\ref{sec:StackedMaps} and of Proposition~\ref{prop:Melo2}. It generalizes the bijection between binary trees and rooted plane trees, and encodes the bicolored cycles of the colored graph.

	 %
	   \begin{prop}
	   \label{prop:MeloBij2}
	   Melonic graphs in $\bG_D$ are in bijection with plane trees with vertices and edges colored in $\llbracket 1,D+1\rrbracket$, and white square vertices, such that 
	   \begin{itemize}
	   \item Edges link colored vertices to white vertices. Edges incident to color-$i$ vertices all have color $i$. 
	   \item White square vertices are of valency $D+1$ and incident edges all have different colors. The ordering of edges around white vertices is that of the colors.
	   \item The color-$ij$ cycles of a colored graph are mapped to the connected components of the submap obtained by keeping only the edges and vertices of color $i$ and $j$.
	   \end{itemize} 
	  \end{prop}
	  
	 %
	 %
	 
	 \
	 
	 The generating function 
	 of connected melonic graphs with one distinguished color-0 edge is
	 \be
	 \label{eqref:MeloGenFunct}
	 \GF
	 (z)=\sum_{\substack{{\G\in\bG}\\{\textrm{melonic}}}} z^{V(\G)/2}.
	 \ee 
	 It satisfies the following equation,
	 \be
	  \GF(z)=\sum_{k\ge0}z^k \GF(z)^{kD}=\frac1{1-z\GF(z)^D},
	  \ee
	 which is rewritten as 
	  \be
	  \label{eqref:GenFunMelo}
	  \GF(z)=1+z\GF(z)^{D+1}.
	  \ee
	 This is a consequence of the bijection between the $D$-ary trees of Prop.~\ref{prop:MeloBij1} and the objects described in Prop.~\ref{prop:MeloBij2}. The solutions to this equation are well known (see e.g. \cite{GoulJack} p.125), and it can be shown that the combinatorial solution (satisfying $\GF(0)=1$) has the expansion
\be
 \GF(z) = \sum_{k\ge0} C_k^{D+1} z^k, \quad\text{where}\quad C_k^{D+1}=\frac 1{(D+1)k+1}\binom{(D+1)k+1}{k},
\ee
the $C_k^{D+1}$ being Fuss-Catalan numbers. 
	  The generating function has a singularity at
	  \be
	  z_c=\frac{D^D}{(D+1)^{D+1}},
	  \ee
	  and the the coefficients $c_n$ behave asymptotically as
	  \be
	  c_n\sim \frac e {\sqrt{2\pi}} \frac {\sqrt{D+1}}{D^{3/2}} z_c^{-n}n^{-3/2}. 
	  \ee
	  The critical exponent $\gamma$ is defined for the first non-vanishing coefficient in the Puiseux expansion in $z-z_c$ of the generating function of rooted elements $\GF(z)$  corresponding to a non-integer exponent:
	   \be
	  \label{eqref:CritExp1}
	  \GF(z)= \sum_{i=0}^k a_i(z_c-z)^{n_i} +   (z_c-z)^{1- \gamma}  + o\bigl((z_c-z)^{1- \gamma} \bigr), \quad \forall i,\ n_i\in\bN,\text{ and }\gamma\in\bQ\setminus\bN.
	  \ee
	  The coefficients $a_i$ may be zero. A classical result (see for instance the famous book \cite{Flajolet}) is that the critical exponent (also called the string susceptibility in physics) can be deduced from the asymptotics of the coefficients,
	  \be
	  \label{eqref:CritExp2}
	  c_n\sim \alpha z_c^{-n}n^{\gamma-2}, 
	  \ee
in the one-root case. We find the corresponding critical exponent
	  \be
	  \gamma=\frac 1 2,
	  \ee
which is characteristic of families of trees, and the corresponding critical behavior for the generating function of rooted melonic graphs near the singularity is 
	  \be
	  \label{eqref:CritBehMelo}
	  \GF(z)= a + b \sqrt{z_c-z} + o(\sqrt{z_c-z}), 
	  \ee
	  with $a,b>0$. 	It was proven in \cite{CritBehavior} that the continuum limit of melonic graphs is the continuous random tree \cite{Aldous}. More precisely, melonic graphs are in bijection with the family of melonic D-balls  \cite{CritBehavior}, which converges uniformly in distribution in the Gromov- Hausdorff topology on compact metric spaces towards the continuum random tree (CRT).   It has Hausdorff dimension 2 and spectral dimension 4/3.
	  It is known in the physics literature as branched polymers. This continuous space gathers the properties of a continuum limit of one-dimensional discrete spaces. For instance, deleting any internal point (not a leaf) of the CRT, raises the number of connected components.  As briefly explained in the introduction, the CRT can therefore not be interpreted as a $D$-dimensional quantum space-time. We report the reader to Section~\ref {sec:QG} for more details on the link to quantum gravity. As the same conclusions were reached numerically for dynamical triangulations (simplices with no colors), it was therefore considered as a fact that because of the predominance of singular discrete space, this Euclidean discrete approach to quantum gravity failed in dimension three and higher. This conclusion however relies on the assumption that the space $\Sb$ introduced in the introduction is the full set of (colored) simplicial pseudo-complexes (Def.~\ref{def:PseudoComp}). We have mentioned in the introduction that by choosing other sets $\Sb$, it was possible to escape this universality class. The aim of this thesis is to develop combinatorial tools to explore possible choices for  $\Sb$. 
	  
	  Furthermore, melonic graphs have received a recent renewed interest in the context of holography and quantum black holes as they are the leading order Feynman graphs of the Sachdev-Ye-Kitaev model. We report the reader to Subsection~\ref{subsec:SYK} for more details.

\section{$p$-Angulations in higher dimension}
\label{sec:pAng}


	In two dimensions, combinatorial maps are obtained by gluing polygons, and $p$-angulations are degree-restricted maps obtained by gluing solely $p$-gons, i.e. discs with a discretized boundary of $p$ edges (1-simplices).  The natural generalization of combinatorial maps in higher dimensions are gluings of polytopes, in which context $p$-angulations generalize to gluings of $p$-topes, i.e. $D$-dimensional balls with a boundary made of $p$ facets. As before for simplicial pseudo-complexes, we will consider colored objects to avoid ambiguities and in order to work with the dual edge-colored graphs. We introduce two ways of doing so. The first one, described in the following section, leads to more singular gluings but is easier to work with and is the framework we will consider for the rest of this work. The second one is very general but will only be introduced shortly in Section \ref{subsec:ColPol}.



	\subsection{Building blocks : bubbles}
	\label{subsec:Bubbles}

\subsubsection{Two dimensions: bubbles and polygons}

In dimension two, a $p$-gon is a disc bounded by $p$ edges.
$p$-Angulations are obtained by taking a certain number of $p$-gons and identifying two by two all of the edges of their boundaries. 
By considering the star subdivision of the $p$-gons, the latter are triangulated, so that one can think of a $p$-angulation as a certain kind of triangulation, with a degree restriction on some vertices (see the top of Fig. \ref{fig:Hexa}). Because they avoid ambiguities and allow us to work with the dual colored graph, we wish to stay in the context of colored triangulations when going to higher dimensions. In an attempt to generalize this idea to higher dimensions, we will consider building blocks which are made up of $D$-simplices which all share one common vertex. These are the elementary building blocks which we will glue along facets of their boundaries, and are called \emph{bubbles}.  

	%
	 \begin{figure}[h!]
 \begin{eqnarray}
	\centering
	\Bc \ \includegraphics[scale=1]{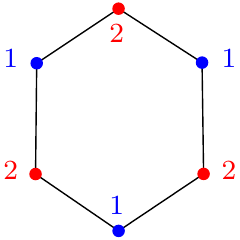}\ & \hspace{1cm}\raisebox{6ex}{\begin{tabular}{@{}c@{}} cone\\$\rightleftarrows$\\ boundary\end{tabular} }\hspace{0.5cm}&\quad\ \ \includegraphics[scale=1]{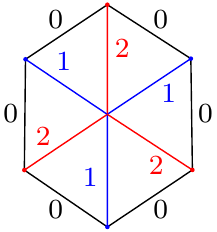}\ \BC  \nonumber\\[+1ex]
	\textrm{duality}\updownarrow  \hspace{1.3cm}&& \hspace{1.65cm} \updownarrow   \nonumber\\[+1ex]
	\raisebox{2.1ex}{\B \quad \includegraphics[scale=0.8]{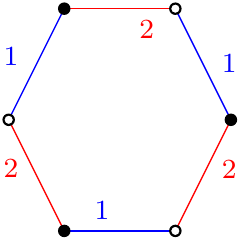}} \quad& \hspace{1cm}\raisebox{7ex}{$\longleftrightarrow$ } \hspace{0.9cm}&\ \ \includegraphics[scale=0.77]{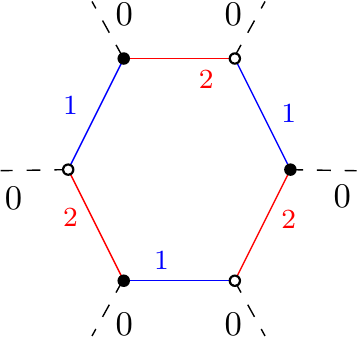}\raisebox{2.1ex}{$ \BB$}\nonumber
	\end{eqnarray}
\caption{On the left is a cycle and its dual colored graph, on the right is the star subdivision of an hexagon and its dual colored graph (with color-0 boundaries). }
\label{fig:Hexa}
	\end{figure}
More precisely, a bubble in dimension 2 is the  cone (Def.~\ref{def:Cone}) of a circle with a colored triangulation. As a 1-simplex is a line between two vertices, a 1-dimensional colored triangulation $\Bc$ of a circle  is just a cycle of even length alternating vertices of colors 1 and 2, and is entirely specified by fixing its size $2p$ (top left of Fig.~\ref{fig:Hexa}). The colored graph $\B$ dual to $\Bc$ is itself a bipartite cycle that alternates edges of color 1 and 2 (bottom left of Fig.~\ref{fig:Hexa}). 
It's two-dimensional  cone is obtained by embedding $\B$ in the plane, adding a vertex in its interior and linking it to every existing vertex without crossings (top right of Fig.~\ref{fig:Hexa}). A bubble in 2D is thus the star subdivision of a polygon. The boundary  of the bubble  $\BC$  is $\Bc$. The coloring of $\Bc$ translates into a coloring of $\BC$ : edges on the boundary are given the additional color 0, and the radial edges are alternatively of colors~1 (resp.~2) if they were added between the central vertex and a vertex of $\Bc$ of color~1 (resp.~2). The colored graph $\BB$ dual to the colored triangulation $\BC$ is therefore precisely $\B$, but to which color-0 half-edges have been added to every vertex (bottom right of Fig.~\ref{fig:Hexa}).

 The gluing of two (non necessarily distinct) bubbles along one edge is done by identifying edges of the boundaries of two bubbles with opposite orientation in the unique possible way (they all have the same color 0). Note that this respects the coloring of vertices so that {\it the resulting map is bipartite}. In the dual picture, it is done by identifying two color-0 half-edges incident to a black and a white vertex.

\subsubsection{Bubbles in higher dimension}

	 \begin{figure}[h!]
 \begin{eqnarray}
	\centering
	\Bc \ \includegraphics[scale=0.55]{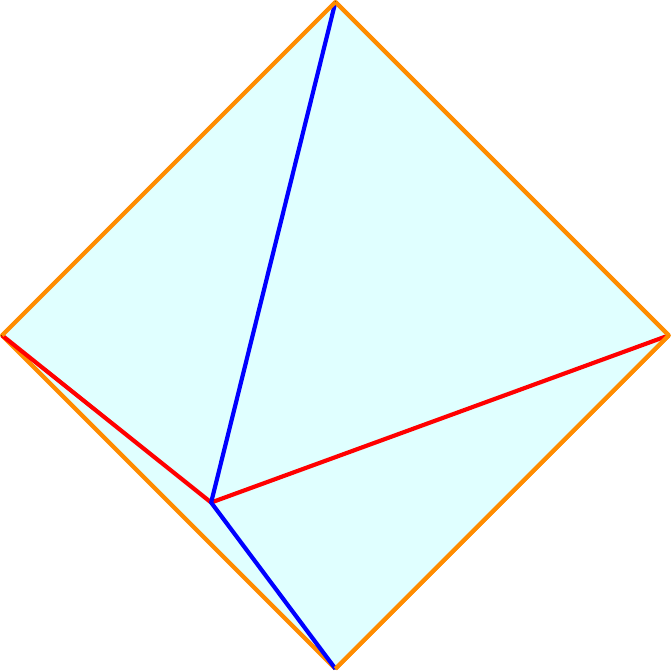}& \hspace{1cm}\raisebox{11ex}{\begin{tabular}{@{}c@{}} cone\\$\rightleftarrows$\\ boundary\end{tabular} }\hspace{1cm}&\includegraphics[scale=0.55]{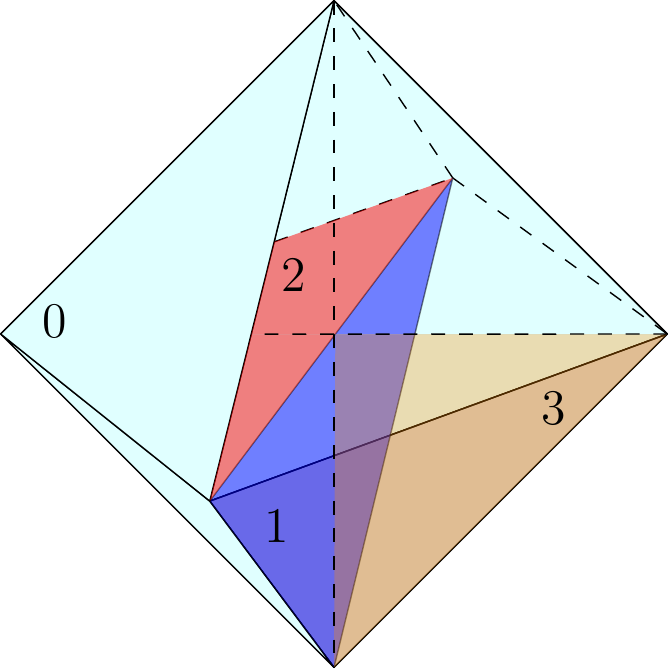}\ \BC \nonumber\\
	\nonumber\\
	\updownarrow \textrm{duality} \qquad&&\qquad\textrm{duality} \updownarrow   \nonumber\\
	\nonumber\\
	\raisebox{2.1ex}{\B \quad \includegraphics[scale=0.65]{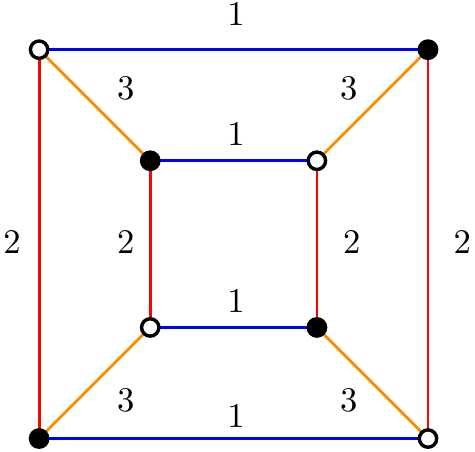}} \quad& \hspace{1cm}\raisebox{11ex}{$\longleftrightarrow$ } \hspace{0.9cm}&\ \ \includegraphics[scale=0.65]{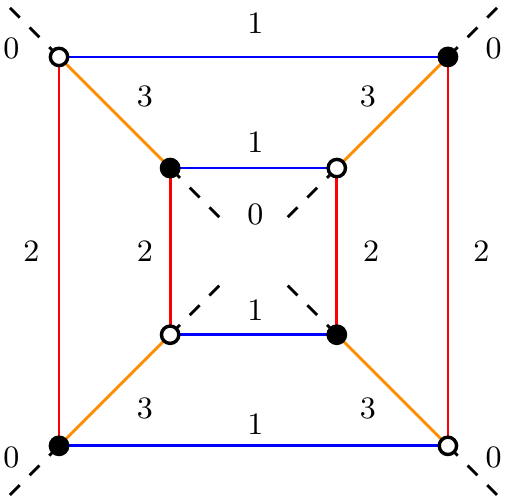}\raisebox{2.1ex}{$\quad \BB$}\nonumber
	\end{eqnarray}
\caption{On the left is a $(D-1)$-dimensional colored triangulation and on the right its $D$-dimensional cone. On the bottom are their dual colored graphs. }
\label{fig:EquiCone}
	\end{figure}
	The building blocks we will consider in most of this work are a generalization of $2p$-gons to dimensions three and higher. They are $D$-dimensional spaces with a single vertex in their interior and  a $(D-1)$-dimensional discrete connected boundary, which we further require to be a colored triangulation of size~$p$. 
	 This is more general than what announced in the preamble of Section~\ref{sec:pAng}, as we don't restrict the topology of the building blocks, which can even have singularities (they may be pseudo-manifolds with a connected boundary). In the more natural case of a $D$ dimensional ball (with a $(D-1)$-colored triangulated sphere as a boundary), it is a simplicial polytope. The notion of {\it cone} was defined in Def.~\ref{def:Cone}. 
	 
	 \begin{definition}[Bubble] 
	 \label{def:bubbles}
	 In dimension $D$, a bubble 
	 is the  cone of a $(D-1)$-dimensional colored triangulation $\Bc$. It is therefore a $D$-dimensional colored triangulation with boundary $\Bc$. We denote $\B\in\bG_{D-1}$ the colored graph dual to $\Bc$.
	 \end{definition}

	Bubbles are the $D$-dimensional objects we glue together along their boundaries. However there is a canonical bijection between the $D$-dimensional object and its $(D-1)$-dimensional boundary, as illustrated in Figure \ref{fig:EquiCone}. 
Therefore, in the rest of this work, {\it bubble} might equivalently refer to one or the other, and we will always denote $\Bc$ its boundary. The boundary has a connected $D$-edge-colored dual graph in $\bG_{D-1}$, which we also call {\it bubble graph}, or simply bubble when there is no ambiguity, and which we will always denote $\B$. The canonical bijection just mentioned is obvious in the dual picture: given a graph $\B\in\bG_{D-1}$ with colors in $\{1,\cdots,D\}$ dual to a $(D-1)$-dimensional triangulation $\Bc$, the graph $\BB$ dual to its cone is obtained by adding a pending half-edge on every vertex. These half-edges do not reach any other vertex, as they are dual to a facet of the boundary. The graph $\B$ is the boundary graph (Def.~\ref{eqref:Boundary}) of $\BB$ : $\B=\partial \BB$. The  canonical bijection in terms of dual graphs is shown in the bottom of Figure \ref{fig:EquiCone}.


\subsubsection{Degree}
	
	
	In dimension $D$, the degree of a bubble is defined as the Gurau degree \eqref{eqref:Deg} of its $(D-1)$-dimensional connected boundary $\Bc$ 
\be
\deltaG(\Bc)=D-1 + \frac{(D-1)(D-2)} 2 n_{D-1}(\Bc) -n_{D-3}(\Bc),
\ee	
The degree of it's interior $\BC$, seen as a $D$-dimensional object with boundary is given by it's Gurau degree, 
\be
\deltaG(\BC)=D + \frac{D(D-1)} 2 n_{D}(\BC) -n_{D-2}(\BC),
\ee
however it is not a quantity we will be interested in throughout this work, as it only differs from $\deltaG(\Bc)$  by a linear function of $n_{D-1}(\Bc)$.
Indeed, denoting $\B$ the colored graph dual to $\Bc$, and $\BB$ that dual to $\BC$, which only differs from $G$ by adding color-0 half-edges incident to every vertex, we have the following relations
\begin{align}
\label{eqref:D2IntBound}
n_{D-3}(\Bc)&=\Phi(\B)=\Phi_{\hat0}(\BB)=n_{D-2}(\BC) \\[+5pt]
n_{D-1}(\Bc)&=V(\B)=\,V(\BB)\, =n_{D}(\BC),
\end{align}
where $\Phi_{\hat0}(\BB)$  is the number of bicolored cycles that do not contain color 0.
 \be
 \Phi_{\hat 0}=\sum_{0<i<j}\Phi_{i,j}.
 \ee


	\subsection{Gluings of bubbles}
	\label{subsec:GluBub}
	
	The discrete objects we study in most of this work are the cell pseudo-complexes obtained by gluing bubbles (Def.~\ref{def:bubbles}) along facets of their boundaries in every possible way.  Facets that lie on the boundary of a bubble with colors $\{1,\cdots,D\}$ are given the additional color~0 (as a convention). 
	\begin{figure}[h!]
	\centering
	\includegraphics[scale=0.4]{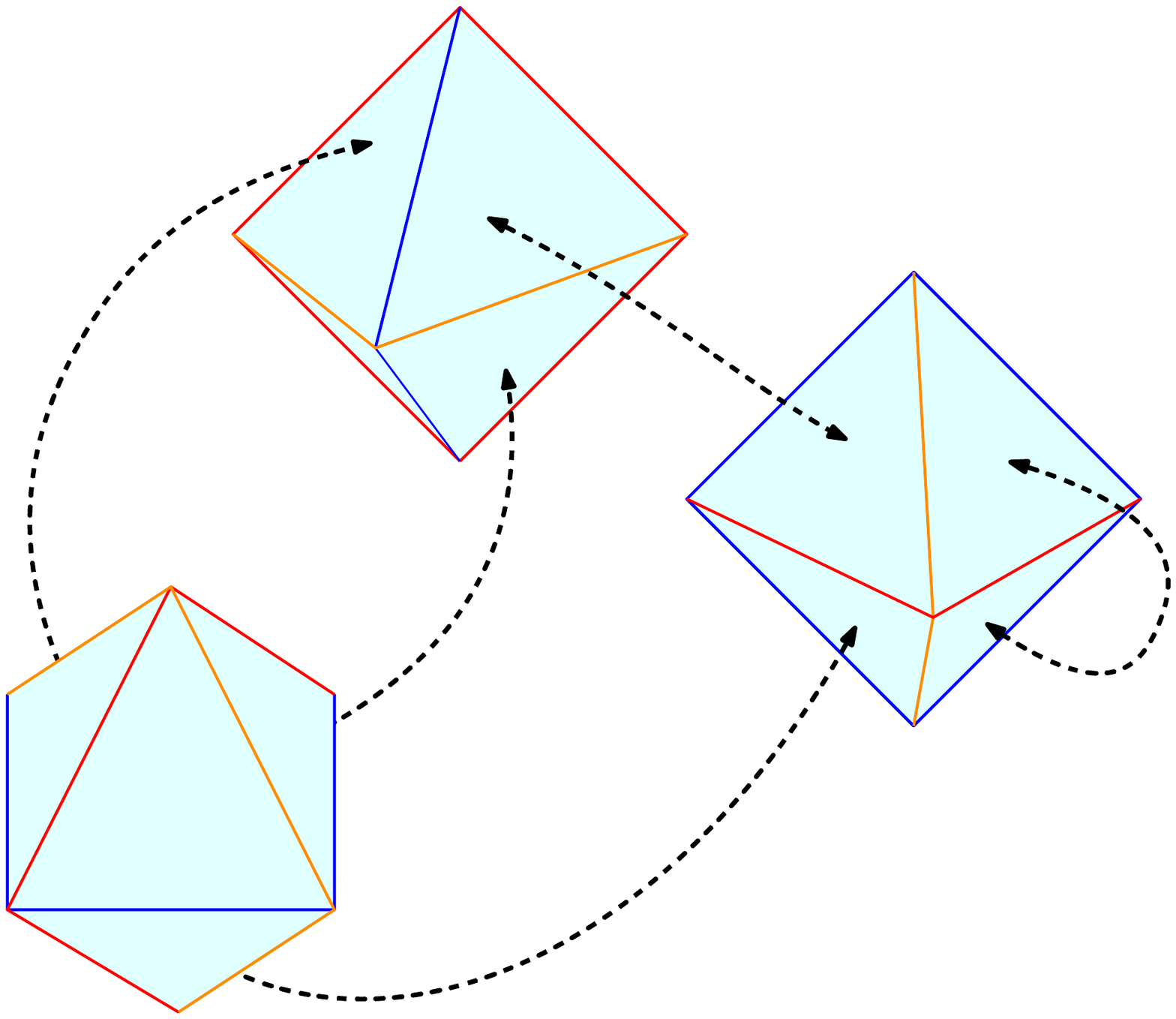}
\caption{Gluings of octahedra in three dimensions. }
\label{fig:GluOcta}
	\end{figure}
Two color-0 facets are glued in a unique way, as described in Section~\ref{sec:Simpl}. Therefore, the gluing of two facets of color~0 belonging to two $D$-simplices (which in turn might belong to the same or to  two different bubbles) is represented as an edge of color~0 between the corresponding vertices in the dual picture. We represent edges of color~0 as dashed, as they play a special role. Non-restricted connected gluings of bubbles in dimension~$D$ are dual to colored graphs in $\bG_D$. Any graph in $\bG_D$ can be interpreted as a gluing of bubbles once a choice is done for the color $i$ playing the role of ``color 0" (one might just exchange the names of the two colors to be consistent with previous and following definitions). The bubbles are then the connected components of the subgraph obtained by deleting all color-0 edges. We can also choose to restrict the possible building blocks :
\begin{definition}[Bubble-restricted gluings]
\label{def:Restricted}
Consider a 
subset of bubbles $\bB\subset \bG_{D-1}$. The connected discrete spaces obtained by gluing only copies of bubbles from $\bB$ are called $\bB$-restricted gluings. We denote $\bG(\bB)$ the set of their dual colored graphs, which we call $\bB$-restricted graphs, and bubble-restricted graphs more generally. 
%
%
In the boundary case, we denote $\bG^q(\bB)$ the set of graphs in $\bG_D^q$ such that, when deleting all color-0 edges, only copies of bubbles in $\bB$ remain.
\end{definition}
Generalized $2p$-angulations are obtained when all bubbles in $\bB$ have (a boundary of) size $2p$. 
Non-orientable gluings can be obtained from orientable bubbles, as done in Subsection~\ref{subsec:LOQuart}, or by gluing non-orientable bubbles, such as the $K_4$ graph  on the right of Fig.~\ref{fig:Ex2D}.
\begin{definition}[Locally-orientable gluings]
\label{def:NORestricted}
Consider a 
subset of bubbles $\bB\subset \bG_{D-1}$. One may decide to glue copies of bubbles from $\bB$ so that the resulting dual colored graphs are connected {\it non-necessarily bipartite} $(D+1)$-regular edge-colored graphs, which set we denote $\tilde \bG(\bB)$. Non-bipartite graphs correspond to non-orientable spaces. One may also consider non-orientable bubbles.
\end{definition}
The dual graphs are such that when deleting every color-0 edge, the connected components all belong to the bubble-set $\bB$. An example of 8-angulation is shown in Figure~\ref{fig:8Ang}.
	\begin{figure}[h!]
	\centering
	\includegraphics[scale=0.5]{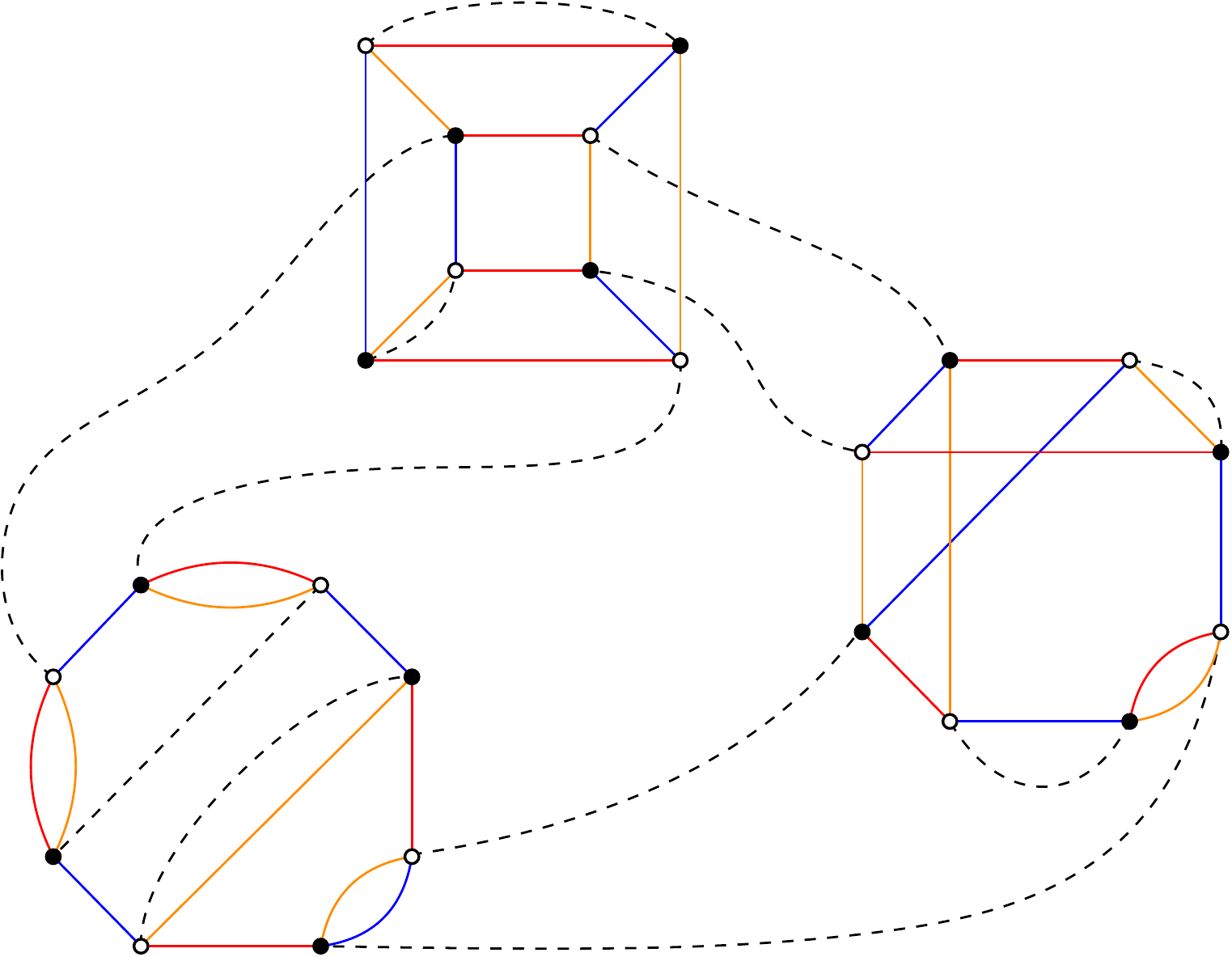}
\caption{Graph dual to an 8-angulation in three dimensions. }
\label{fig:8Ang}
	\end{figure}
Remark that the previous definition allows self-gluings. Bubble-restricted gluings are also simplicial pseudo-complexes in which there is less freedom on how to put simplices together. A first consequence of this is that Theorem~\ref{thm:Gurau} is still valid, i.e. a bubble-restricted gluing  $\C$ is such that 
\be
\label{eq:IneqGurau}
n_{D-2}(\C)\le D + \frac{D(D-1)}4 n_D(\C).
\ee
However, a consequence of Theorem~\ref{thm:Melo} is that \emph{this bound is saturated only for gluings of melonic bubbles}. Indeed, it is saturated for graphs of vanishing Gurau degree, which are melonic, and melonic graphs have melonic bubbles when deleting all the edges of any color~$i$ (consequence of Proposition~\ref{prop:Melo2}). Therefore, melonic graphs cannot be obtained from non-melonic bubbles. 

Another important remark is that $n_{D-2}(\C)$ counts every $(D-2)$-simplex, in particular it also counts $(D-2)$-simplices {\it inside} the bubbles (those that appear in $\BC$ when taking the cone of the boundary $\Bc$). Such  $(D-2)$-simplices are \emph{identified by bicolored cycles which do not contain color 0}, as illustrated on the left of Figure~\ref{fig:EdgesBubble} in three dimensions. As the number of such $(D-2)$-simplices is just 
\be
\label{eqref:D2Int}
n^{\circ}_{D-2}(\C)=\sum_{\B \in \bB} n_{D-2}(\BC) \times n_\B(\C),
\ee
where $n_{D-2}(\BC)=n_{D-3}(\Bc)$ is the number of $(D-2)$-simplices that do not belong to the boundaries $\Bc$ of the bubbles  (\ref{eqref:D2IntBound}), and $n_\B(\C)$ is the number of such bubbles in $\C$. If $\G$ is the graph dual to $\C$, the number of $(D-2)$-simplices inside the bubbles rewrites in terms of its bicolored cycles 
\be
\label{eqref:PhiInt}
n^{\circ}_{D-2}(\C)=\sum_{0<i<j}\Phi_{i,j}(\G) = \Phi_{\hat 0}(\G).
\ee
 The remaining  $(D-2)$-simplices belong to boundaries of bubbles,
 \be
n^\partial_{D-2}(\C)=n_{D-2}(\C) - n^{\circ}_{D-2}(\C). 
\ee
These are the  $(D-2)$-simplices we are interested in counting. They are identified by color-$0i$ cycles, as illustrated on the right of Figure~\ref{fig:EdgesBubble}, and therefore:
\begin{definition}[0-Score]
The 0-score is the total number of bicolored cycles containing color 0:
\be
\label{def:Score0}
 n^\partial_{D-2}(\C)=\sum_{i=1}^D\Phi_{0,i}(\G) = \Phi_{0}(\G).
 \ee
 It is also the number of $(D-2)$-simplices of $\C$ which lie on the boundaries of the bubbles.
 \end{definition}
	\begin{figure}[h!]
	\centering
	\raisebox{+1ex}{\includegraphics[scale=0.69]{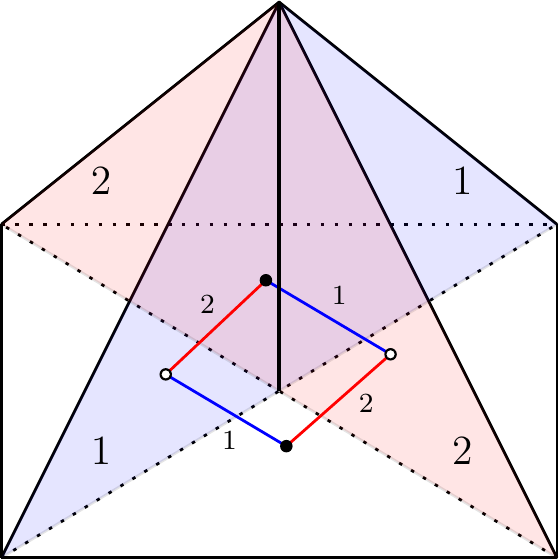}}
	\hspace{2cm}
	\includegraphics[scale=0.9]{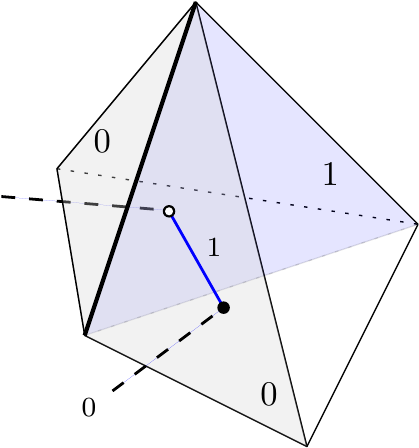}
\caption{Bicolored cycles identifying edges lying respectively in the interior of the bubble and on its boundary. }
\label{fig:EdgesBubble}
	\end{figure}

The number of $D$-simplices can also be expressed in terms of the number of bubbles, 
\be
n_D(\C)=\sum_{\B \in \bB} n_{D-1}(\B) \times n_\B (\C),
\ee
where we used (\ref{eqref:D2IntBound}). For $p$-angulations, it simplifies to 
\be
n_D(\C)=p \times \nb(\C),
\ee
in which $\nb$ denotes the total number of bubbles of the triangulation, 
\be
\label{eqref:NumBub}
\nb(\C)=\sum_{\B \in \bB}  n_\B(\C).
\ee
The linear bound (\ref{eq:IneqGurau}) rewrites 
\be
n^\partial_{D-2}(\C)\le D+ \sum_{\B\in \bB}\biggl[\frac{D(D-1)}4 n_{D-1}(\B) - n_{D-3}(\B)\biggr] n_\B(\C)
\ee
which can be expressed in the dual picture using (\ref{eqref:D2IntBound}),
\be
\label{eqref:Bound31}
\Phi_{0}(\G)\le D+ \sum_{\B\in \bB}\biggl[\frac{D(D-1)}4 V( \B) - \Phi(\B)\biggr] n_\B(\G).
\ee
In order to express this relation in terms of the degree  $\deltaG(\B)$ of the boundary of the bubble, we use that 
\be
\frac{D(D-1)}4=\frac{(D-1)}2 +\frac{(D-1)(D-2)}4,
\ee 
so that \eqref{eqref:Bound31} can be rewritten as
\be
\label{eqref:BoundBoundary}
\Phi_{0}(\G)=n^\partial_{D-2}(\C)\le D+ \sum_{\B\in \bB}\biggl(\deltaG(\B) + (D-1)\bigl(\frac{V(\B)} 2 -1\bigr)\biggr)n_\B(\C),
\ee
and (\ref{eq:IneqGurau}), which is a bound on the number of $(D-2)$-simplices in terms of the number of simplices, is now re-expressed as a bound on the number of $(D-2)$-cells that lie on the boundaries of the bubbles, in terms of the number of bubbles.
%

	\subsection{Bubble-dependent degree}
	\label{subsec:BubDeg}

	As mentioned in the previous section, a consequence of Theorem \ref{thm:Melo} is that the bounds (\ref{eq:IneqGurau}) and (\ref{eqref:BoundBoundary}) are saturated only for gluings of melonic bubbles. It is a (non-trivial) consequence of \cite{GurauSchaeffer} that there are only finitely many gluings of non-melonic bubbles that contribute to each order when they are classified according to Gurau's degree\footnote{With the vocabulary of the paper, the authors show that there are a finite number of schemes. The schemes are obtained by recursively contracting the $D$-dipoles, and by replacing the chains by their minimal realizations. As gluings of non-melonic bubbles can neither have chains of arbitrary length nor melonic contributions of arbitrary size, it implies that there are only finitely many gluings of a non-melonic bubble $\B$ which have the same Gurau degree. This is a particular case of the more general result we prove in Th.~\ref{thm:ExUnicBBD}. }. 
Intuitively,  they produce less $(D-2)$-simplices than gluings of melonic bubbles at fixed number of $D$-simplices, and therefore the term $\frac{D(D-1)}4 n_D$ in Def.~\ref{def:Deg} or equation (\ref{eq:IneqGurau}) is too strong. Its influence has to be softened by replacing the factor 	$\frac{D(D-1)}4$ by some factor 
	\be
	\label{eqref:Bounda}
	a_\B<\frac{D(D-1)}4.
	\ee
For $D$-dimensional gluings $\C$ of a single kind of bubble $\B$, we therefore rather consider 
	\be
	\label{eqref:BubDeg}
	\delta_\B(\C)=D+a_\B \times n_D(\C) - n_{D-2}(\C).
	\ee
The factor $a_\B$ has to be determined so that 
$\delta_\B(\C)\in \bN$
for any gluing $\C$ of bubbles $\B$. This is equivalent to the bound
\be
n_{D-2}(\C)\le D+a_\B \times n_D(\C).
\ee	
For a given $a_\B$, relation (\ref{eqref:BubDeg}) naturally classifies contributions according to the corresponding $\delta_\B\ge 0$.
We further want the set of contributions of at least one given corresponding order to be of infinite cardinality, where we define the order as
\begin{definition}[Order]
\label{def:Order}
For a given $a_\B\in\bR$, the $k^{\rm th}$ order of contribution in the classification of $\B$-restricted graphs $\bG(\B)$ given by the corresponding $\delta_\B$ \eqref{eqref:BubDeg} is the preimage of $\delta_\B$ in $\bG(\B)$, which we denote $\delta_\B^{-1}(k)$. It defines a partition of the $\B$-restricted graphs
\be
 \bG(\B) = \bigsqcup_{k}\delta_\B^{-1}(k).
\ee
We call $k$ the order and we say that a graph $\G\in\delta_\B^{-1}(k)$ is of order $k$ or contributes at order $k$.
\end{definition}
The bubble-dependent degree is then defined as follows
\begin{definition}[Bubble dependent degree] 
\label{def:BubDepDeg}
For $\B\in\bG_{D-1}$, if there exists a scalar $a_\B\in\bR^+$ such that the form 
\be
\label{eqref:BubDeg2}
\begin{array}{rccl}
\delta_\B : & \bG(B) &\longrightarrow& \bN \\
&\G &\longmapsto& D+a_\B V(G) - \Phi(G)
\end{array}
\ee
satisfies the two following conditions
\begin{align}
&\label{eqref:Cond1}\bullet\quad  \delta_\B\in\bN\\
&\label{eqref:Cond2}\bullet\quad \exists k\in\bN {\rm\ such\ that\ Card}\biggl(
\delta_\B^{-1}(k)\biggr)=\infty,
\end{align}
then $\delta_\B$ is said to be well-defined \eqref{eqref:Cond1} and non-trivial \eqref{eqref:Cond2}, and is called a bubble-dependent degree.
\end{definition}
If the last condition is not satisfied, the generating function of contributions to each order are polynomials in the counting parameter. In particular they do not have singularities, and the corresponding theory does not define any continuum limit at a given order (see Section~\ref{subsec:EH}). This is clear intuitively: if the number of gluings contributing to a given order is finite, one cannot have a limit of graphs of the same order with an infinite number of bubbles, each rescaled to have vanishing volume. We say that the corresponding degree is \emph{trivial}. 

The first non-empty order is called the {\it leading order}. One could then just translate (\ref{eqref:BubDeg2}) to have a constant term smaller than $D$. However, we will prove in Theorem~\ref{thm:ExUnicBBD} that if such a value of $a_\B$ exists, then Condition~\eqref{eqref:Cond2} is always satisfied at order 0. In particular, the leading order is always order~0. 
%

As underlined in the definition, the existence of such a value of $a_\B$ is not guaranteed, and it is not excluded for the 0-score of graphs which maximize the 0-score at fix number of vertices (maximal graphs) to have a non-linear dependence in the number of vertices. 
In fact, the simplest non-orientable bubble generates maximal graphs which exhibit a different behavior for even and odd numbers of bubbles. In this case, we can still define the degree with the strongest slope, and exclude the other maximal graphs from the leading order (see the example of the $K_4$ bubble in Section~\ref{subsec:TreeLike} on tree-like families). It can lead however to a non-trivial degree which is non-negative but rational, and therefore defines rational orders. We argue in Section~\ref{sec:K336} that more exotic behaviors are not excluded, although they have never been observed and are not-likely to occur.
%

Moreover, we have stated the definition as if such a value of $a_\B$ was necessarily unique. 
We will prove in Theorem~\ref{thm:ExUnicBBD} that whenever it exists, there is a unique such value.
More precisely, we will show that when choosing a smaller value of $a_\B$, we can easily exhibit an infinite family of graphs for which $\delta_\B\rightarrow-\infty$, and when choosing a higher value of $a_\B$, we only have a finite number of contributions per order.

\

As before for the Gurau degree, we can \emph{express the bubble-dependent degree in terms of the $(D-2)$-cells that lie on the boundaries of the bubbles}, as the $(D-2)$-cells in their interiors is just a constant times the number of bubbles. Making use of relations (\ref{eqref:D2Int} - \ref{eqref:Bound31}), we can rewrite $\delta_\B$ in terms  of the number of boundary $(D-2)$-cells $\Phi_0$ and of the number of bubbles $\nb$,
	\be
	\label{eqref:DeltaTildeA}
	\delta_\B(\G)=D+\tilde a_\B\times \nb(G) - \Phi_0(\G),
	\ee
which only depends on quantities linked to the boundary of the bubbles, and where $\tilde a_\B$ is determined so that conditions (\ref{eqref:Cond1} - \ref{eqref:Cond2}) are satisfied. The relation between $a_\B$ and $\tilde a_\B$ is
\be
\label{eqref:Tildeaa}
\tilde a_\B = a_\B V(\B) - \Phi(\B).
\ee
In practice, for a given bubble $\B$, we rather try to determine $\tilde a_\B$ such that $\delta_\B$ satisfies (\ref{eqref:Cond1} - \ref{eqref:Cond2}). Moreover, for counting results such as in Section \ref{subsec:CountUni}, we are only interested in counting quantities that lie on the boundary of the bubbles. However, comparisons between different models can only be done using relation (\ref{eqref:BubDeg2}), and the coefficient 
\be
a_\B - \frac{D(D-1)}{4} <0
\ee 
gives the \emph{correction to Gurau's degree} (Def.~\ref{def:Deg}). Indeed, we have
\be
\label{eqref:CorrectionGurau}
\Delta_\B=\frac{\deltaG-\delta_\B}V= \frac{D(D-1)}4- a_\B>0.
\ee
According to \eqref{eqref:Tildeaa}, the condition $a<\frac{D(D-1)}4$ for non-melonic bubbles is equivalent to
\be
\label{eqref:BoundAvsBoundTildeA}
a<\frac{D(D-1)}4 \qquad \Leftrightarrow\qquad \tilde a < \deltaG(\B) + (D-1)\bigl(\frac{V(\B)}2 - 1\bigr),
\ee
where $\deltaG(\B)$ is the Gurau degree of the bubble $\B$. This is consistent with \eqref{eqref:BoundBoundary}.

\

This trivially \emph{generalizes to $\bB$-restricted gluings}. The $\bB$-dependent degree of a dual colored graph is\footnote{The notation is rather conflictual as although it is the case for all known examples, it has actually never been proven that the $a_\B$ computed for gluings of a single bubble was necessarily the same $a_\B$ computed for $\bB$-restricted gluings with $\B\in\bB$. A possible candidate could be the  example in Section~\ref{sec:K336}, where we need two conjugate bubbles to produce a graph which has more bicolored cycles than expected when gluing only one or the other bubble. A more accurate notation would therefore be $a_\B^\bB$ in the case of $\bB$-restricted gluings, and the reader should keep in mind that for each one of the examples of $\bB$-restricted gluings treated in this thesis, and for each $\B\in\bB$, it is proven independently that $a_\B^\bB=a_\B$.}
	\be
	\delta_\bB(\G)=D+\sum_{\B\in\bB} a_\B V(B)n_B(G) - \Phi(\G),
	\ee
which can also be written in terms of boundary $(D-2)$-cells 
	\be
	\label{eqref:BubDeg3}
	\delta_\bB(\G)=D+\sum_{\B\in\bB} \tilde a_\B n_\B(\G) - \Phi_{0} (\G).
	\ee
The orders are then defined as
\be
\bG(\bB) = \bigsqcup_{k\in\bN} \delta_\bB^{-1}(k).
\ee

\

Again, as a consequence of Theorem \ref{thm:Melo}, we know that 
for the complete family of melonic bubbles $\bB_{\rm melo}$ in dimension $D$, $a_\B=D(D-1)/4$ does not depend on the size of the bubble, and $\deltaG(\B)=0$ in (\ref{eqref:BoundBoundary}), so that for a melonic bubble $\B_{\rm melo}$ with $V$ vertices,
\vspace{-5pt}
\be
\label{eqref:TildeAMelo}
\tilde a_{\B_{\rm melo}}=(D-1)\bigl(\frac {V} 2 - 1\bigr),
\vspace{-5pt}
\ee
as expected from \eqref{eqref:BoundAvsBoundTildeA}.
In two dimensions, the Gurau degree is twice the genus of triangulations, and for a combinatorial map $M$, $a_\B=1/2$ and $\tilde a_\B=V(\B)/2 - 1$, as for melonic bubbles in higher dimension.

\

For a given bubble $\bB$, our first aim is to identify the maximal gluings:
\begin{definition}[Maximal gluings]
\label{def:Max}
For a given $\bB$, we call maximal the $\bB$-restricted gluings which maximize the number of  $(D-2)$-cells at fixed number of $D$-cells. We will call maximal graphs their dual colored graphs, which maximize the number of bicolored cycles at fixed number of vertices.
\end{definition}
This is done in the dual picture by identifying the colored graphs in $\bG(\bB)$ which maximize the number of bicolored cycles \emph{that contain color 0}, at fixed number of vertices. Indeed, we are interested in the $(D-2)$-cells that lie on the boundaries of the bubbles, and furthermore, the number of remaining bicolored cycles is just a constant times the number of bubbles (\ref{eqref:D2Int}). In most known (orientable) cases they satisfy a relation of the type 
	\be
	\Phi_0(\G_\textrm{max})=D+ \sum_{\B\in\bB}\tilde a_\B n_\B (\G),
	\ee
from which we deduce $\tilde a_\B$ and $a_\B$. If $\tilde a_\B\in\bN$,  the bubble dependent degree defined for this value of $\tilde a$ is an integer, which vanishes only for maximal configurations, and is positive otherwise. It therefore satisfies \eqref{eqref:Cond1} and maximal graphs are precisely the graphs contributing to the leading order thus defined. We will show in Theorem~\ref{thm:ExUnicBBD} that Condition~\eqref{eqref:Cond2} is necessarily satisfied at order 0.
We will calculate the values of $a_\B$ and $\tilde a_\B$ for a certain number of models. A table summarizes these results in Section~\ref{sec:Summary}, and lists the sections where they are treated.


	\subsection{Pairings and coverings}
	\label{subsec:PaCov}

A colored graph in $\bG_D$ always has an even number of vertices: because there is one color-1 edge incident to each vertex and one only, color-1 edges define a partition of the vertices in pairs (a 1-matching).

\begin{definition}[Pairing, covering]
\label{def:Pairing}
For a given bubble $\B\in\bG_{D-1}$, we call pairing a partition of its vertices in pairs of black and white vertices.  Pairings are usually denoted~$\Om$.
We call $\Om$-covering of $\B$ and denote $\BCO\in\bG_D$ the edge-colored graph obtained by adding color-0 edges between the vertices of each pair of $\Om$.
\end{definition}

\begin{definition}[Optimal pairing and covering]
\label{def:OptPair}
We call optimal the coverings which maximize the score (or equivalently the 0-score) among all other coverings of the same bubble. We call optimal the corresponding pairings.
\end{definition}

We will generally denote $\PhiM$ the 0-score of an optimal covering, and $\Opt$ an optimal pairing.
The following is not a sufficient condition for a graph to be melonic. We do not prove it here as we prove a more general statement in Lemma~\ref{lemma:TreesImplieOpt}.

\begin{prop}
Melonic graphs have a unique optimal pairing, its canonical one. 
\end{prop}

\subsection{Colored polytopes in three dimensions}
	\label{subsec:ColPol}
	
	In this subsection we argue that the formalism of colored triangulations can be generalized to study gluings of other kinds of building blocks, not necessarily having colored-triangulated boundary. This is for instance the case of an icosahedron, which boundary is a triangulation which is not properly colorable, but also of polytopes having non-triangular facets, as the cube, or the dodecahedron. We consider the example of the cube. Its boundary is a quadrangulated sphere with 6 facets, which we color with indices from 1 to 6. The edges (resp.~vertices) inherit the color set of the 2 (resp.~3) colors of the facets they belong to. As before, when gluing two facets of the same color, we require that the edges and vertices which have the same color sets are identified. We consider two orientations of the cube, with opposite ordering of colors around the boundaries of the facets. This is shown in Fig.~\ref{fig:Cube}. One can use this formalism to study gluings of bubbles. It would be possible to study gluings of cubes by subdividing them into 3-simplices throughout a barycentric subdivision, but this would require 56 simplices and would allow self-gluings, which are not authorized in the present formalism.
	\begin{figure}[h!]
	\centering
	\includegraphics[scale=0.8]{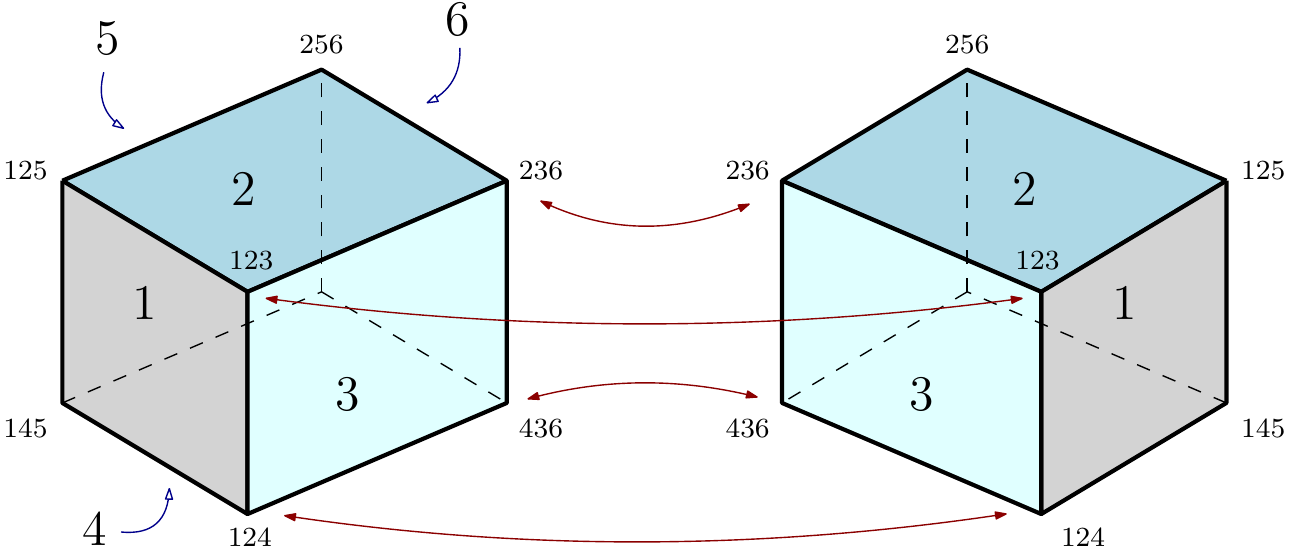}
\caption{Gluing colored cubes. }
\label{fig:Cube}
	\end{figure}

In order to study the combinatorial properties of the discrete spaces obtained by gluing colored cubes, we represent the two kinds of cubes by black and white vertices, and the gluing of two facets of the same color by a colored edge between the two corresponding vertices. The graph we obtain are precisely the same as when gluing tetrahedra in dimension $D=5$, i.e. 6-colored graphs in $\bG_5$. We can use this formalism to classify configurations according to their number of edges at fixed number of cubes.
However, the difference here relies on the fact that \emph{not all bicolored cycles identify edges}. Indeed, with the choice of coloring of Fig.~\ref{fig:Cube}, only the bicolored cycles $12,13,14,15,23,25,26,34,36,45,46,56$ identify edges; the others cycles $16, 24, 35$ do not. We define the score of gluings of colored cubes as
\be
\Phi_\text{cube}(\G\in\bG_6)= \Phi(\G) - \Phi_{1,6}(\G) - \Phi_{2,4}(\G)- \Phi_{3,5}(\G).
\ee	
Let us compute the score of gluings of four cubes. A graph in $\bG_5$ with four simplices is   a cycle alternating $k$ edges and $6-k$ edges. Its usual score is
\be
\Phi^{4,k} = 30 - k(6-k).
\ee
In the case where  $k=1$, e.g. with distinguished color 1, we have one cycle 16, and two cycles for 24 and 35. We find
\be
\Phi_\text{cube}^{4,1}= 30 - 5 - 5 = 20.
\ee
In the case where  $k=2$, there are two cases, shown in Fig.~\ref{fig:FourCubes}. 
	\begin{figure}[h!]
	\centering
	\includegraphics[scale=0.8]{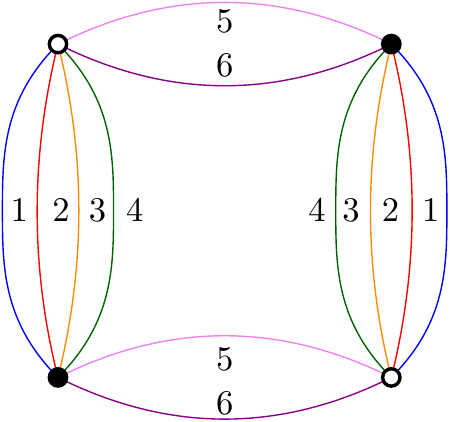}\hspace{2cm}\includegraphics[scale=0.8]{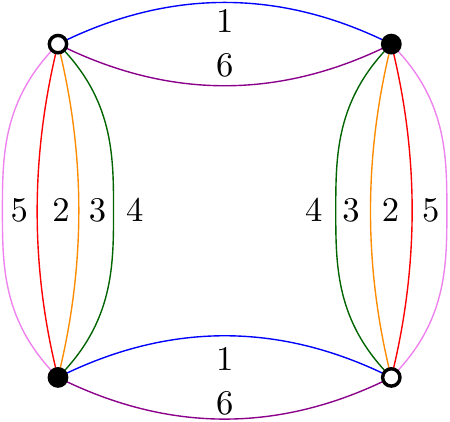}
\caption{Gluing of four cubes. }
\label{fig:FourCubes}
	\end{figure}
	
\noindent They lead to different scores: for the graph on the left, 
\be
\Phi_\text{cube}^{4,2a}= 30-8 - 4 = 18, 
\ee	
and for the graph on the right, 
\be
\Phi_\text{cube}^{4,2b}= 22- 5 = 17.
\ee	
In the case where  $k=3$, there are also two cases, leading to the scores
\be
\Phi_\text{cube}^{4,3a}= 30-9 - 3 = 18, \quad\text{and}\quad \Phi_\text{cube}^{4,3b}= 30-9 - 5 = 16.
\ee
	
We find that $\Phi_\text{cube}^{4,2a}=\Phi_\text{cube}^{4,3a}$. \emph{This is different from the  $D=5$ gluings of tetrahedra corresponding to these graphs, for which}  $\Phi^{4,2}>\Phi^{4,3}$:
\be
\Phi^{4,1}= 25, \qquad \Phi^{4,2}= 22, \quad\text{and}\quad \Phi^{4,3}= 21.
\ee
In this interpretation of the graphs in $\bG_5$, the contribution of the cycle alternating 3 edges has therefore been ``enhanced". Although in this example, $\Phi_\text{cube}^{4,1}$ is still larger, we understand that interpreting graphs in $\bG_5$ as representing gluings of colored cubes could render maximal some graphs which were not. This also motivates the study of discrete spaces obtained by gluing building blocks bigger than tetrahedra.

\section{Quantum gravity and random tensor models}
\label{sec:QG}
	\subsection{Discretized Einstein-Hilbert action}
	\label{subsec:EH}
	
	In dimension $D$, the Euclidean Einstein-Hilbert partition function describes a theory of pure gravity - i.e. without interacting matter - which is classical in the sense that there are no quantum effects 
	\be
\cZ_{EH}=\int D[g]e^{-S_{EH}(g)},
\ee
where the functional integration
is done over all metrics on some manifold $\cM$, and  denoting $\Lambda$ the cosmological constant and $R$ the Ricci scalar, the Einstein-Hilbert action is
\be
\label{eqref:EH}
S_{EH}(g, \cM)=\int_\cM d^Dx\sqrt {\lvert g\rvert} (\Lambda-\frac{1}{16\pi G}R),
\ee
in which $G$ is Newton's constant, and we have set $\hbar=c=1$. A common approach to make sense of this partition function is to discretize the manifold $\cM$, as principally developed by Regge \cite{Regge}.
We consider a triangulation of $\cM$, and provide it with an induced geometry by assuming that all the edges have the same length $l$. For such a choice $\C$ of triangulation, 
the first term of the action is just the volume of $\cM$ - proportional to the number of $D$-simplices of $\C$ - multiplied by $\Lambda$. The second term, which encodes the local curvature, can be shown to be proportional to a sum of deficit angles around $(D-2)$-simplices, which expression is \cite{CritBehavior}
\be
\Delta_{D-2}=2\pi n_{D-2} - \biggl[\frac{D(D+1)} 2 {\rm arccos}\frac 1 D 
 \biggr] n_D,
\ee
where $n_D$ and $n_{D-2}$ denote respectively the number of $D$ and $(D-2)$-simplices, as in the rest of this work. Denoting $v_k=\frac {l^k}{k!} \sqrt{\frac{k+1}{2^k}}$ the volume of a $k$-simplex,
the Einstein-Hilbert action is expressed as a particular case of the Regge action,
\be
S_{Regge}(\Lambda,\C)=\Lambda v_D n_D - \frac 1 {16\pi G}v_{D-2}\Delta_{D-2}.
\ee
In the partition function, we replace the integration over geometries on $\cM$ with a sum over homogeneous triangulations of the manifold $\cM$
\be
\int D[g] \qquad \leftrightarrow \quad \sum_{ {  \substack{\C\ {\rm triangulation\ }}\\{   {\rm of} \ \cM}   }     }.
\ee
We escape this classical background by summing over triangulations belonging to some set $\Sb$, that may contain triangulations of any topology, and consider the discrete partition function 
\be
\label{eqref:PartKD}
\cZ_\Sb(\Lambda)=\sum_{  \substack{{\rm triangulations}\\{ \rm in\ }\Sb}} e^{-\kappa_D n_D} e^{\kappa_{D-2} n_{D-2}},
\ee
where 
\be
\kappa_{D-2}=\frac{v_{D-2}}{8G}, \qquad {\rm and} \qquad \kappa_D= \Lambda v_D + \frac{v_{D-2}}{16\pi G} \frac{D(D+1)} 2 {\rm arccos}\frac 1 D.
\ee
Here, the coefficient $\kappa_{D-2}$ only depends on the choice of ``cut-off" $l$, the length of the edges, and on $G$. In a quantum limit, $l$ is of order 1 and $G\rightarrow 0$, so that $\kappa_{D-2}\rightarrow +\infty$\footnote{$l$ should be thought of as a cut-off which is removed in the continuum limit $l\rightarrow 0$, not as the Planck scale $l_P$, as we would have $\kappa_{D-2}\sim l_P^{D-2}/G \sim 1$ for $D>2$.}. In the continuum limit, $l\rightarrow0$ but we consider a ``mesoscopic" limit in which the action of gravity is still negligible. The two limits may however be of comparable order (see the double scaling limit \cite{DartoisDoubleScaling, 
DoubleScal}). We refer the reader to \cite{Invitation} for a detailed discussion. 
In every case, we are in the limit where 
\be
\label{eqref:LargeKappa}
\kappa_{D-2}\longrightarrow +\infty.
\ee
On the other way, the cosmological constant $\Lambda$ is not fixed, and $e^{-\kappa_D}$ is a variable of the theory. 
The partition function (\ref{eqref:PartKD}) therefore makes sense if the exponent of $e^{\kappa_{D-2}}$ does not become arbitrarily large for large connected triangulations. We therefore rewrite
\be
\label{eqref:PartKD2}
\cZ_\Sb(\Lambda)=\sum_{  \substack{{\rm triangulations}\\{ \rm in\ }\Sb}}(e^{-\kappa_D+a\kappa_{D-2}})^{ n_D} 
(e^{\kappa_{D-2}})^{ n_{D-2}- a n_D},
\ee
and search for some $a\in\bR^+$ and $A\in\bR$ such that for any connected triangulation in $\Sb$, $n_{D-2}- a n_D\le A$. 

\

Another way of understanding this is from statistical physics considerations. $\Sb$ can be seen as a grand-canonical ensemble in which triangulations are micro-states which can exchange energy and whose number of simplices is not fixed, and we wish to identify \eqref{eqref:PartKD} with the grand-canonical partition function of the system, 
\be
\cZ_\Sb(\Lambda)= \sum_{  \substack{{\rm \C\ triangulation}\\{ \rm in\ }\Sb}} e^{\mu n_D(\C)} e^{ - \beta \cE(\C)},
\ee
where $\cE(\C)$ is the energy associated with the triangulation $\C$.
 As $\kappa_{D-2}>0$, we could naively identify $\beta=\kappa_{D-2}$,  $\mu=-\kappa_D$, and $\cE=-n_{D-2}$, however this choice of energy would not be bounded from below. The rescaling \eqref{eqref:PartKD2} is therefore interpreted as a attempt to define an energy 
 \be
 \cE^a_\Sb(\C) = a n_D(\C) - n_{D-2}(\C),
 \ee
 with $a$ chosen such that $\cE^a_\Sb$ is bounded from below for $\C\in\Sb$. 
 
 \

 We furthermore expect the smallest such $A$ to be the dimension $D$. As $n_{D-2}\in\bN$, this implies $an_D\in\bN$ for saturating configurations. In the case where
\be
\label{eqref:QGBubDepDeg}
\exists a_\Sb\in\bQ^+\ {\rm such\ that\ } \delta_\Sb=D+a_\Sb n_D-n_{D-2}\in\bN,
\ee
we recognize condition (\ref{eqref:Cond1}), and $\delta_\Sb $ is the {\it degree} of a connected discrete space. In the most general case, $\delta_\Sb\in\bQ^+$. From previous sections, we know that if $\Sb$ is the full set of colored connected triangulations defined in Section \ref{sec:Simpl} (dual to colored graphs in $\bG_D$), then 
\be
a_{\bG_D}=\frac{D(D-1)}4,
\ee
and $\delta_\Sb$ is  Gurau's degree (Def.~\ref{def:Deg}). Many other choices are possible for the set $\Sb$. Here we are mainly interested in discrete spaces obtained by gluing bubbles (see Sections \ref{subsec:Bubbles} and \ref{subsec:GluBub}), which are ``rigid" blocs, themselves made of several $D$-simplices glued together. We know (Thm.~\ref{thm:Melo}) that the same value $a_\Sb=a_{\bG_D}$  is found if $\Sb$ is the set of spaces obtained by gluing melonic bubbles. For non melonic bubbles, we will show in Theorems~\ref{thm:FinitNumbPosDeg} and~\ref{thm:ExUnicBBD} that this value leads to a finite number of contributions per order. When such a value exists, it is smaller than $D(D-1)/4$ (\ref{eqref:Bounda}). 
Defining 
\be
\lambda=e^{-\kappa_D+a\kappa_{D-2}} \qquad {\rm and} \qquad N=e^{\kappa_{D-2}},
\ee
the equation (\ref{eqref:PartKD2}) can be expressed in the simpler form 
\be
\cZ_\Sb(\lambda)=\sum_{  \C\in\Sb} \lambda^{n_D(\C)} N^{n_{D-2}(\C)- a n_D(\C)}.
\ee
In statistical physics terms, this is the partition function 
of micro-states in $\Sb$, weighted by the Boltzman weight $N^{n_{D-2}(\C)- a n_D(\C)} = e^{-\beta\cE_\Sb^a(\C)}$, and with fugacity $\lambda$.
In the case where the elements in $\Sb$ are not necessarily connected, the free-energy (related to the grand potential),
\be
\cF_\Sb(\lambda)=\ln \cZ_\Sb(\lambda)
\ee
is the sum over {\it connected} discrete spaces. Remark that the object we interpret as the partition function of some non-classical theory of gravity has a sum over geometries of \emph{connected} spaces so that is coincides with the partition function if the elements of $\Sb$ are connected, or rather with the free-energy if the elements in $\Sb$ are not connected. To adjust the notations with the tensor model generating functions of Subsections~\ref{subsec:ColTenMod} and \ref{subsec:Uncolored}, we place ourselves in the second case, and consider the expansion of the free-energy. 
%
%
If \eqref{eqref:QGBubDepDeg} holds, the free-energy writes
\be
\label{eqref:NExp}
\cF_\Sb(\lambda)=N^D\sum_{ \substack{{ \C\in\Sb}\\{\rm connected}}} \lambda^{n_D(\C)} N^{-\delta_\Sb(\C)},
\ee
which is also known as the $1/N$ {\it expansion} in theoretical physics. We call {\it order} (see Def.~\ref{def:Order}) the integer values taken by $\delta_\Sb$ and {\it leading order} the first non vanishing value. We write
\be
\label{eqref:NExp2}
\frac1{N^D}\cF_\Sb(\lambda)=\cF^{(0)}_\Sb(\lambda) + \frac 1 N \cF^{(1)}_\Sb(\lambda) + \frac1{N^2}\cF^{(2)}_\Sb(\lambda) + \cdots
\ee
where $\cF^{(i)}_\Sb(\lambda) $ is the {\it generating function} of elements of $\Sb$ of order $i$, counted with respect to their number of $D$-simplices, 
\be
\cF^{(i)}_\Sb(\lambda) = \sum_{n\ge0} c_n^{(i)} \lambda ^ n,
\ee
where $c_n^{(i)} $ is the number of elements of $\Sb$ of order $i$ with $n$ D-simplices. If there is a least one such space, the leading order corresponds to spaces in $\Sb$ with vanishing degree $\delta_\Sb$ (we will prove in Theorem~\ref{thm:ExUnicBBD} that it is always the case). 
%
Such discrete spaces {\it maximize the number of $(D-2)$-simplices at fixed number of $D$-simplices}. In the general case, they are a subset of {\it maximal spaces} (Def.~\ref{def:Max}). In the {\it large $N$ limit},  which corresponds to the limit (\ref{eqref:LargeKappa}) discussed above, only elements of the leading order survive. If the leading order generating function has a dominant singularity $\lambda_c$, we denote $\gamma$ the {\it critical exponent}, also called the \emph{string susceptibility} in physics\footnote{In practice, we rather count rooted spaces in order not to deal with symmetries. The critical behavior is then given by \eqref{eqref:CritExp1} and \eqref{eqref:CritExp2}. }
\be
\label{eqref:AsymptFree}
\cF^{(0)}_\Sb(\lambda) = \alpha + \beta (\lambda_c - \lambda)^{2-\gamma} + o\bigl((\lambda_c - \lambda)^{2-\gamma}\bigr) ,
\ee
in which case the coefficients $c_n^{(0)} $ asymptotically behave \cite{Flajolet} as
\be
\label{eqref:AsympCoeff}
c_n^{(0)} {\stackbin[ n \rightarrow +\infty]{}{ \sim}} \lambda_c^{-n} n^{\gamma-3}.
\ee
The behavior of the average volume of leading order triangulations at criticality satisfies
\be
<V(\C)>\sim l^D <n_D(\C)> \sim l^D \lambda \frac \dr {\dr \lambda} \ln \cF^{(0)}_\Sb(\lambda),
\ee
where we used that the average number of $D$-simplices of leading order contributions is
\be
<n_D(\C)> = \frac{\sum_{n\ge0} n c_n^{(0)}\lambda^n}{\sum_{n\ge0} c_n^{(0)}\lambda^n} = \lambda \frac \dr {\dr \lambda} \ln \cF^{(0)}_\Sb(\lambda).
\ee
This can be computed using (\ref{eqref:AsymptFree}) and leads to
\be
<V(\C)>{\stackbin[ \lambda \rightarrow \lambda_c]{}{ \sim}} l^D  \frac{ (\lambda_c - \lambda)^{1-\gamma} }{ \alpha + \beta (\lambda_c - \lambda)^{2-\gamma}}.
\ee
Intuitively, the large-scale limit should occur when the behavior does not depend on the details of the discretization. This is precisely the case at the singularity $\lambda_c$, as the partition function only depends on the asymptotics of the coefficients, i.e. on $\lambda_c$ and on the critical exponent $\gamma$. In order to reach a thermodynamical limit $<n_D> \rightarrow +\infty$ at the singularity, the partition function can be normalized\footnote{It can rightfully be argued that the partition function should be normalized otherwise, e.g. by setting it to 1 for $\lambda=0$. This has been noticed by physicists in the 80's. Another possibility is to consider $\Sb$ a set of discrete spaces with a certain number $k\ge 2$ of connected boundaries, in which case the exponent $2-\gamma$ is replaced with the exponent  $ 2-k-\gamma<1$, so that $<n_D>$ indeed diverges at the singularity, and the same argument can then be applied. }  to have $\alpha=\cF^{(0)}_\Sb(\lambda) = 0$. With this choice, at fixed $l$, the average volume diverges at criticality. However, this can be avoided by sending $l\rightarrow 0$ while keeping $\frac{l^D }{\lambda_c-\lambda}$ finite. In this limit, the volume of $D$-simplices goes to zero and a {\it continuum limit} is reached. The resulting continuous space is interpreted as a {\it quantum space-time}, which we would like to characterize. 

In the first place, we will be interested in the critical exponent $\gamma$. In the case of colored triangulations, the continuum limit of maximal configurations - which are the melonic ones (Subsection~\ref{subsec:Melonic}) - is the continuum random tree, or branched polymers in the physics literature \cite{MelonsAreBP}. It is characterized by the critical exponent of trees, $\gamma=1/2$. For two dimensional models, maximal configurations are the planar ones, the critical exponent is $\gamma=-1/2$, and the continuum limit is the Brownian map, as detailed in the following subsection.
The critical exponent does not provide information on the metric properties of the continuum limit. However in general, the exponent  $1/2$ or $-1/2$, are obtained from equations which  suggest a bijection with trees or planar maps, and we expect the continuum limits to be the continuous random trees or the Brownian map from universality arguments. One of the aims of this work is to explicitly describe these bijections. 

It should be noted however that there are cases of statistical systems on maps which lead to the same critical exponent but not to the same theory in the continuum limit. An example is the case of the Ising model on random surfaces, for which the critical exponent is $\gamma=-1/3$ \cite{Kazakov}, corresponding in the continuum to unitary conformal matter coupled to 2D gravity. The multi-critical hard-dimer model on random lattices has the same critical exponent \cite{Staudacher}, but leads to non-unitary conformal matter coupled to 2D gravity in the continuum. The critical exponent is therefore just a first indicator of what we obtain in the continuum.
To characterize the metric properties of the continuum limit with more accuracy, one should calculate other quantities, such as the Hausdorff dimension or the spectral dimension, but this goes beyond the scope of this work.




	\subsection{Two dimensions}
	\label{subsec:2DQG}
	
	In dimension two, the Einstein-Hilbert action (\ref{eqref:EH}) is discretized using the Gauss-Bonet theorem, which states that the curvature term is topological
	\be
	\frac 1 {4\pi} \int_\cM d^2x\sqrt {\lvert g\rvert}R = 2-2g(\cM),
	\ee
$g$ being the genus of the surface $\cM$.	This can be recovered as for a triangulation with equilateral triangles, the Ricci scalar, which encodes the local curvature, relies on the number of triangles around vertices. The triangulation is locally flat at some vertex $v$ if there are precisely 6  equilateral triangles incident to it. If there are less, the curvature is positive, if there are more, it is negative. More precisely,
\be
\int_\cM d^2x\sqrt {\lvert g\rvert}R \qquad\leftrightarrow\quad \sum_{v\in\cV(\C)}4\pi \biggl(1-\frac{\deg(v)} 6\biggr),
\ee
where the sum is taken over vertices $v$ of the triangulation $\C$ and $\deg(v)$ is the valency of $v$, i.e. the number of incident triangles. 
As there are 3 corners per triangle, 
\be
\sum_{v\in\cV(\C)} \deg(v) = 3F(\C), 
\ee
where $F$ is the number of triangles of $\C$, or faces. The discretized Einstein-Hilbert action is therefore 
\be
S_{Regge}(g)=\Lambda v_2 F(\C) - \frac 1 {4 G}v_{0}\biggl(V(\C)-\frac{F(\C)} 2\biggr).
\ee
As for a triangulation, $3F(\C)=2E(\C)$, the Euler characteristics is
\be
2-2g=V-E+F= V-\frac F 2,
\ee
we deduce the coefficient $a=1/2$, and the degree is twice the genus, $\delta=2g$. In dimension two the $1/N$ expansion (\ref{eqref:NExp}) is topological:
\be
\cF_\Sb(\lambda)=N^2\sum_{ \substack{{ \C\in\Sb}\\{\rm connected}}} \lambda^{F(\C)} N^{-2g(\C)}.
\ee
In the large $N$ limit, only maximal triangulations survive, i.e. triangulations of vanishing genus. The leading order partition function is the generating function of planar triangulations of $\Sb$. If $\Sb$ is the full set of triangulations, the coefficients of the generating function of rooted configurations - corresponding to the 2-point function - behave asymptotically as  \cite{TutteTriang}
\be
G^{(0)}_\Sb(\lambda) \sim \frac{1}{16}\sqrt{\frac{3}{2\pi}}n^{-\frac{5}{2}}\biggl(\frac{256}{27}\biggr)^n\propto n^{\gamma-2}\lambda_c^{-n},
\ee
so that $\lambda_c=27/256$ and the critical exponent \eqref {eqref:AsympCoeff} is 
\be
\gamma=-\frac 1 2.
\ee	
If $\Sb$ is the set of colored triangulations, then in the large $N$ limit, only planar colored triangulations survive, which are shown to exhibit the same critical behavior \cite{ ColPlan1, ColPlan2, Rectang, Review}. In the case where $\Sb$ is the set of gluings of bubbles, which in $D=2$ are $p$-gons (with $p$ even), the calculation in terms of their constituting triangles (taking the star subdivision of every $p$-gon) also leads to $a=1/2$ and $\gamma=-1/2$. Stated in terms of the $p$-gons however, $\tilde a = (p-2)/2 $, and $\delta=2g$ is re-expressed as
\be
2g(\C_p)=2+\frac{p-2}2 F(\C_p) - V(\C_p).
\ee
 In the large $N$ limit, only planar $p$-angulations survive. Planar triangulations or $2p$-angulations converge in distribution towards the Brownian map \cite{MarckMierm, MarcMok,  Mie, LeGallUnivers}, sometimes called the Brownian sphere. 
It is a fractal random metric space homeomorphic to the 2-sphere \cite{LGPaul2Sph}, with spectral dimension 2 and Hausdorff dimension 4 \cite{GallHaus}.  Intuitively, it means this space is very creased, but can be continuously deformed into a 2-sphere. 
As the coefficient $a$ and the critical exponent $\gamma$, this continuum limit does not depend of the choice of discretization $p$ \cite{LeGallUnivers}, at least for $p$ even or $p=3$, which are precisely the values we are interested in. 
%
Let us state this in a more precise way. Denote $\C_p^n$ the number of rooted planar $p$-angulations with $n$ faces, $p$ being 3 or any even number, pick a map $\cM_p^n$ uniformly at random in $\C_p^n$ and denote $\cV_n$ its vertex set. See $(\cM_p^n, d_{gr})$ as a random variable, $d_{gr}$ being the graph distance, i.e. the number of edges in the smallest path between two vertices. Denote $d_{GH}$ the Gromov-Hausdorff distance between compact metric spaces, and set $c_3=6^{1/4}$ and $c_{2q}=\bigl(\frac 9{4q(q-1)}\bigr)^{1/4}$. Let $\bK$ be the space of isometry classes of compact metric spaces, which we equip with the distance $d_{GH}$. 
 \begin{theorem}[Universality - Le Gall, 2013 \cite{LeGallUnivers}] 
 \label{thm:LeGall}
 There exist a random compact metric space $(\cM_\infty,D^*)$ called the Brownian map, $D^*$ being a distance on $\cM_\infty$, such that 
 \be
 (\cM_p^n,c_pn^{-1/4}d_{gr}) \quad\substack{{\longrightarrow}\\{n\rightarrow \infty}}\quad (\cM_\infty, D^*),
 \ee
 regardless of $p$ even or equal to 3.
 \end{theorem}
 
The relation between 2D quantum gravity and random planar maps has been studied since the 80's (see \cite{MatrixReview} for a review).  Liouville quantum gravity \cite{Liouv} is a theory of random surfaces with canonical \emph{conformal} structure, which is the effective continuum gravitational theory obtained from coupling conformal matter to 2D gravity. It was introduced by Polyakov as a model to describe the world-sheets in string theory \cite{Polya}. It is a construction of random surfaces a priori different from the Brownian map - or sphere - which is the limit in distribution of random planar maps, and which has a canonical \emph{metric} structure. The link between the Brownian sphere and Liouville quantum gravity has firstly been investigated principally throughout the KPZ relation  \cite{CFTQG2, CFTQG3, DuplSheff0, DuplSheff}, conjectured in \cite{CFTQG1}.   See \cite{GallHaus, DuplHaus} concerning the Hausdorff dimension.  It has been recently shown by Miller and Sheffield \cite{MilSheff} that the two objects could be equipped with the other object's canonical structure and that they are consequently equivalent. 
These continuum fractal random objects are understood as ``quantum 2D space-time". This is what we would like to generalize to higher dimensions, especially dimensions 3 and 4.
 
	\subsection{Guideline in higher dimension}
	\label{subsec:Guideline}

In this subsection,  we summarize our problem and the steps of the study of colored discrete spaces obtained by gluing bubbles (Def.~\ref{def:bubbles}), which are $D$-dimensional elementary building blocks with a $(D-1)$-dimensional boundary which has a colored triangulation. Given a set of bubbles $\bB$, we will in the first place be interested in {\it identifying and counting maximal gluings of bubbles} (Def.~\ref{def:Max}), which maximize the number of $(D-2)$-cells at fixed number of $D$-cells. If in their dual colored graphs, the 0-score (Def.~\ref{def:Score0}) of maximal maps satisfies a relation of the type 
	\be
	\label{eqref:MaximalBound}
	\Phi_0(\G_{\rm max})=D+ \sum_{\B\in\bB}\tilde a_\B n_\B(\G),
	\ee
then we choose this value (and corresponding $a_\B$) for the bubble-dependent degree (\ref{eqref:BubDeg3})
\be
\delta_\bB(\G)=D+\sum_{\B\in\bB} \tilde a_\B n_\B(\G) - \Phi_{0} (\G),
\ee
which vanishes for maximal gluings\footnote{Or a subset of the maximal gluings in the most general case, see the discussion in Subsection~\ref{subsec:TreeLike} on tree-like families} and is positive for other configurations. If $\tilde a_\bB$ is an integer, (\ref{eqref:Cond1}) is satisfied, and to each positive integer corresponds an order of contribution (Def.~\ref{def:Order}). 
If furthermore there are infinitely many leading order configurations, condition (\ref{eqref:Cond2}) is also satisfied, and 
the generating function of (connected) maximal configurations (\ref{eqref:NExp2}) is expected to have a dominant singularity hyperspace. 
When there is only one counting parameter, such as for gluings of a single kind of building block, the asymptotic behavior of the 2-point function (the generating function $\GF^{(0)}_\Sb$ of rooted maximal gluings counted according to their number of $D$-simplices) is, at the singularity (\ref{eqref:CritExp1}), (\ref{eqref:CritExp2}),

\be
c_n^{(0)} \sim \lambda_c^{-n} n^{\gamma-2}\hspace{2cm} \GF^{(0)}_\Sb(\lambda) \sim (\lambda_c - \lambda)^{1-\gamma}.
\ee 
from which we can deduce  the critical exponent $\gamma$. Given a bubble which properties are unknown, we first study gluings of copies of this bubble, to characterize maximal maps and find $\tilde a_\B$. It is not straightforward to characterize maximal maps when there are more bubbles. Then, the generating function is multivariate, and the analysis can be more involved (Subsection~\ref{Subsec:D2CycBub}).\footnote{Also, as mentioned before, the coefficient $a_\B^\bB$ computed for $\bB$-restricted bubbles is not necessarily the same as the coefficient $a_\B$ computed for $\{\B\}$-restricted gluings.}
 At the singularity, the volume of the gluings is kept finite by rescaling the volume of the bubbles to zero, in which case we reach a continuum limit, which we would like to characterize, in the first place by its critical exponent $\gamma$, and then by its Hausdorff dimension, spectral dimension, etc. In the case were the critical exponent is that of trees, $1/2$, or that of planar maps, $-1/2$, we expect it to be a good hint that the continuum limit is the continuous random tree or the Brownian map. In this framework, the emergence of new critical behaviors would be an important step towards establishing a theory describing quantum gravity.
 
 Once the coefficient $a_\bB$ and the asymptotics of the generating function are known, which allow us to find the singularity and the critical exponent, we are interested in the {\it topology of maximal configurations}. This is done using theorems in Sec.~\ref{subsec:Cryst}.
 
In this analysis we only obtained information on the continuum limit of maximal configurations. The next step would be to study how the higher order contributions behave at the singularity (if they have no smaller dominant singularity). It is then possible to compensate their negligible dependence in $N$ with their behavior at the singularity, leading to a double scaling limit \cite{DartoisDoubleScaling, 
DoubleScal}, but this goes beyond the scope of this work.


	\subsection{Colored random tensor model}
	\label{subsec:ColTenMod}
In this subsection, we introduce briefly the colored random tensor model, which was introduced as a non-perturbative approach to quantum gravity and an analytical tool to study random geometries in dimension three and higher. For more details, we refer the reader to the review \cite{Review} or to the recent book \cite{Book}, which is an exhaustive introduction to the subject. 
In dimension $D$, the colored tensor model has a partition function
	\bea
	\cZ(\lambda, \bar\lambda,N)&=& \int \prod_{i=1}^D dT^{(i)}d\bar T^{(i)} e^{-S(\{T^{(i)},\bar T^{(i)}\})},\\
\nonumber	S(\{T^{(i)},\bar T^{(i)}\}) &=& \sum_{i=1}^D T^{(i)}.\bar T^{(i)} + \frac1 {N^{\frac{D(D-1)}4}} \biggl(\lambda \Tr_D(\{T^{(i)}\})+\bar\lambda \Tr_D(\{\bar T^{(i)}\})\biggr),
	\eea
	where $T^{(i)}$ is a rank $D$ tensor with $N^D$ complex entries indexed by $\{a_0,\cdots, a_D\}\setminus a_i$, and where we have denoted
	\be
	T^{(i)}.\bar T^{(i)}=\sum_{\{a_0,\cdots, a_D\}\setminus a_i=1}^N T^{(i)}_{\{a_0,\cdots, a_D\}\setminus a_i}\bar T^{(i)}_{\{a_0,\cdots, a_D\}\setminus a_i}, \\
	\ee
	and $\Tr_D(\{T^{(i)}\})$ is the generalized ``trace" pictured below in the case $D=4$.
	\begin{figure}[h!]
	\centering
	\includegraphics[scale=0.8]{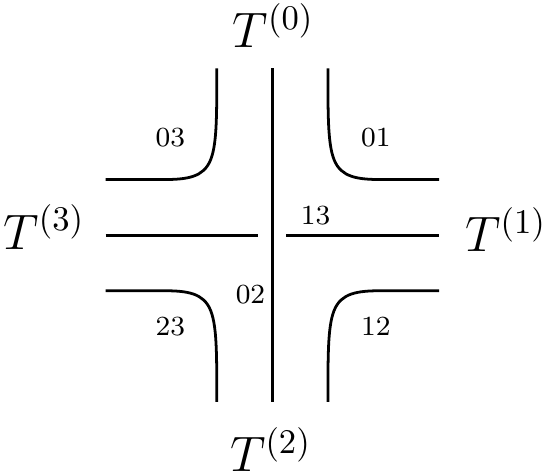} \hspace{2cm}\raisebox{8.5ex}{\includegraphics[scale=0.8]{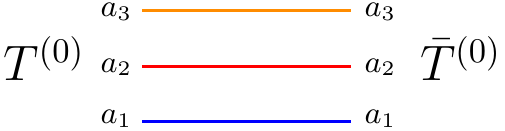} }
\caption{Interaction of the $D=4$ colored tensor model. }
\label{fig:ColTenInt}
	\end{figure}
	Representing interactions of the type $\Tr_D(\{T^{(i)}\})$ by white vertices,  $\Tr_D(\{\bar T^{(i)}\})$ by black vertices, and a propagator of the type $T^{(i)}.\bar T^{(i)}$ by an edge of color $i$, the perturbating expansion is labeled by Feynman graphs which are the non-necessarily connected graphs which connected components are the edge-colored bipartite graphs in $\bG_D$, described in Def.~\ref{def:cG}.	The perturbative expansion of the free energy, 
	\be
	e^\cF(\lambda,\bar\lambda,N)=\cZ(\lambda, \bar\lambda,N)
	\ee
is indexed by \emph{connected} graphs,  in $\bG_D$. 
The amplitude of a graph $\G\in\bG_D$ is proportional to
\be
\cA(\G)\sim g^{V(\G)/2}N^{\Phi(\G) - \frac{D(D-1)}4 V(\G)},
\ee
up to a symmetry factor, $\Phi(\G)$ being the score of $\G$ \eqref{eqref:score}. It rewrites in terms of Gurau's degree (Def.~\ref{def:Deg}),
\be
\label{eqref:AmpCol}
\cA(\G)\sim g^{V(\G)/2}N^{D-\deltaG(\G)}.
\ee	
From Thm.~\ref{thm:Melo}, the graphs which have the strongest scaling in $N$ are the melonic ones, for which 
\be
\cA_{\rm melo}(\G)\sim g^{V(\G)/2}N^{D}.
\ee	
The free energy admits a $1/N$ expansion \eqref{eqref:NExp2}.
In the large $N$ limit (leading order), 
\be 
\cF_{LO}(g,N)=\lim_{N\rightarrow \infty} \frac 1 {N^D} \cF(\lambda,N).
\ee 
only melonic graphs survive. $\cF_{LO}$ is the generating function of melonic graphs, counted according to their number of white vertices (Subsection~\ref{subsec:Melonic}). In this thesis, we are mostly interested in gluing bigger building blocks, the bubbles of Subsection~\ref{subsec:GluBub}. This requires the introduction of uncolored (enhanced) tensor models.


	\subsection{Uncolored random tensor models}
	\label{subsec:Uncolored}

While graphs in $\bG_D$ label the Feynman expansion of colored tensor models, $(D+1)$-edge-colored graphs in $\bG(\bB)$ are generated by the uncolored tensor model with interactions in~$\bB$ \cite{Uncoloring}. We denote 
\be
T \cdot \bar T = \sum_{a_1, \dotsc, a_D} T_{a_1\dotsb a_D} \bar T _{a_1\dotsb a_D},
\ee
and consider the partition function 
\begin{equation}
\label{eqref:ArbTensZ}
\cZ_\bB(\{\lambda_\B\},N) = \exp\,\cF = \int \exp \left( - N^{D-1} \sum_{B\in\bB} N^{s_\B} \lambda_\B \Tr_\B(T, \overline{ T})\right)\, d\mu_0(T, \overline{T}),
\end{equation}
where $d\mu_0( T, \overline{T})$ is the Gaussian measure,
\be
\label{eqref:Gaussianmeasure}
d\mu_0( T, \overline{T})= \exp \left(- N^{D-1} {T}\cdot \overline{T}\right)\ \prod_{a_1, \dotsc, a_D} \frac{dT_{a_1 \dotsb a_D} d\overline{T}_{a_1 \dotsb a_D}}{2i \pi},
\ee
and we require $\Tr_\B( T, \overline{ T})$ to be a $U(N)^{\otimes N}$ invariant, 
\bea
T_{a_1,\cdots, a_D}&\rightarrow&\sum_{b_1,\cdots, b_D} U^{(1)}_{a_1,b_1}\cdots U^{(D)}_{a_D,b_D}T_{b_1,\cdots, b_D}, \\
 \bar T_{a_1,\cdots, a_D}&\rightarrow&\sum_{b_1,\cdots, b_D} \bar U^{(1)}_{a_1,b_1}\cdots \bar U^{(D)}_{a_D,b_D}\bar T_{b_1,\cdots, b_D}.
\eea
 The polynomials $\Tr_\B( T, \overline{ T})$ satisfying these conditions are such that the $i^\textrm{th}$ index of a tensor $T$ is summed with the $i^\textrm{th}$ index of a conjugate tensor $\bar T$.  If we represent the tensors $T$ as white vertices, the tensors $\bar T$ as black vertices, and the summation of the $i^\textrm{th}$ index between a $T$ and a $\bar T$ as an edge of color $i$ between the corresponding black and white vertices, the polynomials $\Tr_\B( T, \overline{ T})$ are encoded into bubbles, which are elements of $\bG_{D-1}$. In the tensor model context, a bubble therefore labels an invariant interaction of the action. For the example of the complete bipartite $K_{3,3}$ bubble in the middle of  Fig.~\ref{fig:Ex2D}, 
 \be
 \includegraphics[scale=1]{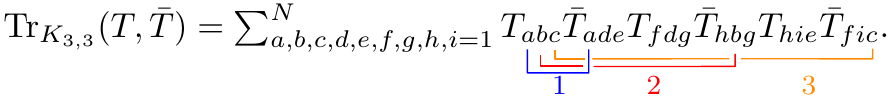}
 \ee
 
We perform a Feynman expansion of $\cZ$. First expand each $\exp(-N^{D-1+s_{\B}} \lambda_\B \Tr_\B(T, \bar T))	$ as a series in $\lambda_\B$ and (illegally) commute the sums with the integral. We are then left with Gaussian moments. By Wick's theorem, they are evaluated as sums over all pairings between $T$'s and $\overline{T}$'s. Those pairings are represented by new edges to which the color 0 is assigned. Therefore, the Feynman expansion of the free energy $\cF = \ln \cZ$ is labeled by the connected $(D+1)$-colored graphs of $\bG(\bB)$, whose bubbles are from the set $\bB$. The amplitude of a Feynman graph as a factor $N^{s_\B}$ per bubble, a factor  $N$ per color-$0i$ cycle, and a factor $N^{-(D-1)}$ per color-0 edge. There are $\sum_{\B\in\bB} \frac{n_\B(\G) V(\B)} 2$ color-0 edges in the graph, so that 
\be
\label{eqref:PertFreeEn}
\cF_\bB(\{\lambda_\B\},N) = \sum_{\G\in\bG(\bB)} c(\G) 
\prod_{\B\in\bB}(-\lambda_\B)^{n_\B(\G)} 
N^{\Phi_0(\G) + \sum_{\B\in\bB} n_\B(\G) \bigl(s_\B + (D-1)(1 - \frac {V(\B)} 2) \bigr)},
\ee
where $c(\G)$ is a symmetry factor. We see that the perturbative expansion of the free energy is well defined if 
\be
D-\delta_\bB(\G)=\Phi_0(\G) + \sum_{\B\in\bB} n_\B(\G) \bigl( s_\B + (D-1)(1 - \frac {V(\B)} 2) \bigr)
\ee
is bounded from above, and the coefficient $s_\B$, which we call \emph{the scaling}, has to been chosen accordingly, if it is possible.\footnote{As explained before for $a_\B$, we should a priori differentiate $s_\B$ computed for $\{\B\}$-restricted gluings and $s_\B^\bB$, computed for $\bB$-restricted gluings.} We recognize the non-negativity of condition~\eqref{eqref:Cond1}, and if we further require \eqref{eqref:Cond2}, we see that $\delta_\bB$ is the bubble-dependent degree (Def.~\ref{def:BubDepDeg}), and we  identify
 \be
 \label{eqref:SFromTildeA}
\tilde a_\B = (D-1)\bigl(\frac {V(\B)} 2 - 1\bigr) - s_\B.
\ee   
If $\tilde a_B\in\bN$, the bubble-dependent degree is a positive or vanishing integer, and the free-energy admits the $1/N$ expansion \eqref{eqref:NExp2}
\be
\frac1{N^D}\cF_\bB\bigl(\{\lambda_\B\},N\bigr)=\cF^{(0)}_\bB\bigl(\{\lambda_\B\}\bigr) + \frac 1 N \cF^{(1)}_\bB\bigl(\{\lambda_\B\}\bigr) + \frac1{N^2}\cF^{(2)}_\bB\bigl(\{\lambda_\B\}\bigr)+ \cdots
\ee
where $\cF^{(i)}_\bB(\{\lambda_\B\},N)$, the free energy of order $i$ contributions, is the {\it generating function} of $\bB$-restricted gluings of bubble-dependent degree $i$, counted with respect to their number of bubbles. 
For melonic bubbles, from \eqref{eqref:TildeAMelo} we know that we should choose $s_\B=0$. From \eqref{eqref:BoundAvsBoundTildeA}, we see that the condition $a<\frac{D(D-1)}4$ for non-melonic bubbles can be translated for $s_\B$
\be
\label{eqref:BoundAvsBoundS}
a_\B<\frac{D(D-1)}4 \qquad \Leftrightarrow\qquad s_\B > - \deltaG(\B).
\ee
The random tensor model obtained choosing the value $s_\B = - \deltaG(\B)$ for the bubbles was originally called the \emph{uncolored tensor model} \cite{Universality, Uncoloring}. The degree is then Gurau's degree, and from what we explained (see \eqref{eqref:Bounda} and preceeding paragraph), there are no leading order contributions of Feynman graphs in $\bG_D$ containing non-melonic bubbles when deleting all color-0 edges, and therefore the leading order comprises solely the melonic contributions of Subsec.~\ref{subsec:Melonic}. We will see in Corollary~\ref{coroll:scaling}, that under certain assumptions, which are satisfied for  all known bipartite examples in dimension $D<6$, $s_\B$ has to be at least 1 for non-melonic colored graphs to contribute to the leading order. For the $K_4$ bubble, $s=1/2$. It is expected that $s_\B$ should always be positive.
The random tensor model obtained choosing the coefficients $\tilde a_\B$ satisfying Conditions~\eqref{eqref:Cond1} and \eqref{eqref:Cond2} is called an \emph{enhanced tensor model} \cite{Enhanced, SigmaReview, SWM, GluOcta, Johannes}\footnote{We discuss in Subsection \ref{subsec:TreeLike} the case where the degree is rational and non-negative.}.
 The degree is then the bubble-dependent degree of Def.~\ref{def:BubDepDeg}, and there are contributions of non-melonic graphs to the leading order.

The generating function of the cumulants is 
\be
\ln \cZ_{B}[\{\lambda_\B\},N;J, \bar J]
=
\ln\int e^{-{\lambda}{N^s} \Tr_\B(T, \bar T)-\bar{J}.T-\bar{T}.J}\frac{d\mu_0(T, \overline{T})}{\cZ_\bB(\{\lambda_\B\},N)}.
\ee
Observables are also polynomials associated to bubbles, and their expectation is 
\begin{equation}
<\Tr_\B( T, \overline{ T})> = \int \Tr_\B( T, \overline{ T}) \exp\left( -\sum_{B\in\bB} N^{s_\B} \lambda_\B \Tr_\B(T, \overline{ T})\right)\, \frac{d\mu_0(T, \overline{T})}{\cZ_\bB(\{\lambda_\B\},N)}.
\end{equation}
Their Feynman expansion is a sum over graphs in $\bG(\bB)$ with a distinguished bubble $\B$, which need not be in $\bB$. We may also be interested in the amplitude of transition between a certain number of  particular triangulated $(D-1)$-pseudo-manifolds
\be
<\Tr_{\B_1}( T, \overline{ T})\Tr_{\B_1}( T, \overline{ T})\cdots >,
\ee
 whose perturbative expansion is a sum over connected graphs in $\bG(\bB)$ with distinguished bubbles $\B_1, \B_2, \cdots$.
The Feynman expansion of the 2-point function, 
\be
\GF^\bB_2(\{\lambda_\B\},N)=<T.\bar T>
\ee
is labeled by colored graphs with a distinguished elementary melon (Fig.~\ref{fig:ElMel}). Contracting that melon, it goes back to distinguishing a particular color-0 edge. The  amplitude of a Feynman graph is therefore as in  \eqref{eqref:PertFreeEn}, without  the  symmetry factor $c(\G)$, as the automorphism group of a rooted graph is trivial. In practice, we therefore compute the leading order 2-point function, in order to determine the critical exponent $\gamma$ from the critical behavior near the dominant singularity.

In this thesis we consider a certain number of examples for $\bB$, for which we compute the scalings $s$ and the critical exponents $\gamma$. The results are summarized in Section~\ref{sec:Summary}.

	\subsection{The Sachdev-Ye-Kitaev model and real tensors}
	\label{subsec:SYK}

	\subsubsection{The SYK model}
	
	Sachdev and Ye proposed a toy-model of $N$ spins with Gaussian-random, infinite-range exchange interactions, of order $D$, in $0+1$ dimensions \cite{sy}. This quantum mechanics model attracted a certain interest within the condensed matter community.  A simple variant of the Sachdev-Ye model was proposed by Kitaev in a series of seminars \cite{kitaev}. The Sachdev-Ye-Kitaev (SYK) model is a quantum mechanical model of $N$ Majorana fermions $\psi_i$ ($i=1,\ldots, N$) living in $0+1$ dimensions with random interactions of order $D$, $D$ being an even integer\footnote{$D$ is usually denoted $q$ in the SYK literature.}. The action writes:
\be
\label{act:syk}
S_{\mathrm{SYK}}=\int d\tau \left( 
\frac 12 \sum_{i=1}^N \psi_i \frac{d} {dt} \psi_i
- \frac{i^{D/2}}{D!}
\sum_{i_1,\ldots, i_D=1}^N
j_{i_1\ldots i_D}\psi_{i_1}\ldots \psi_{i_D}
\right).
\ee
Kitaev proposed this model as a model of holography. The most widely studied version of the SYK model has real fermions, as above. As an example of Feynman graphs obtained by perturbative expansion, we show on the left of Fig.~\ref{fig:melonSYK} a melonic graph, which is a dominant graph in the large $N$ expansion of the SYK model. The dashed lines represent the quenched disorder.
\begin{figure}
\centering
\includegraphics[scale=0.9]{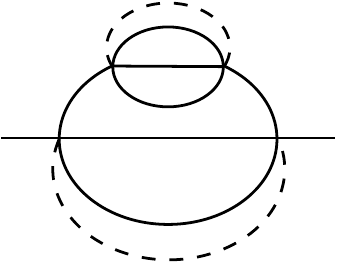}\hspace{2cm} \includegraphics[scale=0.9]{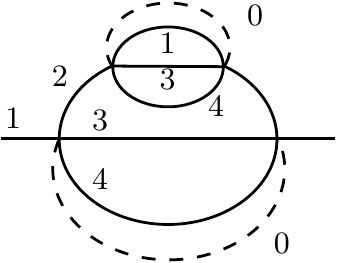}
\caption{\label{fig:melonSYK}Melonic graphs of the SYK and colored SYK models}
\end{figure}

The SYK model has three remarkable features \cite{kitaev, PR, MS}: it is solvable at strong coupling, 
maximally chaotic and, finally, it presents emergent conformal symmetry in the infrared limit. The SYK model is the first model having all these three properties (other known models only have some of these properties, but not all three of them).  The emergent conformal symmetry, here a reparametrization invariance,  is spontaneously broken, and the resulting Goldstone bosons are described by a universal Schwarzian action, which also describes theories of gravity in AdS$_2$. The fact that the system is maximally chaotic, i.e. that the Lyapunov exponent saturates the bound of \cite{BoundChaos}, is a consequence of these Goldstone bosons.  These properties suggest that the holographic dual of the SYK model describes black holes (maximally chaotic objects) in the context of the (near) $\text{AdS}_2\text{/CFT}_1$ holography. The strong coupling solvability, which is a rare and therefore precious property, makes it a toy model to study the quantum properties of black holes.

This attracted a lot of interest within the high energy physics community. Thus, Maldacena and Stanford studied in detail the two- and four-point functions of the model \cite{MS}, Polchinski and Rosenhaus solved the Schwinger-Dyson equation and computed the spectrum of two-particle states \cite{PR}, Fu{\it~et.~al.} proposed a supersymmetric version of the model \cite{Fu}, see also \cite{SYK26}. The bulk dual has been studied in \cite{Gross2, SYK20}, and so on \cite{SYK21, Gross, gurau-ultim, GurauSYK3}.

	\subsubsection{Colored SYK models}

In \cite{Gross}, Gross and Rosenhaus proposed a generalization of the SYK model, in which they have included $f$ flavors (which we rather refer to as colors) of fermions, each occupying $N_a$ sites and appearing with a $q_a$ order in the interaction. We consider a particular case of their proposal where each flavor appears only once in the interaction ($q_a = 1$). It corresponds to a colored SYK model. The coloring makes it possible to use tools developed in the random tensor model literature to study the diagrammatics of the model.  A complex version was first mentioned in \cite{gurau-ultim},  we studied the diagrammatics of the sub-leading orders in \cite{BLT},  and the emergent conformality of the next-to-leading order 2-point function is shown in \cite{StephSYK}. The action writes:
\begin{equation}
\label{act:noi}
S=\int d\tau \left( 
\frac 12 \sum_{f=1}^D\sum_{i=1}^N \psi_i^f \frac{d} {dt} \psi_i^f
- \frac{i^{D/2}}{D!}\sum_{i_1,\ldots, i_D=1}^N
j_{i_1\ldots i_D}\psi_{i_1}^{1}\ldots \psi_{i_D}^{D}
\right).
\end{equation}
Note that we use superscripts to denote the flavor. Moreover, to ease notations, we now work with $D\cdot N$ fermions - we have $N$ fermions of a given flavor.

The Feynman graphs obtained through perturbative expansion of the action \eqref{act:noi} are the edge-colored graphs where the colors are the flavors. At each vertex, each of the $D$ fermionic fields which interact has one of the $D$ flavors, and each flavor is present exactly once. Therefore, at fixed couplings $j_{i_1 \dotsc i_D}$, the Feynman graphs are the $D$-regular non-necessarily bipartite  edge-colored graphs in $\tilde\bG_{D-1}$ (Def.~\ref{def:cG}).
In order to study the $1/N$ expansion, one must average over the disorder with the covariance
\begin{equation}
\label{eqref:Averaging}
\langle j_{i_1 \dotsc i_D} j_{l_1 \dotsc l_D} \rangle \sim \frac1{N^{D-1}} \prod_{k=1}^D \delta_{i_k, l_k}.
\end{equation}
Each graph is thus turned into a sum over Wick pairings which can be represented with edges carrying a new color, say the color $0$. Color-0 edges encode the disorder and are represented as dashed edges.
An example of such a Feynman graph is given on the right of Fig.~\ref{fig:melonSYK}.  Therefore, the Feynman graphs of the real colored SYK model are the elements of $\tilde \bG_D$ which are connected when deleting all the color-0 edges. They correspond to coverings (Def.~\ref{def:Pairing}) of bubbles $\tilde\B^\Om$, where $\tilde\B$ spans $\tilde\bG_{D-1}$.

\

The complex version \cite{gurau-ultim} is obtained by considering the propagation from $\psi$ to $\bar \psi$ and by considering the interacting term in \eqref{act:noi} as well as its complex conjugate:
\begin{equation}
\label{act:noir}
\int d\tau \left( 
\frac 12 \sum_{f=1}^D\sum_{i=1}^N \bar \psi_i^f \frac{d} {dt} \psi_i^f
- \frac{i^{D/2}}{D!}\sum_{i_1,\ldots, i_D=1}^N
j_{i_1\ldots i_D}\psi_{i_1}^{1}\ldots \psi_{i_D}^{D}
- \frac{(-i)^{D/2}}{D!}\sum_{i_1,\ldots, i_D=1}^N
\bar j_{i_1\ldots i_D}\bar \psi_{i_1}^{1}\ldots \bar\psi_{i_D}^{D}
\right).
\end{equation}
The Feynman graphs obtained through perturbative expansion of the complex action have the same structure as the one explained above for the real model \eqref{act:noi}. However, in the complex case, one has two types of vertices, which we can refer to as white and black. Each edge connects a white to a black vertex. The Feynman graphs of \eqref{act:noir} are thus the subset of the Feynman graphs of \eqref{act:noi} which are bipartite. The averaging over the disorder is done throughout the covariance $\langle j_{i_1 \dotsc i_D} \bar{j}_{l_1 \dotsc l_D} \rangle$. The Feynman graphs of the complex colored SYK model are the elements of $ \bG_D$ which are connected when deleting all the color-0 edges. They correspond to coverings (Def.~\ref{def:Pairing})  of bubbles $\B^\Om$, where $\B$ spans $\bG_{D-1}$.

	\subsubsection{$1/N$ expansions}

A Feynman graph $\G=\B^\Om$ has a free sum for each bicolored cycle $0i$ in the graph, which contributes to the amplitude of the graph with a factor $N$. Each color-0 edge contributes with a factor $N^{-(D-1)}$ \eqref{eqref:Averaging}. The amplitude of a Feynman graph $\G=\B^\Om$ is therefore proportional to 
\begin{equation}
\label{SYKdeg}
\cA_N(\G) = N^{\Phi_0(\G) - (D-1)E_0(\G)} A(\G),
\end{equation}
where $\Phi_0$ is the 0-score (Def.~\ref{def:Score0}), and $E_0$ is the number of color-0 (disorder) edges
\be
E_0(\B^\Om)= \frac{V(\B)}2.
\ee
As the graph $\G$ is a covering $\B^\Om$, $b(\G)=1$ and we can rewrite
\be
\Phi_0(\G) - (D-1)E_0(\G) = \Phi_0(\G) - (D-1)(\frac{V(\B)}2 - 1)b(\G) - D + 1 = 1-\delta_0(\G),
\ee
where we recognize a degree $\delta_0$ obtained by choosing the coefficients $\tilde a_\B$ \eqref{eqref:DeltaTildeA} as
\be
\tilde a_\B = (D-1)(\frac{V(\B)}2 - 1), 
\ee
for all bubbles $\B\in\bG_{D-1}$. From \eqref{eqref:SFromTildeA}, this is equivalent to choosing the scaling 
\be
s_\B=0
\ee
 for all bubbles, and we remark that \emph{in zero dimension}, the free energy of the theory would just correspond to 
\be
\label{eqref:ColSYKFreeEn}
\cF(N) = \int \frac{dTd\bar T}{\pi^{N^D}} e^{-N^{D-1}T.\bar T} \sum_{\tilde \B\in\tilde \bG_{D-1}} N^{D-1}\Tr_\B(T,\bar T) = \sum_{\tilde\B\in\tilde\bG_{D-1}} \sum_{\Om(\tilde\B)}N^{1- \delta_0\bigl(\tilde\B^{\Om(\tilde\B)}\bigr)},
\ee
where we used the notations of Subsection~\ref{subsec:Uncolored}. In the case of the one-dimensional SYK model under consideration, the free-energy admits a similar expansion 
\be
\label{eqref:ColSYKFreeEn2}
\cF(N) = \log \cZ (N)  =\sum_{\tilde\B\in\tilde\bG_{D-1}} \sum_{\Om(\tilde\B)}N^{1- \delta_0\bigl(\tilde\B^{\Om(\tilde\B)}\bigr)} A(\tilde\B^\Om),
\ee
in which the $A(\tilde\B^\Om)$ are the amplitudes of the right hand side of \eqref{SYKdeg}. This can be written as the $1/N$ expansion 
\be
\label{eqref:ColSYKFreeEn3}
\cF(N) = \sum_{l\ge 0}N^{1-l} \cF^{(l)},
\ee
where $\cF^{(l)}$ is the free-energy of order $l$ contributions, i.e. of coverings which have degree $\delta_0(\B^\Om)=l$. 

\

{\it The two-point function} $<\psi^c_{i_c}(\tau)\psi^{c'}_{i_{c'}}(\tau')> $ has a perturbative expansion over graphs with one distinguished color $c$ half-edge and one distinguished color $c'$ half-edge. Because of the coloring properties of the graph, we see that this is only possible if $c=c'$, in which case Feynman graphs are coverings $\B_{(c)}^{\Om(\B_{(c)})}$, where the graph $\B_{(c)}\in \tilde\bG^1_{D-1}$ is a bubble with one missing color-$c$ edge (Def.~\ref{def:GraphWithBound}).  In fact, we write
\be
\label{eqref:ColSYK2PT}
<\psi^c_{i_c}(\tau)\psi^{c'}_{i_{c'}}(\tau')>_c = \delta_{c,c'}\delta_{i_c, i_{c'}}G_2(\tau,\tau'),
\ee
where $G_2$ is the normalized 2-point function
\be
\label{eqref:2PtSYK1}
G_2(\tau,\tau') = \frac1{N}\frac{\int d\psi \bigl(\sum_{i_c=1}^N \psi^c_{i_c}(\tau)\psi^{c}_{i_{c}}(\tau')\bigr) e^{-S}}{\int d\psi e^{-S}},
\ee
 $S$ being the action \eqref{act:noi}.
%
The normalized 2-point function admits the $1/N$ expansion
\be
\label{eqref:2PtSYK2}
G_2(\tau,\tau') = \frac 1 N \sum_{\tilde\B_1\in\tilde\bG_{D-1}} \sum_{\Om(\tilde\B)}N^{1- \delta_0\bigl(\tilde\B_1^{\Om(\tilde\B)}\bigr)} A_{\lambda}(\tilde\B_1^{\Om(\tilde\B_1)}),
\ee
where the sum is over coverings with one marked color-$c\neq 0$ edge. Graphs contributing to the normalized 2-point functions are obtained from those contributing to the 2-point function $<\psi^c_{i_c}(\tau)\psi^{c}_{i_{c}}(\tau')>_c$ by gluing the two pending color-$c$ half-edges. This creates precisely one bicolored cycle, thus the factor $1/N$ in \eqref{eqref:2PtSYK1} and \eqref{eqref:2PtSYK2}. 

\

We see from similar arguments, that {\it the $2n$-point functions}
\be
\label{eqref:ColSYK2nPT}
<\psi^{c_1}_{i_{c_1}}(\tau_1)\cdots\psi^{c_{2n}}_{i_{c_{2n}}}(\tau_{2n})> 
\ee
vanish unless there exists a fixed-point free involution $\sigma\in\cS_{2n}$ (a permutation which is the product of disjoint transpositions) such that 
\be
\label{eqref:ColSYK2nPT2}
<\psi^{c_1}_{i_{c_1}}(\tau_1)\cdots\psi^{c_{2n}}_{i_{c_{2n}}}(\tau_{2n})>  = \prod_{c=1}^{2n} \delta_{c, \sigma(c)} G^\sigma_{2n} (\tau_1,\cdots, \tau_n),
\ee
i.e. if the $c_1,\cdots,c_{2n}$ are equal two-by-two. For instance, in the case of the 4-point function, non-vanishing terms have $\{c_1,c_2,c_3,c_4\}=\{i,i,j,j\}$ where $i\neq j$, or $\{c_1,c_2,c_3,c_4\}=\{i,i,i,i\}$ (the sets are unordered). The $2n$-point function has a perturbative expansion over graphs $\tilde\G_n$ which are bubbles with $n$ missing colored edges. We will see in Section~\ref{sec:UniGraphColSYK} how to characterize the contributions to the $1/N$-expansion of the $2n$-point functions at a given order.


\subsubsection{SYK-like models without disorder and tensor models}

	

Recently, Witten proposed a reformulation of the SYK model using real fermionic tensor fields without quenched disorder \cite{Witten}. This comes from the fact that both the SYK model and random tensor models have the same class of dominant graphs in the large $N$ limit ($N$ being, in the tensor model framework, the size of the tensor). They are the melonic graphs described in Subsection~\ref{subsec:Melonic}.  In \cite{GurSyk}, Gurau complemented Witten's results with some modern results of random tensor theory (albeit using a complex version of \cite{Witten}). He gave a classification of the Feynman graphs of the model at all orders in the $1/N$ expansion for the free energy and the 2-point function, based on the Gurau-Schaeffer classification of colored graphs \cite{GurauSchaeffer}. Notice that this classification remains somewhat formal, in the sense that it does not give the graphs which contribute at a given order of the $1/N$ expansion. This SYK-like model is commonly referred to as the Gurau-Witten model. The action of the Gurau-Witten model is
\be
\label{actGW}
S=\int dt \left( 
\frac{\imath}{2}\sum_{f=1}^4 \psi^f\frac{d}{dt}  \psi^f
+\frac \lambda {N^{\frac {D(D-1)} 4}} \psi^1\psi^2\psi^3\cdots \psi^q
\right),
\ee
where we used $q=D+1$ as $D$ is commonly used for random tensor models. Note that the $q=D+1$ fields above $\psi^{f}$, $f=1,\ldots, q$ are now rank $q-1$ tensor fields.
The notation $ \psi^1\psi^2\psi^3\psi^4$ stands for the interaction $ \psi^1\psi^2\psi^3\psi^4$ - note that the contraction of tensor indices leads to different types of interaction terms.
As described in Subsection~\ref{subsec:ColTenMod}, the Feynman graphs of the model are the colored graphs in $\bG_{D}=\bG_{q-1}$. The Gurau-Witten model has been the subject of many publications, among which \cite{SYK22, SYK23, SYK27,  SYK25}.
In \cite{BLT} we compare the diagrams contributing to the sub-leading orders of the colored SYK model and the real colored random tensor model. The emergent conformality of the next-to-leading order 2-point function is shown in \cite{StephSYK}.

Other kinds of SYK-like tensor models have been studied in the literature. Uncolored SYK-like tensor models have for instance been studied in \cite{Kleba}, with the interaction corresponding to the $K_4$ complete graph shown on the right of Fig.~\ref{fig:Ex2D}. This corresponds to an enhanced model, with scaling $s=\frac 1 2$. This uncolored enhanced tensor model had first been studied in \cite{Tan}, together with an order 4 melonic interaction. Some other publications studying this models are \cite{SYK28, SYK27}. A multi-orientable SYK-like tensor model has been introduced in \cite{StephSYK}.

\chapter{Bijective methods}
\label{chap:BijMeth}

As mentioned in the introduction, the aim of this chapter is to develop bijections which would enable a systematic characterization of the discrete spaces obtained by gluing bubbles according of their mean curvature. In particular, as detailed in Section~\ref{sec:QG}, we are interested in identifying and counting the spaces which maximize the number of $(D-2)$-cells at fixed number of $D$-cells. This is summarized in the guideline in Subsection~\ref{subsec:Guideline}. Our criteria for the bijection is therefore that it should keep track of the number of $(D-2)$-cells, and map configurations to recognizable combinatorial families. We come to the general bijection step-by-step, by first looking at a simple case, then applying Tutte's bijection, and then generalizing to any kind of colored graph.

	\section{A first bijection in the simpler case of cyclic bubbles}
	\label{sec:SimplerBij}
	
	We first focus on the simplest case of gluings of a single specific kind of bubble $\bB=\{\B_{k,p}\}$. In dimension $D$ and for $0<k\le\ED$, we define a $k$-cyclic bubble
	of size $2p$ as a bubble $\B_{k,p}$ with $2p$ vertices which alternates $k$ edges 
	and $D-k$ edges as shown in Fig.~\ref{fig:kCycle}. It is determined by fixing $k$ colors. 
	\begin{figure}[h!]
	\centering
	\includegraphics[scale=0.65]{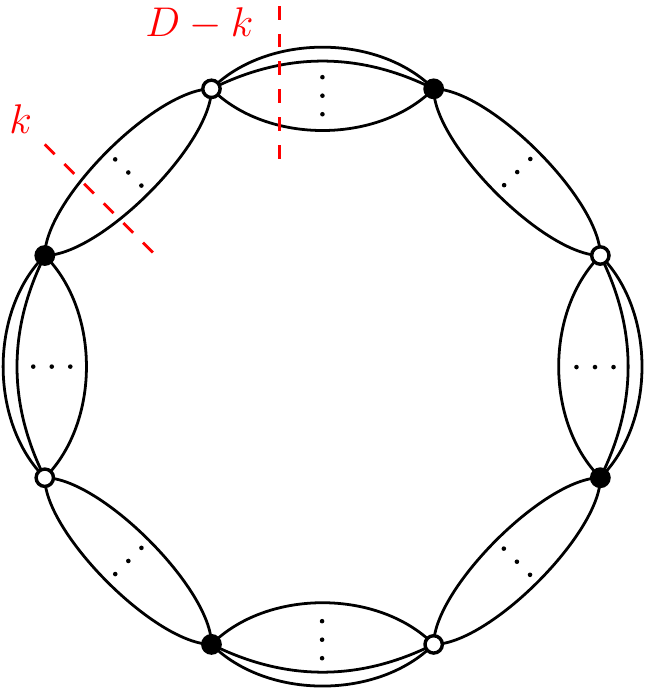}
		\caption{A $k$-cyclic bubble of length 8.}
\label{fig:kCycle}
	\end{figure}
In dimension~2, all bubbles are $1$-cyclic, and a $1$-cyclic bubble of size $2p$ is the colored graph dual to the boundary of a $2p$-gon (with bipartite boundary). As said before, the gluing of two $2p$-gons is done so that the coloring of vertices matches, and therefore the resulting $2p$-angulations  are bipartite. The dual of a $2p$-angulation is a $2p$-valent map. Vertices are mapped to faces and vice-versa, and edges are mapped to edges. From the colored graph representation, the map dual to the $2p$-angulation is intuitively obtained by ``collapsing" each bicolored cycle to a vertex (i.e. contracting all the edges of the cycle), leaving only the color-0 edges. There is an ordering of color-0 edges around the new collapsed vertex, which is that of their appearance around the cycle. 
%
%
This is shown on the left of Fig.~\ref{fig:QuartGlu2D}.
	%
	\begin{figure}[h!]
	\centering
	\raisebox{0ex}{\includegraphics[scale=0.45]{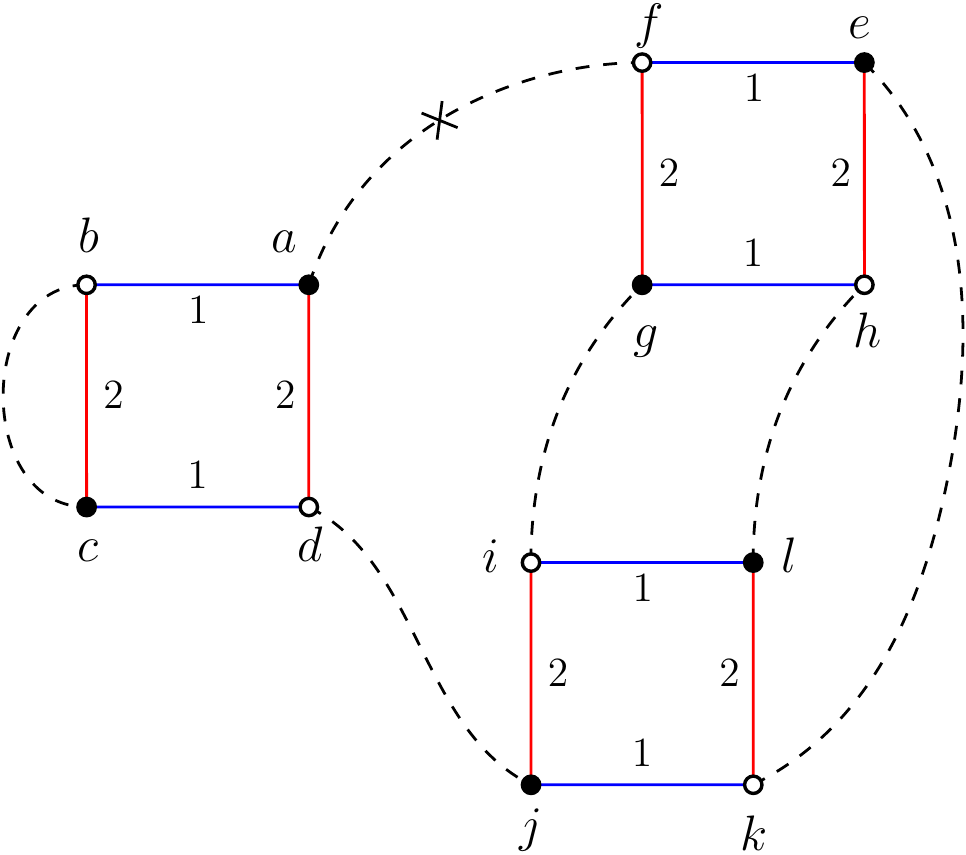}}
	\hspace{0.5cm}\raisebox{10.5ex}{$\leftrightarrow$}\hspace{0.3cm}
	\raisebox{5ex}{\includegraphics[scale=0.8]{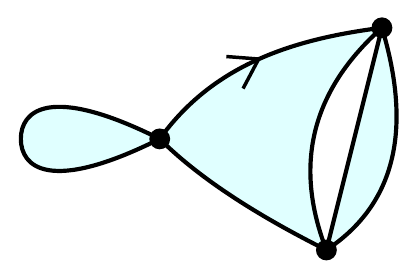}}
	\hspace{0.2cm}\raisebox{10.5ex}{$\leftrightarrow$}\hspace{0.5cm}\includegraphics[scale=0.45]{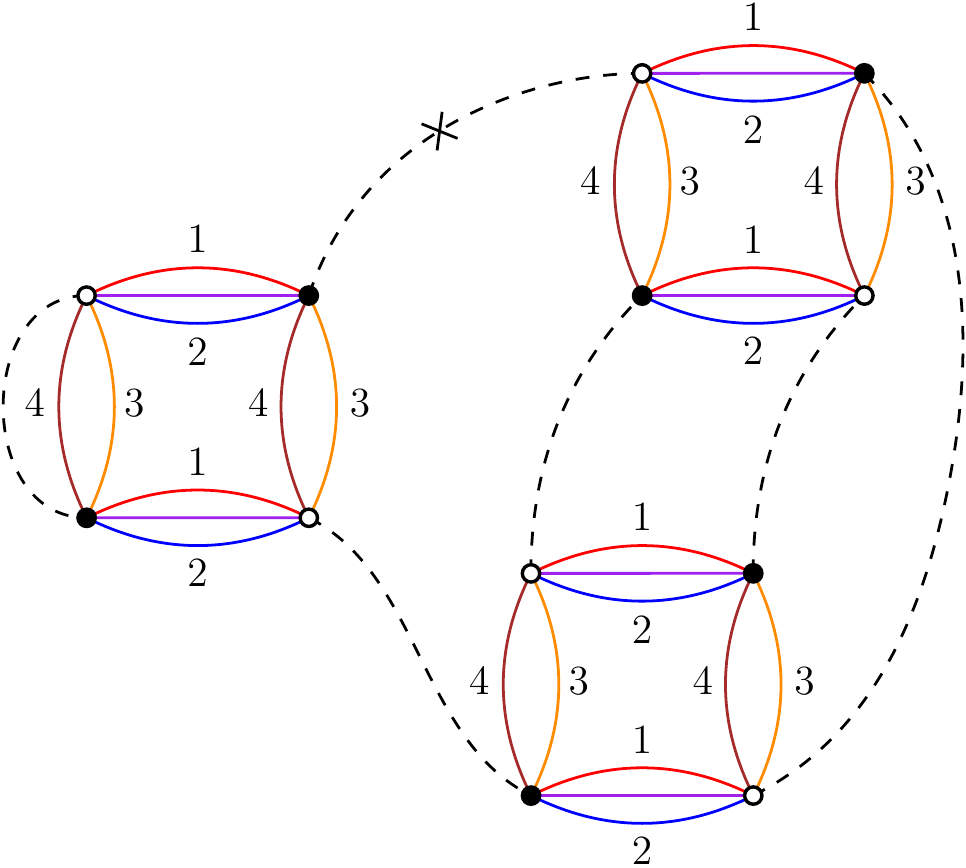}
	\caption{The resulting map $M^*$ does not depend on the cyclic bubble (and on the dimension). }
\label{fig:QuartGlu2D}
	\end{figure}

\begin{prop}
\label{prop:CycBubTopo}
Cyclic bubbles have a spherical topology.
\end{prop}
\prf Consider a $k$-cyclic bubble alternating edges with colors $(i_1,\cdots, i_k)$ and edges with colors 
$(i_k+1, \cdots, i_D)$. For any two permutations $\nu_k\in\cS_k$ and $\nu_{D-k}\in\cS_{D-k}$,
choosing $(i_{\nu_k(1)},\cdots, i_{\nu_k(k)} ; i_{\nu_{D-k}(k+1)}, \cdots, i_{\nu_{D-k}(D)})$ to orient the half-edges around white vertices and $(i_{\nu_k(k)},\cdots, i_{\nu_k(1)} ; i_{\nu_{D-k}(D)}, \cdots, i_{\nu_{D-k}(k+1)})$ to orient the half-edges around black vertices, one gets a planar regular embedding (jacket), and we conclude from Proposition~\ref{prop:PlanJack}. \qed

	%
	
\begin{prop}
There is a bijection between 3-colored graphs $\G$ with one marked color-0 edge and combinatorial maps $M^*$ with bicolored faces and one oriented edge. Bubbles of size $2p$ are mapped to $2p$-valent vertices.
\end{prop}
\prf Of course this follows directly from the duality between colored graphs and colored triangulations, and from the duality between the colored $2p$-angulations and the colored $2p$-valent maps. However, when going to higher dimension,  it is straightforward to use the construction we will detail here.
%
Starting from the $3$-colored dual graph $\G$, 
we define the set of darts of $M^*$ as the set $\cV$ of the vertices of $\G$. Restricting to colors 1 and 2 in $\G$, we are left with a collection of bipartite cycles that alternate edges of colors 1 and 2 (1-cyclic bubbles). We decide to orient them e.g  from black to white on color 1 edges. This naturally organizes the vertices of $\G$ into disjoint sets which are cyclically ordered. We define those sets to be the disjoint cycles of the permutation $\sigma$. 
The edges of color 0 of $\G$ are a set of disjoint unordered pairs 
on $\cV$, which we define to be the disjoint transpositions of $\alpha$. The marked color-0 edge in $\G$ is naturally oriented, e.g. from black to white, which translates into an orientation of the corresponding edge in $M^*$. For the example of Fig.~\ref{fig:QuartGlu2D}, the permutations are
\bea
\sigma&=&(abcd)(efgh)(ijkl)\nonumber\\
\alpha&=&\overrightarrow{(af)}(bc)(dj)(ek)(gi)(hl).
\eea

If we embed the colored graph $\G$ so that the ordering we chose for cycles of colors 12 corresponds to a counter-clockwise ordering and so that color-0 edges are always on the exterior of the cycle, a bicolored cycle 01 arrives and leaves a color-0 edge on  the same side of that edge (this can be seen on  Figure~\ref{fig:QuartGlu2D}). Therefore,
half of the faces of $M^*$ are one-to-one with the bicolored cycles 01
of $\G$, and the other half with the bicolored cycles 02. More precisely, bicolored cycles 01 only share color-0 edges with bicolored cycles 02, so that the corresponding faces in $M^*$ can be colored with colors 1 and 2 in a bipartite way (we recover the fact that it is the dual of a bipartite $2p$-angulation).

On the other way, we start from a combinatorial map $M^*$ with one oriented edge and bipartite faces of colors 1 and 2 and construct the colored graph $\G$. The orientation of the marked edge translates into an orientation of every edge in the map, such that all the edges around a given face have the same orientation. This orientation is opposite for faces of different colors, and around each vertex, incident edges are alternatively in-going and out-going. We define the set of black (resp. white) vertices of $\G$ as the set of outgoing (resp. ingoing) half-edges of $M^*$, and the set of edges of color 0 of $\G$ as the set of disjoint transpositions of $\alpha(M^*)$. Going counter-clockwise around vertices of $M^*$, we encounter either corners (pairs of half-edges) going from an outgoing edge to a ingoing edge or the opposite. We define
 the set of edges of color 1 (resp. 2) of $\G$ as the set of corners of $M^*$
 that start on an outgoing (resp. ingoing) half-edge. 
 
 This describes a bijection between rooted colored graphs with labeled vertices and rooted labeled combinatorial maps. For rooted objects the automorphism groups are trivial, and the bijection extends to unlabeled graphs and maps.
In the non-rooted case, the correspondence we described gives a bijection between colored graphs with labeled vertices and non-rooted labeled combinatorial maps. Since the symmetries of the cyclic bubble are the same as the symmetries of the embedded vertex in a face-bipartite map, the equivalence classes upon relabeling the vertices on one side and the half-edges on the other side coincide.   \qed
 
 \

In higher even dimension, gluings of $k$-cyclic bubbles of size $2p$ are also in bijection with face-bipartite $2p$-valent maps. This bijection is a generalized duality, in which bubbles are mapped to $2p$-valent vertices,  and color-0 facets (or color-0 edges in the colored graph) are mapped to edges. 
\begin{prop}
There is a bijection between $D$-dimensional gluings of $k$-cyclic bubbles - of any size but with the same alternating colors $i_1,\cdots i_{k}$ and  $i_{k+1}, \cdots, i_D$ - and face-bipartite combinatorial maps.\end{prop}
%
%
Restricted gluings of such $k$-cyclic bubbles of size $2p$ with $2p\in \Pc=\{2p_1, \cdots 2p_P\}$ are in bijection with combinatorial maps with allowed vertex valencies $\Pc=\{2p_1, \cdots 2p_P\}$. Gluings of such $k$-cyclic bubbles without restricting the sizes of the bubbles are in bijection with Eulerian combinatorial maps (maps with vertices of even valencies). 
As before, the graph is obtained by collapsing the bubble to a vertex, and the construction in terms of permutations is exactly the same. In fact, one gets the exact same $2p$-valent map from a 3-colored graph and from the $(D+1)$-colored graph obtained by replacing every 12 bicolored cycle with a $k$-cyclic bubble alternating edges of colors $1,i_2,\cdots i_{k}$ and edges of colors $i_{k+1}, \cdots, i_D$ (this is illustrated in Figure~\ref{fig:QuartGlu2D}).
The map has bicolored faces: one can color the faces of the resulting $2p$-valent map $M^*$ with two colors 1 and 2 so that an edge is always incident to two different colors (its dual map is bipartite). However, now, faces of $M^*$ of e.g. color 1 correspond to the bicolored cycles of colors $(0i_1), \cdots ,(0i_k)$, and faces of $M^*$ of color 2 correspond to the  remaining bicolored cycles $(0i_{k+1}), \cdots , (0i_{D-k})$.
The 0-score (Def.~\ref{def:Score0}) of a colored graph $\G$ dual to a gluing of $k$-cyclic bubbles with the same alternating colors $i_1,\cdots i_k$ and  $i_{k+1}, \cdots, i_D$ can therefore be expressed with respect to the faces of the map $M^*$ as follows
\be
\Phi_0(\G)=kF_1(M^*)+(D-k)F_2(M^*)= (2k-D)F_1(M^*)+(D-k)F(M^*)
\ee
in which we have denoted $F_1$ and $F_2$ the number of faces of type 1 and 2 of $M^*$ and $F=F_1+F_2$ the total number of faces of $M^*$. We know that for a generic map,
\be
F=2-2g+E-V\le 2+E-V,
\ee 
with equality if and only if the map is planar. Because the vertices have allowed valencies  in $\Pc=\{2p_1, \cdots 2p_P\}$, we have the following relation
\be
2E(M^*)=\sum_{2p\in \Pc} 2p V_{p}(M^*),
\ee
in which $V_{p}$ is the number of $k$-cyclic bubbles of size $2p$ (or valency of $2p$ vertices in $M^*$), so that 
\be
F(M^*)\le 2+\sum_{2p\in \Pc} (p -1)V_{p}(M^*).
\ee 
There are two cases :

\begin{itemize}
\item If $2k-D = 0$, $\Phi$ is maximal if and only if the corresponding $M^*$ is planar, in which case 
\be
\Phi_0(\G_{\textrm{max}})=D + \frac D 2\sum_{2p\in \Pc} (p -1)V_{p}(M^*),
\ee
which is the linear bound (\ref{eqref:MaximalBound}) we were looking for. The bubble-dependent degree of the dual colored graph is therefore defined as 
\be
\delta_{p,D/2} (\G)= D + \frac D 2\sum_{2p\in \Pc} (p -1)n_{p}(\G)-\Phi_0(\G), 
\ee
where $n_{p}$ is the number of $k$-cyclic bubbles in $\G$ of size $2p$. We deduce the coefficients 
\be
\label{eqref:CoeffsD2Cyc}
\tilde a_{p,D/2}=\frac {D(p-1)} 2,\quad \textrm{and}\quad s_{p,D/2}= (p-1)(\frac D 2 -1)
\ee
from \eqref{eqref:SFromTildeA}. For $k$-cyclic bubbles of size $2p$, we compute the score
\be
\label{eqref:kCycleFaces}
\Phi(\B_{k,p})=p\frac{D(D-1)}2 - k(p-1)(D-k),
\ee
from which we deduce the correction to Gurau's degree (Def.~\ref{def:Deg}) using \eqref{eqref:Tildeaa},
\be
a_{p,D/2}= \frac{D(D-1)}4 - \frac{(p-1)}{2p}\frac{D(D-2)}{4} <  \frac{D(D-1)}4,
\ee
when $p>1$, as expected for a non-melonic bubble.
The counting of their dual planar bipartite $2p$-angulations, or more generally of planar bipartite degree-restricted maps was done in \cite{Ben}. Planar bipartite maps have faces of even sizes. Bender and Canfield showed that the number $c_n^\Pc$ of bipartite maps with $n$ edges, such that the faces have allowed degrees $\Pc=\{2p_1,\cdots, 2p_P\}$ ($P$ may be infinite) behaves asymptotically as
\be
\label{eqref:NumPlanBip}
c_n^\Pc \sim \alpha(\Pc)n^{-5/2}\lambda_c^{-n},
\ee
where $\alpha$ and $\lambda_c$ are given explicitly in terms of the roots of $2=\sum_{i=1}^P (i-1)\binom{2i} i X^i$, and therefore $\gamma=-1/2$. In particular, the number of bipartite planar maps with no restriction on the degree of the faces is 
\be
\frac 1 {2(n+1)(n+2)}\binom {2n}n 2^n.
\ee
It is also possible to count the contributions to higher orders, as done in \cite{Chap4}. As a consequence of Theorem \ref{thm:LeGall}, the continuum limit of maximal maps is the Brownian map.

\item If $2k-D < 0$, the maximal value of $\Phi_0$ is obtained when  $F_1(M^*)=1$ and $g(M^*)=0$, in which case, again,
\be
\Phi_0(\G_{\textrm{max}}) =  D   +(D-k)(E(M^*)-V(M^*)).
\ee
As the dual of $M^*$ is bipartite and $F_1(M^*)=1$ it is easy to see that such maps are in bijection with trees. Indeed, every face of color 2 is a ``cut-face" (see Fig.~\ref{fig:TutteTree}).
The same calculation leads to a bubble-dependent degree
\be
\delta_{\Pc,k} (\G)= D + (D-k)\sum_{2p\in \Pc} (p -1)n_{p}(\G)-\Phi_0(\G), 
\ee
and to the coefficients
\be
\label{eqref:ACycles}
\tilde a_{p,k}=(D-k)(p-1),\ a_{p,k}= \frac{D(D-1)}4 - \frac{(p-1)(D-k)(k-1)}{2p},\ s_{p,k}= (p-1)(k-1).
\ee
We denote $\GF_{k_1, \cdots ,k_P}$ the generating function of gluings of $k$-cyclic bubbles of sizes $k\in \{k_1, \cdots k_P\}$ that have the same alternating colors $i_1,\cdots i_{k}$ and  $i_{k+1}, \cdots,i_D$, that have one marked color-0 facet, counted with respect to their number of edges. The marked facet corresponds to  a distinguished color-0 edge
(which is already oriented from black to white),
and therefore corresponds to distinguishing an oriented edge of $M^*$, or equivalently marking a corner. 
\be
\label{fig:TutteTree}
\begin{array}{c}
\includegraphics[scale=0.8]{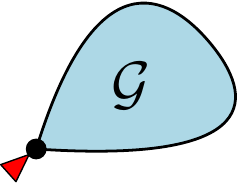}\end{array}\qquad =\qquad \raisebox{-0.5ex}{\includegraphics[scale=0.9]{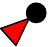}}\qquad +\quad\sum_{k\ge 1}\quad\begin{array}{c} \includegraphics[scale=0.6]{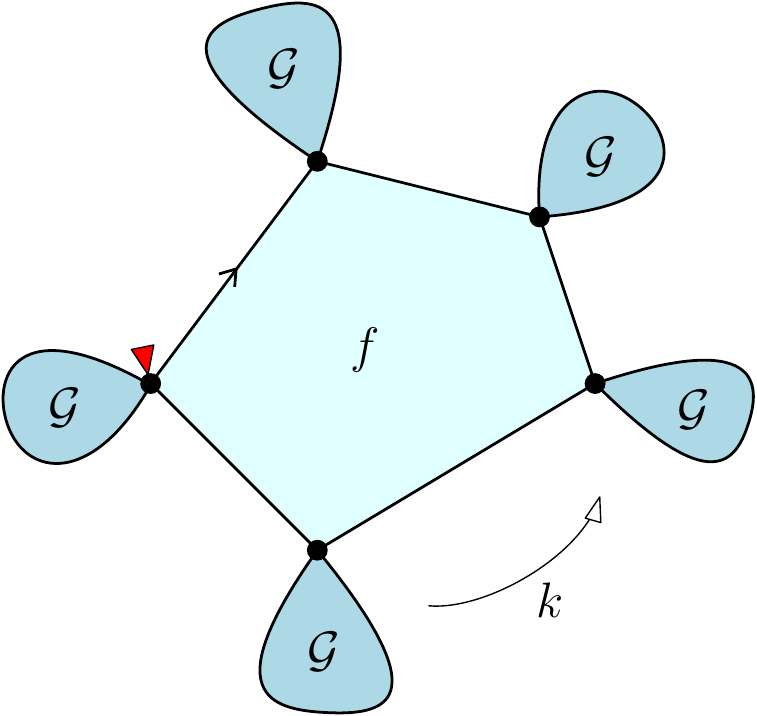}
\end{array}
\ee
The generating function satisfies the well known tree equation,

\be
\GF_{k_1, \cdots ,k_P}(z)=1+ \sum_{k\ge 1 } z^k \GF^k_{k_1, \cdots, k_P}(z),
\ee
or equivalently
\be
\GF_{k_1, \cdots, k_P}(z)=1+z\GF^2_{k_1, \cdots, k_P}(z).
\ee
The tree structure is rendered explicit in the following section. This is done by taking the dual and applying Tutte's bijection. The critical exponent is that of plane trees, $\gamma=1/2$, and the continuum limit is expected to be the continuum random tree (CRT) \cite{Aldous}, also called branched polymers.
\end{itemize}

\

The construction would still hold for gluings of cyclic bubbles alternating different sets of colors, however, in that case, it would no longer allow one to count easily the bicolored cycles of the original colored graph. More precisely, while some bicolored cycles would indeed be mapped to faces of the combinatorial map $M^*$, some others would be mapped to faces with twist factors on certain edges (Fig.~\ref{fig:LOMaps}). 

\

A first conclusion of this section is that unlike in the two dimensional case, for which planar degree-restricted combinatorial maps all fall in the same universality class regardless of the choice of polygons of even length, higher dimensional gluings of polytopes with spherical colored-triangulated boundary may fall in at least two different universality classes, that of plane trees and that of planar maps.

\section{Tutte's bijection and cyclic bubbles}
\label{sec:Tutte}

\subsection{Tutte's bijection}
\label{subsec:Tutte}

Tutte's bijection \cite{TutteBij} maps  bipartite quadrangulations to combinatorial maps without valency restriction. From quadrangulations to maps, a choice of keeping either white or black vertices has to be made, e.g. black vertices here. An edge is then added between the pair of black corners in every face. The white vertices and the edges of the original quadrangulation  are then deleted. It therefore maps faces to edges between black vertices. A planar example is shown in Fig.~\ref{fig:Tutte}. From maps to quadrangulations, a white vertex is added in each face, and edges are added between this vertex and every corner of the corresponding face. The ordering of the corners around the face is opposite to the ordering of edges around the corresponding vertex. 
	\begin{figure}[h!]
	\centering
	\includegraphics[scale=1]{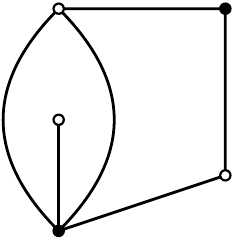}\hspace{2cm}\includegraphics[scale=1]{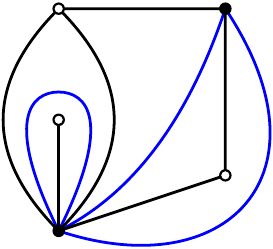}\hspace{2cm}\includegraphics[scale=1]{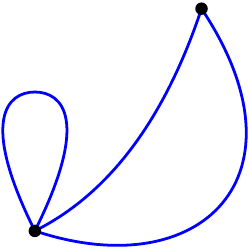}
\caption{Tutte's bijection between bipartite quadrangulations and generic maps. }
\label{fig:Tutte}
	\end{figure}
	%

	
\subsubsection*{Generalization to generic  bipartite maps}

As bipartite graphs have only cycles of even length, the faces of a bipartite map are of even degree. As before, a choice is made of either black or white vertices (e.g. black). 
The bijection now maps bipartite maps (with black and white vertices) to hyper-maps. 
To explain this bijection, we choose to represent hyper-maps as bipartite maps (with two kinds of vertices, white squares and black disks).
Such objects are the Walsh representation of hyper-maps, in which hyper-edges are replaced with ``stars", as shown in Fig.~\ref{fig:Walsh}, or detailed in \cite{Walsh}.
	\begin{figure}[h!]
	\centering
	\includegraphics[scale=0.7]{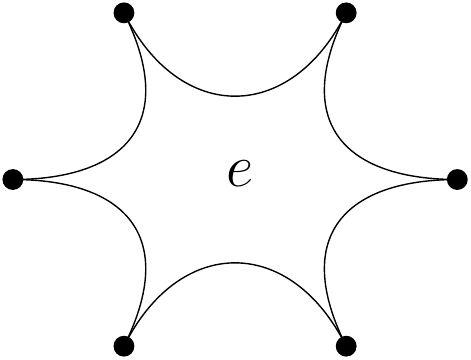}\hspace{2.5cm}\includegraphics[scale=0.7]{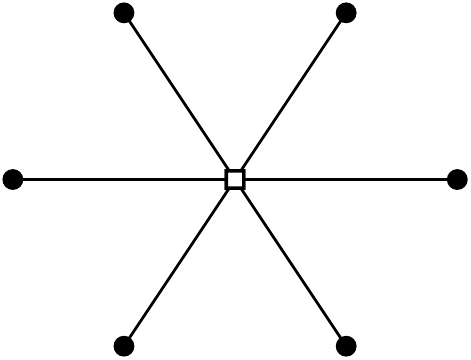}
\caption{Walsh's representation of an hyper-edge. }
\label{fig:Walsh}
	\end{figure}
 The bijection is now quite similar in both ways, and is a particular kind of duality for bipartite maps. A planar example is shown in Fig.~\ref{fig:GenTutte}. Starting from a bipartite map, a square vertex is added in each face, and edges are added between the new vertex and every corner in the corresponding face, so that their ordering around the new vertex is opposite to the ordering of corners around the face. The initial edges and the white vertices are then deleted. Each initial face of length $2p$ has been replaced with a square vertex of valency $p$, and the number of nearest neighbors of black vertices is the number of incident faces in the original map. 
 Similarly, in the other way white vertices are added in each face and edges are added that reach every corner on black vertices. 
	\begin{figure}[h!]
	\centering
	\includegraphics[scale=1]{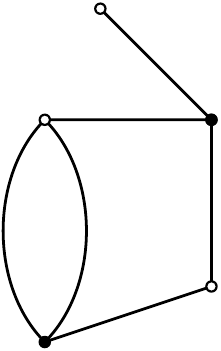}\hspace{2cm}\includegraphics[scale=1]{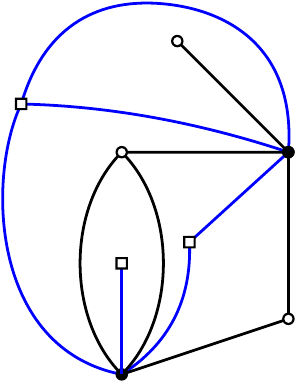}\hspace{2cm}\includegraphics[scale=1]{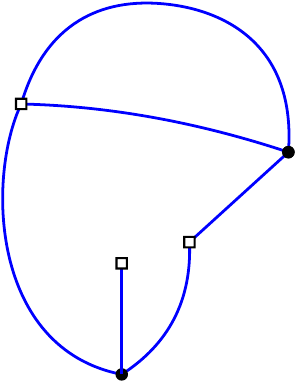}
\caption{Tutte's bijection generalized for bipartite maps. }
\label{fig:GenTutte}
	\end{figure}
In Walsh's representation, a generic bipartite map is therefore mapped to a unique other bipartite map. Moreover, any face-degree restriction is now translated into a degree restriction on the possible valencies of square vertices, so bipartite $2p$-angulations are mapped to bipartite maps such that square vertices are of valency $p$.

\subsection{Cyclic bubbles again}
\label{sec:CycBub2}

Applying this generalized Tutte bijection on the bipartite dual\footnote{As  Eric Fusy suggested, it would actually be more natural to apply the version of Tutte's bijection between bicolored maps and bipartite maps, instead of taking the dual first and applying the version of the bijection between bipartite maps and bipartite maps.} of the map $M^*$ described in Section~\ref{sec:SimplerBij}, we know that there is a bijection between gluings of a single kind of cyclic bubbles and bipartite maps $\Ga$ in which bubbles become white square vertices and black disks encode the way the bubbles are glued together. If the allowed sizes of $k$-cyclic bubbles are $\{2p_1,\cdots,2p_P\}$, then  the allowed valencies of white squares are $\{p_1,\cdots,p_P\}$, which indicates that there is one edge in $\Ga$ per white vertex of the original colored graph $\G$. Depending on the choice of black or white vertices in Tutte's bijection, there are either $D-k$ bicolored cycles per black disk and $k$ bicolored cycles per face of the resulting bipartite map $\Ga$, or the opposite. We know that if $k<D/2$, a maximal  $M^*$ has only one face  corresponding to the $k$ bicolored cycles, so that if we do the right choice in Tutte's bijection, the maximal maps $\Ga$ are planar maps with a single face, i.e. trees. 
In this section, we will describe this bijection between gluings of cyclic bubbles and bipartite maps. Importantly, this bijection holds when gluing cyclic bubbles with different colorings, and will even be generalized to gluings of any kind of bubbles in Section~\ref{sec:StackedMaps}.

\begin{theorem}
\label{thm:BijCycles}
There is a bijection between gluings $\G$ of $k$-cyclic bubbles of allowed length in $\{2p_1,\cdots,2p_P\}$ 
and bipartite combinatorial maps $\Ga$, such that white square vertices have allowed valencies in $\{p_1,\cdots,p_P\}$ and black disk vertices have no degree restriction.
The edges in $\Ga$ are labeled with subsets of $\lDr$ and all edges incident to the same white vertex share the same color set. The bicolored cycles $0i$ in the colored graph $\G$ are mapped to the faces of the combinatorial maps obtained from $\Ga$ by keeping all the black vertices and only the edges which color sets contain color $i$. 
Distinguishing edges of $\G$ corresponds to marking corners on black vertices of $\Ga$.
\end{theorem}

	\begin{figure}[h!]
	\centering
	\includegraphics[scale=0.55]{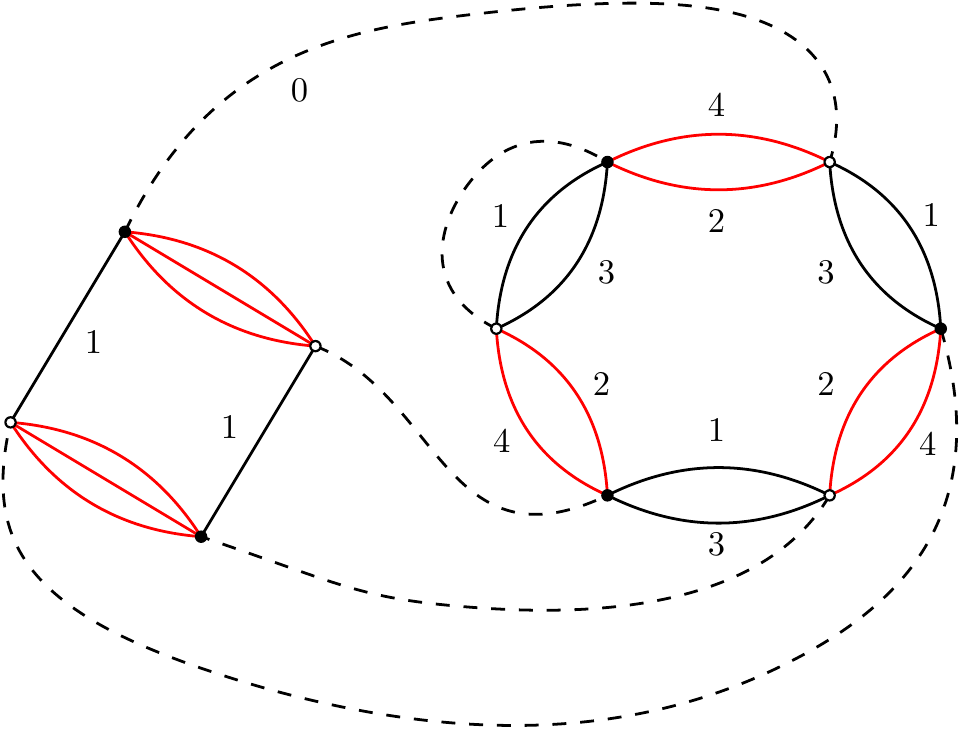}\qquad\ \raisebox{12ex}{$\leftrightarrow$}\qquad\ 
	\raisebox{7ex}{\includegraphics[scale=0.9]{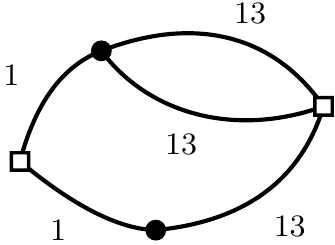}}
	\hspace{0.5cm}
	\caption{Bijection between gluings of cycle bubbles and bipartite maps. }
\label{fig:ExBijCyc}
	\end{figure}

\prf Given sets of integers $\{p_1,\cdots,p_P\}$,  $\{k_1,\cdots,k_P\}$ we consider the corresponding sets of all the $k_i$-cyclic bubbles of length $2p_i$. 
If $k<D/2$, we choose to pair together the black and white vertices that share $D-k$ edges. If $k=D/2$, we choose to pair the black and white vertices which do not share an edge of color 1. As a vertex belongs to a single bubble, these choices translate into  a pairing $\Om_\G$ (Def.~\ref{def:Pairing}) of the vertices of any gluing of cyclic bubbles. We consider such a graph $\G$, we label the resulting pairs from 1 to $V/2$ and we construct the combinatorial map $\Ga$. 
	\begin{figure}[h!]
	\centering
	\includegraphics[scale=0.6]{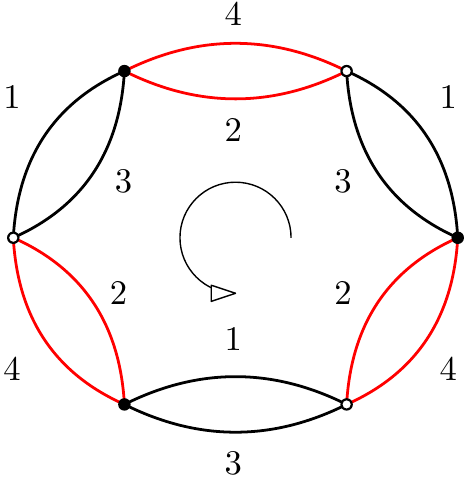}\quad\ \raisebox{8.5ex}{$\leftrightarrow$}\quad\ \raisebox{3ex}{\includegraphics[scale=0.8]{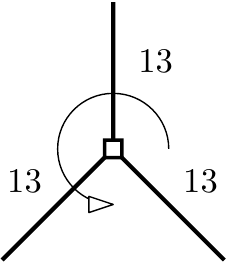}}
	\hspace{1.7cm}
	\raisebox{2ex}{\includegraphics[scale=0.6]{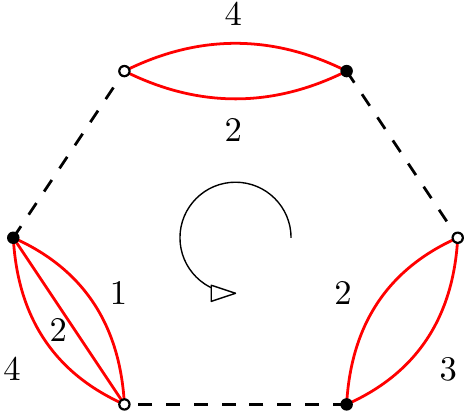}}\quad\ \raisebox{8.5ex}{$\leftrightarrow$}\quad\ \raisebox{3ex}{\includegraphics[scale=0.8]{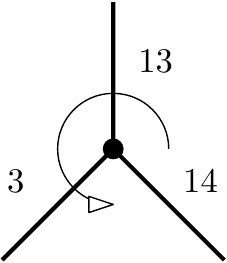}}
	\end{figure}

We first define the set of darts as $\cD=\cD_\diamond\sqcup\cD_\bullet$, where the $i^{\rm th}$ element of both $\cD_\diamond$ and $\cD_\bullet$ corresponds to the $i^{\rm th}$ pair. Each bubble is a cycle which is naturally oriented (e.g. from white to black on the edges between paired vertices in $\Om_\G$, as pictured above), so that to each bubble corresponds a disjoint cyclically ordered set of darts of  $\cD_\diamond$. The disjoint union of these cycles therefore defines a permutation $\sigma_\diamond$ on $\cD_\diamond$, corresponding to white square vertices. 

We now consider the cycles which alternate pairs in $\Om_\G$ and color-0 edges. This time, we choose to orient them from black to white on the  pairs of $\Om_\G$, i.e. from white to black on color-0 edges, so that each cycle defines a disjoint cyclically ordered set of darts from $\cD_\bullet$, and their disjoint union defines a permutation $\sigma_\bullet$ on $\cD_\bullet$ which corresponds to black vertices. 

Each pair corresponds to a unique dart in both  $\cD_\diamond$ and $\cD_\bullet$. The pair of those two elements defines an edge, and the disjoint union of the corresponding transpositions defines the permutation $\alpha$ corresponding to the edges.

Each edge $e$ of $\Ga$ therefore corresponds to a given pair $\pi$ of $\G$ and we color $e$ with the set of all the colors of the edges incident to $\pi$  which do not link its two vertices. All the edges incident to a given white vertex have the same color set. 
Color-0 edges of the original colored graph $\G$ are now corners around black vertices of the bipartite map $\Ga$. An edge is incident to four corners. It has two corners on one side, and two corners on the other side.  Counterclockwise, the white extremity of the color-0 edge in the colored-graph picture is on the side of the edge visited before the corresponding corner in the combinatorial map picture, while its black extremity is on the side of the edge visited after the corner (see the figure below). 

Color-$i$ edges of $\G$ are either corners on white vertices of $\Ga$ (if $i$ belong to the set labeling the incident edges), either half-edges around black vertices (if $i$ does not belong to the set labeling this edge).
Therefore, by keeping only the edges which color set contains $i$, and by deleting all the isolated white square, we obtain a combinatorial map $\Gai$ whose faces correspond to bicolored cycles $0i$ of $\G$ (see the paragraph \emph{color-$i$ submap} below). 
\be
\label{eqref:FacesGa}
\Phi_{0,i}(G)=F(\Gai).
\ee

To go back to $\G$ from $\Ga$, we do the steps in the opposite order. Each step performed here is the inverse operation of the step performed when building $\Ga$ from $\G$, which ensures that this is a bijection. We replace each edge by a pair of black and white vertices, one for each side of the edge, with the convention pictured below. 
	\begin{figure}[h!]
	\centering
	\includegraphics[scale=0.7]{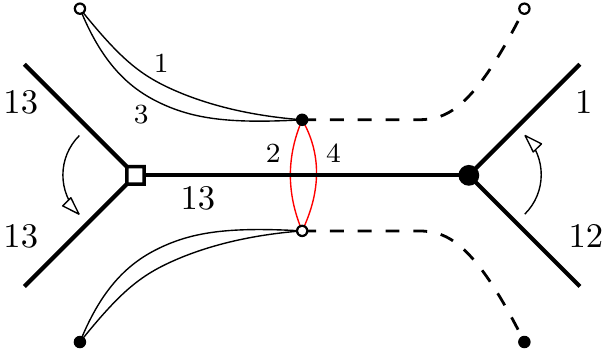}
	\end{figure}
	%
For each corner on a black vertex of $\Ga$ we draw a color-0 edge between the vertices corresponding to the two sides of the edges which are incident to the corner. For each corner on a white square vertex incident to edges with colors $\{i_1,\cdots,i_k\}$, we draw edges of colors $\{i_1,\cdots,i_k\}$ between the corresponding vertices. Then for each edge of $\Ga$,  we draw edges of the missing colors between the two corresponding vertices so that the resulting graph is $(D+1)$-edge-colored.  The two strands of each edge of $\Ga$ therefore correspond to the same pair, one is incident to a cyclic bubble, and the other to a cycle alternating color-0 edges and chosen pairs. 

We consider the equivalence classes upon relabeling the pairs. The cyclic bubbles and the cycles alternating color-0 edges and chosen pairs have the same symmetries, so we get a bijection between equivalence classes.
\qed

\

\emph{Because white squares are incident to edges that all have the same color set, we can specify that color set on the white square itself.}

\subsection*{Color-$i$ submap and score}

The color-$i$ submap $\Gai$ is obtained by keeping all the edges which color set contains color $i$ and deleting isolated square vertices. We refer to them as monochromatic submaps. 
%
Because of relation (\ref{eqref:FacesGa}), and since $\tilde a$ has been calculated before for cycles  (\ref{eqref:ACycles}), the bubble-dependent degree writes
\be
\delta_{\Pc,k} (\G)= D + \sum_{\substack{{2p\in \Pc}\\{k\ge1}}} (D-k)(p -1)V^k_{p,\diamond}(\Ga)-\sum_{i=1}^DF(\Gai), 
\ee
where $V^k_{p,\diamond}$ is the number of $p$-valent white squares whose incident edges carry $k<D/2$ colors. The critical behavior of the generating function of discrete spaces obtained by gluing $D/2$-cyclic bubbles without restriction on their coloring is treated in Section~\ref{Subsec:D2CycBub} using tools developed in the beginning of Chapter~\ref{chap:PropSWM}. We will see it involves three universality classes, characterized by the critical exponents $1/2$, $-1/2$, and $1/3$.

A map with $q$ marked corners on $q$ black vertices corresponds to a colored graph with $q$ marked color-0 edges. Deleting all the marked color-0 edges, we get a colored graph in $\bG^q_D$ with $2q$ degree-$D$ vertices.  The sub-faces on monochromatic submaps which start on a marked corner and end on a marked corner without encountering any other marked corner in the meantime are called \emph{broken faces}. In the colored graph picture,  they correspond to the bicolored paths which start on a degree-$D$ vertex and end on another degree-$D$ vertex. 

\subsection*{Boundary graph}
The boundary graph is obtained from a colored map $\Ga$ with marked corners  the following way.  Faces around the color-$i$  submap $\Gai$ are oriented clockwise (they visit corners counterclockwisely). Keep all the marked vertices of $\Ga$, and whenever a broken face starts on a marked vertex $v_a$ and ends on some marked vertex $v_b$, draw a directed line from $v_a$ to $v_b$. Unmark all the vertices, to obtain the Eulerian graph $\Ga_\circlearrowleft$ (there is no ordering of edges around vertices). The graph $\Ga_\circlearrowleft^{(i)}$ obtained by keeping only the edges of color $i$, is a collection of disjoint directed cycles with no isolated vertices. This implies that every vertex has two color-$i$ incident half-edges, one outgoing and one ingoing. They belong to the same edge if the cycle is of length 1. This is shown in Figure~\ref{fig:BoundMapCyc}.
	\begin{figure}[h!]
	\centering
	\includegraphics[scale=0.9]{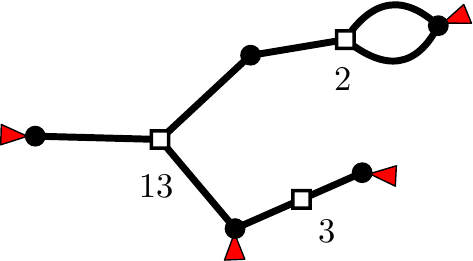}\qquad
	\includegraphics[scale=0.9]{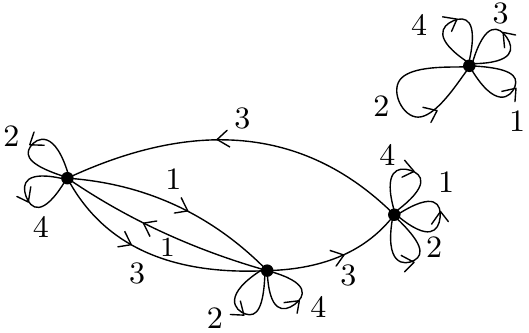}\qquad
	\includegraphics[scale=0.9]{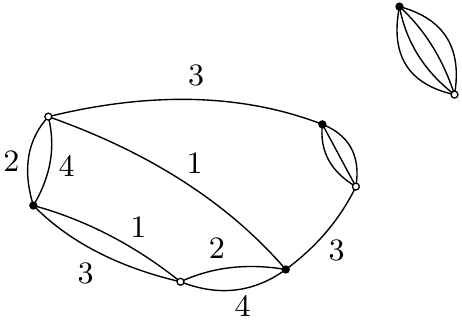}
	\caption{Boundary graph from the bipartite map. }
\label{fig:BoundMapCyc}
	\end{figure}

We now replace each vertex with a pair of black and white vertices and attach all the outgoing (resp.~ingoing) edges to the black (resp.~white) vertex. The corresponding graph $\partial \Ga$ is clearly in $\bG_{D-1}$.  Its pairs of vertices correspond to marked corners of $\Ga$, which in turn correspond to pairs of degree-$D$ vertices of the $(D+1)$ colored graph $\G$ in bijection with $\Ga$. 
Furthermore, $\partial \Ga$ has an edge of color $i$ between two vertices when a broken face joined two marked corners in $\Ga$. Because faces always leave a marked corner of $\Ga$  from a black vertex of $\G$ and arrive on a marked corner of $\Ga$ when they arrive to a white vertex of $\G$, $\partial\Ga$ coincides with the boundary graph of the colored graph in bijection with $\Ga$. 


Importantly, the boundary $\B=\partial\Ga$ naturally comes with a pairing of its vertices (Def.~\ref{def:Pairing}). We say this pairing of $\B$ is \emph{induced} by $\Ga$. Another important remark is that \emph{commuting two half-edges of different colors incident to a marked vertex does not change the boundary graph}, while commuting two half-edges of the same color may change the boundary graph.

\section{Bijection between colored graphs and stacked maps}
\label{sec:StackedMaps}

We have seen several bijections with combinatorial maps. The interest of studying maps instead of colored graphs is because bicolored cycles become faces. In the new representations, we can therefore use results on maps, such as counting formulas, bijections with known families, matrix models... We saw that $k$-cyclic bubbles were in bijection with bipartite maps with prescribed face degree on some vertices (Fig.~\ref{fig:ExBijCyc}). This is Walsh's representation \cite{Walsh} of hyper-maps. In this section, we generalize this bijection for any colored graph. The objects we obtain are superpositions of $D$ hyper-maps - or of bipartite maps in Walsh's representation. 
We have named such objects  {\it stuffed Walsh maps} in \cite{SWM}, ``stuffed" referring to the non-trivial internal structure of the vertices corresponding to hyper-edges.
We develop here a slightly different - but equivalent - point of view, in which the objects we consider are combinatorial maps with colored edges. We call such objects stacked maps.
%

\subsection*{Stacked maps}

Stacked maps have two kinds of vertices, vertices which carry a color $i\in\{0,1,\cdots,D\}$ and white vertices. They have edges which also carry a color $i\in\{0,1,\cdots,D\}$. Edges of color $i$ can only link a white vertex and a vertex of color $i$. Color-$i$ vertices are only incident to edges of color $i$, and white vertices are incident to one edge of each color in $\{0,\cdots,D\}$ and one only. 
There is no cyclic ordering of edges around white vertices (an equivalent choice would be to order the edges around white vertices according to their color).
At most one corner per black vertex may be marked.
We denote $\bS_D$ the set of connected stacked maps with edges colored in $\{0,\cdots,D\}$ and $\bS^q_D$ if $q$ corners are marked. Examples are shown on the right of Figs.~\ref{fig:BijBub} and~\ref{fig:BijSM}.
Before proving the bijection, we need three definitions.

\subsection*{Canonically adding marked edges in the boundary case}

The bijection in the previous section stood only for colored graphs without boundary. However, we described how deleting marked color-0 edges corresponding to marked corners of a map led to a colored graph with boundary. If we wish to extend the bijection to colored graphs with boundaries, starting from a graph in $\bG^q_D$, we need a prescription on how to add back these color-0 edges in order to apply the bijection for graphs without boundary but with distinguished edges. This is the subject of this paragraph. 
  
Consider a generic colored graph $\G_q\in\bG_D^q$  dual to a triangulation with boundary, and a pairing $\Om$ of its vertices. There is a canonical way of adding $q$ color-0 marked edges to obtain a graph $\Gcan\in\bG_D$ without boundary, but with $q$ marked edges. 
%
Indeed, we consider the $q$ paths which alternate pairs of vertices and color-0 edges, and which start and end on degree-$D$ vertices. A color-0 marked edge is then added between the two degree-$D$ vertices for each path, as illustrated in Fig.~\ref{fig:CanAdd}. We therefore obtain $q$ cycles alternating color-0 edges and pairs of $\Om$, each containing a single marked color-0 edge.

	\begin{figure}[h!]
	\centering
	\includegraphics[scale=0.7]{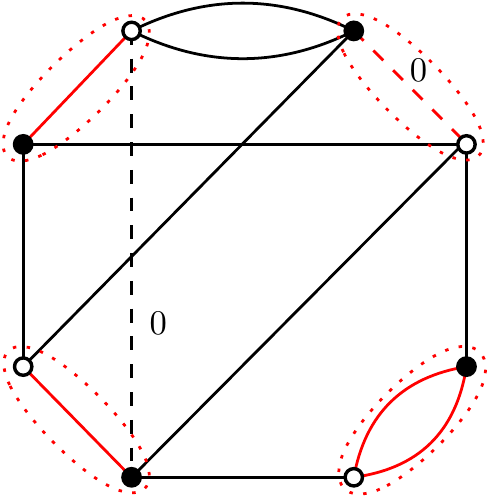}\hspace{2.5cm}
	\includegraphics[scale=0.7]{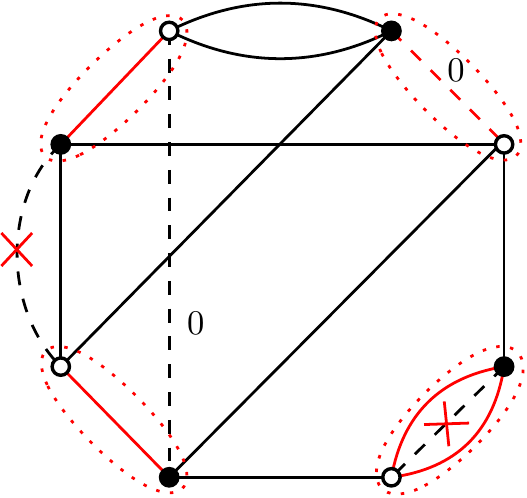}
	\caption{Canonically adding marked color-0 edges. }
\label{fig:CanAdd}
	\end{figure}

\begin{lemma}
\label{lemma:GqG0}
We consider $\G_q\in\bG_D^q$, a pairing $\Om$ and the corresponding $\Gcan$ obtained by canonically adding marked color-0 edges. The $q$ marked edges induce a pairing $\tilde\Om$ of the boundary graph $\partial\G_q$ and we consider the corresponding covering $\partial\G_q^{\tilde\Om}$. Then, the 0-scores satisfies the following identity
\be
\Phi_0(\G_q)=\Phi_0(\Gcan) - \Phi_0(\partial\G_q^{\tilde\Om}).
\ee
and the degree satisfies the identity (regardless of the chosen coefficient $a$):
\be
\delta(\G_q)=\delta(\Gcan) + \Phi_0(\partial\G_q^{\tilde\Om}).
\ee
\end{lemma}

It is therefore possible to classify triangulations with boundary according to their degree by choosing a pairing of the boundary graph and using results and classifications on triangulations without boundary. See the characterizations of sub-leading orders in Subsection~\ref{subsec:BoundQuart} and  Section~\ref{sec:UniGraphColSYK} for applications.


\subsection*{The graph $\GOM$}


%
%
\begin{definition}
\label{def:EulColGraph}
Given a colored graph $\G\in\bG_D^q$ and a pairing $\Om$ of its vertices, the Eulerian graph $\GOM$ is obtained from $\G$ by canonically adding $q$ color-0 marked edges, orienting every edge from black to white and contracting\footnote{In the usual graph theoretical sense.} every pair of $\Om$. 
\end{definition}
	\begin{figure}[h!]
	\centering
	\includegraphics[scale=0.75]{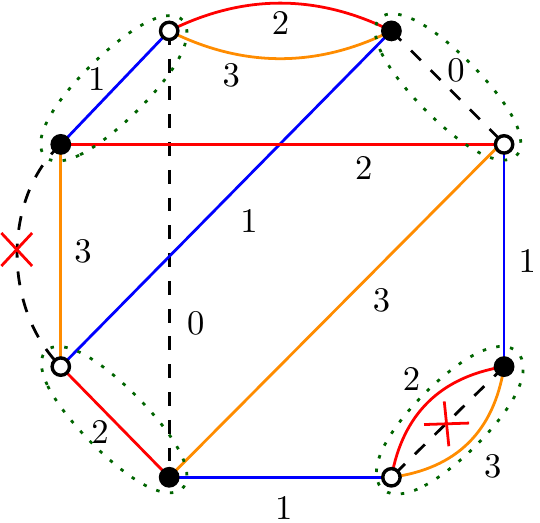}\hspace{2.5cm}
	\includegraphics[scale=0.75]{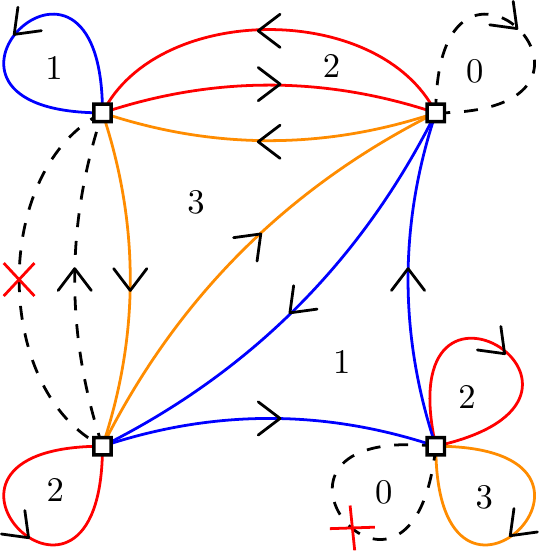}
	\caption{A graph $\G$, a pairing $\Om$, and $\GOM$. }
\label{fig:BOM}
	\end{figure}
Examples are shown in Fig.~\ref{fig:BijBub} for a graph in $\bG_3$ and in Fig.~\ref{fig:BOM} for a graph in $\bG_3^2$. As $\G$ is properly edge-colored, in $\GOM$ each vertex is incident to two color-$i$ half-edges, one ingoing and one outgoing. To recover $\G$ from $\GOM$, replace every vertex with a pair of black and white vertices, and attach all the outgoing (resp.~ingoing) half-edges to the black (resp.~white) vertex. Clearly, $\GOM$ is equivalent to $(\G, \Om)$.  The color-$i$ subgraphs $\GOM^{(i)}$ are collections of disjoint directed cycles with no isolated vertices, and which span all the vertices of $\GOM$. 

Remark that in the case where $\G\in\bG_D^U$ is not connected, but a pair of $\Omega$ contains a black vertex in one connected component and a white vertex in another connected component, then $\GOM$ has less connected components than $\G$.
Also note that if $\partial \Ga=\B\in\bG_{D-1}$ and $\Om$ is the pairing of $\B$ induced by $\Ga$, then
$\Ga_\circlearrowleft$ defined at the end of Section~\ref{sec:CycBub2} coincides with $\BOM$. 
%

\subsection*{Bicolored submaps}

\begin{definition}[Bicolored submap]
\label{def:BicolSubmap}
A bicolored submap $\Gaij$ of a stacked map $\Ga\in\bS_D$ is obtained by keeping all the color-$i$ and color-$j$ edges and vertices. The vertices and edges of color $k\neq i,j$ are deleted. 
We will be mostly interested in the bicolored submap for colors 0 and $i$, which we will denote $\Gai$.
\end{definition}

The bicolored submaps for colors $i$ and $j$ is called the \emph{color-$ij$ submap}.
Some bicolored submaps are shown in Fig.~\ref{fig:BicSub} for the example on the right of Fig.~\ref{fig:BijSM}. Importantly, \emph{bicolored submaps are combinatorial maps} in the usual sense. Indeed, the white vertices all have degree two and the two incident edges therefore have a trivial ordering. In fact, one may contract all the color-0 edges in $\Gai$ to get rid of the valency-two vertices. 

\

Given $i<j$, we also need to define the twisted bicolored submaps $\Ga_\wr^{(ij)}$, obtained from $\Gaij$ by adding a twist (Fig.~\ref{fig:LOMaps}) on every color-$i$ edge. 
\begin{definition}[Twisted bicolored submap]
\label{def:TwistbicolSubmap}
The twisted bicolored submap $\Ga_\wr^{(ij)}$ with $i<j$ is the combinatorial map obtained from $\Gaij$ by adding a twist-factor $(-)$ on every color-$i$ edge while color-$j$ edges carry $(+)$ factors.
\end{definition}
We stress that by doing a local change of orientation (Def.~\ref{def:LocChange}) on every color-$i$ vertex, all the twists are eliminated, so that \emph{the twisted bicolored submaps are orientable}. The reason is that  the white vertices all have degree two and have one edge incident to  a color-$i$ vertex and the other one incident to a color-$j$ vertex, i.e. contracting all the color-$j$ edges, the map is bipartite.

\subsection{The bijection}
\label{subsec:Bij}

We first describe the map $\Ps$ which to each pair $(\G,\Om)$ of a colored graph $\G\in\bG_D^U$ (i.e. whose connected components are in $\bG_D$) with $q$ marked color-0 edges and a pairing of its vertices associates a stacked map $\Ps(\G,\Om)$ which connected components are in $\bS^q_D$.  

\

We consider the Eulerian graph $\GOM$, which is obtained by orienting the edges from black to white and contracting each pair in $\Om$ into a white square vertex.
For each color $i\in\lDr$, the subgraph $\GOM^{(i)}$ obtained by keeping only the color-$i$ edges is a set of disjoint directed cycles $C_1^{(i)}, \dotsc, C_{K_i}^{(i)}$. The subgraph $\GOM^{(0)}$ contains precisely $q$ disjoint cycles containing one marked edge.

Each color-$i$ cycle is replaced with a star-map of color $i$. The cyclic ordering of appearance of the vertices around the cycle is translated into a cyclic ordering of strands around the color-$i$ square vertex. A directed edge of color $i$ is now a directed corner around a star-map of color $i$, i.e. part of a face around the submap of color $i$.  This is shown in Figure.~\ref{fig:CycStar}. Similarly, each color-$0$ cycle is also replaced with a star-map of color $0$. 
\begin{figure}[!h]
\centering
\includegraphics[scale=0.6]{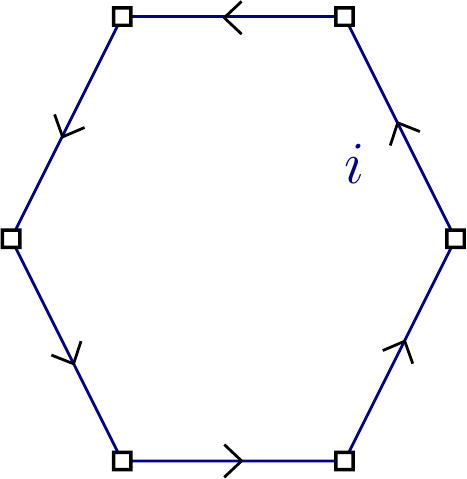}
\hspace{1.5cm}\raisebox{+8ex} {$\leftrightarrow$}\hspace{1.5cm}\includegraphics[scale=0.6]{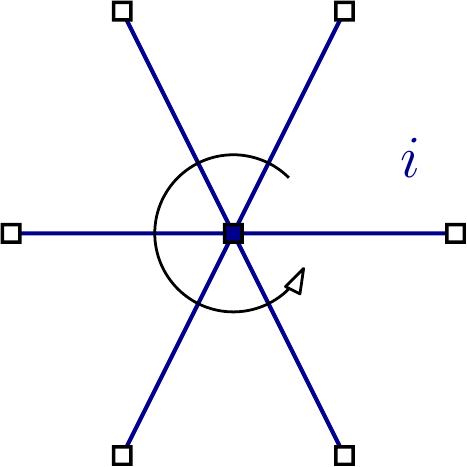}
\caption{Directed cycles are mapped to star-maps of the same color.}
\label{fig:CycStar}
\end{figure}
The cycles which contain one marked edge have a (non-cyclic) ordering of the vertices, which starts and end at the marked edge. The marked edge is mapped to a marked corner of the star-map, and the ordering of vertices around the cycle is translated into an ordering of strands around the color-0 vertex, which starts and ends at the marked corner. This is shown in Figure.~\ref{fig:CycStar0}.
\begin{figure}[!h]
\centering
\includegraphics[scale=0.6]{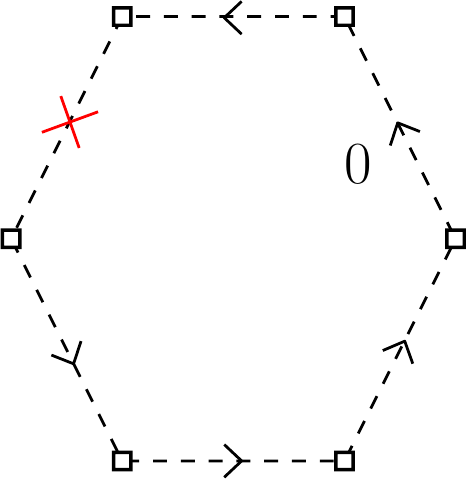}
\hspace{1.5cm}\raisebox{+8ex} {$\leftrightarrow$}\hspace{1.5cm}\includegraphics[scale=0.6]{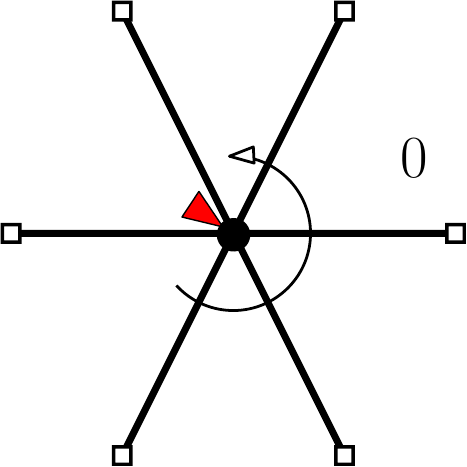}
\caption{Directed cycles with a marked edge become  star-maps with a marked corner.}
\label{fig:CycStar0}
\end{figure}

As they were incident to precisely one cycle per color $i\in\{0,\cdots,D\}$, white squares are now incident to precisely one edge for each color. There is no ordering of edges around white vertices (incident edges are distinguishable because of their color).

Examples are shown in Fig.~\ref{fig:BijBub} (with no marked corner and for colors 1, 2, 3), and in Fig.~\ref{fig:BijSM}.

We stress that the bijection is between non-necessarily connected objects. Indeed, if $\G$ has two connected components, $\GOM$ can be connected in the case where the pairing $\Om$ has at least one pair with one vertex in each connected component (Fig.~\ref{fig:Connect}). 
\begin{figure}[!h]
\centering
\raisebox{+3ex} {\includegraphics[scale=0.8]{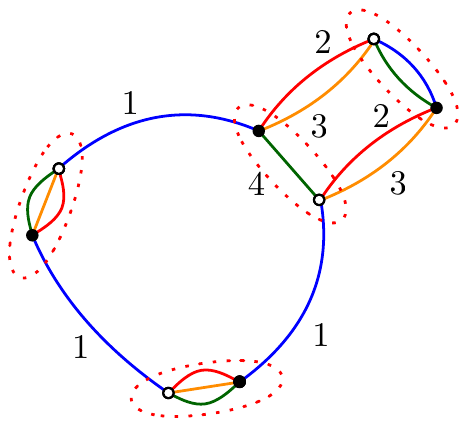}}
\hspace{1cm}
\includegraphics[scale=0.7]{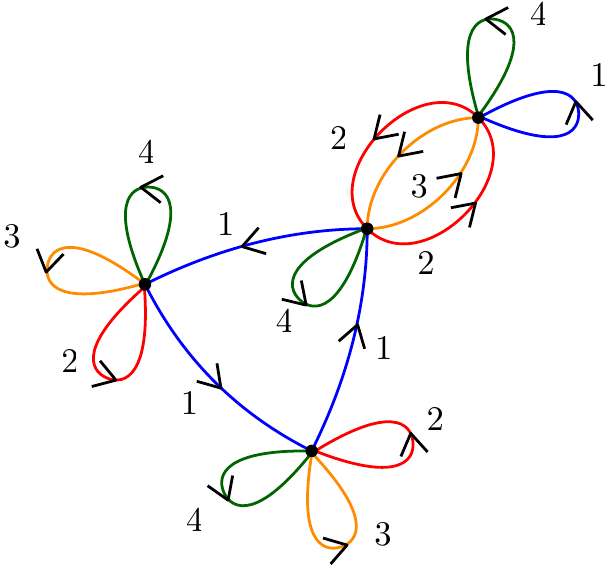}
\hspace{1cm}
\raisebox{+1ex} {\includegraphics[scale=0.8]{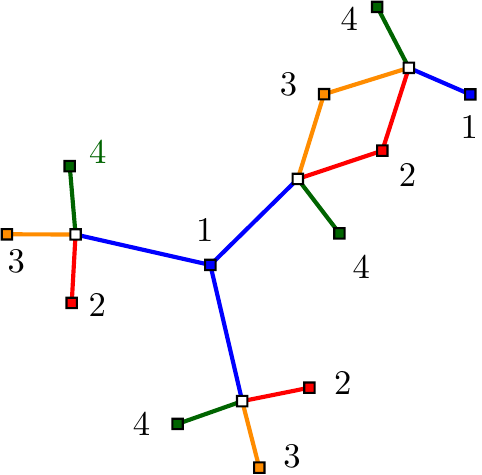}}
\caption{A graph $\B$ with pairing $\Om_\B$, the graph $\BOM$, and the stacked map $\Ps(\B,\Om_B)$.}
\label{fig:BijBub}
\end{figure}

Note that unmarking the corners  in the stacked map picture corresponds to canonically adding unmarked color-0 edges on $(\G,\Om_\G)$ in the colored graph picture.  The reason why there can be at most one marked corner per color-0 vertex is because the $q$ marked edges have been added canonically. Considering stacked maps with more than one mark per color-0 vertex, we would recover cycles alternating color-0 edges and pairs with more than one marked color-0 edge. Deleting the marked edges, we would then recover the same element of $\bG_D^q$ for different stacked maps, corresponding to non-canonical ways of adding $q$ marked color-0 edges.

\

We detail the construction of $\Ps(\G,\Om)$ in terms of permutations : we label the pairs in $\Om$ and define the set of $V(\G)(D+1)$ darts $\cD$ of the map $\Ps(\B,\Om)$ as follows.
For each subgraph $\GOM^{(i)}$ we have a set $\cD^\star_i$  and a set $\cD^\diamond_i$, each one of $V(\G)/2$ darts  of color $i$ (one for each pair of $\G$), and
\be
\cD=\sqcup_{i=1}^{D+1}\bigl(\cD^\star_i\sqcup \cD^\diamond_i\bigr).
\ee
 To each vertex labeled $l$ in $\GOM$ corresponds one dart in $\cD^\star_i$ and one dart in $\cD^\diamond_i$, with the same label $l$.
The subgraph $\GOM^{(i)}$ induces a partition of $\cD^\star_i$ in directed cycles which define a permutation 
 $\sigma^\star_i$ on  $\cD^\star_i$, corresponding to the color-$i$ vertices. 
Some of the cycles of the color-0 subgraph $\GOM^{(0)}$ contain marked edges, and therefore, some of the cycles of the permutation  $\sigma^\star_0$ have a non-cyclic ordering, which starts and ends at the marked edge.
Each vertex in $\GOM$ has a label $l$ and belongs to a single $C_a^{(i)}$ for each $i$, so that to each $l$ corresponds one dart $d^\diamond_{i,l}$ with the same label $l$ in each $\cD^\diamond_i$. This gives a partition of $\sqcup_{i=1}^{D+1} \cD^\diamond_i$ in $l$ unordered sets of $D+1$ darts, one for each color.
We define $\sigma^\diamond$ as the disjoint union of these sets, and 
\be
\sigma=\bigl(\sqcup_{i=1}^{D+1} \sigma^\star_i\bigr) \sqcup \bigl(\sqcup_{l=1}^{V(\G)/2} \sigma^\diamond_l \bigr).
\ee
To each pair $(i,l)$ correspond two darts, $d^\star_{i,l}\in\cD^\star_i$ and $d^\diamond_{i,l}\in\cD^{\diamond}_i$, and we define 
\be
\alpha=\bigsqcup_{(i,l)=(1,1)}^{(D+1,\frac{V(\G)}2)}\bigl(d^\star_{i,l},d^\diamond_{i,l}\bigr).
\ee
We then consider the equivalence classes upon relabeling the pairs. Twisted bicolored submaps have been defined in Def.~\ref{def:TwistbicolSubmap}.

	%
	\begin{figure}[h!]
	\centering
	\includegraphics[scale=0.75]{BijSM2.pdf}
	\hspace{1.5cm}\raisebox{+10ex} {$\leftrightarrow$}\hspace{1.5cm}
	\includegraphics[scale=0.75]{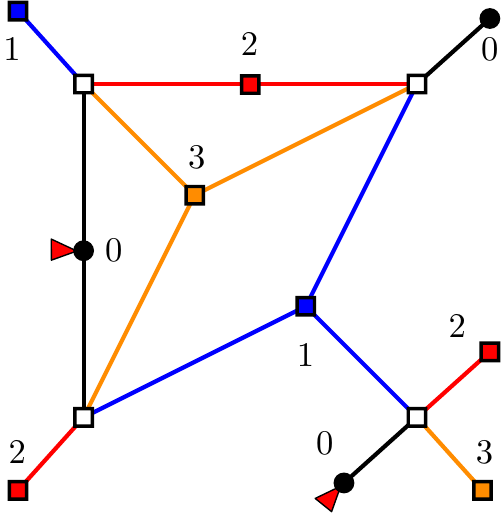}
	\caption{ Graph $\GOM$ and corresponding stacked map $\Ps(\G,\Om)$. }
\label{fig:BijSM}
	\end{figure}

\begin{theorem}
\label{thm:BijSMGen}
The map $\Ps$ is a bijection, which maps color $i<j$ cycles to faces around the twisted bicolored submaps $\Ga_\wr^{(ij)}$.
\end{theorem}

\prf
We describe the inverse map $\Ps^{-1}$ and show that $\Ps^{-1}(\Ps(\G,\Om))=(\G,\Om)$. Starting from a stacked map in $\bS_D^q$, for each counterclockwise  corner $\overrightarrow{(e_1,e_2)}$ around a color $i$ vertex $v_i$, where $e_1=(v_i, v_1^\diamond)$ and $e_2=(v_i, v_2^\diamond)$, we draw a directed edge from $v_1^\diamond$ to $v_2^\diamond$ (which might be the same white square vertex). If the corner is marked, we mark the corresponding color-0 edge. 
Each star-map of color $i$ gives rise to a directed cycle of color $i$ and at this point, each cycle of color-$i$ directed edges is as pictured below.
\begin{figure}[!h]
\centering
\includegraphics[scale=0.55]{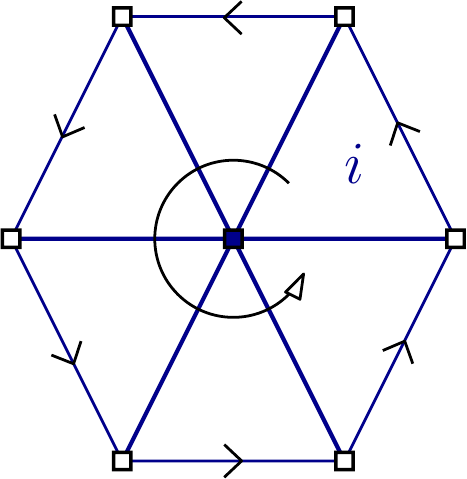}
\end{figure}
At this point, it is clear that if we started from $(\G,\Om)$ and replaced every directed cycle with a star-map, then deleted every directed edge, and then performed the step just described, we recover a directed cycle for each directed cycle of $\GOM$, between the same vertices (corresponding to pairs), and directed the same way. 
From this intermediary diagram, deleting all the color-$i$ vertices and their incident edges, we recover $\GOM$, and then $(\G,\Om)$ as in Def.~\ref{def:EulColGraph}. Therefore, $\Ps^{-1}(\Ps(\G,\Om))=(\G,\Om)$.

Conversely, starting from a stacked-map $\Ga\in\bS_D$, applying the first steps of the map $\Psi^{-1}$, we get an intermediary diagram in which the oriented cycles and star-maps are as in the picture above. Deleting the-star-maps and performing the first steps of $\Ps$, that is, adding a star-map for every oriented cycle so that the counterclockwise ordering of edges around the colored vertices corresponds to the orientation of the cycle, as in Figs.~\ref{fig:CycStar} and \ref{fig:CycStar0}, we recover the star-maps we deleted for every oriented cycle, with the same ordering of edges. Therefore, $\Ps(\Ps^{-1}(\Ga))=\Ga$. 
	\begin{figure}[h!]
	\centering
	\includegraphics[scale=0.75]{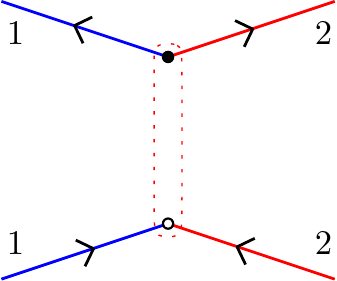}
	\hspace{1.5cm}\raisebox{+6ex} {$\leftrightarrow$}\hspace{1.5cm}
	\includegraphics[scale=1.2]{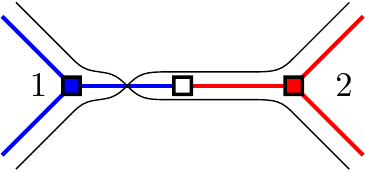}
	\caption{ Color $12$ cycles around a pair of $\Om$, and corresponding faces in $\Ga_\wr^{(12)}$. }
\label{fig:TwistedBicolor}
	\end{figure}

There remains to prove that bicolored cycles are mapped to faces around twisted bicolored submaps $\Ga_\wr^{(ij)}$. In $\GOM$, the color $ij$ cycles alternate edges with opposite orientations: the color-$i$ edges all have the same orientation, which is opposite to that of the color-$j$ edges. Each edge is mapped to a corner, and therefore, the bicolored cycles are mapped to corners around $\Gaij$, which are alternatively clockwise and counterclockwise. By introducing a twist on every color-$i$ edge, the faces precisely visit the corresponding corners in the right order, as shown in Fig.~\ref{fig:TwistedBicolor}. 
\qed

\


\begin{definition}[Score]
The score of a stacked map $\Ga$ is the sum of the unbroken faces around twisted bicolored submaps.
\be
\label{eqref:TwistedScore}
\Phi(\Ga)=\sum_{i=1}^D \Fint (\Ga_\wr^{(ij)}).
\ee
\end{definition}

	\begin{figure}[h!]
	\centering
	\includegraphics[scale=0.6]{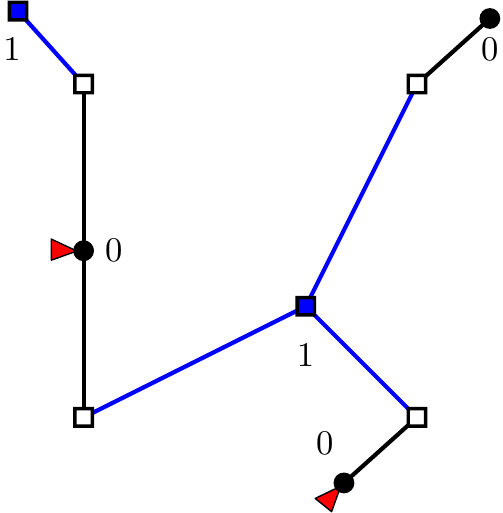}\hspace{1.5cm}
	\includegraphics[scale=0.6]{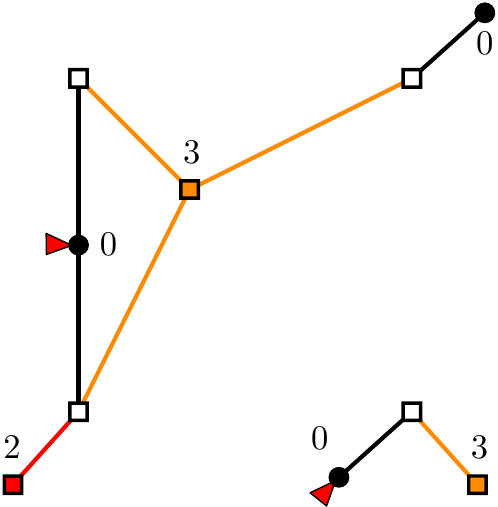}\hspace{1.5cm}
	\includegraphics[scale=0.6]{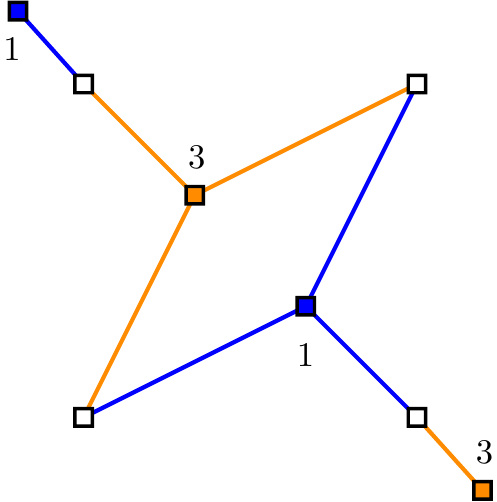}
	\caption{Some bicolored submaps. }
\label{fig:BicSub}
	\end{figure}

The example $\Ga=\Ps(\B,\Om)$ of Figure~\ref{fig:BijSM}, which color 01, 03 and 13 submaps are shown in Fig.~\ref{fig:BicSub} has, when adding twists on all the edges of the color of smaller index, $\Fint(\Ga_\wr^{(01)})=\Fint(\Ga_\wr^{(02)})=0$,  $F(\Ga_\wr^{(12)})=\Fint(\Ga_\wr^{(03)})=1$,  $F(\Ga_\wr^{(13)})=2$ and $F(\Ga_\wr^{(23)})=3$, so that its score is $\Phi(\Ga)=7$, as can be verified on Fig.~\ref{fig:BOM}.

\

The procedure to obtain the {\bf boundary graph of a stacked map with marked corners} is as described at the end of Section~\ref{sec:CycBub2} for gluings of cyclic bubbles. The only difference is that one follows broken faces around twisted color-$0i$ submaps  instead of broken faces around color-$i$ submaps. For instance, the boundary graph of the example in Fig.~\ref{fig:BijSM} is a 1-cyclic bubble of length 4 (quartic melonic graph) alternating color 1 and colors 2,3.

Some parts of the stacked map necessarily correspond to recognizable structures.

\begin{prop}[Tree-contributions]
\label{prop:MelContriBij1}
Tree contributions to a stacked-map $\Ga\in\bS_D$ correspond to melonic contributions to $\G\in\bG_D$, where $(\G,\Om)=\Ps^{-1}(\Ga)$. 
\end{prop}
\prf A tree contribution necessarily has a white vertex incident to $D$ colored leaves. Applying $\Ps^{(-1)}$, we see that this white vertex corresponds to a $D$-pair. We can contract it and proceed recursively: in the stacked map, this goes back to suppressing this white vertex and the incident colored edges and leaves. \qed

\

Note however that some melonic contributions in $\G$ may not correspond to tree-contributions if $\Om$ does not coincide with the canonical pairing of the melonic subgraph (Def.~\ref{def:CanoPairing}). 

\subsection*{Connectedness and face-exploration }

As underlined previously, $\Ps(\G, \Omega)$ might be connected for a non connected $\G\in\bG_D^U$. Here, we provide a mean of determining directly on a stacked map $\Ga$ whether it is the image of a connected colored graph. We focus on the faces around the one-color star-maps. They correspond to oriented edges of $\GOM$, and therefore to edges of $\G$ such that $\Ps(\G,\Om)=\Ga$. Starting from any white square $v^\diamond_0$, we consider all the out-going faces around the $D$ color-$i$ star-maps, and denote $\{v^\diamond_{1;1}, v^\diamond_{1;2}, \cdots\}$ the first white squares they reach. We keep in mind the faces we explored already. For each one of the $v^\diamond_{1,a}$, either we have explored the $D+1$ ingoing faces, or we only have explored a subset of them, in which case we explore the remaining ingoing faces. We denote $\{v^\diamond_{2;1}, v^\diamond_{2;2}, \cdots\}$ the  white squares they first reach. Now, some of the $v^\diamond_{2;a}$ have unexplored outgoing faces, which we explore, denoting $\{v^\diamond_{3;1}, v^\diamond_{3;2}, \cdots\}$ the white vertices we first reach. We continue, alternating the exploration of ingoing and outgoing faces, and stop when we encounter only vertices which have already been visited, and which we reach from faces we have already explored (this necessarily happens as the map is finite). We define the exploration starting from $v^\diamond_0$ as
\be
\bX_\text{out}(v^\diamond_0) = {\biggl [} \{v^\diamond_0\}_\text{out}, \{v^\diamond_{1;1}, v^\diamond_{1;2}, \cdots\}_\text{in}, \{v^\diamond_{2;1}, v^\diamond_{2;2}, \cdots\}_\text{out}, \cdots\biggr].
\ee
In the example on the right of Fig.~\ref{fig:Connect}, we have $\bX_\text{out}(v^\diamond_0) = {\bigl [} \{v^\diamond_0\}_\text{out}, \{v^\diamond_{1;1}, v^\diamond_{1;2}\}_\text{in}, \{v^\diamond_{2;1}\}_\text{out}\bigr]$. We have the following characterization:
\begin{prop}
\label{prop:FaceExpl}
A stacked map $\Ga=\Ps(\G,\Om)$ corresponds to a connected $\G\in\bG_D$ if and only if for any  white square $v^\diamond_0$, every white vertex of $\Ga$ appears precisely twice in $\bX_\text{out}(v^\diamond_0)$. 
\end{prop}
As the vertices $v^\diamond_0$ and $v^\diamond_{1;1}$ appear only once in the exploration on the right of Fig.~\ref{fig:Connect}, we know that the map corresponds to a non-connected colored graph. Furthermore, one of the white vertices does not appear at all.

\

\prf Indeed, it is easy to see that the exploration in $\Ga$ explores the connected component of $\G$ containing the black vertex of $v^\diamond_0$: starting from this black vertex we consider all the incident edges, then we consider all the edges incident to all the white edges they reach, and so on. Each visited white vertex appears exactly once in $\bX_\text{out}(v^\diamond_0)$ if only one vertex of the pair is in the connected component, or twice if both vertices of the pair do: once in a set labeled ``out", and once in a set labeled ``in". In particular, the exploration ignores $\Omega$, and $\G$ is connected if every vertex of $\G$ has been visited, i.e. if each pair of $\Omega$ has been visited twice (once on a black vertex, corresponding to outgoing faces of $\Ga$, once on a white vertex, corresponding to ingoing faces of $\Ga$). \qed

	\begin{figure}[h!]
	\centering
	\includegraphics[scale=0.6]{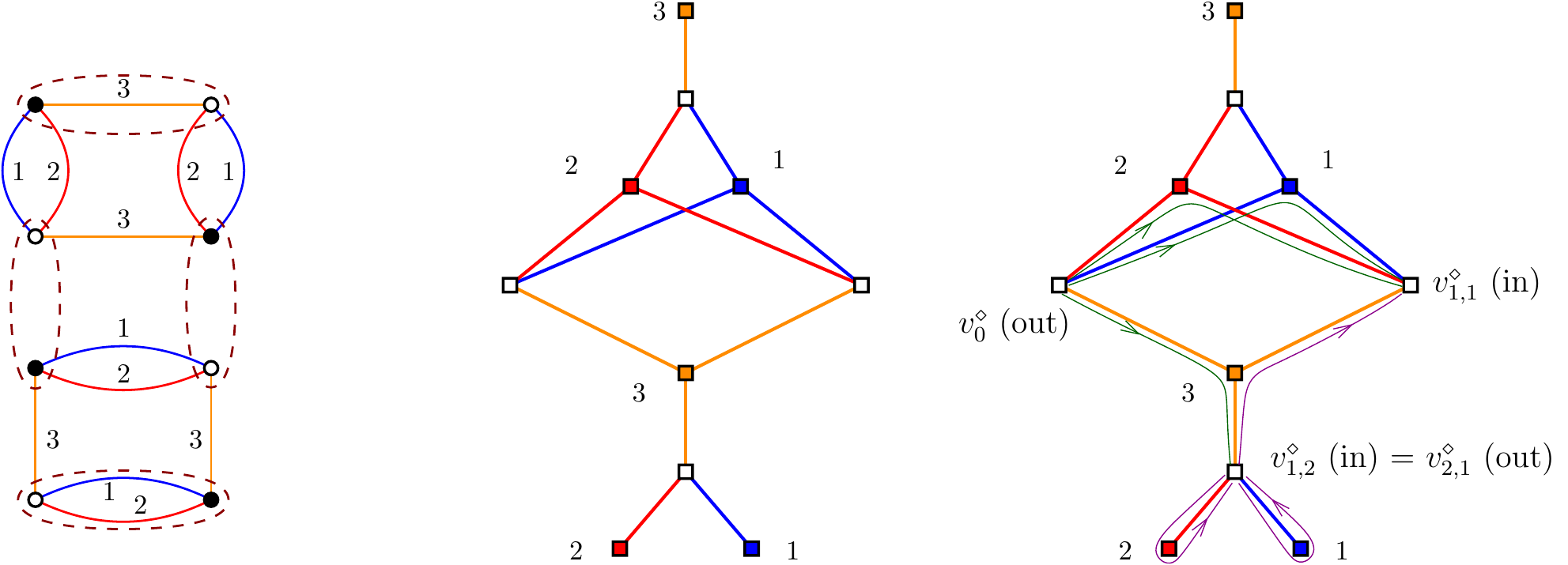}
	\caption{Non-connectedness from the face-exploration: a non-connected $\G$ and $\Om$, the corresponding $\Ps(\G,\Om)$, and an exploration from $v^\diamond_0$. The faces explored at the first step are in green, those explored at the second step are in magenta. }
\label{fig:Connect}
	\end{figure}
\begin{coroll}
\label{coroll:EvenWhiteConnect}
If $\Ga=\Ps(\G,\Om)$ has less connected components than $\G\in\bG_D$, then some face-explorations visit some vertices only once. The number of vertices visited only once in an exploration is even. 
\end{coroll}
\prf This follows from the fact that the vertices visited a single time correspond, in $(\G, \Om)$, to pairs with only one vertex in the component. As the number of vertices in a connected component of $\G$ is even, the number of white vertices visited only once in an exploration is even. \qed

\

This characterization becomes somehow heavy when there are many colors and for large maps. A sufficient condition for a stacked map to be the image of a connected graph is the following.
\begin{lemma}
If a connected stacked map $\Ga$ has a colored leaf incident to every white vertex (but one), then it is the image $\Ga=\Ps(\G,\Om)$ of a connected $\G\in\bG_D$.
\end{lemma}
\prf Every vertex incident to a colored leaf will clearly be visited twice in the exploration, once in an outgoing set, once in an ingoing set. In the case where a single vertex has no incident colored leaf, it also appears twice in the exploration which starts at this white vertex, by applying Coroll.~\ref{coroll:EvenWhiteConnect}. \qed

\

\subsection*{Pairing induced by a color}

One may wonder how to specify a unique pairing per colored graph in order to have a unique associated stacked map.  Also, it would be useful to have a bijection between connected objects. A way of doing so is to choose a particular color $i$, which we take to be different from 0 in the case where  some color-0 edges are marked. We recall that $\G^{\hat i}$ is the graph obtained from $\G$ by deleting all color-$i$ edges. In a graph $\G\in\bG_D$, the color-$i$ edges are a 1-matching which defines a pairing $\Omi$of $\G^{\hat i}$.
%
\begin{theorem} 
\label{thm:BijOneColor}
The following induced map is a bijection
\bea
\Ps^{(i)}: \bG^q_D &\longrightarrow& \bS^q_{D-1}\\
\G\  &\longmapsto&  \Ps\bigl(\G^{\hat i}, \Omi(\G)\bigr). \nonumber
\eea
\end{theorem}
\prf Applying $\Ps$ to $(\G, \Omi(\G))$ for $\G$ connected, one gets a connected map in $\bS_D^q$ such that all color-$i$ vertices are leaves. Every stacked map in $\bS_D^q$ such that all color-$i$ vertices are leaves has a unique antecedent $(\G,\Om)$ with $\G$ connected, and such that color $i$ links the vertices of each pair of $\Om$. Deleting these leaves and their incident edges, one gets a stacked map in $\bS^q_{D-1}$, putting back a color-$i$ leave incident to every white vertex is done in a unique way. \qed


\

The score of $\G\in\bG^q_D$ is that of $\Ga\in\bS^q_{D-1}$ plus the number of color-$j$  submaps (which do not carry marked corners) for each $j$. The 0-score of $\G\in\bG^q_D$ is that of $\Ga\in\bS^q_{D-1}$ plus the number of color-$0$  submaps which do not carry marked corners. 

\

It is also possible to do so choosing color 0, even if some edges are marked. Some color-0 leaves in the stacked map would then be marked. Contracting them, one should just distinguish the incident white vertices. Consequently, 
\be
\label{eqref:BijCol0}
\Ps^{(0)}: \G\  \longmapsto \Ps_q\bigl(\G^{\hat 0}, \Omi[0](\G)\bigr),
\ee
where $\Ps_q(\G, \Omi[0](\G))$ has $q$ distinguished vertices, is also a bijection between colored graph in $\bG_D$ and stacked maps in $\bS_{D-1}$ with $q$ distinguished white square vertices. \emph{The 0-score of a map is then the sum for $j\in\lDr$ of the number of color-$j$ submaps which do not carry distinguished vertices. }

\

Proposition~\ref{prop:MelContriBij1} adapts in the case of an induced pairing for a graph without boundary: tree contributions to a stacked-map $\Ga\in\bS_{D-1}$  correspond to melonic contributions to $\G\in\bG_D$, such that $\G=(\Ps^{(i)})^{-1}(\Ga)$. Again, not all melonic contributions of $\G$ correspond to tree contributions of $\Ga$, this only happens when, in the melonic contributions,  the color-$i$ edges all link vertices of the canonical pairs (Def.~\ref{def:CanoPairing}).

\subsection*{Gluings of 1-cyclic bubbles and boundaries}

The following  proposition states that any connected colored orientable triangulation is the boundary of some colored discrete space obtained by gluing 1-cyclic bubbles in one more dimension. We will state a stronger result in Thm.~\ref{thm:Represent}. 

\begin{prop}
\label{prop:Bound1Cyc}
Every graph in $\bG_{D-1}$ is the boundary of some connected graph in $\bG_D(\bB^1_\text{cyc})$, where $\bB^1_\text{cyc}$ is the set of 1-cyclic bubbles in $\bG_{D-1}$ of any size. 
\end{prop}
\prf This is a simple consequence of Theorems \ref{thm:BijSMGen} and \ref{thm:BijCycles}. Applying Thm~\ref{thm:BijSMGen} to a graph $\bG_{D-1}$ (with colors $\{1,\cdots,D\}$) for any choice of pairing, we get a stacked map in $\bS_{D-1}$. We now embed all the white vertices, e.g. by ordering the incident edges according to their colors, and then mark a corner per white vertex, e.g. that between the edges of color $D$ and 1. 
We obtain a bipartite map $\cM$ with marked corners on every white vertex and $K$ connected components such that edges carry colored labels. 
Because of the construction of $\Ps$, $\cG$ is recovered as the boundary graph of $\cM$ (here, the faces are around monochromatic submaps, as described at the end of Section~\ref{sec:CycBub2}). 
From Thm~\ref{thm:BijCycles}, the corresponding map is mapped bijectively to a colored graph in $\bG_D(\bB^1_\text{cyc})$ (with respect to the conventions of Thm~\ref{thm:BijCycles}, the color-$i$ vertices become the white vertices there, and the white vertices here are the black vertices there). \qed

\subsection{Bubble-restricted gluings }
\label{subsec:BubStacked}

Throughout this work, we are mostly interested in bubble-restricted gluings, i.e. in the case where deleting all color-0 edges in the colored graph, one is left with copies of bubbles from some set $\bB\subset\bG_{D-1}$. Deleting color-0 edges in a colored graph $\G$ corresponds to deleting all color-0 vertices in $\Ps(\G,\Om)$, together with the incident color-0 edges.  Because the construction of $\Ps(\G,\Om)$ is done for each color independently (star-maps only meet at white vertices but are built independently one from another), the construction for the colors $\{1,\cdots,D\}$ is the bijection between colored graphs with connected components in $\bG_{D-1}$ and stacked maps with connected components in $\bS_{D-1}$. Deleting all color-0 edges and vertices in $\Ps(\G,\Om)$, the remaining connected submaps are copies 
of $\Ps(\{\B_a\},\Om_{\lvert\{\B_a\}})$ where $\{\B_a\}$ are copies of bubbles $\B_a\in\bB$ and the pairings $\Om_{\lvert\{\B_a\}}$ are the pairings induced by $\Om$ on this particular set of bubbles  $\{\B_a\}$ in $\G$, such that the ``covering" $\{\B_a\}^{\Om_{\lvert\{\B_a\}}}$ is connected.

\begin{figure}[!h]
\centering
\includegraphics[scale=0.8]{BtoM3.pdf}
\hspace{2cm}
\raisebox{+3ex} {\includegraphics[scale=0.8]{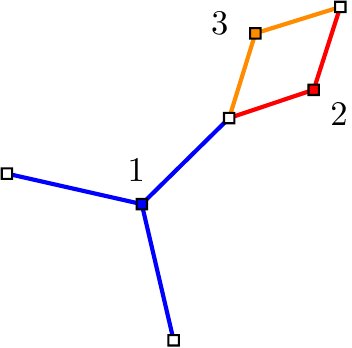}}
\caption{$\Ps(\B,\Om_B)$, and a simpler version with no colored leaf.}
\label{fig:BijBubSimp}
\end{figure}

In the case of restricted gluings, we decide to make a global choice of pairing $\Om_\B$ for all the copies of $\B$.  It translates into a pairing $\Om$ of each graph in $\bG(\bB)$. 
We denote $\bS(\bB,\Om_\bB)$ the set of connected stacked maps built for $\bB$ and a choice of pairing $\Om_\B$ for each bubble $\B\in\bB$, and $\bS^q(\bB,\Om_\bB)$ the set of such objects with $q$ marked corners on $q$ different black vertices. They are stacked maps such that, when deleting all the color-0 edges and vertices, we are left with a collection of submaps isomorphic to $\Ps(\B, \Om_\B)$ for our choices $\Om_\B$, where $\B\in\bB$. If $\bB=\{\B\}$, we rather use $\bS(\B,\Om_\B)$ and $\bS^q(\B,\Om_\B)$. 

\begin{theorem}
\label{thm:BijSM}
For any set $\bB$ of bubbles $\B$, each equipped with a pairing $\Om_\B$, there is a bijection 
\be
\Ps_0: \bG^q(\bB) \longrightarrow \bS^q(\bB,\Om_\bB).
\ee 
The bubbles isomorphic to $\B$ are mapped to the submaps isomorphic to $\Ps(\B, \Om_\B)$, 
and the bicolored cycles $0i$ to the faces around bicolored submaps $0i$, denoted $\Gai$.
\end{theorem}
	\begin{figure}[h!]
	\centering
	\includegraphics[scale=0.75]{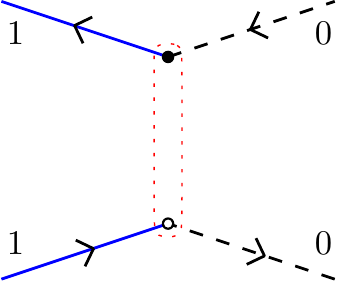}
	\hspace{1.5cm}\raisebox{+6ex} {$\leftrightarrow$}\hspace{1.5cm}
	\includegraphics[scale=1.2]{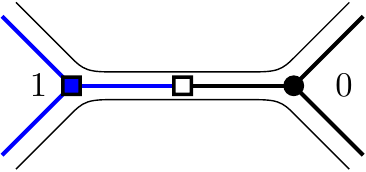}
	\caption{ Color $01$ cycles around a pair of $\Om_\B$, and corresponding faces in $\Ga^{(1)}$. }
\label{fig:BicolorUnt}
	\end{figure}

\prf The map $\Ps$ restricted to colored graphs in $\bG(\bB)$ and pairings inherited from the particular choices $\Om_\B$ is easily seen to be a bijection with elements in $\bS(\bB,\Om_\bB)$. However, because we will only be interested in color-$0i$ cycles, \emph{we decide to orient color-0 edges from white to black and not from black to white, as done for the other colors.} As in the proof for the bijection $\Ps$, we show easily that we have a bijection by looking at the oriented color-$i$ cycles and the corresponding star-maps. The only difference is that here, color-0 edges are oriented from white to black, which has to be taken into account when going from $\GOM$ to $(\G,\Om)$ and conversely, but all the other steps are as for $\Ps$. In particular, the restriction of the bijection to colors in $\lDr$ gives a collection of stacked maps isomorphic to $\Ps(\B, \Om_\B)$ for $\B\in\bB$. Another way of obtaining this bijection from the bijection $\Ps$ is by adding a twist (Fig.~\ref{fig:LOMaps}) on every color-0 edge and then doing a local change of orientation (Def.~\ref{def:LocChange}) on every color-0 vertex.
Bicolored cycles are therefore mapped to faces around bicolored submaps (without twists), as illustrated in Fig.~\ref{fig:BicolorUnt}.
\qed

\

It is this theorem which was proven directly in \cite{SWM}. Examples are shown in Fig.~\ref{fig:BijBub} (bijection $\Ps$ for a bubble $\B$ with pairing $\Om_B$), Fig.~\ref{fig:BijBubSimp} (simplified bubble) and Fig.~\ref{fig:BijSWM} (a graph in $\bG(\B)$ and corresponding stacked map in $\bS(\B, \Om_\B)$). Bicolored submaps have been defined in Def.~\ref{def:BicolSubmap}. We recall that unbroken faces are faces which do not meet any marked vertex. 

\begin{definition}[0-Score]
The 0-score of a stacked map $\Ga\in\bS(\bB,\Om_\bB)$ is the sum of the unbroken faces around color-$0i$ submaps.
\be
\Phi_0(\Ga)=\sum_{i=1}^D \Fint (\Gai).
\ee
\end{definition}

Although this is different from the 0-score for the bijection $\Ps$, which involves twists \eqref{eqref:TwistedScore}, it will not be conflictual as we are interested in bubble-restricted gluings, and we will always use the present definition, which applies in the context of Theorems~\ref{thm:BijCycles},~\ref{thm:BijSM}, and~\ref{thm:BijSimp}. Because of the particular role played by the color 0,  color-0 vertices of stacked maps are represented as black discs, while for other colors, the vertices are represented as colored squares and the edges are colored. 
 Furthermore, if a color-$i$ vertex is a leaf in $\Ps(\B, \Om_\B)$, we can as well not represent it when drawing the maps, as it has no effect on the faces of $\Gai$ (Fig.~\ref{fig:BijBubSimp}).
Note that when a color $i$ is such that color-$i$ square vertices all have valency 1, $\Fint (\Gai)$ is the number of black (disc) vertices without marked corners (e.g. color 4 in Fig.~\ref{fig:BijSWM}). 
We will still call bubbles the submaps $\Ps(\B,\Om_\B)$ and denote the number of bubbles $n_\B$, or $b=\sum_\B n_\B$. We call {\bf maximal maps} in $\bS^q(\bB,\Om_\bB)$ the maps which maximize the 0-score at fixed number of bubbles.

\begin{figure}[!h]
\centering
\includegraphics[scale=0.7]{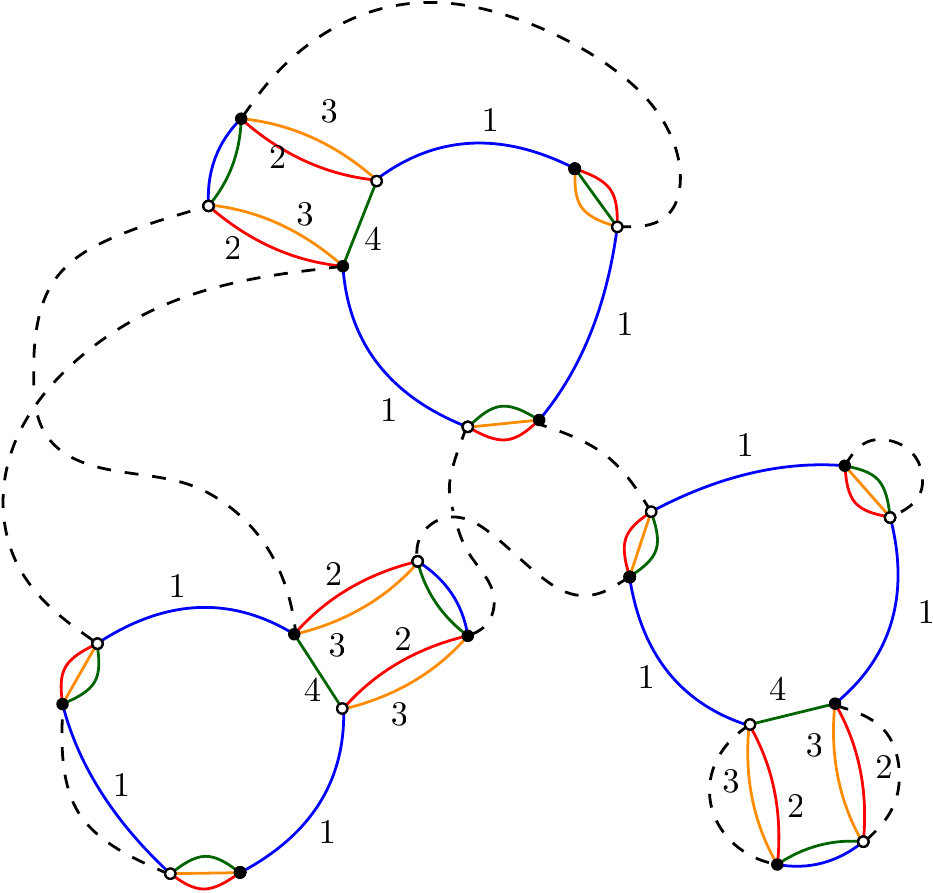}
\hspace{1.5cm}\raisebox{+1.5ex} {\includegraphics[scale=0.8]{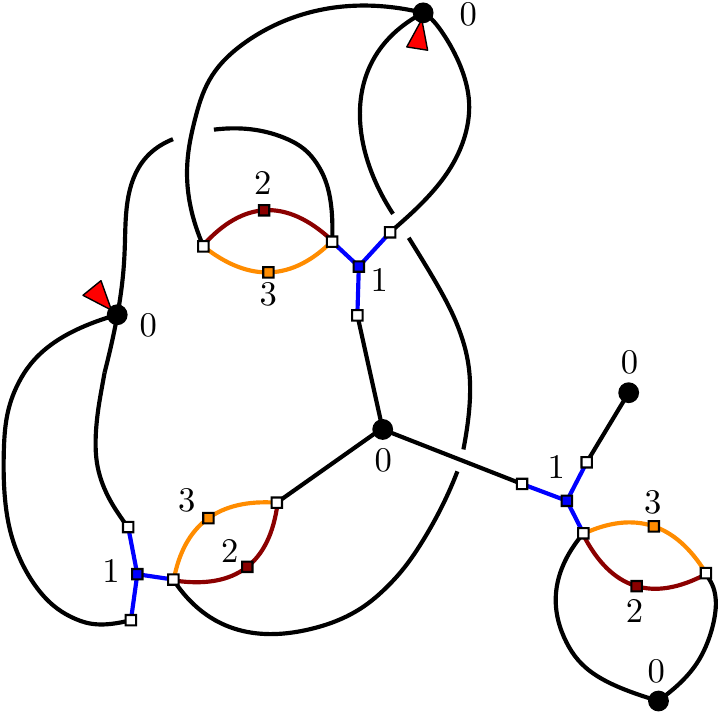}}
\caption{Bijection $\Ps_0$ between edge-colored graphs and stacked maps.}
\label{fig:BijSWM}
\end{figure}

We can compute the 0-score from the stacked map. For the example of Fig.~\ref{fig:BijSWM}, 
 $\Fint(\Ga^{(1)})=2$, $\Fint(\Ga^{(2)})=\Fint(\Ga^{(3)})=3$ and $\Fint(\Ga^{(4)})=3$, so that $\Phi_0(\Ga)=11$. Now consider the map $\Ga_0$ obtained by unmarking the corners. We have  $F(\Ga^{(1)})=3$, $F(\Ga^{(2)})=F(\Ga^{(3)})=4$ and $F(\Ga^{(4)})=5$, so that $\Phi_0(\Ga_0)=16$. 

\

The procedure to obtain the {\bf boundary graph of a stacked map in $\bS^q(\bB,\Om_\bB)$ with $q$ marked corners} is as described at the end of Section~\ref{sec:CycBub2} for gluings of cyclic bubbles. The only difference is that one follows broken faces around color-$0i$ submaps  instead of broken faces around color-$i$ submaps.  Lemma~\ref{lemma:GqG0} can be restated for stacked maps: 
Given a stacked map $\Ga_q\in\bS^q(\Ga(\B, \Om))$ with $q$ marked corners on black vertices and $\Ga_0$ obtained from $\Ga_q$ by unmarking the corners, the relation between the bubble-dependent degree of $\Ga_q$ and that of $\Ga_0$ is
\be
\label{eqref:VacVersusExt}
\delta_\B(\Ga_q)=\delta_\B(\Ga_0)+\Phi_0({\partial\Ga_q}^{\tilde \Om})
\ee
where $\tilde \Om$ is the pairing of $\partial\Ga_q$ induced by $\Ga_q$, and $\partial\Ga_q^{\tilde \Om}$ is the corresponding covering.

\

The submaps $\Ps(\B,\Om_\B)$ can be seen as effective ``vertices" with a non-trivial internal structure, which is a superposition of hyper-edges (in Walsh's representation). This is the reason why stacked maps were called {\bf stuffed Walsh maps} in \cite{SWM}, ``stuffed" referring to the internal structure of vertices. This point of view was motivated by the intermediate field theory  (Section~\ref{sec:IFT}), in which the generating function is expressed as a matrix model with partial traces. The reason why we chose the name stacked maps in the version of the bijection described here is because we are interested in bicolored submaps, and a map in $\bS_D^q$ can be seen as a stacking of $D(D+1)/2$ combinatorial maps. In the case of $\bS^q(\bB,\Om_\bB)$, we are only interested in the $D$ layers of bicolored submaps containing color 0. The edges and vertices of color 0 are common to these $D$ stacked combinatorial maps.

\subsection*{Simpler bijections}

To recover the bijection of Thm.~\ref{thm:BijCycles} for $k$-cyclic bubbles from the bijection of Theorem~\ref{thm:BijSM}, we notice that $\Ps$ applied to a $k$-cyclic bubble in which we have chosen to pair the $h$-pairs (Def.~\ref{def:hPair}) with $h=D-k>D/2$ gives the stacked map pictured on the left, below (without color 0). It has
$k$ star-maps linking all the white vertices in the same cyclic order, and for the $D-k$ remaining colors it has a leaf attached to each white vertex. As underlined before, we can delete all the leafs with colors in $\lDr$ and the incident edges. We then merge the $k$ star-maps (with colors denoted $i_1,\cdots,i_k$) into a single one as pictured below, and label the edges with the color set $\{i_1,\cdots,i_k\}$. 
\begin{figure}[!h]
\centering
\includegraphics[scale=0.7]{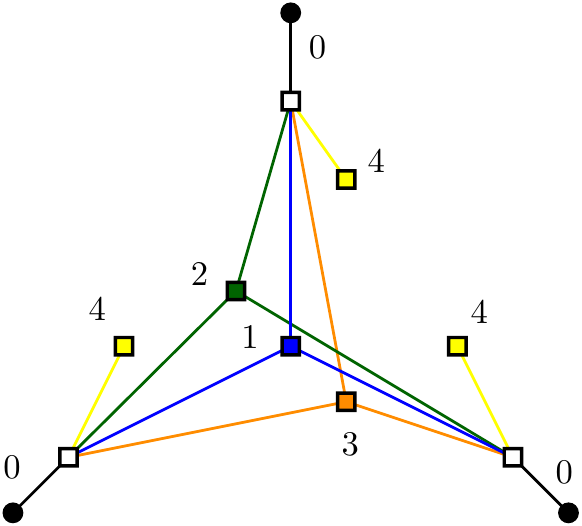}\qquad
\includegraphics[scale=0.7]{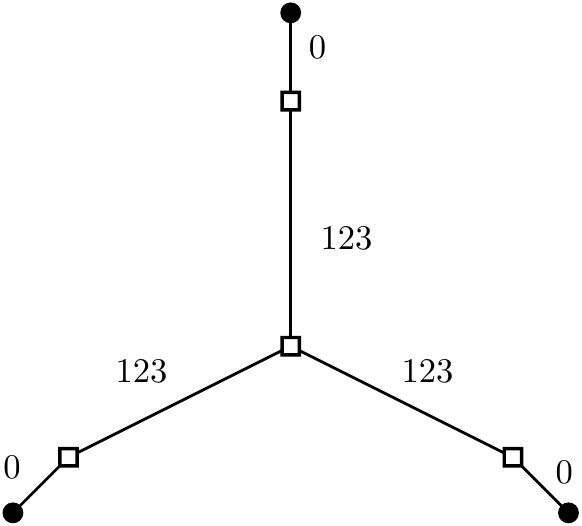}\qquad
\includegraphics[scale=0.7]{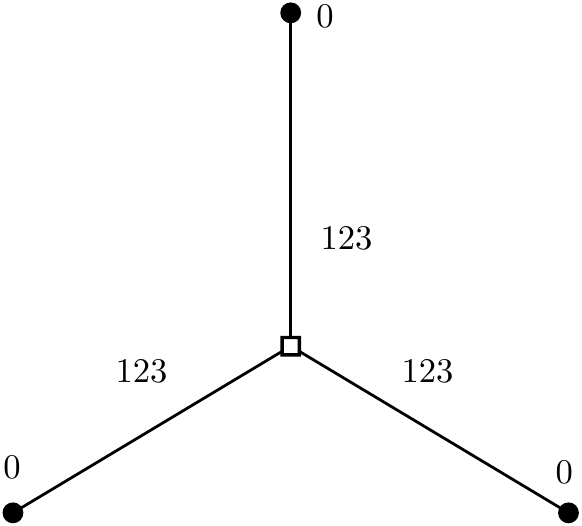}
\end{figure}
After performing this operation for every bubble in a stacked map, white vertices all have degree two. We can contract all the color-0 edges to obtain the bipartite maps described in Thm.~\ref{thm:BijCycles} (the convention on the coloring of vertices is different, as underlined above).
No color-0 edge remains, only color-0 vertices, to which we gave no color in Thm.~\ref{thm:BijCycles}, and the bicolored submaps $0i$ in the stacked map picture become the color-$i$ submaps.
This simplified bijection can be applied for non-cyclic bubbles in some cases, as stated in Thm.~\ref{thm:BijSimp} below.
See, for instance, the examples in Subsection~\ref{subsec:K334}.
The white vertices now have a cyclic ordering of edges around them, except if they have valency two in every color-$i$ submap, in which case the ordering is trivial (see the example of the $K_{3,3}$ bubble in  Subsection~\ref{subsec:K33}).

\begin{theorem}
\label{thm:BijSimp}
If for a bubble $\B\in\bG_{D-1}$ and a pairing $\Om$, $\Ps(\B,\Om)$ has only one or no star-map which is not a leaf for each color, and if the cyclic ordering of incidence of white squares is the same, then there is a simplified bijection with bipartite combinatorial maps. Edges carry color sets and bicolored cycles $0i$ are mapped to faces around color-$i$ submaps.
\end{theorem}
\begin{figure}[!h]
\centering
\includegraphics[scale=0.7]{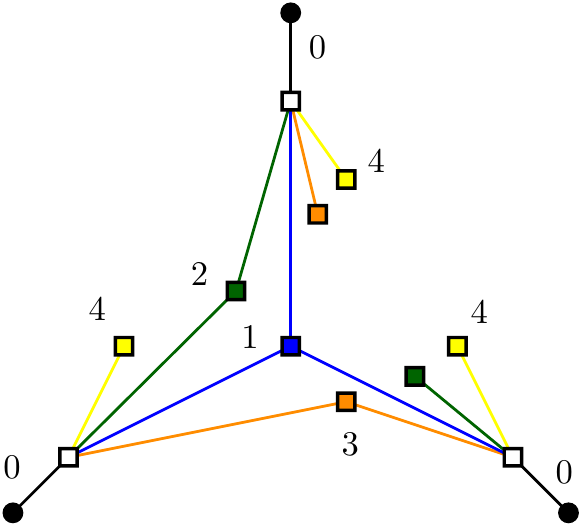}\qquad
\includegraphics[scale=0.7]{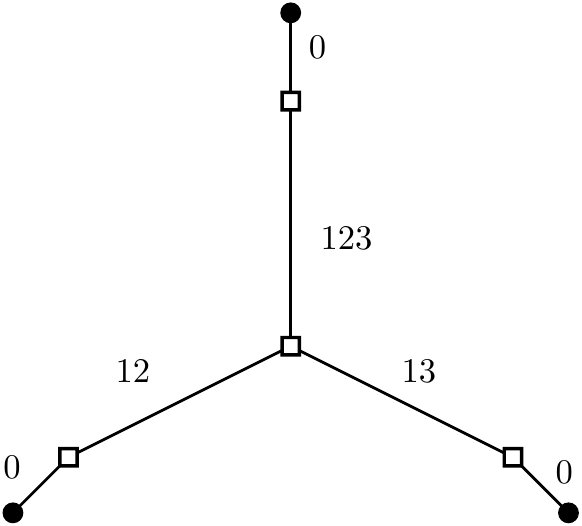}\qquad
\includegraphics[scale=0.7]{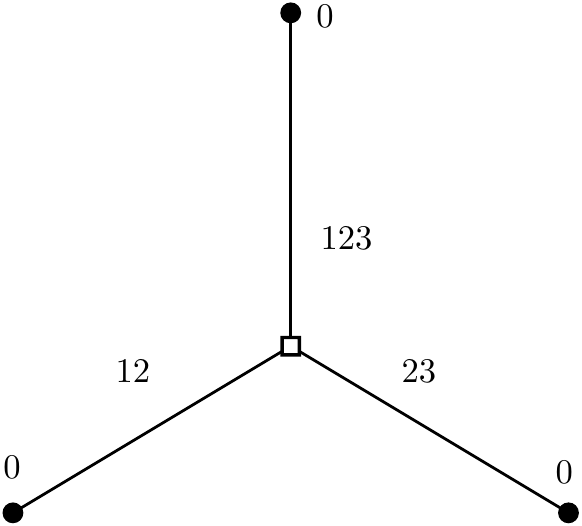}
\caption{Simplifying stacked maps in Thm.~\ref{thm:BijSimp}}
\label{fig:SimpBij}
\end{figure}
\prf The bijective operation performed on every bubble is described in Figure~\ref{fig:SimpBij}. \qed

\

The properties of stacked maps are studied in Chapter~\ref{chap:PropSWM}. Before, we focus on the most simple and well known case of quartic melonic bubbles. We prove a few more involved results in this simple case.

\section{The quartic melonic case}
\label{sec:QuartMelSec}

When the bubbles are 1-cyclic bubbles of length 4 (also called quartic melonic bubbles), we saw in Section~\ref{sec:CycBub2} that there is a bijection between the corresponding edge-colored graphs and combinatorial maps such that edges each have a single color in $\lDr$. The colors of edges correspond to the colors of the simple edges of the 1-cycles. In the case of quartic melonic bubbles,  colored  vertices have degree 2, so that the bijection simplifies. We denote $\bB_4^m$ the set of quartic melonic bubbles (1-cycles of length 4).
\begin{theorem}
\label{thm:BijBoundQuart}
There is a bijection between colored graphs in $\bG^q_D(\bB_4^m)$ and combinatorial maps with edges colored in $\lDr$ and $q$ marked corners, at most one per vertex. We denote $\Mb^q_D$ the set of such maps.
\end{theorem}
Examples are shown in Fig.~\ref{fig:BijQuart}. 

\

\prf  The bijection described in the proof of Theorem~\ref{thm:BijCycles} is trivially generalized to the present case, canonically adding marked color-0 edges as described in the previous section, and then mapping a colored graph with $q$ marked color-0 edges to a bipartite  combinatorial map with $q$ marked corners, such that white vertices all have degree 2. An edge is labeled with the distinguishable color of the corresponding 1-cycle.  By replacing degree 2 white vertices with edges, we obtain a (non-bipartite) combinatorial map.
In the other direction, there is a canonical way to add a white vertex on every edge, and from Theorem~\ref{thm:BijCycles} we obtain a colored graph with $q$ marked color-0 edges which can be deleted canonically. \qed
\begin{figure}[!h]
\centering
\includegraphics[scale=0.77]{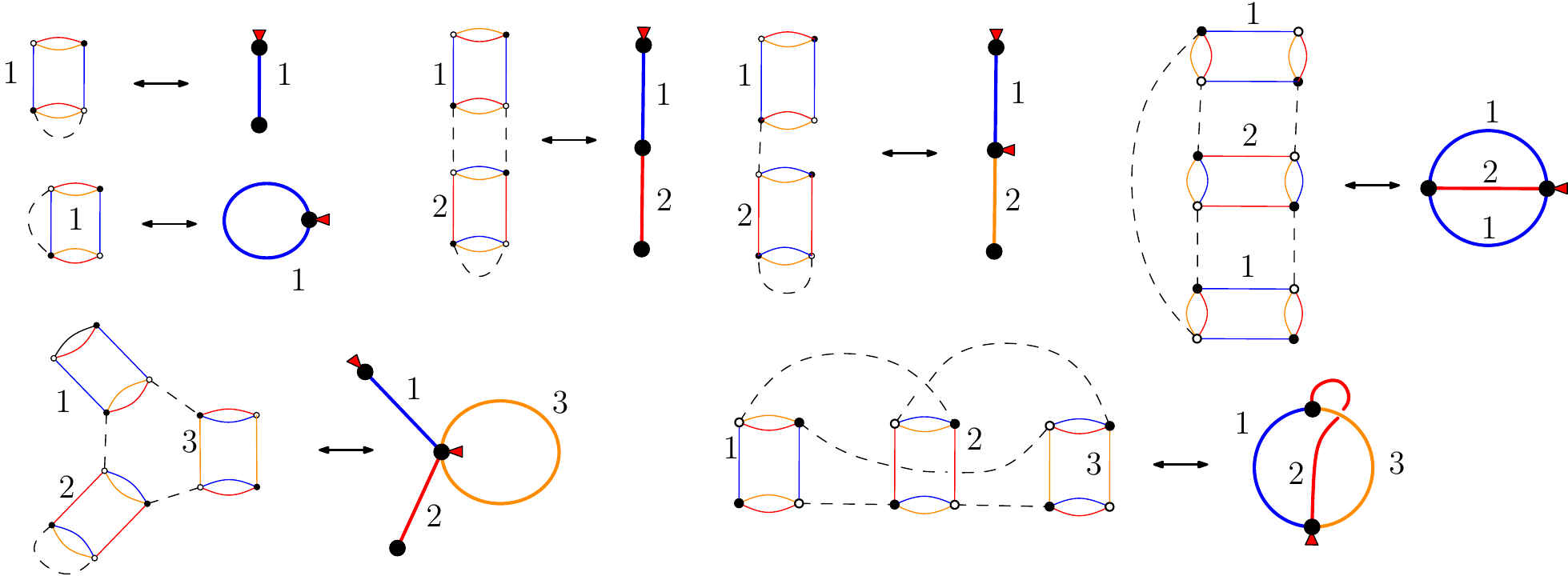}
	\caption{Quartic gluings with boundaries are mapped to combinatorial maps with marked corners. }
\label{fig:BijQuart}
\end{figure}

\

Again, denoting $\Gai$ the combinatorial map obtained by keeping all the vertices and only the edges of color $i$ - which we call {\it color-$i$ submap},
bicolored cycles of color $0i$ are mapped to faces around $\Gai$, and bicolored paths between two degree-$D$ vertices are mapped to faces of $\Gai$ between two marked corners,  called {\it broken faces}.
We only count cycles encountering no marked corner, and denote $\Fint(\Gai)$ the corresponding faces. The bubbles are melonic, so that $\tilde a=D-1$ (\ref{eqref:TildeAMelo}), and the degree is given by 
\be
\delta_{\bB_4^m}(\Ga)=D+(D-1)E(\Ga) - \sum_{i=1}^D \Fint(\Gai).
\ee
As the bubbles are melonic, it coincides with the Gurau degree (Def.~\ref{def:Deg}). Throughout this section, we will denote $\delta=\delta_{\bB_4^m}$ for simplicity.

\subsection{Quartic melonic gluings of positive degree}
\label{subsec:SubDomQuartMel}

For regular colored graphs, which are in bijection with colored combinatorial maps in $\Mb^0_D$, we have the following results.
\begin{prop}
\label{prop:PowVacQuart}
The degree of a map $\Ga\in\Mb^0_D$ can be written as
\be
\label{eqref:PowerVacQuart}
\delta(\Ga) = D\Lc(\Ga)-2\sum_{i=1}^D \Lc(\Gai)+2\sum_{i=1}^D g(\Gai),
\ee
where $\Lc=E-V+1$ is the number of independent cycles
 of a connected graph, $g$ is the genus of a map, given by $2-2g=V-E+F$, and $\Gai$ is the color-$i$ submap.
\end{prop}
 
 \prf The degree of $\Ga\in\Mb^0_D$ is given by 
 \be
 \delta(\Ga)=D+(D-1)E(\Ga) - \sum_{i=1}^D F(\Gai).
 \ee
 Combining the definitions of the genus and of the circuit-rank, 
 \be
 \label{eqref:gXL}
F=2L-2g +V-E.
 \ee
 As edges carry a single color, $\sum_{i=1}^D E(\Gai)=E(\Ga)$. Furthermore, $V(\Gai)=V(\Ga)$, so that 
  \be
 \delta(\Ga)=D(1+E(\Ga)-V(\Ga)) + 2 \sum_{i=1}^D \bigl(g(\Gai)-L(\Gai)\bigr),
 \ee
 which concludes the proof, as the map is connected. \qed

\

The first following result is already known as we showed in Section~\ref{sec:SimplerBij} that maximal gluings of 1-cyclic bubbles of a single kind are trees. However, here we are also able to characterize maps of positive degree.
We call monochromatic cycle in a graph with colored edges a proper cycle (all vertices are distinct) such that all edges have the same color. Similarly, we call polychromatic cycles a proper cycle such that at least two edges have different colors. The total number of independent color-$i$ cycles is the circuit-rank of 
the color-$i$ submap, and the number of independent polychromatic cycles is the circuit-rank of the graph to which we subtract the total number of monochromatic cycles.

\begin{coroll}
\label{coroll:CorollMajTree}
The degree $\delta$ of a map $\Ga\in\Mb^0_D$ satisfies
\be
\label{eqref:CorollMajTree}
\delta(\Ga)\ge 0,
\ee 
and the first non-empty orders can easily be characterized:
\begin{itemize}
\item $\delta=0$: maximal maps are trees
\item $\delta=D-2$: maps with a single monochromatic cycle (left of Fig.~\ref{fig:NLONNLOM0})
\item{ There are 3 cases: 
	\begin{itemize}
	\item If $D>4$, $\delta=D$: maps with a single polychromatic cycle (right of Fig.~\ref{fig:NLONNLOM0})
	\item If $D=4$, $\delta=4$: maps with a single polychromatic cycle,  planar maps with two color-$i$ cycles, or maps with a color-$i$ and a color-$j\neq i$ cycle
	\item If $D=3$, $\delta=2$:  planar maps with two color-$i$ cycles, or maps with a color-$i$ and a color-$j\neq i$ cycle

	\end{itemize}}
\end{itemize}
The following orders can also be characterized using equation $(\ref{eqref:DeltaQuartPos})$, but the number of possible cases increases with the order. Roughly, monochromatic cycles in planar submaps contribute with $D-2$, polychromatic cycles with $D$ and raising the genus of monochromatic submaps by 1 contributes with 2, but implies a certain number of monochromatic cycles. 
\end{coroll}

\begin{figure}[!h]
\centering
\includegraphics[scale=0.6]{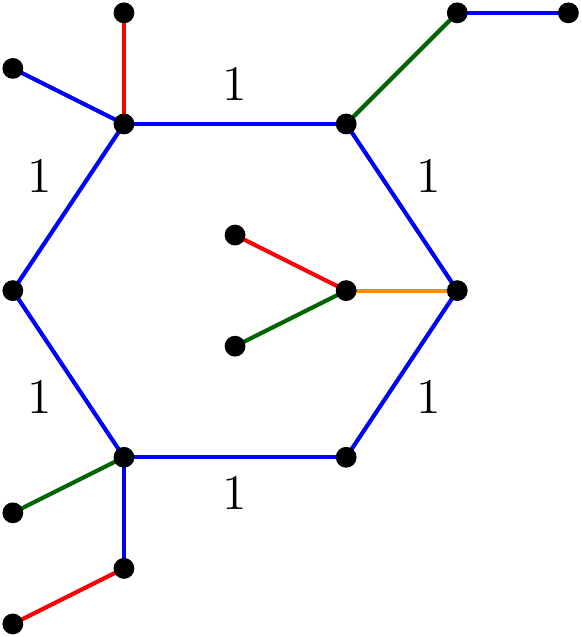}\hspace{3cm}
\includegraphics[scale=0.6]{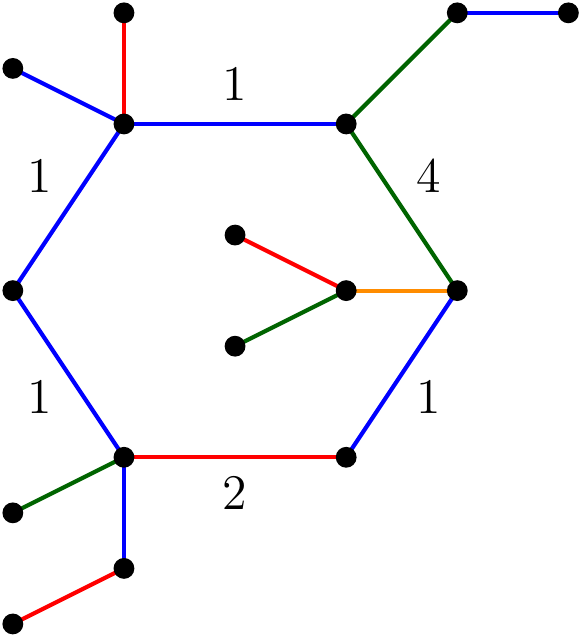}
\caption{Maps contributing to the first two non-empty orders of $\Mb^0_D$.}
\label{fig:NLONNLOM0}
\end{figure}

\prf To prove that result, we use \eqref{eqref:PowerVacQuart} and first show that
\be \label{ColoredCycleBound}
\frac{D}{2}\Lc(\Ga)\ge\sum_{i=1}^D\Lc(\Gai).
\ee
For each $i\in\lDr$, consider a forest $\cT^{(i)}$ spanning $\Gai$. There are $L(\Gai)$ edges in $\Gai\setminus \cT^{(i)}$ and they identify independent cycles. The union $\bigcup_{i\in[D]} \cT^{(i)}$ is connected and spans $\Ga$. A spanning tree $\cT$ of $\bigcup_{i\in\lDr} \cT^{(i)}$ is thus a spanning tree of $\Ga$. Since all the edges in $\bigcup_{i\in\lDr} (\Gai\setminus \cT^{(i)})$ are distinct, 
\begin{equation}
L(\Ga)\ge\sum_{i=1}^DL(\Gai),
\end{equation}
and \eqref{ColoredCycleBound} follows from $D\ge2$. Since $g(\Gai)\ge0$, one obtains \eqref{eqref:CorollMajTree}. 
Furthermore, if $e\not\in\cT$, then either $e\in \cup_{i\in\lDr} \cT^{(i)}$ or $e\not\in\cup_{i\in\lDr} \cT^{(i)}$ and since $\cT$ is a tree spanning $\cup_{i\in\lDr}\cT^{(i)}$, we find
\be
L(\Ga)=L(\cup_{i=1}^D\cT^{(i)})+\sum_{i=1}^DL(\Gai),
\ee
where $L(\cup_{i=1}^D\cT^{(i)})$ is the number of independent polychromatic cycles of $\Ga$. Equation \eqref{eqref:PowerVacQuart} can be rewritten as 
\be
\label{eqref:DeltaQuartPos}
\delta(\Ga)=DL(\cup_{i=1}^D\cT^{(i)})+(D-2)\sum_{i=1}^DL(\Gai)+2\sum_{i=1}^Dg(\Gai),
\ee
which is a sum of positive or vanishing terms. The inequality is saturated if and only if $L(\cup_{i=1}^D\cT^{(i)})=L(\Gai)=g(\Gai)=0$, i.e. if and only if $\Ga$ is a tree. Having genus 1 requires at least two cycles, so that the first positive degree is $D-2$, for maps such that $\sum_{i}L(\Gai)=1$. For $D>4$, the second positive degree is $D$, for maps such that $L(\cup_{i=1}^D\cT^{(i)})=1$ but $\forall i,\ L(\Gai)=0$. This describes maps such that all color-$i$ submaps $\Gai$ are trees, but the union of these trees has a single cycle. If $D=4$, $2D-4=4$ so that at order $D$ we also have contributions of maps satisfying  $\sum_{i}L(\Gai)=2$, but they require $\sum_{i}g(\Gai)=2$. This does not give any restrictions if two monochromatic submaps have one cycle each, but it requires a map with one monochromatic submaps containing two cycles to be planar. If $D=3$, $2D-4=2=D-1$, so that the second positive order is 2, with only contributions from maps satisfying $\sum_{i}L(\Gai)=2$. The same cases apply.
\qed

\

Proposition~\ref{prop:PowVacQuart} and Corollary~\ref{eqref:CorollMajTree} are both true for gluings of $1$-cyclic bubbles of any length.

\subsection{The boundary case}
\label{subsec:BoundQuart}



%
\begin{theorem}
\label{thm:Represent}
Any non-necessarily connected colored bipartite graph $\B\in\bG^U_{D-1}$ with $q$ white vertices is the boundary graph of some connected $\G\in\bG^q_D(\bB_4^m)$ with $q$ degree-$D$ white vertices, or equivalently of some connected colored combinatorial map with $q$ marked corners. 
\end{theorem}
%
 It means that the boundary maps $\partial : \bG^q_D(\bB_4^m) \rightarrow \bG^q_{D-1}$ and $\partial : \Mb^q_D\rightarrow \bG^q_{D-1}$ are surjective.
For any $\B\in\bG^q_{D-1}$ and for any pairing 
we construct in the proof a specific map $\cM(\B, \Om)$ such that
\begin{equation}
\partial \cM(\B, \Om) = \B.
\end{equation}

\begin{figure}[!h]
\centering
\includegraphics[scale=0.8]{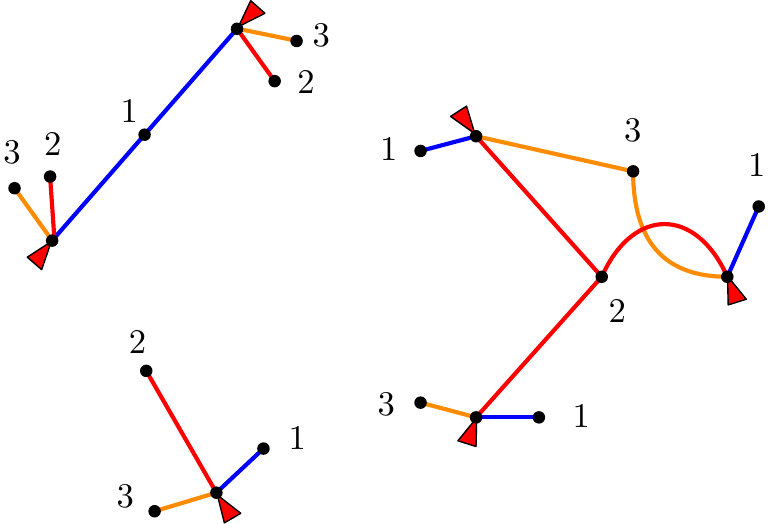}
\hspace{0.7cm}\raisebox{+10ex} {$\rightarrow$}\hspace{0.7cm}
\includegraphics[scale=0.8]{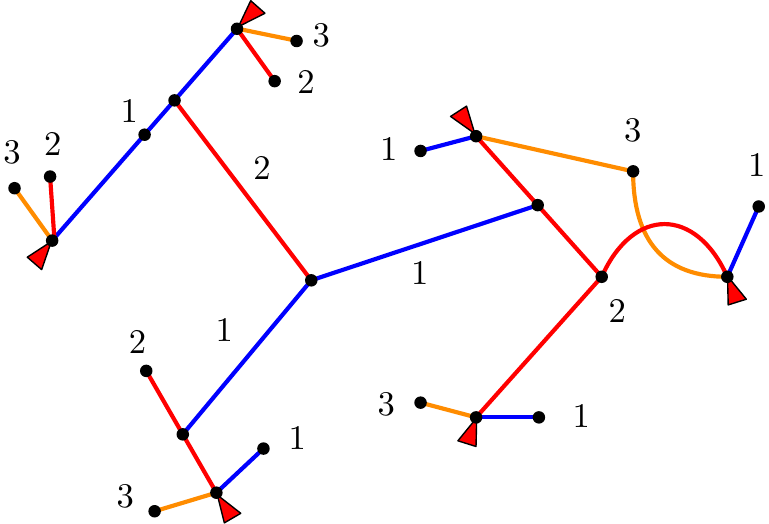}
\caption{Obtaining a connected map with the right boundary graph.}
\label{fig:Represent1Cyc}
\end{figure}

\prf As before in the proof of Prop.~\ref{prop:Bound1Cyc}, we apply Thm~\ref{thm:BijSMGen} to a set $\cG$ of $K$ connected graphs in $\bG_{D-1}$ (with colors in $\lDr$), for any choice of pairing, and get a set of stacked maps in $\bS_{D-1}$. We order the edges incident to white vertices from 1 to $D$  and then mark the corner between the edges of color $D$ and 1 on each white vertex. 
We obtain a combinatorial map $\cM$ with marked corners on every white vertex and $K$ connected components such that each edge carries a colored label, with boundary graph $\cG=\partial\cM$.
We gather these connected components into a single connected map with the right boundary graph.  The procedure is illustrated in Fig.~\ref{fig:Represent1Cyc}: we add one isolated vertex, pick one edge in each component, add a valency two vertex on that edge, and then, if this edge has color $i$, add an edge of color $j\neq i$ between the newly added degree two vertex and the newly added isolated  vertex. The original connected components now interact through a star-map. The newly added vertex is only  incident to unbroken faces, so that the boundary graph is unchanged, and the corresponding connected map suits. \qed

\



Considering a graph $\B\in\bG_{D-1}$,  any map in $\partial^{-1}(\B)$ induces a pairing of its vertices. Denoting $\bM(\B,\Om)$ the family of maps in $\bM_D^{V(\B)/2}$ with boundary graph $\B$ and induced pairing $\Om$, it defines a partition $\partial^{-1}(\B)=\sqcup_{\Om_{i}}\bM(\B, \Om_{i})$, where the union is over inequivalent pairings. Each $\bM(\B,\Om)$ contains infinitely many maps as adding degree two unmarked vertices does not change the boundary graph, and adding contributions of $\Mb^0_D$ to unmarked vertices neither.

 \begin{definition}
 \label{def:MultCycl}
Given a graph $\G$ with colored edges in $\lDr$, we define $\Lc_m(\G)$ as 
\be
\Lc_m(\G)=L(\G)-\sum_{i=1}^D L(\G^{(i)}).
\ee
It is the number of independent polychromatic - or rainbow - cycles of $\G$, i.e. cycles such that at least two edges carry different colors.
 \end{definition}

The 0-score of a covering $\BCO$ can also be computed using the following lemma.
 
 \begin{lemma}
\label{lemma:PhiLm}
The score 
of a covering $\BCO$ of $\B\in\bG_{D-1}$  is related to 
$\Lc_m(\BOM)$ as follows,
\be
\Phi_0(\BCO)=1 + V(\BOM)(D-1)- \Lc_m(\BOM).
\ee 
\end{lemma}

As an example, we see at first glance on the \emph{right} of Fig.~\ref{fig:BijBub} that for $\B$ and $\Om$ in the \emph{left} of Fig.~\ref{fig:BijBub}, there is a single polychromatic cycle. As there are 4 pairs and $D=4$, adding color-0 edges on $\Om$, we get  $\Phi_0(\BCO)=1+4\times 3 -1=12$.

\

\prf This relies on the fact that  the  bicolored cycles $0i$ of $\BCO$ are in one-to-one correspondence with the cycles of $\BOM^{(i)}$, 
\be
\Phi_0(\BCO) = \sum_{i=1}^D L(\BOM^{(i)}).
\ee 
As $\BOM$ is connected, $L(\BOM)=E(\BOM)-V(\BOM)+1$, and as $E(\BOM)=DV(\BOM)$,
\be
L(\BOM)=V(\BOM)(D-1)+1.
\ee
\qed

Recall from Definition~\ref{def:OptPair} that an optimal pairing is one which maximizes $\Phi(\BCO)$. \emph{It is thus equivalent to minimizing} $\Lc_m(\BOM)$. 
 We have the following results.

\begin{lemma}
 \label{lemma:IneqRain}
Let $\Ga$ be an edge-colored map with connected boundary graph $\partial \Ga = \B$, induced pairing $\Om$,
and $\BOM$ as in Def.~\ref{def:EulColGraph}, then
 \be
 \label{eqref:IneqOpt}
\Lc_m(\Ga)\ge \Lc_m(\BOM).
 \ee
 \end{lemma}
 
When each monochromatic submap is a forest, $\Lc_m(\BOM)$ coincides with the number of independent cycles of the map itself.

\
 
\prf
For each color $i$, consider $\cT^{(i)}$ a forest that spans $\Gai$. As in the proof of Corollary~$ \ref{coroll:CorollMajTree}$, the number of polychromatic cycles of $\Ga$ is
\be
L(\Ga)-\sum_{i=1}^DL(\Gai)=L(\cup_{i=1}^D\cT^{(i)}).
\ee
Consider $\cT_{ext}^{(i)}$ the sub-forest of $\cT^{(i)}$ made of its connected components which have at least one marked vertex. 
%
The union $\cup_{i=1}^D\cT_{ext}^{(i)}$ is a subgraph of $\cup_{i=1}^D\cT^{(i)}$, hence
\be
\label{eqref:LmLmExt}
L(\cup_{i=1}^D\cT^{(i)})\ge L(\cup_{i=1}^D\cT_{ext}^{(i)}).
\ee
As $\B$ is connected,  $\cup_{i=1}^D\cT_{ext}^{(i)}$ is connected. Moreover, $E(\cup_{i=1}^D\cT_{ext}^{(i)})=\sum_{i=1}^DE(\cT_{ext}^{(i)})$, so that
\be
\label{eqref:Lm1}
L(\cup_{i=1}^D\cT_{ext}^{(i)})=\sum_{i=1}^DE(\cT_{ext}^{(i)})-V(\cup_{i=1}^D\cT_{ext}^{(i)})+1.
\ee
Since $\cT_{ext}^{(i)}$ is a forest, $E(\cT_{ext}^{(i)})=V(\cT_{ext}^{(i)})-K(\cT_{ext}^{(i)})$, $K$ being the number of connected components. We denote $q$ the number of marked vertices and split the sum over vertices
\be
\sum_{i=1}^D V(\cT_{ext}^{(i)}) = Dq + \sum_{i=1}^D V_{\rm int}(\cT_{ext}^{(i)})= Dq\ + \sum_{\substack{{v\in\cup_{i=1}^D\cT_{ext}^{(i)}}\\ { \text{unmarked}}}} \col(v),
\ee
where $\col(v)$ is the number of colors incident to $v$ in $\cup_{i=1}^D\cT_{ext}^{(i)}$. Equation (\ref{eqref:Lm1}) now writes
\be
\label{eqref:Lm2}
L(\cup_{i=1}^D\cT_{ext}^{(i)})=q(D-1)+ \sum_{\substack{{v\in\cup_{i=1}^D\cT_{ext}^{(i)}}\\ { \text{unmarked}}}} (\col(v)-1)-\sum_{i=1}^DK(\cT_{ext}^{(i)})+1.
\ee
Each connected component of $\cT_{ext}^{(i)}$ gives a disjoint cycle of $\BOM=\Ga_{\circlearrowleft}$, so that 
\be
K(\cT_{ext}^{(i)})=\Phi_{0,i}(\BCO).
\ee
Using Lemma~\ref{lemma:PhiLm}, we can rewrite 
\be
\label{eqref:Lm3}
L(\cup_{i=1}^D\cT_{ext}^{(i)})=\Lc_m(\BOM)\ + \sum_{\substack{{v\in\cup_{i=1}^D\cT_{ext}^{(i)}}\\ { \text{unmarked}}}} (\col(v)-1).
\ee
As there are no isolated vertex, the sum in the right hand side is a sum of positive terms, which concludes the proof. 
We furthermore deduce from (\ref{eqref:Lm3}) that 
\be
\label{eqref:Lm4}
L(\cup_{i=1}^D\cT_{ext}^{(i)})=\Lc_m(\BOM)\ \qquad \Rightarrow \qquad \forall v\in\cup_{i=1}^D\cT_{ext}^{(i)} \text{ unmarked},\  \col(v)=1.
\ee
 \qed

\begin{theorem}
\label{thm:OptMapBound}
Let $\Ga_q\in \Mb^q_D$ with $q$ marked corners, and connected boundary graph $\B = \partial \Ga_q$. Then its degree is bounded from below
\be
\label{eqref:PowDomB}
\delta(\Ga_q)\ge 1+(D-1)(q+\Lc_m(\BOM)),
\ee
$\Om$ being the pairing of $\B$ induced by $\Ga_q$ and $K$ the number of connected components.
Moreover the equality holds if and only if 
\begin{itemize}
\item $\Ga_q^{(i)}$ is a forest for each $i\in\lDr$, and
\item denoting $\cT_{ext}^{(i)}$ the restriction of $\Ga_q^{(i)}$ to its trees which have at least one marked vertex, those trees can meet on marked vertices only, and
\item $\Ga_q\setminus \bigcup_{i\in\lDr} \cT_{ext}^{(i)}$ is a forest whose trees each meet $\bigcup_{i\in\lDr} \cT_{ext}^{(i)}$ at a single vertex.
\end{itemize}
\end{theorem}

\

\prf We start with Lemma \eqref{eqref:VacVersusExt}, $\delta(\Ga_q) = \delta(\Ga_0) + \Phi_0(\BCO)$, and apply Proposition \ref{prop:PowVacQuart} to $\delta(\Ga_0)$ and Lemma \ref{lemma:PhiLm} to $\Phi(\BCO)$. That gives
\begin{equation}
\label{eqref:DegBoundQuartMel}
\delta(\Ga_q) = D\Lc_m(\Ga_0)+(D-2)\sum_{i=1}^D \Lc(\Ga_0^{(i)})+2\sum_{i=1}^D g(\Ga_0^{(i)}) + 1 + q(D-1)- \Lc_m(\BOM).
\end{equation}
As the two sums in the right hand side are sums of positive terms, and because of Lemma~\ref{lemma:IneqRain}, we get (\ref{eqref:PowDomB}). Furthermore, the equality occurs when 
$\forall i\in\lDr, L(\Ga_0^{(i)})=g(\Ga_0^{(i)})=0$, i.e. monochromatic submaps are trees (first condition),
and when $\Lc_m(\Ga_0)=\Lc_m(\BOM)$, which on one hand implies that 
\be
\label{eqref:SatLm}
L(\cup_{i=1}^D\cT^{(i)}) =  L(\cup_{i=1}^D\cT_{ext}^{(i)})
\ee
in (\ref{eqref:LmLmExt}), and on the other hand implies (\ref{eqref:Lm4}). This last condition means that unmarked vertices belong to at most one $\cT_{ext}^{(i)}$ (second condition). If the last condition in Thm.~\ref{thm:OptMapBound} is not satisfied, then (\ref{eqref:SatLm}) is not either. On the other way, a map satisfying the three conditions of Thm.~\ref{thm:OptMapBound} saturates (\ref{eqref:PowDomB}), which concludes the proof. \qed

\

We recall that $\bG_4^m$ is the set of quartic melonic bubbles, i.e. 1-cyclic bubbles of size 4. A consequence of this theorem is that the degree of a colored graph in  $\bG^q_D(\bB_4^m)$ is bounded from below by the optimal bound (that is, the inequality is saturated for some colored graphs):
\be
\label{eqref:BoundDegQuart}
\delta(\G_q\in \bG^q_D(\bB_4^m))\ge 1+(D-1)\bigl(q+\Lc_m(\partial \G_{q/\Opt})\bigr),
\ee
where $\Opt$ is an optimal pairing of the boundary graph $\partial \G_q$. Therefore, we can give the best possible lower bound on the degree of graphs with a given boundary graph:

\begin{coroll}
For a given bubble $\B\in\bG_{D-1}$, any colored graph in $\bG^q_D(\bB_4^m)$ with boundary graph $\B$ has a degree higher or equal to:
\be
\label{eqref:BoundDegQuart2}
\delta\bigl(\G_q\in \partial^{-1}(\B)\bigr)\ge 1+(D-1)\bigl(q+\Lc_m(\B_{/\Opt})\bigr),
\ee
where $\Opt$ is an optimal pairing of the bubble $\B$. This bound is optimal. 
\end{coroll}

Importantly, the map $\cM(\B,\Om)$ built from $\Ps(\B,\Opt)$ in the proof of Thm.~\ref{thm:Represent}  is an example of map saturating (\ref{eqref:BoundDegQuart}). Therefore, for any given bubble $\B$, having found an optimal pairing, it is easy to exhibit a map in  $\partial^{-1}(\B)$ of minimal degree. It is also trivial to compute the degree of the corresponding map, as its circuit-rank coincides with $L_m(\BOM)$. The next important step would be to obtain a relation of the kind 
\be
\deltaG(\B)=k \quad \Rightarrow \quad \Lc_m(\B_{/\Opt})\ge f(k).
\ee
We suggest that this problem could be addressed by studying the maps obtained when applying the bijection $\Ps$ of Thm.~\ref{thm:BijOneColor} on the pairing induced by a covering. This kind of manipulations are used in Section~\ref{sec:UniGraphColSYK} for other purposes. For instance, if $\B$ is not melonic, then $L_m(\B^{\Opt})\ge1$, and 
\be
\label{eqref:BoundDegQuart3}
\deltaG(\B)> 0 \quad\Rightarrow \quad\delta\bigl(\G_q\in \partial^{-1}(\B)\bigr)\ge 1+(D-1)\bigl(q+1\bigr).
\ee
But we will see in Section~\ref{sec:UniGraphColSYK} that coverings with $L_m(\B^{\Opt})= 1$ have a single cycle when applying the bijection of Thm.~\ref{thm:BijOneColor} on color 0 (as on the left of Fig.~\ref{fig:NLOCycle} but with possibly trivial melonic insertions on the edges). From the inverse bijection, it corresponds to a graph such as in Figure~\ref{fig:Handle} or with the two same colors going along the ``ribbon". These graphs have Gurau degree $D$ or $D-2$. Therefore, 
\be
\label{eqref:BoundDegQuart4}
\deltaG(\B)> D \quad\Rightarrow \quad\delta\bigl(\G_q\in \partial^{-1}(\B)\bigr)\ge 1+(D-1)\bigl(q+2\bigr).
\ee
This is an important question, which we leave for future work.

\

From equation \eqref{eqref:DegBoundQuartMel}, we see that one can also deduce the sub-leading orders in $\partial^{-1}(\B)$, as done in Corollary~\ref{coroll:CorollMajTree}. Indeed, it goes back to using Corollary~\ref{coroll:CorollMajTree} for $\Ga_0$, and be sure that the marked corners satisfy the canonical condition to obtain back $\Ga_0$. 
We apply this to the study of some non-maximal maps in $\bM^2_D$. More precisely, 
\be
\bM^2_D=\sqcup_{\B\in\bG^4_D}\partial^{-1}(\B),
\ee
 and we can characterize the first subleading orders of some $\partial^{-1}(\B)$. We do it for a quartic melonic bubble $\B_i\in\bB_4^m$, $i$ being the distinguished color. There are two pairings if $\B_i$, one leading to $L_m(\BOM)=0$ (as the graph is melonic), and one leading to $L_m(\BOM)=D-2$, so that with this pairing the minimal degree of contribution is $1+(D-1)( 2+ D-2)=D^2-D+1$, which is bigger than $D$ for $D>1$. Therefore, 

\begin{coroll}
\label{coroll:CorollMaj4Point}
The degree $\delta$ of a map with quartic melonic boundary graph $\Ga_2\in\partial^{-1}(\B_i)$,   satisfies
\be
\label{eqref:CorollMajTree4PT}
\delta(\Ga_2)\ge 2D-1
\ee 
and the first non-empty orders can easily be characterized:
\begin{itemize}
\item $\delta=2D-1$: maximal maps are trees with two corners marked on the same color-$i$ subtree $\Gai$
\item $\delta=3D-3$: maps with a single monochromatic cycle and two corners marked on the same face in $\Gai$. 
\item{ There are 3 cases: 
	\begin{itemize}
	\item If $D>4$, $\delta=3D-1$: maps with a single polychromatic cycle, with two corners marked on the same color-$i$ subtree $\Gai$
		\item If $D=4$, $\delta=11$: maps with a single polychromatic cycle,  planar maps with two color-$i$ cycles, or maps with a color-$i$ and a color-$j\neq i$ cycle, all with two corners marked on the same face in $\Gai$. 
	\item If $D=3$, $\delta=9$:  planar maps with two color-$i$ cycles, or maps with a color-$i$ and a color-$j\neq i$ cycle, all with two corners marked on the same face in $\Gai$.

	\end{itemize}}
\end{itemize}
\end{coroll}

\prf The proof uses Corollary~\ref{coroll:CorollMajTree} to the map $\Ga_0$ obtained by unmarking the corners, which corresponds to adding color-0 edges canonically. Here, one only has to be sure to recover the right boundary graph, i.e. that the two marked corners are on the same face on the color-$i$ submap $\Gai$. \qed

\

In the case of a non-connected boundary,  gathering  maps of minimal degree from $\partial^{-1}(\B_i)$ as done in the proof of Thm.~\ref{thm:Represent}, where the $\B_i$ are the connected components of the boundary graph, we can easily build a map $\Ga_q$ with boundary $\{\B_i\}$ such that
\be
\delta(\Ga_q)= K(\B)+(D-1)(q+\Lc_m(\BOM)), 
\ee
where $K(\B)$ is the number of connected components. We expect this map to be of minimal degree in  $\partial^{-1}(\{\B_i\})$ but we do not prove it here.

\subsection{Locally-orientable quartic melonic gluings}
\label{subsec:LOQuart}

We can adapt these theorems in the case where we drop the bipartiteness condition and consider non-necessarily orientable gluings of quartic melonic bubbles $\tilde \bG_D(\bB_4^m)$ (Def.~\ref{def:NORestricted}), and also spaces with boundaries $\tilde \bG^q_D(\bB_4^m)$. Most of the previous results still hold in this case. Theorem~\ref{thm:BijBoundQuart} stated that once a choice of orientation (e.g. from white to black on color-0 edges) was done for the cycles alternating edges of color 0 and $(D-1)$-pairs, then we had a bijection between  $\bG_D(\bB_4^m)$ and $\Mb^q_D$. In the non-orientable case, one cannot make such a global choice. An orientation has to be chosen arbitrarily for each such cycle. The resulting (LO) map will have twist factors on the edges (Fig.~\ref{fig:LOMaps}), which depend on this choice. Two different choices of orientation lead to maps which can be obtained one from another by a sequence of local changes of orientation (Def.~\ref{def:LocChange}). 
We consider the equivalence relation between two \LOR maps 
\be
\Ga \sim_\circlearrowleft \Ga' \qquad \Leftrightarrow \qquad
\biggl[
 \substack{{\Ga \text{ and } \Ga' \text{can be obtained one from another}} \\[+1ex]{\text{by a finite sequence of local change of orientations}}}
 \biggr]
\ee
We denote  $\tilde \Mb^q_D$ the set of locally orientable combinatorial maps with edges colored in $\lDr$ and $q$ marked corners on $q$ different vertices, and consider the subset  $\tilde \Mb^q_D(\cP)\subset\tilde \Mb^q_D$ which elements satisfy the property $(\cP)$.
\be
\begin{tabular}{rcl}
$M\in\tilde \Mb^q_D(\cP)$ & $\Leftrightarrow$ &  \emph{The vertices of $M$ can be colored in black and white so that}: \\[+1ex]
&&{\begin{tabular}{l}
$(-)$ edges carrying a $(+)$ factor only link white vertices,\\[+0.5ex]
 $(-)$ edges carrying a $(-)$ factor only link a black and a \\
 \quad white vertex.
\end{tabular}}
\end{tabular}
\ee
We consider the quotient set $\tilde \Mb^q_D(\cP) / _{\sim_\circlearrowleft}$. 

\begin{theorem}
There is a bijection between colored graphs in $\tilde\bG^q_D(\bB_4^m)$ and equivalence classes in $\tilde \Mb^q_D(\cP) / _{\sim_\circlearrowleft}$.
\end{theorem}

\prf The proof is similar to the previous proofs with the difference that here, one cannot make a global choice of orientation of all the cycles which alternate color-0 edges and $(D-1)$-pairs. We detail the constructions. Each $(D-1)$-pair defines an half-edge, and each oriented cycle defines a vertex, with at most one marked corner if the cycle contains a marked color-0 edge. As before, each quartic melonic bubble defines an edge between the two corresponding pairs, colored with the colors not in the $(D-1)$-pairs. Here however, the edge carries a twist factor $(+)$ or $(-)$ (Fig.~\ref{fig:LOMaps}). The cycles are embedded in the plane according to the orientation they were given, and a factor is assigned as in Fig.~\ref{fig:BijLO}. The locally orientable map defined this way satisfies $\cP$. We see that choosing the opposite ordering of the cycle in the colored graph, we reverse the ordering of the half-edges around the corresponding vertex, and we invert all the twist factors. This is precisely the definition of a local change of orientation. Conversely, starting from an element of $M\in\tilde \Mb^q_D(\cP) / _{\sim_\circlearrowleft}$, we replace each vertex by a cycle alternating color-0 edges and $(D-1)$-pairs of edges colored with the colors not on the corresponding edge of $M$, and embed in the plane while respecting the color: countercklockwise (resp. $clockwise$) for cycles corresponding to white (resp. black) vertices. For each color-$i$ edge of $M$, we add the two missing color-$i$ edges between the corresponding $(D-1)$-pairs in accordance with the rule of Fig.~\ref{fig:BijLO}. We then  consider the underlying graph. 
\qed

\begin{figure}[!h]
\centering
\includegraphics[scale=0.6]{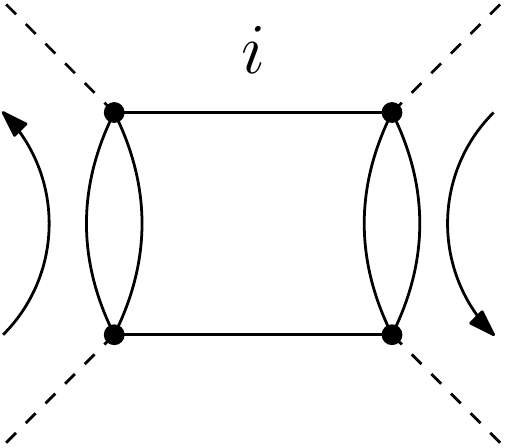}\quad\raisebox{+7.5ex}{$\leftrightarrow$}\quad\raisebox{+7.5ex}{\includegraphics[scale=0.6]{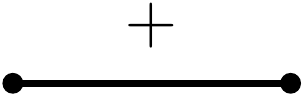}}\hspace{2cm}
\includegraphics[scale=0.6]{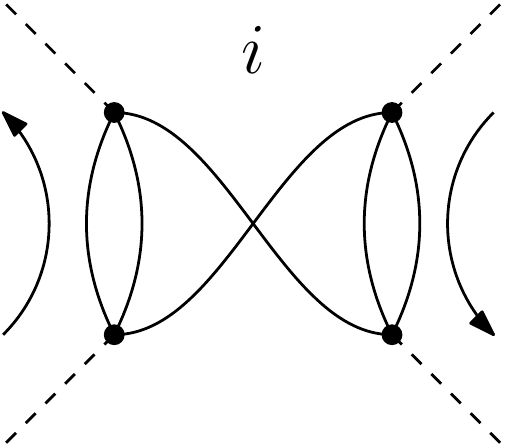}\quad\raisebox{+7.5ex}{$\leftrightarrow$}\quad\raisebox{+7.5ex}{\includegraphics[scale=0.6]{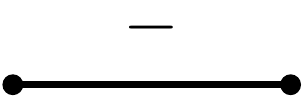}}
\caption{Assigning twist factors in the bijection.}
\label{fig:BijLO}
\end{figure}

\

Proposition~\ref{prop:PowVacQuartLO} trivially generalizes to the locally orientable case.

\begin{prop}
\label{prop:PowVacQuartLO}
The degree of a locally orientable map $\Ga\in\tilde \Mb^0_D$ is
\be
\label{eqref:PowerVacQuartLO}
\delta(\Ga) = D\Lc_m(\Ga)+(D-2)\sum_{i=1}^D \Lc(\Gai)+2\sum_{i=1}^D g(\Gai),
\ee
where $\Lc=E-V+1$ is the number of independent cycles of a connected graph, $g\in\frac 1 2 \bN$ is the genus of a map, given by $2-2g=V-E+F$, and $\Gai$ is the color-$i$ submap.
\end{prop}

We define a twisted cycle as a proper cycle containing an odd number of edges carrying $(-)$ twist factors. Twists only modify the contribution of monochromatic cycles. 

\begin{coroll}
\label{coroll:CorollMajTreeLO}
The degree $\delta$ of a locally orientable map $\Ga\in\tilde\Mb^0_D$ verifies
\be
\label{eqref:CorollMajTreeLO}
\delta(\Ga)\ge 0,
\ee 
and the first non-empty orders can easily be characterized:
\begin{itemize}
\item $\delta=0$: maximal maps are trees
\item $\delta=D-2$: maps with a single monochromatic non-twisted cycle (left of Fig.~\ref{fig:NLONNLOM0})
\item $\delta=D-1$: 
	\begin{itemize}
	\item If $D>3$, maps with a single monochromatic twisted cycle (left of Fig.~\ref{fig:NLONNLOMLO})
	\item If $D=3$,maps with a single monochromatic twisted cycle or two monochromatic cycles (planar)
	\end{itemize}
\item{ $\delta=D$: 
	\begin{itemize}
	\item If $D>4$, $\delta=D$: maps with a single polychromatic cycle (right of Fig.~\ref{fig:NLONNLOM0})
	\item If $D=4$, $\delta=4$: maps with a single polychromatic cycle or two monochromatic cycles (planar)
	\item If $D=3$, $\delta=3$: maps with two monochromatic cycles, and genus 1/2 (right of Fig.~\ref{fig:NLONNLOMLO})

	\end{itemize}}
\end{itemize}
The following orders can also be characterized using equation \eqref{eqref:PowerVacQuartLO}. Roughly, monochromatic non-twisted cycles contribute with $D-2$,  polychromatic cycles with $D$ and raising the genus of monochromatic submaps by 1/2 (resp. 1) contributes with 1 (resp. 2), but implies a certain number of monochromatic cycles. 
\end{coroll}

\begin{figure}[!h]
\centering
\includegraphics[scale=0.6]{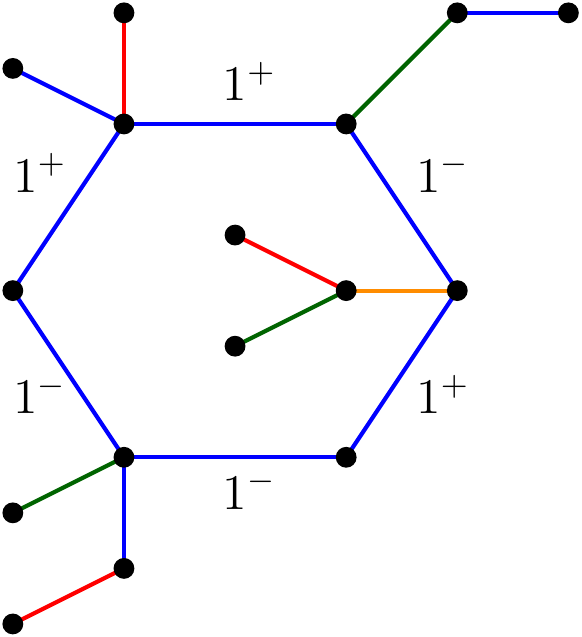}\hspace{3cm}
\includegraphics[scale=0.6]{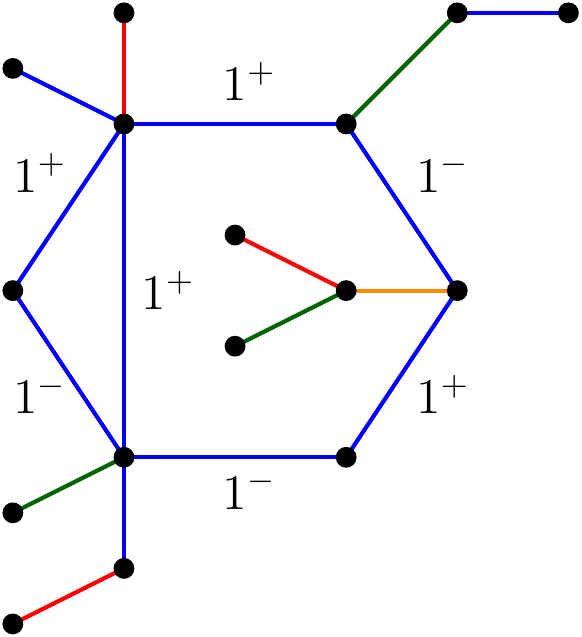}
\caption{A map with a twisted cycle and a map with two monochromatic cycles and genus 1/2.}
\label{fig:NLONNLOMLO}
\end{figure}

 For the first orders, the proof is the same as for Corollary~\ref{coroll:CorollMajTree}, with the difference that the genus can be a half-integer.

\

Although we do not prove it here, Thm.~\ref{thm:Represent}, which states that any non necessarily connected colored triangulation is the boundary of some graph in $\bG^q(\bB^4_m)$  extends here in the case of non-necessarily orientable triangulations.
A map with non-orientable boundary graph $\B$ is constructed from $\BOM$, but one has to be careful when defining the twist factors. 
Relation~\eqref{eqref:VacVersusExt}, Lemma~\ref{lemma:PhiLm} and Lemma~\ref{lemma:IneqRain} still hold, as well as the lower bounds (\ref{eqref:PowDomB}) and \eqref{eqref:BoundDegQuart} on the degree of maps with boundary. These bounds are expected to be saturated but it requires to exhibit a map with the right degree, as was done in the proof of Thm.~\ref{thm:Represent}, and we do not study further details here.

\section{Unicellular graphs and the colored SYK model}
\label{sec:UniGraphColSYK}

\subsection{Complex SYK: bipartite contributions}
\label{subsec:ComplexSYK}


\subsubsection{Partition function}

We recall from \eqref{eqref:ColSYKFreeEn2} and \eqref{eqref:ColSYKFreeEn3}, that the order $l$ free-energy of the colored SYK model is computed for coverings $\tilde \B^{\Om}$ which have degree $\delta_0(\tilde \B^{\Om})=l$, where $\tilde \B\in\tilde\bG_{q-1}$. Here, the pairing $\Om=\Om^{(0)}$ is that defined by the color-0 edges. In this subsection, we focus on bipartite coverings. The amplitude of a covering in the colored SYK model is proportional to
\be
\cA(\BCO)\sim N^{1-\delta_0(\BCO)},
\ee
where $\delta_0$ is a degree defined by the choice 
\be
\tilde a = (D-1)(\frac{V(\B)}2 - 1),
\ee
which does not define a bubble-dependent degree in the sense of Def.~\ref{def:BubDepDeg}. 
This choice corresponds to a scaling \eqref{eqref:SFromTildeA}
\be
s = 0, 
\ee
for every bubble, and to the coefficient 
\be
a = \frac{D(D-1)}4 - \frac{\delta(B)}{V(\B)}.
\ee
As we focus on the case where there is a single bubble, the corresponding degree is
\be
\label{eqref:Delta0Uni}
\delta_0(\BCO)= D +(D-1)( \frac{V(\B)} 2 -1) - \Phi_0(\BCO),
\ee
which, using Lemma~\ref{lemma:PhiLm}, leads to 
\be
\delta_0(\BCO)=L_m(\BOM).
\ee
In the case where $\BCO$ is bipartite, going from $\BOM$ to $\Ps(\B,\Om)$, we replace each monochromatic cycle with a star-map. \emph{This corresponds to applying the bijection $\Ps^{(0)}$ of Thm.~\ref{thm:BijOneColor} on $(\B, \Om^{(0)}(\BCO))$ (where again $\Om=\Om^{(0)}(\BCO)$ is the pairing induced by the color-0 edges)}. The number of polychromatic cycles of $\BOM$ is just the number of independent cycles of $\Ps(\B,\Om)$:
\be
L_m(\BOM)=L\bigl(\Ps(\B,\Om)\bigr).
\ee
The order of contribution of a bipartite unicellular graph $\B^\Om$ is therefore the circuit-rank of the stacked map $\Ps(\B,\Om)$. Identifying order $k$ vacuum contributions goes back to identifying maps of a given circuit-rank, as detailed in the subsection \emph{Identifying diagrams contributing to a given order}.

\

However, importantly, not all stacked maps in $\bS_{D-1}$ represent colored-graphs which are unicellular, meaning that the graph obtained when applying the inverse bijection $(\Ps^{(0)})^{-1}$ of \eqref{eqref:BijCol0} is connected when deleting all color-0 edges. Indeed, 
for any connected stacked map $\Ga\in\bS_{D-1}$, there exists a connected graph $\G\in\bG_D$ such that 
\be
\label{eqref:InvBijSYK}
\Ga = \Ps^{(0)}(\G)  = \Ps\bigl(\G^{\hat 0}, \Om^{(0)}(\G)\bigr)
\ee 
and $\Ga$ is the image of a unicellular graph iff $\G^{\hat 0}$, obtained from $\G$ by deleting all color-0 edges, is connected. Only in that case, we can interpret $\G$ as some covering $\G=\B^{\Om^{(0)}(\G)}$, or equivalently as some contribution to the complex colored SYK model. From \eqref{eqref:InvBijSYK}, $\Ga\in\bS_{D-1}$ corresponds to a unicellular $\G=(\Ps^{(0)})^{-1}(\Ga)$  iff it represents a connected $\G^{\hat 0}$ through $\Ps$, i.e. iff it satisfies the conditions of Prop.~\ref{prop:FaceExpl}.


In the unicellular case however, the exploration can also be simplified, as tree contributions  correspond to melonic contributions of the pre-image. In fact, we can state a stronger version of Proposition~\ref{prop:MelContriBij1}:

\begin{prop}[Melonic contributions in the unicellular case]
\label{prop:MelContriBij2}
If $\Ga=\Ps^{(0)}(\G)$ with $\G^{\hat 0}\in\bG_{D-1}$ connected ($D-1>2$), then tree-contributions to $\Ga$ precisely correspond to melonic contributions to $\G$.
\end{prop}
\prf From Prop.~\ref{prop:Melo2}, the color-0 edges all belong to completely separating cycles in the melonic subgraphs of $\G$. Because of the connectedness of $\G^{\hat 0}$, they therefore link the vertices of the canonical pairs (Def.~\ref{def:CanoPairing}). Therefore, when applying $\Ps^{(0)}$, each such subgraph corresponds to a tree-contributions. The converse was already known from Prop.~\ref{prop:MelContriBij1}. \qed

\subsubsection{ $2$-Point functions}

A covering $\B^\Om$ corresponds to a vacuum contribution of the colored SYK model. Contributions $\G_1$ to the 2-point function \eqref{eqref:ColSYK2PT} have two degree-$D$ vertices, which miss a color-$i$ edge. Adding a marked color-$i$ edge between the two vertices, we obtain a covering ${\B_1}^{\Om}$ of a bubble $\B_1$  with one marked color-$i$ edge. Applying the bijection of Thm.~\ref{thm:BijOneColor}, we obtain a stacked-map with one marked corner on a color-$i$ vertex (Fig.~\ref{fig:SYKA}). The star-map of color $i$ containing the root corner does not contribute to the 0-score of $\B_1^\Om$, and therefore the amplitude of a 2-point contribution in the colored SYK model is proportional to
\be
\cA(\G_1)\sim N^{1-\delta_0(\G_1)} = N^{-L_m(\B_{1;/\Om})}= N^{-L(\Ps(\B_1,\Om))}
\ee
This explains the factor $1/N$ in \eqref{eqref:2PtSYK1}. The order of contribution of a bipartite unicellular graph $\G_1$ with one missing color-$i$ edge to the corresponding 2-point function \eqref{eqref:ColSYK2PT} is therefore the circuit-rank of the stacked map $\Ps(\B_1,\Om)$, where the covering $\B_1^\Om$ is obtained from $\G_1$ by adding a marked color-$i$ edge between the two degree-$D$ vertices. 
Identifying order $k$  contributions  to the 2-point function goes back to identifying maps of a given circuit-rank with one marked color-$i$ corner, as detailed in the subsection \emph{Identifying diagrams contributing to a given order}.

Again, one has to worry about the connectedness of $\B$, which is done throughout a face-exploration in $\Ps(\B_1,\Om)$. In that case however, the exploration does not take into account the face which encounters the marked corner. 


\begin{figure}[!h]
\centering
\raisebox{0ex} {\includegraphics[scale=0.8]{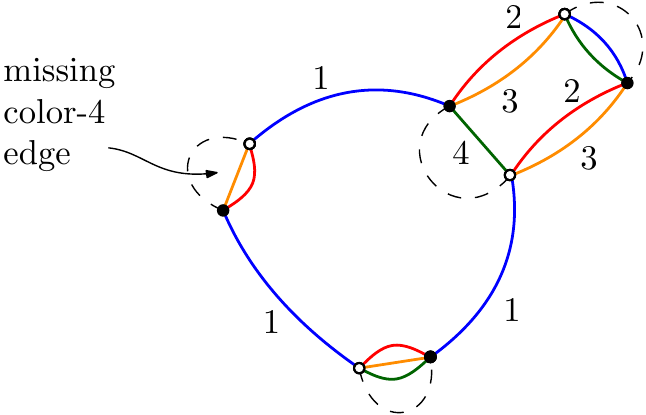}}
\quad\raisebox{+7.5ex}{$\leftrightarrow$}\hspace{1cm}
\includegraphics[scale=0.7]{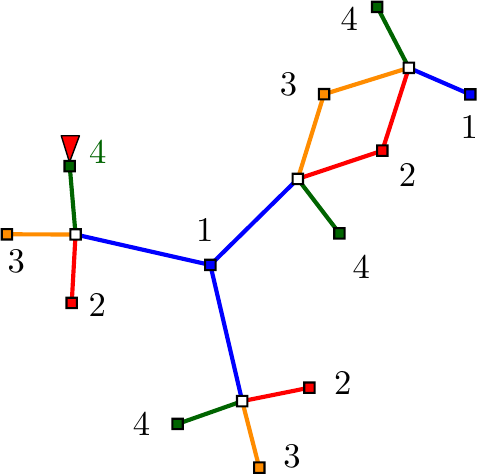}
\caption{Applying the bijection $\Ps$ to a 2-point contribution.}
\label{fig:SYKA}
\end{figure}

\subsubsection{$2n$-Point functions}

We wish to use the same characterization in the case of the $2n$-point functions \eqref{eqref:ColSYK2nPT}, that is, add the missing colored edges in the contributing graphs $\G_n$ to go back to the vacuum case, and use the bijection $\Ps$ on the pairing induced by the color-0 edges. However, in the case where the missing edges do not all have different colors, there are different ways of adding them back to recover a vacuum graph. To avoid the ambiguity, we add the colored edges canonically (Fig.~\ref{fig:CanAdd}) on the pairing $\Om$ induced by the color-0 edges. We recall this procedure here. In $\G_n$,  starting from a vertex which does not have a color-$i$ edge incident to it, we follow the path alternating color-$i$ edges and color-0 edges. It necessarily ends on another vertex which does not have a color-$i$ edge incident to it (as there are no missing edges of color 0). We then add a marked color-$i$ edge between the two endpoint of the path. This leads to a covering $\B_n^\Om$. 
The bicolored cycles of $\B_n^\Om$ containing a marked vertex do not contribute to the 0-score of $\G_n$, and because marked edges have been added canonically, there is at most one marked edge per bicolored cycle, so that
\be
\Phi_0(\G_n)=\Phi_0(\B_n^\Om) - n.
\ee
Applying the bijection of Thm.~\ref{thm:BijOneColor}, a graph $\G_n$ contributing to $<\psi_{i_1}\psi_{i_1}\cdots\psi_{i_n}\psi_{i_n}>$ (where the $i_j$ are not necessarily different) is mapped to a stacked map with $n$ corners marked on $n$ different colored vertices (one on a color $i_1$ vertex, one on a color $i_2$ vertex, and so on). The amplitude of such a graph is proportional to 
\be
\cA(\G_n)\sim N^{1-\delta_0(\G_n)} = N^{1-n-L_m(\B_{n;/\Om})}= N^{1-n-L(\Ps(\B_n,\Om))}.
\ee
Identifying order $k$  contributions  to the $2n$-point function $<\psi_{i_1}\psi_{i_1}\cdots\psi_{i_n}\psi_{i_n}>$ goes back to identifying maps of a given circuit-rank with $n$ corners marked on $n$ different colored vertices (one on a color $i_1$ vertex, one on a color $i_2$ vertex, and so on), as detailed in the following subsection. Again, one has to worry about the connectedness of $\B$, which is done throughout a face-exploration in $\Ps(\B_n,\Om)$. In that case however, the exploration does not take into account the faces which encounter a marked corner. 


\subsubsection{Identifying diagrams contributing to a given order}

\textbf{Pruning.} Considering $\Ps(\B,\Om)$, we recursively contract all the edges incident to unmarked leaves (delete the edge and the isolated vertex), until there are no more unmarked leaves, as shown in Figure~\ref{fig:SYKB}. We call this operation pruning. From Prop.~\ref{prop:MelContriBij2}, it goes back to contracting the melonic contributions in $\B^\Om$, and therefore it does not change the connectedness of $\B$ (if $\B$ was not connected, it still is not, and vice versa). Examples of colored graphs corresponding to pruned stacked maps are shown in Figure~\ref{fig:NLOCycle}. In the pruned stacked map, vertices are either colored vertices incident to edges which all have the same color, or white squares incident to edges which all have different colors (non necessarily $D$ of them). Infinitely many contributions will give the same pruned map. To generate all the stacked maps leading to the same pruned map, firstly, colored leaves have to be inserted incident to white squares, so that the latter have one incident colored edge for each color in $\lDr$. Then, possibly trivial tree contributions have to be added on every colored corner. Considering the marked corner as two corners divided by a cilium (the mark), possibly trivial tree contributions should also be added on each side of the cilium. 

\begin{figure}[!h]
\centering
\includegraphics[scale=0.7]{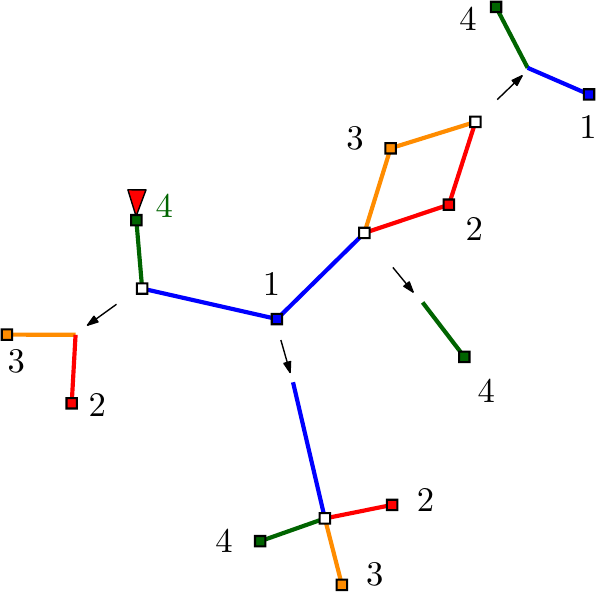}
\qquad\raisebox{+9.5ex}{$\rightarrow$}\hspace{1.5cm}
\raisebox{+5ex}{\includegraphics[scale=1]{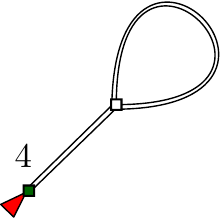}}
\caption{Pruning a map and the corresponding scheme.}
\label{fig:SYKB}
\end{figure}

\

\noindent \textbf{Chain edges.} We now consider maximal sequences of degree-two vertices, which start and end on edges which have one endpoint of higher valency or marked, and alternate colored vertices and white squares.  We replace them with a \emph{chain-edge} between the two vertices of higher valency or marked (which are not necessarily of the same kind, colored or white). Chain edges are represented as double lines.  We are left with a graph which has no valency-two unmarked vertices, and which we call a \textbf{scheme} (see the decompositions in \cite{GurauSchaeffer, 
Fusy} in the colored graph picture). For vacuum graphs, $n=0$, there can be ambiguities, for instance when the sequence of degree-two vertices is a tree, or a cycle. In these two cases we get $D+1$ possible schemes: an isolated vertex or a cycles with one vertex, which is either colored or white. In the vacuum case, different schemes therefore lead to the same stacked maps. This is not a problem in this section as we are only interested in identifying these stacked maps. In Subsection~\ref{subsec:CountUni}, in which we focus on counting stacked maps, we therefore only study rooted cases. 
To go back from schemes to pruned maps, each chain-edge is to be replaced with a sequence of arbitrary length alternating edges and degree two vertices, so that
every edge links a colored vertex to a white vertex.

However, it is important to make sure that the connectedness of $\B$ stays unchanged when replacing the chain-edge with some realization. This is not a problem \emph{as long as there is at least one white vertex in the realization}. The only cases which might cause problems are the minimal realizations in the case of a chain edge between a color-$i$ vertex and a white vertex, and in the case of a chain-edge between two white vertices. For these cases, one has to apply the face-exploration of Prop.~\ref{prop:FaceExpl}.

We stress that the polychromatic nature of all the independent cycles is preserved, even when replacing chain-edges with edges of a single color. If there are two or more vertices of different colors in the cycle it is obvious, if there is one white square or more on the cycle, then necessarily there are two different colors incident to it, and if there are only color-$i$ vertices, then there are necessarily white square between them in every realization.


\

To identify the contributions to the $k^\text{th}$ order of the $2n$-point function $<\psi_{i_1}\psi_{i_1}\cdots\psi_{i_n}\psi_{i_n}>$ of the colored complex SYK model for $n\ge0$, we identify all the schemes $M$ which edges are seen as chain-edges, and which have
\begin{itemize}[label=$-$]
\item $k$ independent cycles; 
\item $n$ corners marked on $n$ different colored vertices: one on a color $i_1$ vertex, one on a color $i_2$ vertex, and so on;
\item colored embedded vertices which have a marked corner or are of valency 3 or more;
\item white non-embedded vertices of valency 3 or more. 
\end{itemize}
To recover all the colored graphs, one has to replace the chain-edges with all possible realizations, i.e. sequences of degree-two vertices alternatively colored and white, such that color-$i$ edges always link a color-$i$ vertex and a white square (the colored vertices of the scheme impose boundary conditions on the colors in the sequence).  The particular cases of minimal realizations of chain edges between a color-$i$ vertex and a white vertex or between two white vertices have to be treated using Prop.~\ref{prop:FaceExpl}.
This procedure is shown in Fig.~\ref{fig:SchemeToSM} for a leading order contributions to the 4-point function. 

\begin{figure}[!h]
\centering
\includegraphics[scale=0.7]{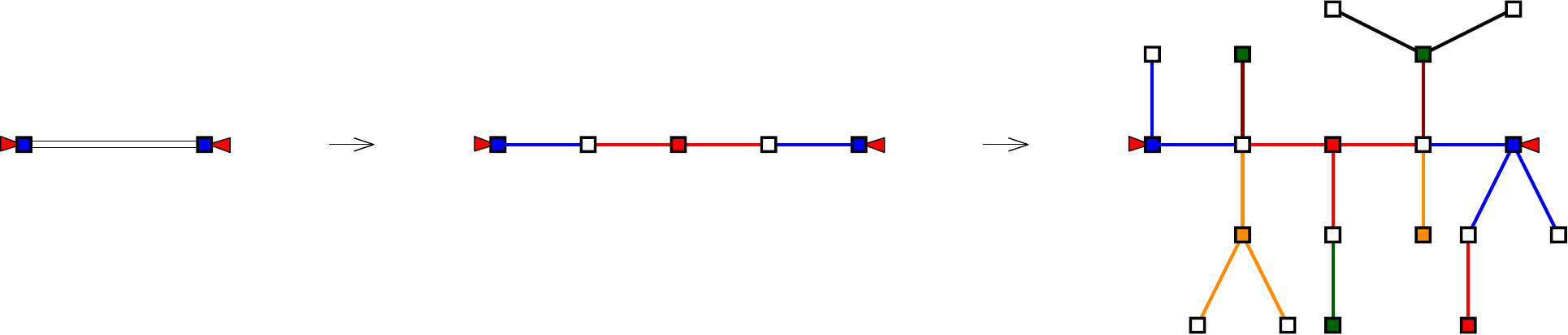}
\caption{From a scheme to a stacked map contributing to the 4-point function ($D=3$).}
\label{fig:SchemeToSM}
\end{figure}

The schemes contributing to the first orders of the partition function, $2$-point function and 4-point function of the complex colored SYK model for $n\le0$ are shown in Figures~\ref{fig:SchemesLO} and~\ref{fig:SchemesNLO}, as well as the leading order of the 6-point function Fig.~\ref{fig:SchemesLO}. Again, one should be careful to the particular cases when replacing chain-edges by their mininmal realizations. It is in principle not complicated to find all the contributions to higher orders or/and to $2n$-point functions with $n>3$, however the number of schemes grows 
extremely fast, and it is in practice very tedious. In these figures, the labels $i,j,k,l$ represent colors in $\lDr$, which are not necessarily distinct. 
\begin{figure}[!h]
\centering
\includegraphics[scale=0.7]{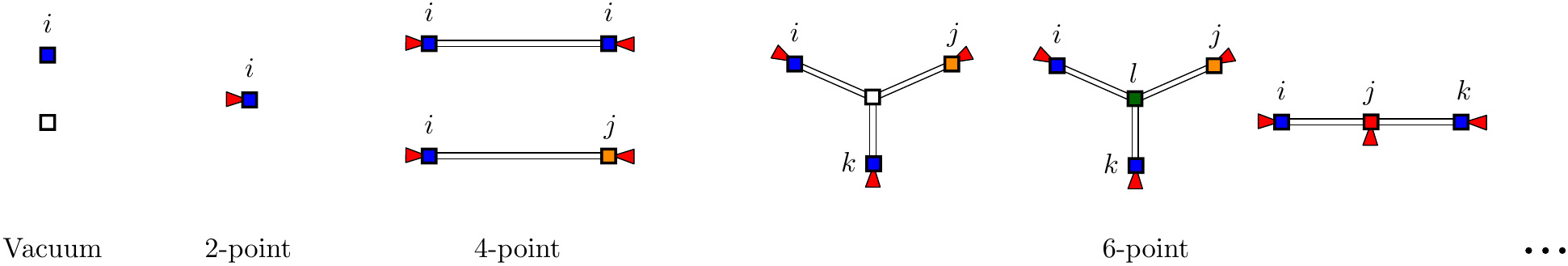}
\caption{Vacuum, 2-point, 4-point, and 6-point leading order schemes.}
\label{fig:SchemesLO}
\end{figure}
\begin{figure}[!h]
\centering
\includegraphics[scale=0.7]{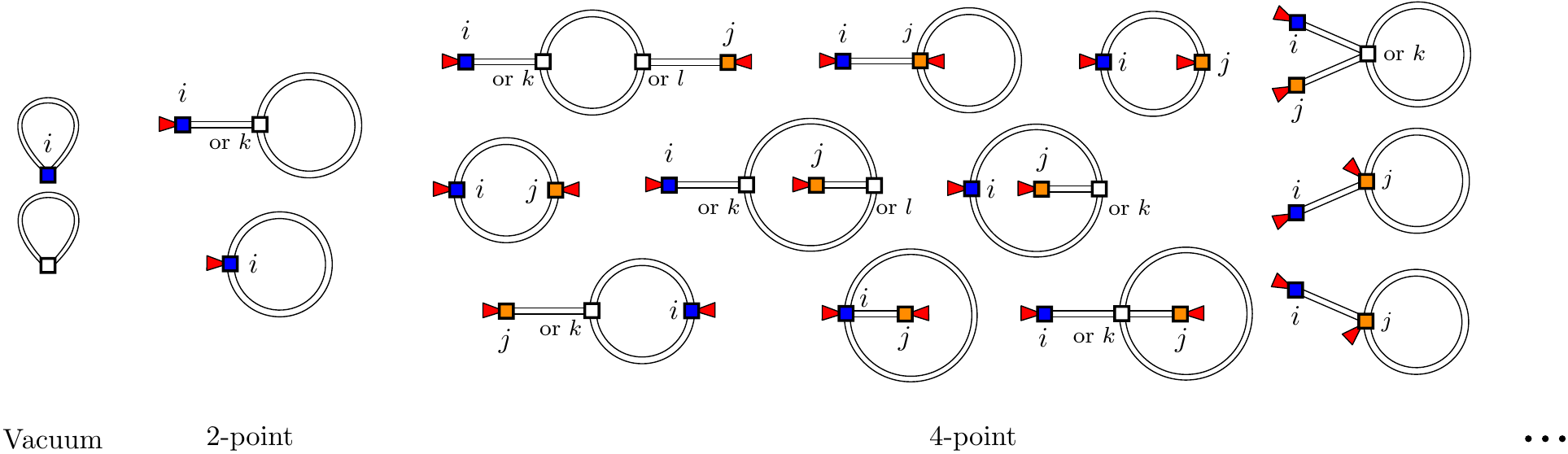}
\caption{Vacuum, 2-point and 4-point next-to-leading order schemes.}
\label{fig:SchemesNLO}
\end{figure}
Examples of coverings contributing to the next-to-leading order and next-to-next-to-leading order of the partition function are shown in Figure~\ref{fig:NLOCycle} in the colored graph picture. These examples have no melonic contributions and therefore correspond to pruned stacked maps. We do not provide the equivalents of all the schemes in the colored-graph pictures, and report the reader to \cite{BLT}, in which  they are listed.

\begin{figure}[!h]
\centering
\raisebox{0.2ex}{\includegraphics[scale=0.5]{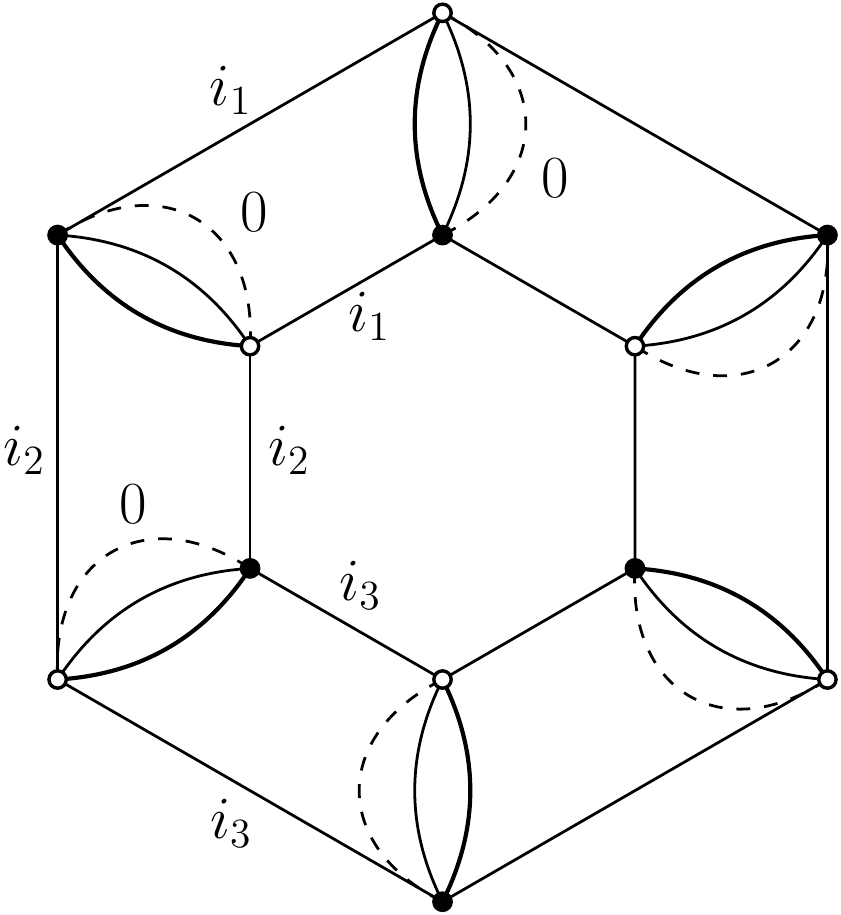}}\hspace{2cm}\includegraphics[scale=0.43]{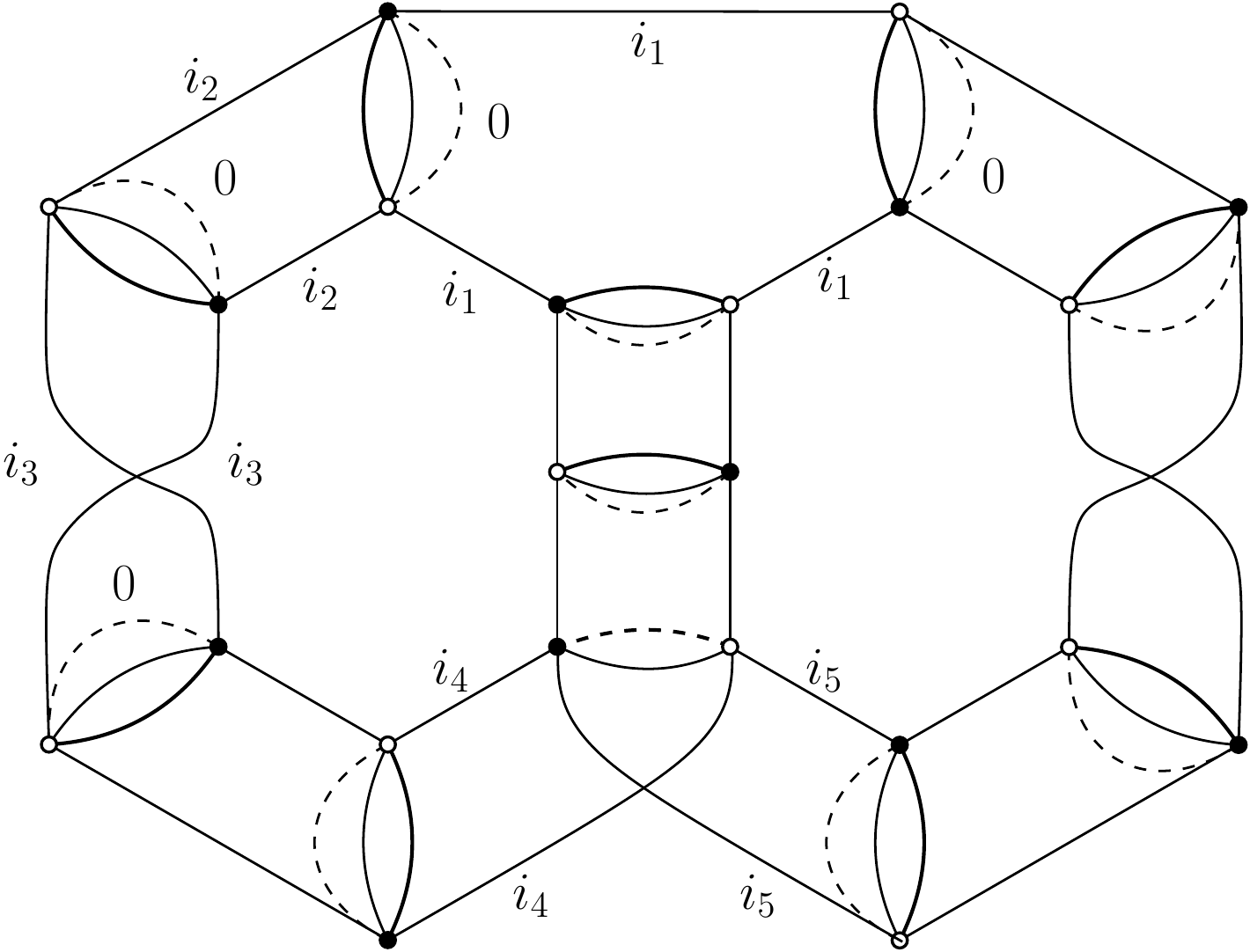}
\caption{Vacuum NLO and NNLO coverings.}
\label{fig:NLOCycle}
\end{figure}

\subsection{Real SYK: non-bipartite contributions}

We do not adapt the bijection $\Ps$ to the non-orientable case, however we explain here how to generalize the previous procedure to non-orientable unicellular coverings, for which we do not need to count bicolored cycles, since here, the degree depends only on the number of polychromatic cycles of $\BOM$. For a non-bipartite bubble $\tilde\B\in\tilde\bG_{D-1}$ and a pairing $\Om$ of its vertices, we contract the pairs and are left with an Eulerian graph for which monochromatic submaps are collections of cycles. 
Importantly, we have lost the information on how which edge was attached to which vertex of the pair, as it was carried by the orientation in the bipartite case. Several $(\tilde\B,\Om)$ give the same $\tilde\B_{/\Om}$. In order to obtain  all the $(\tilde\B,\Om)$ leading to the same $\tilde\B_{/\Om}$, we need to expand each vertex of $\tilde\B_{/\Om}$ into a pair of vertices and \emph{try all possible ways of attaching the two incident color-$i$ edges to one or the other vertex}, for each color $i$. Some of the graphs obtained will be bipartite, some others will be non-bipartite.   

We then replace each oriented cycle with a star-graph, i.e. add a new color-$i$ vertex and color-$i$ edges from that vertex to every vertex of the corresponding cycle. If one of the edges is marked, we distinguish the color-$i$ vertex. The procedure is a \emph{non-embedded} version  of  Fig.~\ref{fig:CycStar} and Fig.~\ref{fig:CycStar0}, leading to a graph $\tilde\Phi({\tilde \B, \Om})$. In particular, the information on which edge of the cycle was marked is lost, and to obtain all possible $\tilde\B_{/\Om}$ leading to the same,  $\tilde\Phi({\tilde \B, \Om})$ we need to consider all possibilities of marking one edge on a cycle corresponding to a distinguished color-$i$ vertex. 
As before in the bipartite case, the amplitude of a $2n$-point covering is
\be
\cA(\tilde \G_n)\sim N^{1-\delta_0(\tilde \G_n)} = N^{1-n-L_m(\tilde \B_{n;/\Om})}= N^{1-n-L(\tilde\Phi(\tilde \B_n,\Om))},
\ee
where $\tilde\B_n^{\Om}$ is obtained from $\tilde G_n$ by adding the missing edges canonically with respect to the pairing induced by color 0 (this only depends on the pairing and can also be done for non-bipartite bubbles).
We stress again that the graph we obtain only keeps the information on $\Phi_0({\tilde \B_n}^{\Om})$ and $L_m(\tilde \B_{n;/\Om})$, which is the information needed in the present case. \emph{The graphs $\tilde\Phi({\tilde \B, \Om})$ contributing to a given order are the underlying graphs of the stacked maps contributing to the same order in the complex (bipartite) case.} The pruning procedure and replacements of chains by chain-edges is done as in the bipartite case. The scheme-graphs contributing to the first orders of the 2, 4, 6-point functions are the underlying graphs of the schemes in 
Figures \ref{fig:SchemesLO} and \ref{fig:SchemesNLO}. For instance, a non-bipartite colored graph leading to the top-left 4-point scheme of \ref{fig:SchemesNLO} is shown on the left of Fig.~\ref{fig:NLOCycleTwist}. An example of next-to-next-to-leading order vacuum contribution is shown on the right of Fig.~\ref{fig:NLOCycleTwist}. All the leading order and next-to-leading order colored graph contributing to the partition function, 2-point and 4-point functions of the real SYK model are listed in \cite{BLT}.

\begin{figure}[!h]
\centering
\raisebox{1ex}{\includegraphics[scale=0.4]{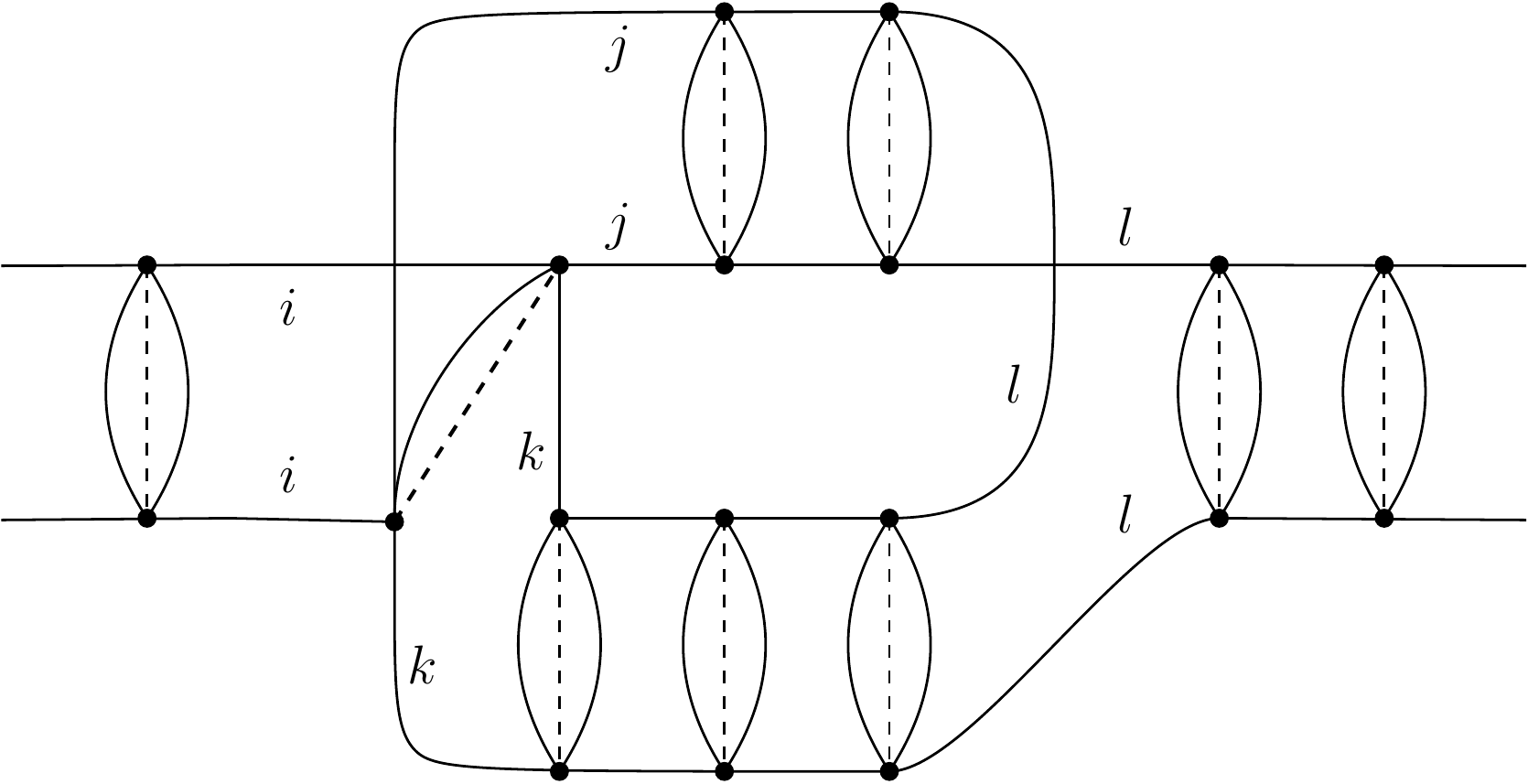}}\hspace{0.8cm}\includegraphics[scale=0.35]{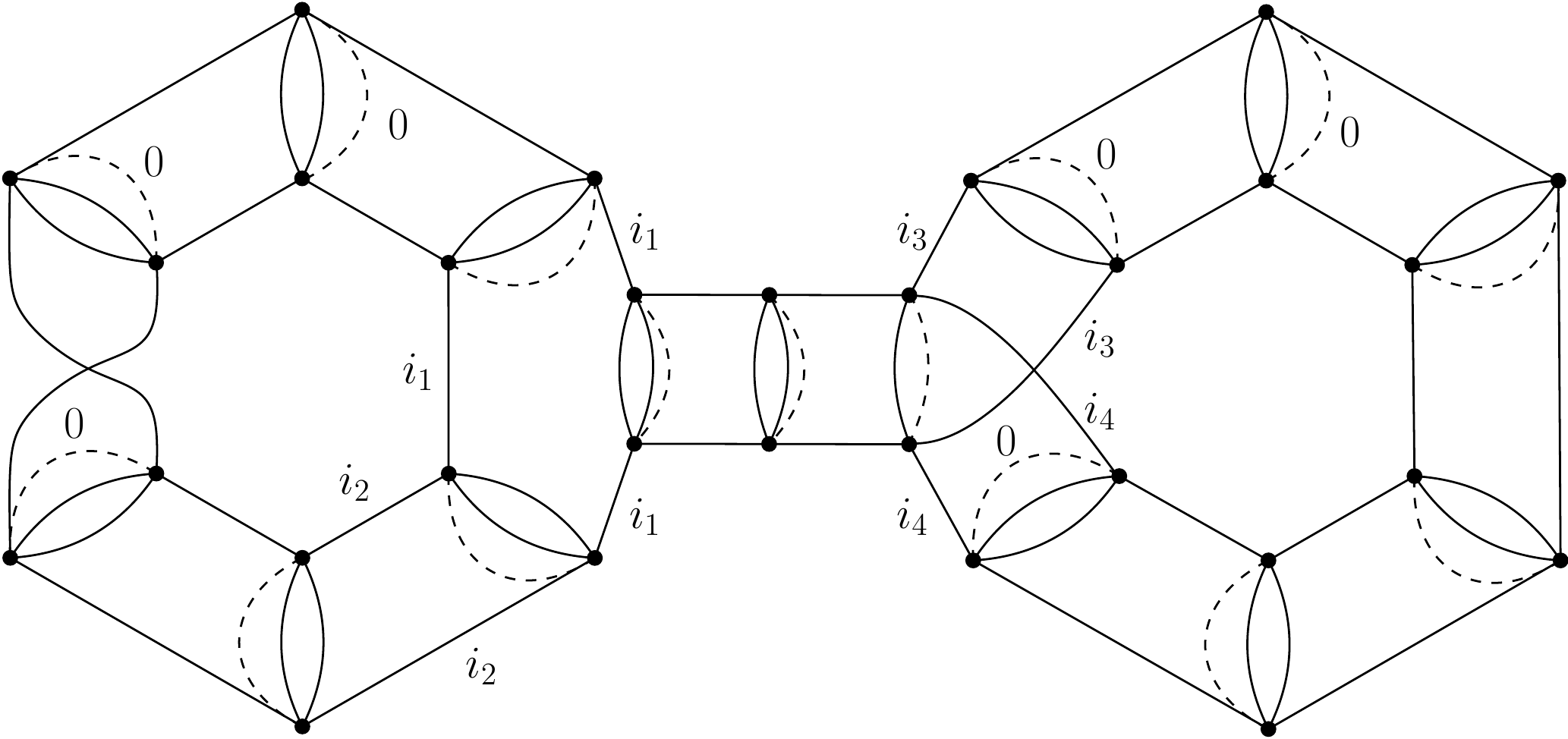}
\caption{4-Point NLO and vacuum NNLO contributions to the real colored SYK model.}
\label{fig:NLOCycleTwist}
\end{figure}

\subsection{Counting unicellular graphs of a given order}
\label{subsec:CountUni}
In dimension two, unicellular maps are maps with a single face. They are obtained by taking a single polygon, and gluing the edges of its boundary in order to obtain a closed surface. Their enumeration and the link with known counting formulas such as the Harer-Zagier formula is the subject of many papers, among which \cite{ChapUnicell, BernardiHZ, ChapFusUnicell}. 
In higher dimension, we define unicellular spaces as self-gluings of a single bubble. In the dual  picture, they correspond to graphs in $\bG_D$ such that, when deleting all the color-0 edges, we are left with a single connected component. \emph{Unicellular graphs are therefore what we have called coverings} (Def.~\ref{def:Pairing}). They are also the objects we have studied in the context of the SYK model, in the previous subsections. In Subsection~\ref{subsec:Unicell}, and more generally in the following chapters, we will be interested in certain properties of unicellular graphs $\B^\Om$ for $\B\in\bG_{D-1}$ a given chosen bubble. Determining precisely the 0-score of all the coverings of a bubble without trying them all (i.e. by only knowing its Gurau degree and its number of facets, for instance) is a difficult question which remains unsolved. In this subsection, we tackle another simpler problem, which is a first step in solving this question: we adapt the decomposition of the previous section to the characterization of unicellular graphs which have a given 0-score, without restricting the possible bubbles. 
We stress that the choice 
\be
\tilde a = (D-1)(\frac{V(\B)}2 - 1),
\ee
is an obvious choice to classify and compare coverings of bubbles (unicellular graphs) with respect to the number on $(D-2)$-cells lying on the boundary of the bubbles, not worrying about the number of internal $(D-2)$-simplices, or equivalently about the Gurau-degrees $\deltaG$ of the bubbles. Indeed, the degree $\delta_0$ \eqref{eqref:Delta0Uni} thus defined only depends on the number of facets and  $(D-2)$-cells which do not lie in the interior of the bubbles, i.e. on the number of vertices of bubble-graphs and on the 0-score of the coverings.  The degree $\delta_0$ defines orders (Def.~\ref{def:Order}) restricted to unicellular graphs. Among bubble-graphs with the same number of vertices $V(\B)$,
coverings contributing to the same order share the same number of boundary $(D-2)$-cells. Among coverings with the same degree $\delta_0$, the number of boundary $(D-2)$-cells is a linear function of the number of vertices of the bubble-graphs, with slope $(D-1)/2$.
In the orientable case, the coverings are bipartite and we can use the bijection~$\Ps$ and the decomposition in pruned maps and schemes as described in Subsection~\ref{subsec:ComplexSYK}.
Once the schemes contributing to a given order have been characterized, we can try to compute the range of possible Gurau degrees of the bubbles inside the coverings, and see if for a given degree we can determine the set  of allowed 0-scores, depending on the number of vertices. We leave this for future work, and focus on counting the maps contributing to a given order of $\delta_0$.
 The generating function of $D$-colored stacked trees rooted on a color-$i$ corner and counted according to their number of white squares is 
 \be
 \GF_T(z)= 1+\sum_{k\ge 1} z^k  \GF_T(z)^{k(D-1)},
\ee
which is the generating function of melonic graphs in $\bG_{D-1}$ \eqref{eqref:MeloGenFunct}
\be
 \GF_T(z) = \sum_{k\ge0} C_k^{D} z^k, \quad\text{where}\quad C_k^{D}=\frac 1{Dk+1}\binom{Dk+1}{k},
\ee
the  $C_k^D$ being Fuss-Catalan numbers. To recover all the maps leading to a given pruned map, colored leaves have to be added in that pruned map so that white squares are incident to one colored vertex for each color in $\lDr$, and then the generating function $\GF_T(z)$ has to be added in every colored corner. Again, planting or pruning tree contribution does not change the connectedness of 
\be
\label{eqref:Bubble0Unicell}
\G^{\hat 0},\text{ where }\G=(\Ps^{(0)})^{-1}(\Ga\in\bG_{D-1}).
\ee
We compute the generating function of chains, counted according to their number of white squares ($z_\circ$) and colored squares ($z_\bullet$). There are six types of chain-edges, each replaced with a different generating function in the schemes, corresponding to chains between:
\begin{itemize}[label=-]
\item two color-$i$ vertices,  $\GF_{\bullet\bullet}^{ii}$;
\item a vertex of color $i$ and a vertex of color $j\neq i$,  $\GF_{\bullet\bullet}^{ij}$;
\item two white vertices with incident edges of color $i$, $\GF_{\circ\circ}^{ii}$;
\item  two white vertices with incident edges of color $i$ and $j\neq i$, $\GF_{\circ\circ}^{ij}$;
\item  a white vertex with incident edge of color $i$ and a color-$i$ vertex, $\GF_{\circ\bullet}^{ii}$;
\item a white vertex with incident edge of color $i$ and a color-$j\neq i$ vertex, $\GF_{\circ\bullet}^{ij}$.
\end{itemize}

We may also need the generating functions of chains without their minimal realization, in the particular cases where replacing  chain edges between a color-$i$ vertex and a white vertex or between two white vertices  by their minimal realizations changes the connectedness of $\G^{\hat 0}$ \eqref{eqref:Bubble0Unicell}.
\begin{itemize}[label=-]
\item two white vertices with incident edges of color $i$, without minimal realization, $\GF_{\circ\circ}^{ii\star}$;
\item  a white vertex with incident edge of color $i$ and a color-$i$ vertex, without minimal realization, $\GF_{\circ\bullet}^{ii\star}$.
\end{itemize}

The generating function of chains between a colored vertex and a white square, starting with color $i$ and never encountering color $i$ again is 
\be
\GF_{\bullet\circ\setminus i}^{ij}(z_\circ ,z_\bullet) = \sum_{k\ge 1} (D-1)(D-2)^{k-1} y^k = \frac {(D-1)y}{1-(D-2)y},
\ee
where we denoted 
\be
y=z_\circ z_\bullet.
\ee

A chain between two color-$i$ vertices starts with an half-edge of color $i$, then as an arbitrarily long chain without any color-$i$ edge $(.)$, then has a white square, which either is the last white square of the chain, either is incident to a color-$i$ vertex followed by a smaller chain between two color-$i$ vertices $[.]$.
\be
\includegraphics[scale=0.7]{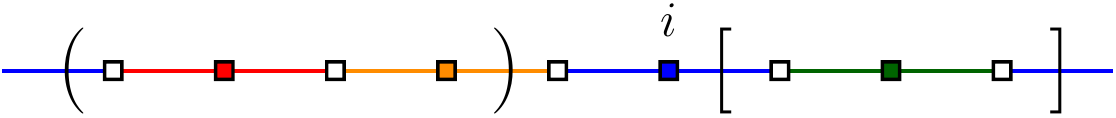}
\ee
Therefore, 
\be
\GF_{\bullet\bullet}^{ii} (z_\circ ,z_\bullet) = \GF_{\bullet\circ\setminus i}^{ij}(z_\circ ,z_\bullet) \times z_\circ \times \bigl(1+z_\bullet\GF_{\bullet\bullet}^{ii} (z_\circ ,z_\bullet) \bigr),
\ee
which leads to 
\be
\GF_{\bullet\bullet}^{ii} (z_\circ, z_\bullet ) =\frac 1 {z_\bullet} \frac{(D-1)y^2}{\bigl(1+y \bigr)\bigl(1-(D-1)y\bigr)}.
\ee

A chain between a vertex of color $i$ and a vertex of color $j\neq i$ is either a single white square between a color $i$ and a color $j$ half-edges, either is that same white square followed by a color-$j$ vertex and a chain between two color-$j$ vertices, either starts with a white square  followed by a color-$k$ vertex, followed by a chain between a color-$k$ vertex and a color-$j$ vertex:
\be
\GF_{\bullet\bullet}^{ij} (z_\circ, z_\bullet ) = z_\circ\bigl(1+z_\bullet\GF_{\bullet\bullet}^{jj} (z_\circ, z_\bullet)  \bigr)+ (D-2)y \GF_{\bullet\bullet}^{kj} (z_\circ, z_\bullet),
\ee
which leads to 
\be
\GF_{\bullet\bullet}^{ij} (z_\circ, z_\bullet ) =\frac{z_\circ}{\bigl(1+y \bigr)\bigl(1-(D-1)y\bigr)}.
\ee
Linear combinations of these two generating functions are often needed in the following: 
\be
\label{eqref:LinCombiGenF}
a\GF_{\bullet\bullet}^{ii} (z_\circ ,z_\bullet) + b\GF_{\bullet\bullet}^{ij} (z_\circ, z_\bullet )  = \frac1{z_\bullet}\frac{a(D-1)y^2 + by}{\bigl(1+y \bigr)\bigl(1-(D-1)y\bigr)}.
\ee
We obtain the other generating functions from these ones:
\bea
\GF_{\circ\circ}^{ij}(z_\circ, z_\bullet)&=&z_\bullet^2\GF_{\bullet\bullet}^{ij} (z_\circ, z_\bullet),\\
\GF_{\circ\circ}^{ii} (z_\circ, z_\bullet )  &=& z_\bullet (1+ z_\bullet\GF_{\bullet\bullet}^{ii} (z_\circ, z_\bullet))\\
\GF_{\circ\bullet}^{ij} (z_\circ, z_\bullet )  &=& z_\bullet \GF_{\bullet\bullet}^{ij} (z_\circ, z_\bullet),\\
\GF_{\circ\bullet}^{ii} (z_\circ, z_\bullet )  &=& 1 + z_\bullet \GF_{\bullet\bullet}^{ii} (z_\circ, z_\bullet),\\
\GF_{\circ\circ}^{ii\star}(z_\circ, z_\bullet)&=& z_\bullet^2  \GF_{\bullet\bullet}^{ii} (z_\circ, z_\bullet),\\
\GF_{\circ\bullet}^{ii\star} (z_\circ, z_\bullet )  &=&  z_\bullet \GF_{\bullet\bullet}^{ii} (z_\circ, z_\bullet).
\eea



\

We apply this to the enumeration of rooted coverings (at least one marked corner) which  have the same degree
\be
\delta_0(\B^\Om)=L_m(\BOM)=L\bigl(\Ps(\B,\Om)\bigr).
\ee
Let us first consider examples. The 2-point scheme in Figure~\ref{fig:SchemesLO} has one colored marked vertex. Following the procedure, this is just the generating function of rooted trees $\GF_T$.
The two 4-point schemes in Fig.~\ref{fig:SchemesLO} give back $\GF_{\bullet\bullet}^{ii}$ and $\GF_{\bullet\bullet}^{ij}$. Inside each realization of the chain, we need to add $D-2$ generating functions of trees $\GF_T$ per white vertex, and  two per colored vertex. Moreover, the colored marked vertices are not included in the realization of the chain-edge, which gives four more factors  $\GF_T$. If we suppose that we can distinguish the two roots, we get
\be
\GF_\text{4}^{(0)} (z)= \GF_T(z)^4 \biggl[D \GF_{\bullet\bullet}^{ii}\bigl(z \GF_T(z)^{D-2},   \GF_T(z)^{2}  \bigr) + D(D-1)  \GF_{\bullet\bullet}^{ij}\bigl( z \GF_T(z)^{D-2},   \GF_T(z)^{2} \bigr)\biggr].
\ee
From \eqref{eqref:LinCombiGenF},
\be
D \GF_{\bullet\bullet}^{ii}(z_\circ, z_\bullet) + D(D-1) \GF_{\bullet\bullet}^{ij}(z_\circ, z_\bullet ) 
=
\frac{1}{ z_\bullet} \frac{D(D-1)y}{1-(D-1)y},
\ee
and $\GF_\text{4}^{(0)}$ simplifies to 
\be
\GF_\text{4}^{(0)} (z)= \frac {D(D-1) z\GF_T(z)^{D+2}}{1-(D-1)z\GF_T(z)^D},
\ee
which is the generating function of two-rooted leading order contributions, i.e. the leading order 4-point function, for the degree $\delta_0$.
The second and the third 6-point contributions are computed similarly as products of two or three  $\GF_{\bullet\bullet}^{ii} $ and $\GF_{\bullet\bullet}^{ij}$, where all non-equivalent choices of $i, j, k$ and $l $. The first 6-point schemes gives products of $\GF_{\circ\bullet}^{ij}$ and $\GF_{\circ\bullet}^{ii}$. 
%

\

The generating functions of the next-to-leading order contributions to the 2-point function (one-root schemes) are similar to those computed  in \cite{GurauSchaeffer}. The difference relies on the fact that here, we do not need to worry about whether two colors pass along the cycle or more. The second scheme in Fig.~\ref{fig:SchemesNLO} is simply
\be
\label{eqref:NLO2PT1}
 D\GF_T(z)^{3}  \GF_{\bullet\bullet}^{ii}\bigl(z \GF_T(z)^{D-2},   \GF_T(z)^{2}\bigr) =   D\GF_T(z)\biggl[ \frac{(D-1)y^2}{\bigl(1+y \bigr)\bigl(1-(D-1)y\bigr)}\biggr]_{y=z\GF_T(z)^{D}}.
\ee
The first drawing refers to two schemes, depending on whether the valency 3 vertex is white or of some color-$i$. If it is white, either the bridge chain-edge is replaced with  $\GF_{\bullet\circ}^{ii}$, in which case there are $\binom{D-1}2$ ways of choosing the two other colors $j,k$ incident to the white vertex. The other chain-edge is then replaced with $\GF_{\circ\circ}^{jk}$. We carefully verify that the minimal realization of $\GF_{\bullet\circ}^{ii}$ still corresponds to a connected bubble, which is the case here. If the bridge chain-edge is replaced with  $\GF_{\bullet\circ}^{ij}$, there are $D-1$ choices for $j$, and then again $\binom{D-1}2$ ways of choosing the two other colors incident to the white vertex. Up to some factors, we find a linear combination 
\bea
&&\GF_{\circ\circ}^{jk}(z_\circ, z_\bullet)\bigl[ \GF_{\circ\bullet}^{ii}(z_\circ, z_\bullet)  + (D-1)  \GF_{\circ\bullet}^{il}(z_\circ, z_\bullet) \bigl] \hspace{7.5cm}\\
\nonumber&&\hspace{4cm}= z_\bullet^2\GF_{\bullet\bullet}^{jk}(z_\circ, z_\bullet)\bigl[1+ z_\bullet \GF_{\bullet\bullet}^{ii}(z_\circ, z_\bullet)  + (D-1) z_\bullet \GF_{\bullet\bullet}^{il}(z_\circ, z_\bullet) \bigl] 
\eea
Using \eqref{eqref:LinCombiGenF}, we rewrite
\be
1+ z_\bullet \GF_{\bullet\bullet}^{ii}(z_\circ, z_\bullet)  + (D-1) z_\bullet \GF_{\bullet\bullet}^{il}(z_\circ, z_\bullet)
= \frac{1}{1-(D-1)y}
 \ee
One has to add an extra $\GF_T(z)^{2}$ for the marked vertex, and an extra $z\GF_T(z)^{D-3}$ for the white square.  We obtain the contribution for this scheme
\be
\label{eqref:NLO2PT2}
\frac{D(D-1)(D-2)}2 \biggl[ \frac{y^2}{\bigl(1+y \bigr)\bigl(1-(D-1)y\bigr)^2}
\biggr]_{y=z\GF_T(z)^{D}}.
\ee
 If the valency-3 vertex is colored, then either it is of color $i$, in which case the contribution is $(\GF_{\bullet\bullet}^{ii})^2$, either it is of some other color $k$, in which case the contribution is $(D-1)\GF_{\bullet\bullet}^{ik}\GF_{\bullet\bullet}^{kk}$. Using \eqref{eqref:LinCombiGenF}, we obtain the contribution 
 \be
 D\frac 1 {z_\bullet} \frac{(D-1)y^2}{\bigl(1+y \bigr)\bigl(1-(D-1)y\bigr)}
\frac{(D-1)y}{1-(D-1)y}.
 \ee
 We replace $z_\bullet$ per $\GF_T(z)^2$, $y$ per $z\GF_T(z)^D$, and add 2 factors $\GF_T(z)$ for the marked vertex and three for the valency-3 vertex, obtaining the contribution
  \be
  \label{eqref:NLO2PT3}
 D(D-1)^2 \GF_T(z)^3  \biggl[ \frac{y^3}{\bigl(1+y \bigr)\bigl(1-(D-1)y\bigr)^2}\biggr]_{y=z\GF_T(z)^{D}}.
 \ee
 Summing the three contributions \eqref{eqref:NLO2PT1},  \eqref{eqref:NLO2PT2} and  \eqref{eqref:NLO2PT3}, we obtain the \emph{generating function of rooted order one schemes (next-to-leading order 2-point function)}:
   \be
  \label{eqref:NLO2PTtot}
 \GF_2^{(1)}(z)= \frac{D(D-1)} 2  \biggl[ y^2\frac{2\GF_T(z) +D-2+2y\GF_T(z)(D-1)(\GF_T(z)^2 - 1  \bigr)}{\bigl(1+y \bigr)\bigl(1-(D-1)y\bigr)^2}\biggr]_{y=z\GF_T(z)^{D}}.
 \ee
This can in theory be applied to count the contributions to any order of the degree $\delta_0$. However, the number of schemes becomes rapidly impossible to handle.

\section{Intermediate field theory}
\label{sec:IFT}

The intermediate field theory is a way of rewriting the generating functions of bubble-restricted gluings as matrix models. Hopefully, we can then apply matrix models methods, for instance throughout their eigenvalue decomposition. Of course, this is very common in two dimensions. The fact that $\Ps_0$ (Thm.~\ref{thm:BijSM}) is a bijection with stacked combinatorial maps suggests that this is possible. 

Let us first give an example in the case of the simpler bijection of  Section~\ref{sec:SimplerBij}.  It maps bijectively  gluings of a single kind of $k$-cyclic bubble to four-valent combinatorial maps which faces are bicolorable, i.e. which dual maps are the bipartite quadrangulations. It is easily seen that the generating function of such maps can be written as a rectangular matrix model 
\be
\cZ(\lambda,N)=\int \exp\bigl\{-N^{D-1}\Tr\bigl(B.B^\dagger - \frac\lambda2 N^{(k-1)(p-1)} (B.B^\dagger) .(B.B^\dagger) \bigr)\bigl\} \frac{dBdB^\dagger}{\cZ_0},
\ee
where $B$ is a $N^k\times N^{D-k}$ complex matrix, and the scaling $s=(k-1)(p-1)$ has been computed from \eqref{eqref:ACycles}.  The left and right indices of the matrices are distinguished and give rise to two kind of faces, which are colored with 1 and 2.  The maps are face-bipartite and faces of different colors do not contribute with the same power of $N$. We refer the reader to the paper of Di~Francesco \cite{Rectang} on rectangular matrix models for further developments. 

\

In the case of quartic melonic gluings, the generating function can be rewritten as a multi-matrix model throughout the bijection of Thm.~\ref{thm:BijBoundQuart} with combinatorial maps whose edges carry colors in $\lDr$. We need $D$ size-$N$ matrices - one for each color, and as the valency of vertices is not restricted, the potential is a infinite sum of all the possible traces of products of copies of these $D$ matrices. Each monomial with $k$ matrices has to be rescaled with a symmetry factor $1/k$ so the potential should look like $\ln(\un - \sum_{i=1}^D M_i)$. In the quartic case, these intermediate field generating functions can be derived throughout the Hubbard-Stratonovich transformation applied to the random tensor models, as detailed in the following subsections.

But first let us insist on the fact that the intermediate field representation has proven a very powerful tool in the quartic case. The first application is in deriving analyticity results for the corresponding matrix or tensor models. This representation is used to prove the analyticity of correlation functions in the Borel summability domain of the perturbative expansion at the origin \cite{BeyondPert,Mutiscale, Analyticity,ConstrQuart}. Using this method, R. Gurau and T. Krajewski showed in  \cite{Analyticity} that the planar sector is the large $N$ limit of quartic one-matrix models \emph{beyond perturbation theory}. Attempts have been made by V.~Rivasseau and myself to apply intermediate field technics in order to obtain analyticity results for non-quartic matrix and tensor models with positive interactions \cite{PosTens}. We could prove the non-uniform Borel-Leroy summability, however the domain of convergence shrinks to zero in the large $N$ limit. As mentioned in the introduction, this thesis focuses on geometric and combinatorial properties of discrete spaces, so we do not present these developments here. In the quartic melonic case, the intermediate field representation has also been used in \cite{DartEyn} to apply eigenvalue methods, or the topological recursion in \cite{TopoTens}. In  \cite{PhaseTrans, SymBreak}, it is used to study the effective theory of fluctuation around the melonic vacuum, the phase transition that occurs at criticality, and the breaking of the symmetry of the model. We detail in Section~\ref{sec:Fluctuations} the very first steps leading to a generalization of these results.

	\subsection{The Hubbard-Stratonovich transformation}
	
	The Hubbard-Stratonovich transformation \cite{Hubbard, Strato} re-expresses the scalar $\Phi^4$ theory as a theory with interactions of any order. The first step towards the intermediate field representation is to develop the vertices using a new real scalar field $\sigma$. This new field splits the $\phi^4$ vertex into two $\phi^2\sigma$ vertices, which is the reason why it is called \emph{intermediate} (see Fig.~\ref{fig:SplitVert}). We  consider the generating function of cumulants of the complex $\phi^4$ theory,
\be
\cZ[J,\bar{J};\lambda]={\frac 1 {\cZ(\lambda)}}\int_{\bC} d\mu_{G}(\phi, \bar{\phi})\exp\{ -\frac{\lambda} 2 (\bar{\phi}\phi)^2+\bar{J}\phi +J\bar{\phi}\},
\ee
where $\cZ(\lambda)=\cZ[0,0;\lambda]$. We denote $d\mu(\sigma)=\frac{1}{\sqrt{2\pi}}d\sigma e^{-\frac{\sigma^2}{2}}$. The first step relies on the identity
\be
\forall \lambda\in\mathbb{R}^{+},\, X\in \mathbb{R},\; e^{-\frac{\lambda} 2 X^2}=\int_{\mathbb{R}}d\mu(\sigma)\exp\{ -i\sqrt{\lambda}X \sigma\}
\ee
applied to $X=\bar{\phi}\phi$. It leads to 	
\be
\cZ[J,\bar{J};\lambda]=\int_{\bC\otimes\bR} d\mu_{G}(\phi, \bar{\phi})d\mu_{G}(\sigma)\exp\{ -i\sqrt{\lambda} \bar{\phi}\phi\sigma+\bar{J}\phi +J\bar{\phi}\}.
\ee
 This is now a Gaussian integral on $\phi$ and $\bar \phi$. Performing the integration over the initial fields, 
\be
\int _\bC e^{-\phi\bar\phi(1+i\sigma\sqrt{\lambda}) +\bar{J}\phi +J\bar{\phi}}\frac{d\phi d\bar\phi} \pi = \frac 1 {1+i\sigma\sqrt{\lambda}} e^{\frac{J\bar{J}}{1+i\sqrt{\lambda}\sigma}},
\ee
and the partition now writes 
\be
\label{eqref:generatingscalar}
\cZ[J,\bar{J};\lambda]={1\over{\cZ(\lambda)}}\int_{\bR} d\mu(\sigma)\exp\biggl( - \ln(1+i\sqrt{\lambda}\sigma)+\frac{J\bar{J}}{1+i\sqrt{\lambda}\sigma}\biggr).
\ee	

As mentioned previously, this transformation can be understood on the Feynman graphs, as shown in Fig.~\ref{fig:SplitVert}. The vertex is first split in two throughout a $\sigma$ field. In the new graphs, the half-edges corresponding to the $\phi$ fields are now gathered into oriented cycles, which are then contracted into vertices of the new $\sigma$ theory.
\begin{figure}[!h]
\centering
\includegraphics[scale=0.9]{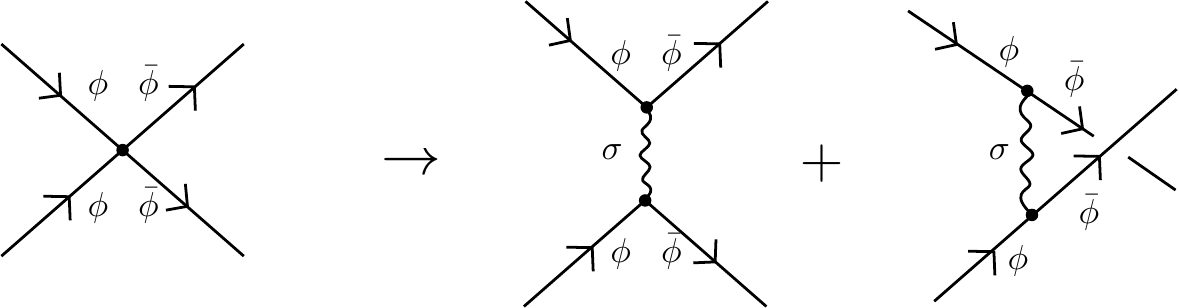}
\caption{Splitting of the quartic vertex}
\label{fig:SplitVert}
\end{figure}

The transformation generalizes to quartic matrix models, e.g. for the complex model
\be
\cZ[J,\bar{J};\lambda, N] = \int \frac{dMd\bar M}{\cZ(\lambda,N)}\exp\bigl[-N\Tr\bigl(MM^\dagger +\frac{\lambda} 2 MM^\dagger MM^\dagger + JM^\dagger + MJ^\dagger \bigr) \bigr],
\ee
where $dMd\bar M=\pi^N \prod_{i,j=1}^N d\text{Re}(M_{ij})d\text{Im}(M_{ij})$. The first step is
\be
e^{-\frac \lambda 2 N\Tr MM^\dagger MM^\dagger}=\int d\nu(A) \exp\biggl( i\sqrt{\lambda N}\Tr(M^\dagger A M)\biggr), 
\ee
where the variable $A$ is an Hermitian $N\times N$ matrix, and $d\nu(A)=\frac {dA}{\cZ_0} e^{-\frac 12 \Tr A^2}$ is normalized. We are left with a Gaussian integral which can be performed
\bea
&&\int  e^{-N\Tr\bigl[MM^\dagger(\un-iA\sqrt{\frac{\lambda} N})+J^\dagger M +JM^\dagger\bigr] }\frac{dMd\bar M}{\cZ(\lambda,N)}\hspace{7cm} \\
&&\nonumber \hspace{5cm}= \frac 1 {\det\bigl[(\un-iA\sqrt{\frac{\lambda} N})\otimes \un\bigr]} e^{N\Tr J  \bigl[(\un-i A \sqrt{\frac{\lambda} N})\otimes \un \bigr]^{-1}  J^\dagger  }.
\eea
As $\det\bigl[(\un-iA\sqrt{\frac{\lambda }N})\otimes \un\bigr] = \det(\un-iA\sqrt{\frac{\lambda} N})^N$, we obtain the intermediate field representation
\be
\cZ[J,\bar{J};\lambda, N] = \int d\nu(A) \exp-N\Tr\biggl[\ln(1-iA\sqrt{\frac{\lambda} N})   - J  (1-i A \sqrt{\frac{\lambda} N})^{-1}J^\dagger  \biggr].
\ee
Changing $A$ to $N^{-1/2}A$, we get the expression
\be
\cZ[J,\bar{J};\lambda, N] = \int \frac{dA}{\cZ(\lambda,N)}\exp-N\Tr\biggl[\frac{A^2} 2 + \ln(1-iA\sqrt{\lambda} )   + J  (1-i A \sqrt{\lambda} )^{-1}J^\dagger  \biggr].
\ee

The perturbative expansion of the intermediate field theory is over combinatorial maps with vertices of any valency. They are obtained from the maps of the Feynman expansion of the original quartic model by applying Tutte's bijection (Subsection~\ref{subsec:Tutte}) to the dual quadrangulations. 
	
	\subsection{Matrix models for quartic tensor models}
	
	The development for quartic (non-necessarily melonic) tensor models is similar to the previous section. For $k$-cyclic bubbles of size 4 \eqref{eqref:ACycles}, the scaling $s$ is 
	\be
	s=k -1.
	\ee
If $\cI_{D-k}$
and $\cI_{k}$
are respectively the sets of the $D-k$ and $k$ colors which alternate on the considered $k$-cyclic bubble,  we denote the corresponding invariant quartic polynomial 
\be
(T._{\cI_{D-k}}\bar T)._{\cI_k}(T._{\cI_{D-k}}\bar T) = \Tr_{\cI_k}\biggl[\Tr_{\cI_{D-k}}\bigl[T\otimes\bar T\bigr].\Tr_{\cI_{D-k}}\bigl[T\otimes\bar T\bigr]\biggr].
\ee
The generating function for cumulants is 
\be
\cZ_{k}[J,\bar{J};\lambda, N] 
=\int_{\bC^D}e^{-N^{D-1}\bigl(T.\bar T + \frac \lambda 2 N^{k-1}(T._{\cI_{D-k}}\bar T)._{\cI_k}(T._{\cI_{D-k}}\bar T)  + T.\bar J + J.\bar T\bigr)}\frac{dTd\bar T}{\cZ(\lambda,N)}.
\ee
We choose to pair the vertices which share $D-k\ge D/2$ edges (this is an optimal pairing), and as before, we write the quadratic term as 
\be
e^{-\frac\lambda2 N^{D+k-2}(T._{\cI_{D-k}}\bar T)._{\cI_k}(T._{\cI_{D-k}}\bar T)}=\int d\nu(M_{\cI_k}) \exp\biggl( i\sqrt{\lambda N^{D+k-2}}\Tr((T._{\cI_{D-k}}\bar T).M_{\cI_k})\biggr), 
\ee
where $M_{\cI_k}$ is an Hermitian $N^k\times N^k$ matrix. We rewrite 
\be
\Tr((T._{\cI_{D-k}}\bar T).M_{\cI_k}) = \Tr\biggl([T\otimes\bar T].[M_{\cI_k}\otimes\un_{\cI_{D-k}}]\biggr),
\ee
and see that the covariance is $N^{D-1}\un^{\otimes D} -i\sqrt{\lambda N^{D+k-2}}M_{\cI_k}\otimes\un_{\cI_{D-k}}$. Integrating over $T$ and $\bar T$, we get 
\be
\frac 1 {\det\bigl[\un^{\otimes D} -i\sqrt{\lambda N^{k-D}}M_{\cI_k}\otimes\un_{\cI_{D-k}}\bigr]}
\Tr \bigl(N^{D-1}[{ J}\otimes{ \bar{J}}].(\un^{\otimes D} -i\sqrt{\lambda N^{k-D}}M_{\cI_k}\otimes\un_{\cI_{D-k}})^{-1}\bigl),
\ee
which, tracing on the identity factors and rescaling $M_{\cI_k}$ to $N^{-k/2}M_{\cI_k}$ leads to the intermediate field representation
\bea
&&\cZ_{k}[J,\bar{J};\lambda, N] =\int\frac{ dM_{\cI_k}}{\cZ_{k,0}}e^{-\frac{N^k}2\Tr M_{\cI_k} ^2- N^{D-k}\Tr\ln\bigl[\un^{\otimes k} -i\sqrt{\frac \lambda {N^{D-2k}}}M_{\cI_k}\bigr]}\hspace{4.5cm} \\
&&\nonumber\hspace{5.5cm}\times\ e^{+N^{D-1}\Tr \bigl([{ J}\otimes{ \bar{J}}].(\un^{\otimes D} -i\sqrt{\frac \lambda {N^{D-2k}}}M_{\cI_k}\otimes\un_{\cI_{D-k}})^{-1}\bigl)  }.
\eea
The partition function  for the full quartic random tensor model in dimension $D$ is
\be
\cZ(\lambda, N) =\int e^{-\Tr\ln\bigl[   \un^{\otimes D} -i\sum_{\cI_k\subset \lDr}\sqrt{\frac \lambda {N^{D-k}}}M_{\cI_k}\otimes\un_{\cI_{D-k}}\bigr]\ }\frac{\prod_{\cI_k\subset \lDr}d\nu(M_{\cI_k})}{\cZ_0}. 
\ee
The Feynman perturbative expansion has been studied at leading order in Section~\ref{sec:SimplerBij} for the case of a single set $\cI_k$, in Section~\ref{sec:QuartMelSec} for the case of quartic melonic bubbles, and in Section~\ref{Subsec:D2CycBub}  for the case of 2-cyclic bubbles in dimension 4.

	\subsection{Matrix models for generic tensor models}

The matrix model we are going to present has multi-trace interactions. It has a Gaussian measure, while its potential is a sum of two distinct terms. The first one stands for the interaction bubble $\Ps(\B,\Om_B)$, while the second one, which does not depend on $\B$ and corresponds to the color-0 edges of the stacked maps, is an infinite series in the conjugate matrices which performs all the possible gluings of interaction bubbles.
%
%

\subsubsection{Matrix bubbles}

Given a bubble $\B$ with $V$ vertices and an arbitrary labeling of both its black and white vertices, there exist $D$ permutations\footnote{The permutation of color $i$ is made explicit in $\B$ by keeping only edges of color $i$.}
 $\tau_1,...,\tau_D$ of $\llbracket1,\cV\rrbracket$ such that $\B$ can be written in the following form
\be
\Tr_\B({ T}, { \bar T})= 
\sum_{  \substack {  { i^1_{1},..,i^1_{D},\ \   }   \\ {...\ \ \  }  \\ { i^{\cV}_{1},..,i^{\cV}_{D}=1  }  }   }^N\ 
 \sum_{  \substack {  { j^1_{1},..,j^1_{D},\ \   }   \\ {...\ \ \ }  \\ { j^{\cV}_{1},..,j^{\cV}_{D}=1  }  }   }^N  
 \prod_{a,b=1}^{\cV}\ T^{}_{i^a_1 ... i^a_D}\bar T^{}_{j^b_{1}...j^b_{D}}
 \prod_{k=1}^D\delta_{i_k^a, j_k^{\tau_{k}(a)}}.
\ee
Since the key ingredient to go from $\B$ to the edge-colored map $\Ps(\B, \Om)$ which encodes $\B$ through its broken faces is a pairing $\Om$ of $\B$, we re-organize the sums over tensor indices according to $\Om$. The pairing  can also be thought of as a permutation $b$ that identifies each white vertex $a$ with a particular black vertex $b(a)$. For each color $i$, we define the induced permutation $\tilde\tau_i$ so that this expression reorganizes as
\be
\label{eqref:bubble}
\Tr_\B({ T}, { \bar T})=
\sum_{  \substack {  { i^1_{1},..,i^1_{D},\ \   }   \\ {...\ \ \ }  \\ { i^{\cV}_{1},..,i^{V}_{D}=1  }  }   }^N\ 
 \sum_{  \substack {  { j^1_{1},..,j^1_{D},\ \   }   \\ {...\ \ \ }  \\ { j^{\cV}_{1},..,j^{V}_{D}=1  }  }   }^N
 \prod_{a=1}^{V}\ \ \bigl[ T_{i^a_1 ... i^a_D}\bar T_{j^{b(a)}_{1}...j^{b(a)}_{D}}\bigr]
 \prod_{k=1}^D\delta_{i_k^a, j_k^{\tau_{k}(a)}}.
	\ee
The {\it matrix bubble} equivalent to the pairing $\Omega=\{(a,b(a)\mid a$ white vertex$\}$ on the bubble $\B$ is then obtained from ($\ref{eqref:bubble}$) by replacing each pair  $(T^{(a)},T^{(b(a))})$ with a $N^D\times N^D$ complex matrix $M$ :
\be
\label{eqref:matrixbubble}
V_{\B, \Omega}(M)=
\sum_{  \substack {  { i^1_{1},..,i^1_{D},\ \   }   \\ {...\ \ \ }  \\ { i^{\cV}_{1},..,i^{\cV}_{D}=1  }  }   }^N\ 
 \sum_{  \substack {  { j^1_{1},..,j^1_{D},\ \   }   \\ {...\ \ \ }  \\ { j^{\cV}_{1},..,j^{\cV}_{D}=1  }  }   }^N
 \prod_{a=1}^{\cV}M_{(i^a_1 ... i^a_D); (j^{b(a)}_{1}...j^{b(a)}_{D})    } 
  \prod_{k=1}^D\delta_{i_k^a, j_k^{\tau_{k}(a)}}.
  \ee
Graphically, each one of the $n$ pairs is identified with a copy of the matrix $M$, its first $D$ sub-indices being the edges reaching the white vertex of the pair, while its $D$ second half of sub-indices are those reaching the black vertex. The contraction scheme between the $n$ copies of $M$ is then obtained by contracting sub-indices of color $i$ according to the edges of the bubble. 
\begin{figure}[!h]
\centering
\includegraphics[scale=0.65]{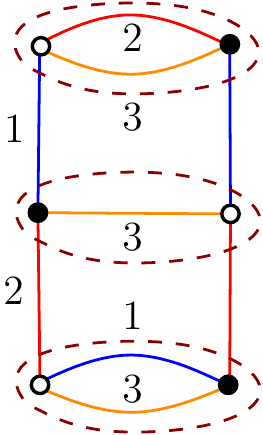}\hspace{2cm}\includegraphics[scale=0.65]{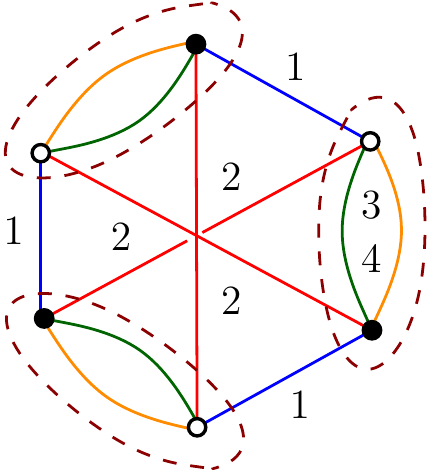}
\caption{ Two bubbles of order six together with their optimal pairings. }
\label{fig:BubEx}
\end{figure}

\noindent The matrix bubble associated to the examples in Figure \ref{fig:BubEx} are 
\be
\sum_{\substack{{i_1,j_1}\\ {i_2, j_2}}}(\Tr_{23} M)_{i_1; j_1}(\Tr_3 M)_{j_1,j_2 ; i_1,i_2}(\Tr_{13} M)_{i_2;j_2} =\Tr\bigl[(\Tr_{23} M)\otimes(\Tr_{13} M)\bigr].(\Tr_3 M) 
\ee
for the left example, and for that on the right, 
\be
\sum_{\substack{{i_1,j_1, k_1}\\ {i_2, j_2, k_2}}}
(\Tr_{34} M)_{i_1,i_2 ; j_1,j_2}(\Tr_{34} M)_{k_1,j_2 ; i_1,k_2}(\Tr_{34} M)_{j_1,k_2 ; k_1,i_2}.
\ee
It is clear in these examples that a smaller matrix can be considered if a subset of colors is always traced on $M$. However, we intend to give general results, and our choice of matrix further allows to consider a theory with infinitely many interactions of a well chosen family. Thanks to the following theorem, proven in \cite{SWM}, we can rewrite the correlation functions of any random tensor model as that of a multi-trace matrix model. However, the resulting interactions are very complicated in general. 

\subsubsection{Intermediate field representation}

\begin{theorem}[Intermediate field representation] 
\label{thm:IFT}
Any tensor model has an intermediate field representation involving a complex matrix of size $N^D\times N^D$. More precisely, for any pairing $\Om$ of the $2n$ vertices of the bubble $\B$, the associated partition function and generating function of the cumulants formally\footnote{
i.e. as a formal power series in the coupling constant $\lambda$.
}
 rewrite :
\be
\label{eqref:TensorMatrixEquality}
 \cZ_{\B}(\lambda, N)
=\int_{\bC^D}e^{-\frac \lambda n N^{s} \Tr_\B({ T}, { \bar{T}})}d\mu({ T}, { \bar{T}})
=\int e^{ - \frac{\lambda}{n}{N^{s-n(D-1)}} V_{\B,\Om}(M)-\Tr\ln\bigl(\un^{\otimes D}+
\bar M\bigr)
  }d\nu(M, \bar M),
\ee
where $d\nu(M,  \bar M)=\frac{dM d\bar M}{\pi^{N^{2D}}}e^{-\Tr(M\bar M)}$ and $V_{\B,\Om}$ is the matrix bubble equivalent to $\B$ with pairing $\Om$, and the generating function of the cumulants is
\be
\label{eqref:genfunc}
 \cZ_{\B}[\lambda, N;{ J, \bar J}]
=\int e^{ - \frac{\lambda}{n}{N^{s-n(D-1)}} V_{\B,\Om}(M)-\Tr\ln\bigl(\un^{\otimes D} +
\bar M\bigr)-\Tr \bigl[({ J}\otimes{ \bar{J}}).(\un^{\otimes D} +\bar M)^{-1}\bigl]
  }d\nu(M, \bar M).
\ee
\end{theorem}

We stress however that if in $V_{\B,\Om}(M)$, a subset $I$ of colors is always traced for each $M$, as mentioned earlier, the latter can be replaced by a matrix of size $\lvert I\rvert $ in the previous theorem, provided that every $\un^{\otimes D}-
\bar M$ term is replaced with $\un^{\otimes D}-
\bar M\otimes \un^{\hat I}$, where $\hat I=\llbracket1,D\rrbracket\setminus I$. In general, the cost of reducing the sized of the matrices is that instead of a  $N^D\times N^D$ one-matrix model, we get a multi-matrix model with $N^{\lvert\cI\rvert}\times N^{\lvert \cI \rvert}$ matrices. This second approach is the one developed on \cite{SWM}. Depending on the interaction bubble, it can be an improvement or not.
The previous theorem is stated for any choice of pairing. In practice, we always choose an optimal pairing.
In the case of bubbles with a single optimal pairing, the subscript $\Om$ is forgotten.

\

 The strategy to prove the theorem is simply to expand the exponentials of the non-quadratic terms and prove the equalities order by order in $\lambda$. The proof then relies on the one-dimensional case which we present as a lemma. We denote 
$d\nu(z, \bar z)=\frac{dz d\bar z}{\pi}e^{-z\bar z}$.

\begin{lemma} 
For all $a\in\bC$ and $p\in\bN$, the following relation holds
\be
\label{eqref:relationprf}
\langle z^p\,e^{-a\bar{z}}\rangle_0 \equiv \int_{\mathbb C} z^p \ e^{-  a \bar z }d\nu(z, \bar z) = a^p.
\ee
\end{lemma}
\prf 
\bea
\nonumber \int_{\bC} z^p \ e^{-  a\bar z -z\bar z}\frac{dz d\bar z}{\pi}
&=&\biggl[ e^{\frac{\partial}{\partial z}\frac{\partial}{\partial \bar z}} e^{a\bar z}z^p\biggr]_{z=\bar z=0}\\
&=&\sum_{k\ge0}\frac{1}{k!}\bigl[ \bigl(\frac{\partial}{\partial z}\bigr)^k z^p\bigr]_{z=0}\bigl[ \bigl(\frac{\partial}{\partial \bar z}\bigr)^k e^{a\bar z}\bigr]_{\bar z=0}\\
\nonumber&=&\sum_{k\ge0}\frac{1}{k!}(p!\delta_{p,k})(a^k)=a^p. 
\eea
\qed

This is easily extended to products of multivariate polynomials.
\begin{lemma} 
\label{polprod}
For any $K\in\bN,\ \{a_i\}\in\bC^K$, and given  $P[z_1,\dotsc,z_K]$ and $Q[z_1,\dotsc,z_K]$ two complex polynomials, the following relation holds
\begin{equation}
\begin{aligned}
\langle P(z_1,\dotsc,z_K) Q(z_1,\dotsc,z_K)\ e^{- \sum_i a_i\bar z_i } \rangle_0 &\equiv \int_{\mathbb C ^K} P(z_1,\dotsc,z_K) Q(z_1,\dotsc,z_K)\ e^{- \sum_i a_i\bar z_i }\prod_id\nu(z_i, \bar z_i)\\
&= P(a_1,\dotsc, a_K) Q(a_1,\dotsc, a_K).
\end{aligned}
\end{equation}	
\end{lemma}

\prf The generalization of \eqref{eqref:relationprf} to $K$ complex variables $z_1,..,z_K$ is the obvious relation
\be
\label{eqref:monomial}
\int_{\mathbb C ^K} \prod_{i=1}^K z_i^{p_i} \ e^{- \sum_i a_i.\bar z_i }\prod_id\nu(z_i, \bar z_i)=\prod_{i=1}^Ka_i^{p_i},
\ee
so that by linearity, for two polynomials $P(z_1,\dotsc,z_K)=\sum_{m}\alpha_m\prod_i z_i^{p_{m,i}}$ and $Q(z_1,\dotsc,z_K)=\sum_{n}\beta_n\prod_i z_i^{q_{n,i}}$, 
\bea
\nonumber \langle P(z_1,\dotsc,z_K) Q(z_1,\dotsc,z_K)\ e^{- \sum_i a_i\bar z_i } \rangle_0
&=&\sum_{m,n}\alpha_m\beta_n\langle \prod_i z_i^{p_{m,i}+q_{n,i}}e^{-\vec{a}.\vec{\bar z}}\rangle_0\\
\nonumber&=&\sum_{m,n}\alpha_m\beta_n\langle \prod_i a_i^{p_{m,i}+q_{n,i}}e^{-\vec{a}.\vec{\bar z}}\rangle_0\\
&=&P(a_1,..,a_K)Q(a_1,..,a_K).\hspace{2cm}
\eea
\qed

\noindent{\bf Proof of Theorem \ref{thm:IFT}.} Given a bubble $\B$, $\Om$ a pairing of its vertices, and $q\in\bN$, we now prove that the following equality holds
\be
\label{bubblepower}
\bigl[ \Tr_\B({ T},{ \bar T}) \bigr]^p
= 
\int  \bigl[V_{\B,\Omega}(M)\bigr]^p\ \exp\biggl( - \Tr\ [{ T\otimes \bar T}]M^\dagger  \biggr)d\nu(M,M^\dagger),
\ee
where $[T\otimes\bar T]$ is seen as a $N^D\times N^D$ matrix, 
\be
\label{eqref:TBarTM}
\Tr\ [{ T\otimes \bar T}]M^\dagger = \sum_{ \substack {   {i_1,\dotsc, i_D} \\ {j_1,\dotsc, j_D}  } }T_{i_1\dotsb i_D}\bar T_{j_1\dotsb j_D}\bar M_{(i_1,\dotsc, i_D);(j_1,\dotsc, j_D)}.
\ee
We use expression \eqref{eqref:bubble} for the bubble $\B$. Relation \eqref{eqref:monomial} applied to the products over the vertices of matrix elements of ${ T\otimes\bar  T}$ gives
\be
 \prod_{a=1}^{V}\ \ \bigl[ T_{i^a_1 ... i^a_D}\bar T_{j^{b(a)}_{1}...j^{b(a)}_{D}}\bigr]
=
\int  \prod_{a=1}^{\cV}\ M_{(i^a_1 ... i^a_D); (j^{b(a)}_{1}...j^{b(a)}_{D})}  
\ e^{ - \Tr\bigl(  [{ T\otimes \bar T}]M^\dagger \bigr)  }
d\nu(M,M^\dagger).
\ee

Inserting this relation into \eqref{eqref:bubble} and using expression \eqref{eqref:matrixbubble}, one gets relation \eqref{bubblepower} for $p=1$. The relation for $p\ge0$ then follows from applying Lemma $\ref{polprod}$.
Using \eqref{eqref:TBarTM}, we can integrate $T$ and $\bar T$
\be
\int d\mu_{0}({ T}, { \bar{T}})e^{ - [{ T\otimes \bar T}]M^\dagger}
=\frac{1}{\det\bigl[ \un^{\otimes D}+\bar M \bigr]}
=\exp\bigl[ -\Tr \ln (\un^{\otimes D}+\bar M ) \bigr].
\ee

Changing $T\rightarrow N^{\frac{D-1}2}T$ and $\bar T\rightarrow N^{\frac{D-1}2}\bar T$ in the tensor model, the interaction becomes $\Tr_\B(T,\bar T)\rightarrow N^{-n(D-1)}\Tr_\B(T,\bar T)$, and the terms of the perturbative expansion of the tensor and matrix models match:
\be
\label{eqref:pBubbles}
\frac 1{p!} \biggl(\frac{-\lambda N^{s}}{n}\biggr)^p\int \bigl[ \Tr_\B({ T},{ \bar T}) \bigr]^p d\mu_{0}({ T}, { \bar{T}})
=
\frac 1{p!} \biggl(\frac{-\lambda N^{s}}{nN^{n(D-1)}}\biggr)^p\int  \bigl[V_{\B,\Omega}(M)\bigr]^p\ e^{ -\Tr \ln (\un^{\otimes D}+\bar M) }
d\nu(M,\bar M).
\ee
Equation \eqref{eqref:genfunc} works similarly with sources.
\qed

\subsubsection{Feynman diagrams are stacked maps}

We now perform the Feynman expansion of the matrix model \eqref{eqref:TensorMatrixEquality}. There are two types of non-quadratic terms in the action which give rise to two types of vertices. 
\begin{itemize}
\item The logarithm is expanded as an infinite sum of monomials,
\be
\Tr\ln(\un^{\otimes D}+\bar M)=\Tr\sum_{k>0}\frac{(-1)^{k+1}}{k}\bar M^k
\ee
 To each monomial $\Tr\bar M^k$, one associates a \emph{black} vertex of degree $k$. Incident edges are cyclically ordered (counter-clockwise) according to their order in the trace.
Since a matrix $\bar M$ represents a pair of vertices of $\B$, and since the summation of all the indices of two tensors is represented as a color-0 edge in the colored graph picture, it is clear that $\Tr\bar M^k$ corresponds to a cycle of a colored graph which alternates pairs of vertices and edges of color 0 (represented as corners around the vertex). One thus recovers the color-0 vertices of stacked maps.

\item The potential $V_{\B,\Om}$ corresponds to the stacked map $\Ps(\B,\Om)$, with one color-0 half-edge on each white square, corresponding to $M$. As color-0 does not appear in $\Ps(\B,\Om)$, we can interpret it as a stacked map with $D+1$ colors, but with a boundary. The boundary graph of the stacked map gives the contraction rules between the indices of the copies of $M$.

\item The quadratic term of the action propagates a $M$ to a $\bar M$, connecting black vertices to  bubbles.
\end{itemize}

\chapter{Properties of stacked maps}
\label{chap:PropSWM}


In this chapter, we aim at providing general results which would make it possible to easily characterize and count maximal maps for a given set of bubbles. We saw in Section~\ref{sec:SimplerBij} that we could recover the universality classes of trees and that of planar maps, respectively called branched polymers and 2D pure gravity in physics, and we wish to understand which properties of the bubbles will ineluctably lead to known combinatorial families when restricting to maximal maps. Another of our goals is to obtain general properties for the coefficient $\tilde a_\B$ and the scaling $s_\B$ leading to well-defined and non-trivial bubble-dependent degrees, and to consistent $1/N$ expansions: Do they exist? Are they uniquely defined? 

We recall that maximal gluings are those which maximize the number of $(D-2)$-cells at fixed number of $D$-cells, or equivalently, stacked maps which maximize the 0-score at fixed number of bubbles. If the dependence in the number of bubbles of the 0-score of maximal maps is linear \eqref{eqref:MaximalBound}, we can deduce the coefficients $\tilde a$ and the appropriate bubble-dependent degree \eqref{eqref:BubDeg3}, the coefficient $a$ \eqref{eqref:Tildeaa}, which gives the correction to Gurau's degree, and the scaling $s$ of the corresponding enhanced tensor model \eqref{eqref:ArbTensZ}. We refer the reader to the guideline in Subsection~\ref{subsec:Guideline}, in which references to most of the notions defined in the first chapter are listed. We also recall the reader that the list of symbols gives references to the definitions. 

Again, we are interested in the topology of maximal gluings, and wish to understand how the results of Section~\ref{subsec:Cryst} apply in the context of stacked maps. In Subsection~\ref{subsec:Unhook}, we adapt the notion of edge-deletion to stacked maps (whose white vertices all have degree $D+1$), and study how the 0-score  and the topology vary when performing that operation. 

In Subsection~\ref{sec:MaxMaps}, we discuss whether trees belong to maximal maps, as in that case there is a simple formula for the coefficients $\tilde a$, $a$, and $s$, leading to a well-defined and non-trivial  $1/N$ expansion.  Theorem~\ref{thm:ExUnicBBD} gives a sufficient and necessary condition of existence of an appropriate bubble-dependent degree, and states that  when it exists, is is uniquely defined, as well as the coefficients $\tilde a$, $a$, and $s$. We also prove results which give general rules to determine which aspects of bubbles will modify the critical behavior of maximal maps or not.

We generally use the stacked-map picture and the bijection $\Ps_0$ of Thm.~\ref{thm:BijSM}, however we stress here that all results are also true using, when it is possible,  the simplified bijection of Thm.~\ref{thm:BijSimp}, in which bubbles are white vertices, and bicolored submaps are monochromatic submaps. 
In the rest of this thesis, when referring to stacked maps or simpler versions, black vertices/edges are the color-0 vertices/edges, which play a special role when considering bubble-restricted gluings, as the structure of the bubbles $\Ps(\B,\Om_\B)$ for $\B\in\bB$ is rigid. By face of a stacked map, we refer to a face around a color-$0i$ submap. 
In general, we study gluings of a single kind of bubble, in order to determine the corresponding coefficients $\tilde a$, $s$.... We may then try to characterize maximal maps obtained when considering sets of bubbles.

\section{Trees and the bound on $(D-2)$-cells}
\label{sec:TreeBound}

\subsection{Projected maps and trees}
\label{subsec:ProjXTrees}

\begin{definition}[Projected map]
The projected map $\Ga^\star$ of a stacked map $\Ga\in\bS(\bB,\Om_\bB)$ is obtained by replacing each bubble $\Ps(\B,\Om_\B)$ with a non-embedded vertex. It is bipartite and we color the resulting non-embedded vertices in white.
\end{definition}

The same projected map may be associated to several stacked maps. 
This is not a problem, as we are mostly interested in the circuit-rank of the projected map. The projected map on the left of Fig.~\ref{fig:ProjMap} is that of the map in Fig.~\ref{fig:BijSWM}.


\begin{figure}[!h]
\centering
\includegraphics[scale=0.6]{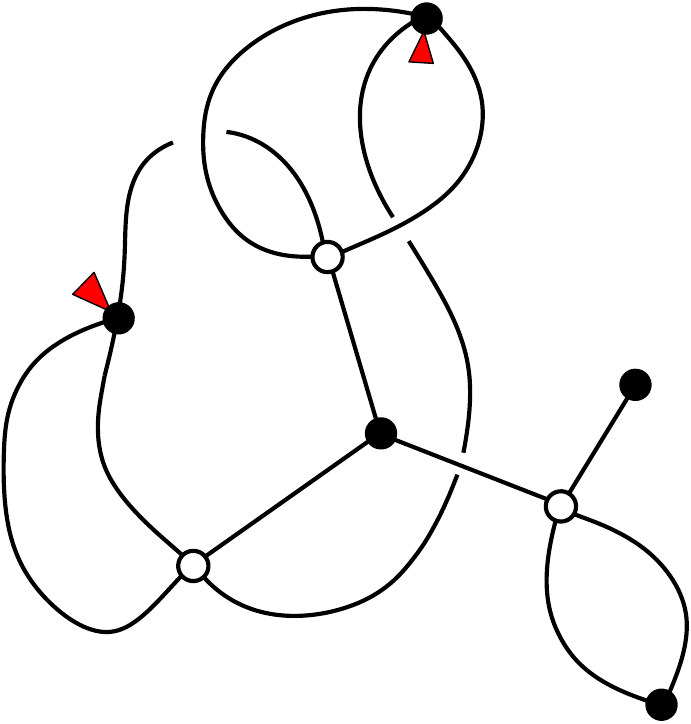}\hspace{2cm}\raisebox{+1.7ex}{\includegraphics[scale=0.6]{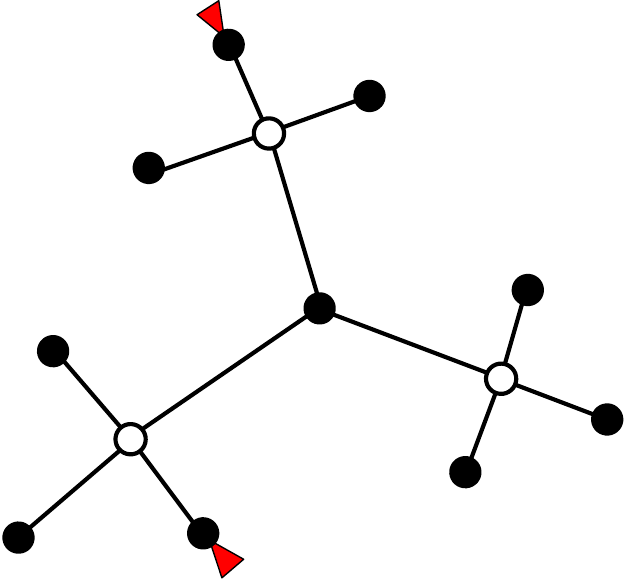}}
\caption{Projected map and projected tree.}
\label{fig:ProjMap}
\end{figure}

\begin{definition}[Stacked trees]
A stacked tree $\Ga$ is such that its projected map $\Ga^\star$ is a tree. 
\end{definition}
When there is no ambiguity, we will just call them trees. In general, by number of independent cycles of a stacked map, we refer to the circuit-rank of its projected map. Remark that in the case of the simpler bijection of Thm.~\ref{thm:BijSimp}, the circuit-rank of the map coincides with the circuit-rank of the projected map. 

\subsection{Generalized unicellular maps}
\label{subsec:Unicell}

Stacked maps with a single bubble generalize unicellular maps, as explained in Subsection~\ref{subsec:CountUni}. Here however, we are interested in the case where we have chosen a given bubble $\B$. The set of generalized unicellular colored graph is the set of coverings of $\B$ $\{\B^\Om\}_{\Om}$. It corresponds to the two-dimensional case of unicellular maps with prescribed number of edges. 

 If a bubble $\B$ has $V/2$ pairs, then, building the bijection of Section~\ref{subsec:BubStacked} with a given choice of pairing $\Om_\B$, there are $(V/2)!$ different unicellular maps with a single bubble, each corresponding to a covering $\B^{\Om'}$ where $\Om'$ differs from $\Om_\B$.  Choosing another pairing to build the bijection, we would obtain $(V/2)!$ different unicellular maps. However, we would recover the same $(V/2)!$ projected maps, counted with multiplicity.
The map $\Ps_0(\B^{\Om_\B}, \Om_\B)$ is the only unicellular tree. The number of independent cycles of any other unicellular map $\Ps_0(\B^{\Om'}, \Om_\B)$ is the minimal number of pairs of color-0 edges that have to be switched to recover $\B^{\Om_\B}$ from $\B^{\Om'}$ in the colored graph picture (see the definition of a $\rho$-pair switching \ref{def:RhoHSwitch}).

Importantly, the choice of pairing $\Om_\B$ breaks the symmetry on equivalent pairings of $\B$. Indeed, if there are $p$ equivalent pairings of $\B$ (i.e. such that the corresponding graphs $\BOM$ are isomorphic),
then they give rise to $p$ different unicellular maps. This can be seen on the example of the $K_{3,3}$ bubble Fig.~\ref{fig:UniMaps}, where all three unicellular maps on the bottom are maximal (they corresponding to choosing three edges of different colors in the bubble). There are three maps which are not isomorphic, but correspond to equivalent pairings.
%
\begin{figure}[!h]
\centering
\includegraphics[scale=0.7]{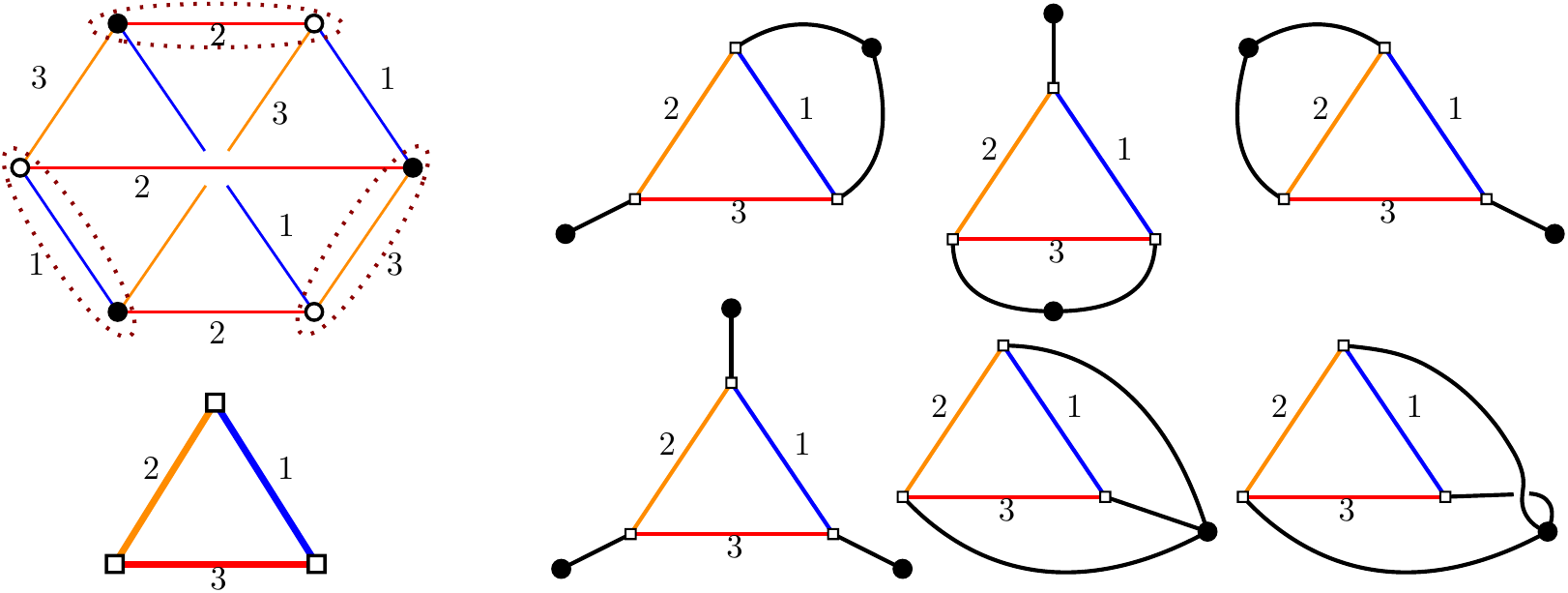}
\caption{On the top left is the $K_{3,3}$ bubble and a pairing. On the bottom left is the corresponding map, without colored vertices. On the right are the six unicellular maps. }
\label{fig:UniMaps}
\end{figure}

If one chooses an optimal pairing (Def.~\ref{def:OptPair}) to build the bijection of Sec.~\ref{subsec:BubStacked}, then the only unicellular tree is maximal among unicellular maps. If there are other optimal pairings, there are other unicellular maps which are maximal. If one chooses a non-optimal pairing to build the bijection, then the only unicellular tree is not maximal. We will see in the Section~\ref{subsec:TreeDeg} that with such a choice, no tree is maximal. In particular,
\begin{lemma}
If stacked trees belong to maximal maps in $\bS(\bB,\Om_\bB)$, then the pairing $\Om_B$ is optimal for the bubble $\B$.
\end{lemma}

Consider a bubble with more than one optimal pairing, and choose any optimal pairing to build the bijection. Then there is at least one unicellular map which is not a tree but is maximal. Consequently,
\begin{lemma}
\label{lemma:TreesImplieOpt}
If stacked trees are the only maximal maps in $\bS(\B,\Om)$, then $\B$ has only one optimal pairing.
\end{lemma}
%

\subsection{Edge-unhooking}
\label{subsec:Unhook}

The following operation replaces the usual edge-deletion, which is central in obtaining Tutte's equations for families of combinatorial maps.

\begin{figure}[!h]
\centering
\includegraphics[scale=0.75]{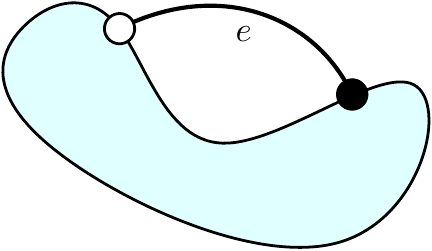}\qquad\raisebox{+4ex}{$\rightarrow$}\qquad\includegraphics[scale=0.75]{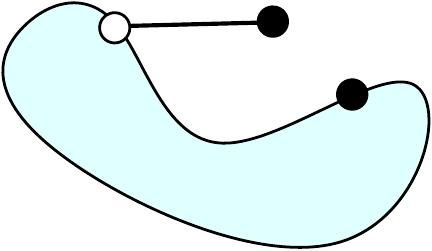}
\caption{Edge unhooking.}
\label{fig:Unhook}
\end{figure}

\begin{definition}[Edge unhooking]
Given an edge $e$ in a stacked map $\Ga$ incident to a black vertex, one may unhook it from its black endpoint, and attach the pending end to some newly added black vertex, thus obtaining another  stacked map $\Gae$ (Figure~\ref{fig:Unhook}).
\end{definition}

In practice, we do not unhook edges incident to marked corners. We stress that the unhooking may raise the number of connected components. Edge-unhooking is defined the same way for the simpler bijections of Thm.~\ref{thm:BijSimp}. We call \emph{bridge} or cut-edge an edge $e$ such that $\Gae$ has more  connected components than $\Ga$.
\begin{definition}[cut-bubble]
\label{def:CutBub}
A cut-bubble $\Ps(\B,\Om)$ is such that every incident edge is a bridge. 
\end{definition}

The two corners incident to the edge $e$ correspond to two color-0 edges 
of the corresponding edge-colored graph $\G\in\bG(\bB)$. In the colored graph picture, unhooking $e$ goes back to switching these two color-0 edges (Def.~\ref{def:RhoHSwitch}).

\subsubsection{Edge-unhooking and score}

Before switching them, the two color-0 edges belong to a certain number 
\be
\label{eqref:I1I2}
\cI_1(e)=D-\I(e)
\ee
 common bicolored cycles of $\G$. The unhooking is therefore a $\rho_{\cI_1}$-pair switching. 
 In the stacked map picture, the definition of $\I$ is 
\be
\label{eqref:I2E}
\I(e)=\Card\{i\in\lDr \mid \textrm{Two different faces run along }e\ {\rm in }\ \Gai\}.
\ee
If $f$ and $f'$ are the two exchanged edges in the colored graph picture, or the two corresponding corners in the stacked map, we will also need to denote $\I(f,f')$ this quantity.

\begin{prop}
\label{prop:FaceUnhook}
When unhooking an edge $e$ of a map $\Ga$, the 0-score of the resulting map $\Gae$ is 
\be
\label{eqref:FaceUnhook}
\Phi_0(\Gae)=\Phi_0(\Ga) + D-2\I(e).
\ee
\end{prop}
\prf  In each bicolored submap, there are either 1 or 2 faces running along $e$. In the first case, the unhooking splits this face into two faces, while in the second case, the two faces merge into a single face.  Therefore, 
\be
\Phi_0(\Gae)=\Phi_0(\Ga) + \cI_1(e)-\I(e),
\ee
and we conclude with (\ref{eqref:I1I2}). \qed

\

 In particular, as we wish to determine maximal maps, we will be interested in knowing whether the stacked map obtained after unhooking or hooking an edge as a higher 0-score
\be
\Phi_0(\Gae)\ge\Phi_0(\Ga)\quad \Leftrightarrow \quad\I(e)\le \frac D 2.
\ee
In the case where $\Gae$ and $\Ga$ have the same number of connected components ($e$ is not a bridge), the bubble-dependent degree of $\Gae$ is
\be
\delta_\bB(\Gae)=\delta_\bB(\Ga)-D+2\I(e).
\ee
If $e$ is a bridge, i.e. if the unhooking disconnects the component of $e$ into two components $\Gae=\Gae^1\sqcup\Gae^2$,
\be
\Phi_0(\Gae^1)+\Phi_0(\Gae^2)=\Phi_0(\Ga) +D,
\ee
and as the number of connected components is increased by one, the degree \eqref{eqref:BubDeg} satisfies
\be
\label{eqref:ConSumDeg}
\delta_\bB(\Gae^1)+\delta_\bB(\Gae^2)=\delta_\bB(\Ga).
\ee
$h$-Pairs were defined in Definition~\ref{def:hPair}.

\begin{prop} 
\label{prop:MoreD2}
If $\B$ contains an $h$-pair with $h>D/2$, then the corresponding pair of vertices belongs to every optimal pairing of $\B$.
\end{prop}
\prf In the colored graph picture, consider some $\Om$-covering $\BCO$ in which the vertices of this pair, which we denote $\pi$, are incident to two different color-0 edges $e_1$ and $e_2$, and then switch these two color-0 edges.  Doing so, we get a $\Om'$-covering in which a color-0 edge links the two vertices of $\pi$. The variation of the 0-score is 
\be
\Phi(\B^{\Om'})-\Phi(\BCO)=D-2\I,
\ee 
where $\I$ is the number of colors $i$ for which $e_1$ and $e_2$ belong to two different bicolored cycles $0i$ in  $\BCO$. Because $\pi$ is an $h$-pair, $\I\le D-h<D/2$, and 
\be
\Phi(\B^{\Om'})>\Phi(\BCO),
\ee
so that $\BCO$ is not optimal. \qed

\subsubsection{Edge-unhooking and topology}

If the pair corresponding to an edge $e$ is an $h$-dipole (Def.~\ref{def:Dipole}), then $\I(e)=D-h$, and we know how the 0-score varies upon unhooking $e$. The converse is not true, an $h$-pair with $\I(e)=D-h$ is not necessarily a $h$-dipole.
We have the following properties:

\begin{itemize}
\item If the edge-unhooking disconnects the graph into two connected components $\G_1$ and $\G_2$ representing two pseudo-manifolds $\cM_1$ and $\cM_2$,  and if the color-0 edges belong to spheres in every $\Gi[1]$ and $\Gi[2]$ for $i\neq0$, then $\G$ represents the (topologically unique) connected sum $\cM_1\#\cM_2$ (Prop.~\ref{prop:DirSum2}). The condition is always satisfied for PL-manifolds. If $\G_2$ represents  a sphere, then $\G$ represents $\cM_1$.
\item If the pair of $\Om$ corresponding to the edge $e$ is a proper $h$-dipole with $h\ge 1$, then the edge-unhooking is  a flip  in the colored graph picture (Def.~\ref{def:Flip}),  and the topology is unchanged (Prop.~\ref{prop:FlipBlop}).
\end{itemize}

An application of the first property is the following corollary.
\begin{coroll} (of Prop.~\ref{prop:DirSum2})
\label{coroll:TreeTopo}
If $\BCO$ represents a PL-manifold $\cB^\Om$, then the triangulation $\C$ 
represented by a tree $\cT$ of $\bS(\B,\Om)$ with $b(\cC)$ bubbles is homeomorphic
 to the direct sum of $b(\cC)$ copies of $\cB^\Om$
\be
\C\cong_{PL}\#_{b(\cC)} \cB^\Om.
\ee
\end{coroll}

If a pair of vertices is a combinatorial handle in $\G$ (Def.~\ref{def:CombHandle}) belonging to the pairing $\Om$, then the corresponding white square vertex in $\Ps_0(\G,\Om)$ is incident to $D-1$ colored leafs $\lDr\setminus\{i,j\}$, and a single face goes around the two other incident color $i$ and $j$ edges  in $\Ga^{ij}$, and conversely. In the case where $j=0$, then in $\Ps(\B,\Om)$, the white square is incident only to colored leaves, apart for the color $i$. Remark that from Prop.~\ref{prop:MoreD2}, for $D\ge 3$ such a pair belongs to every optimal pairing of $\B$.
Consider an edge $e$ of a stacked map of $\bS(\B,\Om)$, which is incident to a white square, itself incident to $D-1$ leaves. If $\I(e)=1$, then the corresponding pair is a proper $h$-dipole, and unhooking $e$, the topology is unchanged.  If in the contrary $\I(e)=0$, then, unhooking $e$,
\begin{itemize}
\item if $\Gae$ is connected and  represents a PL-manifold $N$, then $\G$ represents $N\#(\cS^{D-1}\times\cS^1)$,
\item If $\Gae$ is not connected, $\G$ represents the connected sum of the two components of $\Gae$.
\end{itemize}
Indeed, as we focus on the orientable case, the handle $\bH$ in Thm.~\ref{thm:CombHandle} is the orientable bundle over $\cS^1$.

\

As an application of these results, we can specify the {\bf topology of the contributions to the first non-vanishing orders in the quartic melonic model} which are  detailed in Corollary~\ref{coroll:CorollMajTree}. Indeed:
\begin{itemize}[label=-]
\item Maximal maps are trees of cyclic bubbles, which represent $D$-spheres $\cS^D$, and a connected sum of spheres is a sphere (it confirms something we already new, as the corresponding colored graphs are melonic). 

\item Maps of degree $D-2$ have a single monochromatic cycle. Any edge has $\I(e)=1$, and therefore unhooking it does not change the topology. Indeed, the corresponding pair is a proper $(D-1)$-dipole, once we have contracted all the edges corresponding to tree contributions, which again, does not change the topology. As it leads to a tree,  contributions of degree $D-2$ represent $D$-spheres. With the same reasoning, maps with only monochromatic cycles and planar monochromatic submaps represent $D$-spheres. 

\item In $D>4$, maps with one polychromatic cycle are such that $\I(e)=0$, and the pairs corresponding to half-edges are $(D-1)$-pairs once the tree contributions are contracted. These pairs are therefore combinatorial handles. Unhooking any edge, we get a sphere. As we are in the orientable case, degree $D$ contributions in $D>4$ represent $\cS^{D-1}\times\cS^1$.
\end{itemize}

\subsubsection{Vertex splitting/merging}
\begin{figure}[!h]
\centering
\hspace{1cm}\includegraphics[scale=0.7]{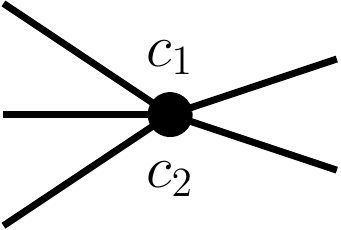} \hspace{1cm}\raisebox{4ex}{\begin{tabular}{@{}c@{}} splitting\\$\rightleftarrows$\\ merging\end{tabular} }\hspace{1cm} \includegraphics[scale=0.7]{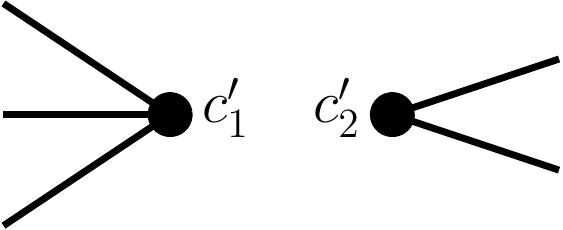} 
\caption{Vertex splitting and merging.}
\label{fig:VertSplit}
\end{figure}

The edge-unhooking is a particular case of a $\rho$-pair switching (Def.~\ref{def:RhoHSwitch}) where the two color-0 edges are incident to the same element of the pairing $\Om$. More generally, in the stacked map picture, a $\rho$-pair switching corresponds to picking two corners and:
\begin{itemize}[label = -]
\item If they are on the same vertex, splitting the black vertex into two black vertices along these two corners as shown in Figure~\ref{fig:VertSplit}

\item  If they are on two different black vertices, merging the two vertices along the two corners.
\end{itemize}
\noindent  Note that this is done without ambiguity, respecting the fact that corners oriented counterclockwise correspond to color-0 edges going from white to black vertices in the colored graph picture (Thm.~\ref{thm:BijSM}).
The edge unhooking/hooking corresponds to the case where the corners $c_1$ and $c_2$ are incident to the same edge. 
The variation of the 0-score is the same as when unhooking an edge: if we denote $\Ga_{c_1,c_2}$ the map obtained from $\Ga$ by splitting the two corners $c_1$ and $c_2$, then,
\be
\Phi_0(\Ga_{c_1,c_2})=\Phi_0(\Ga) + D-2\I(c_1,c_2),
\ee
 where we have denoted $\I(c_1,c_2)$ the number of colors $i$ such that the corners $c_1$ and $c_2$ belong to two different faces in $\Gai$. The topological properties are also as for the edge-unhooking. In particular, if the splitting raises the number of connected components and these connected components represent PL-manifolds,
 then $\Ga$ is a connected sum of the two connected components, and if one of the new corners defines a proper $h$-dipole, then the topology is unchanged. 
In general, we only need to unhook/hook edges, as any two maps differ one from another by a finite sequence of edge unhooking/hookings. However, vertex splitting/merging turn out useful for topology proofs, and in the case of tree-like families (Subsection~\ref{subsec:TreeLike}).

\subsection{Score of trees and the coefficient $a_\B$}
\label{subsec:TreeDeg}

\begin{prop}[Score of trees]
\label{prop:TreeDeg}
Consider a stacked tree $\cT\in\bS(\B,\Om)$ with $b(\cT)$ bubbles.  Each of its color $0i$ subtree satisfies
\be
\label{eqref:TreeDegI}
F(\cT^{(i)}) = 1+ b(\cT)\bigl(K(\BOM^{(i)}) - 1\bigr),
\ee
where $K(\BOM^{(i)})$ is the number of disjoint cycles of color $i$ in $\BOM$.
The 0-score of $\cT$ is
\be
\label{eqref:TreeDeg}
\Phi_0(\cT) = D+ b(\cT)\bigl(\Phi_0(B^\Om) - D\bigr).
\ee
\end{prop}
\prf  We prove this recursively on the number of bubbles $b$. There is only one unicellular tree, and it has  
$\Phi_{0,i}(B^\Om)=K(\BOM^{(i)})$ faces of color $i$. Now consider a stacked tree $\cT$ with $b>1$. It necessarily has an edge $e$ between a bubble and a vertex of valency greater than one. Unhooking $e$, we get two trees $\cT_1$ (resp. $\cT_2$) with $b_1$ (resp. $b_2$) bubbles, such that $b_1+b_2=b$. The same face goes around both sides of $e$ in $\cT^{(i)}$, and it is duplicated after the unhooking. Therefore, 
\bea
F(\cT^{(i)})&=&-1+F(\cT^{(i)}_1)+F(\cT^{(i)}_2)\\
&=&1 + (b_1+b_2)(K(\BOM^{(i)}) - D),
\eea
which proves (\ref{eqref:TreeDegI}). Equation (\ref{eqref:TreeDeg}) follows from summing over all colors.\qed

\

Consequently, the bubble-dependent degree (Def.~\ref{def:BubDepDeg}) of a tree is 
\be
\label{eqref:Phi0TreesTildeA}
\delta_\B(\cT)= b(\cT)(\tilde a_\B  -\Phi_0(B^\Om) + D).
\ee
To satisfy condition~(\ref{eqref:Cond1}), $\tilde a_\B$ therefore has to be chosen such that 
\be
\label{eqref:Cond1TildeA}
\tilde a_\B \ge \Phi_0(B^\Om) - D,
\ee
as otherwise the family of stacked-trees has un-bounded negative degree. In particular, the $1/N$ expansion (\ref{eqref:NExp}) would not be defined. This is true for any pairing 
$\Om$ of $\B$, so that we must choose
\be
\tilde a_\B \ge \PhiM - D,
\ee
where $\Opt$ is  an optimal pairing (Def.~\ref{def:OptPair}). \emph{In practice we will always build the bijection of Thm.~\ref{thm:BijSM} with an optimal pairing.} Doing so, the degree of stacked trees is precisely 
\be
\label{eqref:DegTildeA}
\delta_\B(\cT)= b(\cT)(\tilde a_\B  -\PhiM + D).
\ee
 Choosing $\tilde a_\B = \PhiM - D$, all the trees have vanishing degree. We have just proven the following result
\begin{coroll}
\label{coroll:TreeMaxAB}
If, choosing an optimal pairing to build the bijection, stacked trees are maximal 
(Def.~\ref{def:Max}), 
then the smallest possible choice for $\tilde a_\B$ is 
\be
\label{eqref:TildeABTree}
\tilde a_\B = \PhiM - D,
\ee
This choice leads to a well-defined (\ref{eqref:Cond1}) and non-trivial (\ref{eqref:Cond2}) bubble-dependent degree 
\be 
\delta_\B(\Ga)= D+ b(\Ga)(\PhiM - D) - \Phi_0(\Ga).
\ee
\end{coroll}

\

The only bipartite examples we know which do not satisfy this condition are in dimension 6 (see Section~\ref{sec:K336}). For all other known bipartite cases, this is the most appropriate definition of the bubble dependent degree. We can reformulate this degree in terms of $a_\B$ (\ref{eqref:Tildeaa})
\be
a_\B = \frac{\PhiM +\Phi(\B)- D} {V(\B)},
\ee
which we rewrite in terms of the score of $\BCO$ (not only counting bicolored cycle $0i$)
\be
\label{eqref:TreeMaxAB2}
a_\B = \frac{\PhijM - D} {V(\B)},
\ee
the bubble-dependent degree therefore writes 
	\be
	\label{eqref:BubDegTr}
	\delta_\B(G)=D+ \bigl(\PhijM - D\bigr)\frac{V(G)}{V(\B)} - \Phi(G).
	\ee
The corresponding scaling for the enhanced random tensor model is the following.
\begin{coroll}
\label{coroll:scaling}
The strongest scaling one can choose to have a defined $1/N$ expansion for enhanced tensor models associated to $\B$ such that trees belong to maximal maps in $\bS(\B,\Opt)$  is 
\be
s_\B=1+\frac{V(\B)} 2 (D-1)-\PhiM.
\ee
From Lemma.~\ref{lemma:PhiLm}, the scaling writes in terms of the number of polychromatic cycles of any optimal covering $\LmM$
\be
\label{eqref:ScalingCircuit}
s_\B=\LmM \ge 0,
\ee
with equality iff the bubble is melonic.
\end{coroll}

\prf We just need to prove the last statement, which is trivial as $\Phi_0(\BCO)\le 1+\frac{V(\B)} 2 (D-1)$ with equality iff the bubble is melonic. \qed

\

For $\bB$-restricted gluings, the 0-score of a stacked tree $\cT\in\bS(\bB,\Om_\bB)$ with $n_\B(\cT)$ bubbles $\Ps(\B,\Om_\B)$ is
\be
\label{eqref:TreeDegbB}
\Phi_0(\cT) = D+ \sum_{\B\in\bB}n_\B(\cT)(\Phi_0(B^\Om) - D).
\ee

If stacked trees belong to maximal maps in $\bS(\bB,\Om_\bB)$, we obtain the same coefficient $\tilde a_\B = \tilde a_\B^\bB$ for all $\B\in\bB$.

\subsection{Maps of positive degree and the choice of $a_\B$ }
\label{subsec:FiniteNumContr}
	
	
	In the previous subsection, we showed that when trees were maximal, there was a lower bound on the coefficient $\tilde a$ for it to satisfy Condition~\ref{eqref:Cond1}, i.e. for the $1/N$ expansion  \eqref{eqref:NExp} to be well-defined. We showed that the choice 
\be
\tilde a_\B = \PhiM - D,
\ee
also satisfied Condition~\ref{eqref:Cond2}, and therefore the bubble-dependent degree thus defined led to a non-trivial $1/N$ expansion. In this subsection, we show that this value is an upper bound on $\tilde a$ for Condition~\ref{eqref:Cond2} to be satisfied.
A consequence of \eqref{eqref:DegTildeA} is that when choosing 
\be
\tilde a_\B > \PhiM - D,
\ee
there are only finitely many trees per order (Def.~\ref{def:Order}). In Theorem~\ref{thm:FinitNumbPosDeg}, we prove that when trees belong to maximal maps, such a choice in fact leads to a finite number of contributions per order. 
	
	\begin{theorem}
	\label{thm:FinitNumbPosDeg}
	 If trees belong to maximal maps in $\bS(\B,\Opt)$ 
	 and if the degree is built choosing the coefficient $\tilde a_\B\in\bR^+$ such that
	\be
	\label{eqref:TildeASmall}
	\tilde a_\B > \PhiM - D,
	\ee
	then  there are finitely many contributions per order $\delta_B^{-1}(k)$, $k\in\bR^+$.
	\end{theorem}
	\prf Because the degree of a tree $\cT$ with $b$ bubbles is \eqref{eqref:Phi0TreesTildeA}
	\be
	\label{eqref:ScoreTreesFinite}
	\delta_\B(\cT)= b(\cT)\bigl(\tilde a_\B  -\Phi_0(B^\Om) + D\bigr),
	\ee
 	if $\tilde a_\B$ is as in \eqref{eqref:TildeASmall}, then there are a finite number of trees contributing to every order.  If $\cT$ and $\cT'$ are two trees, we have from \eqref{eqref:ScoreTreesFinite}
	\be
	\delta_B(\cT)<\delta_\B(\cT') \qquad \Rightarrow \qquad b(\cT)<b(\cT').
	\ee	
Furthermore,  if $\cT_\Ga$ and $\Ga$ have the same number of bubbles, as trees are maximal, 
\be
\label{eqref:BoundingTrees}
\delta_\B(\cT_\Ga)\le\delta_\B(\Ga)<\delta_\B(\cT') \quad \Rightarrow  \qquad \ b(\Ga)<b(\cT').
\ee	
Consider $k\in\bR^+$ and $\Ga$ contributing at order $k$. From \eqref{eqref:ScoreTreesFinite}, the degree of trees can be arbitrarily large, so that there exists a tree $\cT_k$ such that $k<\delta_\B(\cT_k)$, and from \eqref{eqref:BoundingTrees}, $b(\Ga)< b(\cT_k)$. Therefore, 
\be
\text{Card}[\delta_B^{-1}(k)] \le \text{Card}\{ \Ga\in\bS(B) \mid b(\Ga)<b(\cT_k)\}.
\ee
But because the number of edges of a map is proportional to the number of bubbles
\be
E(\Ga)=n b(\Ga),
\ee
there is a  finite number of maps with less or as much bubbles as $\cT_k$, which concludes the proof. \qed

\begin{coroll}
	\label{coroll:FinitNumbPosDeg}
	 If trees belong to maximal maps of $\bS(\B,\Opt)$, then the only value of $\tilde a$ for which the $1/N$ expansion \eqref{eqref:NExp} is well defined (Condition~\eqref{eqref:Cond1}) and not trivial (Condition~\eqref{eqref:Cond2}) is that of \eqref{eqref:TildeABTree}.
It corresponds to the bubble-dependent degree of \eqref{eqref:BubDegTr}, to the scaling \eqref{eqref:BubDegTr}, and the correction to Gurau's degree is \eqref{eqref:TreeMaxAB2}.
	\end{coroll}
	%


\section{Maximal 
maps}
\label{sec:MaxMaps}

There is a great diversity of bubbles of any size in any dimension. We seek to  find their  coefficients $a_\B$, to count maximal maps, and to deduce their critical behavior near the dominant singularity. Because our aim would be to find new continuum limits, which should be characterized by different critical exponents,
we would like to understand which properties of bubbles trivially lead to known  critical behaviors, in order  to focus on bubbles which might have more interesting properties.
Because of the results of the previous section,  one of the first questions  is whether trees belong to maximal maps. When this is the case, we can compute $a_\B$ using (\ref{eqref:TildeABTree}). Proposition~\ref{prop:PhivsTrees} compares the score of maps and trees, and is very helpful in simple cases.
Most of the other results in this section will allow us to restrict the set of bubbles which might lead to interesting maximal maps. Subsection~\ref{subsec:TreeLike} addresses the cases where trees do not belong to maximal maps when choosing an optimal pairing. We treat a number of examples in Section~\ref{sec:Examples}. Some of them are simple applications of the results in this section, and some more involved examples require a careful study. In the present section, we often refer to examples of Section~\ref{sec:Examples}, in order to illustrate our results. First of all, there are simple but yet very useful consequences of Prop.~\ref{prop:FaceUnhook}.

\begin{prop}
\label{prop:eD2}
Every edge $e$ of a maximal map of $\bS(\bB,\Om_\bB)$ is either a bridge, either satisfies $\I(e)\ge \frac D 2$.
\end{prop}
\prf Otherwise, unhooking $e$ which is not a bridge and satisfies $\I(e) < \frac D 2$ leads to a connected map, which from Prop.~\ref{prop:FaceUnhook} has a higher 0-score. \qed

\begin{prop}
\label{prop:hD2Trees}
If every pair (but one) in the pairings $\Om_\B$ of the bubbles $\B\in\bB$ is an $h$-pair with $h\ge D/2$, then
trees belong to maximal maps.
\end{prop}
\prf In the submaps $\Ps(\B,\Om_\B)$ corresponding to the bubbles, the white square vertices are incident to at least D/2 colored leaves. If an edge is not a bridge, unhooking it does not raise the number of connected components, and does not increases the 0-score. At best, the 0-score remains unchanged, but this requires $h=D/2$ and $\I(e)=D/2$. There is at most one edge incident to each bubble that corresponds to an $h$-pair with $h<D/2$, its other endpoint being a black vertex. The subset of such edges is therefore a collection of star-maps. We can complete this forest  into a tree spanning the map. Unhooking the $L(\Ga^\star)$ edges defining independent cycles, which all correspond to $h$-pairs with $h\ge D/2$,  one obtains a tree with higher (or equal) 0-score and with the same number of bubbles. \qed

\subsection{Large $h$-pairs}
\label{subsec:hPair}

We recall that an $h$-pair, defined in  \ref{def:hPair},  is a pair of vertices linked by $h$ parallel edges. By large $h$-pair, we mean that $h > D/2$. The importance of the results in this subsection is that they are local, and consequently they do not depend on the set $\bB$. 
%
%
From Prop.~\ref{prop:MoreD2}, we know that an $h$-pair $\pi$ with $h>D/2$ parallel edges belongs to every optimal pairing, and Prop.~\ref{prop:eD2} implies that when building the bijection with any optimal pairing, edges corresponding to the pair $\pi\in\Om(\B)$ are bridges in every maximal map, regardless of the details of the rest of the map. This is a local property. 
%

%
\begin{coroll}
\label{coroll:MoreD2Trees}
An edge which corresponds to an $h$-pair of $\B\in\bB$ with $h>D/2$ is a bridge in every maximal map of $\bS(\bB,\Om_\bB)$.
If this is the case for every pair but one in the pairing $\Om_\B$ of the bubble $\B$, then every edge incident to a submap $\Ps(\B,\Om_\B)$ is a bridge in every maximal map.
\end{coroll}
\prf The first statement is because such an edge necessarily satisfies $\I(e)>D/2$. The second statement is proved exactly as in the proof of Prop.~\ref{prop:hD2Trees}, with the difference that the edges not in the spanning tree all have $\I(e)>D/2$, so that the 0-score strictly increases at each edge-unhooking. \qed

\

This applies to  gluings of $k$-cyclic bubbles with $k<\frac D 2$. It is stronger than proving that maximal such gluings are in bijection with bipartite trees: it means that the corresponding bubbles behave the same way, whatever the other bubbles they are glued to. Cut-bubbles were defined in Def.~\ref{def:CutBub}.
\begin{definition} We say that $(\B,\Om_\B)$ satisfies the cut-bubble property if
\be
\label{CutBubbleProp}
\forall\bB\supset\B, \forall \Om_{\bB\setminus\B}, \quad \Ga\textrm{ maximal in }\bS(\bB,\Om_\bB) \ \Rightarrow\  \Ps(\B,\Om_\B)\textrm{ are cut-bubbles in }\Ga.
\ee
\end{definition}

Being a bridge in every maximal map means, in the colored graph picture, that in every maximal map, either a color-0 edge links the two vertices of the corresponding pair, or they are incident to two color-0 edges which form a 2-bond (deleting them separates the graph). 
\be
\label{fig:hPair}
\begin{array}{c}\includegraphics[scale=0.6]{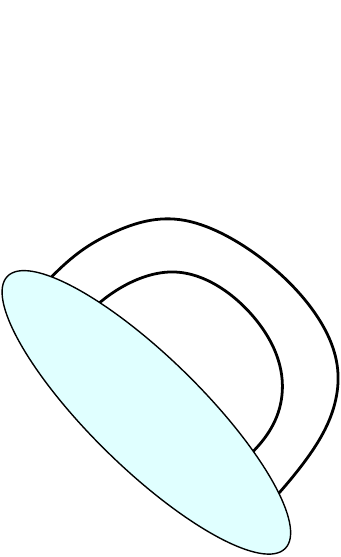}\end{array}\qquad\rightarrow\qquad
\begin{array}{c}\includegraphics[scale=0.6]{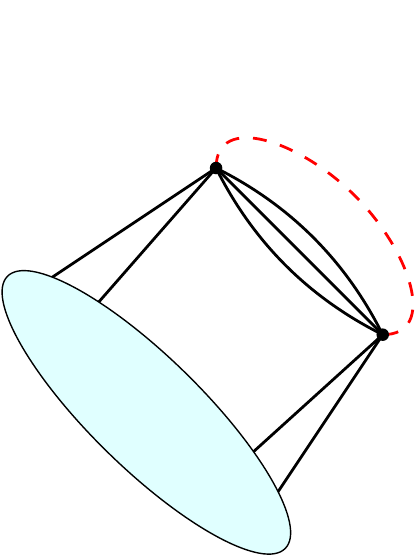}\end{array}\qquad\text{or}\qquad
\begin{array}{c}\includegraphics[scale=0.6]{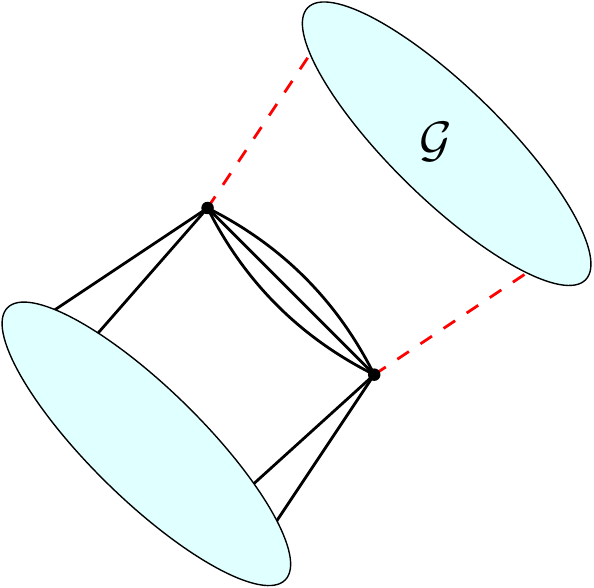}\end{array}
\ee
If $\pi$ is an $h$-pair of $\B$ with $h>D/2$, we know how the corresponding pair will be in every maximal map. Contracting this pair as pictured in Figure~\ref{fig:PairIns}, we obtain a smaller bubble $\Bpi$.  To recover the maximal graphs in $\bG(\B)$ from the maximal graphs in $\bG(\Bpi)$, one inserts the pair $\pi$ in every bubble and adds color-0 edges so that each pair $\pi$ either has an
 edge of color 0 between its two vertices, either the two incident color-0 edges form a 2-cut leading to a maximal map of $\bG^1(\B)$ (a leading order 2-point function, in the physics vocabulary). This is illustrated in (\ref{fig:hPair}), in which $\GF$ is the generating function of maximal maps in $\bG^1(\B)$ with one marked color-0 edge. 
 
 \

In the stacked map picture, contracting the pair corresponds to deleting the corresponding edge, white square, and the colored edges incident to it inside $\Ps(\B,\Om_\B)$. The contraction induces a pairing $\Om_{\Bpi}=\Om\setminus \pi$ of $\Bpi$. 
To recover maximal maps in $\bS(\B,\Om_\B)$, we can study maximal maps in $\bS(\Bpi,\Om_{\Bpi})$, and then replace $\Ps(\Bpi,\Om_{\Bpi})$ with $\Ps(\B,\Om_\B)$ and add bridges between the newly added white squares, and black vertices in various combinatorial maps of $\bS(\Bpi,\Om_{\Bpi})$. This is better illustrated in the  simple example of the first section of Sec.~\ref{subsec:K334}. 
We might be interested in the optimality of the corresponding pairing, for instance if maximal maps are trees. Because a large $h$-pair is in every optimal pairing, we know that if $\Om_{\Bpi}$ is optimal for $\Bpi$, then $\Om_\B$ is optimal for $\B$. The converse is also true:
\begin{lemma}
\label{lemma:LargeHPairPairing}
With these notations, an optimal pairing $\Om_\B$ induces an optimal pairing $\Om_{\Bpi}$ of $\Bpi$.
\end{lemma}
\prf Denote $e_1,\cdots,e_{D-h}$ the edges of $\Bpi$ obtained by contracting $\pi$. Given any covering $\Bpi^{\Om_{\bar \pi}}$ of $\Bpi$, we can insert back the pair $\pi$ on $e_1,\cdots,e_{D-h}$ directly on the covering, obtaining a covering $\BCO$ of $\B$ with an edge of color 0 between the vertices of $\pi$ for which 
\be
\label{eqref:PhiBpi}
\Phi_0(\BCO)=\Phi_0(\Bpi^{\Om_{\bar \pi}}) + h.
\ee
Comparing $\Opt[,\bar \pi]$ with some other pairing $\Om_{\bar \pi}$:
\be
\Phi_0(\Bpi^{\Opt[,\bar \pi]})-\Phi_0(\Bpi^{\Om_{\bar \pi}}) = \Phi_0(\BCO[opt])-\Phi_0(\BCO) \ge 0.     \qed
\ee

\

If trees belong to maximal maps of $\bS(\Bpi,\Om_{\Bpi})$, we can deduce the coefficients  $\tilde a_\Bpi$, $a_\Bpi$ and $s_\Bpi$ from those for $\B$. From Corollary~\ref{coroll:TreeMaxAB} and relation \eqref{eqref:PhiBpi}, 
\be
\label{eqref:TildeABpi}
\tilde a_\B= \tilde a_\Bpi + h.
\ee
Using (\ref{eqref:TreeMaxAB2}) and 
\be
\Phi(\B)=\Phi(\Bpi) + \frac{h(h-1)} 2, 
\ee
we see that 
\be
\label{eqref:ABpi}
 a_\B=a_\Bpi\bigl(1-\frac 2{V(\B)}\bigr) + \frac{h(h+1)} {2V(\B)},
\ee
and from  \ref{coroll:scaling}, 
\be
\label{eqref:SBpi}
s_\B=s_\Bpi + D-1-h.
\ee

\

\noindent We can give a stronger version of Corollary~\ref{coroll:MoreD2Trees}

\begin{coroll}
Consider $\B\in\bB$ and a pairing $\Om_\B$, $\tilde\B$ obtained from $\B$ by contracting a certain number of pairs in a given order, and $\Om_{\tilde\B}$ the pairing induced by $\Om_\B$. 
\begin{itemize}
\item If $\pi$ was an $h$-pair in $\Om_\B$ with $h\le D/2$ but is now an $h$-pair in $\tilde\Om_\B$ with $h> D/2$, then the corresponding edge is a bridge in every maximal map of $\bS(\bB,\Om_\bB)$.
\item If $\tilde\B$ is the elementary melon (Fig.~\ref{fig:ElMel}), then $(\B,\Om_\B)$ satisfies the cut-bubble property: every edge incident to a submap $\Ps(\B,\Om_\B)$ is a bridge in every maximal map of $\bS(\bB,\Om_\bB)$.
\end{itemize}
\end{coroll}

\begin{figure}[!h]
\centering
\includegraphics[scale=0.6]{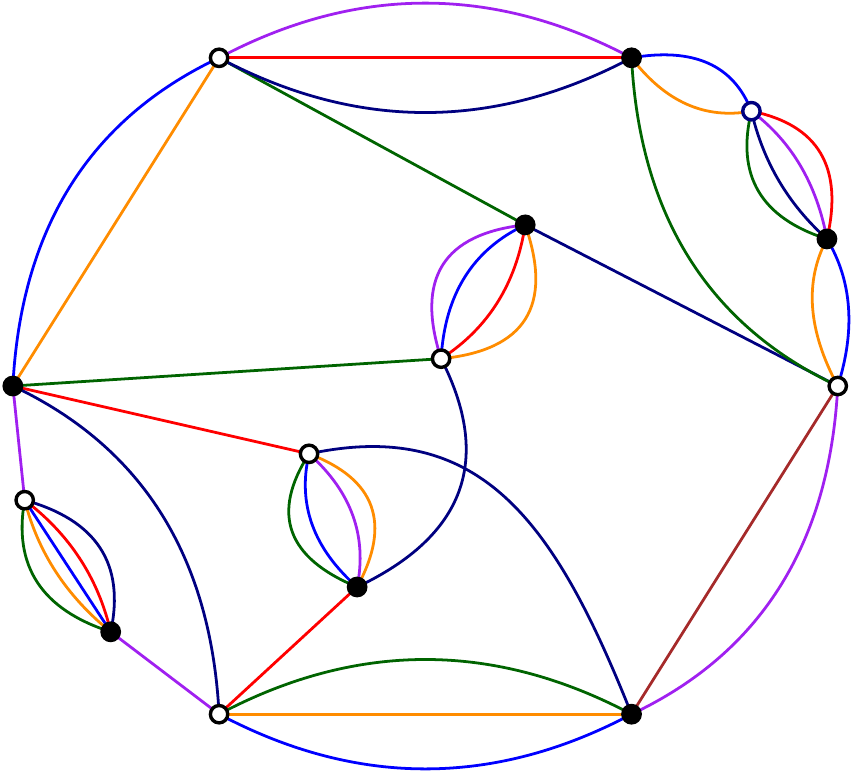}\hspace{2cm}\includegraphics[scale=0.6]{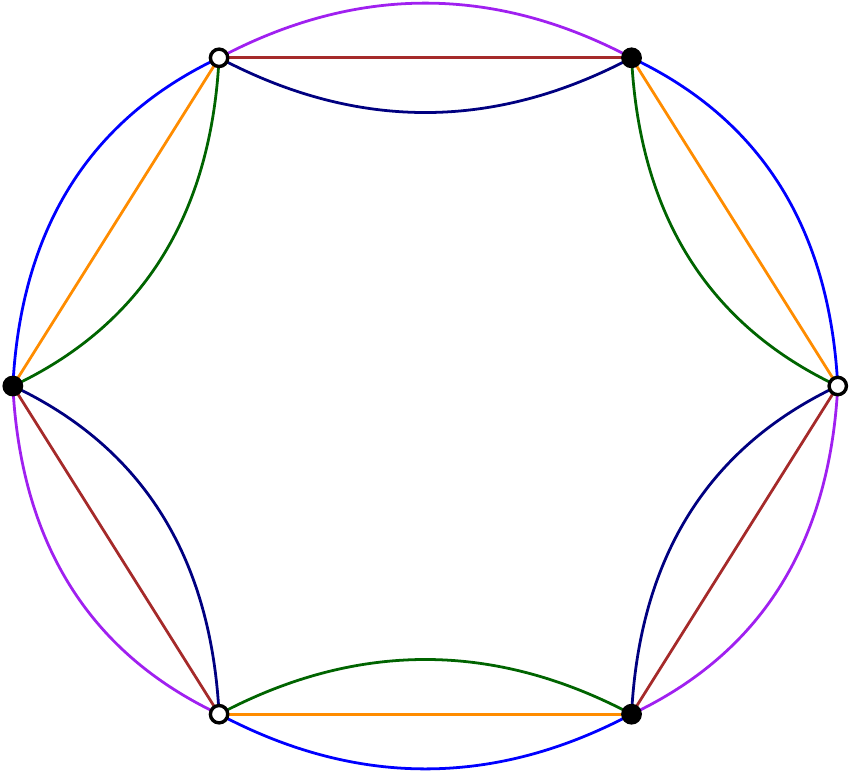}\caption{A bubble $\B$ and simpler bubble $\tilde\B$ obtained by contracting large $h$-pair.}
\label{fig:LargehPair}
\end{figure}

This includes all the melonic graphs, but also less trivial example, such as the one in Fig.~\ref{fig:LargehPair}. On the left of Fig.~\ref{fig:LargehPair} is shown an example of bubble $\B$ that contains large $h$-pairs, and on the right is a bubble $\tilde\B$ obtained from $\B$ by contracting all the large $h$-pairs. It is enough to study maximal maps in $\bS(\tilde \B,\Om_{\tilde\B})$ to understand those in $\bS(\B,\Om_\B)$. We can deduce the coefficients $\tilde a_\B$, $a_\B$ and $s_\B$ from \eqref{eqref:TildeABpi} \eqref{eqref:ABpi} and \eqref{eqref:SBpi}, knowing that for a 3-cyclic bubble in dimension $D=6$, we have \eqref{eqref:CoeffsD2Cyc}: $\tilde a_{3,3}=6,\ a_{3,3}=11/2,\ {\rm and}\ s_{3,3}= 4$. We obtain 
\be
\tilde a_{\B}=23
\quad{\rm and}\quad s_{ \B}= 7.
\ee
A simple but similar example is treated in the first section of Sec.~\ref{subsec:K334}. \emph{We restrict our study to bubbles $\B\in\bG_{D-1}$ which only have $h$-pairs with $h\le \frac D 2$}.

\subsection{Comparing maps and trees}

Proposition~\ref{prop:PowVacQuart} generalizes for stacked maps with no marked corner. It compares the 0-scores of stacked maps and of trees that have the same number of bubbles (with no other restriction). 
\begin{prop}
\label{prop:PhivsTrees}
Consider a stacked map $\Ga\in\bS(\bB,\Om_\bB)$ and a stacked tree $\cT\in\bS(\bB,\Om_\bB)$ with the same number of bubbles (both connected). Then,
\be
F(\cT^{(i)}) -F(\Gai)= L(\Ga^\star) -2L(\Gai)+ 2 g(\Gai),
\ee
and summing over the colors $i$, 
\be
\label{eqref:PhivsTrees}
\Phi_0(\cT) -\Phi_0(\Ga)= DL(\Ga^\star) -2\sum_{i=1}^D L(\Gai)+ 2\sum_{i=1}^D g(\Gai).
\ee
\end{prop}

\

\prf  As in the proof of Prop.~\ref{prop:PowVacQuart}, we write the number of faces of color $0i$ submaps as  
 \be
F(\Gai)=2L(\Gai)-2g(\Gai) + V(\Gai) -E(\Gai).
 \ee
The vertices of $\Gai$ are all the black vertices ($V_\bullet(\Ga)$), the white square vertices, and the color $i$ squares:
\be
V_\diamond(\Ga)=\sum_{\B\in\bB} V(\B)/2 \times n_\B(\Ga), \qquad\text{and}\qquad V_i(\Ga)=\sum_{\B\in\bB} \Phi_{0,i}(\BCO)\times n_\B(\Ga). 
\ee
The edges of $\Gai$ are all the edges incident to black vertices ($E_\bullet(\Ga)$), and all the edges incident to color $i$ square vertices, both equal to $V_\diamond$. Therefore, 
\be
V(\Gai) -E(\Gai)= V_\bullet(\Ga) -E_\bullet(\Ga) +\sum_{\B\in\bB} \Phi_{0,i}(\BCO)n_\bB(\Ga).
\ee
Inserting the expression of $L(\Ga^\star)$,
\be
L(\Ga^\star)=E_\bullet(\Ga)-V_\bullet(\Ga)- b(\Ga) + 1,
\ee
leads to 
\be
V(\Gai) -E(\Gai)=-L(\Ga^\star)+ 1+\sum_{\B\in\bB} n_\B(\Ga)(\Phi_{0,i}(\BCO)-1),
\ee
and (\ref{eqref:TreeDegbB}) concludes the proof.\qed

\

The degree of a map in $\bS_D(\B)$ is therefore
\be
\delta_\B(\Ga)=b(\Ga)\bigl(\tilde a_\B-\Phi_0(\B^\Om)+D\bigr) + DL(\Ga^\star) -2\sum_{i=1}^D L(\Gai)+ 2\sum_{i=1}^D g(\Gai).
\ee
In particular, \emph{if trees belong to maximal maps in $\bS_D(\B)$, from \eqref{eqref:TildeABTree}, the degree of a map is}
\be
\label{eqref:DegGenTree}
\delta_\B(\Ga)= DL(\Ga^\star) -2\sum_{i=1}^D L(\Gai)+ 2\sum_{i=1}^D g(\Gai).
\ee
Proposition~\ref{prop:PhivsTrees} also holds when using the simpler bijection of Thm.~\ref{thm:BijSimp}. In that case,  $L(\Ga^\star)=L(\Ga)$, and trees are so in the usual sense. 
As the right hand side of \eqref{eqref:PhivsTrees}  is a sum of positive or vanishing terms (which vanishes for trees) it provides a  sufficient condition for trees to be maximal.

\begin{coroll}
\label{Coroll:LstarLiTree}
If for any $\Ga\in\bS(\bB,\Om_\bB)$, the following inequality is satisfied
\be
\sum_{i=1}^D L(\Gai)\le \frac D 2 L(\Ga^\star),
\ee
then stacked trees are maximal and other maximal maps are such that
\be
\label{eqref:CharMax}
\sum_{i=1}^D L(\Gai) = \frac D 2 L(\Ga^\star)\quad\text{ and }\quad \forall i\in\lDr,\ g(\Gai)=0.
\ee
\end{coroll}

The opposite is not always true. Indeed, this quantity may be negative for maximal maps with bicolored submaps of positive genus. 
It is sometimes easy to determine if trees are maximal, e.g. using Prop.~\ref{prop:hD2Trees}. This allows us to compute the coefficient $\tilde a_\B$ using Corollary~\ref{coroll:TreeMaxAB}. The next step is to characterize the full set of maximal maps. The following corollary provides a characterization.

\begin{coroll}
\label{coroll:SuffCharMax}
If stacked trees are maximal, and for every map $\Ga\in\bS(\B,\Om)$, there exists another map $\Ga'\in\bS(\B,\Om)$ such that their projected maps and every bicolored submaps have the same circuit-ranks, and $\Ga'$ has planar bicolored submaps, 
$\forall i\in\lDr,\ g(\Ga'^{(i)})=0$, then 
\be
\forall \Ga\in\bS(\B,\Om),\ \sum_{i=1}^D L(\Gai)\le \frac D 2 L(\Ga^\star),
\ee
and other maximal maps are such that 
\be
\sum_{i=1}^D L(\Gai) = \frac D 2 L(\Ga^\star)\quad\text{ and }\quad \forall i\in\lDr,\ g(\Gai)=0.
\ee
\end{coroll}

\


The example of $D/2$-cyclic bubbles of different kinds is treated in Section~\ref{Subsec:D2CycBub}, and illustrates the use of the results stated in this section. The example of  Sec.~\ref{subsec:K334_2}, \emph{a size 6 bubble in} $D=4$ is also a rather simple application.  We recall that $\B^{\hat i}$ is obtained from $\B$ by deleting every edge of color $i$. An example of the following result is illustrated in the first section of Sec.~\ref{subsec:K334}.


\begin{prop}
\label{prop:AddColOpt}
Consider a bubble $\B\in\bG_{D-1}$.  If there exists a color $i\in\lDr$ such that $\B^{\hat i}$ is connected, if edges of color $i$ define an optimal pairing $\Om_i$ of $\B^{\hat i}$, and if trees are maximal among maps in $\bS(\B^{\hat i},\Om_i)$, then 
stacked trees are the only maximal maps in $\bS(\B,\Om_i)$. 
\end{prop}
\prf  We apply Theorem~\ref{prop:PhivsTrees}, which compares the 0-score of trees and maps: 
\bea
\Phi_0(\cT) -\Phi_0(\Ga)&=& L(\Ga^\star) -2 L(\Gai)+ 2g(\Gai)+ \Phi_0(\cT^{\hat i}) -\Phi_0(\Ga^{\hat i})\\
&\ge& L(\Ga^\star) -2 L(\Gai)+ 2g(\Gai),
\eea
as  trees are maximal among maps in $\bS(\B^{\hat i},\Om_i)$.
Furthermore, as there is an edge of color $i$ on every pair of $\Om_i$, all color $i$ square vertices in $\Ga(\B,\Om_i)$ are leaves. Consequently, $L(\Gai)=g(\Gai)=0$, and 
\be
\Phi_0(\cT) -\Phi_0(\Ga)\ge L(\Ga^\star)
\ee
which is positive if $\Ga$ is not a tree. 
\qed

\subsection{One-cycle maps}
%
\begin{lemma}
\label{Lemma:OptD2B}
Consider an optimal covering $\B^{\Opt}$ of $\B$, and two particular color-0 edges. They cannot belong to more than $D/2$ common bicolored cycles.
 \end{lemma}
\prf Then the two color-0 edges $e$, $e'$ would satisfy $\I(e,e')<D/2$. Exchanging them leads to another covering $\B^{\Om}$ with $\Phi_0(\B^{\Om})=\Phi_0(\B^{\Opt}) + D-2\I(e,e')>0$, which contradicts the optimality of $\Opt$. \qed
 
 \
 
 Translated in the stacked map picture, this lemma states that unicellular maps in $\bS(\B,\Opt)$ with one (projected) cycle have a 0-score  which is smaller or equal to that of the only unicellular tree. In fact, this property is true for stacked maps of any size:

\begin{prop}
Consider $\Om_\B$ an optimal pairing for each $\B\in\bB$, and $\Ga$ a stacked map in $\bS(\bB,\Om_\bB)$ satisfying $L(\Ga^\star)=1$. Denote $n_\B$ the number of bubbles  $\B\in\bB$. Then any tree $\cT$ whith the same numbers of bubbles $n_\B$ as $\Ga$ for each $\B\in\bB$ has a highest or equal 0-score.
 \end{prop}
\prf Consider a one-cycle map. Consider a bubble $B$ of this cycle (i.e. which is not only incident to bridges). It has precisely two edges $e$ and $e'$ which are not bridges. Consider the corresponding unicellular tree with two marked leaves,  at the extremities of $e$ and $e'$. It is almost a covering of $\B$, with the difference that the pairs $\pi_e$ and $\pi_{e'}$ corresponding to $e$ and $e'$ have no incident color-0 edge.  The boundary graph of this unicellular tree is a $k$-cyclic bubble of size 4, with $k=\Card\cI(e,e')$ colors going from $\pi_e$ to $\pi_{e'}$, $\cI(e,e')$ being the set of such colors. 
Because we have chosen an optimal pairing of $\B$, Lemma~\ref{Lemma:OptD2B} implies $k\le D/2$. As $\I(e)=\I(e')\le k$, from Prop.~\ref{prop:FaceUnhook}, unhooking $e$ leads to a tree with the same number of bubbles $n_\B$ as $\Ga$, and which has a highest or equal 0-score. All trees with the same number of bubbles $n_\B$ have the same 0-score, which concludes the proof. \qed

\

We can be more precise. The set $\I(e)$ is the same for all the edges which are not bridges, as unhooking any one of them leads to a tree with the same 0-score, and we denote it $\I$. The colors which belong to $\I$ are those for which $L(\Gai)=1$ (a combinatorial map with one cycle has two faces), i.e. those for which color $i$ is one of the colors in $\cI(e,e')$, which go between $\pi_e$ and $\pi_{e'}$ for every bubble of the cycle $\cC$. Therefore, 
\be
\I=\cap_{e,e' \in \cC}\ \cI(e,e').
\ee
The 0-score of the one-cycle map is that of the tree obtained when unhooking one of the edges if and only if $\I=D/2$, i.e. if all sets $\cI(e,e')$ have precisely the same colors. If all bubbles in $\bB$ and all pairs in their optimal pairings have $\cI(e,e')<D/2$, then one-cycle maps have a lower 0-score than trees. On the contrary, the following theorem shows that if at least two pairs in the optimal pairing of one of the bubbles has $\cI(e,e')=D/2$, then maximal maps contain a subset in bijection with planar combinatorial maps.

\begin{theorem}
Consider a bipartite bubble $\B$ and build the bijection with an optimal pairing $\Opt$. If there exists a maximal unicellular map $\Ga_1$ with $L(\Ga_1^\star)=1$, then $D$ is even and a subset of maximal stacked maps are in bijection with planar combinatorial maps.
\end{theorem}
\prf We denote $e$ and $e'$  the two edges which are not bridges in $\Ga_1$, and $\pi$ and $\pi'$ the two corresponding pairs in $\Opt$. As $\Opt$ is an optimal pairing of $\B$,  the only unicellular tree is maximal (see Subsec.~\ref{subsec:Unicell}).
Unhooking either $e$ or $e'$, we therefore get that 
\be
\label{eqref:ProofL1Plan}
0=D-2\I(e)=D-2\I(e'),
\ee
so that $D$ is even and $\I(e)=\I(e')=D/2$. We will now consider colored graphs in $\bG(\B)$ such that every pair apart from $\pi$ and $\pi'$ has a color-0 edge between its two vertices. Adding color-0 edges on the pairs of $\B$ in $\Opt$ but not on $\pi$ and $\pi'$ and then considering the boundary graph, we obtain an effective quartic bubble $\B(\pi,\pi')$. Because of \eqref{eqref:ProofL1Plan}, we know that $D/2$ colors  cross from $\pi$ to  $\pi'$, and because the bubble is bipartite, we know that the boundary graph $\B(\pi,\pi')$ is a $D/2$-cyclic quartic bubble. We know from our first study of $D/2$-cyclic bubbles in Sec.~\ref{sec:SimplerBij},  that maximal gluings of $D/2$-cyclic bubbles of size 4 are in bijection with planar combinatorial maps, as  the white vertices all have valency two. The maps in $\bS(\B, \Opt)$ for which all the pairs which do not correspond to $e$ and $e'$ are leaves are therefore in bijection with planar combinatorial maps (contracting all the leaves or adding them back is done in a unique way). \qed


%


\subsection{Connected sum}
\label{subsec:ConecSum}

In this subsection, we are interested in the case where a bubble $\B$ is obtained as the connected sum of two smaller bubbles $\B_1$ and $\B_2$, as described in Prop.~\ref{prop:DirSum}. Can we deduce properties on the maximal gluings of $\B$ knowing the properties of the maximal gluings of $\B_1$ and $\B_2$? Can we, conversely, deduce the properties of the maximal gluings of $\B_1$ and $\B_2$, knowing that of $\B$? We begin  by stating a simple but useful lemma. By contracting a color-0 edge, we mean contracting its two-endpoints as an $h$-pair (Figure~\ref{fig:Meander0}). 
\begin{figure}[h!]
\centering
\includegraphics[scale=.55]{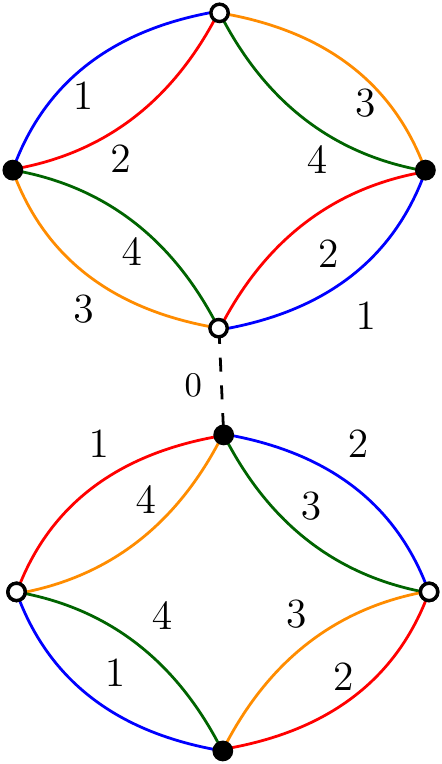}\hspace{1.5cm}\raisebox{+12ex}{$\leftrightarrow$}\hspace{1.5cm}\raisebox{+0.5ex}{\includegraphics[scale=.6]{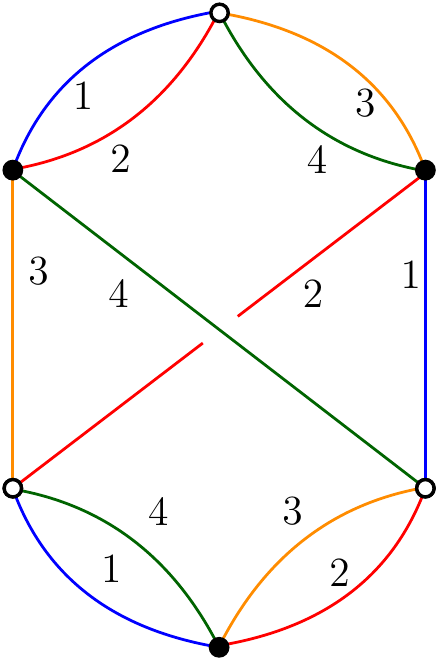} }
\caption{
Contracting a color-0 edge between two bubbles.
}
\label{fig:Meander0} 
\end{figure}
\begin{lemma}
\label{lemma:ContractScore}
Contracting a color-0 edge between two distinct bubbles in a graph does not change the 0-score. The score satisfies
\be
\label{eqref:ScoreCon}
\Phi(\B_1\#\B_2)=\Phi(\B_1)+\Phi(\B_2)-\frac{D(D-1)}2.
\ee
\end{lemma}
\prf This operation just reduces by one the length of the color $0i$ cycles passing through that edge. The only bicolored cycles which would be suppressed are those with a single color-0 edge, but there are none because the edge is between two distinct bubbles. As for the score, we see that
one bicolored cycle is suppressed for each of the $2\times\binom  D2$ bicolored cycles incident to the contracted vertices. \qed

\

The operation suppresses one vertex $v_i$ in each bubble. We call $\cE$ the set of $D$ edges, one of each color in $\lDr$, which results from the contraction of $(v_1,v_2)$.
Lemma~\ref{lemma:ContractScore} also conversely implies that inserting back a pair of vertices on $\cE$ inside each bubble $\B=\B_1\#\B_2$ of a graph in $\bG(\B)$, one obtains a graph of $\bG(\B_1,\B_2)$ with the same 0-score. 

Consider a pairing $\Om_1$ (resp. $\Om_2$) of $\B_1$ (resp. $\B_2$). We study the effect of the operation on the stacked maps. The partner of $v_i$ in $\Om_i$, denoted $v'_i$, so that $\pi_i=(v_i,v'_i)$,  is therefore left alone when contracting the pair. We define the induced pairing of $\B=\B_1\#\B_2$ as the pairing containing all the pairs of each $\Om_i$ but $\pi_i$, and the additional pair $\pi_\#=(v'_1,v'_2)$. We denote $\Om_\#$ that pairing. 

As the operation creates a pair with one vertex from each bubble, in the stacked maps it will merge two white squares, one from each $\Ps(\B_i,\Om_i)$, and each one of the colored edges and vertices incident to it. 
Indeed, for each color $i$, the contraction merges two oriented cycles (Fig.~\ref{fig:CycStar}), $(\pi_1, \pi_a, \pi_b, \cdots )$ of ${\B_1}_{/\Om_1}$  and $(\pi_2, \pi'_a, \pi'_b, \cdots )$  of ${\B_2}_{/\Om_2}$ into a single oriented cycle. If $v_1$ is a white  vertex, then $v'_1$ is black and the ordering of appearance of the pairs along the merged oriented cycle is $(\pi_\#, \pi_a, \pi_b, \cdots, \pi'_a,\pi'_b,\cdots )$. The resulting operation in the stacked map picture is illustrated in Figure~\ref{fig:ConSumStack}.

\begin{figure}[h!]
\centering
\raisebox{1.8ex}{\includegraphics[scale=.9]{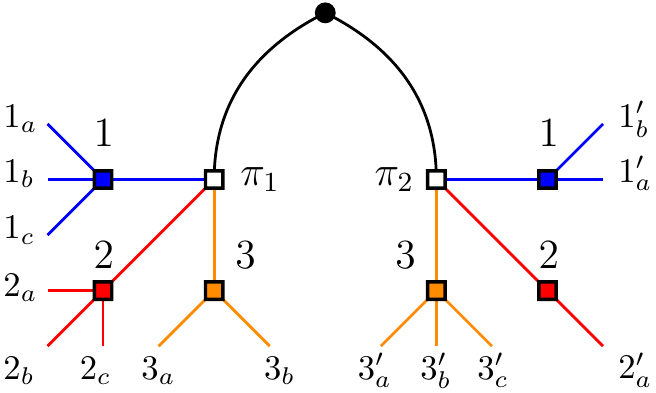}}
\hspace{1cm}
\includegraphics[scale=.9]{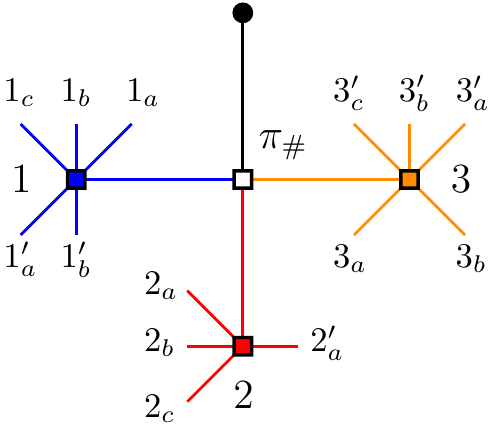} 
\caption{Local effect of the contraction on the stacked maps.}
\label{fig:ConSumStack}
\end{figure}

In a map of $\bS(\B_1,\B_2; \Om_1,\Om_2 )$, locally around each pair $\B_1$, $\B_2$,  two  black edges and two bubbles are merged, so that $L(\Ga^\star)$ remains unchanged. Likewise, two color-$i$ vertices and two color-$i$ edges are merged, so that $L(\Gai)$ remains unchanged. As, from Lemma~\ref{lemma:ContractScore}, the 0-score remains the same, we know from \eqref{eqref:PhivsTrees} comparing the 0-score of trees and maps that $g(\Gai)$ does too, as we can also verify directly on the stacked maps.

\begin{lemma} 
\label{lemma:CircConSum}
Consider a map $\Ga_\#\in\bS(\B_1\#\B_2, \Om_\# )$ and denote $\Ga$ the map of   $\bS(\B_1,\B_2; \Om_1,\Om_2 )$ obtained by inserting a pair on $\cE$ in every $\B=\B_1\#\B_2$. Then 
\be
L(\Ga)=L(\Ga_\#), \qquad L(\Gai)=L(\Gai_\#), \quad\textrm{and}\quad g(\Gai)=g(\Gai_\#).
\ee
In particular,   $\Ga$ is a tree if and only if $\Ga_\#$ is a tree. Similarly, if the simpler bijection of Thm.~\ref{thm:BijSimp} has been used, $\Ga$ is planar if and only if $\Ga_\#$ is planar. 
\end{lemma}

In a more general way,  Cor.~\ref{coroll:SuffCharMax} and 
Cor.~\ref{Coroll:LstarLiTree} in $\bS(\B_1,\B_2; \Om_1,\Om_2 )$ give close characterizations in $\bS(\B_1\#\B_2, \Om_\# )$. We denote $V_i=V(\B_i)$ and $V_\#=V(\B_1\#\B_2)$. 
For statement $(c)$ of the following proposition, we define, for $\bB_\#\supset\B_1\#\B_2$,  
\be
\bB=\bigl(\bB_\#\setminus\B_1\#\B_2\bigr)\cup\{\B_1, \B_2\}.
\ee

\begin{prop}
With the notations of Lemma~\ref{lemma:CircConSum}, we have the following properties:
\begin{enumerate}[label=(\alph*)]
\item If trees belong to maximal maps in $\bS(\B_1,\B_2; \Om_1,\Om_2 )$, then trees belong to maximal maps in $\bS(\B_1\#\B_2, \Om_\# )$, the pairing $\Om_\#$ is optimal for $\B_1\#\B_2$, 
\bea
\label{eqref:TildeACon}
\tilde a_{\B_1\#\B_2}&=&\tilde a_{\B_1}+\tilde a_{\B_2}, \\
\label{eqref:ACon}
a_{\B_1\#\B_2}&=&\frac1 {V_\#} \bigl(a_{\B_1}V_1+a_{\B_2}V_2-  \frac{D(D-1)} {2}\bigr),\\
\label{eqref:SCon}
s_{\B_1\#\B_2}&=&s_{\B_1}+s_{\B_2},
\eea
and a map $\Ga_\#$ is maximal in $\bS(\B_1\#\B_2, \Om_\# )$ if $\Ga$ is maximal in $\bS(\B_1,\B_2; \Om_1,\Om_2 )$.
\item If trees are the only maximal maps in $\bS(\B_1,\B_2; \Om_1,\Om_2 )$,  then the same property holds in $\bS(\B_1\#\B_2, \Om_\# )$.
\item If the following property is satisfied for any $\bB_\#$,
\be
\label{eqref:PropConCutBub}
\biggl[\Ga_\#\textrm{ maximal in }\bS(\bB_\#, \Om_{\bB_\#})\biggr]\ \Rightarrow\ \biggl[\Ga\textrm{ maximal in }\bS(\bB, \Om_{\bB})\biggr],
\ee

\noindent then if  $(\B_1, \Om_1)$ and $(\B_2,\Om_2)$ satisfy the cut-bubble property \eqref{CutBubbleProp}, so does $(\B_1\#\B_2,\Om_\#)$.
\item If in the simpler bijection of Thm.~\ref{thm:BijSimp}, planar maps belong to maximal maps in $\bS(\B_1,\B_2; \Om_1,\Om_2 )$, then planar maps belong to maximal maps in $\bS(\B_1\#\B_2, \Om_\# )$.
\end{enumerate}
\end{prop}
\prf  $(b)$, $(d)$, and the first statement of $(a)$ are proven similarly. Consider a map $\Ga_\#$ and another map $\Ga'_\#$ with the same number of bubbles. Then with the notations of Lemma~\ref{lemma:CircConSum}, $\Ga$ has the same number of bubbles as $\Ga'$ 
\be
n_{\B_1}(\Ga')=n_{\B_2}(\Ga')=n_{\B_1\#\B_2}(\Ga'_\#).
\ee
Suppose that $\Ga'_\#$ is a tree, or a planar map with the simpler bijection, then from Lemma~\ref{lemma:CircConSum}, $\Ga'$ has the same property in $\bS(\B_1,\B_2; \Om_1,\Om_2 )$. If in addition trees (or planar maps) belong to maximal maps in $\bS(\B_1,\B_2; \Om_1,\Om_2 )$, from Lemma~\ref{lemma:ContractScore}, 
\be
\Phi_0(\Ga'_\#)-\Phi_0(\Ga_\#) =  \Phi_0(\Ga')-\Phi_0(\Ga) \ge 0
\ee
which proves the characterizations of $(a)$, and $(d)$. If trees are the only maximal maps in $\bS(\B_1,\B_2; \Om_1,\Om_2 )$, the  inequality is strict, and the same property holds in $\bS(\B_1\#\B_2, \Om_\# )$, proving $(b)$. We compute the coefficients $\tilde a$, $a$ and $s$. If $T_\#$ is a tree in  $\bS(\B_1\#\B_2, \Om_\# )$, then $T$ is a tree in $\bS(\B_1,\B_2; \Om_1,\Om_2 )$, and
\be
 \Phi_0(T_\#)=\Phi_0(T)=D+a_{\B_1}n_{\B_1}(T)+a_{\B_2}n_{\B_2}(T)= D+(a_{\B_1}+a_{\B_2})n_{\B_1\#\B_2}(\Ga'_\#),
\ee
which proves \eqref{eqref:TildeACon}. Using \eqref{eqref:TreeMaxAB2} and \eqref{eqref:ScoreCon}, \eqref{eqref:ACon} comes easily. Using that $V\#=V_1+V_2-2$ and  \eqref{eqref:TildeACon}, we obtain the scaling \eqref{eqref:SFromTildeA}
\be
s_{\B_1\#\B_2}=\bigl(\frac{V_\#}2 -1\bigr)(D-1) - \tilde a_{\B_1\#\B_2} = s_{\B_1}+s_{\B_2}.
\ee 
We prove the last statement of $(a)$: if trees belong to maximal maps in both $\bS(\B_1\#\B_2, \Om_\# )$ and $\bS(\B_1,\B_2; \Om_1,\Om_2 )$, then 
\be
\Ga_\#\textrm{ maximal }\ \Leftrightarrow\ \Phi_0(\Ga_\#)=\Phi_0(T_\#)\ \Leftrightarrow\ \Phi_0(\Ga)=\Phi_0(T)\ \Leftrightarrow\ \Ga \textrm{ maximal }.
\ee

It remains to prove $(c)$. Given $\bB_\#$,  $\Om_{\bB_\#}$ such $\B_1\#\B_2$ has pairing $\Om_\#$, and $\Ga_\#$ maximal in $\bS(\bB_\#, \Om_{\bB_\#})$, then from \eqref{eqref:PropConCutBub}, $\Ga$ is maximal in $\bS(\bB, \Om_{\bB})$ and both $\Ps(\B_1, \Om_1)$ and $\Ps(\B_2, \Om_2)$ are cut-bubbles in $\Ga$. Therefore, they must locally be as on the left of Fig.~\ref{fig:CutBubCon}. Contracting the color-0 edge, we see on the right of Fig.~\ref{fig:CutBubCon}, that  $\Ps(\B_1\#\B_2, \Om_\#)$ is a cut-bubble. \qed

\begin{figure}[h!]
\centering
\includegraphics[scale=.9]{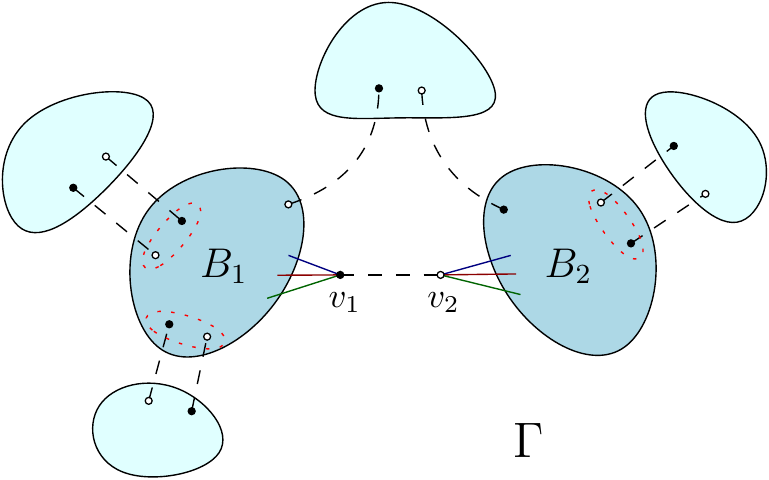}
\hspace{0.5cm}
\includegraphics[scale=.9]{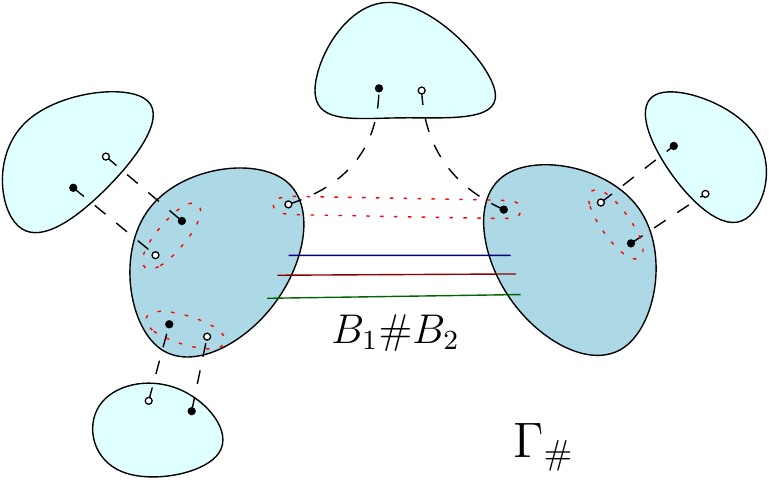} 
\caption{If $\Ps(\B_i,\Om_i)$ are cut bubbles in $\Ga$, $\Ps(\B_1\#\B_2,\Om_\#)$ is a cut-bubble in $\Ga_\#$.}
\label{fig:CutBubCon}
\end{figure}
%


\begin{prop}
If $\B_1$ or $\B_2$ represents a $(D-1)$-sphere, then $\Ga_\#$ and $\Ga$ represent the same $D$ dimensional pseudo-manifold. 
\end{prop}
\prf As $v_1$ and $v_2$ belong two different bubbles, deleting all the color-0 edges, they belong to different connected components. If furthermore one of the bubbles represents a sphere, then $(v_1,v_2)$ is a proper 1-dipole, and from Thm.~\ref{thm:GagliaProper}, contracting it does not change the topology. \qed

\

The results of this section imply that, given a bubble which contains an edge-cut of $D$ edges of different colors, we may just insert a pair of vertices on this edge-cut and study the two smaller bubbles. The example of Fig.~\ref{fig:Meander0} is treated in \emph{a size 6 bubble in $D=4$}, in Subsection~\ref{subsec:K334}.
When the properties of maximal maps in $\bG(\bB)$ are known, the results translate to the infinite family of bubbles which are connected sums of elements of $\bB$. For a detailed description, we report the reader to \cite{Enhanced}, where the family of connected sums of 2-cyclic bubble sin $D=4$ is studied. We stress  that \emph{all the examples we treat in Section~\ref{sec:Examples} extend to the infinite families of bubbles obtained by doing the connected sums of the studied bubbles in all possible ways.}

\

Conversely, it is more difficult to characterize maximal maps in  $\bS(\B_1,\B_2; \Om_1,\Om_2 )$, knowing the properties of maximal maps in $\bS(\B_1\#\B_2, \Om_\# )$. The reason is that in a generic map of $\bS(\B_1,\B_2; \Om_1,\Om_2 )$, there is not necessarily a color-0 edge between the two vertices $v_1$ and $v_2$ for pairs of bubbles $\B_1$ and $\B_2$. In particular, there can be an odd number of bubbles. A careful study makes it sometimes possible to characterize maps in  $\bS(\B_1,\B_2; \Om_1,\Om_2 )$, knowing the properties of maximal maps in $\bS(\B_1\#\B_2, \Om_\# )$. This is the case when $\B_1\#\B_2$ satisfies the cut-bubble property (Def.~\ref{def:CutBub}). Indeed, if ever there is a color-0 edge in a maximal graph between a bubble $\B_1$ and a bubble $\B_2$ in $\bG(\B_1,\B_2; \Om_1,\Om_2 )$, then contracting it does not change the 0-score and we have a bubble $\B_1\#\B_2$ in a maximal graph. It satisfies the cut-bubble property, so the bubbles are as on the left of Fig.~\ref{fig:CutBubCon}, and we may study the rest of the graph. For instance, if $\B_1=\B_2=\B$, there is always one such edge. 
Another approach that sometimes makes it possible to make strong deductions is the following. 

\

In the colored graph picture, consider a graph $\G$ in $\bG(\bB)$, and the abstract graph which has a vertex for each bubble and an edge between two vertices if there is a color-0 edge between the corresponding bubbles. Choose a spanning tree $\cT$, and contract each of the corresponding color-0 edge in $\G$. Denoting $\bB=\{\B_1,\B_2,\cdots\}$,\emph{ the resulting graph $\G^\cT$ is a covering  of a  bubble} 
\be
\B^\cT\cong( \#_{n_{\B_1}(\G)} \B_1)\#( \#_{n_{\B_2}(\G)} \B_2)\#\cdots
\ee
Each contraction preserves the 0-score and, if trees belong to maximal maps, the bubble-dependent degree. Therefore, if we restrict the possibilities for $\G^\cT$, it can restrict the possibilities for $\G$. 
This idea is applied in  \cite{BLT} to identify colored graphs of positive Gurau degree. 

We have studied bubbles obtained by contracting a vertex in each pair. But  is a bubble which has an edge-cut of $D$ edges of different colors, always the graph connected sum of two smaller bubbles? Here are certain consequences to the existence of such an edge-cut. 

\begin{lemma} Be $\B$ a bubble which has an edge-cut $\cE$ of $D$ edges of different colors. Deleting the cut, we obtain a graph $\B_{\setminus\cE}$ with two connected components, and if $e_i\in\cE$, we denote $e_i=(v^\bullet_i,v^\circ_i)$ where $v^\bullet_i$ (resp. $v^\circ_i$) is the black (resp. white) extremity of $e_i$. The following properties are true:
\begin{enumerate}[label=(\alph*)]
\item There is an odd number of vertices on each side of the cut. 
\item All the $v^\bullet_i$ (resp. $v^\circ_i$) are in the same component of $\B_{\setminus\cE}$.
\item Inserting a pair of vertices on $\cE$ preserves the bipartiteness. 
\end{enumerate}
\end{lemma}

\prf The statement $(a)$ follows from considering the edges of color 1. There is one in $\cE$, which as one extremity in each one of the two components of  $\B_{\setminus\cE}$, which we denote $\B_{\setminus\cE}^1$ and  $\B_{\setminus\cE}^2$. For the rest, 
there are $E_i$ edges of color 1 in $\B_{\setminus\cE}^i$. The number of vertices in $\B_{\setminus\cE}^i$ is therefore $2E_i+1$.
To prove $(b)$, we focus on the bicolored cycles $1i$ in $\B_{\setminus\cE}^1$. The edges which were incident to edges of $\cE$ are now $D-1$-valent  while the others are $D$-valent. Consider the vertex which was incident to the color-1 edge of $\cE$. There is a color $12$ path starting at this vertex, and which necessarily ends at the only vertex which does not have an incident color-2 edge. Because the path alternates edges of color 1 and 2, there is an even number of vertices in the path, and because the vertices in the path are alternatively black and white, its two endpoints have the same color (black or white). Repeating this for each $i\neq 1$ proves $(b)$.
%
If $\B$ is bipartite, inserting a pair on $\cE$ does not change this property apart for the vertices incident to the edge-cut. Because of $(b)$, it is enough to color the vertex incident to the $\{v_i\bullet\}$ in white and conversely. \qed


\subsection{Non-connected bubbles}
\label{subsec:NonConBub}

In two dimensions, whatever set $\bB$ of connected bubbles we consider, maximal maps are always the planar ones, and unless the counting parameters are tuned to reach multi-critical behaviors, the critical exponent will be $-1/2$. It is still possible however to reach other critical behaviors by considering non-connected bubbles. A non-connected bubble in $2D$ is a  collection of polygons. Gluings of non-connected polygons are generated by multi-trace matrix models \cite{IndianBaby,  Korchem, BabyCrit, Baby}. 
Maximal maps are such that connected components have to be planar. We build an abstract graph by associating to each planar discrete surface a white vertex, to each non-connected bubble a black vertex, and for each connected component of a bubble, if it belongs to a discrete planar surface, we draw an edge between the corresponding black and white vertices. Then one shows that for the map to be maximal, this graph has to be a tree.
Therefore, to each non-connected bubble made of $k$ polygons will correspond a  nodal point between $k$ planar discrete surfaces. At the nodal point, the surfaces are ``close", but don't ``touch". In this subsection, we show that the situation is very similar  in higher dimension. 

\

We consider a set $\bB_\sqcup$ with $\B\in\bB_\sqcup$ made of $k$ non-necessarily connected components $\B_\sqcup=\B_1\sqcup\cdots\sqcup\B_k$,
and consider a maximal graph $\G\in\bG(\bB_\sqcup)$. It is considered as connected when non-connected bubbles are considered as  connected objects (i.e. if for any pairing of non-connected bubbles, the projected map is connected).
We denote $\bB$ the set of bubbles 
\be
\bB=\bigl(\bB_\sqcup\setminus \B_\sqcup\bigr)\cup\{\B_1,\cdots, \B_k\}.
\ee
To study the influence of the non-connectivity of $\B_\sqcup$, we still consider all the bubbles in $\bB_\sqcup\setminus \B_\sqcup$ as connected objects and decide not to consider the bubble $\B_\sqcup$ as connected anymore, but rather as $k$  connected bubbles $\B_1\cdots\B_k$. With this choice, the graphs in $\bG(\bB_\sqcup)$ which were considered as connected are not seen as connected anymore when changing our point of view for $\B_\sqcup$,  and we denote $\{K_\alpha\}_{\alpha\in\{1,\cdots,A\}}$ its $A$ corresponding ``connected" components in $\bG(\B_\sqcup)$.\footnote{It is not excluded for other bubbles in $\bB$ to be non-connected.}
We consider the following abstract graph of incidence relations. As explained above, if $\G\in\bG(\bB_\sqcup)$, we represent each $K_\alpha$ as a white vertex and each $\B_\sqcup$ as a black vertex. For each connected component of $\B_\sqcup$, if it belongs to $K_\alpha$, we draw an edge between the corresponding black and white vertices, obtaining a graph $H$. We denote $V_j=V(\B_j)$, and $V_\sqcup=\sum_{j=1}^k V_j$.

\begin{prop}
If maximal graphs of $\bG(\bB)$ satisfy \eqref{eqref:MaximalBound}, that is if their 0-score has a linear dependance in the number of bubbles, then a graph in $\bG(\bB_\sqcup)$ is maximal if and only if $H$ is a tree and the $K_\alpha$ are maximal in $\bG(\bB)$, and
\bea
\label{eqref:TildeASq}
\tilde a_{\sqcup_{j=1}^k \B_j}&=&D+\sum_{j=1}^k\tilde a_{\B_j}, \\
\label{eqref:ASq}
a_{\sqcup_{j=1}^k \B_j}&=&\frac1 {V_\sqcup} \bigl((k-1)D + \sum_{j=1}^k a_{\B_j}V_j\bigr),\\
\label{eqref:SSq}
s_{\sqcup_{j=1}^k \B_j}&=&1-k+ \sum_{j=1}^k s_{\B_j}.
\eea
\end{prop}

\prf If  $\G\in \bG(\bB_\sqcup)$, we consider $\tilde\G$, whose connected components are in $\bG(\bB)$. It is the same colored graph, but the notion of connectivity is different. In particular, $\G$ and $\tilde\G$ have the same 0-score. 
\be
\Phi_0(\G)=\sum_{\alpha=1}^A\Phi_0(K_\alpha).
\ee
From the hypothesis, the $K_\alpha$ satisfy
\be
\Phi_0(K_\alpha)\le D+\sum_{\B\in\bB} \tilde a_\B n_\B(K_\alpha),
\ee
with equality iff the $K_\alpha$ are maximal. Therefore, 
\be
\Phi_0(\G)\le  \sum_{\alpha=1}^A(D+\sum_{\B\in\bB} \tilde a_\B n_\B(K_\alpha)) = AD+\sum_{\B\in\bB} \tilde a_\B n_\B(\G).
\ee
This is maximal if and only if $A$ is maximal, and the $K_\alpha$ are maximal. The graph $H$ has $k \times n_{\B_\sqcup}$ edges, $A+n_{\B_\sqcup}$ vertices, and is connected, so its circuit-rank  satisfies 
\be
(k-1)n_{\B_\sqcup}-A+1=L(H),
\ee
and at fixed $n_{\B_\sqcup}$, $A$ is maximal if and only $H$ is a tree, in which case, $A=(k-1)n_{\B_\sqcup}+1$. This proves the first statement. By looking at the coverings of $\B_\sqcup$, we see it implies that an optimal pairing of  $\B_\sqcup$ is obtained by taking an optimal pairing of each of the $\B_j$. To find the coefficients, we suppose that $\bB_\sqcup=\{\B_\sqcup\}$. A maximal graph $\G_M\in \bG(\B_\sqcup)$ satisfies 
\be
\Phi_0(\G_M)=((k-1)n_{\B_\sqcup}+1)D + \sum_{j=1}^k \tilde a_{\B_j} n_{\B_j}(\G_M)=D+\bigl((k-1)D+ \sum_{j=1}^k \tilde a_{\B_j}\bigr)n_{\B_\sqcup}(\G_M),
\ee
as for any $j$, $n_{\B_j}(\G_M)=n_{\B_\sqcup}(\G_M)$. Therefore, $\B_\sqcup$ also satisfies \eqref{eqref:MaximalBound}, and we deduce  \eqref{eqref:TildeASq}. From \eqref{eqref:Tildeaa}, 
\be
a_{\B_\sqcup}=\frac{\sum_{j=1}^k\Phi(\B_j) + \tilde a_{\B_\sqcup}}{V_\sqcup},
\ee
and \eqref{eqref:ASq} follows from \eqref{eqref:TildeASq}. The coefficient $s$ is deduced from  \eqref{eqref:TildeASq} using  \eqref{eqref:SFromTildeA},
\bea
s_{\B_\sqcup} &=& \bigl(\sum_{j=1}^k \frac {V_j}2 - 1\bigr)(D-1) - \tilde a_{\B_\sqcup}\\
&=& \sum_{j=1}^k (\frac {V_j}2 - 1)(D-1) + (k-1)(D-1)  - ((k-1)D+\sum_{j=1}^k \tilde a_{\B_j}).  
\eea

\subsection{Tree-like families and the bound on $(D-2)$-cells}
\label{subsec:TreeLike}

This section contains important results, however it is not needed to understand the examples in a first reading.  The notion of tree depends on the choice of pairing. As detailed in the previous subsections, we are interested in knowing if trees belong to maximal maps, which requires that the bijection is built with optimal pairings. If a bubble $\B$ has more than one optimal pairing, then, choosing one of them, $\Opt$,  to build the bijection, we know (Subsec.~\ref{subsec:Unicell}) that there are maximal unicellular maps which are not trees. Consider one of these unicellular maps $\Ga_1$, corresponding to an other optimal covering $\B^{\Om'}$. The corners around black vertices correspond to color-0 edges in the colored graph picture. Picking two corners incident to black vertices of two copies of $\Ga_1$, and merging the black vertices as detailed in Figure~\ref{fig:VertSplit}, we obtain a connected stacked map with two bubbles, which is not a tree, but has the 0-score  of a tree with two bubbles. Repeating this vertex merging in a tree-like way, we can generalize this to arbitrarily many bubbles. The maps built this way are not trees but share all their properties. In our context, we call them tree-like maps. 

\begin{definition}[Tree-like family of stacked maps]
\label{def:TreeLikeStack}
An infinite family $\bF$ of stacked maps of $\bS(\bB,\Om_\bB)$ is said to be tree-like if there exists a finite subset  $\bT$ of $\bS(\bB,\Om_\bB)$ such that for any map in the family, there exists a map $\Ga_T\in\bT$, a black vertex $v_\bullet$ and two corners $c$ and $c'$ around $v_\bullet$, such that when splitting $v_\bullet$ along $c$ and $c'$, we obtain two connected components, one of them being $\Ga_T$, and the other one being in $\bF$. We call $\bF$ a $\bT$-tree-like family.
\end{definition}

\begin{figure}[h!]
\centering
\includegraphics[scale=.5]{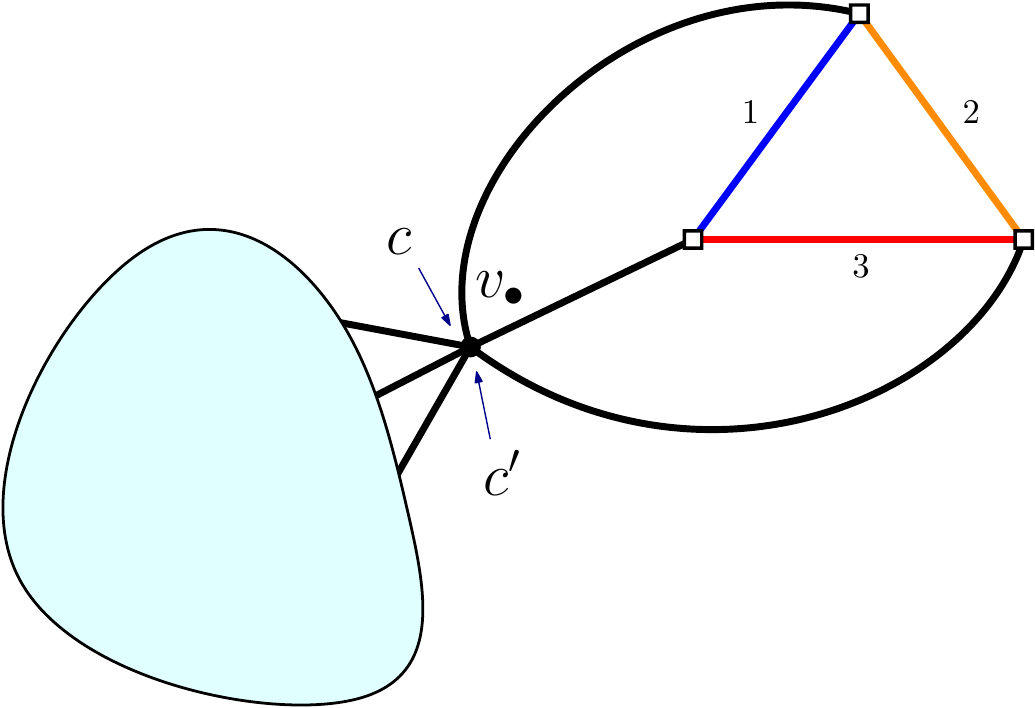}
\hspace{0.5cm}
\includegraphics[scale=.5]{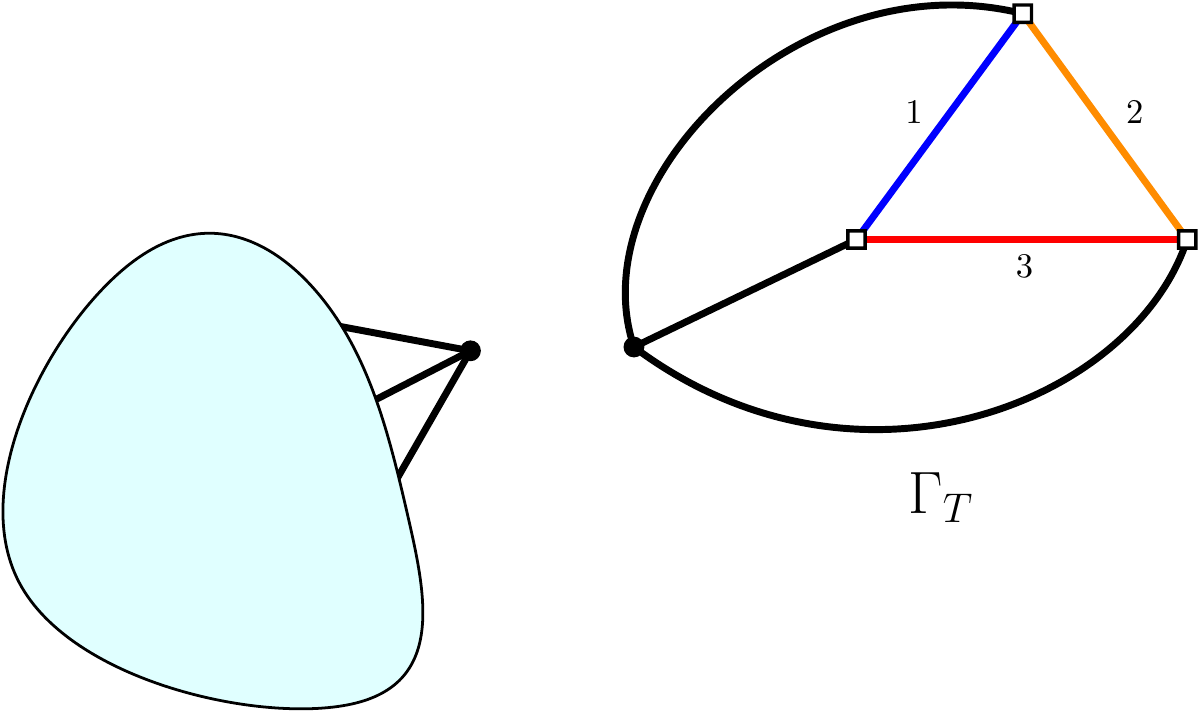} 
\caption{Recursive definition of a tree-like family in the stacked map picture.}
\label{fig:TreeLikeEx}
\end{figure}

This is illustrated in Fig.~\ref{fig:TreeLikeEx}. Put more simply, it is precisely like defining trees as  connected graphs for which there exists a leaf, and when deleting the edge incident to the leaf, we are left with a vertex and a smaller tree. In this definition, all the corners of a $\Ga_T$ can be merged to another element of $\bF$ to build a bigger one. We could have decided to allow only mergings for certain corners of the $\Ga_T$ and not the others, but such families are not needed in our context. 
For the description to be well defined, we further require that an element of $\bT$ cannot itself be split into a smaller element of $\bT$ and a smaller element of $\bF$. We can state the same definition in the colored graph picture:

\begin{definition}[Tree-like family of colored graphs]
An infinite family $\bF$ of colored graphs of $\bG_D$ is said to be tree-like if there exists a finite subset  $\bT$ of $\bG_D$ such that for any graph in the family, there exists a graph $\G_T\in\bT$, a pair of color-0 edges $f$ and $f'$, such that when switching  $f$ and $f'$, we obtain two connected components, one of them being $\G_T$, and the other one being in $\bF$. We call $\bF$ a $\bT$-tree-like family.
\end{definition}
An example is shown in \ref{fig:TreeLikeEx2}. Choosing the dotted pairing to build the bijection, we recover precisely the example of Figure~\ref{fig:TreeLikeEx} (see Figure~\ref{fig:UniMaps}).
\begin{figure}[h!]
\centering
\includegraphics[scale=.5]{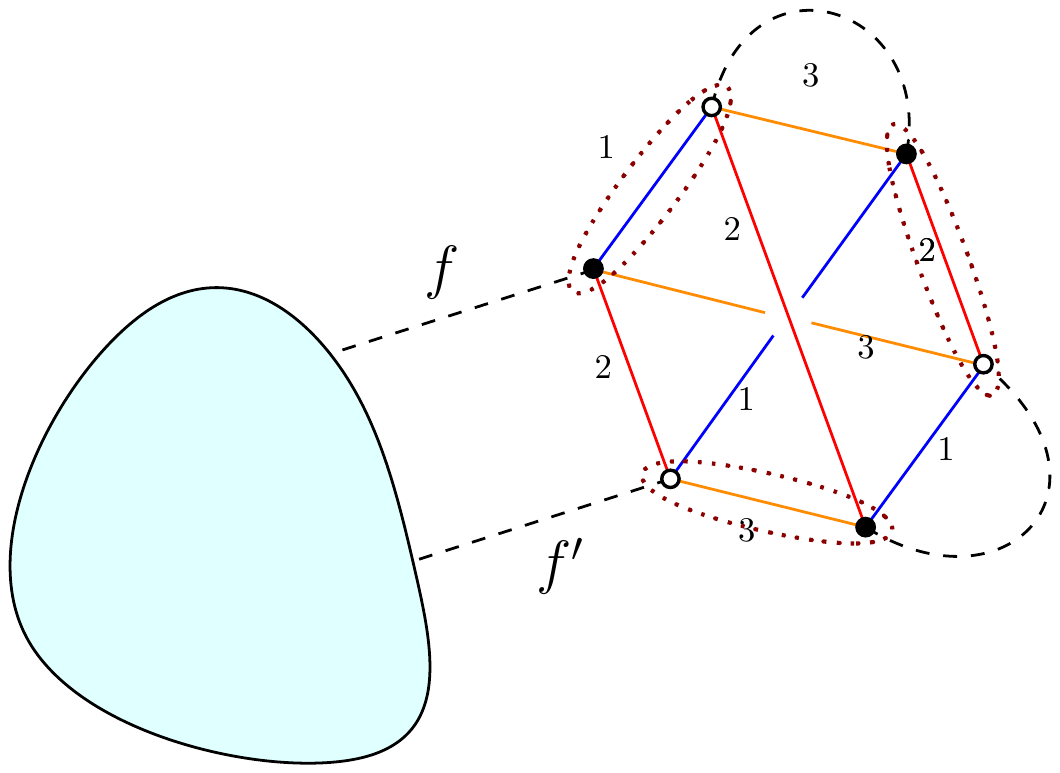}
\hspace{0.5cm}
\includegraphics[scale=.5]{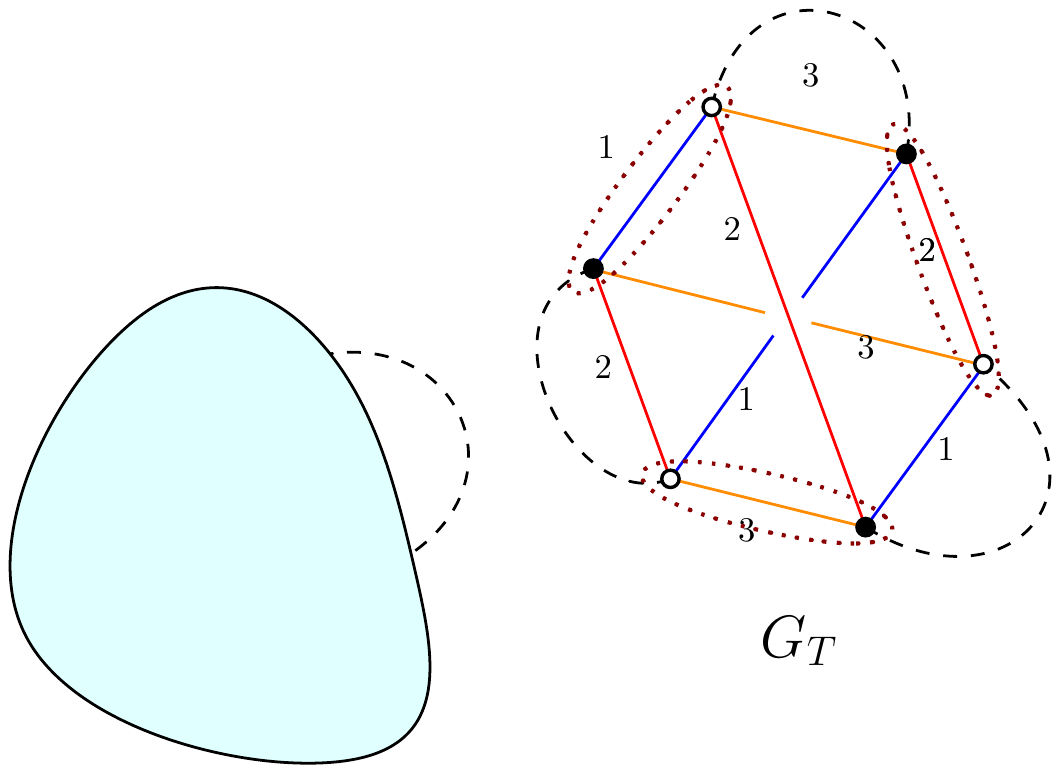} 
\caption{Recursive definition of a tree-like family in the colored graph picture.  }
\label{fig:TreeLikeEx2}
\end{figure}

In the case of graphs in $\bG(\B)$, $\G_T$ is a set of bubbles $\{\B_a\}_{a\in A}$ with color-0 edges added in a connected way. $\G_T$ induces a pairing $\Om_T$ of this set of bubbles $\{\B_a\}_a$, so that $\G_T$ coincides with the ``covering" $\{\B_a\}_a^{\Om_T}$.  The set $\{\B_a\}_a$ is seen as an \emph{effective bubble}, but unlike in the case of non-connected bubbles (previous section), here, the color-0 edges have to be added so that $\{\B_a\}_a^{\Om_T}$ is connected. 
A more global characterization of $\bF$ in the colored graph picture, illustrated in Figure~\ref{fig:TreeFamily}, is that a graph $\G$ belongs to a $\bT$-tree-like family if any of its vertices belongs to a subset $\{\B_a\}_{a\in A}$ with pairing $\Om_T$ of its vertices, such that    for each pair in $\Om_T$, either a color-0 edge links the two vertices of the pair, either  the two incident color-0 edges form a 2-cut. In addition there must be no subset $A'\subsetneq A$ for which the property is satisfied for $\{\B_a\}_{a\in A'}$. 


\begin{figure}[h!]
\centering
\raisebox{0ex}{\includegraphics[scale=.9]{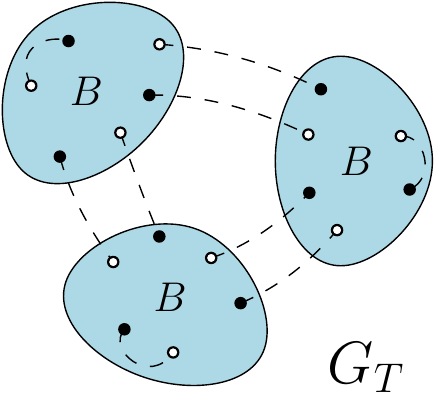}} \hspace{1cm}
\includegraphics[scale=.9]{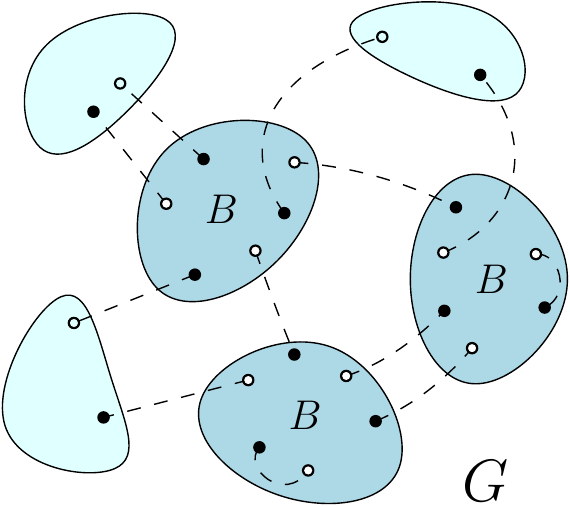} 
\caption{Local decomposition of a tree-like family.}
\label{fig:TreeFamily}
\end{figure}

\

In the case where there are more than one optimal pairings, the notion of tree is too narrow.
In particular, if trees belong to maximal maps, then \emph{the tree-like family with $\bT$ the set of maximal unicellular maps is a subset of maximal maps}. Sometimes, they are the only maximal maps (see the example of gluings of octahedra in Subsection~\ref{subsec:BiPyr}, for which $\bT$ comprises the patterns in~\eqref{eqref:MaxOcta} with empty blobs, and an example is shown in Fig.~\ref{fig:Dominant}). Sometimes maximal maps are a larger $\bT$-tree-like family (see the example of the $K_{3,3}$ bubble in Subsection~\ref{subsec:K33}, for which $\bT$ comprises the patterns in Fig.~\ref{fig:ExDomOrdK33} and Fig.~\ref{fig:K33Deg3Vert} - with empty blobs - and an example of tree-like map is shown in Fig.~\ref{fig:ExDomOrdK33}). And sometimes, tree-like families are just part of a larger set of maximal maps (as in the example of $D/2$-cyclic bubbles, see Subsection~\ref{Subsec:D2CycBub}).

\subsubsection{Bijection with trees}

There is always a bijection between $\B$-restricted gluings $\bG(\B)$ for which maximal colored graphs are a $\bT$-tree-like family, and a set of stacked maps $\bS_\bT$ such that maximal graphs of $\bG(\B)$ are mapped to trees of $\bS_\bT$.  We describe it briefly. 
Choose a pairing for $\B$,
and consider a vertex. If it belongs to a subset $\{\B_a\}_{a\in A}$ with $\Om_T$ a pairing of its vertices, as described above, then we choose this pairing to pair the vertices in the subgraph  $\{\B_a\}_{a\in A}$, and continue exploring the components which are attached to $\{\B_a\}_{a\in A}$ by 2-cuts of color-0 edges. If not, we take the pairing we chose initially for the bubble $\B$ it belongs to, and pick another vertex not in the same bubble. Having explored all the graph, we have built a partition of the vertices in pairings of subgraphs $\{\B_a\}_{a\in A}$, which translates into a pairing $\Om_\sqcup(\G)$ of the whole graph. We then consider  the map
\be
\label{eqref:BijTreeLike}
\G\in\bG(B) \longmapsto \Ps_0(\G,\Om_\sqcup(\G)).
\ee 

The images of $\bG(\bB)$ under the bijection have submaps $\Ps(\{\B_a\}_{a},\Om_T)$ (in one less dimension) which are ``cut-bubbles", i.e. which are incident only to bridges, and have submaps  $\Ps(\B,\Om_\B)$, which can be cut-bubbles or not. However, there are a certain number of prohibited  patterns, which correspond to the maps $\Ps_0(\{\B_a\}_{a}^{\Om_T},\Om_B)$. For instance, the patterns corresponding to other optimal coverings $\B^{\Opt'}$ with (possibly trivial) map insertions on the corner around black vertices are forbidden, if $\Opt'$ is not the optimal pairing used to build the bijection. This is illustrated for the gluings of octahedra in Subsection~\ref{subsec:BiPyr}. The patterns on the right of \eqref{eqref:MaxOcta} are prohibited, and are replaced with the two new cut-bubbles in Fig.~\ref{fig:NewVertOcta}.


\subsubsection{Counting tree-like families}

We denote $p_T$ the number of pairs in $\G_T$. Consider the generating function $\GF_\bT$ of rooted $\bT$-tree-like maps counted according to their number of ``cut-bubbles" isomorphic to $\G_T$, It satisfies the tree equation
\be
\GF_\bT(\{z_{T}\})=1 + \sum_{\G_T\in\bT} z_{T} \GF_\bT(\{z_T\}) ^{p_T}.
\ee
As $\bT$ is finite, the expected critical exponent of the generating function at the dominant singularity is that of trees, 1/2. In the case of a $\bT$-tree-like family for which all  $\G_{T}$ have the same $n_T$, as it is the case if $\bT$ is restricted to the $k$ optimal coverings of $\B$, then taking all $z_T$ equal to $z$, the equation simplifies to 
\be
\label{eqref:CountTreeLike}
\GF_\bT(z)=1 +k z \GF_\bT(z) ^{p_\B}.
\ee
We refer the reader to the examples in Sections \ref{subsec:BiPyr} and \ref{subsec:K33} for concrete examples.

\subsubsection{Score}

 Because vertex-splittings present the same properties as edge-unhookings, \eqref{eqref:ConSumDeg} is true for vertex splittings which disconnect the map: the sum of the degrees of the connected components is the degree of the graph before the splitting. As for trees, the structure of tree-like families enables us to deduce their 0-score. 
 
\begin{prop}
\label{prop:TreeLikeScore}
If $\bF\subset\bG(\bB)$ is a $\bT$-tree-like family, the 0-score and the bubble-dependent degree of an element $\G\in\bF$  with $n_T(\G)$ submaps isomorphic to $\G_T\in\bT$ are
\bea
\label{eqref:PhiTreeLike}\Phi_0(\G)&=&D+ \sum_{\G_T\in\bT} n_T(\G)(\Phi_0(\G_T)-D),\\
\label{eqref:DeltaTreeLike}\delta_\bB(\G)&=&\sum_{\G_T\in\bT} n_T(\G)\delta_\bB(\G_T).
\eea
\end{prop}


In the case where trees are not maximal (or just a finite number of trees belong to maximal maps, as the unicellular tree is always maximal), then  we cannot use the results of Subsection~\ref{subsec:TreeDeg} to determine the coefficient $\tilde a$. If however maximal maps are ultimately a $\bT$-tree-like family $\bF$, then the 0-score of graphs in $\bF$ which contain the same number of bubbles $\B$ is the same. Denoting $n_\B(\G_T)$ the number of bubbles in $\G_T$, 
\be
b(\G)=\sum_{\G_T\in\bT} n_T(\G)n_\B(\G_T).
\ee
The degree of graphs in $\bF$ is
\be
\label{eqref:DeltaTTreeLikeFam}
\delta(\G) =\sum_{\G_T\in\bT} n_T(\G)[ n_\B(\G_T) \tilde a_\B - \Phi_0(\G_T)+D].
\ee

If  $n_\B(\G_k) \tilde a_\B - \Phi_0(\G_k)+D<0$ for some $\G_k\in\bT$, then elements of $\bF(\bT)$ containing only effective bubbles $\G_k$ have degree 
\be
\delta\bigl(\G\in\bF(\G_k)\bigr) = n_k(\G)[ n_\B(\G_k) \tilde a_\B - \Phi_0(\G_k)+D],
\ee
and we can find infinite families with unbounded negative degree. Therefore, to satisfy Condition~\eqref{eqref:Cond1}, we need to choose $\tilde a_\B$ satisfying 
\be
\label{eqref:TileATreeLike1}
\forall \G_T\in\bT,\ n_\B(\G_T) \tilde a_\B - \Phi_0(\G_T)+D\ge0.
\ee
In the case where the 0-score of maximal maps is a linear function of the number of bubbles for maximal graphs ultimately 
\be
\label{eqref:LinearDepTreeLike}
\exists K\in\bN,\ \exists\tilde a>0, \quad\text{ such that }\quad \biggl\{{\begin{tabular}{@{}c@{}}  $b(\G_\text{max})\ge K$\\$\G_\text{max}$ maximal \end{tabular}} \quad \Rightarrow \quad\Phi_0(\G_\text{max})=D+\tilde a_\B b(\G_\text{max}), 
\ee
then as \eqref{eqref:PhiTreeLike}, for $\G_\text{max}$ large enough, we have the relation
\be
\sum_{\G_T\in\bT} n_T(\G_\text{max})[n_\B(\G_T) \tilde a_\B - \Phi_0(\G_T)+D] = 0,
\ee
which from \eqref{eqref:TileATreeLike1} leads to 
\be
\forall \G_T\in\bT,\quad n_\B(\G_T) \tilde a_\B - \Phi_0(\G_T)+D=0,
\ee
and choosing any $\G_T\in\bT$,  we can define 
\be
\tilde a_\B  = \frac{ \Phi_0(\G_T)-D}{ n_\B(\G_T) }.
\ee

\subsubsection{Non-linear 0-score} 

It is not forbidden to have tree-like maximal graphs, whose 0-score does not satisfy a linear dependence in the number of bubbles  \eqref{eqref:LinearDepTreeLike}.  This happens for instance if there exists a graph $\G_2$ with two bubbles such that maximal maps are a $\bT$ tree-like family with 
\be
\bT=\{\G_1=\B^{\Opt}, \G_2\}, \quad\text{and}\quad  \Phi_0(\G_2) - D > 2 (\Phi_0(\G_1) - D),
\ee
i.e. $\G_2$ has two bubbles but its  0-score is higher than that of trees with two bubbles. Then elements of the sub-tree-like family $\bF(\G_2)$  have a stronger 0-score than elements of $\bF\setminus\bF(\G_2)$ with even number of bubbles. However, among graphs with odd number of bubbles, maximal maps  belong to $\bF\setminus\bF(\G_2)$. In that case, a choice 
\be
\tilde a_\B  = \frac{ \Phi_0(\G_2)-D}{ n_\B(\G_2) },
\ee
leads to a non-trivial \eqref{eqref:Cond2} bubble-dependent degree which vanishes for $\bF(\G_2)$ but is positive otherwise. In particular, maximal maps with an odd number of bubbles  are excluded from the leading order: the degree of tree-like graphs is \eqref{eqref:DeltaTTreeLikeFam}
\bea
\delta_\B(\G\in\bF)&=&n_1(\G)[\tilde a_\B - \Phi_0(\G_1)+D] + n_2(\G)\bigl[\overbrace{2\tilde a_\B - \Phi_0(\G_2)+D}^{=\ 0} \bigr] \nonumber\\[+2ex]
&=&n_1(\G)\bigl[\underbrace{\frac{\Phi_0(\G_2)-D} 2 - \Phi_0(\G_1)+D}_{>\ 0}\bigr].
\eea

\

\noindent\emph{Example of the $K_4$ bubble} 

\

An example of a family which illustrates this behavior is that of maximal maps in $\bG(\B_{K_4})$, where $\B_{K_4}$ is the complete graph with 4 vertices, shown in Fig.~\ref{fig:Ex2D}. It is shown in \cite{Tan} that the choice
\be 
\delta_{\B_{K_4}}(\G)=3+ \frac 3 2 b(\G) - \Phi_0(\G)
\ee
satisfies almost all the conditions of a bubble-dependent degree (Def.~\ref{def:BubDepDeg}), and that the leading order thus defined is a $\{\G_2\}$-tree-like family, where $\G_2$ is represented in Fig.~\ref{fig:TK4G2}. However this only includes graphs with an even number of bubbles, so that we know that maps which maximize the 0-score at fixed but odd number of bubbles have an 0-score smaller than $3+3b/2$. Therefore, the 0-score of maximal maps is not linear in the number of bubbles, although it is for maximal maps with an odd number of bubbles. It is easy to see that a $\{\B_{K_4}^\Om, \G_2\}$-tree-like graphs with a single $\B_{K_4}^\Om$ are maximal among maps with an even number of bubbles. The only condition which we required for a well-defined bubble-dependent degree, and which $\delta_{\B_{K_4}}$ does not satisfy, is that it takes value in $\frac 1 2\bN$ instead of $\bN$.\footnote{In two dimensions, the degree reduces to twice the genus, so that it still takes integer values for non-orientable maps.}

\begin{figure}[h!]
\centering
\includegraphics[scale=.85]{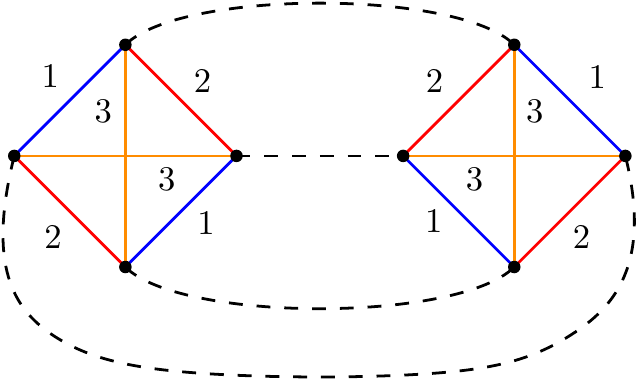} 
\caption{The graph $\G_2$ with two bubbles $\B_{K_4}$.}
\label{fig:TK4G2}
\end{figure}

The major part of the results stated in the previous sections of this chapter under the assumption that trees were part of the maximal maps generalize to the case where a $\bT$-tree-like family $\bF$ is (ultimately) part of maximal maps. The choice 
\be
\tilde a_\B = \max_{\G_T\in\bT} \frac{ \Phi_0(\G_T)-D}{ n_\B(\G_T) }
\ee
leads to non-trivial \eqref{eqref:Cond2} bubble-dependent degree taking values in $\bQ^+$. This is very general, as we show in Thm.~\ref{thm:ExUnicBBD} below that whenever $\tilde a$ can be defined, there necessarily exists a tree-like family in the leading order. intersection of $\bF$ and the  leading order contains only the $\bT_\text{max}$-tree-like family $\bF_M$ where $\bT_M$ is the subset of elements of $\bT$ saturating this choice of $\tilde a$
\be
\delta_\B(\G\in\bF)= \sum_{\G_T\in\bT} n_T(\G)n_\B(\G_T)\bigl[\tilde a_\B -  \frac{ \Phi_0(\G_T)-D}{ n_\B(\G_T) } \bigr] = \biggl\{ {\begin{tabular}{@{}c@{}}  0 if $\G\in\bF_M$ \\[+1ex] $>0$ otherwise\end{tabular} }.
\ee
From this we can deduce the coefficients $s$ and $a$, add large $h$-pairs, study connected sums of bubbles, and so on. Corollary~\ref{coroll:TreeMaxAB} gave a sufficient condition of existence of the coefficient $\tilde a_B$ for the corresponding bubble-dependent degree to be well-defined \eqref{eqref:Cond1} and non-trivial. Theorem~\ref{thm:FinitNumbPosDeg} proved the unicity of  this value. We saw however that is was possible - at least for non-orientable bubbles - to have only finitely many maximal trees (one, in the case of the $K_4$ bubble). 
The following results enlarge the sufficient condition of existence of a non-trivial and non-negative rational bubble-dependent degree to more general cases. The case which is not included is discussed in Sec.~\ref{sec:K336}. Intuitively, it states that $\tilde a$ exists if there is ultimately a subsequence of the 0-score of maximal graphs which is stronger than the other terms and which  is ultimately linear in the number of bubbles.  It gives the corresponding value of $\tilde a_\B$, and therefore of $a_\B$ and $s_\B$, and further shows that whenever they exist, they are uniquely defined. 

\begin{lemma}
\label{lemma:ExUnicBBD1}
Given a non-necessarily bipartite bubble $\B\in\tilde \bG_{D-1}$, the supremum
\be
\label{eqref:DefSup}
\alpha ^M_\B = \sup_{\G\in\tilde \bG(\bB)} \frac{\Phi_0(\G) - D}{b(\G)}
\ee
exists, and we have the following bound for the coefficient $\tilde a_\B$ to have a well-defined bubble-dependent degree ($\tilde a_\B b - \Phi_0$ is bounded from below),
\be
\label{eqref:Supremum}
\tilde a_\B \ge \alpha^M_\B.
\ee
\end{lemma}

\prf The bound on the 0-score is given by melonic graphs \eqref{eqref:Bound31}: For any $\G\in\tilde \bG(\B)$,
\be
\Phi_0(\G) \le \bigl[ \frac{D(D-1)}4 - \Phi(\B)\bigr]b(\G) + D \quad\Leftrightarrow \quad \frac{\Phi_0(\G) - D}{b(\G)} \le  \frac{D(D-1)}4 - \Phi(\B).
\ee
The set $\bigl\{\frac{\Phi_0(\G) - D}{b(\G)}\bigr\}_{\G\in\tilde\bG(\B)}$ is therefore a non-empty subset of $\bR$ with an upper bound, thus the existence of the supremum. 

Now suppose that \eqref{eqref:Supremum} does not hold, then $ \tilde a_\B < \alpha^M_\B$, so that there exist $\G_T\in\tilde\bG(\B)$ such that  $\tilde a_\B - \frac{\Phi_0(\G_T)-D}{b(\G_T)}<0$. Consider the $\{\G_T\}$-tree-like family $\bF(\G_T)$. A graph $\G\in\bF(\G_T)$ has 0-score \eqref{eqref:PhiTreeLike}, and therefore satisfies
\be
\tilde a_\B b(\G) - \Phi_0(\G) = \bigl(\tilde a_\B - \frac{\Phi_0(\G_T)-D}{b(\G_T)}\bigr) b(\G) + D.
\ee
As $\tilde a_\B - \frac{\Phi_0(\G_T)-D}{b(\G)}<0$, this family has unbounded negative degree and this choice for $\tilde a_\B$ does not lead to a well-defined degree. \qed

\begin{lemma}

\label{lemma:ExUnicBBD2}
Given a non-necessarily bipartite bubble $\B\in\tilde \bG_{D-1}$, choosing 
\be
\label{eqref:Supremum2}
\tilde a_\B > \alpha^M_\B.
\ee
leads to a trivial bubble-dependent degree (for any $k$, the equation $\tilde a_\B b(\G) -\Phi_0(\G) = k$ has only finitely many solutions).
\end{lemma}
\prf For any $\G\in\tilde\bG(\B)$, we have the following bound
\be
\Phi_0(\G) \le \alpha^M_\B b(\G) + D.
\ee
If $\tilde a_\B b(\G) -\Phi_0(\G) = k$, we can bound the number of bubbles of $\G$,
\be
\tilde a_\B b(\G) -  \alpha^M_\B b(\G) - D \le k \quad \Rightarrow\quad b(\G) \le \frac{D+k}{\tilde a_\B-  \alpha^M_\B}.
\ee
But there is a  finite number of graphs in $\bG(\B)$ with less or as much bubbles as $\frac{D+k}{\tilde a_\B-  \alpha^M_\B}$, which concludes the proof. \qed

\begin{theorem}[Existence and unicity of the bubble-dependent degree]

\label{thm:ExUnicBBD}

 For $\B\in\tilde\bG(\B)$, the only possible choice which might lead to a non-trivial and non-negative rational bubble-dependent degree is 
\be
\label{eqref:Supremum3}
\tilde a_\B = \alpha^M_\B.
\ee
If the supremum $\alpha^M_\B$ is reached (i.e. if it is a maximum), then the choice \eqref{eqref:Supremum3} indeed leads to a non-trivial and non-negative rational bubble-dependent degree
\be
\delta_\B\bigl(\G\in\tilde\bG(\B)\bigr) = D+  \alpha^M_\B b(\G) - \Phi_0(\G) \in \bQ^+,
\ee
with infinitely many solutions to $D+  \alpha^M_\B b(\G) - \Phi_0(\G) =0$. In particular, the leading order is the order 0, and $N^D$ is always the right way of rescaling the free-energy \eqref{eqref:NExp2} to extract the contributions to the leading order.
\end{theorem}

If on the contrary $\alpha^M_\B$ is not a maximum, then $D+  \alpha^M_\B b(\G) - \Phi_0(\G) \in \bQ^{+,\star} $ has no order-0 contributions. This case is discussed in Section~\ref{sec:K336}.

\

Note that such a degree satisfies Condition~\eqref{eqref:Cond2}  at order 0 but does not a priori satisfy Condition \eqref{eqref:Cond1} as it does not necessarily take values in $\bN$, as in the case of the $K_4$ bubble. A non-integer $\tilde a$ leads to rational orders, 
\be
\bG(\B)=\bigsqcup_{q\in\bQ^+}\delta_\B^{-1}( q ),
\ee
and therefore to a $1/N$ expansion \eqref{eqref:NExp} with rational powers in $1/N$, but apart from that, it still defines a well-behaved theory. From 
\be
\tilde a_\B = \max_{\G\in\tilde\bG(\B)}  \frac{\Phi_0(\G) - D}{b(\G)},
\ee
 the coefficients $a_\B$ and $s_\B$ are deduced uniquely, using \eqref{eqref:Tildeaa} and \eqref{eqref:SFromTildeA}.
If for instance, one finds a graph $\G_4\in\bG(\B)$ with 4 bubbles in 4 dimensions, with 0-score 15, and such that $ \forall \G\in\bG(\B),\ \Phi_0(\G)\le 4 + \frac {11} 4 b(\G)$, 
then $\tilde a_\B = 11/4$ and the bubble-dependent degree is defined as $\delta_B=4+\frac {11} 4 b - \Phi_0$ and takes values in $\frac14\bN$, and the scaling is $s=3/2V(\B) - 23/4$. 

\

\prf The fact that \eqref{eqref:Supremum3} is the only choice possibly leading to a well-behaved bubble-dependent degree follows from Lemma~\ref{lemma:ExUnicBBD1} and \ref{lemma:ExUnicBBD2}.

In the case where $\alpha^M_\B$ is a maximum, then there exist $\G_T\in\tilde\bG(\B)$ such that 
\be
\alpha^M_\B=\frac{\Phi_0(\G_T) - D}{b(\G_T)} \quad \Leftrightarrow\quad \delta_\B(\G_T)=D+  \alpha^M_\B b(\G_T) - \Phi_0(\G_T) = 0.
\ee
In particular, from the left hand side, $\alpha^M_\B\in\bQ$. From the definition of $\alpha^M_\B$ \eqref{eqref:DefSup}, the degree is non-negative. Furthermore, any graph $\G$ in the $\{\G_T\}$-tree-like family $\bF(\G_T)$ has degree \eqref{eqref:DeltaTreeLike}
\be
\delta_\B\bigl(\G\in\bF(\G_T)\bigr) = n_T(\G) \delta_\B(\G_T) = 0,
\ee
so that there are infinitely many contributions to the order 0. \qed

\subsubsection{Topology}

Because of their definition, tree-like families decompose into connected sums of the graphs in $\bT$, throughout a sequence of vertex splittings. Because vertex-splittings present the same properties as edge-unhookings, Corollary~\ref{coroll:TreeTopo} generalizes here:

\begin{coroll} (of Prop.~\ref{prop:DirSum2})
\label{coroll:TreeLikeTopo}
If the elements of $\bT$ represent PL-manifolds, then the triangulation $\C$ 
represented by a tree-like map $\Ga\in\bF$ with $n_T(\C)$ submaps isomorphic to $\G_T\in\bT$ is homeomorphic
 to the direct sum of $n_T(\C)$ copies of each $\G_T$. Denoting arbitrarily $\bT=\{\G_{T_1}, \G_{T_2}, \cdots \}$,
\be
\C\cong_{PL}\bigl(\#_{n_{T_1}(\C)} \G_{T_1}\bigr)\#\bigl(\#_{n_{T_2}(\C)} \G_{T_2}\bigr)\#\cdots.
\ee
This is also true in the case of pseudo-manifolds, upon other conditions (see the subsection \emph{Edge-unhooking and topology} in \ref{subsec:Unhook}). 
\end{coroll}

\section{Examples}
\label{sec:Examples}
		
\subsection{$D/2$-cyclic bubbles of different kinds}
\label{Subsec:D2CycBub}
%

We consider restricted gluings of $D/2$-cyclic bubbles of any kinds and sizes, and choose to pair the edges not linked by color 1. Using the bijection of Thm.~\ref{thm:BijCycles}, we can study the corresponding bipartite maps instead, which have edges carrying sets of $D/2$ colors in $\lDr$, and black and white vertices, each one of the latter being incident to edges which all carry the same color set. From Prop.~\ref{prop:hD2Trees}, as edges have $D/2$ colors, trees are maximal. 
The score of a $D/2$-cyclic bubble of size $2p$, and the 0-score of the covering obtained by adding color-0 edges parallel to color-1 edges are respectively 
\be
\Phi(p)=\frac {D  \bigl((p+1)D -2p\bigr)} 4 , \qquad\text{and}\qquad\Phi_0(p)=(p+1)\frac D 2, 
\ee
According to Corollaries~\ref{coroll:TreeMaxAB} and \ref{coroll:scaling} and to (\ref{eqref:TreeMaxAB2}), we can deduce the following  coefficients for a bubble of size $2p$ (which we had already obtained in \eqref{eqref:CoeffsD2Cyc})
\be
\label{eqref:CoeffsD2Cyc2}
\tilde a_p=\frac{(p-1) D} 2, \quad a_p=\frac{D(D-1)} 4 -\frac {p-1} p\times\frac{D(D-2)} 8, \quad\text{and}\quad s_p = (p-1)(\frac D 2-1),
\ee
and if all bubbles have size $2p$, the bubble-dependent degree of a colored graph can be defined as
	\be
	\delta_p(\G)=D\bigl(1+\frac{(p-1) } 2\nb(G)\bigr) - \Phi_0(\G),
	\ee
it vanishes for maximal configurations, which we want to characterize, and is positive otherwise.

\subsubsection{Single kind of $D/2$ cyclic bubble}
%

If there is only one kind of $D/2$ cyclic bubble, i.e. if $D/2$ cyclic bubbles alternate the same color $i_1,\cdots, i_{D/2}$ and $i_{D/2+1},\cdots, i_{D}$, then $L(\Gai[i_1])=\cdots= L(\Gai[i_{D/2}])=L(\Ga^\star)$ and the other $L(\Gai)$ vanish, so that using \eqref{eqref:PhivsTrees} which compares the 0-score of trees and maps with the same number of bubbles,
\be
0\le \Phi_0(\cT) -\Phi_0(\Ga)= DL(\Ga^\star) -2\sum_{i=1}^D L(\Gai)+ 2\sum_{i=1}^D g(\Gai)= 2Dg(\Ga)
\ee
and we recover the fact that maximal maps are the planar bipartite maps, (see Section~\ref{sec:SimplerBij}).

\subsubsection{2-Cyclic bubbles in $D=4$}
%

\noindent{\it Maximal maps}

\

\noindent If there are all allowed  2-cyclic bubbles in $D=4$, as color 1 appears in every color set, $\Gai[1]=\Ga$, and 
\bea
\Phi_0(\cT) -\Phi_0(\Ga)&=& 4L(\Ga) -2\sum_{i=1}^D L(\Gai)+ 2\sum_{i=1}^D g(\Gai)\\
\label{eqref:TvsGD2} &=&2\bigl(L(\Ga) -\sum_{i\neq1} L(\Gai)\bigr)+ 2\bigl(g(\Ga) + \sum_{i\neq1} g(\Gai)\bigr),
\eea
where we recognize the number of polychromatic cycles $L_m(\Ga)=L(\Ga) -\sum_{i\neq1} L(\Gai)$ (Def.~\ref{def:MultCycl}). 
If $\cT^{(i)}$ is a tree spanning $\Gai$, then $\cup_{i=2}^D \cT^{(i)}$ is a connected graph spanning $\Ga$, which misses $\sum_{i=2}^D L(\Gai)$ edges, each defining an independent cycle of $\Ga$. Therefore, 
\be
L(\Ga) -\sum_{i\neq1} L(\Gai) = L(\cup_{i=2}^D \cT^{(i)}) \ge 0,
\ee
and (\ref{eqref:TvsGD2}) vanishes when the map is planar, and $L(\cup_{i=2}^D \cT^{(i)})=0$, i.e. when there is no cycle containing edges $1i$ and $1j$ for $i\neq j$. The maximal maps are ``trees" of planar color $1i$ components (the incidence relations between planar connected bicolored components are tree-like). A schematic example is shown in Fig.~\ref{fig:Cactus}. 
\begin{figure}[!h]
\centering
\includegraphics[scale=0.5]{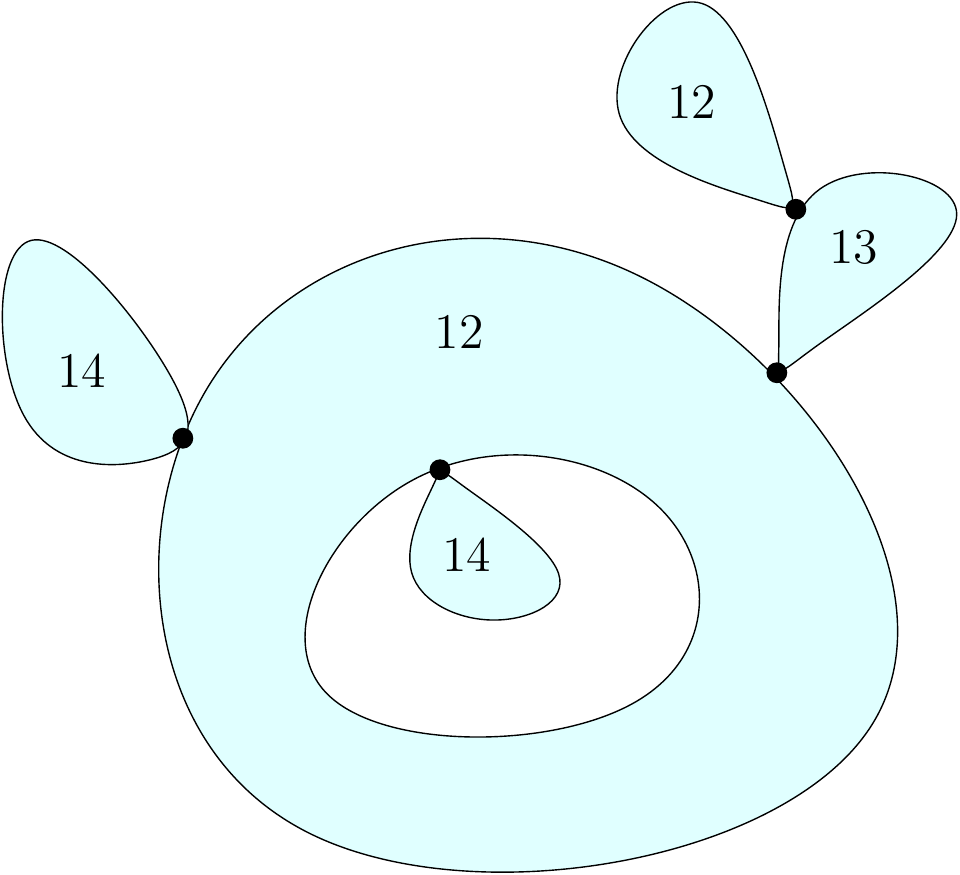}
\caption{Structure of a maximal maps for three 2-cyclic bubbles in $D=4$ (the blobs are planar components).}
\label{fig:Cactus}
\end{figure}

\

\noindent{\it Topology of maximal maps}

\

Before counting maximal maps, we prove that they have the topology of the 4-sphere. We first consider a planar component $1i$. In the colored graph picture, an edge corresponds to a 2-pair $\pi$ with internal colors $j$ and $k$ both different from $i$. Deleting every edge of color $j$ and $k$, we are left with a collection of cycles alternating color-0 edges and parallel edges of color 1 and $i$. As the component is planar, we know that $\I(e)=2$, which means that the vertices of $\pi$ belong to two different connected components (which are both 2-spheres). The pair $\pi$ is therefore an $h$-dipole, and the unhooking $e$ is a flip in the colored graph, which leaves the topology unchanged. Unhooking edges we end up with a tree, which, combining  Prop.~\ref{prop:CycBubTopo} and Cor.~\ref{coroll:TreeTopo}, has the topology of the 4-sphere. We now consider a maximal map. As the incidence relations between the planar components $1i$ are tree-like, there is one such components which  only interacts with the rest of the map at a  black vertex. There exist two corners around that black vertex such that when splitting this vertex as described in Figure~\ref{fig:VertSplit}, we obtain two connected components, one of them being a planar map with edge colors $1i$. The maximal map is therefore the connected sum of a 4-sphere with a smaller maximal map, and recursively, this proves that maximal maps have the topology of the 4-sphere.

\

\noindent{\it Generating function}

\

We denote $\GF_{p_1, \cdots, p_k}(z_2,z_3,z_4)$ the generating function of rooted maximal maps, counted according to their number of bubbles of type $1i$ for each $i$, and such that  white vertices have allowed valencies $p_1, \cdots, p_k$. We denote  $\GP_{p_1, \cdots, p_k}(\lambda_1, \cdots, \lambda_k)$ the generating function of rooted planar bipartite maps such that  white vertices have allowed valencies $\{p_i\}=\{p_1, \cdots, p_k\}$ and no face visits twice the same black vertex, counted according to their number of white vertices of each kind. Using the generalized version of Tutte's bijection (Subsection~\ref{subsec:Tutte}), this is also the generating function of rooted bipartite combinatorial maps with no cycle of length two, such that faces have allowed degrees $2p_1, \cdots, 2p_k$, counted according to their number of faces of each kind. They may be computed using the literature on degree-restricted maps \cite {Ben, BernFusBij, CEAMatrix, BernFusHyp}. 
The root edge belongs to a maximal planar component such that no face visits twice the same black vertex, in which all the edges have the same color set $1i$, and such that on each corner on each black vertex is a (possibly trivial) rooted maximal map. \emph{The generating function of maximal maps therefore satisfies}
\be
\GF_{\{p_i\}}(z_2, z_3, z_4) = 1+ \sum_{i=2}^4\biggl[  \GP_{\{p_i\}}\biggl( z_i \bigl(\GF_{\{p_i\}}(z_2, z_3, z_4)\bigr)^{p_1}, \cdots,  z_i\bigl(\GF_{\{p_i\}}(z_2, z_3, z_4)\bigr)^{p_k}  \biggr) - 1 \biggr].
\ee
If only white vertices of valency $p$ are allowed, then this simplifies to 
\be
\label{eqref:GenFuncNeck}
\GF_{p}(z_2, z_3, z_4) = 1+ \sum_{i=2}^4\biggl[  \GP_{p}\biggl( z_i \bigl(\GF_{p}(z_2, z_3, z_4)\bigr)^{p} \biggr) - 1 \biggr].
\ee

\

\noindent{\it 4D quadrangulations and proliferation of baby universes}

\

Solving such equations can in general be somehow involved. Here we treat the particular case where $p=2$, corresponding to 2-cycles of size 4. In that case, $\GP$ is just the generating function $\GP$ of non-separable planar maps counted according to their number of edges. Non-separable maps are maps such that no face has two corners incident to the same vertex \cite{GoulJack}. 
$\GP$ is given by the system \cite{GoulJack}
\bea
\label{eqref:NonSep}
z&=&u(1-u)^2\\
\GP&=&(3u+1)(1-u).
\eea
 We furthermore consider only two such bubbles, with counting parameters $z_1$ and $z_2$, as the critical behavior is the same.
When $z_1=0$ or when $z_2=0$, maximal maps are precisely the usual planar maps, which are well known, leading to the critical exponent $\gamma=-1/2$. However, there is a narrow interval around $z_1=z_2$ for which there is a ``destructive interference", the critical behavior being that of trees, with critical exponent $\gamma=1/2$.
More precisely, we find in \cite{Johannes} that the tree regime extends  from
\be
\label{eqref:sNeck}
\alpha:=\frac{\lambda_1}{\lambda_2} \in 
\left]\alpha_{c,1},\alpha_{c,2}\right[
,\qquad
(\alpha_{c,1},\alpha_{c,2})=\left(
\frac{5}{4} \left(3-\sqrt{5}\right),\frac{1}{5} \left(\sqrt{5}+3\right)\right)
\,\approx\,(0.95,1.047).
\ee
The generating function has the following expansion near its dominant singularity for $\alpha\in]\alpha_{c,1}, \alpha_{c,2}[$,
\be
\GF(t,\alpha) = \frac{4 \left(-2 \alpha^2+2 \alpha -2 +\sqrt{2 \alpha}(\alpha+1) \right)}{\alpha^2+1}
-\left(\frac{1}{32 \left(\alpha- \sqrt{2\alpha}+1\right)} -t \right)^{1/2}.
\ee
At $\alpha_{c,1}$ and $\alpha_{c,2}$,  the generating function behaves as
\be
\GF(t,\alpha_{c,1}) = \frac{4}{3} \left(\sqrt{5}-1\right) 
-48 \left(3 \left(17 \sqrt{5}-38\right)\right)^{1/3} \left(\frac{\sqrt{5}+3}{96} -t\right)^{2/3}
+ \mathcal{O}\left(\frac{\sqrt{5}+3}{96} -t\right),
\ee
and
\be
\GF(t,\alpha_{c,2}) = \frac{4}{3} \left(\sqrt{5}-1\right) 
-\frac{48 \left(6 \left(5 \sqrt{5}-11\right)\right)^{1/3}}{5^{2/3}} \left(\frac{5}{96} -t\right)^{2/3}
+ \mathcal{O}\left(\frac{5}{96} -t\right).
\ee
We find an \textbf{intermediary critical regime}, with critical exponent
\be
\gamma_{BU}=\frac 1 3,
\ee
which differs from the tree critical exponent $\gamma_T=1/2$ or multi-critical $\gamma_{T,n}=1-1/n$, and from the planar exponent  $\gamma_P=-1/2$ or multi-critical $\gamma_{P,m}=-1/m$. It is argued in \cite{Enhanced} and \cite{SigmaReview} that this exponent is characteristic of the regime for which \textbf{baby-universes proliferate} \cite{Baby}. The $1/3$ exponent has also been found in the context of planar graphs with given 2-connected or 3-connected components \cite{NOY}\footnote{I thank Eric Fusy for pointing this out.}. This regime is obtained in the two-dimensional case using multi-matrix models \cite{IndianBaby,  Korchem, BabyCrit}. As explained in Subsection~\ref{subsec:NonConBub}, this corresponds to gluing non-connected polygons. We stress that in our case, this regime is recovered while selecting maximal graphs obtained by gluing connected bubbles. In this framework,  we see three universality classes in dimension 4 while only the universality class of planar maps appears in dimension 2. 
This regime is recovered in \cite{Enhanced, SigmaReview} by considering a 1-cyclic bubble with color $i$ and a 2-cyclic bubble with colors $1i$, both of length 4. The graphs of this model are in bijection with combinatorial maps with edges carrying color $i$ or $1i$. From Cor.~\ref{coroll:MoreD2Trees},  the edges of color $i$ must be bridges in every maximal maps, and therefore the 2-cyclic bubbles $\B_{1i}$ behave as in $\bS(\B_{1i}, \Om_{1i})$. Consequently, maximal maps are planar and such that color-$i$ edges are bridges, and the generating function satisfies
\be
\GF(z_i,z_{1i})= \GP\bigl(z_{1i}\GF(z_i,z_{1i})^2\bigr) + z_i \GF(z_i,z_{1i})^2.
\ee
When $z_i<<z_{1i}$, it leads to a planar regime with critical exponent $-1/2$, when  $z_i>>z_{1i}$, it leads to a tree regime with critical exponent $1/2$, and for $z_i=3z_{1i}$, it leads to the same intermediary exponent $1/3$.
 To my knowledge, this exponent was first mentioned in \cite{IndianBaby}. In \cite{BabyCrit}, it is argued  that by tuning the theory, one can obtain the more general critical exponents $\gamma=\frac m {m+n+1}$.


\

\subsection{A bubble with toroidal boundary in $D=3$}
\label{subsec:K33}

We consider the complete bipartite graph $K_{3,3}$ with a proper 3-edge-coloring, represented in Figure \ref{fig:CompBip}. It has six different pairings: either vertices inside pairs are all linked by the same color $i=1, 2, 3$, or each one of the three pairs contains a different color. The last three are optimal and symmetrically equivalent. We choose an optimal pairing $\Opt$ and consider the simplified bijection of Thm.~\ref{thm:BijSimp}. The white  vertex corresponding to the bubble is shown on the right of \ref{fig:CompBip}, it is not embedded, as the  colored cycles of $\BOM$ are of length 1 or 2. 
\begin{figure}[!h]
\centering
\includegraphics[scale=.7]{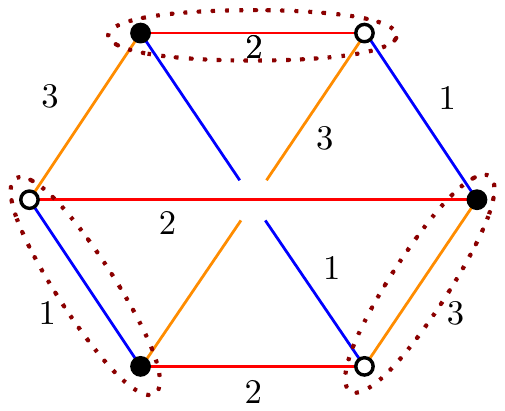} \hspace{3cm}\raisebox{2.7ex}{\includegraphics[scale=.7]{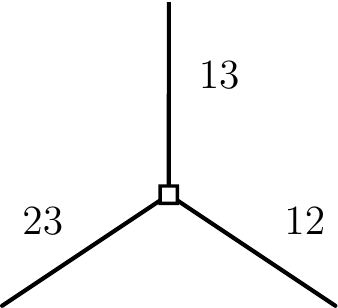} }
\caption{
The $K_{3,3}$ bubble and an optimal pairing, and the corresponding vertex.
}
\label{fig:CompBip} 
\end{figure}

Here, every pair has less than $D/2$ colors, so that Lemma~\ref{prop:hD2Trees} does not apply. We thus focus on Corollary~\ref{Coroll:LstarLiTree} instead (with $L(\Ga)=L(\Ga^\star)$), and try to figure out whether
\be
3L(\Ga)-2\sum_{i=1}^D L(\Gai)\ge 0.
\ee
The union of two monochromatic submaps for any two colors $i\neq j \in\{1,2,3\}$ covers $\Ga$,
\begin{equation}
\Gai\cup\Gai[j]=\Ga.
\end{equation}
The union of two forests respectively spanning $\Gai$ and $\Gai[j]$ is a map which spans $\Ga$. We choose two such forests in the following way: first observe that $\Gai\cap\Gai[j]$ is a forest (a cycle would have both colors $(i, j)$ all along which is impossible), which can thus be completed into both a forest $\cT^{(i)}$ spanning $\Gai$ and a forest $\cT^{(j)}$ spanning $\Gai[j]$. This ensures that $\Gai\setminus\cT^{(i)}$ and $\Gai[j]\setminus\cT^{(j)}$ do not have any common edges. This choice of spanning forests gives us the following inequalities,
\be
\label{LiLj}
\forall i\neq j \qquad L(\Ga)=L(\cT^{(i)}\cup\cT^{(j)})+L(\Gai)+L(\Gai[j])\geq L(\Gai)+L(\Gai[j]).
\ee

Summing the three different relations for $i\neq j\in\{1, 2, 3\}$, we obtain that
\be
3L(\Ga)-2\sum_{i=1}^3 L(\Gai)\geq0, 
\ee
so that from Cor.~\ref{Coroll:LstarLiTree}, trees are maximal, and other maximal maps satisfy 
\be
\label{K33DomOrd}
\left\{
\begin{aligned}
&3L(\Ga)=2\sum_{i=1}^3 L(\Gai) 
\\
&g(\Ga^{(1)})=g(\Ga^{(2)})=g(\Ga^{(3)})=0. \\
\end{aligned}
\right.
\ee
From Corollaries~\ref{coroll:TreeMaxAB} and \ref{coroll:scaling} and from (\ref{eqref:TreeMaxAB2}), we \emph{deduce the sought coefficients}
\be
\tilde a_{K_{3,3}}=3, \quad a_{K_{3,3}}=1, \quad\text{and}\quad s_{K_{3,3}} = 1,
\ee
and the bubble-dependent degree of gluings of bubbles with $K_{3,3}$ toroidal boundary is defined as
	\be
	\delta_{K_{3,3}}(\G)=3(1+\nb(G)) - \Phi_0(\G),
	\ee
it is positive and vanishes for maximal configurations, which we want to characterize.

\

We know focus on \emph{characterizing the full set of maximal maps} of $\bS(\B_{K_{3,3}},\Opt)$. Because of \eqref{LiLj}, the first constraint in \eqref{K33DomOrd} is equivalent to $L(\Gai\cup \Gai[j])=0$ for all $i\neq j$ which from \eqref{LiLj} leads to the system
\be
\label{K33VarCond}
L(\Ga^{(1)}) + L(\Ga^{(2)}) = L(\Ga), \qquad L(\Ga^{(1)}) + L(\Ga^{(3)}) = L(\Ga), \qquad L(\Ga^{(2)}) + L(\Ga^{(3)}) = L(\Ga),
\ee
whose solution is
\be
\label{K33VarVarCond}
2L(\Ga^{(1)}) = 2L(\Ga^{(2)}) = 2L(\Ga^{(3)}) = L(\Ga).
\ee
We now study the case $L(\Ga)=2$, for which \eqref{K33VarVarCond} rewrites
\be
L(\Ga^{(1)})=L(\Ga^{(2)})=L(\Ga^{(3)})=1,
\ee
and identify the bridgeless solutions. There are three distinct cycles $C_1, C_2, C_3$, one for each color, but only two independent cycles. It means that every cycle is the symmetric difference of the other two. $\Ga$ thus has the structure of a Theta graph, i.e. two nodes with three segments between them (middle and right of Fig.~\ref{fig:DomCells}). Each segment must be part of two cycles, meaning that the edges of a segment all have the same couple of colors. Bipartiteness and the structure of the white vertex prevent any map from having a chain with more than two consecutive edges with the same couple of colors. Therefore, each segment has one or two edges, with the same colors. The allowed maps are thus restricted to those shown in Figure \ref{fig:DomCells}.

\begin{figure}[!h]
\centering
\includegraphics[scale=1.3]{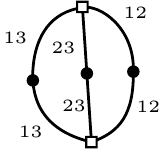}\hspace{3cm}\includegraphics[scale=1.5]{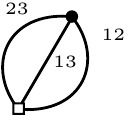} \hspace{3cm}\includegraphics[scale=1.5]{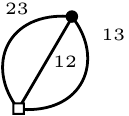} 
\caption{Bridgeless dominant maps with two fundamental cycles.}
\label{fig:DomCells} 
\end{figure}

We now prove by induction on the circuit-rank $L$ that any map $\Ga$ which has no submap homeomorphic to one of those of Figure \ref{fig:DomCells} and which is not a tree verifies
\be
\label{eqref:InequK33}
3L(\Ga)-2\sum_{i=1}^3 L(\Gai)>0,
\ee
and is therefore not maximal.
We saw that this property is true for $L(\Ga)\leq 2$. We now consider $L>2$ and $\Ga\in\bS(\B_{K_{3,3}},\Opt)$ such that $L(\Ga)=L$. From Prop.~\ref{prop:eD2}, all edges which are not bridges are such that $|\cI_2(e)| = 2$. Let $e_{ij}$ be one of them, with colors $\{i,j\}$. It is not a bridge in any of the two monochromatic submaps it is contained in. Let us unhook that edge, which leads to a map $\Ga'$ with 
\be
L(\Ga') = L(\Ga) - 1,\qquad\text{ and }\qquad \sum_{i=1}^3 L(\Ga'^{(i)}) = \sum_{i=1}^3 L(\Gai) - 2,
\ee
 so that 
 \be
 3 L(\Ga) - 2 \sum_{i=1}^3 L(\Gai) = 3 L(\Ga') - 2\sum_{i=1}^3 L(\Ga'^{(i)}) - 1.
 \ee

As $e_{ij}$ is not a bridge in $\Gai$ and $\Gai[j]$, the edges $e_{jk}, e_{ik}$ incident to the same white vertex are not bridges respectively in $\Gai$ and $\Gai[j]$. Indeed, $e_{ij}$ belongs to two distinct cycles, one containing $e_{ik}$ and whose edges all contain the color $i$ and one containing $e_{jk}$ and whose edges all contain the color $j$. Let us denote those two cycles $C_i=(e_{ik}, \dotsc, e_{ij})$, and $C_j=(e_{jk}, \dotsc, e_{ij})$, and $e'_{ij}$ the first edge they have in common (which might be $e_{ij}$). Then the concatenation of the two chains $(e_{ik}, ...,e'_{ij})\subset C_i $ and $(e_{jk}, ...e'_{ij})\subset C_j $, with $e'_{ij}$ excluded, is also a cycle in $\Ga'$. This implies that $e_{ik}$ and $e_{jk}$ are not bridges in $\Ga'$. Furthermore, $e_{ij}$ is now attached to a leaf in $\Ga'$, so that $e_{ik}$ and $e_{jk}$ are respectively bridges in $\Ga'^{(i)}$ and $\Ga'^{(j)}$. When unhooking, say, $e_{ik}$ from its black extremity, we get a map $\Ga''$ such that $L(\Ga'')=L(\Ga')-1$ and $\sum_{i=1}^3 L(\Ga''^{(i)})=\sum_{i=1}^3 L(\Ga'^{(i)})-\epsilon$, with $\epsilon\in\{0,1\}$. It comes that
\bea
3L(\Ga)-\sum_{i=1}^3 L(\Gai) = 3 L(\Ga') - 2\sum_{i=1}^3 L(\Ga'^{(i)}) - 1 &=& 3L(\Ga'')-2\sum_{i=1}^3 L(\Ga''^{(i)})+2(1-\epsilon) \nonumber\\
&\ge& 3L(\Ga'')-2\sum_{i=1}^3 L(\Ga''^{(i)}).
\eea
Notice that $\Ga''$ is not a tree as $L(\Ga)>2 \Rightarrow L(\Ga'')>0$. Moreover, unhooking $e_{ij}$ and $e_{ik}$ cannot create a submap homeomorphic to one in Figure~\ref{fig:DomCells}. The induction hypothesis thus applies to $\Ga''$ from which we conclude that $3L(\Ga)-\sum_{i=1}^3 L(\Gai) >0$.
\begin{figure}[!h]
\centering
\includegraphics[scale=0.7]{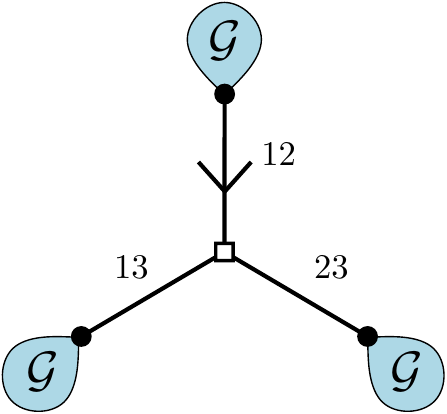}\hspace{2cm}
\raisebox{1.7ex}{\includegraphics[scale=0.75]{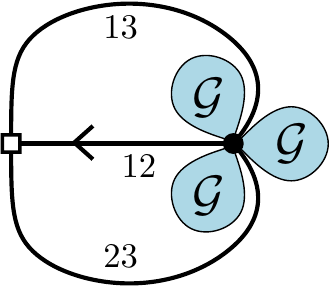} \hspace{2cm}
\includegraphics[scale=0.75]{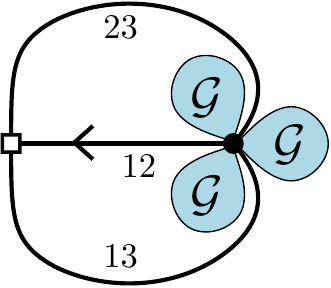} }
\caption{ Size one submaps.}
\label{fig:K33Deg3Vert}
\end{figure}

\

The submaps of Figure~\ref{fig:DomCells} cannot be arranged in any possible way. In order to satisfy the second constraint of \eqref{K33DomOrd}, a bubble is locally as in the six cases pictured in Fig.~\ref{fig:K33Deg3Vert}  and \ref{fig:K33Deg6Vert}. Indeed, it is clear for big Theta submaps, as a color-$i$ submap around a black vertex is just a collection of embedded cycles sharing a black vertex (a ``rosette") which is planar because $g(\Gai[1])=0$.
Concerning  small Theta submaps, suppose that around a black vertex, an edge $e^1$ of a small Theta submap $\Theta_1$ is between two edges $e_a$ and $e_b$ of some other small Theta submap $\Theta_2$. These two edges share a color, e.g. 1. As $g(\Gai[1])=0$,  $\Gai[1]$ is locally as a planar rosette with tree insertions (which we don't need to worry about). If $e^1$ also has color 1, the other edge of $\Theta_1$ has to be in the same part of the plane delimited by $e_a$ and $e_b$ in $\Gai[1]$ as $e^1$. Applying this to the three pairs of edges of $\Theta_2$, the three edges of $\Theta_1$ cannot intersect an edge of $\Theta_2$, so that Theta submaps are as in Fig.~\ref{fig:K33Deg3Vert}. If $e^1$ does not have color 1, then $\Theta_1$ and $\Theta_2$ should be as in Fig.~\ref{fig:K33Deg3Vert}.
As a consequence, a maximal map $\Ga$ of $\bS(\B_{K_{3,3}},\Opt)$ has its monochromatic submaps planar and is  a $\bT$-tree-like family (Subsection~\ref{subsec:TreeLike}), where $\bT$ comprises the maps obtained from Fig.~\ref{fig:K33Deg3Vert}  and \ref{fig:K33Deg6Vert} by putting a trivial $\GF=1$ for each $\GF$.

\begin{figure}[!h]
\centering
\includegraphics[scale=0.7]{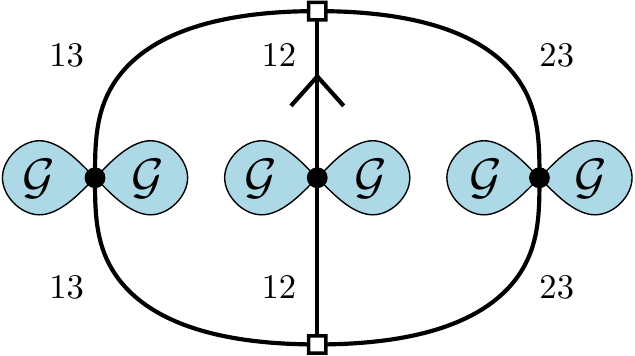}\hspace{1cm}
\includegraphics[scale=0.75]{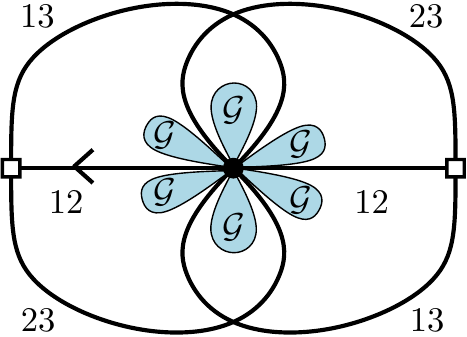} \hspace{1cm}
\includegraphics[scale=0.75]{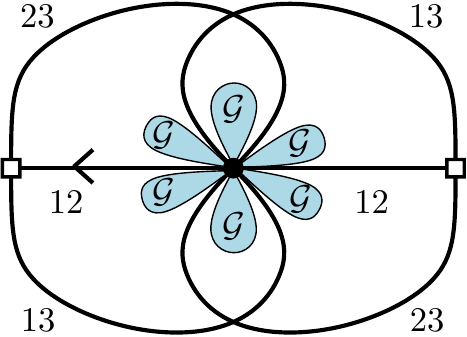} 
\caption{Size two submaps.}
\label{fig:K33Deg6Vert} 
\end{figure}

As explained in the section detailing tree-like families, there is a simple bijection with trees with 3 kinds of valency 3 vertices and 3 kind of valency 6 vertices. It maps the example in \ref{fig:ExDomOrdK33} to a tree with respectively (5, 1, 1) vertices coming from the  (left, middle, right) of Fig.~\ref{fig:K33Deg3Vert}, and (2, 1,0) vertices from the  (left, middle, right) of Fig.~\ref{fig:K33Deg6Vert}.
The bijection extends to non-maximal maps by forbidding the patterns of Figs.~\ref{fig:K33Deg3Vert} and~\ref{fig:K33Deg6Vert}.

\begin{figure}[!h]
\centering
\includegraphics[scale=1.2]{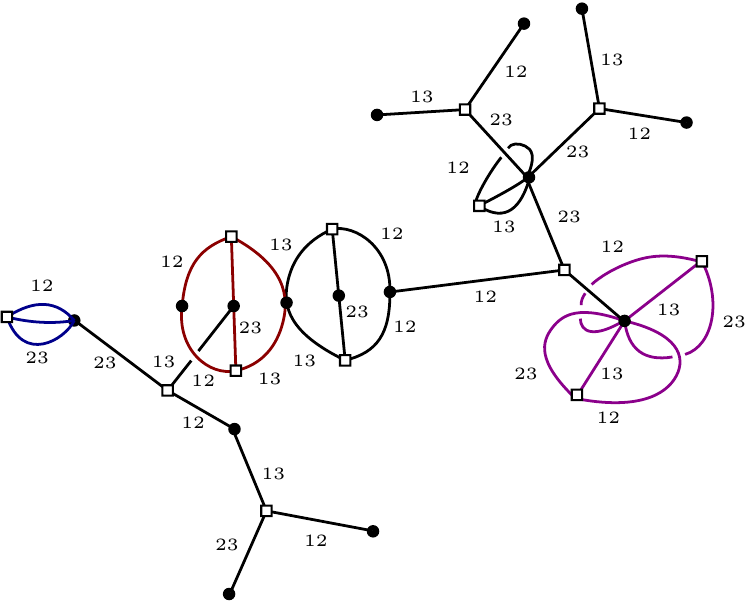} 
\caption{This is an example of a maximal map.}
\label{fig:ExDomOrdK33} 
\end{figure}

To count maximal maps, we decide (in the colored graph picture) to mark a color 0 edge such that its white extremity is the color-3 edge of the optimal pairing, or equivalently, in the map picture, to orient a color 12 edge from a black vertex to a bubble. The generating function of maximal maps rooted in this way and counted according to their number of bubbles satisfies the tree relation
\be
\GF(z)=1+ 3z \GF(z)^3 + 3z^2 \GF(z)^6.
\ee
The generating function for any marked color-0 edge is recovered by replacing $z$ with $3z$. We computed the solution to this equation in \cite{Johannes}. It satisfies $\GF(0)=1$ has the square-root expansion
\be 
\GF(z) = c - \alpha \sqrt{1-z/{z_c}}+O(1-z/{z_c})
\ee
which characterizes tree-regimes, with critical exponent $\gamma=1/2$. We can compute 
\bea
z_c &=& \frac{1}{162} \biggl[164 + \sqrt{2 \left(1334 -939528 \sqrt[3]{2}/Q +9\ 2^{2/3} Q\right)}\nonumber\\
&& -\sqrt{2 \left(2668 + 939528 \sqrt[3]{2}/Q -9\ 2^{2/3} Q+219784 \sqrt{\frac{2}{1334 -939528 \sqrt[3]{2}/Q +9\ 2^{2/3} Q }}\right)} \biggr]\nonumber\\
&\approx& 0.0144
\eea
where $Q=\sqrt[3]{1293487 \sqrt{377}-7870587}$ and 
\be
c \approx 3.14604 \quad,\quad \alpha \approx 0.378154
\ee
can be given only numerically as they involve sixth roots.

\

 {\bf The other order 6 bipartite bubbles in $D=3$ are all melonic}, and therefore are cut-bubbles (Def.~\ref{def:CutBub}) in every maximal map, when building the bijection with optimal pairings. There are three 1-cyclic bubbles, and three other melonic bubbles (similar to that on the left of Fig.~\ref{fig:ExSext4D}, which only differ by the coloring. The generating function of $D=3$ hexangulations with one marked color-0 edge counted according to their number of bubbles therefore satisfies
\be
\GF(z)=1+ 27z \GF(z)^3 + 9z^2 \GF(z)^6.
\ee

\subsection{Sextic bubbles in $D=4$}
\label{subsec:K334}


In this section, we characterize maximal maps for every size 6 bubbles in dimension four. The critical behaviors obtained are the same as for the size 4 interactions in $D=4$.

\subsubsection{Simple examples}

The order 6 bipartite bubbles in $D=4$ include two kind of melonic bubbles, the 1-cyclic ones (4 different colorings) and the kind on the left of Fig.~\ref{fig:ExSext4D}. When building the bijections with optimal pairings, melonic bubbles are cut-bubbles (Def.~\ref{def:CutBub}) in every maximal map. There is a 2-cyclic bubbles of size 6, which maximal maps have been characterized in Sec.~\ref{Subsec:D2CycBub}. There is a bubble obtained by adding edges parallel to an optimal pairing of the $K_{3,3}$ bubble, shown in the middle of Fig.~\ref{fig:ExSext4D}. Because of Proposition~\ref{prop:AddColOpt}, we know that trees are the only maximal maps. For this bubble, the coefficients $\tilde a$, $a$, and $s$, are respectively  5, 7/3, and 1, and the bubble-dependent degree is defined as $\delta=4+5b-\Phi_0$.
\begin{figure}[h!]
\centering
\raisebox{3ex}{\includegraphics[scale=.6]{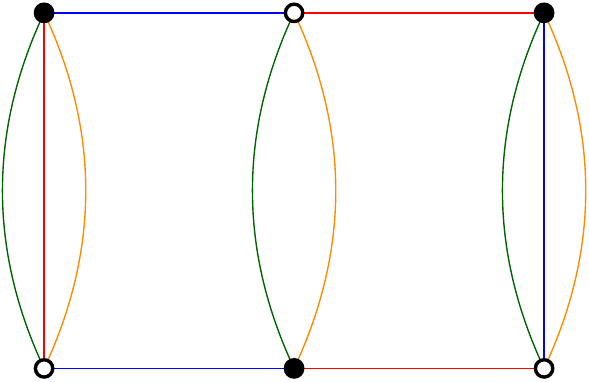}} \hspace{1.5cm}\includegraphics[scale=.75]{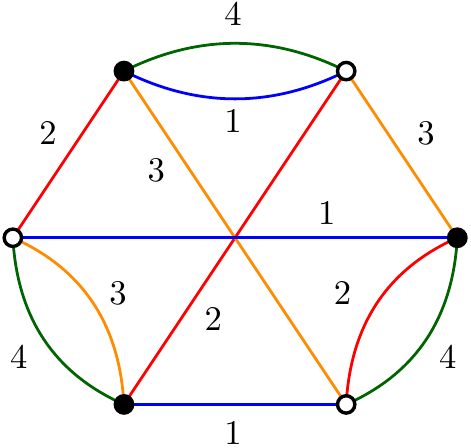} 
\hspace{1.5cm}\includegraphics[scale=.7]{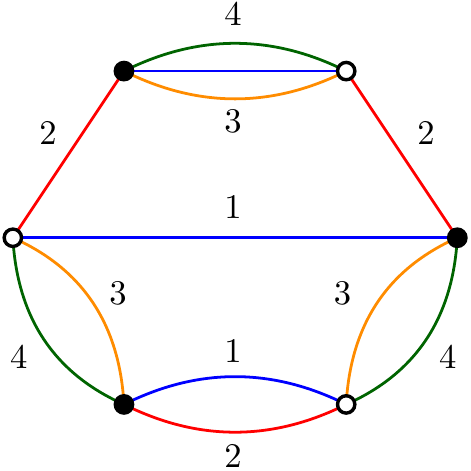} 
\caption{
Order 6 melonic and non-melonic bubbles, and a 2-cyclic bubble with additional 3-pair.}
\label{fig:ExSext4D}
\end{figure}

Then, there are the bubbles as on the right of Fig.~\ref{fig:ExSext4D}. They contain a 3-pair, and we can therefore proceed as explained in the section on large $h$-pairs. When contracting the 3-pair of a bubble $\B_6$ of this kind, a 2-cyclic bubble $\B_4$ of size 4 is recovered. From \eqref{eqref:TildeABpi}, \eqref{eqref:ABpi} and \eqref{eqref:SBpi}, we can deduce the coefficients $\tilde a$, $a$, or $s$ from those of $\B_4$, which are 2, 5/2, and 1; obtaining 5, 8/3, and 2. On the example of Fig.~\ref{fig:ExSext4D}, we choose an optimal pairing $\Om_{\B_6}$, e.g. the one which pairs the vertices linked by colors 3 and 4. This induces an optimal pairing $\Om_{\B_4}$ of $\B_4$ (here it is obvious, however we have proven it in Lemma~\ref{lemma:LargeHPairPairing}). We know that maximal maps in $\bS(\B_4,\Om_{\B_4})$ are planar combinatorial maps, with edges carrying color 2 and 4. To recover the maximal maps in $\bS(\B_6,\Om_{\B_6})$, we need to add a white vertex on each edge, and a color 2 bridge between this edge and some other maximal component. There are 2 ways of doing so, as there are two sides on the color 24 edge. 
For the counting, we root maximal maps at a color 24 edge oriented from a black to a white vertex.
A maximal map of $\bS(\B_6,\Om_{\B_6})$ decomposes as a non-separable \eqref{eqref:NonSep} planar map of color 34 containing the root, with one generating function per black corner and one generating function per edge, that can be added in two possible ways.
The generating equation of maximal maps in $\bS(\B_6,\Om_{\B_6})$, rooted as stated, and counted according to their number of bubbles therefore satisfies 
\be
\GF(z)=\GP(2z\GF(z)^3),
\ee
where $\GP$ is the generating function of non-separable rooted planar maps, counted according to their number of edges. The critical behavior is expected to be that of planar maps. The treatment is similar when considering all kinds of colorings of this bubble together, leading to a relation similar to the one obtained for $D/2$ cyclic bubbles of different kinds \eqref{eqref:GenFuncNeck}.

\subsubsection{A size 6 bubble in $D=4$}
\label{subsec:K334_3}
%
\begin{figure}[h!]
\centering
\includegraphics[scale=.7]{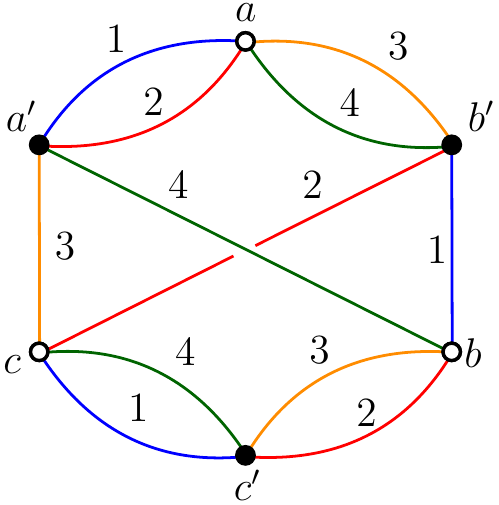} \hspace{3cm}\raisebox{3ex}{\includegraphics[scale=.7]{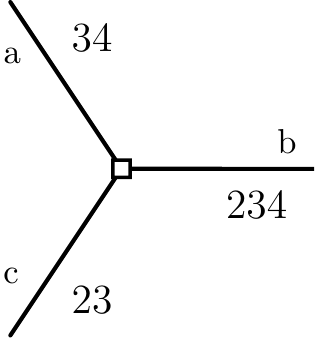} }
\caption{
An order 6 bubble, and the corresponding vertex.
}
\label{fig:Meander} 
\end{figure}
We consider the bubble $\B$ in Figure \ref{fig:Meander}. It has four different optimal pairings, among which $\Om = \{(a,a'),(b,b'),(c,c')\}$ ($\Phi_0(\BCO)=8$), which we choose to build the bijection. It satisfies the conditions of the bijection of Thm.~\ref{thm:BijSimp}, and we therefore study the corresponding bipartite maps, which edges have color sets 23, 34, and 234. White squares have valency 3, and the cyclic ordering of edges around white squares is given by color 3, which appears on every edge, and is $(e_{34}, e_{23}, e_{234})$ (A color-3 edge oriented from  black to white is a counterclockwise corner around a white square vertex). 
Using Prop.~\ref{prop:hD2Trees}, as edges have $D/2$ colors or more, trees belong to maximal maps, and according to Corollaries~\ref{coroll:TreeMaxAB} and \ref{coroll:scaling} and to (\ref{eqref:TreeMaxAB2}),
\be
\tilde a_\B=4, \qquad a_\B=\frac 7 3, \quad\text{and}\quad s_\B = 2,
\ee
where we used $\Phi(\B)=10$. Again, the bubble-dependent degree is defined as 
	\be
	\delta_{\B}(\G)=4(1+\nb(G)) - \Phi_0(\G).
	\ee
We wish to characterize other maximal maps. We could study directly the circuit-ranks and genera of bicolored submaps, as done in \cite{SWM}. However, there is a simpler way of solving this example by noticing that there is a 4-cut of edges with 4 different colors. The bubble is therefore the connected sum of two necklaces, which we denote $\B_1$ and $\B_2$ (Fig.~\ref{fig:Meander0}). Note that regarding our choice of pairing, it is color 3 which is never included in the pairs, and which plays the role of color 1 in the section on 2-cyclic bubbles \ref{Subsec:D2CycBub}. We apply results from Section~\ref {subsec:ConecSum} on bubbles which are connected sums of two smaller bubbles (we could have derived the coefficients $\tilde a$, $a$, and  $s$ from \eqref{eqref:TildeACon}, \eqref{eqref:ACon}, and \eqref{eqref:SCon}. From a map in $\bS(\B,\Om)$, we recover a map of $\bS(\B_1, \B_2,\Om_1, \Om_2)$ by performing the operation of Fig.~\ref{fig:Meander2} on each white vertex corresponding to a bubble. We obtain maps with edges either carrying color set  34, or 23, and such that there is a partition of the edges in pairs of edges sharing a corner, with counterclockwise order 34, 23, and with the same 0-score. Conversely, any map satisfying this property gives back a map in $\bS(\B_1, \B_2,\Om_1, \Om_2)$ with the same 0-score. 
\begin{figure}[h!]
\centering
\includegraphics[scale=.7]{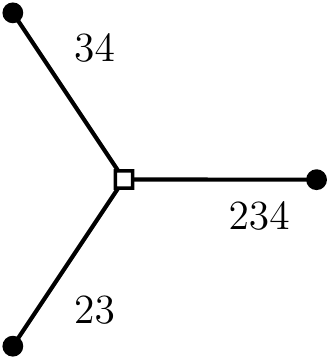}\hspace{1.5cm}\raisebox{+7ex}{$\rightarrow$}\hspace{1.5cm}\includegraphics[scale=.7]{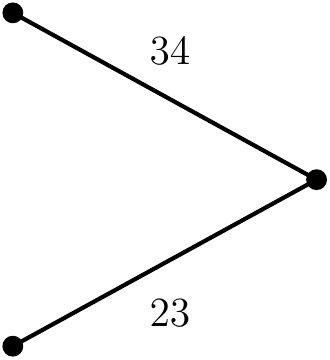} 
\caption{
Splitting each bubble into two necklaces.
}
\label{fig:Meander2} 
\end{figure}
We know  that the 0-score of a map in $\bS(\B_1, \B_2,\Om_1, \Om_2)$ is maximal iff the map is planar, and such that there is no cycle with both an edge 34 and an edge 23.
The set of maximal maps in $\bS(\B,\Om)$ is therefore the preimage of the subset of maximal maps in $\bS(\B_1, \B_2,\Om_1, \Om_2)$ for which there is a partition of the edges in pairs of edges sharing a corner, with counterclockwise order 34, 23. Indeed, such maps exist, and the  0-score of such a map in  $\bS(\B_1, \B_2,\Om_1, \Om_2)$ is higher or equal to that of any other map in  $\bS(\B_1, \B_2,\Om_1, \Om_2)$ such that there is a partition of the edges in pairs of edges sharing a corner, with counterclockwise order 34, 23. In particular, it is strictly higher than the 0-score of the images of maps in $\bS(\B,\Om)$  which do not belong to maximal maps in $\bS(\B_1, \B_2,\Om_1, \Om_2)$.
We do not count maximal maps explicitly here. However, the characterization of maximal maps is very close to that of 2-cyclic bubbles in $D=4$. More precisely, it corresponds to the case  $\lambda_1=\lambda_2$ ($\alpha=1$) in \eqref{eqref:sNeck}, with a hard dimer configuration on the color-0 edges: in the colored graph picture, the color-0 edges resulting from the pair insertions on the edge-cuts inside each bubble can be thought as \emph{occupied}. Here however, every bubble should be incident to a dimer, and dimers can only link the two different 2-cyclic bubbles. The $\alpha=1$ case in \eqref{eqref:sNeck} leads to a tree behavior $\gamma=1/2$, and coupling the system to hard-dimers is expected to lead to a tree-like behavior, or, if the system is fine-tuned, to multi-critical behaviors $\gamma=\frac{n-1}n$.

\subsubsection{Another size 6 bubble in $D=4$}
\label{subsec:K334_2}
%
\begin{figure}[!h]
\centering
\includegraphics[scale=.7]{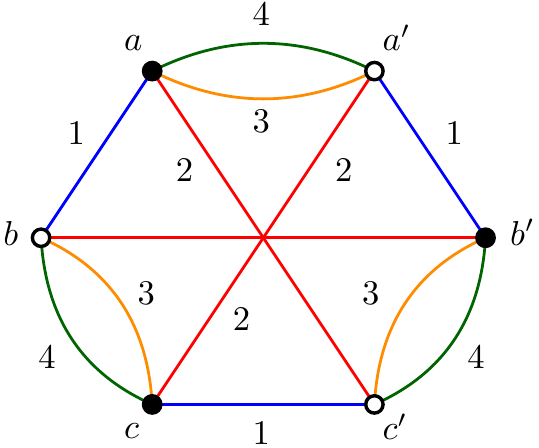} \hspace{3cm}\raisebox{2.7ex}{\includegraphics[scale=.7]{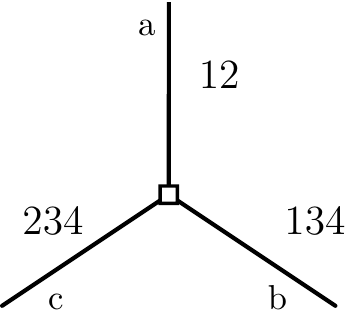} }
\caption{
An order 6 bubble, and the corresponding vertex.}
\label{fig:K334_2} 
\end{figure}
We consider the bubble $\B$ on the left of Figure~\ref{fig:K334_2}. It has four optimal pairings, and we choose $\Om = \{(a,a'),(b,b'),(c,c')\}$ ($\Phi_0(\BCO)=8$) to build the bijection. It satisfies the conditions of the bijection of Thm.~\ref{thm:BijSimp}, and consequently we  focus on the corresponding bipartite maps. The simplified vertex is shown on the right of Figure~\ref{fig:K334_2}: the edges have color sets 12, 234, and 134, white squares have valency 3, and we do not embed these vertices, as all colored cycles of $\BOM$ are of length 1 or 2. 
Here we cannot use Prop.~\ref{prop:hD2Trees} directly. However, the trick here will be to notice that deleting the edges of color 4 in the colored graph we recover the $K_{3,3}$ bubble, and to use the  results of Sec.~\ref{subsec:K33}. We know that when picking any two colors in $\{1,2,3\}$ or in $\{1,2,4\}$, \eqref{LiLj} is satisfied. The quantity $L(\cT^{(i)}\cup\cT^{(j)})$ does not depend on the chosen spanning trees, and we denote
\be
L_{\overline{ij}}(\Ga)=L(\Ga)-L(\Gai)-L(\Gai[j]).
\ee
In particular, we will use the two following properties:
\be
\label{eqref:L12B}
L(\Ga)\ge L(\Gai)+L(\Gai[j]),
\ee
for any $i\neq j\in\{1,2,3,4\}$; and for $i\neq j\in\{1,2,3\}$,
\be
\label{eqref:L12C}
L_{\overline{ij}}(\Ga)>0\quad \Rightarrow \quad \biggl[\substack{{\textrm{There exist some edge in } \Ga \textrm{ which is a cut-edge in }}\\{\textrm{both  }\, \Gai {\rm\ and\ } \Gai[j] \textrm{ but is not a cut-edge in } \Ga}}\biggr].
\ee
As edges containing color 3 or 4 are the same edges, $\Gai[3]=\Gai[4]$. Using that $L(\Gai[3])=L(\Gai[4])$, we rewrite the part of \eqref{eqref:PhivsTrees} depending on the circuit-ranks
\be
4L(\Ga)-2\sum_{i=1}^4L(\Gai) = 2L_{\overline{13}}(\Ga)+2L_{\overline{23}}(\Ga)\ge0.
\ee
From Cor.~\ref{Coroll:LstarLiTree}, trees are maximal, and other maximal maps satisfy 
\be
\label{eqref:CondK334}
L_{\overline{13}}=L_{\overline{23}}=0,\quad \textrm{ and }\quad \forall i,\ g_i=0.
\ee
From Corollaries~\ref{coroll:TreeMaxAB} and \ref{coroll:scaling} and from (\ref{eqref:TreeMaxAB2}), we compute 
\be
\tilde a_\B=4, \qquad a_\B=2, \quad\text{and}\quad s_\B = 2,
\ee
where we used $\Phi(\B)=8$. The bubble-dependent degree is defined as 
	\be
	\delta_{\B}(\G)=4(1+\nb(G)) - \Phi_0(\G),
	\ee
which from \eqref{eqref:DegGenTree} rewrites
\be
\delta_{\B}(\G)=2L_{\overline{13}}(\Ga)+2L_{\overline{23}}(\Ga).
\ee
We focus on characterizing the other maximal maps. 
There are two cases. If in addition $L_{\overline{12}}=0$, then \eqref{K33VarCond} is satisfied for colors 1, 2 and 3, and we know from  Sec.~\ref{subsec:K33} that the map is a maximal map of $\bS(\B_{K_{3,3}}, \Opt)$, so we know how it is. If not, $L_{\overline{12}}(\Ga)>0$, which means that there exists an edge $e$ which is not a cut-edge, but is so in both $\Gai[1]$ and $\Gai[2]$. Unhooking $e$, we obtain a new diagram $\Gae$ such that, 
\be
L(\Gae)=L(\Ga)-1,\qquad
L(\Gae^{(1)})=L(\Gai[1]),\quad \text{and}\quad
L(\Gae^{(2)})=L(\Gai[2]).
\ee
Using (\ref{eqref:L12B}) and (\ref{eqref:CondK334}), 
\be
\label{eqref:L13}
0\le L(\Gae)-L(\Gae^{(1)})-L(\Gae^{(3)})=L(\Ga)-1-L(\Gai[1])-L(\Gae^{(3)})=L(\Gai[3])-L(\Gae^{(3)})-1.
\ee
Furthermore,  $L(\Gai[3])-L(\Gae^{(3)})$ is 0 or 1, but the first value contradicts (\ref{eqref:L13}), so that 
\be
L(\Gae^{(3)})=L(\Gai[3])-1,
 \ee
 so that $\Ga$ is obtained from $\Gae$ by hooking an edge of color $i34$ to create a cycle such that $\Gai[3]=\Gai[4]$ stays planar, and such that it is a cut-edge in $\Gai$. We still have $L(\Gae)\ge L(\Gae^{(1)}+L(\Gae^{(2)})=L(\Gai[1])+L(\Gai[2])$, so there are two cases. Either $L(\Gae)> L(\Gae^{(1)})+L(\Gae^{(2)})$, in which case we repeat the same steps, either $L(\Gae)= L(\Gae^{(1)})+L(\Gae^{(2)})$, in which case $\Gae$ is a tree. From this tree, $\Ga$ is obtained by hooking a certain number of edges $i34$ so that every edge of color $12$ stays a cut-edge and such that the overall map remains planar.
 
 \
 
Maximal maps are quite similar to that of the example in Figure~\ref{fig:ExSext4D}, with the difference that the tree part has three valency 3 vertices and three valency 6 vertices. We root maximal maps at a color 134 or 234 edge oriented from a black to a white vertex. 
%
We denote $\GF$ the generating function of maximal maps in $\bS(\B,\Opt)$, rooted as stated and counted according to their number of bubbles.
  There are two cases. In the first case, the root is either a bridge of color 134 or of color 234, and  it belongs either to one of the three degree 3 vertices or to one of the three valency 6 vertices. The contribution is $6 z\GF(z)^3  + 6 z^2 \GF(z)^6$. In the other case,  the root is not a bridge, in which case it belongs to a non-separable planar map of color 34 containing the root \eqref{eqref:NonSep}, and we need to add one generating function per black corner of the non-separable component and one generating function per edge - corresponding to the color-12 bridge -  that can be added in two possible ways. Indeed, a white vertex is added on every edge, and a color-12 bridge is added between that white square and a maximal submap, and there are two ways of deciding which one of the two color 34 segments has color set 234 and which one has colors 134. The orientation of the color 34 root edge translates into an orientation of either the  color 234 edge or the color 134 edge, so that the two orientations coincide and  so that the edge is oriented from a black to a white vertex. If $\GP$ is the generating function of rooted non-separable planar maps counted according to their number of edges, this second possibility contributes as $\GP\left(2 z \GF ^3\right)$. It generates a tree part $2z\GF^3$ which has to be subtracted from the overall tree part.
The generating function $\GF$ therefore satisfies 
\be
\GF(z) = \GP\left(2 z \GF ^3\right)+4 z\GF(z)^3  + 6 z^2 \GF(z)^6 
\ee

\begin{figure}[!h]
\centering
\includegraphics[width=7.5cm]{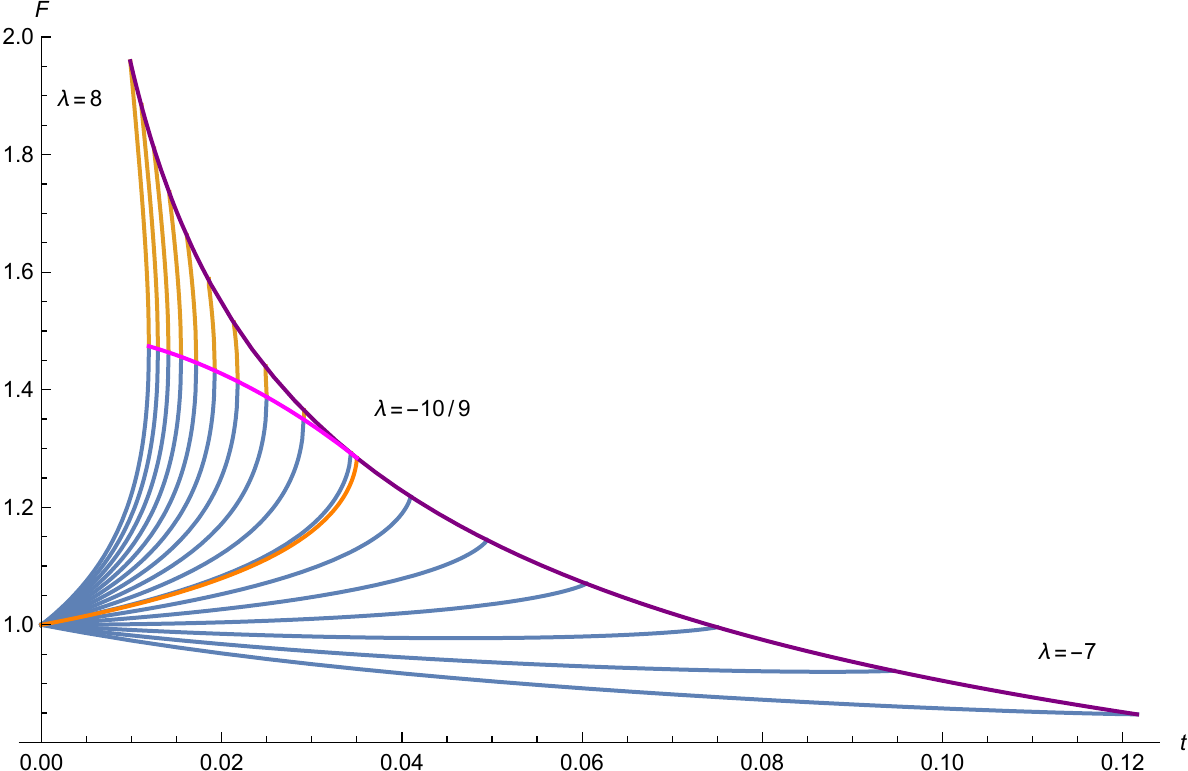}
\caption{Generating functions $\GF(t,\lambda)$ for $\lambda=-7,-6,...,7,8$ and for $\lambda=\lambda_c$. The purple curve shows the loci of planar critical points valid for $\lambda<\lambda_c$ while the magenta curve indicates the critical points of the tree regime for $\lambda>\lambda_c$.
}
\label{fig:4K33mel}
\end{figure}

We have counted maximal maps of $\bS(\{\B, \B_{\text{melo}}\},{\Opt}_{\bB})$  in \cite{Johannes}, where $\B_{\text{mel}}$ is a 1-cyclic bubble of size 6. We denote $\lambda_\text{mel}$ the corresponding counting parameter and $t$ the parameter counting the total number of edges. Adding the melonic bubble simply modifies the cubic term
\be
\GF(z) = \GP\left(2 t \GF ^3\right)+4 \lambda t \GF(z)^3  + 6 t^2 \GF(z)^6,
\ee
where $\lambda=4+\lambda_3$. The explicit system of equations is
\bea
2 (F t)^3 &=& u (1-u)^2\\
F &=&(3 u+1) (1-u) + \frac{\lambda}{2}  \left(u (1-u)^2\right)+ \frac{3}{2} u^2 (1-u)^4
\eea
The solutions are roots of polynomial equations of order 16. The system becomes singular when the Jacobian vanishes,
\bea
\nonumber
\det \left(
\begin{array}{cc}
 6 t^3 F^2 & 2 (1-u) u-(1-u)^2 \\
 1 & -3 u (1-u)^4+6 u^2 (1-u)^3-\lambda  (1-u)^2+2 \lambda  u (1-u)-3 (1-u)+3 u+1 \\
\end{array}
\right) = 0
\eea
The critical coupling is positive for the non-melonic interaction but negative for the melonic part, 
\be
{\lambda_c}=-\frac{10}{9}  \quad , \quad   t_c(\lambda_c)= \frac{19683}{562432}.
\ee

Expanding the generating function around the singularity for $\lambda = \lambda_c$ gives 
\be
F(t,\lambda_c) = \frac{104}{81} - \frac{104}{729} 13^{\frac2 3}( 1 - t/t_c(\lambda_c))^{\frac2 3} + \mathcal{O}( 1 - t/t_c(\lambda_c))
\ee
with the critical  exponent $2/3$ characterizing the proliferation of baby universes.
For $\lambda > \lambda_c$, explicit results are difficult to obtain since higher order roots are involved.
For $\lambda < \lambda_c$, 
the series expansion at the critical point $t_{c,1} = \frac{531441}{4 (9 \lambda +166)^3}$
is
\bea
&&F(t,\lambda) = \frac{2(166 + 9 \lambda) }{243} 
+ \frac{(9 \lambda +62) (9 \lambda +166)}{243 (9 \lambda +10)}\left( 1 -\frac{t}{t_{c,1}}\right)\hspace{6cm}\\
\nonumber&&\hspace{6cm}+ \frac{\sqrt{2 (9 \lambda +166)^5}}{27\sqrt{-3(9 \lambda +10)^5}} \left( 1 -\frac{t}{t_{c,1}}\right)^{3/2}
+ \mathcal{O}\left( 1 -\frac{t}{t_{c,1}}\right)
\eea

At $\lambda_3=0$, the behavior is tree-like. The results are shown in Fig.~\ref{fig:4K33mel}.


\

\subsection{Gluings of bi-pyramids and more}
\label{subsec:BiPyr}

%
\begin{figure}[!h]
\centering
\includegraphics[scale=.5]{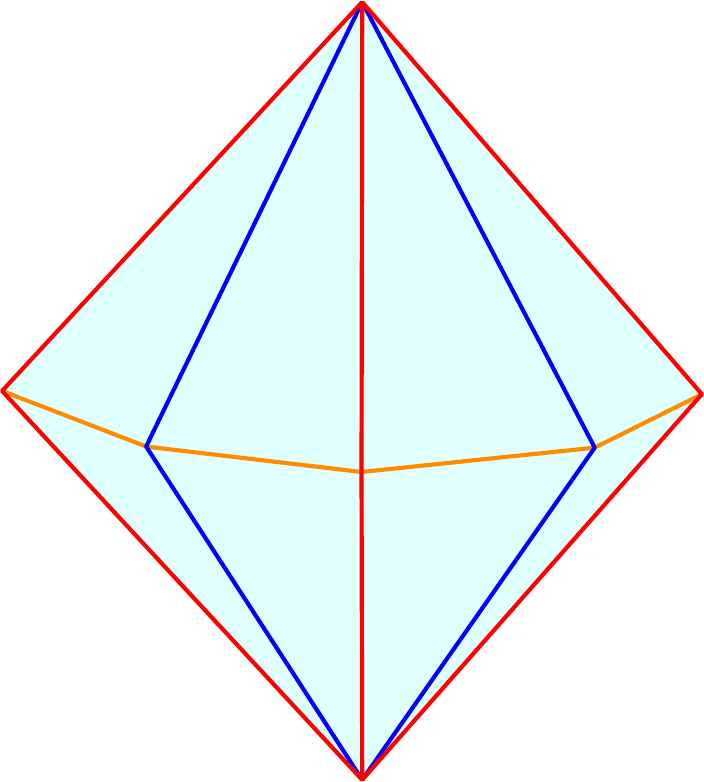} 
\hspace{1cm}\raisebox{0ex}{\includegraphics[scale=.45]{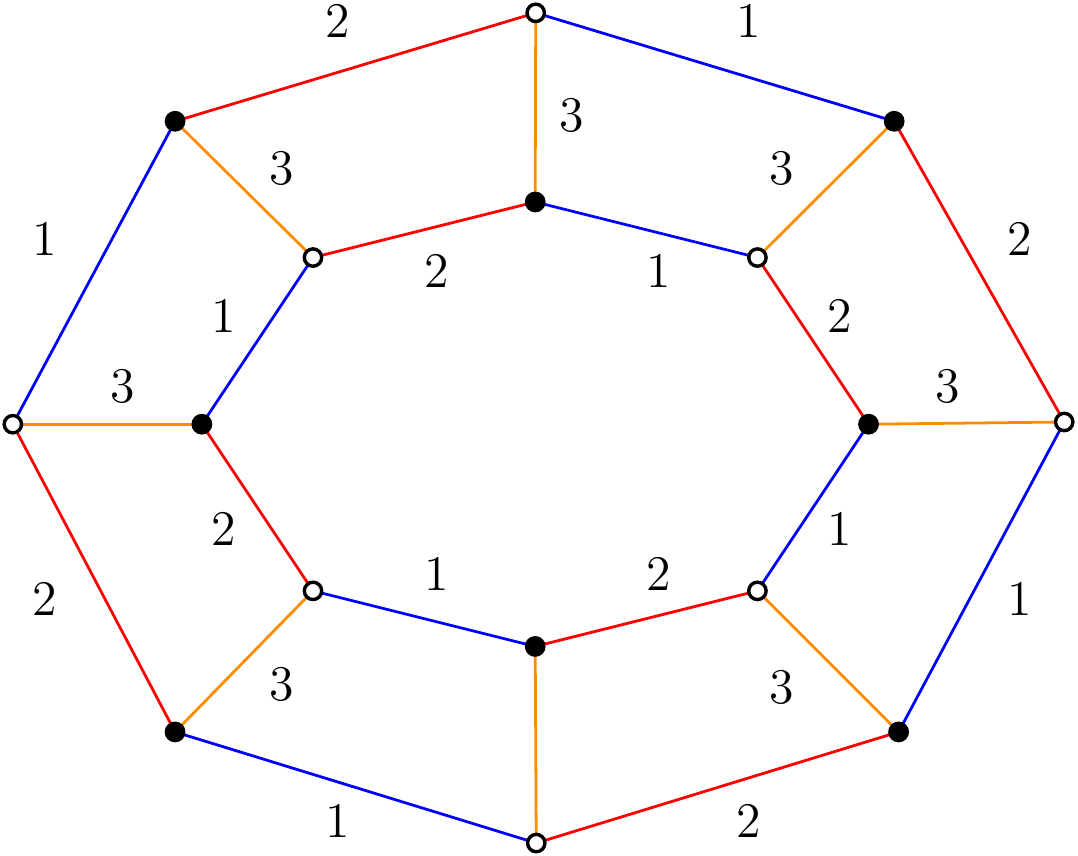} }
\hspace{1cm}\raisebox{3.7ex}{\includegraphics[scale=.53]{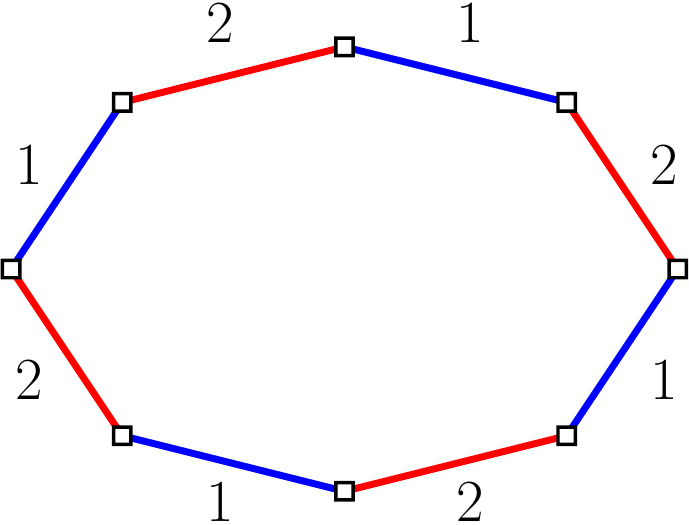} }
\caption{
A bi-pyramid with 16 facets, the dual colored graph $\B_4$, and $\Ps(\B_4,{\Opt}^4)$ without colored vertices.}
\label{fig:BiPyr} 
\end{figure}

We consider the bi-pyramidal bubbles which decompose into two pyramids with a $2p$-gonal basis (left of Fig.~\ref{fig:BiPyr}). The dual colored graphs are shown in the middle of Fig.~\ref{fig:BiPyr}. The case $p=2$ corresponds to the octahedron of Fig.~\ref{fig:EquiCone}. The octahedral bubble (bottom left of Fig.~\ref{fig:EquiCone}) has three optimal pairings, which are defined by choosing all edges of a given color, 1, 2 or 3. Whatever optimal pairing we choose to build the bijection, it therefore has three maximal unicellular maps, and we know from Subsection~\ref{subsec:TreeLike} on tree-like families, that the tree-like maps obtained by gluing those unicellular maps in a tree-like way have the same 0-score as trees. We will show that they are the only maximal configurations. Bi-pyramids of bigger size have a single optimal pairing, defined by the color-3 edges in Fig.~\ref{fig:BiPyr}. We will show that when building the bijection with this pairing, trees are the only maximal maps.

\

We consider a bi-pyramidal bubble $\B_p$ with $4p$ vertices, $p\ge2$, and the  optimal pairing $\Opt^p$ defined by the edges dual to the $2p$-gonal basis, which we take to have color 3. We consider a simplified version of the map $\Ps(\B_p,\Opt^p)$ with no degree-two colored vertex and no colored leaf, as shown on the right of Fig.~\ref{fig:BiPyr}.  We suppose that  $\Ga\in\bS(\B_p,\Opt^p)$ is a maximal map, and we first focus on a particular bubble $\Ps(\B_p,\Opt^p)$. Let $e$ be an edge incident to it. Because $\Ga$ is maximal, from Prop.~\ref{prop:eD2}, either it is a bridge, either $\I(e)=2$ (it cannot be 3 as $\Gai[3]$ is a collection of star-maps). From Proposition~\ref{prop:FaceUnhook}, if $e$ is not a bridge, unhooking it leads to a connected map $\Gae$ with
\be
\label{eqref:Phi0Gae}
\Phi_0(\Gae)=\Phi_0(\Ga)-1.
\ee

\begin{figure}[!h]
\centering
\raisebox{4.5ex}{\includegraphics[scale=.7]{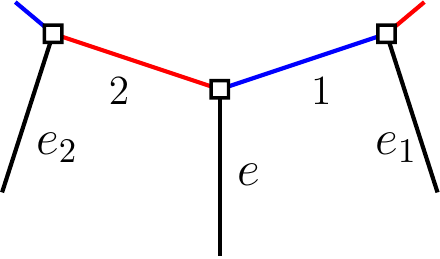} }
\hspace{1cm}\includegraphics[scale=.7]{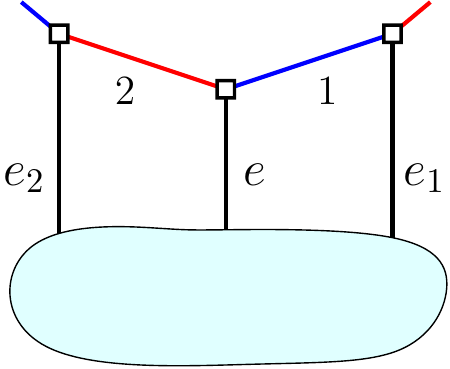} 
\hspace{1cm}\raisebox{3.1ex}{\includegraphics[scale=.7]{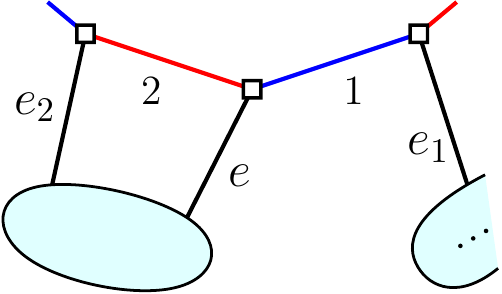} }
\caption{
In a maximal map, the edges $e$, $e_1$ and $e_2$ form an edge-cut.}
\label{fig:BiPyrE} 
\end{figure}

The edge $e$ satisfies the following property :

\

$\cP^1_e\quad:\quad$\emph{If $e$ is a bridge, then all edges incident to the same bubble are also bridges.}

\

Indeed, we consider the edges $e_1$ and $e_2$ as on the left of Fig.~\ref{fig:BiPyrE}. As $e$ is a bridge, $\I(e_1)$ and $\I(e_2)$ are both smaller or equal to one. As $\Ga$ is maximal, from Proposition~\ref{prop:eD2}, $e_1$ and $e_2$ are both bridges. By induction, we obtain $(\cP^1_e)$. We now suppose that $e$ is not a bridge. 
It satisfies the property.

\

$\cP^2_e\quad:\quad$\emph{Unhooking $e$, $e_1$ and $e_2$ raises the number of connected components}

\

Indeed, if $(\cP^2_e)$ is not satisfied, we first unhook $e$, obtaining $\Gae$, with \eqref{eqref:Phi0Gae}. But in $\Gae$, $\I(e_1)$ and $\I(e_2)$ are both smaller or equal  to one, and this is still the case for $e_2$ after unhooking $e_1$. Unhooking both edges leads to a map $\Ga''$ with 
\be
\Phi_0(\Ga'')\ge\Phi_0(\Gae)+ 6-2\I(e_1)-2\I(e_2)\ge \Phi_0(\Gae) + 2 > \Phi_0(\Ga),
\ee
which contradicts the maximality of $\Ga$. We remind that a $k$-bond is a minimal edge-cut comprised of $k$ edges, i.e. a set $S$ of edges such that unhooking all of them disconnects a connected graph into two connected components while unhooking the edges of any proper subset of $S$ does not. The property $(\cP^2_e)$ implies that  $\{e, e_1, e_2\}$ is  an edge-cut, and therefore either it is a 3-bond - middle of Fig.~\ref{fig:BiPyrE}, either it contains a 2-bond - right of Fig.~\ref{fig:BiPyrE} (the blobs represent connected components). In the first case, unhooking $e$ and then $e_1$, we obtain a connected map $\Ga'$ with 
\be
\label{eqref:Phi0Gaee1}
\Phi_0(\Ga')=\Phi_0(\Gae)+ 3-2\I(e_1)\ge \Phi_0(\Gae) + 1 \ge \Phi_0(\Ga).
\ee
The equality cannot be strict as $\Ga$ is maximal, and therefore $\Ga'$ is a maximal map and $e$ is a bridge, so from $(\cP^1_e)$, all the edges incident to the bubble in $\Ga'$ are bridges. There are $2p\ge 4$ edges incident to the bubble so that there is at least one another edge $e_4$, which is a bridge in $\Ga'$ and therefore was also a bridge in $\Ga$ since unhooking $e$ and $e_1$ did not affect this. From $\cP^1_e$, $\{e, e_1, e_2\}$ cannot be a 3-bond. In the second case, suppose that $\{e, e_2\}$ is a 2-bond. We first unhook $e$ and then  $e_1$, as it  is not a bridge in $\Gae$. This leads to a connected map $\Ga'$ for which \eqref{eqref:Phi0Gaee1} is also true. As previously, it is an equality, which means that all the edges incident to the bubble are bridges in $\Ga'$. The only possibility not to arrive at a contradiction is that there are only four edges incident to the bubble. Therefore,

\begin{lemma}
\label{lemma:2bond}
For $p>2$, the only possibility in a maximal map is that all edges incident to a bubble are bridges. For $p=2$, the three possibilities in Fig.~\ref{fig:2Bond} are allowed, which correspond to the three optimal pairings. 
\end{lemma}
%
\begin{figure}[!h]
\centering
\includegraphics[scale=.37]{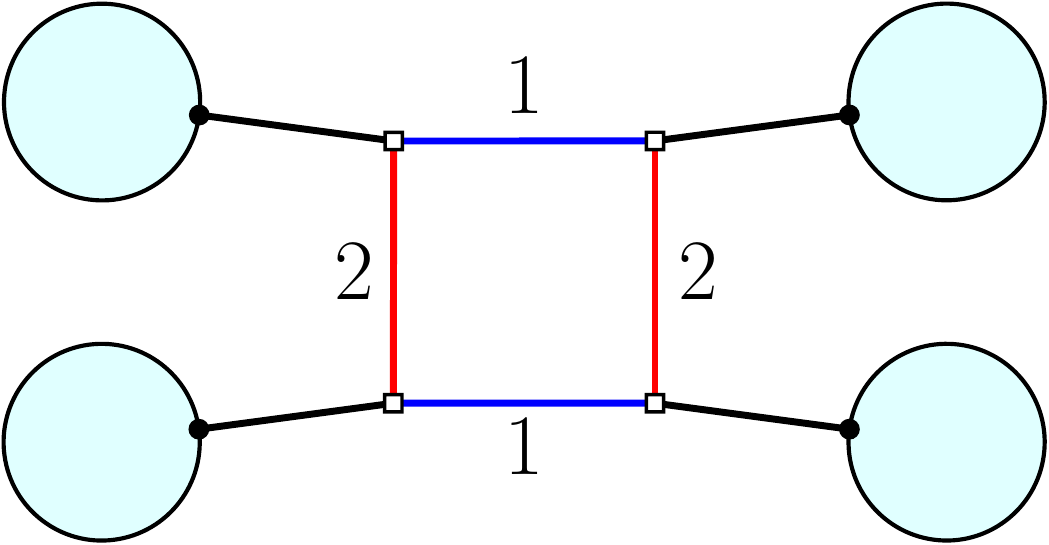}\hspace{2cm} \includegraphics[scale=.45]{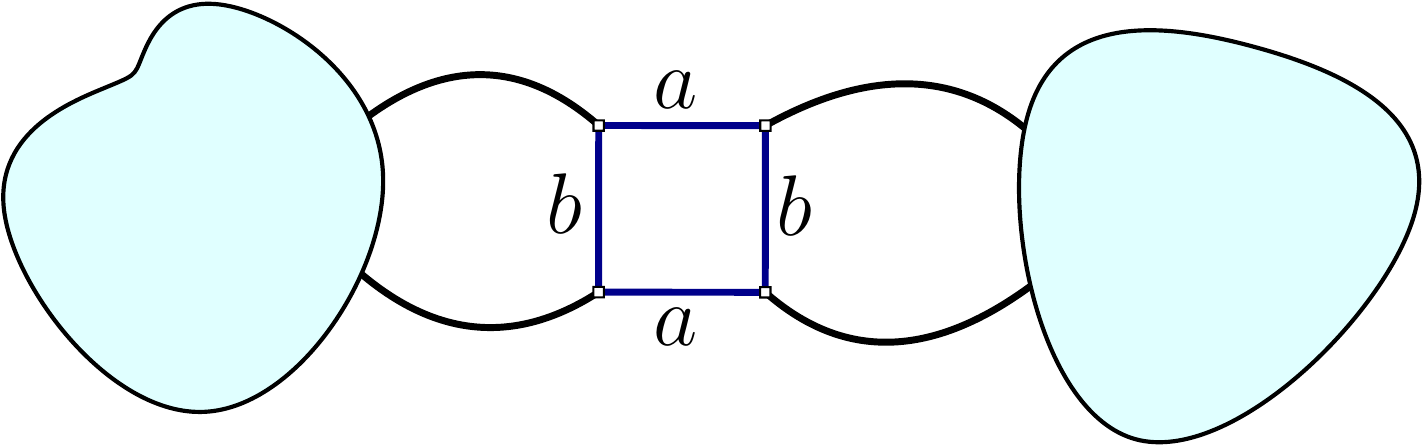} 
\caption{Edges incident to a bubble are either bridges, or form 2-bonds of this sort.}
\label{fig:2Bond} 
\end{figure}

In the case of a bi-pyramidal bubble $\B_p$ with $p>2$,  it implies that maximal maps of $\bS(\B_p,\Opt^p)$ are trees. However we have proven something even stronger: 

\begin{coroll}
$(\B_p, \Opt^p)$ satisfies the cut-bubble property \eqref{CutBubbleProp}: whatever the set $\bB$ of bubbles containing $\B_p$, a bubble $\Ps(\B_p,\Opt^p)$ in a maximal map of $\bS(\bB,\Opt^\bB)$ is a cut-bubble, i.e.  all incident edges are bridges. 
\end{coroll}

The 0-score of an optimal pairing of $\B_p$ is $\Phi_0(\BCO)=4p$, so that from Corollaries~\ref{coroll:TreeMaxAB} and \ref{coroll:scaling}, for $p>2$, 
\be
\label{eqref:TildeASBip}
\tilde a_{\B_p}=4p-3,  \quad\text{and}\quad   s_{\B_p} = 1, 
\ee
and the appropriate bubble dependent degree is, for $\Ga\in\bS(\B_p,\Opt^p)$,
\be
\label{eqref:DegBip}
\delta_{\B_p}(\Ga) = 3+(4p-3)b(\Ga) - \Phi_0(\Ga).
\ee
Computing 
\be
\Phi(\B_p)=2(p+1),
\ee
we deduce from \eqref{eqref:TreeMaxAB2} the coefficient $a$,
\be
\label{eqref:ABip}
a_{\B_p}=\frac 3 2 - \frac 1 {4p},
\ee
and therefore, if $\G$ is the colored graph corresponding to $\Ga$, the correction to Gurau's degree (Def.~\ref{def:Deg}) is 
\be
  \frac{\deltaG(\G) - \delta_{\B_p}(\G)}{V(\G)} =\frac 1{4p}.
\ee

Notice that the optimal covering of a bi-pyramidal bubble has a planar jacket. From Prop.~\ref{prop:PlanJack}, it therefore represents a 3-sphere. Applying Prop.~\ref{coroll:TreeTopo} on the topology of trees, \emph{maximal gluings of bi-pyramids with $p>2$ have the topology of the 3-sphere.} For $p>2$, we denote $\GF_p$ the generating function of rooted maximal configurations of $\bS(\B_p,\Opt^p)$
counted according to their number of bubbles, 
\be
\GF_p(z)=1+z\GF_p(z)^{2p}.
\ee

\subsubsection{Gluings of octahedra}

We now focus on the case where $p=2$. We have just proven that in a maximal map, edges incident to a bubble are as in Fig.~\ref{fig:2Bond}. We denote $\bS_{\max}$ the set of maximal maps in $\bS(\B_2,\Opt^2)$, and call face of a stacked map a face around one of its bicolored submaps. We prove the following property:

\begin{lemma}
\label{lemma:black}
In a maximal map, two edges incident to a bubble and forming a 2-bond are incident to the same black vertex.
\end{lemma}

\prf Indeed, an edge might either be a bridge, or be part of a 2-bond connecting a black vertex to a bubble (i.e. minimal set of two parallel edges), or neither of these two kinds. We denote $p(v)$ the number of the last type of edges incident to the black vertex $v$. They can be part of 2-bonds incident to different bubbles and $k$-bonds for $k>2$. We now prove by induction that for a maximal $\Ga$, $p(v)=0$ for any black vertex $v$. This way, the edges in $\Ga\in\bS_{\max}$ around black vertices are either bridges or form 2-bonds incident on the same bubbles.
 
For some black vertex $v$, the $p(v)$ considered edges $(e_1,...,e_{p(v)})$ form an edge-cut, since the other edges attached to $v$ are all either bridges or pairs of edges that form 2-bonds. Let $p_a\le p_b\le ...$ be the numbers of edges in each bond of the unique decomposition of the edge-cut $(e_1,...,e_{p(v)})$.

Notice that there are no bridges in the bond decomposition of $(e_1,...,e_{p(v)})$, as we excluded them by definition. Therefore $ p(v)=0$ or $\forall a,\ p_a>1$. In particular, $e_1$ is always part of a $k$-bond with $k>1$ and $p(v)=1$ is impossible.
 
\begin{figure}
\includegraphics[scale=0.42]{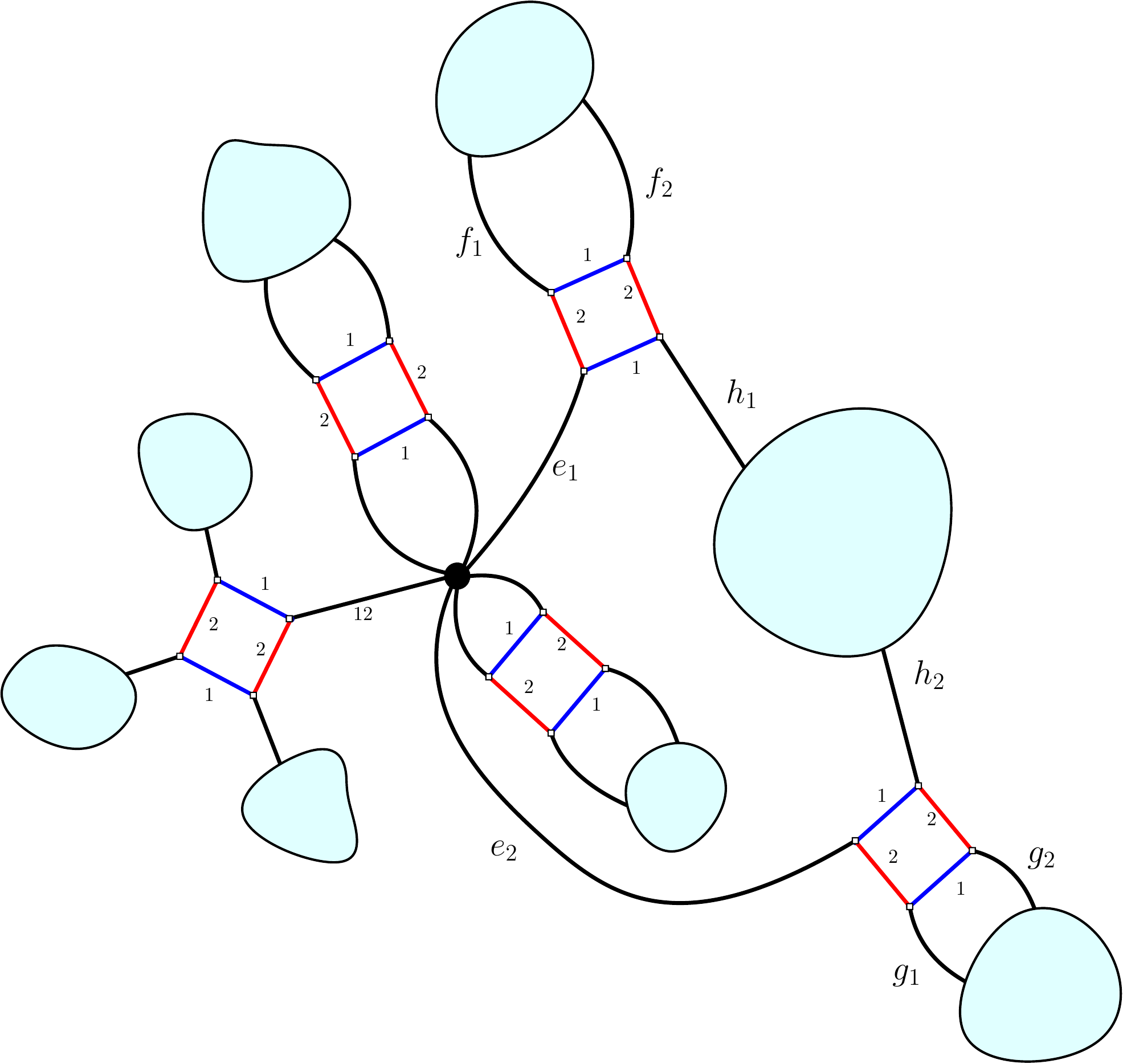}\hspace{0.1cm}\includegraphics[scale=0.45]{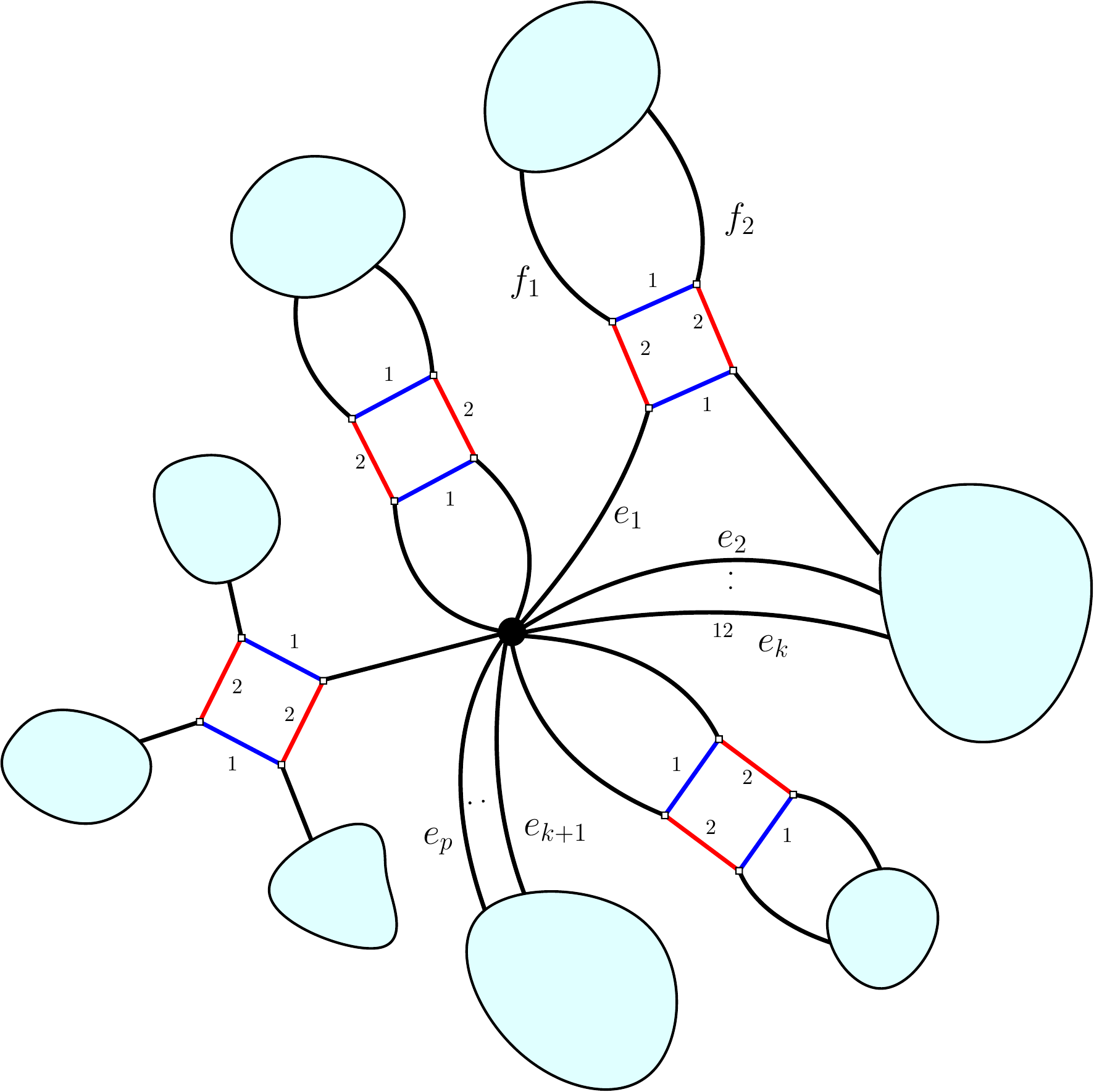}
\caption{On the left is a map with $p(v)=2$ and on the right one with $p(v)=k>2$ for some black vertex.}
\label{fig:Recursion} 
\end{figure}

Let us look at the case $p(v)=2$, i.e. two edges $e_1$ and $e_2$ which form a 2-bond but reach two different bubbles. From Lemma~\ref{lemma:2bond}, we know that to each of them is attached another 2-bond which does not contain $e_1$ or $e_2$. We denote them $(f_1,f_2)$ and $(g_1,g_2)$ as in the left of Figure~\ref{fig:Recursion} (note that the case in Fig.~\ref{fig:Recursion} is the most general case for which $(e_1,h_1)$ and $(e_2,h_2)$ form two 2-bonds). From \eqref{eqref:Phi0Gae}, after unhooking $e_1$, one obtains $\Ga'$ with one face less than $\Ga$ and in which both $e_1$ and $e_2$ are bridges. This implies that $f_1$ and $g_1$ are bridges in at least $\Ga'^{(1)}$ or $\Ga'^{(2)}$, so that unhooking them both brings two additional faces (according to Prop.~\ref{prop:FaceUnhook}). One thus obtains a new element $\Ga''$ with more faces than $\Ga$, which is impossible.
 
Now suppose that for $q>1$, it is proven that $\Ga\in\bS_{\max}$ has no black vertex $v$ for which $1<p(v)\leq q $, and let $\Ga\in\bS_{\max}$ with $p(v)=q+1$ for some black vertex $v$. Since $e_1$ is not a bridge, $\Ga'$ obtained by unhooking $e_1$ has one face less than $\Ga$. From Lemma~\ref{lemma:2bond}, $e_1$ is incident to a bubble to which another pair of edges $(f_1, f_2)$ is attached and form a 2-bond as shown on the right of Fig.~\ref{fig:Recursion}. After detaching $e_1$, we may also detach one of these edges, e.g. $f_1$. As $e_1$ is unhooked, $f_1$ is now a bridge in either $\Ga'^{(1)}$ or $\Ga'^{(2)}$. Let us choose the case $\Ga'^{(2)}$ as in Fig.~\ref{fig:Recursion}. Since $\Ga\in\bS_{\max}$, $f_1$ was not a bridge in $\Ga^{(1)}$ and had two distinct incident faces of color 1, and this is still the case after unhooking $e_1$ (because $e_1$ and $f_1$ belonged to two different connected components in $\Ga^{(1)}$). Unhooking $f_1$ therefore gives a graph $\Ga''$ with one more face than $\Ga'$, hence $\Phi_0(\Ga'') = \Phi_0(\Ga)$ and $\Ga''\in\bS_{\max}$. However $\Ga''$ has a vertex $v$ with $p(v)=q$ which contradicts our hypothesis. Notice that if $p(v) = 3$ in $\Ga$ with the edges $e_1, e_2, e_3$ forming a 3-bond, then $e_2$ and $e_3$ cannot be incident to the same bubble (else they would form a 2-bond). This ensures that $p(v) = 2$ in $\Ga''$ (as the quantity $p(v)$ counts 2-bonds if their edges are not incident to the same bubble). This concludes the induction. \qed

\begin{definition}
Assume that the edges incident to a bubble form two 2-bonds like in Figure \ref{fig:2Bond}. The vertical cut of the bubble consists in removing the inner edges of color $a$ of the bubble which connect both 2-bonds.

Let $\Ga\in\bS_{\max}$ without bridges, and thus satisfying the properties Lemma~\ref{lemma:2bond} and Lemma~\ref{lemma:black} without bridges. The vertical cut of $\Ga$ is obtained by performing the vertical cut of each bubble. It leads to a map $V\Ga$ which has a single black vertex per connected component. We say that $\Ga$ is planar if $V\Ga$ is a planar map.
\end{definition}

\

\begin{lemma}
\label{lemma:planar}
If $\Ga\in\bS_{\max}$ without bridges, then $\Ga$ is planar.
\end{lemma}

\prf From Propositions \ref{lemma:2bond} and \ref{lemma:black}, we know that a bubble in $\Ga\in\bS_{\max}$ without bridges is adjacent to exactly two distinct black vertices, and the vertical cut of the bubble separates $\Ga$ into two connected components,
\begin{equation} \label{SquareVertexDominant}
\begin{array}{c} \includegraphics[scale=.55]{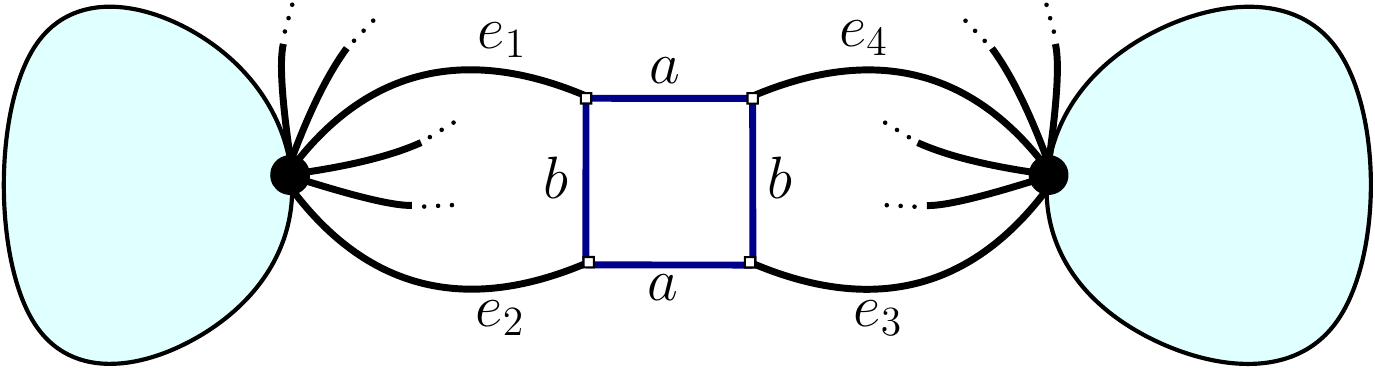} \end{array}
\end{equation}
In the map $\Ga^{(a)}$, one deletes the inner edges of color $b$, and the other way around for $\Ga^{(b)}$. In $\Ga^{(a)}$, the edges $e_1$ and $e_4$ are merged into a single edge, as well as $e_2$ with $e_3$, while in $\Ga^{(b)}$, the edges $e_1$ and $e_2$ are merged into a single loop, as well as $e_3$ with $e_4$. This turns the edges incident to the bubble into a pair of parallel edges in $\Ga^{(a)}$ and a pair of loops in $\Ga^{(b)}$,
\begin{equation}
\Ga^{(a)} = \begin{array}{c} \includegraphics[scale=.35]{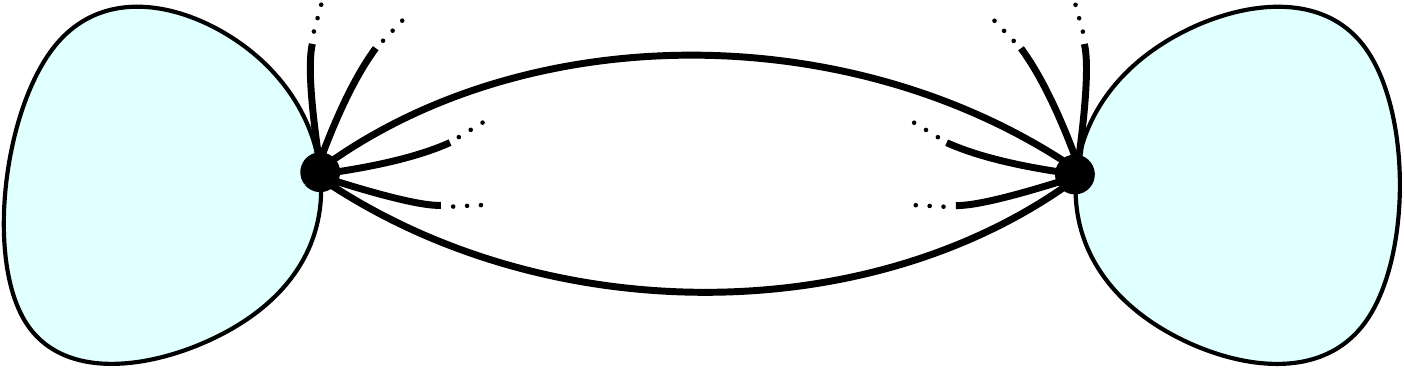} \end{array} \qquad
\Ga^{(b)} = \begin{array}{c} \includegraphics[scale=.35]{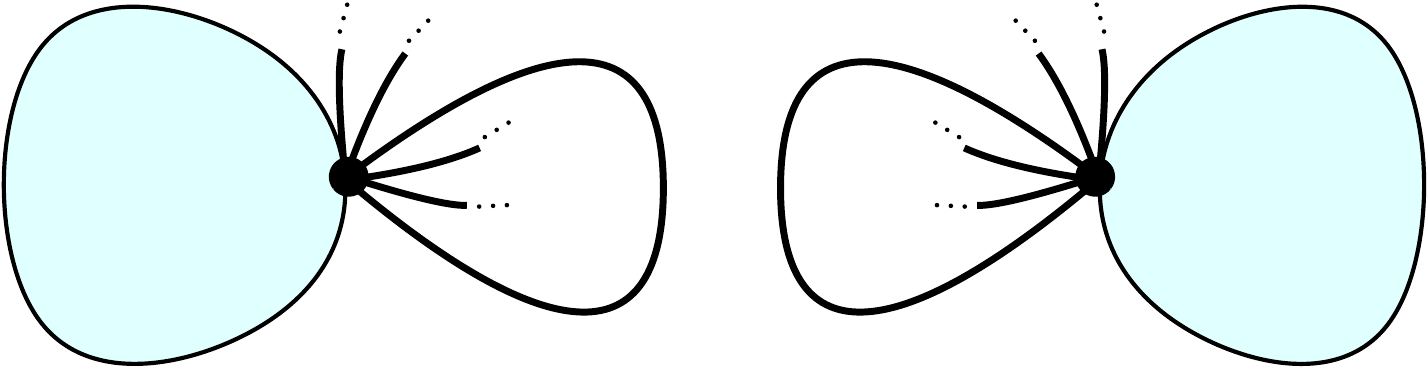} \end{array}
\end{equation}
From Euler's formula, the number of faces of $\Ga^{(c)}$, $c=1, 2$ is
\begin{equation}
\Phi_0(\Ga^{(c)}) = 2b - V + 2(k^{(c)} - g^{(c)}),
\end{equation}
where we have used the fact that the number of edges of $\Ga^{(c)}$ is $2b$ ($b$ the number of bubbles) and the number of vertices of $\Ga^{(c)}$ is $V$ the number of black vertices of $\Ga$. Moreover, $k^{(c)}, g^{(c)}$ respectively denote the number of connected components and the genus of $\Ga^{(c)}$.

One can easily turn $\Ga^{(1)}$ and $\Ga^{(2)}$ into planar maps by permuting the order of the edges around the black vertices. For instance, one can make parallel edges occupy consecutive corners. This way, when edges are parallel in $\Ga^{(a)}$, they become a pair of disjoint loops in $\Ga^{(b)}$ (and the other way around) and there is no edges sitting at the corner inside each loop. This permuting of edges around black vertices does not change the number of connected components of $\Ga^{(c)}$, but only its genus. This thus maximizes $\Phi_0(\Ga^{(c)})$ and it can be concluded that $\Ga\in\bS_{\max}$ without bridges has planar maps $\Ga^{(1)}$ and $\Ga^{(2)}$.

Let $v$ be a black vertex in $\Ga$. The faces of $\Ga^{(c)}$, $c=1, 2$, can be partitioned as those which go through $v$ (there are ${\Phi_{0, v}^{(c)}}$ of them) and those which do not (there are $\Phi_{0,\hat v }^{(c)}$ of them),
\begin{equation}
\Phi_0(\Ga^{(c)}) = \Phi_{0, v }^{(c)} + \Phi_{0,\hat v }^{(c)}.
\end{equation}
Since $\Ga^{(1)}$ and $\Ga^{(2)}$ are planar, there are well defined notions of outside and inside the faces which are delimited by either parallel edges or loops. The face which leaves the black vertex $v$ along the external (resp. internal) side of an edge which is part of a 2-bond in $\Ga^{(a)}$ comes back to $v$ for the first time along the external (resp. internal) side of the edge in the same 2-bond,
\begin{equation}
\begin{array}{c} \includegraphics[scale=.5]{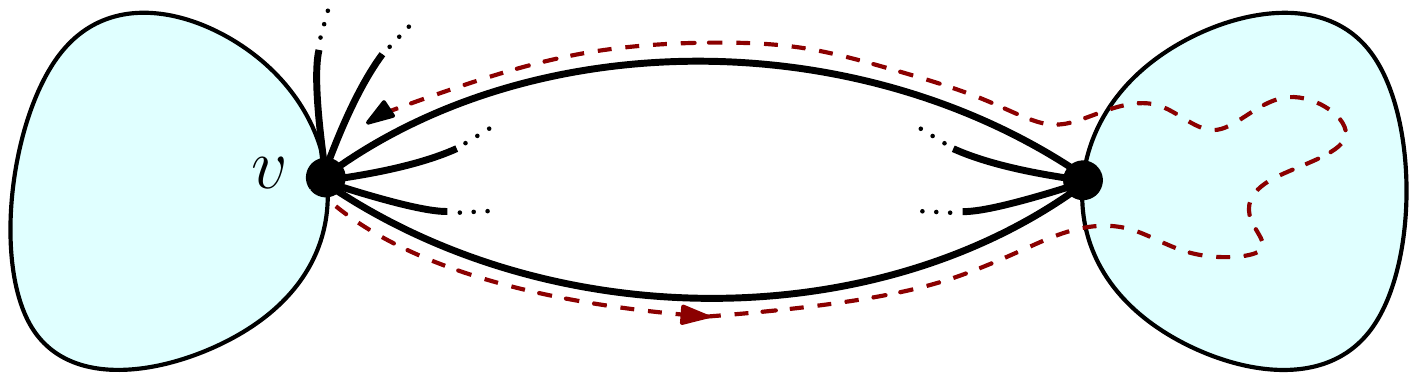} \end{array}
\end{equation}
This property is obvious for $\Ga^{(b)}$ as one travels outside or inside a loop. With some abuse of notation, it can be visualized on $\Ga$ itself, by saying that the external (internal) face leaving $v$ along $e_2$ returns to $v$ for the first time along the external (internal) side of $e_1$,
\begin{equation}
\begin{array}{c} \includegraphics[scale=.5]{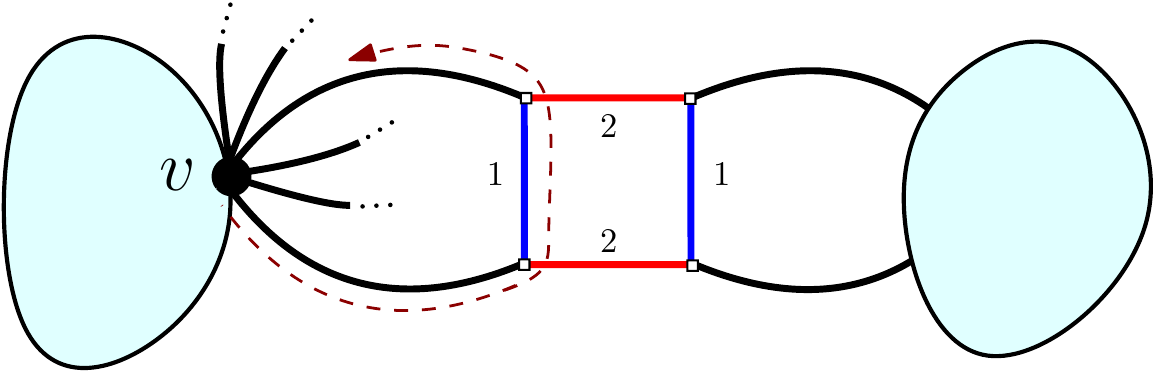} \end{array} \qquad \begin{array}{c} \includegraphics[scale=.5]{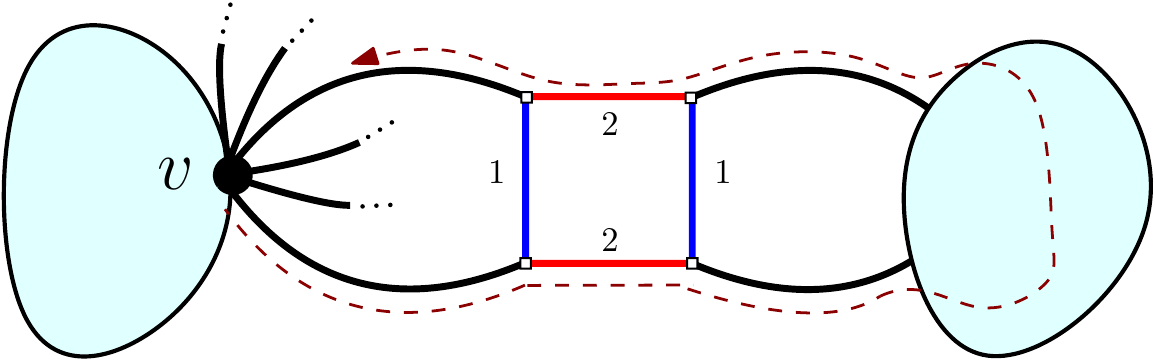} \end{array}
\end{equation}
This shows that there is a bijection between the faces which go through $v$ in $\Ga^{(1)}$ and $\Ga^{(2)}$, and $\Phi_{0,v}^{(1)} = \Phi_{0,v}^{(2)}$.

Let $\Ga_0(v)$ be the connected component of the vertical cut of $\Ga$ which contains $v$. For the bubble represented in \eqref{SquareVertexDominant}, one deletes the inner edges of color $a$ and gets two loops. In fact each bubble provides the vertical cut with two loops. Therefore, there is a bijection between the faces of $\Ga_0(v)$ and the faces of $\Ga^{(1)}$ (and $\Ga^{(2)}$ as well) and thus
\begin{equation}
\Phi_0(\Ga^{(c)}) = \Phi_0(\Ga_0(v)) + \Phi_{0,\hat v }^{(c)}
\end{equation}
At fixed $\Phi_{0,\hat v }^{(c)}$, one maximizes $\Phi_0(\Ga^{(c)})$ by maximizing $\Phi_0(\Ga_0(v))$. Since $\Ga_0(v)$ is a 1-vertex map, this is done by selecting any planar configuration for $\Ga_0(v)$. This reasoning applies to any black vertex of $\Ga$ and shows that the vertical cut has to be planar. \qed

\begin{prop} \label{prop:Dominant}
The set $\bS_{\max}$ is defined by the Propositions \ref{lemma:2bond}, \ref{lemma:black} and \ref{lemma:planar}. It is a tree-like family with $\bT$ the three patterns in \eqref{eqref:MaxOcta} (with empty blobs). The 0-score of an element $\Ga\in\bS_{\max}$ with $b(\Ga)$ bubbles is  $\Phi_0(\Ga) = 5b(\Ga) + 3$.
\end{prop}

\noindent An example of a generic maximal map is given in Fig.~\ref{fig:Dominant}.

\prf Let $\bS_0$ be the set of elements of $\bS(\B_2,\Opt)$ satisfying the criteria given in Propositions \ref{lemma:2bond}, \ref{lemma:black}, \ref{lemma:planar}. From those propositions, we already know $\bS_{\max} \subset \bS_0$. It is therefore sufficient to show that all elements of $\bS_0$ have the same 0-score, $\Phi_0=5b(\Ga) + 3$.
The criteria of Propositions \ref{lemma:2bond}, \ref{lemma:black}, \ref{lemma:planar} show that around a bubble of $\Ga\in\bS_0$, $\Ga$ takes any one of the following forms
\begin{equation}
\label{eqref:MaxOcta}
\begin{array}{c} \includegraphics[scale=.35]{4Bridges.pdf} \end{array} \qquad \begin{array}{c} \includegraphics[scale=.55]{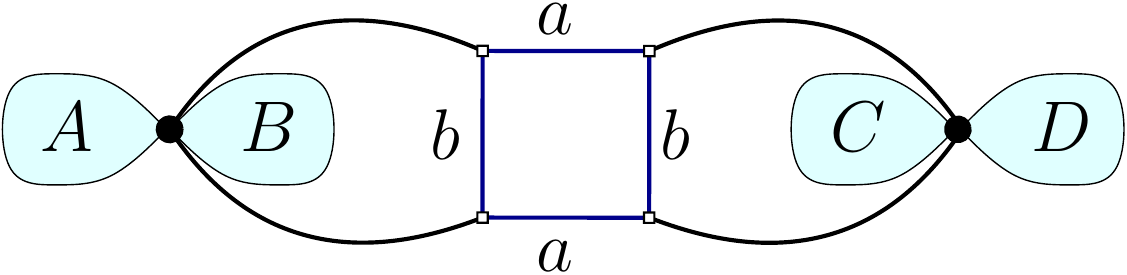} \end{array}
\end{equation}
where the blobs reproduce the same pattern. In the case where the bubble is incident to two 2-bonds, one can unhook an edge in each 2-bond. We saw when treating the general case of bi-pyramids that this led to a map with the same 0-score. Doing so for all bubbles incident to 2-bonds, the number of faces is preserved and $\Ga$ is transformed into a tree for which it is known from Proposition \ref{prop:TreeDeg} that $\Phi_0(\Ga) = 5b(\Ga) + 3$. \qed

\begin{figure}
\centering
\includegraphics[scale=.55]{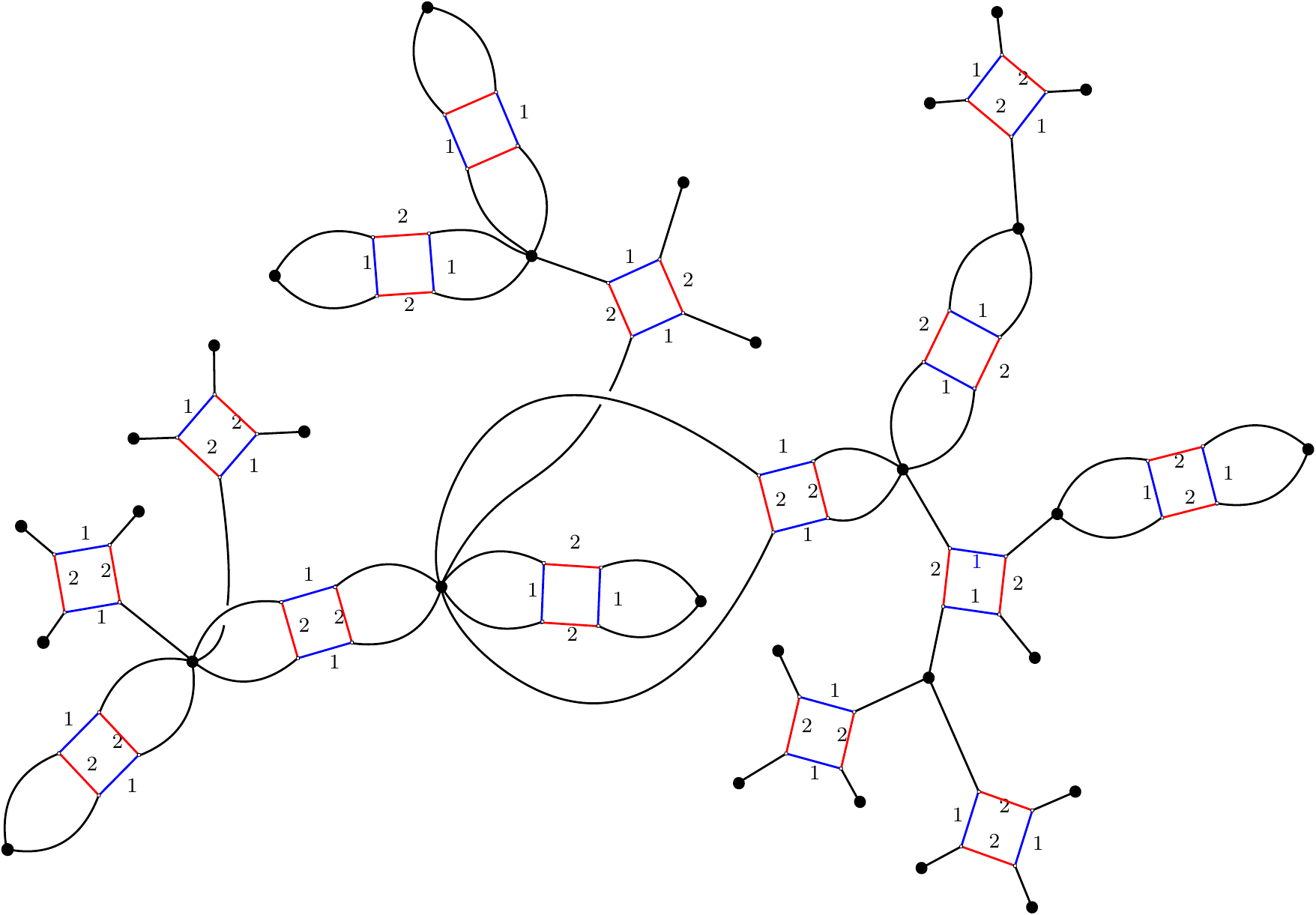}
\caption{A generic example of a maximal map.}
\label{fig:Dominant} 
\end{figure}

\

As trees are dominant, the coefficient $\tilde a$ and the scaling are as in \eqref{eqref:TildeASBip} for $p=2$, as well as the bubble-dependent degree  \eqref{eqref:DegBip} and the correction to Gurau's degree  \eqref{eqref:ABip}. From Corollary~\ref{coroll:TreeLikeTopo} on the topology of tree-like families, maximal maps also have the topology of the 3-sphere. Denoting $\GF_p$ the generating function of rooted maximal gluings of octahedra counted according to their number of bubbles. It satisfies  \eqref{eqref:CountTreeLike}
\be
\GF_2(z)=1+3z\GF_2(z)^{4}.
\ee
Expanding $\GF_2(z)$ as a power series $\GF_2(z) = \sum_{k\geq 0} a_k z^k$, this is equivalent to the recursion
\begin{equation}
a_{n+1} = 3 \sum_{\substack{k_1, k_2, k_3, k_4\\ k_1+k_2+k_3+k_4 = n}} a_{k_1}a_{k_2}a_{k_3}a_{k_4}
\end{equation}
with $a_0=1$. One finds $(1, 3 , 36, 594, \dotsc)$.
The singularity analysis of $\GF_2(z)$ is completely straightforward. Singular points $(\GF_{2,c}, z_c)$ are solutions of $\Phi(\GF_2, z) \equiv 1 - \GF_2 + 3z \GF_2^4 = 0$ and $\partial_{\GF_2} \Phi(\GF_2, z) = -1 + 12z \GF_2^3 = 0$. The second equation gives $z_c = 1/(12 \GF_{2,c}^3)$ which after being plugged into the first equation leads to $\GF_{2,c} = 4/3$ and $z_c = 9/256$. Expanding $1-\GF_2(z)+3z\GF_2(z)^4 = 0$ around that point gives the critical behavior
\begin{equation}
\GF_2(z) = \frac{4}{3} - \sqrt{\frac{2048}{243} \Bigl(\frac{9}{256} - z\Bigr)} + o\Bigl(\sqrt{\frac{9}{256} - z}\Bigr).
\end{equation}
which lies in the universality class of trees. As underlined \eqref{eqref:BijTreeLike} in the subsection on tree-like families, there is a simple bijection with trees with 3 kind of valency 4 vertices, in which the pattern in \eqref{eqref:MaxOcta} where $a$ is 1 and $b$ is 2, is replaced with that on the left of Fig.~\ref{fig:NewVertOcta}, and the pattern where $a$ is 2 and $b$ is 1 is replaced with that on the right of Fig.~\ref{fig:NewVertOcta}. To arrange the blobs $A,B,C$ and $D$, one has to follow the faces around bicolored submaps, remembering that corners oriented counterclockwise on black vertices are, in the colored graph picture, color-0 edges going from a black to a white vertex. 
The bijection holds for non-maximal gluings, as long as the patterns on the right of \eqref{eqref:MaxOcta} are prohibited. This gives a bijection in which maximal maps are precisely trees. 

\begin{figure}[!h]
\centering
\includegraphics[scale=0.45]{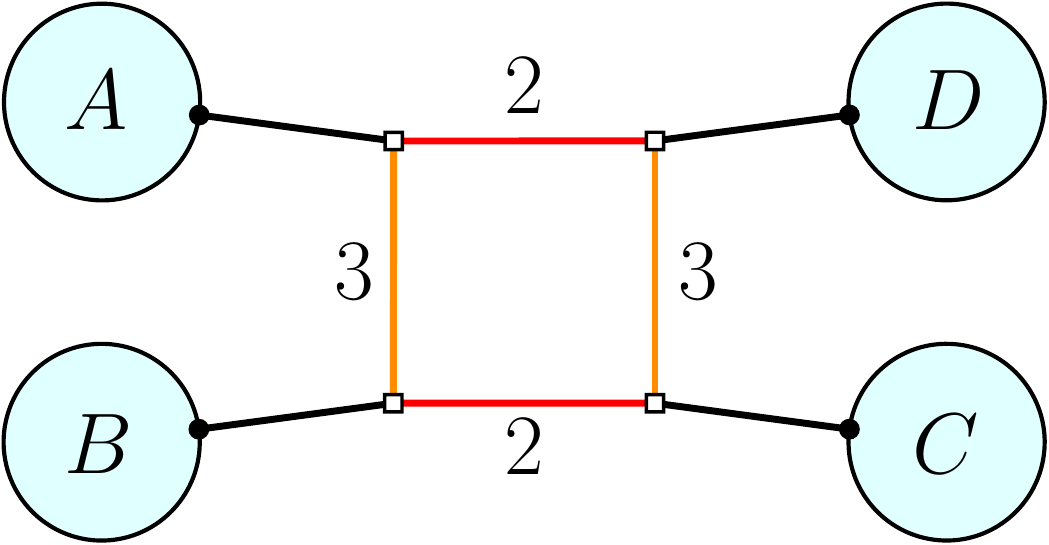}\hspace{2cm}\includegraphics[scale=0.45]{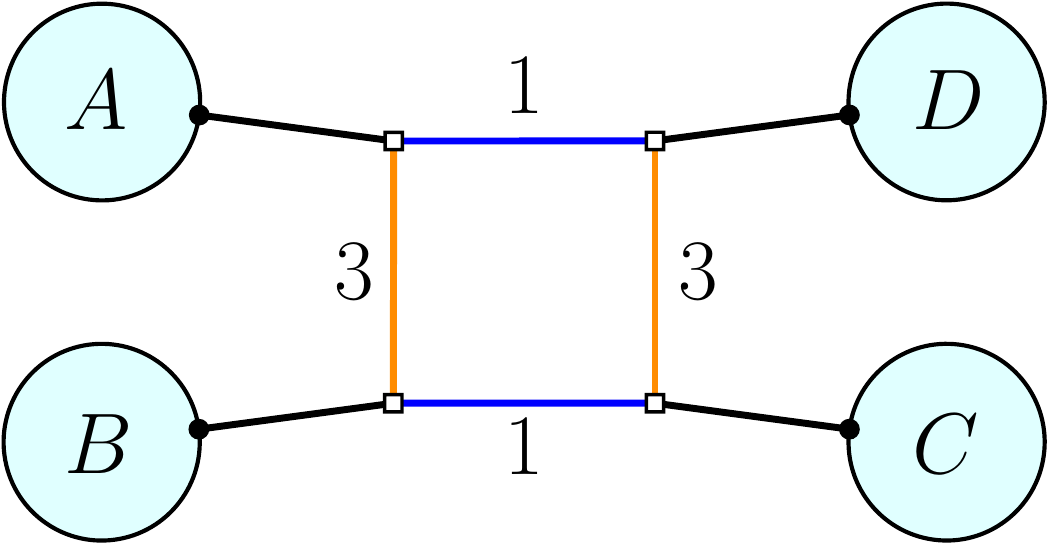}
\caption{Effective bubbles for the other maximal unicellular maps.}
\label{fig:NewVertOcta} 
\end{figure}

Although we obtain the same critical behavior as for bi-pyramids of bigger size, there is a big difference here: the proof is tedious and relies on the details of the bubbles. In particular, we have not proven that octahedral bubbles would always behave as in \eqref{eqref:MaxOcta} when in a maximal map of $\bS(\bB,\Om_\bB)$, where $\B_2\in\bB$.

\subsubsection{Gluings of larger toroidal bubbles}

%
\begin{figure}[!h]
\centering
\includegraphics[scale=.5]{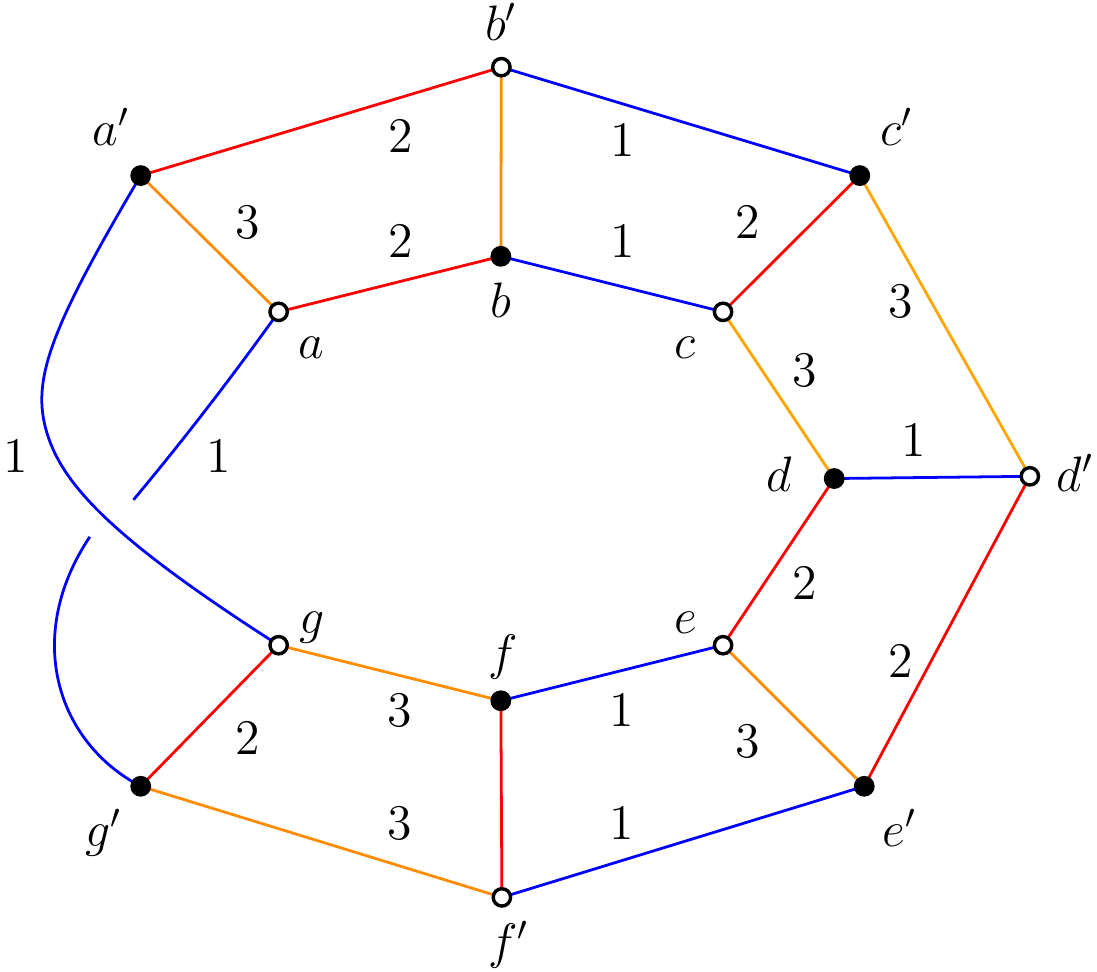} 
\hspace{0.7cm}\raisebox{7.3ex}{\includegraphics[scale=.5]{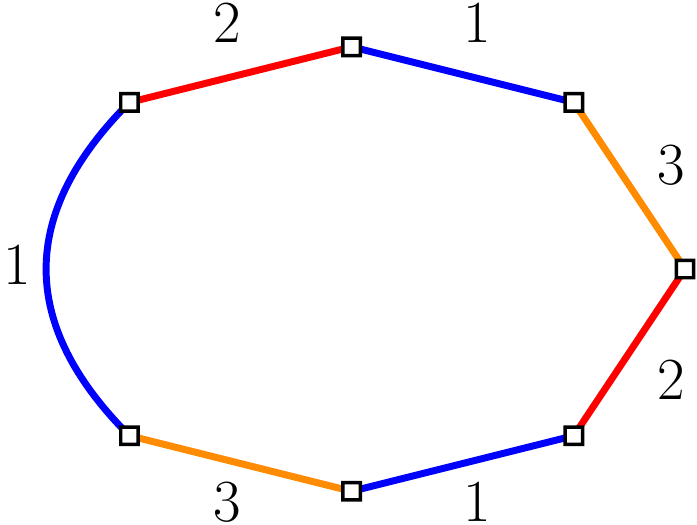} }
\hspace{1cm}\raisebox{8ex}{\includegraphics[scale=.5]{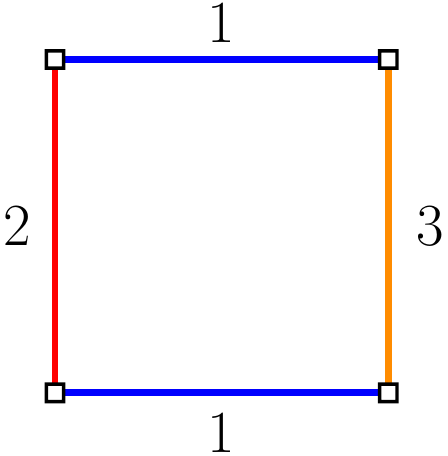} }
\caption{
An example of toroidal bubble and simplified stacked map with $2q=14$, and example of simplified stacked map in the case $q=4$.}
\label{fig:TorBub} 
\end{figure}

In Section~\ref{subsec:K33}, we studied the example of the $K_{3,3}$ bubble, which has a toroidal boundary. It is the smallest bubble of an infinite family of bubbles $K_q$ of size $2q$ ($q\ge3$) with toroidal boundaries 
and with a structure which is very similar to that of bi-pyramidal bubbles, as pictured in Figure~\ref{fig:TorBub}. The difference is that all three colors go around the ``ribbon".
Pairing every letter with the corresponding primed letter in Figure~\ref{fig:TorBub} is an optimal pairing. The stacked map corresponding to the bubble is in the middle of Fig.~\ref{fig:TorBub}.
It is easily seen that $(\cP_e^1)$ and $(\cP_e^2)$ are also satisfied, as they only rely on the fact that two neighboring edges share a single color. We see that the 3-bond case can only work when there are only three edges, which is the case of the $K_{3,3}$ bubble ($q=3$). The two 2-bonds decomposition can only apply if there are four edges, which is the case for the corresponding 8-vertices bubbles $K_4$ ($q=4$, right of Fig.~\ref{fig:TorBub}). For bubbles of bigger size in this family ($q>4$),  the same conclusions apply, and we are left with trees only. 
In the case of the size 8 toroidal bubbles, there are only two optimal pairings, and we get generalized trees with two kind of valency 4 edges, leading to the equation   \eqref{eqref:CountTreeLike}
\be
\GF_4(z)=1+2z\GF_4(z)^{4},
\ee
for $\GF_q$ the generating function of rooted maximal gluings of size 8 toroidal bubble counted according to the number of bubbles. 
The 0-score of an optimal pairing of $K_q$ is $\Phi_0(K_p^{\Opt})=2q$, so that from Corollaries~\ref{coroll:TreeMaxAB} and \ref{coroll:scaling}, 
\be
\label{eqref:TildeASKq}
\tilde a_{K_q}=2q-3,  \quad\text{and}\quad   s_{K_q} = 1, 
\ee
and the appropriate bubble dependent degree is, for $\Ga\in\bS(K_q,\Opt^q)$,
\be
\label{eqref:DegKq}
\delta_{\B_p}(\Ga) = 3+(2q-3)b(\Ga) - \Phi_0(\Ga).
\ee
Computing 
\be
\Phi(K_q)=q,
\ee
we deduce from \eqref{eqref:TreeMaxAB2} the coefficient $a$,
\be
\label{eqref:AKq}
a_{K_q}=\frac 3 2 \bigl(1 - \frac 1 q\bigr),
\ee
and therefore, if $\G$ is the colored graph corresponding to $\Ga$, the correction to Gurau's degree (Def.~\ref{def:Deg}) is 
\be
  \frac{\deltaG(\G) - \delta_{K_q}(\G)}{V(\G)} =\frac 3{2q}.
\ee

\subsubsection{Higher dimensional generalizations}

In dimension $D>3$, we consider the bubble $\B^D_q$ shown in Fig.~\ref{fig:RibbonBub}, and denote $2q$ the number of its vertices. We distinguish two cases. If all edges of one of the colors, say $i$, belong to the ``lateral" $(D-2)$-pairs, then this color has been added on an optimal pairing of the $D-1$ dimensional bubble $\B^{\hat i}$ obtained by deleting all color-$i$ edges. From Prop.~\ref{prop:AddColOpt}, choosing the only optimal pairing of $\B^D_q$, trees are the only maximal maps. If not, then all colors go along the ``ribbon". Choosing to pair the $(D-2)$-pairs, we obtain the simplified stacked map on the right of Fig.~\ref{fig:RibbonBub}. 
\begin{figure}[!h]
\centering
\includegraphics[scale=.55]{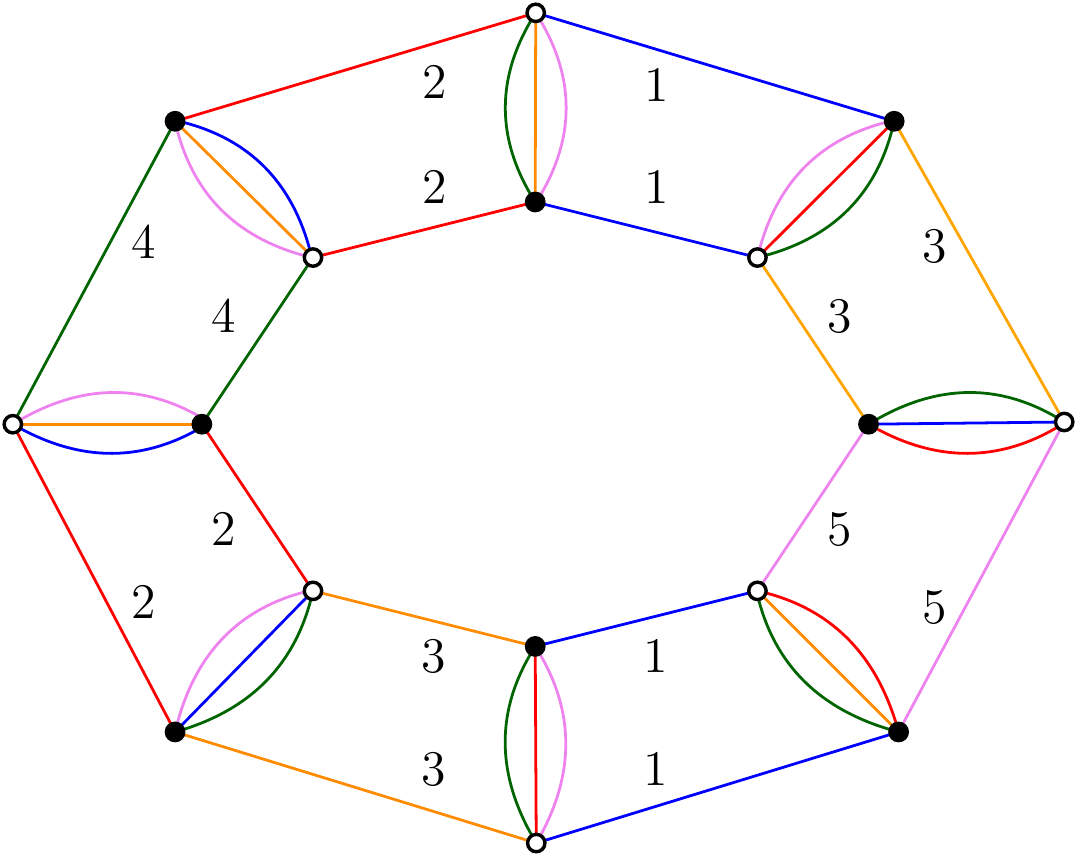} 
\hspace{1.5cm}\raisebox{6ex}{\includegraphics[scale=.55]{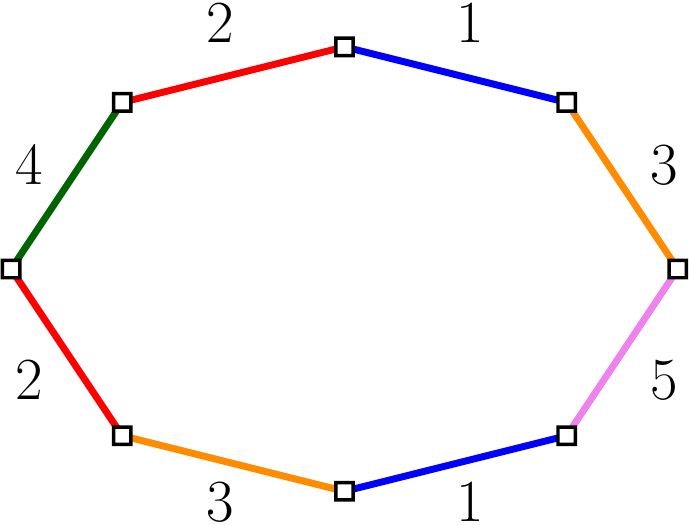} }
\caption{
Bubble representing $\cS^4\times\cS^1$, with $2q=16$, and simplified stacked map.}
\label{fig:RibbonBub} 
\end{figure}

Again, it is easily seen that $(\cP_e^1)$ and $(\cP_e^2)$ are also satisfied. The 3-bond case does not apply: all colors appear around the only cycle of $\Ps(\B^D_q, \Om)$ and there are at least four colors.  The two 2-bonds decomposition can only apply if there are four edges incident to the bubble, which only happens in dimension 4. In that case, we see that since all colors around the cycle of $\Ps(\B^4_4, \Om)$ are different, the two 2-bonds case is also non-maximal. We conclude that higher dimensional generalizations all have only tree maximal maps. 

The 0-score of the optimal pairing is $\Phi_0\bigl((\B^D_q)^\Om\bigr) = q(D-1)$, from which we deduce the coefficients $\tilde a^D_q$ and $s^D_q$ 
\be
\tilde a^D_q = q(D-1) -D,\quad\text{and}\quad s^D_q = 1
\ee
(the scaling is computed trivially from \eqref{eqref:ScalingCircuit}, as $L_m\bigl((\B^D_q)_{/\Om}\bigr)=1$). The score of the bubble depends on whether the two same colors appear on the cycle of $\Ps(\B^D_q, \Om)$:
\begin{itemize}
\item if it does,  the score is 
\be
\Phi(\B^D_{q,a})= 2+q\frac{(D-1)(D-2)}2,
\ee
 and the coefficient $a$ is 
\be
a^D_{q,a}=\frac{D(D-1)}4 - \frac{D-2}{2q}.
\ee
 In that case, we see that every $(D-2)$-pair is a $D-2$ dipole, so that switching two edges of one of the colors going along the cycle is a flip (Def.~\ref{def:Flip}), which does not change the topology and leads to a melonic graph. The topology of the graph is therefore that of $\cS^D$. From Prop.~\ref{coroll:TreeTopo}, as maximal maps are trees, they too have the topology of the sphere $\cS^D$.
 \item If 3 colors or more go along the cycle of $\Ps(\B^D_q, \Om)$, then every $(D-2)$-pair is a combinatorial handle (Def.~\ref{def:CombHandle}), and contracting it leads to a melonic graph. From Thm.~\ref{thm:CombHandle}  the graph represents the connected sum of a sphere and $\cS^{D-1}\times\cS^1$, and therefore represents $\cS^{D-1}\times\cS^1$ itself. Because the bubbles do not represent the sphere, maximal maps do not represent  PL-manifolds (Prop.~\ref{prop:Manifolds}). The score differs from that of the previous case by $D-1$, 
 \be
\Phi(\B^D_{q,b})=q\frac{(D-1)(D-2)}2,
\ee
 and the coefficient $a$ is 
\be
a^D_{q,b}=\frac{D(D-1)}4 - \frac{D}{2q}.
\ee
 \end{itemize}

\section{Fluctuations around non-melonic vacuums}
\label{sec:Fluctuations}

In this section we give an application of the intermediate field representation of the enhanced tensor models to the study of the sub-leading orders, when trees are the only maximal maps. By translating the theory around the leading order, we are able to factorize leading order contributions and isolate the partition function of the sub-leading orders. It is then possible to study the eigenvalues of the mass matrix and to show that the leading order describes a stable vacuum of the theory, and that the effective theory of subleading orders describes a theory of fluctuations around that tree vacuum.

	
	
\subsection{Leading order free energy and trees}

We consider a bubble $\B$, and an optimal pairing of its vertices $\Om_\B$, such that trees belong to maximal maps. From Prop.~\ref{coroll:scaling} and Thm~\ref{thm:FinitNumbPosDeg}, the coefficient $s_\B$ as to be taken as follows in order to have a well-defined and non-trivial $1/N$ expansion,
\be
s_\B=1+ \mu_\B(D-1) - \PhiM,
\ee
where $\mu_\B=\frac{V(\B)}2$. We denote $\Phi_0^\B= \PhiM$. Here, we change the sign $M\rightarrow -M$ and $\bar M \rightarrow -\bar M$, to have positive signs in front of the interactions, and we choose the sign of the coupling constant $\lambda$ to have a positive sign in front of the matrix-bubble, and  rescale it by $\mu_\B$. These manipulations are done so that we can compare the perturbative expansion  with combinatorial results.  We consider the partition function of the corresponding intermediate field theory (Thm.~\ref{thm:IFT}), 
\be
 \cZ_{B}(\lambda, N)
=\int_{\bC^D}e^{\frac{\lambda}{\mu_\B}{N^{s_\B}} \Tr_\B(T , \bar T)}d\mu(T , \bar T)
=\int e^{  \frac{\lambda}{\mu_\B}{N^{D-\Phi^\B_0}} V_{\B}(M)-\Tr\ln(\un^{\otimes D}- \bar M )  }d\nu(   M, \bar M),
\ee
in which $d\nu(   M, \bar M)=\frac{dM d\bar M}{\pi^{N^2}} e^{-\Tr(M\bar M)}$. Motivated by results from \cite{DartEyn} and \cite{PhaseTrans}, 
we do the replacements 
\be
M=x\un^{\otimes D}+M', \quad\text{and}\quad \bar M=y\un^{\otimes D}+\bar M',
\ee
and denote $\mathcal{V}^{(k)}_{\B}(M')$ the sum of the $\binom  \mu k$ interactions obtained from $V_{\B}(M')$ by replacing $k$ copies of $M'$ with $\un^{\otimes D}$ in every possible way. 
%
The partition function then writes
\bea
 \cZ_{B}(\lambda, N)
&=&e^{-N^{D}\bigl(xy - \frac{\lambda}{\mu} x^n+\ln(1-y)\bigr)}\int e^{-\Tr(x\bar M'+yM')}  \\
&& \times e^{+ \frac{\lambda}{\mu}{N^{D-\Phi^\B_0}} \bigl(x^{\mu-1}N^{\Phi^\B_0-D}\Tr(M')+\sum_{k= 0}^{\mu-1}x^k\mathcal{V}^{(k)}_{\B}(M')\bigr)-\Tr\ln(\un^{\otimes D}- \frac{\bar M'}{1-y})  }d\nu(  M', \bar M').\nonumber
\eea

\begin{theorem}
If trees are the only maximal maps in $\bS(\B,\Om)$ the resulting integral is subdominant and describes an effective theory on the fluctuations. The large $N$ free-energy therefore only receives contributions from the exponential factor before it,
\be
\cF_\infty=-xy+\frac{\lambda}{\mu} x^\mu-\ln(1-y),
\ee
where $\mu$ is the number of pairs of the bubble $\B$, and where $a$ and $y $ are the unique solutions of the equations
\be
\label{eqref:xyPartFunct}
x=\frac{1}{1-y}, \qquad\mathrm{and}\qquad
y = \lambda x^{\mu-1}.
\ee
We identify $x$ with the leading order two-point function, which is also the generating function of trees with one marked corner, as it satisfies 
\be
\GF_2=1+\lambda \GF_2^\mu.
\ee
Furthermore, for large $N$, $y=O(1)$ and $x=O(1)$. The partition function becomes
\be
\cZ_{\B}(\lambda, N)
=e^{-N^{D}\cF_\infty(\lambda)}
\int \exp\biggl[ \frac{\lambda}{\mu}{N^{D-\Phi^\B_0}} \sum_{k=0}^{\mu-2}x^k\mathcal{V}^{(k)}_{\B}(M')-\Tr\widetilde\ln(\un^{\otimes D}- x\bar M')  \biggr]d\nu(  M', \bar M'),
\ee
where we have denoted $\tilde\ln(1-\alpha)=\ln(1-\alpha)+\alpha$.
\end{theorem}

\proof We first prove the large $N$ behavior of $x$ and $y$. On one hand, $<\Tr M>=xN^D+<\Tr M'>$ and $<\Tr \bar M>=yN^D+<\Tr \bar M'>$, and $<\Tr M'>$ is at most in $N^D$. On the other hand, we can compute explicitly their first order in $N$. Indeed, at large $N$, $<\Tr \bar M>$ only selects dominant maps with one marked black leaf. In particular, they scale at $N^D$, so that $y=O(1)$. Similarly, at large $N$, $<\Tr M>$ selects only trees with one additional bubble of valency one. It therefore has one more propagator than usual dominant maps, which does not change the factor $N$, so that $<\Tr M>=O(N^D)$, and again $x=O(1)$. 

We now focus on the fluctuation theory. Making the choice \eqref{eqref:xyPartFunct}, two cancellations occur: the term  $\frac{\Tr \bar M'}{1-y}$ coming from the logarithm is cancelled by the $-x\Tr \bar M'$ term, and the terms from $\mathcal{V}^{(\mu-1)}_{\B}$ which are of the form $\lambda x^{\mu-1}\Tr M'$ are in turn canceled by $-y\Tr M'$. 
The interactions are either $V_{\B}$, either terms of the sum $\sum_{k>0}\cV^{(k)}_{\B}(M')$. Regarding the number of faces, the latter behave as if the matrices which have been replaced with identity factors were leaves. The maps of the fluctuation theory therefore have at most as many faces as dominant maps of the initial theory. The assumption that dominant maps are trees also applies, so that a dominant map of the fluctuation theory should be a tree, which must have at least one leaf.  There might be two kind of such leaves,  the $\frac{\Tr \bar M'}{1-y}$ term and those of the form of the form $\lambda x^{\mu-1}\Tr M'$, which have both been cancelled. The fluctuation integral therefore generates no tree, and therefore is at most of order $O(N^{D-1})$. \qed

\

The change of variable only extracts trees. In the case where trees are not the only maximal maps in $\bS(\B,\Om_\B)$, he resulting integral is not subdominant.
The previous result generalizes for the partition function of a family of interaction $\bB$,
\be
 \cZ_{\bB}(\lambda, N)
=\int_{\bC^D}e^{\sum_{\B\in\bB}{\frac{\lambda_\B}{\mu_\B}}{N^{s_\B}} \Tr_\B(T , \bar T)}d\mu(T , \bar T)
=\int e^{ \sum_{\B\in\mathbb{B}}\frac{\lambda_\B}{\mu_\B}{N^{D-\Phi_0^\B}} V_{\B}(M)-\Tr\ln(\un^{\otimes D}- \bar M )  }d\mu(   M, \bar M).
\ee
Changing variables, it becomes
\bea
 \cZ_{\bB}(\lambda, N)
&=&e^{-N^D\bigl(xy - \sum_{\B\in\bB}\frac{\lambda_\B}{\mu_\B}x^{\mu_\B}+\ln(1-y)\bigr)}\\
\nonumber&&\times\int e^{-\Tr(x\bar M'+yM') + \sum_{\B \in \bB}\frac{\lambda\B}{\mu_\B}{N^{D-\Phi_0^\B}} \sum_{k=0}^{\mu_\B}x^k\mathcal{V}^{(k)}_{\B}(M')-\Tr\ln(\un^{\otimes D}- \frac{\bar M'}{1-y})  }d\nu(  M', \bar M').
\eea

\begin{theorem}\label{thm:LOmultibubble}
If trees are the only maximal maps in $\bS(\bB,\Om_\bB)$, the resulting integral is subdominant and describes an effective theory on the fluctuations. The large $N$ free-energy therefore only receives contributions from the exponential factor before it,
\be
\cF_{\bB, \infty}=-xy+\sum_{\B\in\bB}\frac{\lambda_\B}{\mu_\B}x^{\mu_\B}-\ln(1-y),
\ee
where $\mu_\B$ is the number of pairs of the bubble $\B$, and where $x$ and $y $ are the unique solutions of the equations
\be
x=\frac{1}{1-y}, \qquad\mathrm{and}\qquad
y= \sum_{\B\in\bB}{\lambda_\B}x^{\mu_\B-1}.
\ee
We identify $x$ with the leading order two-point function, which is also the generating function of trees with one marked corner, as it verifies 
\be
\GF_{2}=1+\sum_{\B\in\bB}{\lambda_\B}\GF_{2}^{\mu_\B}.
\ee
Furthermore, for large $N$, $y=O(1)$ and $x=O(1)$. The partition function becomes
\be
\label{eqref:FluctB}
\cZ_{\bB}(\lambda, N)
=e^{-N^{D}\cF_{\bB, \infty}(\lambda)}\int \exp\biggl[N^D \sum_{\B\in\bB} \frac{\lambda_\B}{\mu_\B N^{\Phi_0^\B}}\sum_{k=0}^{\mu_\B-2}x^k\mathcal{V}^{(k)}_{\B}(M')-\Tr\widetilde\ln(\un^{\otimes D}- x\bar M')  \biggr]d\nu(  M', \bar M').
\ee

\end{theorem}

\subsection{Effective theory of fluctuations: pruning and schemes}

The Feynman diagrams of the perturbative expansion of the effective theory of fluctuations
\be
\label{eqref:EffPartFunct2}
\cZ^\text{eff}_{\bB}(\lambda, N)= \int \exp\biggl[N^D \sum_{\B\in\bB} \frac{\lambda_\B}{\mu_\B N^{\Phi_0^\B}}\sum_{k=0}^{\mu_\B-2}x^k\mathcal{V}^{(k)}_{\B}(M')-\Tr\widetilde\ln(\un^{\otimes D}- x\bar M')  \biggr]d\nu(  M', \bar M').
\ee
 are similar to that of the full theory, which are stacked maps in $\bS(\bB,\Om_\bB)$,  with the difference that there are no tree contributions: they are obtained from the elements of $\bS(\bB,\Om_\bB)$ by the pruning procedure described in Fig.~\ref{fig:SYKB}. Instead, the generating function $x$ of rooted trees is added on every corner around the black vertices generated by the logarithm (corresponding to color-0 edges in the colored graph picture). In the case where a tree contribution was attached to a white square of the bubble $\Ps(\B,\Om_\B)$, it has been pruned, and the ``leaf" incident to the bubble which is a remainder of the pruned tree - the $x \un^{\otimes D}$ insertion -  also carries a generating function $x$. 

\

We now focus on the quadratic terms in the equation. In \eqref{eqref:EffPartFunct2}, the quadratic terms are either $-\Tr M'\bar M'$, either $+\frac{x^2}{2}\Tr(\bar M'^2)$ from the logarithm, either terms in $\cV^{(n_\B-2)}_{\B}(M')$. The latter are obtained from $V_{\B}$ by inserting $\mu_\B-2$ identities, thus leaving only two copies of $M'$, and are of the form 
\be
N^{\Phi_0^\B-2D+\lvert I \rvert}  \Tr_{I}[\Tr_{\widehat I}(M')^2],
\ee
where $\widehat I =\llbracket 1,D\rrbracket \setminus I$, $I$ being the set of crossing colors. This is pictured on the left of Figure~\ref{fig:TraceI} for $\Tr_{13}[\Tr_{24}(M')^2]$. The contribution of these terms to the quadratic part is
\begin{figure}[!h]
\centering
\includegraphics[scale=0.9]{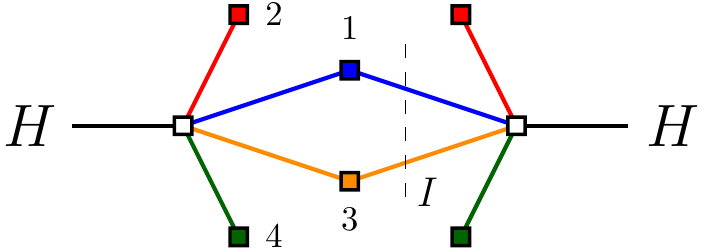}\hspace{2cm}\includegraphics[scale=0.9]{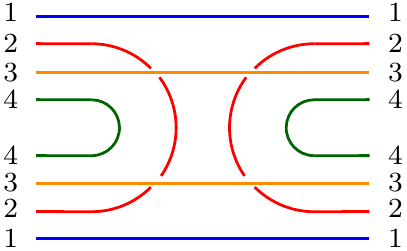}
\caption{Quadratic partial trace and corresponding contraction operator.}
\label{fig:TraceI} 
\end{figure}
\bea
N^D \sum_{\B\in\bB} \frac{\lambda_\B}{\mu_\B N^{\Phi_0^\B}}x^{\mu_\B-2}\mathcal{V}^{(\mu_\B-2)}_{\B}(M')&=&N^{-D} \sum_{\B\in\bB} \frac{\lambda_\B}{\mu_\B } x^{\mu_\B-2}\sum_{p,q\in\Omega}   N^{\lvert I_{p,q} \rvert}  \Tr_{I_{p,q}}[\Tr_{\widehat I_{p,q}}(M')^2]\nonumber\\
&=&\frac{1}{2x^{2}}\sum_{I}K_{I,\bB}N^{\lvert I \rvert-D} \Tr_{I}[\Tr_{\widehat I}(M')^2],  
\eea 
in which $p$, $q$ are pairs in $\Omega$, which is the unique optimal pairing of $\B$ (Lemma~\ref{lemma:TreesImplieOpt}), and denoting $k_{I,\B}$ the number of optimal pairs of $\B$ with crossing colors $I$,
\be
K_{I,\bB}=
\sum_{\B\in\bB} 2k_{I,\B}\frac{\lambda_\B}{\mu_\B }x^{\mu_\B}.
\ee
Remark that from Lemma~\ref{Lemma:OptD2B}, the sum is for $I$ with no more than ${\lfloor\frac{D}{2}\rfloor} $ crossing colors. We denote 
\be
H =(M', \bar M'),
\ee
and gather all the quadratic terms, and rewrite them using a quadratic operator acting on $H$:
\be
\label{eqref:QuadrPart}
-<H; \cO H>=-\Tr M'\bar M'+\frac{x^2}{2}\Tr(\bar M'^2)+\frac{1}{2x^{2}}\sum_{I}K_{I,\bB}N^{\lvert I \rvert-D} \Tr_{I}[\Tr_{\widehat I}(M')^2] .
\ee
We introduce the contraction operator $e_I$ associated to the quadratic partial traces
\be
<M',e_I M'>=\Tr_{I}[\Tr_{\widehat I}(M')^2], 
\ee
with which we can express the operator $\cO$ introduced in (\ref{eqref:QuadrPart}) and its inverse $\cO^{-1}$
\begin{equation}
{\renewcommand{\arraystretch}{2.5}
\cO=
\left(
\begin{array}{c|c}
 -\frac{1}{x^{2}}(\un^{\otimes 2D}-C) \  &\quad\un^{\otimes 2D} \\[+1ex]
\hline
\un^{\otimes 2D} &\quad -x^2\un^{\otimes 2D} \\[+0.75ex]
\end{array}
\right), \qquad
\cO^{-1}=
\left(
\begin{array}{c|c}
\ \  x^2C^{-1}  \quad & \ C^{-1}\\[+1ex]
\hline
\ C^{-1}  \quad & \ \frac{1}{x^{2}}(C^{-1}-\un^{\otimes 2D})\\[+0.75ex]
\end{array}
\right),}
\end{equation}
in which $\un^{\otimes 2D}$ is the $N^{2D}\times N^{2D}$ identity $\un^{\otimes 2D}=\tilde e_{\llbracket 1,D\rrbracket}$, and we have denoted 
\be
C=\un^{\otimes 2D}-\sum_{\substack{{I\subset\lDr}\\{\lvert I \rvert \le \lfloor \frac D 2 \rfloor}}}K_{I,\bB}N^{\lvert I \rvert-D}e_I.
\ee


The action of the effective theory can thus be rewritten using the operator  $\cO$:
\begin{equation}
\label{eqref:EffAction}
S[M',\bar M']=-\frac 1 2 \langle H;\cO H\rangle +N^D \sum_{\B\in\bB} \frac{\lambda_\B}{\mu_\B N^{\Phi_0^\B}}\sum_{k=0}^{\mu_\B-3}x^k\mathcal{V}^{(k)}_{\B}(M')-\Tr\widetilde{\widetilde\log}(\un^{\otimes D}- x\bar M') ,
 \end{equation}
where  $\widetilde{\widetilde\log}$ has no linear and quadratic terms,
\be
\widetilde{\widetilde\log}(1-z)=\log(1-z)+z+z^2/2.
\ee

The Feynman diagrams labeling the perturbative expansion of the theory with effective action \eqref{eqref:EffAction} have no valency one or two vertices. Interpreting the quadratic terms as part of the propagator and not as valency two vertices, we have contracted the sequences of valency two vertices as described in Fig.~\ref{fig:SYKB}. By studying the algebra of the $e_I$, we show in \cite{Stephane} that the propagators in $\cO^{-1}$ are chain-edges which carry the generating function of those sequences of valency two vertices. We further show that for $\lambda_\B$ smaller than the dominant singularity $\lambda_\B^c$, all the eigenvalues of the mass-matrix $\cO$ are positive, so that for $\lambda_\B<\lambda_\B^c$, the non-melonic tree vacuum is the stable vacuum of the theory.  When $\lambda_\B$ reaches the dominant singularity, one of the eigenvalues vanishes, and the theory undergoes a phase transition.

\chapter{Summary and outlooks}
\label{chap:SummaryEnd}

\section{Summary of the studied building blocks}
\label{sec:Summary}

In this table, we summarize the results we obtained with respect to the bubbles studied throughout this work. Given a bubble $\B$, we specify the coefficient $\tilde a_\B$, which, from Thm.~\ref{thm:FinitNumbPosDeg}, is the only value allowing to define a bubble-dependent degree (Def.~\ref{def:BubDepDeg}) which  provides a well-defined \eqref{eqref:Cond1} and non-trivial  \eqref{eqref:Cond2} $1/N$ expansion. 
It provides the smallest bound on the number of $(D-2)$-cells, which for the cases we studied is linear in the number of bubbles of the cell pseudo-complex, 
\be
n^\partial_{D-2}(\C)\le D+ \tilde a_\B \times n_\B(\C),
\ee
saturated only for maximal configurations. 
For non-melonic bubbles, we provide the corresponding value of $a_\B$ \eqref{eqref:Tildeaa}, which gives the smallest bound on the number of $(D-2)$-simplices, and the correction to Gurau's degree:
\be
\Delta_\B=\frac{\deltaG-\delta_\B}V= \frac{D(D-1)}4- a_\B>0.
\ee
We also give the unique scaling $s_\B$ \eqref{eqref:SFromTildeA}, which properly defines an enhanced random tensor model (Subsection~\ref{subsec:Uncolored}). We provide the critical exponent $\gamma_\B$, given by the critical behavior of the free energy of maximal configurations with one distinguished color-0 facet near its dominant singularity \eqref{eqref:CritExp1} and \eqref{eqref:CritExp2}.  We denote 
\be
p_\B=\frac {V(\B)} 2,
\ee
which is the number of pairs in the bubble. For $D\ge 3$, the third column from the left indicated the genus (for $D=3$) or the degree (for $D>3$) of the bubble, and its topology. By ``(sing)", we mean that the represented space is not a piecewise-linear manifold.  The topology of maximal gluings is indicated in the column ``max. top.".  The column ``sec." gives the references to the sections where the corresponding example is treated. For dimension two, we obtain the following values.



\

\noindent\begin{tabular}{|c|c|c|c|c|c|c|c|c|}
\hline
{\bf Bubble} $\B$ & {\bf size} $V_\B/2$ &$\quad\tilde a_\B\quad$ & $\quad a_\B\quad$ & $\quad\Delta_\B\quad$ & $\quad s_\B\quad $ & $\quad\gamma_\B\quad$ & {\bf top.} & {\bf sec.}  \\
\hline
{\begin{tabular}{@{}c@{}}  $p$-gon\\[+1ex]\includegraphics[scale=0.5]{Hexa.pdf}\end{tabular} }& $p$ & $p - 1$ & $\frac 1 2$ & 0 &  0  & $-\frac 1 2 $&$\cS^2$& \ref{subsec:2DQG} \\
\hline
\end{tabular}

\newpage

\subsection{Dimension three}

\

\noindent\begin{tabular}{|c|c|c|c|c|c|c|c|c|c|}
\hline
%
{\bf Bubble} $\B$ &$\ V_\B/2\ $ & {\begin{tabular}{@{}c@{}}  $g(\B)$\\[+2ex] $\B\cong ...$ \end{tabular}}&$\quad\tilde a_\B\quad$ & $\quad a_\B\quad$ & $\ \Delta_\B\ $ & $\  s_\B\  $ & $\ \gamma_\B\ $ &{\begin{tabular}{@{}c@{}}  {\bf max.}\\  {\bf top.} \end{tabular}}& {\bf sec.} 
\\
\hline
{\begin{tabular}{@{}c@{}}  melonic \\[+1ex]\includegraphics[scale=0.5]{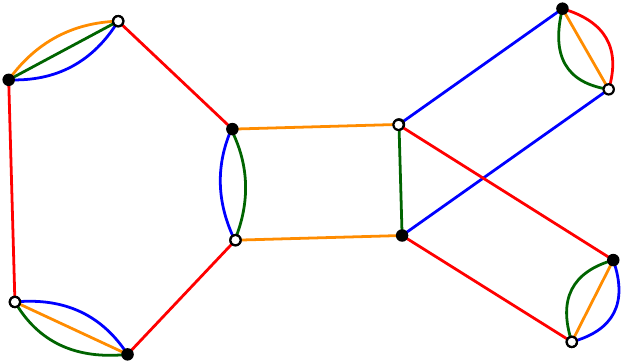}\end{tabular} }& $p$ & {\begin{tabular}{@{}c@{}}  0\\[+2ex] $ \cS^2$ \end{tabular}} &$2(p-1) $ & $\frac 3 2$ & 0 &  0  & $ \frac 1 2 $& $\cS^3$&\ref{subsec:Melonic}\\
\hline
{\begin{tabular}{@{}c@{}}  $K_{3,3}$ \\[+1ex]\includegraphics[scale=0.5]{K33.pdf}\end{tabular} }& 3 &{\begin{tabular}{@{}c@{}}  1\\[+2ex] $ \cT^2$ \end{tabular}}& 3& 1& $\frac 1 2$ &  1  & $ \frac 1 2 $&{\begin{tabular}{@{}c@{}} ? \\ \text{(sing)} \end{tabular}}  &\ref{subsec:K33}\\
\hline
{\begin{tabular}{@{}c@{}}  Bi-pyramid \\[+1ex]\includegraphics[scale=0.3]{BiPyrBub.pdf}\end{tabular} }& $p$ even & {\begin{tabular}{@{}c@{}}  0\\[+2ex] $ \cS^2$ \end{tabular}} &$2p-3$& $\frac{3p-1}{2p}$ & $\frac 1 {2p}$ &  1  & $ \frac 1 2 $& $\cS^3$&\ref{subsec:BiPyr}\\
\hline
{\begin{tabular}{@{}c@{}}  Toroidal \\[+1ex]\includegraphics[scale=0.3]{TorBub0.pdf}\end{tabular} }& $p$ & {\begin{tabular}{@{}c@{}}  1\\[+2ex] $\cT^2$ \end{tabular}} & $2p-3$& $\frac{3p-3}{2p}$ & $\frac 3 {2p}$ &  1  & $ \frac 1 2 $&{\begin{tabular}{@{}c@{}} ? \\ \text{(sing)} \end{tabular}}  &\ref{subsec:BiPyr}\\
\hline
\end{tabular}

\

where we have denoted the 2-Torus 
\be
\cT^2 \leftrightarrow \cS^1\times\cS^1.
\ee

\

\subsection{Dimension  four}

\noindent
\begin{tabular}{ |c|c|c|c|c|c|c|c|c|c|}
\hline
%
{\bf Bubble} $\B$ &$\ V_\B/2\ $ & {\begin{tabular}{@{}c@{}}  $\deltaG(\B)$\\[+2ex] $\B\cong ...$ \end{tabular}}&$\quad\tilde a_\B\quad$ & $\quad a_\B\quad$ & $\ \Delta_\B\ $ & $\  s_\B\  $ & $\ \gamma_\B\ $ & {\begin{tabular}{@{}c@{}}  {\bf max.}\\  {\bf top.} \end{tabular}}& {\bf sec.} \\
\hline
{\begin{tabular}{@{}c@{}}  melonic \\[+1ex]\includegraphics[scale=0.5]{melo2.pdf}\end{tabular} }& $p$ & {\begin{tabular}{@{}c@{}}  0\\[+2ex] $ \cS^3$ \end{tabular}} &$3(p-1) $ & 3 & 0 &  0  & $ \frac 1 2 $& $\cS^4$&\ref{subsec:Melonic}\\
\hline
{\begin{tabular}{@{}c@{}}  $2$-cyclic \\[+1ex]\includegraphics[scale=0.5]{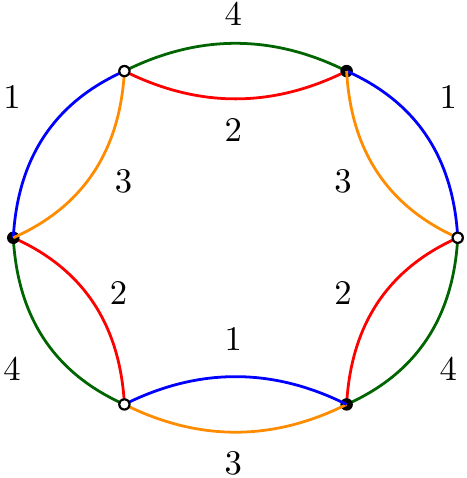}\end{tabular} }& $p$ & {\begin{tabular}{@{}c@{}}  $p-1$\\[+2ex] $ \cS^3$ \end{tabular}}
&$2(p-1) $ & $2+\frac 1 p$ & $1-\frac 1 p$  &  $p-1$  &  {\begin{tabular}{@{}c@{}}  $  -1/ 2$\\[+2ex]$  1/ 2$\\[+2ex]1/3 \end{tabular}}
& $\cS^4$& \ref{sec:SimplerBij}\\
\hline
{\begin{tabular}{@{}c@{}} \includegraphics[scale=0.5]{K334_1.pdf}\end{tabular} }& 3 & {\begin{tabular}{@{}c@{}}  3\\[+2ex] $ \cS^2\times\cS^1$ \end{tabular}} &5 & 7/3 & 2/3 &  1  & $ \frac 1 2 $&   {\begin{tabular}{@{}c@{}} ? \\ \text{(sing)} \end{tabular}}  & \ref{subsec:K334}\\
\hline
{\begin{tabular}{@{}c@{}}\includegraphics[scale=0.5]{Ord6NeckLike.pdf}\end{tabular} }& 3 & {\begin{tabular}{@{}c@{}}  1 \\[+2ex] $ \cS^3$ \end{tabular}} &5 & 8/3 & 1/3 &  1  & $ -\frac 1 2 $& $\cS^4$ & \ref{subsec:K334} \\
\hline
{\begin{tabular}{@{}c@{}}\includegraphics[scale=0.5]{K334_2.pdf}\end{tabular} }& 3 & {\begin{tabular}{@{}c@{}}  4 \\[+2ex] $\cT^2\times I$ \end{tabular}} \footnotemark[1]
&4 & 2 & 1 &  2  & $ \frac 1 2 $&  {\begin{tabular}{@{}c@{}} ? \\ \text{(sing)} \end{tabular}}  & \ref{subsec:K334} \\
\hline
{\begin{tabular}{@{}c@{}}\includegraphics[scale=0.5]{Meander1.pdf}\end{tabular} }& 3 & {\begin{tabular}{@{}c@{}}  2 \\[+2ex] $ \cS^3$ \end{tabular}}  &4 & 7/3 & 2/3 &  2  & $ \frac 1 2 $?& $\cS^4$ & \ref{subsec:K334} \\
\hline
{\begin{tabular}{@{}c@{}}  ``ribbon" \\[+1ex]\includegraphics[scale=0.25]{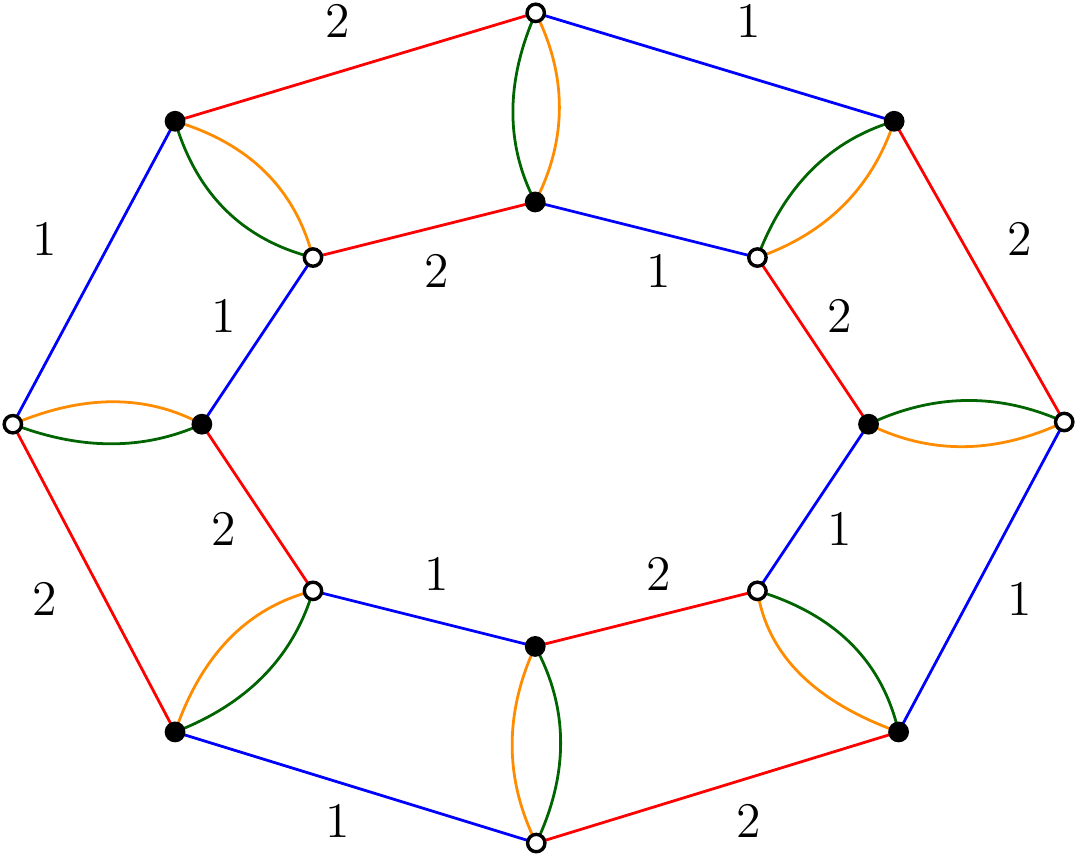}\end{tabular} }& $p$ even & 
{\begin{tabular}{@{}c@{}}  
1
\\[+2ex] $ \cS^{3}$\end{tabular}} &$3p-4 $ &$3-\frac 1 p$ & $\frac{1}{p}$ &  $1$  & $\frac 1 2$ & $\cS^4$& \ref{subsec:BiPyr} \\
\hline
{\begin{tabular}{@{}c@{}}  handle \\[+1ex]\includegraphics[scale=0.25]{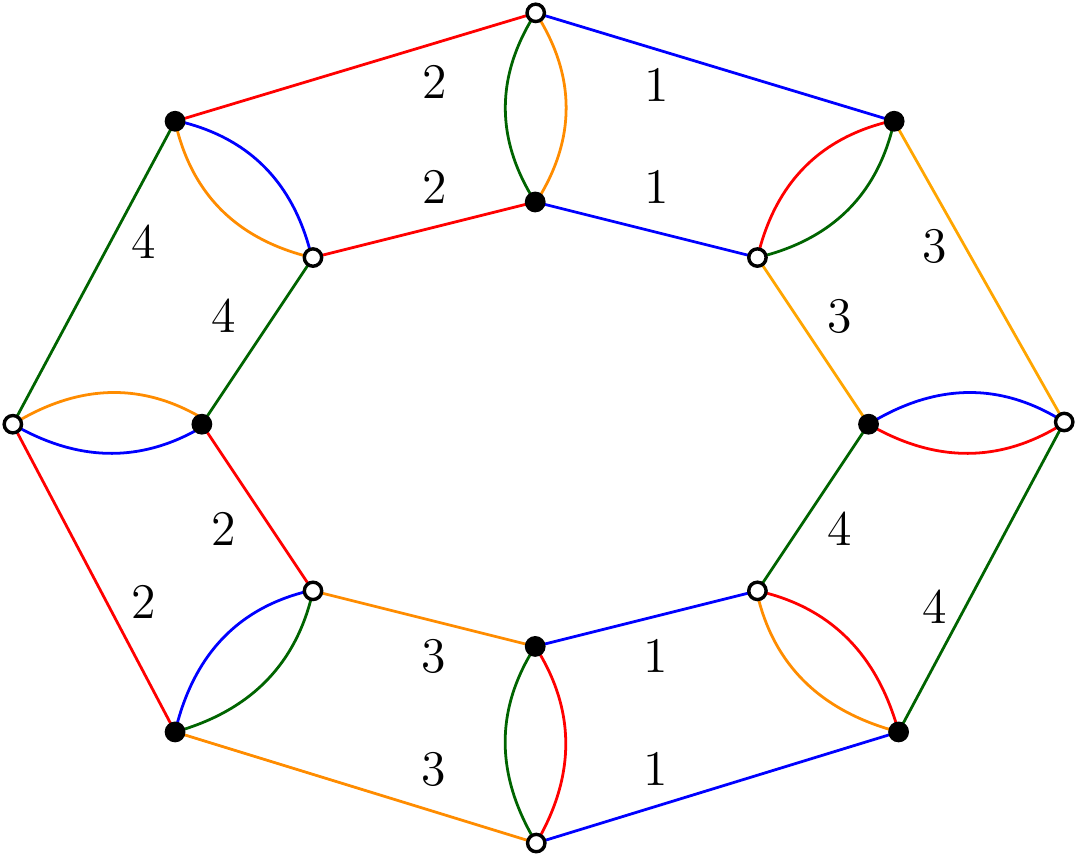}\end{tabular} }& $p$ & 
{\begin{tabular}{@{}c@{}}  
3 \\[+2ex]  $\cS^{2}\times \cS^1$\end{tabular}} &$3p-4 $ & $3-\frac{2}{p}$ & $\frac{2}{p}$ &  $1$  & $\frac 1 2$ &  {\begin{tabular}{@{}c@{}} ? \\ \text{(sing)} \end{tabular}} & \ref{subsec:BiPyr} \\
%
%
\hline
\end{tabular}

\footnotetext[1]{\label{note1}According to \cite{Cristo}, the bubble represents the trivial I-bundle over the torus.}

\subsection{Dimension  $D$}

\

\noindent
\begin{tabular}{ |c|c|c|c|c|c|c|c|c|}
\hline
%
{\bf Bubble} $\B$ &$\frac{V_\B} 2$ & {\begin{tabular}{@{}c@{}}  $\deltaG(\B)$\\[+2ex] $\B\cong ..$ \end{tabular}}&$\tilde a_\B$ & 
$\ \Delta_\B\ $ & $s_\B$ & $\ \gamma_\B\ $ & {\begin{tabular}{@{}c@{}}  {\bf max.}\\  {\bf top.} \end{tabular}}& {\bf sec.} \\
\hline
{\begin{tabular}{@{}c@{}}  melonic \\[+1ex]\includegraphics[scale=0.4]{melo2.pdf}\end{tabular} }& $p$ & {\begin{tabular}{@{}c@{}}  0\\[+2ex] $ \cS^{D-1}$ \end{tabular}} &$(D-1)(p-1) $& 0
 & 0 & $ \frac 1 2 $& $\cS^D$&\ref{subsec:Melonic}\\
\hline
{\begin{tabular}{@{}c@{}}  $k$-cyclic \\[+1ex]\includegraphics[scale=0.4]{kCycle0.pdf}\end{tabular} }& $p$ & 
{\begin{tabular}{@{}c@{}}  
\eqref{eqref:deltakCyc}
\\[+4ex] $ \cS^{D-1}$ \end{tabular}} &$(D-k)(p-1) $ 
& \eqref{eqref:DeltakCyc} &  $(p-1)(k-1)$  &  {\begin{tabular}{@{}c@{}}  $  -1/ 2$\\[+2ex]$  1/ 2$\\[+2ex]1/3 \end{tabular}} & $\cS^D$&\ref{sec:SimplerBij}\\
\hline
{\begin{tabular}{@{}c@{}}  ``ribbon" \\[+1ex]\includegraphics[scale=0.25]{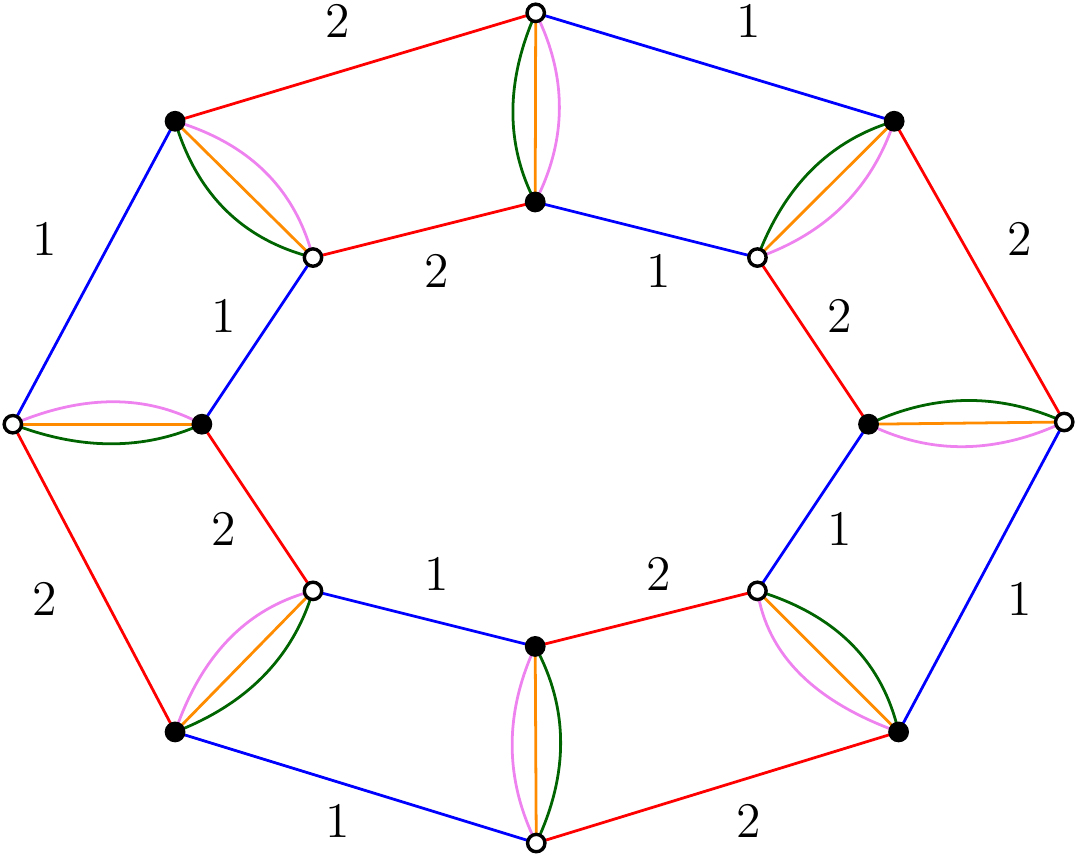}\end{tabular} }& {\begin{tabular}{@{}c@{}}  $p$\\[+2ex] even \end{tabular}}& 
{\begin{tabular}{@{}c@{}}  
$D-3$
\\[+2ex] $ \cS^{D-1}$\end{tabular}} &$p(D-1) - D $ &  $\frac{D-2}{2p}$ &  $1$  & $\frac 1 2$ & $\cS^D$& \ref{subsec:BiPyr} \\
\hline
{\begin{tabular}{@{}c@{}}  handle \\[+1ex]\includegraphics[scale=0.25]{RibbonBub.pdf}\end{tabular} }& $p$ & 
{\begin{tabular}{@{}c@{}}  
$D-1$ \\[+2ex]  $\cS^{D-2}\times \cS^1$\end{tabular}} &$p(D-1) - D $ &  $\frac{D}{2p}$ &  $1$  & $\frac 1 2$ &  {\begin{tabular}{@{}c@{}} ? \\ \text{(sing)} \end{tabular}} & \ref{subsec:BiPyr} \\
%
%
\hline
\end{tabular}

\vspace{1cm}

where we have denoted 
\bea
\label{eqref:deltakCyc}
\delta_{p,k,D}&=&(p-1)[k(D-k)-D+1],\\
\label{eqref:DeltakCyc}
 \Delta_{p,k,D}& =& \frac{(p-1)(D-k)(k-1)}{2p}.
\eea

\

We stress that from these examples, we can study the infinite families obtained by doing the connected sums of any number of copies of these examples, and add any number of $h$-pairs on the resulting bubbles. Using the results in Subsections~\ref{subsec:hPair} and \ref{subsec:ConecSum}, we can deduce the values of $\tilde a$ and $s$ for these infinite families  of bubbles and characterize maximal maps, obtain the critical exponent, and determine their topology.

\newpage

\section{Towards more interesting behaviors?}
\label{sec:K336}

Throughout this thesis, we have mentioned a number of times that all known examples of bipartite bubbles in dimension $D<6$ were such that choosing an optimal pairing to build the bijection, trees were part of the set of maximal maps. In $D=6$, however, we can exhibit a bubble $\B$ for which this is not the case\footnote{We found this bubble together with Thierry Monteil.}. It is illustrated in \ref{fig:K336}. 
\begin{figure}[!h]
\centering
\includegraphics[scale=.7]{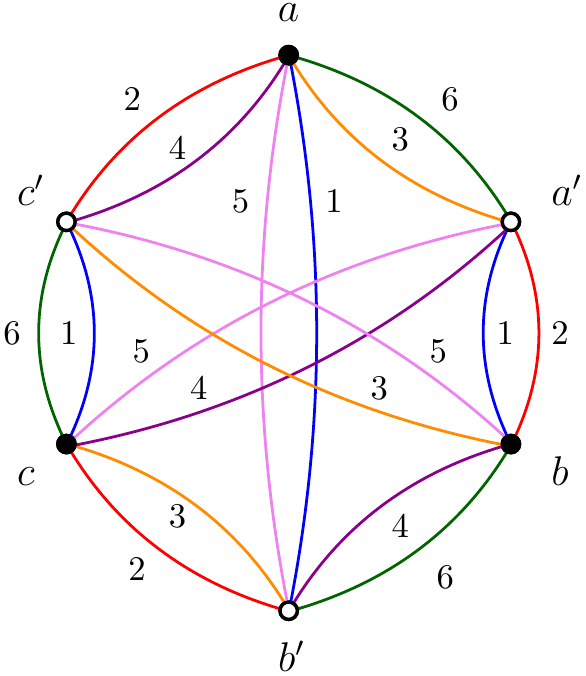} 
\hspace{2cm}
\includegraphics[scale=.55]{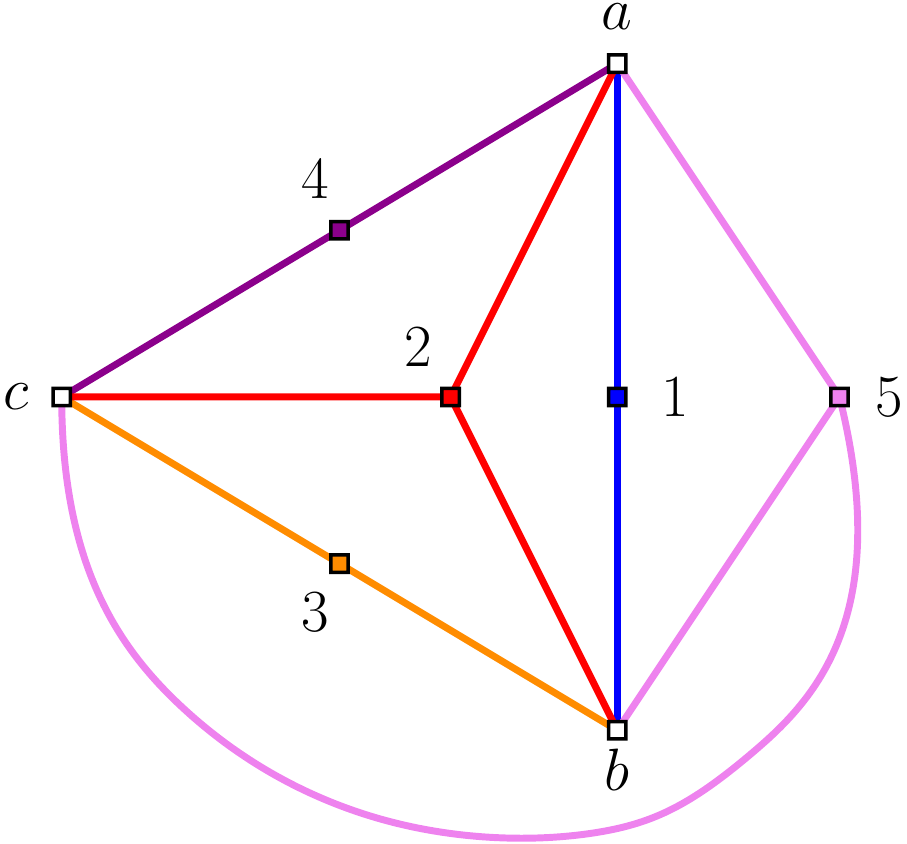} 
\caption{A size 6 bubble in $D=6$ and stacked map without colored leaves.}
\label{fig:K336} 
\end{figure}

All of its pairings are optimal. We choose the pairing $\Opt=\{(a,a'), (b,b'), c,c')\}$. The stacked map $\Ps(\B,\Opt)$ is pictured on the right of Fig.~\ref{fig:K336} (without the color-6 leaves). 
This bubble is not invariant under exchanging the colors of the black and the white vertices. We denote $\bar\B$ the corresponding bubble, and also choose to pair the vertices linked by color 6. The resulting stacked map is similar to that of Fig.~\ref{fig:K336}, with the roles of colors 2 and 5 permuted. 
The 0-score of the corresponding coverings is $\PhiM=11$, and the 0-score of trees in $\bG(\B,\bar\B)$ is therefore 
\be
\Phi_0(\cT)=6+5b(\cT).
\ee
In particular, the 0-score of a tree with two and four bubbles would be 
\be
\Phi_0^{(2)}=16, \qquad\text{and}\qquad \Phi_0^{(4)}=26.
\ee
\begin{figure}[!h]
\centering
\includegraphics[scale=.65]{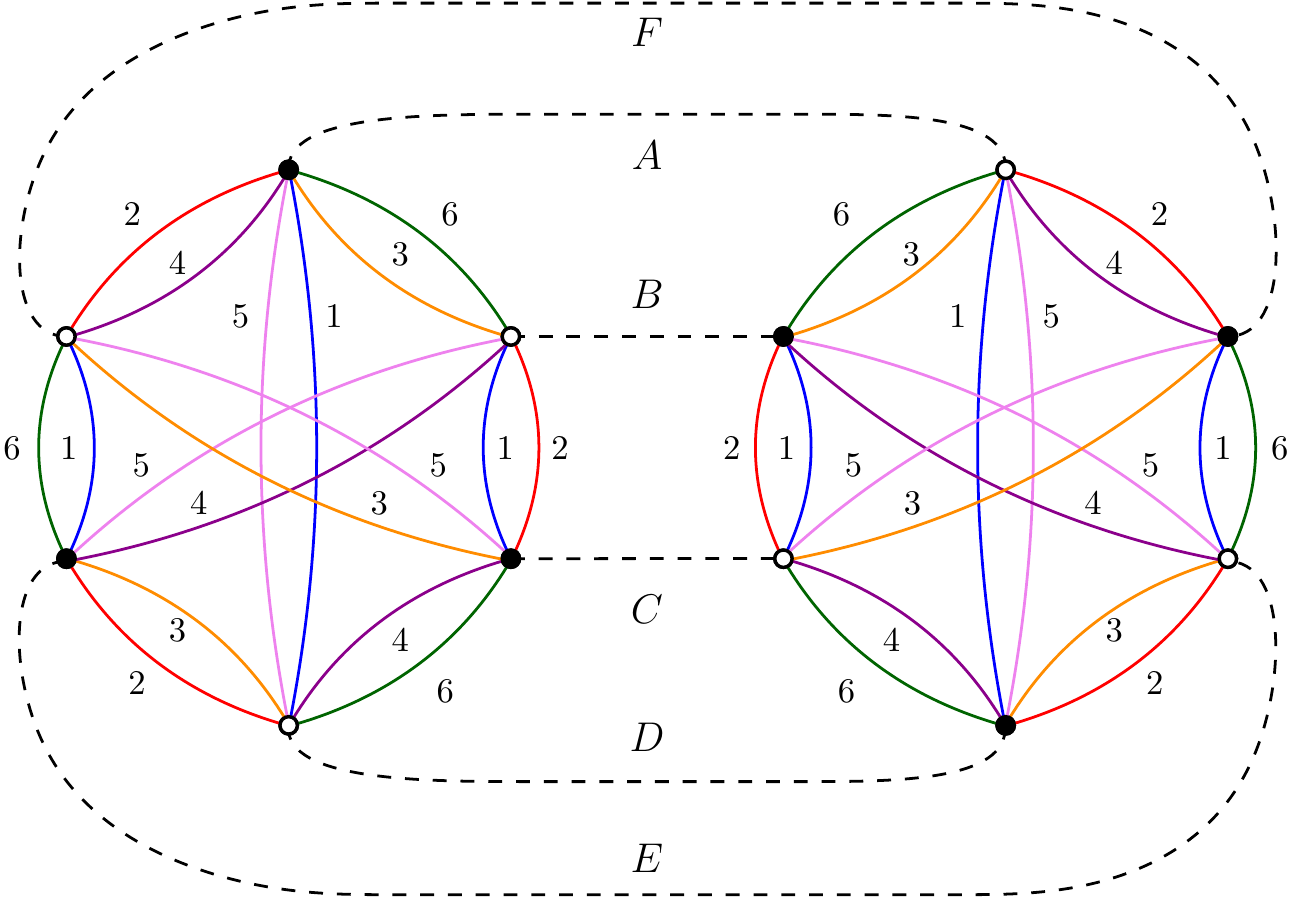} 
\caption{A graph which 0-score is higher than that of a tree.}
\label{fig:K3362Bub} 
\end{figure}

We now exhibit a graph of  $\bG(\B,\bar\B)$ with two bubbles, which 0-score is stronger than $\Phi_0^{(2)}$: the 0-score of the graph $\G_2$ pictured in Fig.~\ref{fig:K3362Bub} has 3 color $0i$ cycle for each $i$, and therefore its 0-score satisfies
\be
\Phi_0(\G_2)=18 >\Phi_0^{(2)}.
\ee
This is already a strong property, as it implies that not all sets of bubbles are such that trees belong to maximal maps. In particular, we can build a $\{\G_2\}$-tree-like family $\bF_2$ (Fig.~\ref{fig:TreeFamily}), and from Prop.~\ref{prop:TreeLikeScore}, the 0-score of a tree-like graph $\G$ with $n_2$ bubbles $\G_2$ is 
\be
\Phi_0(\G\in\bF_2)=6+12 \times n_2(\G).
\ee
In order to compare this with other graphs in $\bG(\B,\bar\B)$, we convert back in terms of $b=n_\B+n_{\bar\B} = 2 n_2$:
\be
\label{eqref:ScoreG2}
\Phi_0(\G\in\bF_2)=6+6 \times b(\G).
\ee
It implies that trees with an odd number of bubbles do not belong to maximal maps.  A consequence of this fact is that the results of  Section~\ref{sec:TreeBound} on the coefficients $\tilde a$, $a$ and $s$ do not hold anymore, as they relied on the fact that trees were part of maximal maps. However, as detailed in Subsection~\ref{subsec:TreeLike}, as long as the sequence $\{\Phi_0^b\}_{b\in\bN}$ of the 0-scores of maximal maps satisfies
\be
\exists n\in\bN,\ \forall\ b\in\bN,\quad \frac{\Phi_0^b-D}b\le \frac{\Phi_0^n-D}n,
\ee
then the unique coefficient $\tilde a$ leading to a non-negative rational and non-trivial bubble-dependent degree is
\be
 \tilde a_\B =  \frac{\Phi_0^n-D}n,
\ee
and the coefficients $a_\B$ and $s_\B$ are uniquely defined accordingly.

\

We do not continue here the precise study of the graphs in $\bG(\B,\bar \B)$. The behavior for the first terms is very similar to that of the $K_4$ non-orientable bubble, and we expect the same behavior to occur, which is non-linear, but is linear for an even number of bubbles. In that case, $\bK=2\bN$, and $\tilde a_B = 6$ from \eqref{eqref:ScoreG2}.
We argue however that it is not \emph{a priori} impossible for a set of bubbles $\bB$ to generate maximal maps which 0-score has a non-linear increasing dependence in the number of bubbles, although it is not likely to occur. If $\bF(\G_k)$ is a ${\G_k}$-tree-like family, where $\G_k$ comprises $k$ bubbles $\B$ and $\bar\B$, the 0-score of elements of $\bF(\G_k)$ is linear in $b(\G)$, and we denote 
\be
\label{eqref:a'k}
a'_{\G_k}=\frac{\Phi_0(\G_k)-D} k 
\ee
the corresponding slope,
\be
\Phi_0(\G\in\bF_k)=6+ a'_{\G_k} \times b(\G),
\ee
where $a'_{\G_k}$  is not necessarily an integer. From \eqref{eq:IneqGurau}, we obtained the bound \eqref{eqref:BoundAvsBoundTildeA}, which applies for $a'_{\G_k}$:
\be
a'_{\G_k}<\deltaG(\B)+(D-1)\bigl(\frac{V(\B)}2 - 1\bigr) = a'_\text{max}.
\ee
We compute $\Phi(\B)=23$, and therefore $\deltaG(\B)=12$, and $a'_\text{max}=22$
\be
a'_{\G_k}<22.
\ee
%
%
Non-linear increasing behaviors could happen for $\bB$-restricted graphs satisfying
\be
\label{eqref:PropNonLin}
\forall k\in\bN,\ \forall A \in  ]0,a'_\text{max}[,\      \exists \G \in\bG(\bB) \text{ such that }  \biggl\{ {\begin{tabular}{@{}c@{}} $b(\G)\ge k $ \\ $a'_{\G}>A$ \end{tabular}}.
\ee
For instance, in the case of  $\bB=\{\B,\bar\B\}$,  starting from $\G_1=\B^{\Opt}$, we can exhibit a graph $\G_2$ such that $b(\G_2)=2$ and $a'_{\G_1}<a'_{\G_2}$.
If ever the property \eqref{eqref:PropNonLin} was satisfied for $\bB=\{\B,\bar\B\}$, then we could find a graph $\G_k$ with $k\ge 3$ bubbles, such that $a'_{\G_2}<a'_{\G_k}<22$, and recursively build an infinite sequence $\{\G_1,\G_2,\G_k\cdots\}$ 
such that 
\be
a'_{\G_1}<a'_{\G_2}<a'_{\G_k}<\cdots<22.
\ee
We stress that this is not a priori forbidden, as $a'_{\G_k}\in\bQ$ \eqref{eqref:a'k}, although it is very likely not to occur. As a consequence, \emph{maximal maps would not have a 0-score linear in the number of bubbles, and it would not be possible to find $\tilde a$ satisfying \eqref{eqref:Cond1} and \eqref{eqref:Cond2}}. Indeed, choosing $\tilde a$ smaller than 22, there would exist a graph $\G$ with $a'_\G> \tilde a$, from which we could build the infinite family of arbitrarily large $\G_n$-tree-like graphs $\bF_n$ with degree
\be
\delta(\G \in\bF_n)=(\tilde a - a'_{\G_n}) \times b(\G) \xrightarrow[b\rightarrow + \infty]{} - \infty, 
\ee
so that the $1/N$ expansion \eqref{eqref:NExp} would not be defined, and choosing $\tilde a$ larger or equal to 22, there would only be a finite number of contributions per order.
If maximal graphs in $\bG(\B,\bar\B)$ satisfy a relation of the type 
\be
\Phi_0(\G_\text{max})=D+f(b(\G_\text{max})),
\ee
where $f$ is a strictly increasing non-linear function taking rational values on integers  and such that 
\be
\lim_{x\rightarrow+\infty} f(x)\le22,
\ee
then a proper definition for the degree would be 
\be
\delta(\G)=D+f(b(\G)) - \Phi_0(\G).
\ee
It would lead to a $1/N$ expansion with rational orders. The leading order contributions would be those of vanishing degree, i.e. maximal configurations. It would not, however, be possible to define a tensor model generating the corresponding $1/N$ expansion.

\end{document}